\documentclass[]{thinking_with_video}

\usepackage[utf8]{inputenc}
\usepackage{latexsym}
\usepackage{inconsolata}
\usepackage{tabularx}
\usepackage{enumitem}
\usepackage{pifont}
\usepackage{listings}
\usepackage{url}
\usepackage{titletoc} 
\usepackage{tikz}
\usepackage{needspace} 

\definecolor{c2cblue}{HTML}{2E5AA8}
\definecolor{c2ccyan}{HTML}{24A7B5}
\definecolor{c2cpurple}{HTML}{7657C8}

\definecolor{Chart2CodeGreen}{rgb}{0.33,0.42,0.18}
\definecolor{Chart2CodeRed}{RGB}{190,45,45}
\providecommand{\cmark}{\textcolor{Chart2CodeGreen}{\ding{51}}}
\providecommand{\xmark}{\textcolor{Chart2CodeRed}{\ding{55}}}

\definecolor{Chart2CodeTextRed}{HTML}{D62728}
\providecommand{\red}[1]{\textcolor{Chart2CodeTextRed}{#1}}
\providecommand{\green}[1]{\textcolor{Chart2CodeGreen}{#1}}

\providecommand{\bench}{Chart2Code}

\definecolor{chart}{HTML}{1F77B4}
\newtcolorbox{example}[1][]{
  breakable,
  colback=chart!5!white,
  colframe=chart,
  title=\centering #1
}

\NewTColorBox{wronganswer}{ O{} }{%
  colback=red!5!white,
  colframe=red!75!black,
  coltitle=black,
  title={Wrong Answer},
  breakable,
  #1
}
\NewTColorBox{correctanswer}{ O{} }{%
  colback=Chart2CodeGreen!5!white,
  colframe=Chart2CodeGreen!75!black,
  coltitle=black,
  title={Correct Answer},
  breakable,
  #1
}

\newenvironment{wideexample}[1]
{
    \clearpage
    \begin{example}[#1]
    \footnotesize
}
{
    \end{example}
    \clearpage
}

\newcommand{\appendixplatetitle}[1]{%
  {\Large\bfseries\textcolor{seedblue}{#1}\par}%
  \vspace{0.6mm}%
  {\color{seedblue!45}\rule{0.34\linewidth}{0.7pt}\par}%
}

\newcommand{\errorcaseseparator}{%
  \par\vfill
  {\color{seedblue!28}\hrule height 0.55pt}%
  \vfill
}

\newenvironment{evaluationcase}
{%
    \clearpage
    \noindent\begin{minipage}{\linewidth}
    \begingroup
    \captionsetup{type=figure,font=footnotesize,skip=3pt}
    \centering
    \setlength{\parskip}{0pt}
}
{%
    \par\endgroup
    \end{minipage}
    \clearpage
}

\newenvironment{compacterrorpage}
{%
    \clearpage
    \noindent\begin{minipage}[t][0.89\textheight][s]{\linewidth}
    \begingroup
    \captionsetup{type=figure,font=footnotesize,skip=3pt}
    \centering
    \setlength{\parskip}{0pt}
}
{%
    \par\endgroup
    \end{minipage}
    \clearpage
}

\hypersetup{
  colorlinks=true,
  linkcolor=seedblue,
  citecolor=seedblue,
  urlcolor=seedblue
}

\makeatletter

\renewcommand{\@toptitlebar}{%
  \noindent
  \begin{tikzpicture}
    \shade[
      left color=c2cblue,
      right color=c2cpurple
    ]
    (0,0) rectangle (\linewidth,2.5pt);
  \end{tikzpicture}
  \par
  \vskip 2mm
}

\renewcommand{\@bottomtitlebar}{%
  \vskip 5mm
  \noindent
  \begin{tikzpicture}
    \shade[
      left color=c2cblue,
      right color=c2ccyan
    ]
    (0,0) rectangle (\linewidth,1pt);
  \end{tikzpicture}
  \par
}

\makeatother

\definecolor{gA}{HTML}{078CF0}
\definecolor{gB}{HTML}{079EEE}
\definecolor{gC}{HTML}{08AFE8}
\definecolor{gD}{HTML}{0DBFD5}
\definecolor{gE}{HTML}{14C6B5}
\definecolor{gF}{HTML}{1CC395}
\definecolor{gG}{HTML}{24B976}
\definecolor{gColon}{HTML}{333333}

\newcommand{\charttocodetitle}{%
  \textcolor{gA}{Fr}%
  \textcolor{gB}{om}%
  \hspace{0.27em}%
  \textcolor{gB}{Ch}%
  \textcolor{gC}{ar}%
  \textcolor{gD}{ts}%
  \hspace{0.27em}%
  \textcolor{gD}{t}%
  \textcolor{gE}{o}%
  \hspace{0.27em}%
  \textcolor{gE}{Co}%
  \textcolor{gF}{d}%
  \textcolor{gG}{e}%
  \textcolor{gColon}{:}%
}

\title{
  \charttocodetitle\\[1.5mm]
  A Hierarchical Benchmark for Multimodal Models
}
\renewcommand{\authorfont}{\fontsize{10}{10.5}\selectfont}
\renewcommand{\affiliationfont}{\fontsize{9.2}{11.2}\selectfont}

\author{
  \mbox{Jiahao Tang\textsuperscript{1,$\dagger$}},
  \mbox{Henry Hengyuan Zhao\textsuperscript{2,*}},
  \mbox{Lijian Wu\textsuperscript{1,$\dagger$}},
  \mbox{Zijian Zhang\textsuperscript{1}},
  \mbox{Yifei Tao\textsuperscript{3}},
  \mbox{Dongxing Mao\textsuperscript{1}},
  \mbox{Yang Wan\textsuperscript{1}},
  \mbox{Jingru Tan\textsuperscript{1}},
  \mbox{Min Zeng\textsuperscript{1}},
  \mbox{Min Li\textsuperscript{1}},
  \mbox{Alex Jinpeng Wang\textsuperscript{1,$\S$}}
}

\renewcommand{\affiliationlist}{%
{\affiliationfont
$^{1}$CSU-JPG, Central South University
\qquad
$^{2}$National University of Singapore\\[0.6mm]
$^{3}$Nanyang Technological University
}%
}

\abstract{
We introduce \bench{}, a new benchmark for evaluating the natural language to chart code generation capabilities of large multimodal models (LMMs). 
\bench{} is explicitly designed from a user-driven perspective, capturing diverse real-world scenarios and progressively increasing task difficulty.
It consists of three levels: \textbf{Level 1 (Chart Reproduction)} reproduces charts from a reference figure and user query; \textbf{Level 2 (Chart Editing)} involves complex modifications such as changing chart types or adding elements; and \textbf{Level 3 (Long-Table to Chart Generation)} requires models to transform long, unprocessed tables into faithful charts following user instructions.
To our knowledge, this is the first hierarchical benchmark that reflects practical chart2code usage while systematically scaling task complexity.
In total, \bench{} contains 2,186 tasks across 22 chart types, paired with multi-level evaluation metrics that assess both code correctness and the visual fidelity of rendered charts. 
We benchmark 29 state-of-the-art (SoTA) LMMs, including both proprietary and the latest open-source models such as GPT-5.2, Qwen3-VL, InternVL3/3.5, MiMo-VL, and Seed-1.6-VL. Experimental results demonstrate that even the SoTA model GPT-5.2 averages 72.21 on code-based evaluation and only 33.41 on chart-quality assessment across the editing tasks, underscoring the difficulty of \bench{}.
We anticipate this benchmark will drive advances in multimodal reasoning and foster the development of more robust and general-purpose LMMs. The code and data are available on \href{https://csu-jpg.github.io/Chart2Code.github.io/}{Chart2Code}
}

\checkdata[Project Page]{\url{https://csu-jpg.github.io/Chart2Code.github.io}}
\checkdata[Code]{\url{https://github.com/CSU-JPG/Chart2Code}}
\checkdata[Data]{\url{https://huggingface.co/datasets/CSU-JPG/Chart2Code}}
\begin{document}

\maketitle

\renewcommand{\thefootnote}{}
\footnotetext{\textsuperscript{$\dagger$} Equal contribution.}
\footnotetext{\textsuperscript{*} Project lead.}
\footnotetext{\textsuperscript{$\S$} Corresponding author.}
\renewcommand{\thefootnote}{\arabic{footnote}}

\vspace{-1.0em}

\section{Introduction}

Charts are one of the most powerful tools in scientific publications and business reports, they distill large amounts of structured data into clear and persuasive visuals. 
With the rapid progress of large multimodal models (LMMs) \citep{gpt5, Claude4}, it becomes increasingly realistic to envision AI systems that not only interpret visual charts \citep{wang2024charxiv} but also generate executable plotting code, a task we refer to as chart-to-code (chart2code). Such capabilities can significantly enhance productivity by automating visualization creation, enabling reproducibility.

However, exisiting chart to code benchmarks exsist the performance saturation issues such as ChartMimic \citep{chartmimic}, Plot2Code \citep{plot2code}. Previous models GPT-4o already reached 82.2\% scores in ChartMimic. With the rapid advancement of frontier models such as GPT-5.2 \cite{gpt5.2}, ChartMimic can no longer serve as a sufficiently challenging or discriminative benchmark for evaluating current or future models.

Additionally, there is a lack of a systematic chart-to-code benchmark that covers more comprehensive and realistic human–AI interactive chart-coding scenarios. Users rarely stop at simple chart reproduction that ChartMimic does, they need to edit figures to improve their work productivity avoiding boring repetitive works. For non-expert users who are not data analyst, they often work with long and raw tables that must be distilled into interpretable plots. However, when applied to these more common and demanding scenarios, the exisiting state-of-the-art (SOTA) models often struggle, revealing substantial gaps in their practical ability (refer to Appendix~\ref{sec:appendix_case_study} for examples). 
This discrepancy \textit{creates a mismatch between reported benchmark performance and real-world utility, highlighting the need for a challenging and relistic benchmark that more comprehensively reflects everyday chart2code challenges}.

\begin{figure*}[t]
    \centering
    \includegraphics[width=0.93\linewidth]{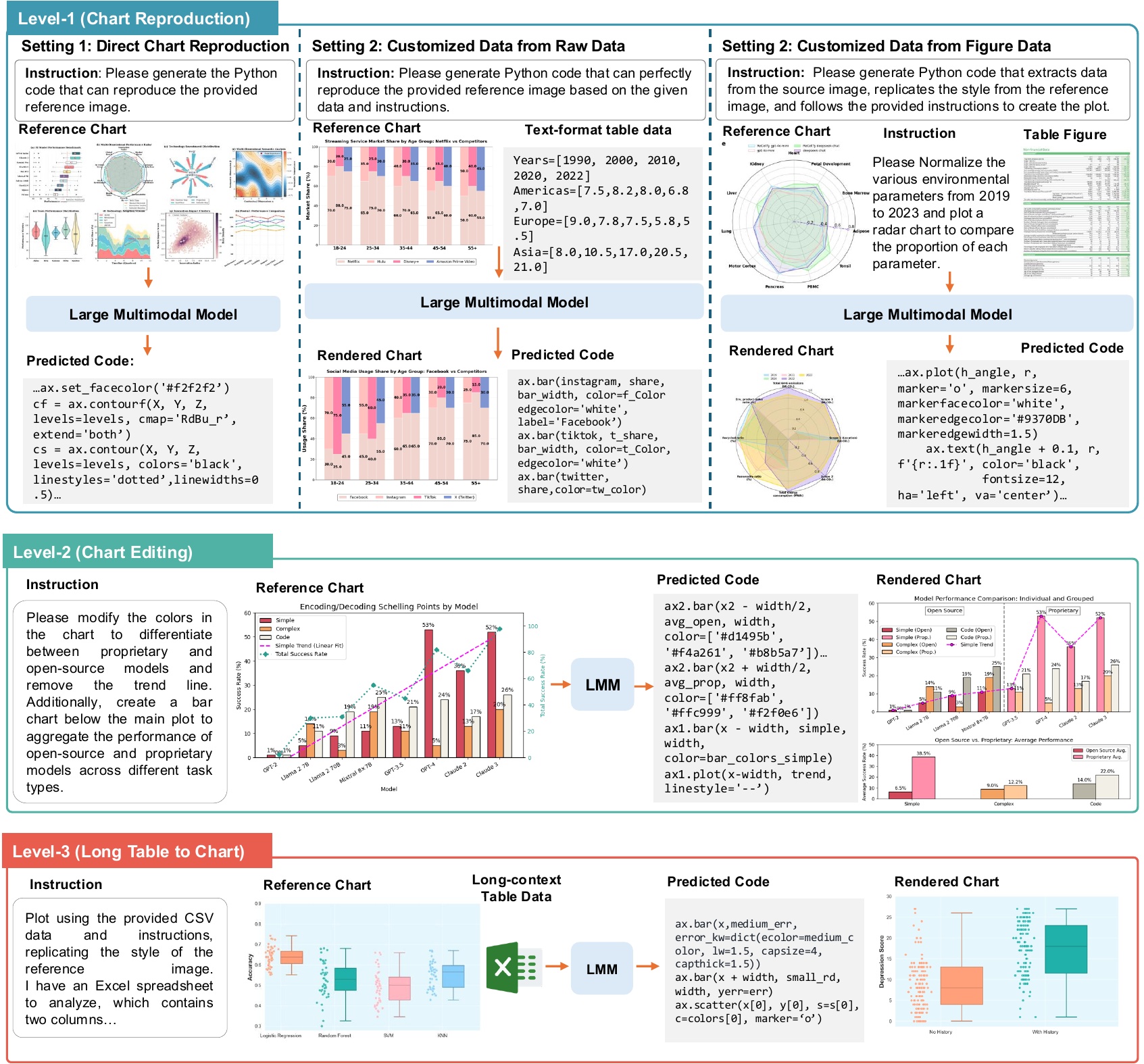}
    \caption{\textbf{\bench{} covers three progressively challenging levels}: reproduction, editing, and long-table to chart generation. 
    It provides a user-driven and diverse benchmark that better reflects real-world chart2code demands.
}
    \label{fig:taskoverview}
\end{figure*}

Motivated by these observations, we introduce \bench{} (Figure~\ref{fig:taskoverview}), a new benchmark designed to rigorously evaluate chart generation capabilities of LMMs under progressively challenging conditions. 
\bench{} consists of three difficulty levels: \textbf{Level 1 (Chart Reproduction)} targets mimicking a reference figure and instruction; \textbf{Level 2 (Chart Editing)} requires complex and precise editing, such as changing chart types or adding new elements; \textbf{Level 3 (Long-Table to Chart Generation)} presents the most demanding setting, where models must convert long, unprocessed data tables into faithful charts from user instructions. 
This hierarchical design reflects real-world usage while progressively increasing difficulty, and its distinctions from prior benchmarks are highlighted in Table~\ref{tab:compare_related_works}.

We comprehensively benchmark 29 state-of-the-art LMMs, including both proprietary and open-weight models, across the three levels of \bench{}.
Our results show that while LMMs demonstrate promising capabilities on simple reproduction tasks, their \textbf{performance deteriorates sharply on complex editing and long-context data-to-chart generation}. 
Together, these findings reveal the \textit{unsolved challenges of chart2code generation and point to future directions for building more reliable visualization assistants}. 
\begin{itemize}
    \item We present \bench{}, the first hierarchical benchmark targeting chart2code generation with progressively more challenging tasks. 
    \item We propose multi-level evaluation protocols that jointly assess code executability and visual fidelity, offering a comprehensive lens on model performance.
    \item We provide an extensive empirical study across 29 mainstream LMMs, yielding new insights into their strengths, weaknesses, and design trade-offs for chart generation.
\end{itemize}

\begin{table*}[t]
\centering
\caption{\textbf{Comparison of existing chart-to-code benchmarks.} 
Ref. Fig.: Reference Figure; Instr.: Instruction; Text Data: Text-format data; Fig. Data: Figure-format data; NL: Natural language.}
\resizebox{1.0\textwidth}{!}{
\begin{tabular}{l|cccc|ccc|c|cc|c}
\toprule
\multirow{2}{*}{\textbf{Benchmark}} 
& \multicolumn{4}{c|}{\textbf{Input Type}} 
& \multicolumn{3}{c|}{\textbf{Task Category}} 
& \multirow{2}{*}{\textbf{Output}} 
& \multicolumn{2}{c|}{\textbf{Metric}}
& \multirow{2}{*}{\begin{tabular}{c}\textbf{Datasets}\\\textbf{Size}\end{tabular}} \\
\cmidrule(lr){2-5} \cmidrule(lr){6-8} \cmidrule(lr){10-11}
& Ref. Fig. & Instr. & Text Data & Fig. Data 
& Chart Reproduction & Chart Editing & Long-table to Chart 
&  
& Rule-based & GPT-score
&  \\
\midrule
CharXiv \citep{wang2024charxiv}      & \xmark & \cmark & \xmark & \xmark & \xmark & \xmark & \xmark & NL   & \cmark & \xmark & 2323 \\
Plot2Code \citep{plot2code}          & \cmark & \cmark & \xmark & \xmark & \cmark & \xmark & \xmark & Code & \xmark & \cmark & 132 \\
AcademiaChart \citep{academiachart}  & \cmark & \cmark & \xmark & \xmark & \cmark & \xmark & \xmark & Code & \cmark & \cmark & 2525 \\
Chartmimic \citep{chartmimic}        & \cmark & \cmark & \xmark & \xmark & \cmark & \xmark & \xmark & Code & \cmark & \cmark & 4800 \\
ChartEdit \citep{chartedit}          & \cmark & \cmark & \xmark & \xmark & \xmark & \cmark & \xmark & Code & \xmark & \cmark &  233\\
\midrule
\bench{} (Ours)                      & \cmark & \cmark & \cmark & \cmark & \cmark & \cmark & \cmark & Code & \cmark & \cmark & 2186 \\
\bottomrule
\end{tabular}
}
\label{tab:compare_related_works}
\end{table*}

\section{Related Work}

\paragraph{Large Multimodal Models.}
Thanks to the success of proprietary LMMs such as GPT-5 \citep{gpt5.2}, Gemini-3-Pro \citep{gemini3}, and Claude-Sonnet-4 \citep{Claude4}, we see the dawn of building AI agents for addressing realistic applications. In the academic community, we see enormous excellent open-source models: MiMo-VL \citep{mimo}, QwenVL-series \citep{qwen2.5vl, qwen2vl}, and InternVL-series \citep{internvl3_5, internvl3}, MolMo \citep{molmo}, Kimi-VL \citep{kimiteam2025kimivltechnicalreport} LLaVA-series \citep{llavaonevision, llava1.5, llavanext}, Deepseek-VL \citep{deepseekvl}, and GLM-4V \citep{GLM-4V}.

\paragraph{Agentic Benchmarks.}
The rapid progress of foundation LLMs and LMMs has motivated the creation of diverse agentic benchmarks, spanning GUI automation \citep{xie2024osworld, zhao2025worldguiinteractivebenchmarkdesktop, videogui, visualwebarena}, agentic coding \citep{swebench, yang2024swebenchmultimodal}, tool use \citep{yao2025taubench}, AI research assistance \citep{nathani2025mlgym}, and chart reasoning \citep{wang2024charxiv}. 
We focus on chart2code, a practical task central to everyday workflows for researchers and professionals. Despite progress, even the best proprietary LMMs still fail to generate faithful charts from long, raw tables, underscoring the need for future modeling advances.

\paragraph{Chart Understanding to Code Generation.}

Chart understanding has evolved through a series of benchmarks that progressively expand task complexity. ChartQA \citep{chartqa} first established large-scale visual question answering over charts, combining logical and visual reasoning. ChartXiv \citep{wang2024charxiv} extended this line by introducing scientific charts with expert-designed questions, further exposing the gap between multimodal models and human performance. 

Moving beyond QA, Chart2Code benchmarks address faithful chart generation. ChartMimic \citep{chartmimic} formalized this by requiring code synthesis from chart images and instructions, while ChartEdit\citep{chartedit} and ChartM$^3$\citep{ChartM3} emphasize interactive modification, where models edit chart-rendering code following natural-language instructions. 

Extending chart generation, StarVector \citep{Starvector} proposes a vision-language approach to produce scalable vector graphics from visual or textual inputs. FigEdit\citep{figedit} highlights the gap between chart editing and general figure editing. ChartBench \citep{chartbench} focuses on chart question answering (VQA), related to CharXiv \citep{wang2024charxiv}, while BigDocs \citep{bigdocs} targets multimodal document understanding with tasks such as Screen2HTML, Image2SVG, and Image2Flow. ChartGalaxy~\citep{chartgalaxy} introduces a large-scale dataset for infographic chart generation, focusing on complex visual layouts and design variations. However, none of these works directly address the chart-to-code formulation studied in this paper.

Although GPT-4o achieves high scores on ChartMimic (83.2) and ChartEdit (93.6), it still struggles with realistic chart2code tasks, motivating a more challenging benchmark for reliable evaluation.

\section{\bench{}: From Visual Charts to Code}
\label{method}


\subsection{Task Definition of \bench{}}
\label{task definition}
The \bench{} can be formulated as:
$C=f(R,I,D)$
where, $R$ is the visual reference chart (e.g., chart screenshot), $I$ is the user's natural language instruction and $C$ is the executable Python code generated by LMM ($f$). $D$ represents the optional underlying data source provided to ground the generation. To accommodate real-world scenarios, we design $D$ to support three modalities: raw text, visual data tables (e.g., screenshots of spreadsheets), and structured files (e.g., Excel).
To ensure rigor and comprehensiveness, we designed three tasks of increasing difficulty.
\subsection{Hierarchical task design}
\paragraph{Level 1 (Chart Reproduction):} This foundational level assesses the model's fidelity in visual mimicry and is divided into two scenarios.

  \begin{itemize}
  \item \textbf{Visual-Only Reproduction ($R + I, D=\emptyset$):} 
  The LMM must generate code to reproduce the visual appearance and underlying data of the reference chart $R$ purely from the image. We denote this setting as \textbf{DR} (\textbf{D}irect \textbf{R}eproduction). This setting primarily evaluates the model's visual perception and understanding capabilities.

  \item \textbf{Style Transfer with Data ($R + D + I$):} 
The LMM is provided with an external data source $D$ and must generate code to visualize it following the style and layout of $R$. We define two variants: \textbf{CRD}(\textbf{C}ustomize \textbf{R}aw \textbf{D}ata), where $D$ is raw tabular data, and \textbf{CFD}(\textbf{C}ustomize \textbf{F}igure \textbf{D}ata), where $D$ is structured figure-level data. This setting reflects real-world workflows in which users aim to replicate chart styles given available data \citep{chartmimic, plot2code,academiachart}.

\end{itemize}

\paragraph{Level 2 (Chart Editing):} At this level, LMM edits the reference chart ($R$) as instructed, performing basic operations such as style changes, type swaps, and color adjustments. These tasks, similar to the editing functions in \citep{chartedit}, primarily test the LMM’s fundamental chart understanding ability. Going a step further, Level 2 involves more sophisticated editing operations. These include calculating correlation coefficients based on chart data, plotting trend lines, generating data analysis tables, splitting charts by data category, incorporating new data analyses and so on. Beyond mere understanding, these tasks further stress-test the LMM’s inferential reasoning. This setting reflects a sophisticated real-world scenario where users demand extended analytical insights and complex visual synthesis, rather than the simplistic localized edits addressed in previous benchmarks like \citep{chartedit}.

\paragraph{Level 3 (Long-Table to Chart):} The final level of our benchmark is also the hardest task setting. It reflects the most common usage in real-world chart creating scenarior. To summarize, the challenges of Level 3 requires the model to (1) absorb long-context inputs, (2) perform reasoning and information retrieval, (3) conduct mathematical calculation, (4) generate executable code by following the diverse user queries. So the Level 3 is distinct from Level 1 in task definition. This task also reflects the scenario that a non-expert user uses the LLM/LMM for data visualization with raw tabular data. This is also a domain that has not been addressed in previous works\citep{chartmimic, chartedit,chartqa}.

\begin{figure*}[t]
    \begin{minipage}[t]{0.53\linewidth}
    \caption{\textbf{Collected charts distribution.}}
    \label{fig:benchdistpiechart}
    \centering
    \vspace{0pt}
    \includegraphics[width=\linewidth]{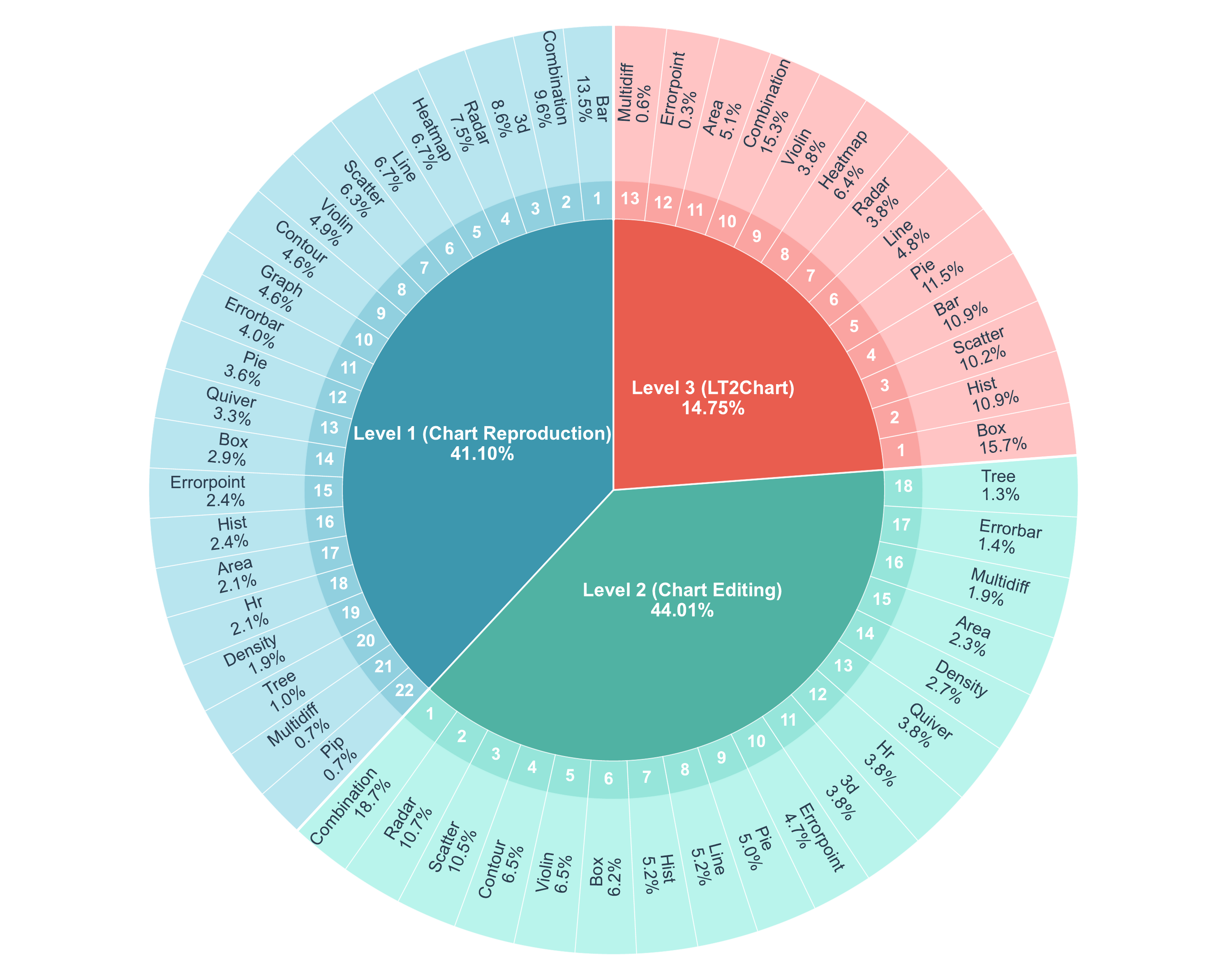}
    \end{minipage}%
  \begin{minipage}[t]{0.458\linewidth}
    \captionsetup{type=table}
    \caption{\textbf{Detailed data statistic.}}
    \label{tab:datastattable}
    \centering
    \vspace{0pt}
    \setlength{\tabcolsep}{6pt}
    \resizebox{0.99\linewidth}{!}{\begin{tabular}{lr}
    \toprule
    \textbf{Statistic} & \textbf{Number} \\
    \midrule
    \multicolumn{2}{l}{\textbf{GT Charts}} \\
    Total charts & 2,186 \\
    - Level 1 / 2 / 3 charts & 863 / 1{,}010 / 313 \\
    Unique charts & 967 \\
    - Unique Level 1 / 2 / 3 charts & 719 / 0 / 248 \\
    \midrule
    \multicolumn{2}{l}{\textbf{Instructions}} \\
    Total instructions & 2,222 \\
    - Level 1 / 2 / 3 instructions & 899 / 1{,}010 / 313\\
    Unique instructions & 1,468 \\
    - Unique instructions - Level 1 / 2 / 3  & 145 / 1{,}010 / 313 \\
    Maximum instruction length - Level 1/ 2 / 3  & 522 / 578 / 390 \\
    Average instruction length - Level 1 / 2 / 3 & 140.5 / 267.3 / 120.8 \\
    \midrule
    \multicolumn{2}{l}{\textbf{GT Code (Lengths/Lines)}} \\
    Maximum code length - Level 1 / 2 / 3 & 96{,}563 / 1{,}1860 / 790{,}130 \\
    Average code length - Level 1 / 2 / 3 & 2{,}714.98 / 3{,}063.2 / 11{,}754.3 \\
    Maximum code lines - Level 1 / 2 / 3 & 842 / 342 / 388 \\
    Average code lines - Level 1 / 2 / 3 & 71.6 / 85.0 / 69.5 \\
    \midrule
    \multicolumn{2}{l}{\textbf{Extremely Long-Table Data}} \\
    Total Excel files & 71 \\
    Average lines per file & 2,647 \\
    Maximum lines & 30{,}427 \\
    Average data entries & 810{,}329.3 \\
    Maximum data entries & 72{,}705 \\
    \bottomrule
    \end{tabular}}
  \end{minipage}
\end{figure*}

\subsection{Data Statistics and Analysis}
\label{datastatistic}

\bench{} comprises 2,186 tasks across three levels--863/1,010/313 for L1/L2/L3--spanning 22/19/12 chart families (e.g., radar, heatmap, scatter, box, tree, error-bar, pie, violin; see Fig.~\ref{fig:benchdistpiechart}).
To maximize diversity, Level 1 emphasizes unique charts (719 unique). 
Level 2 reuses Level 1 charts with at least one edit instruction per chart, resulting in 1,010 unique, non-duplicated edits. 
Level 3 (LT2Chart) includes 245 charts and 313 instructions derived from web-sourced long tables, making annotation and ground-truth code especially challenging. 
As summarized in Tab.~\ref{tab:datastattable}, the instruction scale and substantial code lengths highlight the breadth and difficulty of \bench{}. The data source and annotation details are provided in Appendix \ref{data curataion}

\subsection{Evaluation}
\label{evaluation metric}
\subsubsection{Overall}

To comprehensively evaluate model performance on this benchmark, we first establish the Code \textbf{executable rate, Exec.Rate} as the primary metric. This metric directly measures the model's capability to generate executable code, with its definition and calculation detailed in Appendix \ref{evaluation overall}. Furthermore, we introduce a multi-level and multi-dimensional evaluation framework to evaluate models at both the code-level and chart-level. At the code-level, we propose rule-based and LLM-based evaluation frameworks to compare the similarity in visual outcomes between the ground-truth and generated code. At the chart-level, we employ LMMs to evaluate the visual fidelity of the predicted charts.

\subsubsection{Code-Level}
\paragraph{Base Evaluation Metrics.} To strictly evaluate visual fidelity at the code level, we propose a rule-based base evaluation method, referred to as \texttt{Base}. Leveraging insights into commonly used scientific plotting libraries—such as Matplotlib, Seaborn, Wordcloud, and Networkx—we parse the underlying plotting object \texttt{Figure}. We isolate eight critical dimensions that are fundamental to the evaluation of visual fidelity (detailed in Appendix \ref{base evaluation}).

Compared to the four-dimension heuristic evaluation in \citep{chartmimic}, which relies on function trackers and code injection, our method offers broader dimensional coverage, stronger applicability, faster execution, and higher accuracy (detailed comparison in Appendix \ref{base evaluation}).

\paragraph{LLM Evaluation Metrics.} Recognizing that rule-based matching remains suboptimal for comprehensively evaluating visual fidelity at the code level (see Appendix 1), we leverage GPT-5-mini to evaluate the code, referred to as \texttt{LLM}, thereby exploring its potential to enhance code-level visual fidelity evaluation. By employing more stringent prompts, we evaluate the similarity across various chart dimensions from a code-level perspective and further analyze the performance gap between the code-level and chart-level assessments (see Appendix \ref{LLM vs LMM}).

\subsubsection{Chart-Level}
\paragraph{LMM Evaluation Metrics.} Due to the inherent discrepancy between code-level and chart-level representations, evaluating visual fidelity solely from a code-based perspective is insufficient. Following prior work \cite{chartmimic, chartedit}, we utilize GPT-5-mini to perform multi-dimensional similarity scoring between ground-truth and generated charts, referred to as \texttt{LMM}.

\paragraph{Human Evaluation.}
Given that current LMMs exhibit certain biases in chart visual perception (see Appendix \ref{sec:appendix_case_study}), we also designed a human evaluation system to analyze the consistency between LMM and human assessments. we conducted a human evaluation study involving 20 undergraduate students from the School of Computer Science. Participants assessed both the ground truth (GT) charts and the generated charts across multiple dimensions, aiming for a comprehensive evaluation of the reproduction quality. The results demonstrate a strong correlation between human scores and LMM scores (see Appendix \ref{human_evaluation}).

\section{Experiments}

\subsection{Experiments Setup}
\textbf{Models.}
We conducted tests on 29 widely-used open-source models and proprietary models to evaluate their performance on our benchmark. 
For the open-source models, we selected 12 representative vision-language models, with total parameters ranging from 7B to 72B, including: Qwen2-VL (7B, 72B), Qwen2.5-VL (7B, 72B), Deepseek-VL (7B), Kimi-VL (7B), MiMo-VL-SFT (7B), MiMo-VL-RL (7B), InternVL-2.5 (8B, 38B), InternVL-3 (8B, 38B), InternVL-3.5 (8B, 38B), GLM-4V (9B), LLAVA-onevision-si (7B), LLAVA-onevision-ov (7B), Molmo (7B),Qwen3-VL. 
For proprietary models, we selected the five most popular multimodal large models, including: Gemini-3-pro, Claude-sonnet-4, GPT-5.2, Seed-1.5-VL, and Seed-1.6-VL.

\textbf{Configuration.} 
All experiments were conducted on NVIDIA V100 GPUs.
Models in the 7B–9B range were evaluated using two GPUs with model parallelism; detailed per-level configurations are provided in Tab.~\ref{tab:run_configurations}.  
Empirically, non-thinking models required only 4,096 tokens for inference, with negligible truncation except for InternVL-3.5-38B and the Qwen3-VL series.
To preserve visual fidelity, images were fed at their native resolution, and the maximum input pixel setting supported by each model was used to ensure complete processing of chart details.

\begin{table*}[t]
\small
\centering
\caption{\textbf{Evaluation results on Chart Reproduction (Level 1) with various LMMs.} Each task includes a reference chart as input. \textbf{\textit{DR}}: input without the any data. \textbf{\textit{CRD}}: input with customized text-format table data. \textbf{\textit{CFD}}: input with customized figure-format table data. \textbf{\texttt{Exec.Rate}}: execution rate(\%); \textbf{\texttt{Base}}: the multi-dimensional weighted Base-Score. The Base score uses the same dimension-wise weighting scheme as the LLM-score(\textbf{\texttt{LLM}}). More detail Metrics in Appendix \ref{sec: Analysis}.   We use GPT-5-mini as the base model for both LLM-score and LMM-score(\textbf{\texttt{LMM}}); \textbf{\texttt{Base}}, \textbf{\texttt{LLM}}, \textbf{\texttt{LMM}}: scores on a 0--100 scale (higher is better).}
\label{tab:level1main}
\resizebox{\textwidth}{!}{
\begin{tabular}{l|cccc|cccc|cccc}
\toprule
\multirow{2}{*}{\textbf{Model}}  &  \multicolumn{4}{c}{\textbf{Direct Reproduction(DR)}}  & \multicolumn{4}{c}{\textbf{Customize Raw Data(CRD)}} & \multicolumn{4}{c}{\textbf{Customize Figure Data(CFD)}}   \\

& \textbf{\shortstack{Exec.Rate}} & \textbf{Base} & \textbf{LLM} & \textbf{LMM} &\textbf{\shortstack{Exec.Rate}} & \textbf{Base} & \textbf{LLM} & \textbf{LMM} & \textbf{\shortstack{Exec.Rate}} & \textbf{Base} & \textbf{LLM} & \textbf{LMM }  \\
\midrule
\multicolumn{13}{c}{\it{\textbf{Proprietary}}} \\
\midrule
Gemini-3-Pro & \textbf{97.50} & 78.65 & 75.86 & \textbf{45.42} & \textbf{100.0} & 69.23 & \textbf{69.76} & \textbf{40.72} & \textbf{99.07} & 70.78 & 71.12 & 32.85 \\
Claude-Sonnet-4 & 96.52 & 65.60 & 70.69 & 32.36 & \textbf{100.0} & 61.46 & 55.03 & 40.63 & 93.52 & 65.27 & 65.99 & 26.44 \\
GPT-5.2 & 97.08 & \textbf{79.91} & \textbf{77.88} & 43.73 & 97.22 & 66.31 & 65.51 & 39.26 & \textbf{99.07} & \textbf{73.02} & \textbf{71.42} & \textbf{35.40} \\
Seed-1.5-VL & 87.34 & 63.85 & 53.84 & 26.40 & 97.22 & 65.76 & 58.73 & 34.74 & 79.63 & 65.58 & 64.19 & 19.53 \\
Seed-1.6-VL & 82.61 & 62.46 & 50.85 & 26.16 & 94.44 & 60.69 & 57.87 & 28.44 & 79.63 & 66.22 & 63.12 & 25.51 \\
\midrule
\multicolumn{13}{c}{\it{\textbf{Open-Source LMMs (non-thinking)}}} \\
\midrule
LLaVA-OV-Qwen2-7B-SI & 82.48 & 31.33 & 16.81 & 2.58 & 19.44 & 49.73 & 35.93 & 15.14 & 0.00 & - & - & - \\
LLaVA-OV-Qwen2-7B-OV & 72.60 & 34.33 & 54.85 & 3.42 & 8.33 & 32.68 & 15.67 & 6.00 & 0.00 & - & - & - \\
DeepSeek-VL-7B & 36.44 & 37.49 & 22.02 & 3.37 & 55.56 & 50.73 & 29.28 & 9.25 & 4.63 & 30.37 & 25.20 & 5.60 \\
kimi-VL-A3B & 70.79 & 52.76 & 40.36 & 14.15 & 66.67 & 50.29 & 46.85 & 33.04 & 64.81 & 53.62 & 45.36 & 16.41 \\
Qwen2-VL-7B & 62.45 & 40.99 & 28.74 & 7.02 & 77.78 & 52.21 & 44.50 & 19.61 & 31.48 & 51.02 & 41.03 & 8.65 \\
Qwen2-VL-72B & 74.97 & 50.74 & 38.27 & 12.75 & 86.11 & 56.02 & 46.89 & 28.23 & 64.81 & 59.66 & 56.41 & 15.81 \\
InternVL-2.5-8B & 67.73 & 42.76 & 27.49 & 7.80 & 69.44 & 54.16 & 42.80 & 25.88 & 37.04 & 51.58 & 38.81 & 12.85 \\
InternVL-2.5-38B & 85.67 & 51.97 & 39.83 & 15.18 & 86.11 & 55.74 & 47.98 & 24.71 & 77.78 & 59.86 & 55.51 & 22.88 \\
InternVL-3-8B & 29.49 & 46.58 & 55.68 & 9.20 & 36.11 & 53.47 & 31.38 & 6.23 & 12.96 & 50.73 & 35.61 & 3.14 \\
InternVL-3-38B & 85.26 & 53.57 & 42.85 & 16.68 & 86.11 & 58.17 & 51.10 & 34.00 & 36.11 & 60.17 & 60.42 & 18.56 \\
GLM-4V-9B & 54.24 & 32.50 & 23.12 & 4.80 & 75.00 & 50.35 & 42.91 & 19.70 & 41.67 & 31.64 & 12.84 & 1.00 \\
GLM-4.6v-Flash & 54.52 & 66.67 & 55.38 & 31.70 & 61.11 & 63.44 & 52.75 & 32.45 & 55.56 & 65.64 & 60.23 & 22.57 \\
Intern-VL-3.5-8B & 72.18 & 52.01 & 41.29 & 15.41 & 69.44 & 37.87 & 14.58 & 1.68 & 49.07 & 53.18 & 48.85 & 10.87 \\
Intern-VL-3.5-38B & 84.01 & 55.78 & 42.54 & 19.78 & 86.11 & 60.66 & 53.16 & 38.83 & 24.07 & 61.15 & 60.69 & 21.84 \\
MiMo-VL-7B-RL & 43.12 & 64.41 & 49.82 & 21.43 & 66.67 & 59.73 & 52.21 & 25.38 & 42.59 & 64.39 & 57.68 & 21.44 \\
MiMo-VL-7B-SFT & 51.74 & 63.99 & 49.07 & 22.61 & 75.00 & 53.34 & 37.80 & 22.74 & 32.41 & 64.68 & 54.74 & 22.83 \\
Qwen2.5-VL-7B & 69.26 & 48.20 & 35.46 & 10.54 & 83.33 & 55.42 & 46.72 & 20.14 & 44.44 & 50.30 & 41.78 & 7.81 \\
Qwen2.5-VL-72B & 63.84 & 59.05 & 47.95 & 20.97 & 94.44 & 58.12 & 50.97 & 27.53 & 47.22 & 63.95 & 61.37 & 24.48 \\
Molmo-7B-D & 4.31 & 25.26 & 11.48 & 2.42 & 2.78 & \textbf{70.30} & 40.00 & 4.00 & 1.85 & 45.58 & 24.00 & 0.00 \\
Qwen3-VL-30B-A3B & 72.18 & 63.09 & 50.84 & 15.03 & 69.44 & 55.84 & 52.26 & 23.28 & 74.07 & 63.02 & 60.23 & 23.56 \\
Qwen3-VL-32B & 76.91 & 63.10 & 54.87 & 18.44 & 77.78 & 60.49 & 52.64 & 36.79 & 80.56 & 66.64 & 65.27 & 25.84 \\
\midrule
\multicolumn{13}{c}{\it{\textbf{Open-Source LMMs (thinking)}}} \\
\midrule
MiMo-VL-7B-RL & 21.14 & 70.45 & 60.00 & 30.04 & 75.00 & 64.58 & 57.96 & 24.96 & 37.96 & 68.05 & 64.38 & 29.83 \\
MiMo-VL-7B-SFT & 41.03 & 65.49 & 54.33 & 16.95 & 77.78 & 61.33 & 53.89 & 29.93 & 37.96 & 64.04 & 56.04 & 23.29 \\
Qwen3-VL-30B-A3B & 76.08 & 62.37 & 55.48 & 23.83 & 61.11 & 64.78 & 57.63 & 35.00 & 80.56 & 66.57 & 62.86 & 26.34 \\

\bottomrule
 \end{tabular}}
 \vskip -0.1in
\end{table*}

\begin{table*}[t]
\small
\centering
\caption{\textbf{Evaluation results on Chart Editing (Level 2) with various LMMs.}}
\label{tab:level2main}
\resizebox{0.98\textwidth}{!}{
\begin{tabular}{l|r|cccccccccc|c}
\toprule
\multirow{2}{*}{\textbf{Model}} & \multirow{1}{*}{\textbf{\shortstack{Exec.\\Rate}}} & \multicolumn{10}{c}{\textbf{Code-Level}} & \multicolumn{1}{c}{\textbf{Chart-Level}}  \\

 &  & \textbf{Color} & \textbf{Grid} & \textbf{Layout} & \textbf{Legend} & \textbf{Visual} & \textbf{Data} & \textbf{Text} & \textbf{Type} & \textbf{Base} & \textbf{LLM} & \textbf{LMM} \\
\midrule
\multicolumn{13}{c}{\it{\textbf{Proprietary}}} \\
\midrule
Gemini-3-Pro & \textbf{97.23} & \textbf{52.32} & 75.31 & \textbf{86.45} & 63.33 & 81.58 & 62.75 & 77.16 & 93.86 & 70.78 & 72.21 & \textbf{33.41} \\
Claude-Sonnet-4 & 90.20 & 47.17 & 65.29 & 55.32 & 56.51 & 81.50 & 54.88 & 80.52 & 93.29 & 63.65 & 66.45 & 25.40 \\
GPT-5.2 & 96.04 & 58.44 & 80.83 & 61.51 & 58.16 & \textbf{84.65} & \textbf{64.66} & \textbf{83.77} & \textbf{94.52} & \textbf{70.93} & \textbf{75.66} & 33.03 \\
Seed-1.5-VL & 60.20 & 44.39 & 69.37 & 52.06 & 52.80 & 79.85 & 52.02 & 77.21 & 92.66 & 61.67 & 65.45 & 18.30 \\
Seed-1.6-VL & 70.00 & 43.61 & 69.37 & 52.25 & 51.32 & 79.70 & 53.11 & 79.46 & 92.32 & 61.77 & 65.40 & 17.43 \\

\midrule
\multicolumn{13}{c}{\it{\textbf{Open-Source LMMs (non-thinking)}}} \\
\midrule
LLaVA-OV-Qwen2-7B-SI & 2.57 & 15.35 & 59.38 & 53.85 & 30.79 & 56.34 & 22.59 & 52.70 & 55.33 & 38.43 & 37.40 & 6.40 \\
LLaVA-OV-Qwen2-7B-OV & 1.68 & 15.37 & 63.48 & 38.24 & \textbf{64.56} & 50.15 & 18.13 & 48.81 & 58.43 & 39.07 & 35.44 & 7.35 \\
DeepSeek-VL-7B & 30.46 & 15.32 & 42.58 & 34.44 & 26.24 & 52.95 & 22.36 & 51.56 & 68.43 & 35.16 & 26.87 & 4.20 \\
kimi-VL-A3B & 50.69 & 26.22 & 64.50 & 40.68 & 39.93 & 66.08 & 34.05 & 64.97 & 83.03 & 47.96 & 44.43 & 9.01 \\
Qwen2-VL-7B & 24.55 & 18.72 & 57.60 & 41.87 & 32.73 & 58.65 & 25.41 & 55.07 & 76.38 & 41.06 & 34.82 & 5.27 \\
Qwen2-VL-72B & 57.23 & 27.82 & 64.07 & 42.86 & 40.59 & 69.53 & 36.83 & 64.35 & 84.76 & 49.55 & 39.88 & 10.02 \\
InternVL-2.5-8B & 24.06 & 23.86 & 62.25 & 41.21 & 36.70 & 65.03 & 32.72 & 62.46 & 81.14 & 46.20 & 41.14 & 7.48 \\
InternVL-2.5-38B & 30.30 & 30.40 & 68.25 & 50.34 & 48.97 & 72.92 & 37.38 & 70.78 & 87.96 & 53.48 & 52.90 & 10.67 \\
InternVL-3-8B & 4.55 & 23.22 & 62.68 & 57.03 & 31.34 & 66.79 & 29.34 & 58.63 & 84.49 & 46.41 & 38.02 & 5.48 \\
InternVL-3-38B & 69.80 & 35.87 & 68.36 & 48.66 & 49.30 & 75.86 & 45.33 & 72.28 & 90.53 & 56.74 & 56.98 & 13.32 \\
GLM-4V-9B & 10.50 & 19.71 & 61.79 & 48.11 & 34.54 & 59.98 & 26.53 & 57.80 & 65.80 & 42.05 & 34.78 & 3.87 \\
GLM-4.6V-Flash & 18.22 & 48.06 & 69.23 & 50.13 & 51.20 & 81.56 & 54.78 & 77.90 & 91.86 & 62.76 & 63.25 & 17.91 \\
InternVL-3.5-8B & 37.52 & 29.55 & 68.21 & 43.87 & 41.56 & 69.85 & 39.58 & 67.41 & 83.82 & 51.30 & 49.52 & 9.71 \\
InternVL-3.5-38B & 4.26 & 44.11 & \textbf{92.54} & 67.22 & 60.24 & 72.35 & 40.38 & 76.53 & 88.56 & 62.64 & 57.48 & 8.28 \\
MiMo-VL-7B-RL & 17.23 & 42.72 & 70.76 & 54.30 & 46.81 & 74.50 & 49.23 & 74.57 & 89.75 & 59.42 & 56.95 & 14.54 \\
MiMo-VL-7B-SFT & 30.40 & 37.36 & 67.37 & 48.40 & 47.24 & 75.98 & 46.53 & 70.80 & 88.04 & 56.56 & 53.97 & 15.23 \\
Qwen2.5-VL-7B & 31.98 & 26.87 & 67.22 & 45.41 & 38.75 & 65.34 & 37.21 & 63.68 & 83.65 & 49.22 & 44.35 & 9.44 \\
Qwen2.5-VL-72B & 74.55 & 41.32 & 67.44 & 52.16 & 50.04 & 77.74 & 48.65 & 74.72 & 91.12 & 59.31 & 49.02 & 16.69 \\
Molmo-7B-D & 0.59 & 16.53 & 44.44 & 41.67 & 50.00 & 49.70 & 29.14 & 41.57 & 54.44 & 37.32 & 29.92 & 15.50 \\
Qwen3-VL-30B-A3B & 38.91 & 40.07 & 70.02 & 51.69 & 50.06 & 76.81 & 48.01 & 75.52 & 92.39 & 59.26 & 45.77 & 17.46 \\
Qwen3-32B & 63.76 & 42.90 & 71.98 & 50.76 & 52.24 & 77.96 & 52.35 & 77.43 & 92.09 & 61.30 & 64.15 & 18.27 \\
\midrule
\multicolumn{13}{c}{\it{\textbf{Open-Source LMMs (thinking)}}} \\
\midrule
MiMo-VL-7B-RL & 26.14 & 42.47 & 69.05 & 48.47 & 45.96 & 80.76 & 50.41 & 74.80 & 90.56 & 59.53 & 61.04 & 17.81 \\
MiMo-VL-7B-SFT & 30.50 & 40.56 & 68.22 & 48.10 & 46.65 & 78.42 & 48.68 & 73.48 & 89.86 & 58.32 & 59.33 & 16.38 \\
Qwen3-VL-30B-A3B & 13.76 & 44.00 & 65.02 & 48.92 & 49.19 & 78.19 & 50.18 & 76.18 & 89.60 & 59.55 & 63.04 & 14.37 \\
\bottomrule
\end{tabular}}
\end{table*}





\begin{table*}[t]
\small
\centering
\caption{\textbf{Evaluation results on Long-Table to Chart task (Level 3) with various LMMs.}}
\label{tab:level3main}
\resizebox{\textwidth}{!}{
\renewcommand\arraystretch{1.1}
\begin{tabular}{l|r|cccccccccc|c}
\toprule
\multirow{2}{*}{\textbf{Model}} & \multirow{2}{*}{\textbf{\shortstack{Exec.\\Rate}}} & \multicolumn{10}{c}{\textbf{Code-Level}} & \multicolumn{1}{c}{\textbf{Chart-Level}}  \\

 &  & \textbf{Color} & \textbf{Grid} & \textbf{Layout} & \textbf{Legend} & \textbf{Visual} & \textbf{Data} & \textbf{Text} & \textbf{Type}  & \textbf{Base} & \textbf{LLM} & \textbf{LMM} \\
\midrule
\multicolumn{13}{c}{\it{\textbf{Proprietary}}} \\
\midrule
Gemini-3-Pro & 30.03 & \textbf{56.29} & \textbf{83.71} & 92.55 & 68.39 & \textbf{84.00} & \textbf{62.26} & \textbf{80.45} & \textbf{96.60} & \textbf{74.28} & \textbf{77.90} & \textbf{35.97} \\
Claude-Sonnet-4 & \textbf{46.65} & 35.43 & 68.97 & 84.08 & 61.77 & 74.44 & 46.73 & 67.77 & 89.95 & 61.13 & 62.04 & 16.60 \\
GPT-5.2 & 20.13 & 34.46 & 82.14 & 80.87 & 52.64 & 78.34 & 43.56 & 76.09 & 93.81 & 61.99 & 68.40 & 16.29 \\
Seed-1.5-VL & 10.86 & 44.55 & 76.89 & 93.94 & 69.30 & 73.52 & 51.75 & 73.57 & 95.96 & 67.58 & 75.59 & 22.85 \\
Seed-1.6-VL & 26.52 & 44.26 & 82.28 & \textbf{95.48} & \textbf{74.73} & 75.42 & 52.62 & 77.82 & 86.47 & 68.6 & 76.35 & 25.77 \\

\bottomrule

\end{tabular}}
\end{table*}

\subsection{Main Experimental Results}

\subsubsection{Level-wise Comparison of Models}
\paragraph{Level 1.} As shown in Tab.~\ref{tab:level1main}, proprietary models lead across \textbf{DR}, \textbf{CRD}, \textbf{CFD} achieving high executability but only moderate visual fidelity---for example, Gemini-3-Pro reaches 97.5/100/99.1\% ER on DR/CRD/CFD while LMM-Scores stay around 33.85--45.42. 
CRD is ``easy to run'' (e.g., Gemini, GPT and Qwen2.5-VL-72B at $\approx$100\% ER) yet still low-fidelity ($\approx$27.53--40.72), confirming execution $\neq$ fidelity. 
CFD is the hardest: top proprietary models keep $\geq$99\% ER but LMM-Scores remain $\approx$32.85--35.40, and many open-source models drop sharply (some 0 ER). 
Larger open-source backbones (Qwen3-VL-32B/MOE/Dense, InternVL-2-38B) close part of the execution gap but not the fidelity gap.
A notable outlier is Molmo-7B-D with DR Base-Score $\approx$70.3, which is primarily attributable to statistical bias caused by its extremely low ER.

\paragraph{Level 2.} The results are presented in Tab.~\ref{tab:level2main}. 
Proprietary models sustain $\sim$90\% ER (Gemini 97.23, Claude 90.20, GPT-5.2 96.04) and excel on code-level subskills, especially Layout/Type $\approx$86.45--94.52---yet chart-level remains modest ($\sim$17.43--33.41), evidencing a persistent gap between syntactic compliance and rendered-image fidelity. 
Strong open-source systems improve executability (e.g., Qwen2.5-VL-72B 74.55\%), but chart-level still lags (16.69). Smaller backbones struggle (e.g., LLaVA-OV-Qwen2-7B variants $\leq$2.57\% ER). 
``Thinking'' helps procedure more than pixels: MiMo-VL-7B-RL ER improves 17.23$\rightarrow$26.14, and MiMo-VL-7B-SFT chart-level nudges 15.23$\rightarrow$16.38, but absolute fidelity remains low. Notably, the thinking variant of Qwen3-VL-30B-A3B suffers from a low execution rate due to extensive truncation issues, further highlighting the substantial challenges this task poses for open-source models.

\paragraph{Level 3.} Tab.~\ref{tab:level3main} presented the results.
Coverage is limited because the benchmark is very hard: only a couple of open-source models could even complete inference, and on the proprietary side, five models were run, but overall ER is still <50\%, primarily due to long-context inputs exceeding the maximum input limits. 
Among those that ran, ER drops to 10–46\% (e.g., Seed-1.5-VL 10.86\%), while code-level stays strong (Base $\approx$61--74 LLM $\approx$68--78), indicating structurally plausible code under long context. 
However, chart-level fidelity collapses (Gemini 35.97, Claude 16.60, GPT-5.2 16.29; Seed-1.5/1.6-VL 22.85/25.77), showing that turning lengthy raw tables into pixel-accurate charts is the main bottleneck; this is mainly because current models are unable to simultaneously handle extremely long textual inputs and reference images, and extract the data processing and plotting requirements from the instructions.

\subsubsection{Analysis}
\label{sec:Analysis}


\textbf{Execution vs. Complexity:} 
From level 2 to Level 3, ER for proprietary systems drops from ~90\% in Tab.~\ref{tab:level2main} to 10–30\% on Level 3 (Gemini 30.03, Claude 46.65, GPT-5.2 20.13 in Tab.~\ref{tab:level3main}). 
This mirrors the jump in reasoning load (long-context/table parsing, multi-constraint edits), showing that being able to run code at level 2 does not translate to robust end-to-end success at Level 3.
We concluded \textbf{execution success declines steeply with task complexity, even for top proprietary models}.

\textbf{Code vs. Visual Fidelity:} 
On level 2 (Tab.~\ref{tab:level2main}), proprietary models achieve moderate Base-Score and LLM-Score. (e.g., Gemini 70.78/72.21, Claude 63.65/66.45, GPT-5.2 70.93/75.66), yet chart-level GPT-Score is only ~17.43–33.41 (Gemini 33.41, Claude 25.40, GPT-5.2 33.03). On Level 3 (Tab.~\ref{tab:level3main}), the gap widens: code-level remains moderate to good (e.g., LLM-Score 62.04--77.90 across models), but LMM-Score collapses (Gemini 35.97, Claude 16.60, GPT-5.2 , Seed1.5/1.6-VL 18.30/17.43). 
This demonstrates that \textbf{while code-level compliance is generally high, it does not guarantee pixel-level visual correctness, making chart-level fidelity the primary bottleneck}.

\textbf{Chart Reproduction Challenge: } 
In Tab.~\ref{tab:level1main}, proprietary models still execute but with lower fidelity (e.g., Gemini CFD ER 99.07 with LMM-Score $\approx$32.85; Claude 93.52/26.44; GPT-5 99.07/35.40). 
Open-source models suffer larger drops (e.g., InternVL-3-8B 49.07/10.87, Qwen3-VL-30B-A3B 74.07/23.56; several models hit $\approx$0 ER). 
Compared to DR/CRD in the same table, CFD exposes weaknesses in Base-Score, LLM-Score and LMM-Score.
We concluded \textbf{reproducing existing charts (CFD) is the hardest subtask in Level1}.

\begin{figure*}
    \centering
    \includegraphics[width=1.0\linewidth]{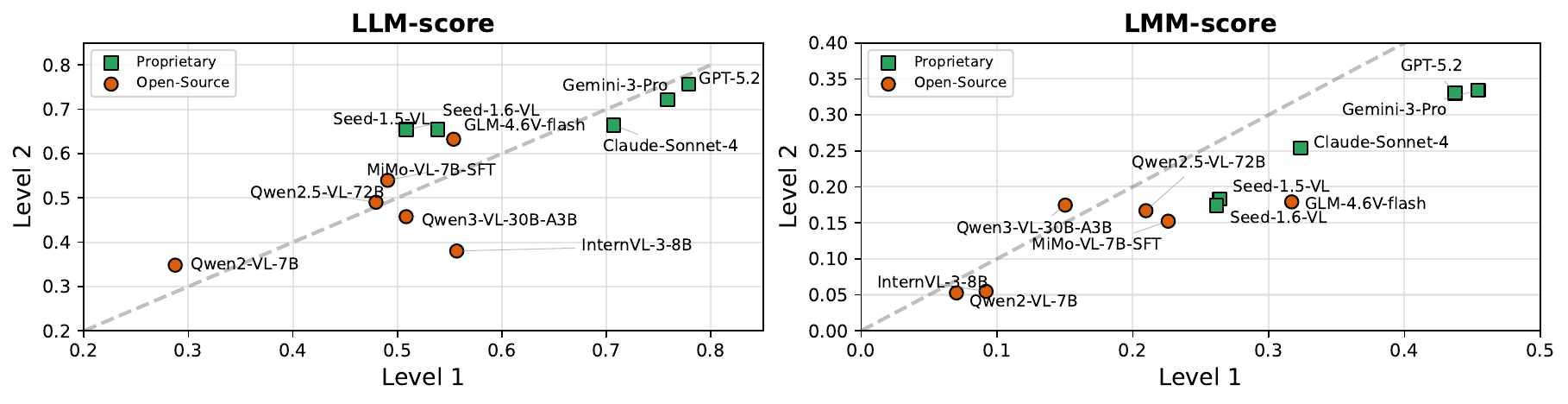}
    \caption{\textbf{Left}: Most proprietary and open-source models generalize well on Level 1 and Level 2 tasks when calculating the LLM-score for predicted code assessment. \textbf{Right}: Proprietary models tend to obtain higher LMM-scores on the Level 1 task rather than the Level 2, while open-source models perform poorly on both tasks (scores are lower than 0.5).}
    \label{fig:scattermodelslevel12}
\end{figure*}

\textbf{Scaling Open-Source Backbones: }
In Tab.~\ref{tab:level2main}, Qwen2.5-VL-72B reaches 74.55 ER with strong code-level, yet chart-level is only 16.69; InternVL-3-38B shows 69.80 ER and similar code-level strength (Base-Score 56.74, LLM-Score 56.98), but chart-level remains 13.32. 
This contrasts with proprietary models’ $\sim$90\% ER and still-low chart-level ($\approx$17.43--33.41), underscoring that fidelity, not executability, is the persistent gap.
These result shows \textbf{larger open-source backbones close part of the execution gap on level 2, but chart-level fidelity gains are modest}.

\textbf{Thinking Helps Procedural Compliance: }
On level 2 (Tab.~\ref{tab:level2main}), MiMo-VL-7B-RL ER rises from 17.23 → 26.14 when enabling thinking; MiMo-VL-7B-SFT nudges 30.40 → 30.50. 
LLM-side (code-level GPT-Score) also improves slightly. 
However, chart-level remains low or mixed (e.g., MiMo-SFT 15.23 → 16.38; MiMo-RL thinking row lacks chart-level). 
The net effect suggests that chain-of-thought/planning aids procedural compliance, yet post-render pixel-level exactness requires additional mechanisms (e.g., render-then-verify loops).
This indicates ''Thinking'' variants help instruction following and executability, but visual fidelity improvements are inconsistent.

\textbf{Metric Analysis:}
From the results in level1, 2, 3(Tab.~\ref{tab:level1main}~\ref{tab:level2main}~\ref{tab:level3main}), we observe that Base-Score and LLM-Score are highly consistent with each other(e.g.,in Tab.~\ref{tab:level2main}, Gemini 70.78/72.21  Qwen2.5-vl-7B 49.22/45.35), yet exhibit a substantial gap compared to LMM-Score. This indicates that\textbf{ both rule-based evaluation and LLM-based code-only assessment correlate poorly with actual visual similarity, revealing a significant mismatch between code-level correctness and pixel-level visual fidelity}.
To further examine the validity of using LMMs for image similarity evaluation, we additionally conduct human comparison studies; detailed experimental results are reported in Appendix \ref{LLM vs LMM}.

\textbf{Table-to-Chart Gap:} 
On Level 1 CRD (Tab.~\ref{tab:level1main}), multiple models achieve very high ER (e.g., Gemini 100; Cluade 100), yet LMM-Score remains low (~40.72–40.63 across models). On level 3 (Tab.~\ref{tab:level3main}), code-level Visual/Text/Type scores are solid for leading models (e.g., Gemini 84.00/80.45/96.60, GPT-5.2 78.34/76.09/93.81), but chart-level stays around 35.97–16.29, highlighting the gap between semantic correctness and visual exactness.
\textbf{Table to chart is relatively “easy to execute” but still hard to render faithfully}.

\textbf{Failure Mode Analysis:}
We further analyze model failures across all levels (see Appendix~\ref{error_case_study}). Errors are mainly caused by data extraction inaccuracies, visual style mismatches, layout inconsistencies, and axis-related issues. \textbf{The dominant bottleneck shifts from perception (Level 1) to reasoning (Level 2) and long-context understanding (Level 3), reflecting the increasing complexity of the tasks}.
\subsection{Discussion.}

\paragraph{Model Performance Across Manually Defined Difficulty Levels.}
In this experiment, we ask the human labeler to split each level into easy, medium and hard, in total three levels, and each subset contains 30 samples.
As shown in Fig.~\ref{fig:difficultscoreheatmap}, model performance exhibits a clear correlation with manually annotated difficulty levels across all benchmark stages. 
On Level 1, proprietary models (e.g., GPT-5.2, Gemini-3-Pro, Claude-Sonnet-4) maintain relatively strong scores across Easy, Medium, and Hard subsets, though the overall fidelity remains moderate. 
In contrast, most open-source models show low scores and struggle particularly on harder cases. 
On Level 2, performance declines noticeably even for proprietary models, with overall scores dropping to $\sim$17.43--33.41 and sharper degradation from Easy to Hard, indicating sensitivity to increased editing complexity. 
By Level 3, almost all models fail regardless of difficulty level: LMM-scores drop to a low range, showing that long-context table-to-chart generation overwhelms current models. 
These trends suggest that \textbf{while models can partially track difficulty scaling on simpler tasks, the hardest scenarios effectively collapse their ability to produce faithful visualizations}.

\paragraph{Code Generalization Holds, Visual Fidelity Lags.}
As shown in Fig.~\ref{fig:scattermodelslevel12}, the performance trends differ substantially when measured by LLM-score versus LMM-score. 
On the left, both proprietary and open-source models generalize reasonably well from Level 1 to Level 2 when evaluated with LLM-score, indicating that code-level syntax and structure can often be preserved across tasks. 
On the right, however, the LMM-score reveals a sharper divide: proprietary models achieve relatively higher visual fidelity on Level 1 than on Level 2, whereas open-source models perform poorly on both levels, with most scores remaining below 0.5. 
This contrast highlights that while models can maintain code-level compliance, translating such compliance into pixel-level faithful renderings remains a key unsolved challenge, particularly for open-source systems.

\section{Limitations}

Overall, \bench{} has two key limitations. 
First, all tasks are currently in English, leaving multilingual chart2code as an important direction for future work. 
Second, our evaluation relies on LLM-based judges to assess code correctness and visual fidelity. 
While this enables scalable evaluation, it may introduce biases and inaccuracies compared to human assessment, especially for subtle visual details. 
Future work will explore multilingual expansion and more robust evaluation protocols to improve the benchmark’s coverage and reliability.

\section{Conclusion} 
We presented \bench{}, a hierarchical benchmark for chart-to-code generation that spans three progressively challenging levels: chart reproduction, chart editing, and long-table–to–chart generation.
Through a large-scale evaluation of 29 state-of-the-art LMMs, we observe a clear and consistent trend: although current models perform reasonably well on simple chart reproduction, their performance degrades substantially when faced with complex editing operations and long-context visualization tasks, thereby revealing significant gaps in practical capability.
Taken together, these findings highlight the persistent and unsolved challenges in chart-to-code generation, and emphasize the need for future models with stronger reasoning abilities, better generalization, and improved robustness in real-world visualization scenarios.



\section{Acknowledgement}
This project is supported by the National Natural Science Foundation of China under Grant No 62502544. 

\bibliographystyle{plainnat}
\bibliography{main}

\clearpage
\beginappendix
\startcontents[app]
\printcontents[app]{l}{1}{}

\clearpage

\section{LLM Usage Statement} 
\label{app:llm}
We disclose the use of Large Language Models (LLMs) in this research in several capacities.

First, during the preparation of this manuscript, we utilized an LLM for grammatical correction and stylistic refinement to improve the paper's readability.

Second, and central to our methodology, multiple LLMs served as the subjects of our experiments to test our proposed benchmark. Furthermore, the evaluation metrics for our benchmark involved using an LLM to assess the comprehensive quality of the results.

We explicitly state that we have never relied on LLMs to generate core research ideas, methodologies, experimental designs, or conclusions. All technical contributions and analyses presented herein are the original work of the authors.





\section{User-Centric Case Studies}
\label{sec:appendix_case_study}

In this section, we present representative examples that reflect scenarios commonly encountered by real users. 
Each example includes the input instruction, structured data, reference visualization, and model-generated outputs, allowing for a direct comparison between correct and erroneous results.

The first example corresponds to a Level 2 task ("Error Sample"), where the model is required not only to generate chart code but also to perform data editing and conditional grouping based on the instruction. 
Specifically, the model must distinguish between sufficient and insufficient inventory categories and reflect this grouping through appropriate visual encoding (e.g., color separation) while preserving the overall chart style. 
This requires jointly reasoning over data values, instruction constraints, and visual design.

As shown in the figure, most LMMs fail in multiple aspects. 
Common errors include: (1) incorrect data analysis, where models misinterpret numerical relationships or fail to apply the required grouping logic; 
(2) visual style inconsistencies, such as failing to reproduce the reference color scheme or incorrectly using a pie chart instead of a donut chart; 
(3) missing or incomplete components, including absent sub-figures or incorrect legend configurations. 
These errors indicate that models struggle to maintain consistency between data, instructions, and visual representation.

We observe that even strong models often fail on this seemingly routine setting, which \textbf{highlights their difficulty in handling tasks that are trivial for humans}. 
In contrast, the correct example demonstrates that solving this task requires precise alignment between data extraction, reasoning, and visual rendering.

Moreover, as illustrated in the subsequent cases ("LLM capability exploration"), existing LMMs frequently produce incorrect outputs even for basic perception tasks, such as identifying chart elements or extracting key information. 
These failures suggest that current models lack robust visual grounding and fine-grained reasoning abilities.

Overall, these observations indicate that \textbf{if models cannot reliably handle such user-level scenarios, they are unlikely to succeed in more complex chart2code tasks involving long-context data and compositional instructions}.

\section{Fine-Grained Visual Perception Analysis}
\label{sec:appendix_perception}

In this section, we analyze the fine-grained visual perception capabilities of LMMs through controlled examples ("LLM capability exploration"), focusing on key attributes such as titles, legends, grid patterns, and axis details.

Each example isolates a specific visual attribute, removing complex reasoning and generation requirements, to directly probe the models’ low-level visual grounding ability. Despite this simplification, even strong models exhibit inconsistent performance. For example, models may confuse title structures (e.g., single-line vs. multi-line), misidentify legend positions, or fail to distinguish visual styles such as dashed versus solid grid lines. They also frequently overlook subtle attributes, including tick label rotation and axis orientation.

These results suggest that current LMMs lack robust fine-grained visual perception, particularly when precise spatial and stylistic details are required. Such limitations are critical, as accurate chart-to-code generation depends on reliably extracting these low-level attributes. Improving visual grounding at this level is therefore essential for advancing chart2code performance.

\begin{example}[Error Sample]
\footnotesize

\textbf{Instruction:}  
Analyze inventory distribution by category.  
- Highlight sufficient inventory in 'Grooming Tools' and 'Kids' Clothing'  
- Highlight insufficient inventory in 'Toys \& Games' and 'Books \& Stationery'  
- Use separate colored sections in the chart to distinguish these two groups  
Generate runnable Python code matching the uploaded image style.

\vspace{0.5em}
\textbf{Data text:}
\begin{verbatim}
{
"Grooming Tools": {"in_stock": 15.2, "out_of_stock": 14.8},
"Kids' Clothing": {"in_stock": 12.5, "out_of_stock": 13.2},
"Toys & Games": {"in_stock": 8.3, "out_of_stock": 9.1},
"Books & Stationery": {"in_stock": 7.1, "out_of_stock": 8.2},
"Health & Wellness": {"in_stock": 6.8, "out_of_stock": 7.4},
"Cameras & Accessories": {"in_stock": 6.5, "out_of_stock": 7.0},
"Beauty & Personal Care": {"in_stock": 6.2, "out_of_stock": 6.7},
"Men's Clothing": {"in_stock": 5.9, "out_of_stock": 6.3},
"Women's Clothing": {"in_stock": 5.4, "out_of_stock": 6.0},
"Shoes & Footwear": {"in_stock": 5.1, "out_of_stock": 5.8}
}
\end{verbatim}

\vspace{0.5em}
\begin{center}
\begin{minipage}{0.46\textwidth}
    \centering
    \textbf{Reference Figure}\\[0.3em]
    \includegraphics[width=0.85\linewidth]{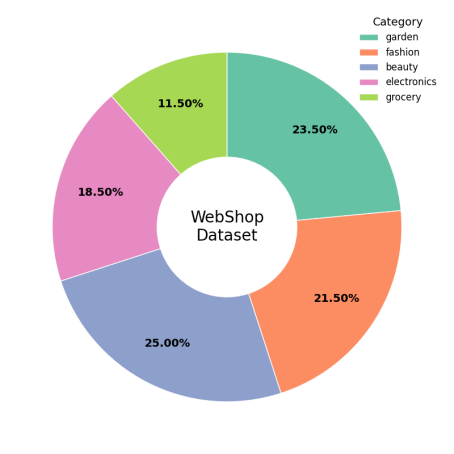}
\end{minipage}
\hfill
\begin{minipage}{0.46\textwidth}
    \centering
    \textbf{GT Figure}\\[0.3em]
    \includegraphics[width=0.85\linewidth]{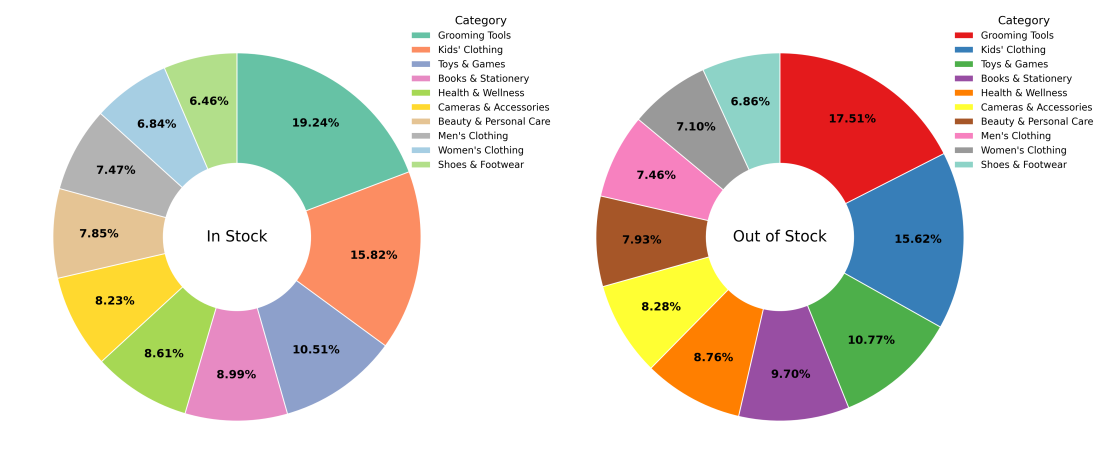}
\end{minipage}
\end{center}

\vspace{1em}
\begin{wronganswer}
\textbf{Doubao-Seed-1.6}

\medskip
\noindent
\begin{minipage}[c]{0.42\linewidth}
  \centering
  \includegraphics[width=0.95\linewidth]{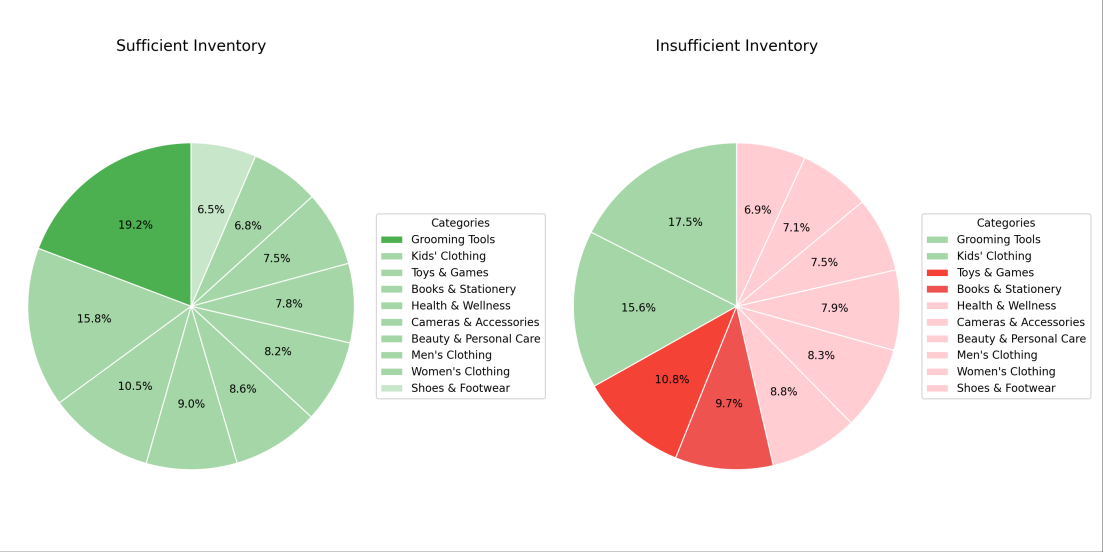}
\end{minipage}\hfill
\begin{minipage}[c]{0.54\linewidth}
  \small
  \red{Data analysis} statistics are incorrect; 
  \red{colors} do not properly mimic the reference style; 
  and the pie chart should be rendered as a \red{donut chart}.
\end{minipage}
\end{wronganswer}

\vspace{0.6em}
\begin{wronganswer}
\textbf{Claude Sonnet 4}

\medskip
\noindent
\begin{minipage}[c]{0.42\linewidth}
  \centering
  \includegraphics[width=0.95\linewidth]{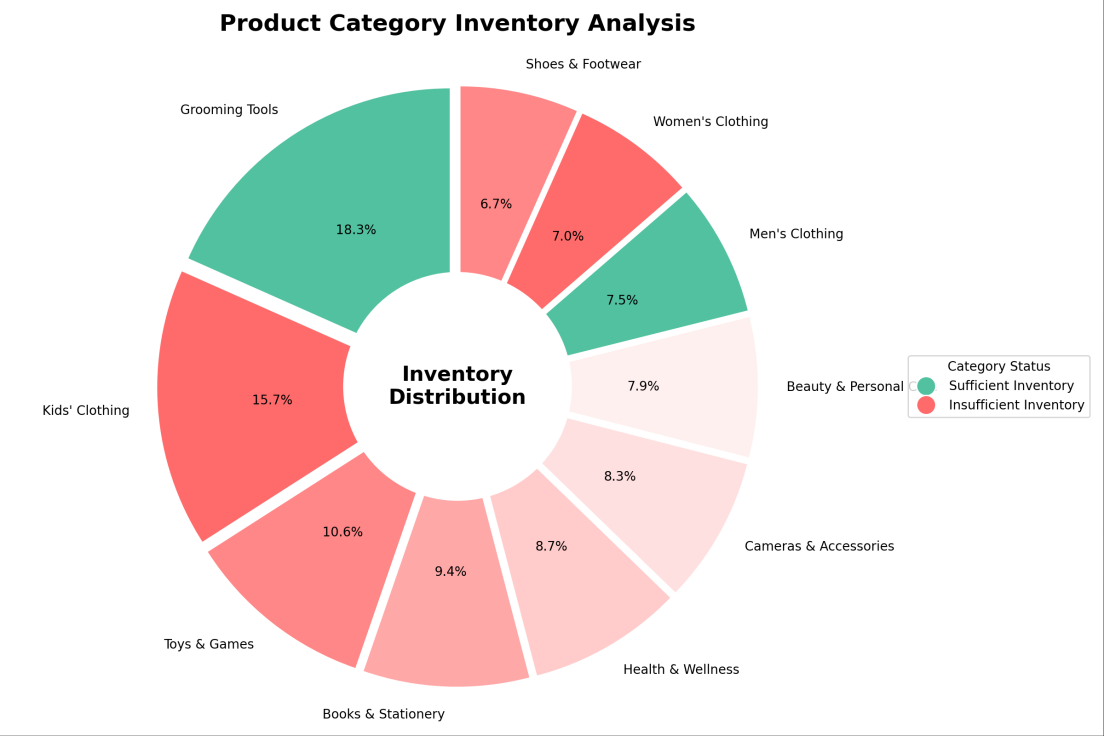}
\end{minipage}\hfill
\begin{minipage}[c]{0.54\linewidth}
  \small
  One \red{sub-figure} is missing; the \red{colors} do not follow the reference style; 
  the \red{data extraction} is incorrect; and the \red{legend style} is wrong.
\end{minipage}
\end{wronganswer}

\vspace{0.6em}
\begin{wronganswer}
\textbf{GPT-5}

\medskip
\noindent
\begin{minipage}[c]{0.42\linewidth}
  \centering
  \includegraphics[width=0.95\linewidth]{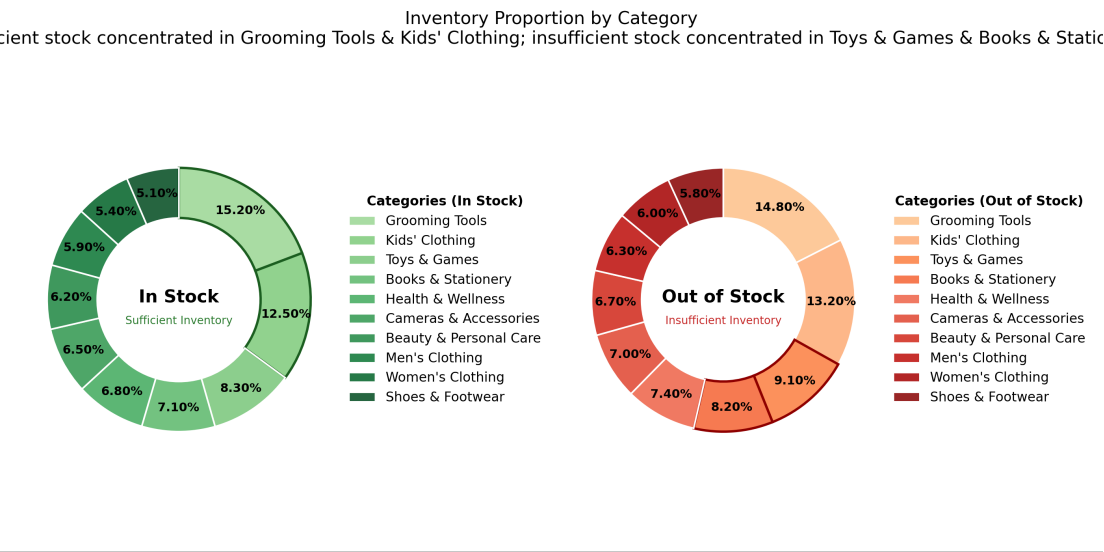}
\end{minipage}\hfill
\begin{minipage}[c]{0.54\linewidth}
  \small
  \red{Color style} error; \red{data analysis} error.
\end{minipage}
\end{wronganswer}

\vspace{0.6em}
\begin{correctanswer}
\textbf{Gemini-2.5-pro}

\medskip
\noindent
\begin{minipage}[c]{0.42\linewidth}
  \centering
  \includegraphics[width=0.95\linewidth]{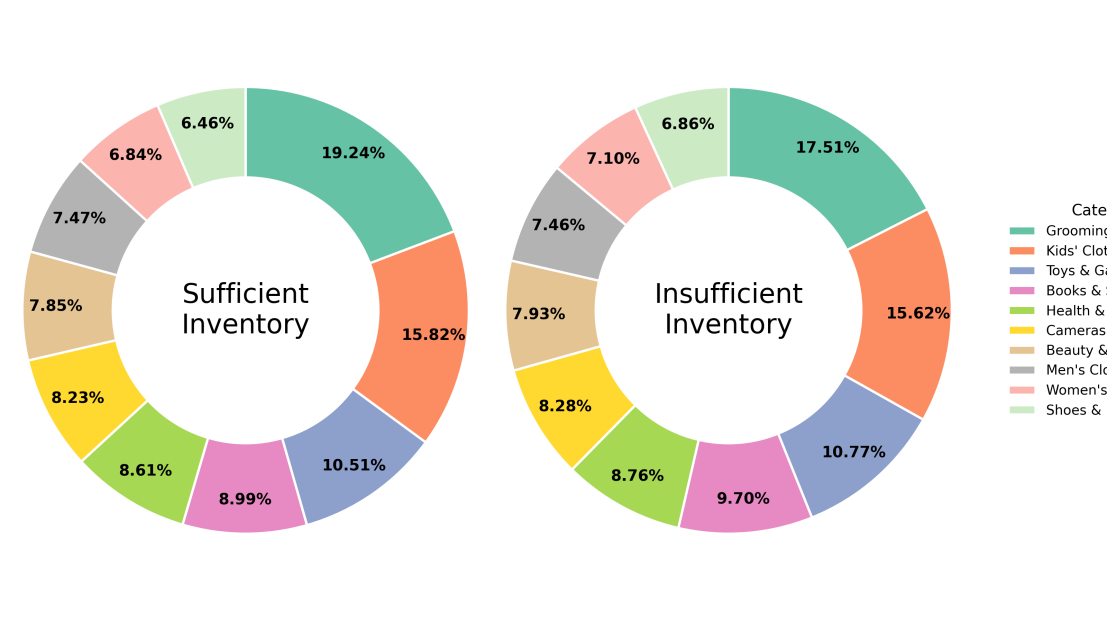}
\end{minipage}\hfill
\begin{minipage}[c]{0.54\linewidth}
  \small
  The image is highly \green{faithful}; the data analysis is \green{correct}; 
  there are minor flaws, but the overall result remains \green{acceptable}.
\end{minipage}
\end{correctanswer}

\end{example}

\clearpage

\clearpage
\begin{example}[LLM capability exploration]
\footnotesize

\textbf{Question}: 
Please, based on this image, tell me if there is any title text in the picture? If so, what is its position, is it single or multiple lines, and what are its size and color?

\vspace{0.5em}
\textbf{Figure}: 
\begin{center}
    \includegraphics[width=0.45\linewidth]{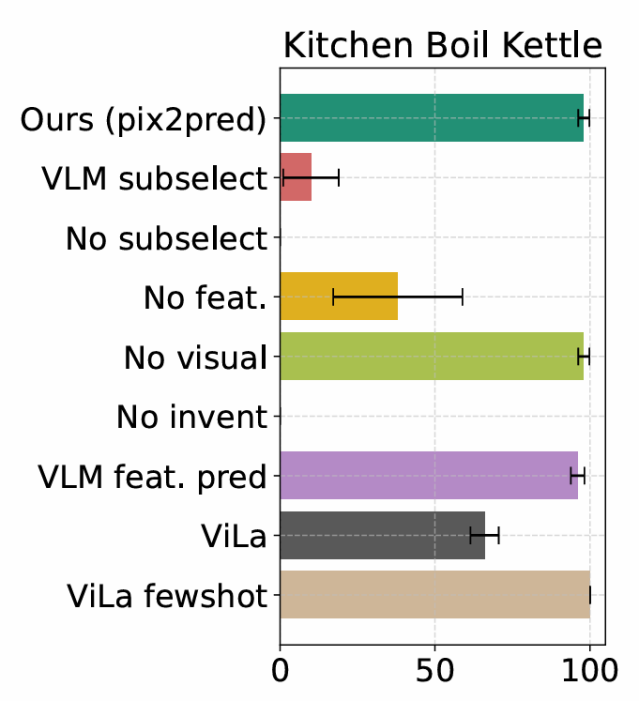}
\end{center}

\vspace{0.8em}
\begin{correctanswer}
\textbf{Answer:} Yes; Upper; Single; fontsize=12; Black;
\end{correctanswer}

\vspace{0.4em}
\begin{correctanswer}
\textbf{Doubao-1.5-thinking-vision-pro:} Yes; Upper; Single; fontsize=14; Black;
\end{correctanswer}

\vspace{0.4em}
\begin{correctanswer}
\textbf{Qwen2.5-VL-72B:} Yes; Center top; Single; fontsize=14-16; Black;
\end{correctanswer}

\vspace{0.4em}
\begin{correctanswer}
\textbf{Claude-opus-4:} Yes; Center top; Single; Large; Black;
\end{correctanswer}

\vspace{0.4em}
\begin{wronganswer}
\textbf{Gemini-2.5-pro:} Yes; Center top; Multiple; Large; Black;
\end{wronganswer}

\vspace{0.4em}
\begin{correctanswer}
\textbf{GPT-5:} Yes; Center top; Single; fontsize=16; Black;
\end{correctanswer}

\end{example}

\clearpage

\begin{wideexample}{LLM capability exploration}

\textbf{Question}: 
Please describe the arrangement and relative position of the legend in this chart: Is it arranged horizontally, vertically, or in a grid? Which side of the main graph is it located on (top, bottom, left, right, or embedded within)?

\vspace{0.5em}
\textbf{Figure}: 
\begin{center}
    \includegraphics[width=0.45\linewidth]{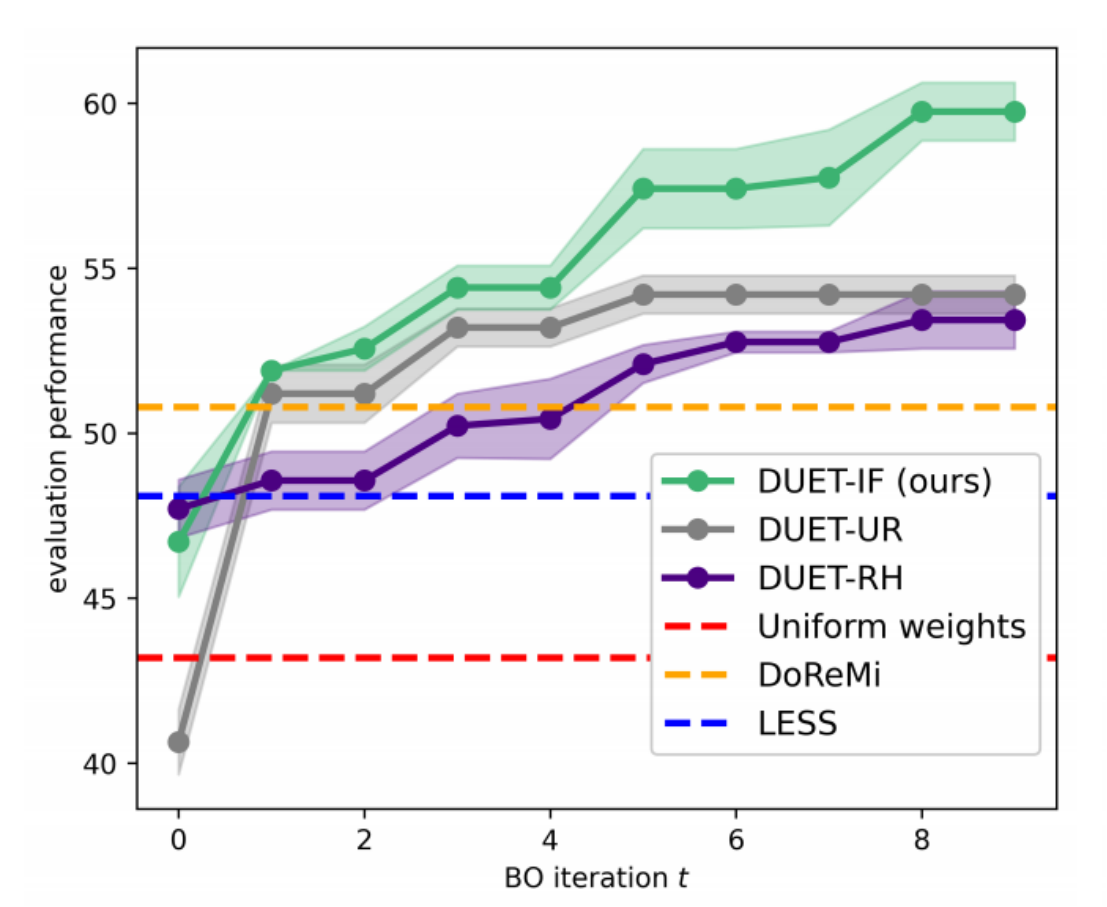}
\end{center}

\vspace{0.8em}
\begin{correctanswer}
\textbf{Answer:} Horizontally(N*1); Lower right;
\end{correctanswer}

\vspace{0.4em}
\begin{wronganswer}
\textbf{Doubao-1.5-thinking-vision-pro:} Horizontally(N*1); Upper right;
\end{wronganswer}

\vspace{0.4em}
\begin{correctanswer}
\textbf{Qwen2.5-VL-72B:} Horizontally(N*1); Lower right;
\end{correctanswer}

\vspace{0.4em}
\begin{wronganswer}
\textbf{Claude-opus-4:} Horizontally(N*1); Middle right;
\end{wronganswer}

\vspace{0.4em}
\begin{wronganswer}
\textbf{Gemini-2.5-pro:} Horizontally(N*1); Middle right;
\end{wronganswer}

\vspace{0.4em}
\begin{wronganswer}
\textbf{GPT-5:} Horizontally(N*1); Right;
\end{wronganswer}

\end{wideexample}

\begin{wideexample}{LLM capability exploration}
\textbf{Question}: 
Please describe the grid lines in this chart: Are they horizontal, vertical, or both? Are the lines dashed or solid?

\textbf{Figure}: 
\begin{center}
    \includegraphics[width=0.7\linewidth]{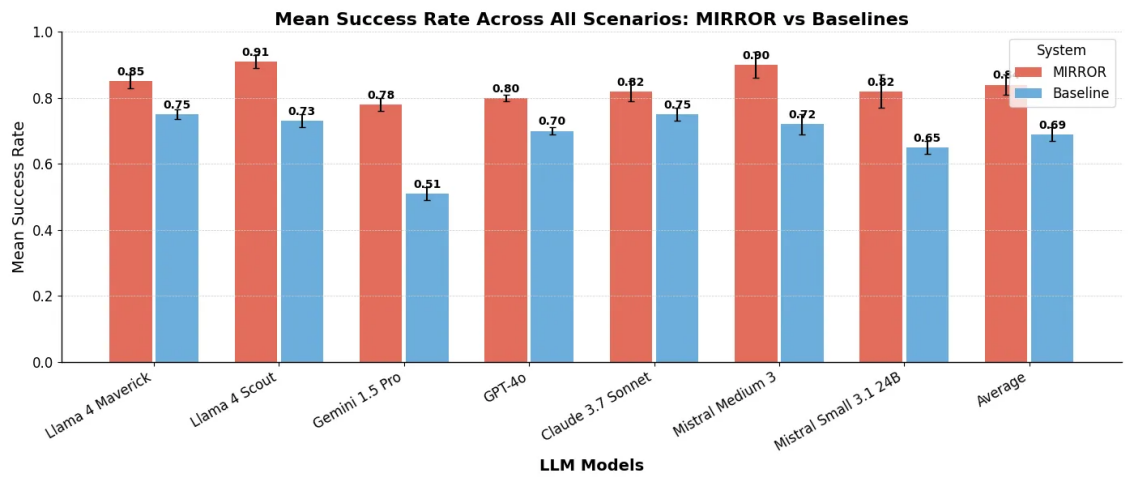}
\end{center}

\begin{correctanswer}
\textbf{Answer:} Only horizontal grid lines; Dashed line;
\end{correctanswer}

\begin{wronganswer}
\textbf{Doubao-1.5-thinking-vision-pro:} Only horizontal grid lines; Solid line;
\end{wronganswer}

\begin{correctanswer}
\textbf{Qwen2.5-VL-72B:} Only horizontal grid lines; Dashed line;
\end{correctanswer}

\begin{wronganswer}
\textbf{Claude-opus-4:} Only horizontal grid lines; Solid line;
\end{wronganswer}

\begin{correctanswer}
\textbf{Gemini-2.5-pro:} Only horizontal grid lines; Dashed line;
\end{correctanswer}

\begin{correctanswer}
\textbf{GPT-5:} Only horizontal grid lines; Dashed line;
\end{correctanswer}

\end{wideexample}

\begin{wideexample}{LLM capability exploration}
\textbf{Question}: 
Please describe the primary tick marks on the axes of this chart: whether they exist, their thickness and orientation (facing outward or inward), as well as the position and rotation angle of the tick labels.

\textbf{Figure}: 
\begin{center}
    \includegraphics[width=0.7\linewidth]{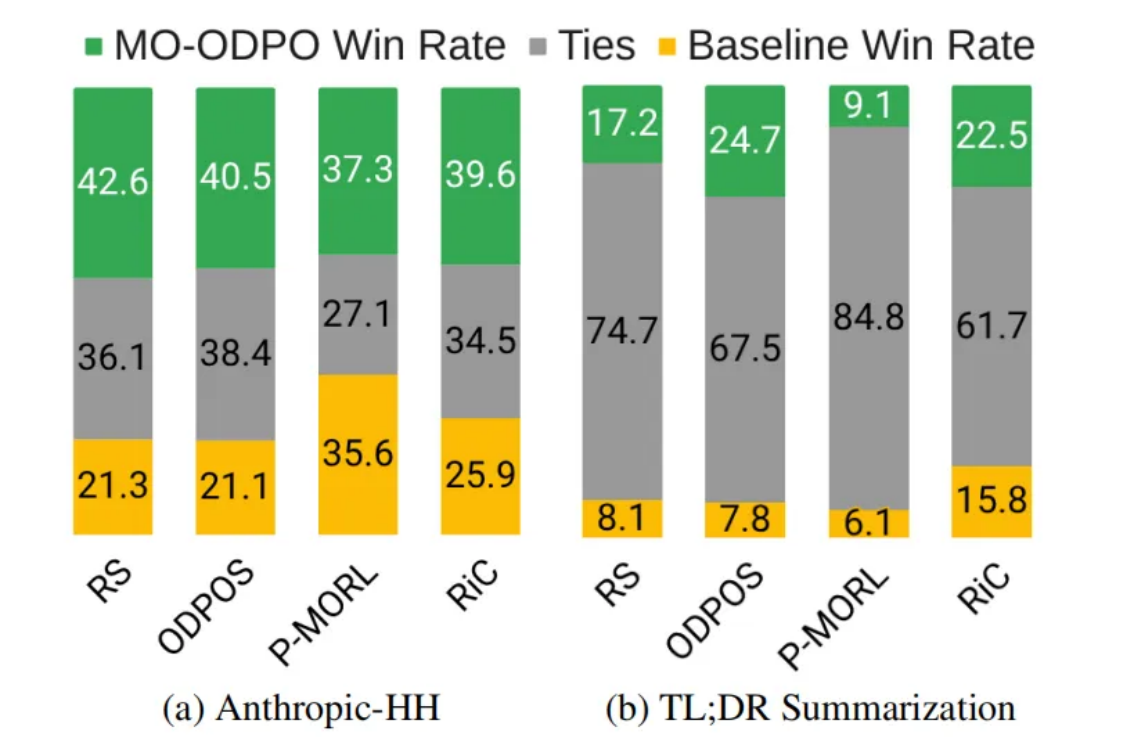}
\end{center}

\begin{correctanswer}
\textbf{Answer:} No; Lower; 45 degrees.
\end{correctanswer}

\begin{wronganswer}
\textbf{Doubao-1.5-thinking-vision-pro:} Implied; Lower; 0 degrees.
\end{wronganswer}

\begin{wronganswer}
\textbf{Qwen2.5-VL-72B:} No; Lower; 0 degrees.
\end{wronganswer}

\begin{wronganswer}
\textbf{Claude-opus-4:} No; Lower; 0 degrees.
\end{wronganswer}

\begin{wronganswer}
\textbf{Gemini-2.5-pro:} No; Lower; 0 degrees.
\end{wronganswer}

\begin{wronganswer}
\textbf{GPT-5:} No; Lower; 0 degrees.
\end{wronganswer}

\end{wideexample}

\section{Data Curation}

To construct a comprehensive and challenging chart benchmark, we collected a rich dataset of chart images and their corresponding raw data from multiple sources.

\subsection{Data Curation and Annotation}
\label{data curataion}

\subsubsection{Data curation}
\textbf{Chart Data:} Our chart figure sources primarily consist of three aspects. 
First, we collected approximately 5,000 paper charts from Arxiv, spanning from January 2024 to July 2025, covering various fields such as CSEE, Physics, Statistics, and Economics, to ensure diversity and modernity in the chart types. 
Second, we gathered 1,000 example charts from function libraries such as Matplotlib, Seaborn, WordCloudX, Scipy, as well as Matlab plotting example tutorials. Finally, we filtered 150 difficult charts from the ChartMimic \citep{chartmimic} dataset.

\textbf{Raw Data:} Our benchmark collects raw data from sources such as Kaggle, Annual Reports, and publicly available data from various company websites. The raw data includes Excel spreadsheets, figures, text, and other formats, covering multiple domains such as corporate financial reports, flight route data, weather data, GDP data, and car sales figures. Additionally, we have intentionally selected data of varying lengths to test the LLM's ability to analyze and process long text data.

\subsubsection{Data filtering}

\textbf{Chart Data:} 
We propose a "gather-distribution" data selection process. 
First, we gather data from various sources into a chart pool, which is then roughly filtered by 10 undergraduate computer science students based on chart type and information complexity. 
Based on this initial selection, we reduce the data to 3,000 charts to ensure that the resulting dataset contain a diverse range of visual elements and chart types. 
Next, the gathered data is distributed by category to 5 experts with many years of experience in Python plotting for independent evaluation. 
The evaluation criteria are refined into three dimensions: data complexity, visual complexity, and programming complexity. Each dimension is independently assessed to select more valuable charts as part of the benchmark data. Finally, the charts from various categories are aggregated to form the 719 unique reference figures in the benchmark.

\textbf{Raw Data:} Since the raw data we collected contains various data formats, we first use automated scripts to filter out the raw data that exhibits rich numerical performance and is suitable for plotting. After that, we conduct manual checks to preserve the diversity of the raw data as much as possible. The final selection includes 71 Excel files, 108 raw data figures, and 36 raw data text files.

\subsubsection{Data Annotation}

During the data annotation process for the three-level tasks, we employed an interactive data annotation method based on Python scripts and agents, which we refer to as the human-AI interactive annotation process. Specifically, in the level 1 data annotation process, annotators, with the assistance of the LMM, recreate the selected data by writing Python code. The data generated here directly serves as the first setting of the Level 1 task. Subsequently, based on the 719 scripts, annotators select and modify suitable chart types using the data from the 108 raw table figures and 36 raw table text files, resulting in 108+36 customized entries for the second setting of the task.

In the Level 2 annotation process, annotators first categorize and summarize chart editing operations commonly encountered in real-world scenarios. 
They then modify the code with the help of prompt engineering and Python code injection, leveraging the programming capabilities of LLM. 
While the LLM may lack proficiency in the chart2code task, its programming ability is exceptional. Through this process, we obtained over 4,700 edited and modified scripts, which were further filtered through the data selection process, ultimately yielding 1,010 high-quality Level 2 data entries.


For Level 3 data annotation, annotators first analyzed the contents of 71 diverse data tables to formulate specific statistical requirements. Leveraging LLMs to generate corresponding preprocessing scripts, they performed systematic data cleaning and transformation. Subsequently, target data aligned with the predefined statistical needs were extracted to facilitate the LLM in generating final plotting code. This rigorous pipeline ultimately yielded 313 high-quality Level 3 data entries.

\subsection{Chart Image Data}
\textbf{Data Scource:} Our chart image library is primarily composed of three parts, designed to cover a wide range of chart types, visual styles, and information densities.

\begin{figure}[t]
    \centering
    \includegraphics[width=\linewidth]{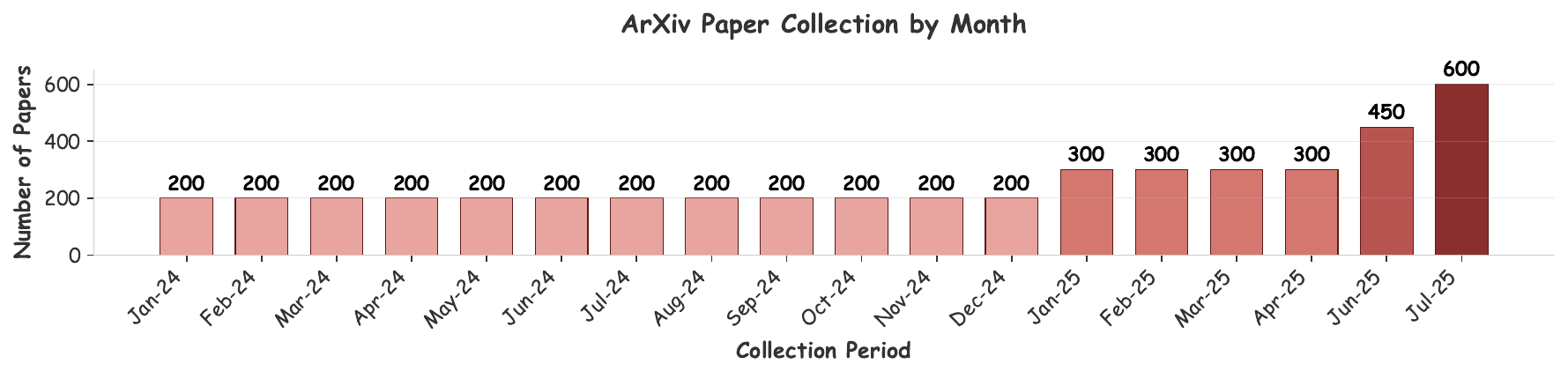}
    \vskip -0.1in
    \caption{Timestamps distribution of chart sources from arxiv preprint.}
    \vskip -0.2in
    \label{fig:arivcollectbar}
\end{figure}

\begin{itemize}[leftmargin=*, label=\textbullet, itemsep=5pt]
    \item \textbf{Charts from Academic Literature:} We extracted chart images from approximately 5,000 PDF documents by crawling and parsing papers from the preprint server arXiv using automated scripts. These publications span from January 2024 to July 2025 and cover multiple disciplines, including computer science, physics, statistics, and economics, timestamps distribution of chart sources from arxiv preprint ~Fig.\ref{fig:arivcollectbar}. This ensures that our dataset not only includes common statistical charts but also covers the highly customized and information-dense visualizations frequently found in academic research, guaranteeing both diversity and state-of-the-art relevance. Due to their superior data quality and structural complexity, academic charts were prioritized for our dataset construction. Following a final filtration stage, 548 charts were curated from arXiv papers. These instances represent the core of our benchmark, providing the necessary depth for evaluating sophisticated chart-to-code generation.


    \item \textbf{Examples from Programming Communities and Tutorials:} To enrich the diversity of our dataset, we curated 1,000 example charts from the official documentation and tutorials of mainstream visualization libraries, including Matplotlib, Seaborn, Plotly, WordCloudX, Scipy, and MATLAB, as well as various active developer forums. By applying systematic data processing and augmentation techniques, we further expanded the variety and complexity of these instances. This integration significantly broadens the taxonomic coverage of our benchmark, ensuring a more comprehensive evaluation of chart-to-code generation across diverse styles and formats. After final data screening, 87 instances were selected from various programming communities and tutorials. The inclusion of these charts substantially enriches our task categories, ensuring that the benchmark covers a more extensive range of practical visualization patterns.
    
    \item \textbf{Existing Chart Datasets:} Additionally, 150 high-quality instances were sourced from ChartMimic\citep{chartmimic} for inclusion in our initial pool. After a final round of filtering, 50 samples were retained for the final benchmark dataset
\end{itemize}

\textbf{Preliminary Collection and Deduplication:}
First, we gathered all charts from the three aforementioned sources into a unified database. We then performed preliminary automated deduplication and format standardization.

\textbf{Coarse Filtering:}
We recruited 10 senior undergraduate students majoring in computer science to conduct an initial screening of the chart pool. The screening criteria were primarily based on the clarity of the chart type (i.e., whether it is a common chart type) and its information complexity (e.g., the number of data series, density of text labels). This stage aimed to quickly eliminate ambiguous, overly simplistic, or low-quality images, reducing the dataset size from approximately 6,300 to 3,000.

\textbf{Expert Evaluation and Annotation:}
We invited five doctoral students and researchers, each with over three years of experience in data visualization, to serve as experts for a fine-grained evaluation of the filtered charts. We assigned the charts to the experts by category (e.g., line charts, bar charts, scatter plots) and asked them to independently score each chart from 1 to 5 across three dimensions:
\textbf{Data Complexity:} Refers to the dimensional and structural complexity of the underlying data required for the chart.
\textbf{Visual Complexity:} Refers to the richness of visual elements in the chart, such as markers, colors, annotations, and dual axes.
\textbf{Programming Complexity:} Refers to the programming skills and volume of code required to reproduce the chart, such as the need for complex layouts or custom functions.
\textbf{Final Adjudication:}
We selected charts that achieved a high composite score across the three dimensions and had high inter-rater agreement ($> 0.8$). For charts with disagreements, two core researchers made the final decision. Through this process, we finalized a set of 719 high-quality reference charts.

\begin{table}[ht]
\centering
\caption{Statistics of Annotation Tasks per Annotator}
\label{tab:annotator_workload}
\begin{tabular}{|l|p{5cm}|p{5cm}|}
\hline
\textbf{Annotator} & \textbf{Raw Data } & \textbf{Chart Data} \\ \hline
Annotator 1 & 20 diverse data tables and Excels. & 50 charts. \\ \hline
Annotator 2 & 20 diverse data tables and Excels. & 50 charts. \\ \hline
Annotator 3 & 50 diverse data figures and Excels. & 50 charts. \\ \hline
Annotator 4 & 50 diverse data figures and Excels. & 50 charts. \\ \hline
Annotator 5 & 50 diverse data tables and Excels. & 150 charts. \\ \hline
Annotator 6 & 50 diverse data tables and Excels. & 150 charts. \\ \hline
Annotator 7 & 40 diverse data tables and Excels. & 150 charts. \\ \hline
Annotator 8 & 40 diverse data tables and Excels. & 150 charts. \\ \hline
Annotator 9 & check data. & check charts. \\ \hline
Annotator 10 & check data. & check charts. \\ \hline
Annotator 11 & check data. & check charts. \\ \hline
Annotator 12 & check data. & check charts. \\ \hline
\end{tabular}
\end{table}

\subsection{Raw Data Filtering}

 \textbf{Automated Preprocessing:} We developed automated scripts to parse raw data files in various formats (e.g., Excel, CSV, TXT, JSON). These scripts prioritized the selection of data tables that contain abundant numerical, time-series, or categorical information suitable for visualization.

 \textbf{Manual Verification and Diversity Preservation:} Subsequently, we manually reviewed the data filtered by the scripts, discarding any incomplete or poorly formatted data. During this process, we placed special emphasis on preserving the diversity of data sources and domains to ensure the final dataset was not biased towards any specific field. Ultimately, we constructed a raw database containing 71 Excel files, 108 structured data figures, and 36 semi-structured text files.

\subsection{Human Annotation Details and Ethical Considerations}

\textbf{Annotation Process and Instructions}. To ensure high-quality data labeling, we employed a human-in-the-loop approach. Due to the dynamic and iterative nature of the task, static instruction documents were not the primary mode of guidance. Instead, annotators were guided through interactive training sessions and real-time feedback loops to align with the specific research goals. Consequently, the full text of static instructions or screenshots is not included in this report. The annotation criteria focused on high-quality chart and human natural language instruction, and ambiguity was resolved through consensus discussions among the annotators and the authors.

\textbf{Recruitment and Compensation}. The annotation task was performed by undergraduate student. The participants were recruited based on their expertise in coding. They were compensated at a rate that exceeds the local minimum wage and is commensurate with standard industry rates for similar tasks. Some annotators were the authors of this paper, and thus no monetary compensation was involved.

\textbf{Consent and Ethics}. All participants were fully informed of the purpose of the study and the usage of the annotated data. Since the task involved standard data processing with no intervention in the participants' personal lives and utilized non-sensitive public data, formal approval from an Institutional Review Board (IRB) was not sought. The study adheres to the ethical guidelines regarding human subject participation, ensuring anonymity and the right to withdraw from the task at any time.

\section{More Analysis}
\label{sec: Analysis}
\subsection{Details Evaluation results}

The detail Base-Score and LLM-Score results on DR, CRD and CFD as show in  Tab.~\ref{tab:level1_direct}, Tab.~\ref{tab:level1_direct_LLM}, Tab.~\ref{tab:level1_customize}, Tab.~\ref{tab:level1_customize_LLM} and Tab.~\ref{tab:level1_figure}, Tab.~\ref{tab:level1_figure_LLM}
Fig.~\ref{fig:level 1 direct}, Fig.~\ref{fig:level 1 customize}, Fig.~\ref{fig:level 1 figure}, Fig.~\ref{fig:level 2}, and Fig.~\ref{fig:level 3} present the comparative results of human scores versus model scores.

\begin{table}[p]
\small
\centering
\caption{Details Evaluation results on level-1 Direct mimic result(Base-Score Details)}
\label{tab:level1_direct}
\resizebox{\linewidth}{!}{
\renewcommand\arraystretch{1.1}
\begin{tabular}{l|r|cccccccccc|c}
\toprule
\multirow{1}{*}{\textbf{Model}} & \multirow{1}{*}{\textbf{\shortstack{Exec.\\Rate}}} & \multicolumn{10}{c}{\textbf{Code-Level}} & \multicolumn{1}{c}{\textbf{Chart-Level}}  \\

 &  & Color & Grid & Layout & Legend & Visual & Data & Text & Type & Base & LLM & LMM \\
\midrule
\multicolumn{13}{c}{\it{\textbf{Proprietary}}} \\
\midrule
Gemini-3-Pro & 97.50 & 61.92 & 83.49 & 84.14 & 87.17 & 83.78 & 68.47 & 93.59 & 93.60 & 78.65 & 75.86 & 45.42 \\
Claude-Sonnet-4 & 96.52 & 43.69 & 51.02 & 77.15 & 80.39 & 74.97 & 55.45 & 85.66 & 88.75 & 65.60 & 56.65 & 32.36 \\
GPT-5.2 & 97.08 & 65.77 & 84.94 & 83.96 & 87.93 & 85.21 & 69.25 & 93.35 & 93.85 & 79.91 & 77.88 & 43.73 \\
Seed1.5-VL & 87.34 & 37.85 & 65.79 & 76.16 & 75.22 & 71.77 & 53.27 & 81.05 & 86.40 & 63.85 & 53.84 & 26.40 \\
Seed1.6-VL & 82.61 & 36.71 & 61.25 & 76.87 & 77.18 & 69.39 & 50.62 & 84.69 & 80.49 & 62.46 & 50.85 & 26.16 \\
\midrule
\multicolumn{13}{c}{\it{\textbf{Open-Source LMMs (non-thinking)}}} \\
\midrule
LLaVA-OV-Qwen2-7B-SI & 82.48 & 7.84 & 46.00 & 58.76 & 29.25 & 41.43 & 17.06 & 44.72 & 43.36 & 31.33 & 16.81 & 2.58 \\
LLaVA-OV-Qwen2-7B-OV & 72.60 & 7.61 & 48.06 & 65.10 & 36.51 & 45.29 & 19.41 & 50.00 & 44.27 & 34.33 & 54.85 & 3.42 \\
DeepSeek-VL-7B & 36.44 & 10.87 & 49.77 & 64.69 & 35.77 & 49.06 & 22.62 & 47.18 & 61.49 & 37.49 & 22.02 & 3.37 \\
kimi-VL-A3B & 70.79 & 26.93 & 58.47 & 69.63 & 61.68 & 63.57 & 38.50 & 64.72 & 78.73 & 52.76 & 40.36 & 14.15 \\
Qwen2-VL-7B & 62.45 & 13.16 & 42.16 & 69.30 & 46.14 & 53.60 & 25.63 & 56.09 & 65.04 & 40.99 & 28.74 & 7.02 \\
Qwen2-VL-72B & 74.97 & 23.44 & 57.84 & 68.96 & 61.70 & 60.72 & 35.52 & 64.78 & 75.42 & 50.74 & 38.27 & 12.75 \\
InternVL-2.5-8B & 67.73 & 16.46 & 49.69 & 68.14 & 46.18 & 54.48 & 28.58 & 57.76 & 61.26 & 42.76 & 27.49 & 7.80 \\
InternVL-2.5-38B & 85.67 & 24.22 & 56.99 & 71.47 & 61.88 & 62.49 & 38.12 & 65.69 & 76.47 & 51.97 & 39.83 & 15.18 \\
InternVL-3-8B & 29.49 & 17.69 & 48.13 & 74.59 & 53.51 & 58.84 & 33.35 & 59.50 & 69.13 & 46.58 & 55.68 & 9.20 \\
InternVL-3-38B & 85.26 & 27.20 & 54.62 & 71.38 & 60.99 & 64.89 & 41.15 & 68.68 & 78.44 & 53.57 & 42.85 & 16.68 \\
GLM-4V-9B & 54.24 & 6.01 & 32.64 & 63.56 & 39.80 & 44.56 & 18.11 & 48.69 & 47.50 & 32.50 & 23.12 & 4.80 \\
GLM-4.6v-flash & 54.52 & 40.68 & 70.91 & 78.12 & 77.96 & 75.47 & 55.51 & 80.94 & 90.95 & 66.67 & 55.38 & 31.70 \\
Intern-VL-3.5-8B & 72.18 & 26.02 & 55.78 & 69.38 & 56.38 & 65.72 & 39.38 & 64.86 & 77.15 & 52.01 & 41.29 & 15.41 \\
InternVL-3.5-38B & 84.01 & 31.37 & 56.41 & 71.04 & 62.63 & 66.82 & 43.84 & 70.42 & 80.06 & 55.78 & 42.54 & 19.78 \\
MiMo-VL-7B-RL & 43.12 & 40.56 & 61.70 & 76.42 & 76.89 & 73.56 & 52.02 & 81.04 & 89.30 & 64.41 & 49.82 & 21.43 \\
MiMo-VL-7B-SFT & 51.74 & 39.96 & 59.75 & 75.99 & 75.20 & 73.44 & 53.71 & 80.75 & 87.66 & 63.99 & 49.07 & 22.61 \\
Qwen2.5-VL-7B & 69.26 & 21.55 & 49.85 & 71.02 & 53.45 & 59.62 & 35.68 & 61.97 & 71.63 & 48.20 & 35.46 & 10.54 \\
Qwen2.5-VL-72B & 63.84 & 35.69 & 56.65 & 74.20 & 68.78 & 69.69 & 46.60 & 74.20 & 82.36 & 59.05 & 47.95 & 20.97 \\
Molmo-7B-D & 4.31 & 10.77 & 33.49 & 45.16 & 32.72 & 31.72 & 8.33 & 34.02 & 37.31 & 25.26 & 11.48 & 2.42 \\
Qwen3-VL-30B-A3B & 72.18 & 39.22 & 59.22 & 76.61 & 73.06 & 73.32 & 51.40 & 78.99 & 88.61 & 63.09 & 50.84 & 23.83 \\
Qwen3-32B & 76.91 & 39.61 & 55.76 & 76.55 & 78.05 & 71.72 & 50.82 & 83.03 & 85.46 & 63.10 & 54.87 & 18.45 \\
\midrule
\multicolumn{13}{c}{\it{\textbf{Open-Source LMMs (thinking)}}} \\
\midrule
MiMo-VL-7B-RL & 21.14 & 42.06 & 74.48 & 88.16 & 82.21 & 78.56 & 59.78 & 83.82 & 93.57 & 70.45 & 60.00 & 30.04 \\
MiMo-VL-7B-SFT & 41.03 & 37.29 & 69.56 & 83.56 & 79.00 & 73.29 & 53.67 & 80.76 & 86.79 & 65.49 & 54.33 & 16.95 \\
Qwen3-VL-30B-A3B & 76.08 & 38.53 & 58.93 & 75.88 & 73.04 & 72.41 & 50.00 & 78.75 & 87.95 & 62.37 & 55.48 & 15.03 \\

\bottomrule
\end{tabular}}
\end{table}

\begin{table}[p]
\vspace{0.2cm}
\small
\centering
\caption{Details Evaluation results on level-1 Direct mimic result(LLM-Score Details)}
\label{tab:level1_direct_LLM}
\resizebox{\linewidth}{!}{
\renewcommand\arraystretch{1.1}
\begin{tabular}{l|r|cccccccccc|c}
\toprule
\multirow{1}{*}{\textbf{Model}} & \multirow{1}{*}{\textbf{\shortstack{Exec.\\Rate}}} & \multicolumn{10}{c}{\textbf{Code-Level}} & \multicolumn{1}{c}{\textbf{Chart-Level}}  \\

 &  & Color & Grid & Layout & Legend & Visual & Data & Text & Type & Base & LLM & LMM \\
\midrule
\multicolumn{13}{c}{\it{\textbf{Proprietary}}} \\
\midrule
Gemini-3-Pro & 97.50 & 75.54 & 87.78 & 77.15 & 88.02 & 67.22 & 58.05 & 84.86 & 86.36 & 78.65 & 75.86 & 45.42 \\
Claude-Sonnet-4 & 96.52 & 53.41 & 69.56 & 61.63 & 76.64 & 50.57 & 34.23 & 70.69 & 62.12 & 65.60 & 56.65 & 32.36 \\
GPT-5.2 & 97.08 & 76.95 & 88.58 & 80.08 & 91.06 & 70.60 & 59.80 & 85.75 & 89.20 & 79.91 & 77.88 & 43.73 \\
Seed1.5-VL & 87.34 & 48.15 & 71.25 & 56.46 & 72.23 & 44.90 & 32.40 & 67.67 & 64.78 & 63.85 & 53.84 & 26.40 \\
Seed1.6-VL & 82.61 & 44.69 & 67.97 & 55.13 & 74.31 & 42.34 & 28.93 & 67.09 & 54.45 & 62.46 & 50.85 & 26.16 \\
\midrule
\multicolumn{13}{c}{\it{\textbf{Open-Source LMMs (non-thinking)}}} \\
\midrule
LLaVA-OV-Qwen2-7B-SI & 82.48 & 15.92 & 35.59 & 14.67 & 25.83 & 11.79 & 2.50 & 34.87 & 8.52 & 31.33 & 16.81 & 2.58 \\
LLaVA-OV-Qwen2-7B-OV & 72.60 & 50.68 & 67.82 & 58.52 & 76.04 & 49.30 & 31.10 & 69.34 & 63.94 & 34.33 & 54.85 & 3.42 \\
DeepSeek-VL-7B & 36.44 & 24.41 & 46.58 & 20.73 & 36.32 & 16.95 & 4.18 & 34.22 & 8.19 & 37.49 & 22.02 & 3.37 \\
kimi-VL-A3B & 70.79 & 41.94 & 60.95 & 42.52 & 58.49 & 31.85 & 18.16 & 54.24 & 35.37 & 52.76 & 40.36 & 14.15 \\
Qwen2-VL-7B & 62.45 & 28.82 & 50.61 & 27.27 & 44.90 & 22.22 & 9.91 & 47.36 & 17.59 & 40.99 & 28.74 & 7.02 \\
Qwen2-VL-72B & 74.97 & 34.97 & 60.72 & 36.60 & 59.18 & 29.96 & 17.24 & 56.57 & 35.20 & 50.74 & 38.27 & 12.75 \\
InternVL-2.5-8B & 67.73 & 27.61 & 49.61 & 24.36 & 44.78 & 20.55 & 9.55 & 46.05 & 15.25 & 42.76 & 27.49 & 7.80 \\
InternVL-2.5-38B & 85.67 & 37.61 & 60.33 & 39.60 & 60.02 & 32.22 & 18.09 & 55.46 & 39.27 & 51.97 & 39.83 & 15.18 \\
InternVL-3-8B & 29.49 & 51.83 & 68.48 & 60.00 & 76.09 & 49.42 & 31.67 & 49.42 & 65.32 & 46.58 & 55.68 & 9.20 \\
InternVL-3-38B & 85.26 & 39.86 & 61.51 & 42.66 & 61.75 & 34.26 & 22.32 & 58.01 & 45.91 & 53.57 & 42.85 & 16.68 \\
InternVL-3.5-38B & 84.01 & 38.52 & 60.64 & 41.42 & 59.30 & 34.12 & 23.61 & 61.58 & 44.30 & 55.78 & 42.54 & 19.78 \\
GLM-4V-9B & 54.24 & 26.33 & 39.44 & 18.47 & 38.35 & 16.92 & 7.63 & 39.97 & 10.13 & 32.50 & 23.12 & 4.80 \\
GLM-4.6v-Flash & 54.52 & 52.26 & 75.31 & 58.02 & 75.24 & 46.41 & 32.26 & 67.76 & 62.08 & 66.67 & 55.38 & 31.70 \\
Intern-VL-3.5-8B & 72.18 & 40.52 & 61.23 & 40.89 & 56.97 & 33.34 & 22.37 & 56.55 & 38.16 & 52.01 & 41.29 & 15.41 \\
MiMo-VL-7B-RL & 43.12 & 45.85 & 68.55 & 52.92 & 71.35 & 41.71 & 26.15 & 67.41 & 52.27 & 64.41 & 49.82 & 21.43 \\
MiMo-VL-7B-SFT & 51.74 & 46.01 & 69.97 & 51.94 & 68.70 & 41.22 & 26.19 & 64.60 & 49.87 & 63.99 & 49.07 & 22.61 \\
Qwen2.5-VL-7B & 69.26 & 32.16 & 55.19 & 34.80 & 55.03 & 28.31 & 16.46 & 53.78 & 30.24 & 48.20 & 35.46 & 10.54 \\
Qwen2.5-VL-72B & 63.84 & 46.10 & 64.36 & 47.97 & 68.66 & 40.05 & 26.83 & 62.88 & 49.67 & 59.05 & 47.95 & 20.97 \\
Molmo-7B-D & 4.31 & 11.61 & 22.90 & 10.48 & 26.45 & 6.45 & 1.13 & 22.10 & 0.97 & 25.26 & 11.48 & 2.42 \\
Qwen3-VL-30B-A3B & 72.18 & 47.46 & 68.09 & 54.27 & 70.81 & 44.26 & 28.21 & 65.92 & 54.64 & 63.09 & 50.84 & 23.83 \\
Qwen3-VL-32B & 76.91 & 50.73 & 67.54 & 58.69 & 76.17 & 48.46 & 31.06 & 69.51 & 64.79 & 63.10 & 54.87 & 18.45 \\
\midrule
\multicolumn{13}{c}{\it{\textbf{Open-Source LMMs (thinking)}}} \\
\midrule
MiMo-VL-7B-RL & 21.14 & 54.41 & 78.59 & 64.11 & 77.14 & 47.66 & 40.41 & 72.24 & 70.59 & 70.45 & 60.00 & 30.04 \\
MiMo-VL-7B-SFT & 41.03 & 49.06 & 74.81 & 57.98 & 74.57 & 44.13 & 31.72 & 68.33 & 63.79 & 65.49 & 54.33 & 16.95 \\
Qwen3-VL-30B-A3B & 76.08 & 51.87 & 68.34 & 59.39 & 76.65 & 49.08 & 32.24 & 69.14 & 63.97 & 62.37 & 55.48 & 15.03 \\
\bottomrule
\end{tabular}}
\end{table}

\begin{table}[p]
\small
\centering
\vspace{-1.2cm}
\caption{Details Evaluation results on level-1 Customize Raw Data result(Base-Score Details)}
\label{tab:level1_customize}
\resizebox{\linewidth}{!}{
\renewcommand\arraystretch{1.1}
\begin{tabular}{l|r|cccccccccc|c}
\toprule
\multirow{1}{*}{\textbf{Model}} & \multirow{1}{*}{\textbf{\shortstack{Exec.\\Rate}}} & \multicolumn{10}{c}{\textbf{Code-Level}} & \multicolumn{1}{c}{\textbf{Chart-Level}}  \\

 &  & Color & Grid & Layout & Legend & Visual & Data & Text & Type & Base & LLM & LMM \\
\midrule
\multicolumn{13}{c}{\it{\textbf{Proprietary}}} \\
\midrule
Gemini-3-Pro & 100.00 & 47.55 & 81.25 & 80.56 & 54.34 & 78.54 & 66.00 & 76.10 & 94.44 & 69.23 & 69.76 & 40.72 \\
Claude-Sonnet-4 &  100.00  & 39.06 & 50.52 & 85.19 & 32.70 & 74.51 & 65.22 & 67.70 & 95.37 & 61.46 & 55.03 & 40.63 \\
GPT-5.2 &  97.22  & 45.29 & 78.93 & 82.86 & 38.79 & 78.83 & 64.71 & 72.25 & 91.43 & 66.31 & 65.51 & 39.26 \\
Seed1.5-VL &  97.22  & 34.89 & 67.86 & 87.14 & 45.53 & 74.32 & 72.14 & 71.55 & 97.14 & 65.76 & 58.73 & 34.74 \\
Seed1.6-VL &  94.44  & 30.76 & 59.44 & 83.82 & 33.47 & 74.55 & 63.89 & 72.24 & 94.12 & 60.69 & 57.87 & 28.44 \\
\midrule
\multicolumn{13}{c}{\it{\textbf{Open-Source LMMs (non-thinking)}}} \\
\midrule
LLaVA-OV-Qwen2-7B-SI &  19.44  & 30.49 & 23.21 & 85.71 & 0.00 & 71.74 & 49.76 & 56.12 & 100.00 & 49.73 & 35.93 & 15.14 \\
LLaVA-OV-Qwen2-7B-OV &  8.33  & 22.92 & 0.00 & 66.67 & 13.98 & 45.32 & 23.22 & 41.94 & 66.67 & 32.68 & 15.67 & 6.00 \\
DeepSeek-VL-7B &  55.56  & 16.67 & 67.50 & 77.50 & 28.58 & 64.17 & 52.16 & 45.18 & 86.67 & 50.73 & 29.27 & 9.25 \\
kimi-VL-A3B &  66.67  & 22.59 & 36.98 & 81.25 & 29.28 & 67.21 & 46.50 & 55.56 & 94.44 & 50.29 & 46.85 & 33.04 \\
Qwen2-VL-7B &  77.78  & 20.25 & 63.39 & 85.71 & 36.01 & 64.19 & 48.49 & 53.13 & 82.14 & 52.21 & 44.50 & 19.61 \\
Qwen2-VL-72B &  86.11  & 22.22 & 62.90 & 90.32 & 37.34 & 68.12 & 55.42 & 53.79 & 92.47 & 56.02 & 46.89 & 28.23 \\
InternVL-2.5-8B &  69.44  & 16.61 & 69.00 & 96.00 & 26.42 & 69.19 & 52.77 & 52.92 & 89.33 & 54.16 & 42.80 & 25.88 \\
InternVL-2.5-38B &  86.11  & 22.85 & 65.32 & 88.71 & 29.53 & 70.70 & 55.40 & 58.51 & 88.17 & 55.74 & 47.98 & 24.71 \\
InternVL-3-8B &  36.11  & 12.86 & 73.08 & 76.92 & 37.02 & 62.84 & 51.94 & 65.45 & 89.74 & 53.47 & 31.38 & 6.23 \\
InternVL-3-38B &  86.11  & 23.00 & 67.74 & 82.26 & 35.38 & 69.99 & 60.67 & 61.15 & 97.85 & 58.17 & 51.10 & 34.00 \\
GLM-4V-9B &  75.00  & 15.14 & 58.33 & 83.95 & 27.78 & 66.36 & 48.62 & 54.36 & 85.19 & 50.35 & 42.91 & 19.70 \\
Glm-4.6v-flash & 61.11 & 33.54 & 84.66 & 81.82 & 31.68 & 73.35 & 71.08 & 61.22 & 92.42 & 63.44 & 52.75 & 32.45 \\
Intern-VL-3.5-8B &  69.44  & 4.05 & 69.73 & 65.33 & 11.50 & 52.01 & 28.69 & 25.33 & 89.33 & 37.87 & 14.58 & 1.68 \\
InternVL-3.5-38B & 86.11 & 23.66 & 70.97 & 88.71 & 44.07 & 70.77 & 63.20 & 64.79 & 93.55 & 60.66 & 53.16 & 38.83 \\
MiMo-VL-7B-RL &  66.67  & 34.85 & 51.14 & 75.00 & 46.20 & 73.42 & 67.99 & 57.95 & 87.88 & 59.73 & 52.21 & 25.38 \\
MiMo-VL-7B-SFT &  75.00  & 25.82 & 37.50 & 71.88 & 47.54 & 70.34 & 55.06 & 61.05 & 83.33 & 53.34 & 37.80 & 22.74 \\
Qwen2.5-VL-7B &  83.33  & 22.51 & 69.58 & 88.33 & 35.67 & 68.37 & 52.26 & 60.51 & 82.22 & 55.42 & 46.72 & 20.14 \\
Qwen2.5-VL-72B &  94.44  & 26.14 & 67.89 & 80.88 & 34.56 & 68.85 & 61.67 & 59.25 & 94.12 & 58.12 & 50.97 & 27.53 \\
Molmo-7B-D &  2.78  & 50.07 & 100.00 & 100.00 & 0.00 & 71.24 & 78.29 & 75.00 & 100.00 & 70.30 & 40.00 & 4.00 \\
Qwen3-VL-30B-A3B &  69.44  & 32.93 & 67.50 & 76.67 & 32.26 & 65.35 & 54.57 & 56.23 & 85.33 & 55.84 & 52.26 & 23.28 \\
Qwen3-VL-32B & 77.78 & 34.69 & 70.09 & 80.36 & 36.22 & 72.69 & 60.29 & 65.07 & 90.48 & 60.49 & 52.64 & 36.79 \\
\midrule
\multicolumn{13}{c}{\it{\textbf{Open-Source LMMs (thinking)}}} \\
\midrule
MiMo-VL-7B-RL &  75.00  & 37.93 & 63.43 & 88.89 & 48.64 & 72.28 & 67.62 & 66.43 & 95.06 & 64.58 & 57.96 & 24.96 \\
MiMo-VL-7B-SFT &  77.78  & 27.23 & 70.98 & 85.71 & 39.14 & 72.44 & 66.65 & 63.25 & 94.05 & 61.33 & 53.89 & 29.93 \\
Qwen3-VL-30B-A3B &  61.11  & 33.61 & 71.59 & 88.64 & 39.56 & 74.41 & 69.98 & 69.46 & 96.97 & 64.78 & 57.63 & 35.00 \\
\bottomrule
\end{tabular}}
\end{table}

\begin{table}[p]
\vspace{0.2cm}
\small
\centering
\caption{Details Evaluation results on level-1 Customize Raw Data result(LLM-Score Details)}
\label{tab:level1_customize_LLM}
\resizebox{\linewidth}{!}{
\renewcommand\arraystretch{1.1}
\begin{tabular}{l|r|cccccccccc|c}
\toprule
\multirow{1}{*}{\textbf{Model}} & \multirow{1}{*}{\textbf{\shortstack{Exec.\\Rate}}} & \multicolumn{10}{c}{\textbf{Code-Level}} & \multicolumn{1}{c}{\textbf{Chart-Level}}  \\

 &  & Color & Grid & Layout & Legend & Visual & Data & Text & Type & Base & LLM & LMM \\
\midrule
\multicolumn{13}{c}{\it{\textbf{Proprietary}}} \\
\midrule
Gemini-3-Pro & 100.00 & 65.56 & 85.06 & 64.58 & 72.50 & 61.81 & 76.06 & 56.67 & 73.75 & 69.23 & 69.76 & 40.72 \\
Claude-Sonnet-4 & 100.00 & 44.03 & 64.58 & 52.08 & 54.86 & 48.19 & 64.44 & 46.67 & 66.94 & 61.46 & 55.03 & 40.63 \\
GPT-5.2 & 97.22 & 60.71 & 82.14 & 62.29 & 68.00 & 61.43 & 65.71 & 62.00 & 66.43 & 66.31 & 65.51 & 39.26 \\
Seed1.5-VL & 97.22 & 37.00 & 72.14 & 55.43 & 61.43 & 45.29 & 76.00 & 52.57 & 74.43 & 65.76 & 58.73 & 34.74 \\
Seed1.6-VL & 94.44 & 39.41 & 72.94 & 58.97 & 59.26 & 49.85 & 67.65 & 55.44 & 68.09 & 60.69 & 57.87 & 28.44 \\
\midrule
\multicolumn{13}{c}{\it{\textbf{Open-Source LMMs (non-thinking)}}} \\
\midrule
LLaVA-OV-Qwen2-7B-SI & 19.44 & 30.00 & 40.00 & 35.00 & 30.00 & 34.29 & 40.00 & 43.57 & 36.43 & 49.73 & 35.93 & 15.14 \\
LLaVA-OV-Qwen2-7B-OV & 8.33 & 13.33 & 33.33 & 13.33 & 26.67 & 16.67 & 6.67 & 26.67 & 0.00 & 32.68 & 15.67 & 6.00 \\
DeepSeek-VL-7B & 55.56 & 21.00 & 67.50 & 29.75 & 31.75 & 26.50 & 23.50 & 19.25 & 29.00 & 50.73 & 29.27 & 9.25 \\
kimi-VL-A3B & 66.67 & 37.92 & 61.46 & 45.83 & 51.88 & 38.12 & 52.71 & 43.12 & 46.88 & 50.29 & 46.85 & 33.04 \\
Qwen2-VL-7B & 77.78 & 31.25 & 74.64 & 39.11 & 48.04 & 32.32 & 54.11 & 41.61 & 38.57 & 52.21 & 44.50 & 19.61 \\
Qwen2-VL-72B & 86.11 & 33.87 & 64.52 & 43.87 & 44.84 & 37.90 & 60.48 & 42.10 & 46.94 & 56.02 & 46.89 & 28.23 \\
InternVL-2.5-8B & 69.44 & 29.20 & 69.80 & 40.80 & 40.40 & 31.80 & 54.40 & 38.00 & 40.00 & 54.16 & 42.80 & 25.88 \\
InternVL-2.5-38B & 86.11 & 29.03 & 73.23 & 44.03 & 46.61 & 37.90 & 61.45 & 46.45 & 50.65 & 55.74 & 47.98 & 24.71 \\
InternVL-3-8B & 36.11 & 20.00 & 73.85 & 29.62 & 38.46 & 26.54 & 25.00 & 26.15 & 29.23 & 53.47 & 31.38 & 6.23 \\
InternVL-3-38B & 86.11 & 34.68 & 72.90 & 47.58 & 51.45 & 40.97 & 67.10 & 45.00 & 49.52 & 58.17 & 51.10 & 34.00 \\
InternVL-3.5-38B & 86.11 & 34.35 & 71.13 & 54.35 & 51.45 & 43.23 & 67.74 & 46.61 & 60.65 & 60.66 & 53.16 & 38.83 \\
GLM-4V-9B & 75.00 & 32.78 & 70.74 & 35.37 & 44.63 & 32.78 & 56.11 & 38.33 & 29.44 & 50.35 & 42.91 & 19.70 \\
GLM-4.6v-Flash & 61.11 & 37.73 & 87.05 & 52.27 & 51.82 & 41.14 & 56.14 & 44.55 & 62.95 & 63.44 & 52.75 & 32.45 \\
Intern-VL-3.5-8B & 69.44 & 13.00 & 51.80 & 14.00 & 14.20 & 14.80 & 0.00 & 3.60 & 21.40 & 37.87 & 14.58 & 1.68 \\
MiMo-VL-7B-RL & 66.67 & 43.12 & 69.17 & 45.62 & 59.17 & 46.67 & 59.79 & 42.29 & 53.33 & 59.73 & 52.21 & 25.38 \\
MiMo-VL-7B-SFT & 75.00 & 28.33 & 61.30 & 40.93 & 46.30 & 34.81 & 37.41 & 32.22 & 30.93 & 53.34 & 37.80 & 22.74 \\
Qwen2.5-VL-7B & 83.33 & 29.33 & 74.00 & 43.33 & 50.33 & 36.33 & 57.00 & 44.50 & 46.00 & 55.42 & 46.72 & 20.14 \\
Qwen2.5-VL-72B & 94.44 & 36.18 & 73.09 & 44.12 & 52.35 & 42.21 & 63.97 & 45.29 & 52.35 & 58.12 & 50.97 & 27.53 \\
Molmo-7B-D & 2.78 & 25.00 & 100.00 & 40.00 & 0.00 & 10.00 & 30.00 & 40.00 & 100.00 & 70.30 & 40.00 & 4.00 \\
Qwen3-VL-30B-A3B & 69.44 & 39.40 & 75.00 & 51.20 & 51.60 & 43.60 & 61.60 & 40.20 & 59.00 & 55.84 & 52.26 & 23.28 \\
Qwen3-Dense-32B & 77.78 & 40.89 & 80.00 & 51.43 & 54.64 & 43.75 & 58.93 & 48.21 & 48.75 & 60.49 & 52.64 & 36.79 \\
\midrule
\multicolumn{13}{c}{\it{\textbf{Open-Source LMMs (thinking)}}} \\
\midrule
MiMo-VL-7B-RL & 75.00 & 43.15 & 73.89 & 56.30 & 62.78 & 45.56 & 72.78 & 47.59 & 61.67 & 64.58 & 57.96 & 24.96 \\
MiMo-VL-7B-SFT & 77.78 & 35.36 & 77.50 & 51.79 & 51.79 & 42.32 & 65.54 & 47.32 & 66.43 & 61.33 & 53.89 & 29.93 \\
Qwen3-VL-30B-A3B & 61.11 & 47.27 & 79.23 & 55.68 & 60.00 & 47.05 & 64.32 & 50.91 & 60.23 & 64.78 & 57.63 & 35.00 \\
\bottomrule
\end{tabular}}
\end{table}

\begin{table}[p]
\small
\centering
\vspace{-1.2cm}
\caption{Details Evaluation results on level-1 Customize Figure Data result(Base-Score Details)}
\label{tab:level1_figure}
\resizebox{\linewidth}{!}{
\renewcommand\arraystretch{1.1}
\begin{tabular}{l|r|cccccccccc|c}
\toprule
\multirow{1}{*}{\textbf{Model}} & \multirow{1}{*}{\textbf{\shortstack{Exec.\\Rate}}} & \multicolumn{10}{c}{\textbf{Code-Level}} & \multicolumn{1}{c}{\textbf{Chart-Level}}  \\

 &  & Color & Grid & Layout & Legend & Visual & Data & Text & Type & Base & LLM & LMM \\
\midrule
\multicolumn{13}{c}{\it{\textbf{Proprietary}}} \\
\midrule
Gemini-3-Pro & 99.07 & 52.32 & 75.31 & 86.45 & 63.33 & 81.58 & 62.75 & 77.16 & 93.86 & 70.78 & 71.12 & 32.85 \\
Claude-Sonnet-4 & 93.52 & 48.84 & 52.64 & 85.48 & 58.65 & 79.04 & 57.41 & 71.10 & 93.30 & 65.27 & 65.99 & 26.44 \\
GPT-5.2 & 99.07 & 61.93 & 78.66 & 85.51 & 61.32 & 83.12 & 62.66 & 75.46 & 96.95 & 73.02 & 71.42 & 35.40 \\
Seed1.5-VL & 79.63 & 48.02 & 59.30 & 87.21 & 57.02 & 74.28 & 57.13 & 75.32 & 92.36 & 65.58 & 64.19 & 19.53 \\
Seed1.6-VL & 79.63 & 45.99 & 64.97 & 83.14 & 64.99 & 76.02 & 57.72 & 74.38 & 91.28 & 66.22 & 63.12 & 25.51 \\
\midrule
\multicolumn{13}{c}{\it{\textbf{Open-Source LMMs (non-thinking)}}} \\
\midrule
LLaVA-OV-Qwen2-7B-SI & 0.00 & - & - & - & - & - & - & - & - & - & - & 0.00 \\
LLaVA-OV-Qwen2-7B-OV & 0.00 & - & - & - & - & - & - & - & - & - & - & 0.00 \\
DeepSeek-VL-7B & 4.63 & 1.15 & 40.00 & 80.00 & 20.00 & 52.25 & 17.00 & 35.15 & 40.00 & 30.37 & 25.20 & 5.60 \\
kimi-VL-A3B & 64.81 & 35.14 & 47.80 & 80.71 & 40.63 & 66.55 & 43.79 & 58.94 & 83.76 & 53.62 & 45.36 & 16.41 \\
Qwen2-VL-7B & 31.48 & 20.76 & 53.31 & 79.41 & 45.90 & 68.20 & 41.07 & 58.34 & 81.37 & 51.02 & 41.03 & 8.65 \\
Qwen2-VL-72B & 64.81 & 38.86 & 52.38 & 87.86 & 58.18 & 71.23 & 47.31 & 65.35 & 89.24 & 59.66 & 56.41 & 15.81 \\
InternVL-2.5-8B & 37.04 & 25.92 & 54.69 & 86.25 & 37.60 & 66.06 & 39.15 & 53.18 & 87.92 & 51.58 & 38.81 & 12.85 \\
InternVL-2.5-38B & 77.78 & 37.53 & 61.90 & 84.13 & 50.74 & 73.02 & 50.31 & 63.71 & 89.44 & 59.86 & 55.51 & 22.88 \\
InternVL-3-8B & 12.96 & 23.97 & 43.75 & 82.14 & 37.88 & 69.88 & 38.55 & 48.62 & 100.00 & 50.73 & 35.61 & 3.14 \\
InternVL-3-38B & 36.11 & 37.10 & 66.99 & 87.18 & 51.74 & 69.71 & 49.32 & 68.27 & 84.96 & 60.17 & 60.42 & 18.56 \\
InternVL-3.5-38B & 24.07 & 38.85 & 61.70 & 90.38 & 58.57 & 74.83 & 46.18 & 68.27 & 87.69 & 61.15 & 60.69 & 21.85 \\
GLM-4V-9B & 41.67 & 5.57 & 40.83 & 75.56 & 17.78 & 47.17 & 21.21 & 28.22 & 53.33 & 31.64 & 12.84 & - \\
GLM-4.6v-Flash & 55.56 & 47.47 & 61.94 & 90.00 & 61.25 & 78.09 & 52.89 & 70.49 & 93.89 & 65.64 & 60.23 & 22.57 \\
Intern-VL-3.5-8B & 49.07 & 29.86 & 53.85 & 83.02 & 43.76 & 63.91 & 40.18 & 60.74 & 86.42 & 53.18 & 48.85 & 10.87 \\
MiMo-VL-7B-RL & 42.59 & 51.85 & 62.23 & 79.35 & 55.70 & 78.59 & 53.98 & 70.77 & 85.58 & 64.39 & 57.68 & 21.44 \\
MiMo-VL-7B-SFT & 32.41 & 44.99 & 58.47 & 75.81 & 59.95 & 81.79 & 57.63 & 68.12 & 97.42 & 64.68 & 54.74 & 22.83 \\
Qwen2.5-VL-7B & 44.44 & 27.43 & 57.12 & 73.61 & 38.66 & 61.61 & 38.58 & 53.06 & 86.94 & 50.30 & 41.78 & 7.81 \\
Qwen2.5-VL-72B & 47.22 & 45.18 & 59.56 & 84.31 & 56.47 & 77.20 & 55.32 & 68.95 & 91.96 & 63.95 & 61.37 & 24.48 \\
Molmo-7B-D & 1.85 & 5.68 & 62.50 & 100.00 & 0.00 & 65.44 & 40.01 & 36.49 & 100.00 & 45.58 & 24.00 & 0.00 \\
Qwen3-VL-30B-A3B & 74.07 & 47.15 & 53.18 & 85.42 & 53.22 & 74.28 & 52.70 & 72.34 & 92.04 & 63.02 & 60.23 & 26.34 \\
Qwen3-Dense-32B & 80.56 & 51.43 & 64.80 & 86.21 & 60.36 & 75.79 & 56.32 & 73.16 & 90.65 & 66.64 & 65.27 & 25.84 \\
\midrule
\multicolumn{13}{c}{\it{\textbf{Open-Source LMMs (thinking)}}} \\
\midrule
MiMo-VL-7B-RL & 37.96 & 57.15 & 58.23 & 90.65 & 53.93 & 77.89 & 59.95 & 73.84 & 91.79 & 68.05 & 64.38 & 29.82 \\
MiMo-VL-7B-SFT & 37.96 & 48.44 & 56.61 & 91.06 & 60.84 & 74.02 & 49.44 & 71.76 & 90.41 & 64.04 & 56.04 & 23.29 \\
Qwen3-VL-30B-A3B & 80.56 & 49.78 & 62.02 & 87.36 & 53.82 & 78.44 & 58.00 & 72.61 & 95.94 & 66.57 & 62.86 & 23.56 \\
\bottomrule
\end{tabular}}
\end{table}

\begin{table}[p]
\vspace{0.2cm}
\small
\centering
\caption{Details Evaluation results on level-1 Customize Figure Data result(LLM-Score Details)}
\label{tab:level1_figure_LLM}
\resizebox{\linewidth}{!}{
\renewcommand\arraystretch{1.1}
\begin{tabular}{l|r|cccccccccc|c}
\toprule
\multirow{1}{*}{\textbf{Model}} & \multirow{1}{*}{\textbf{\shortstack{Exec.\\Rate}}} & \multicolumn{10}{c}{\textbf{Code-Level}} & \multicolumn{1}{c}{\textbf{Chart-Level}}  \\

 &  & Color & Grid & Layout & Legend & Visual & Data & Text & Type & Base & LLM & LMM \\
\midrule
\multicolumn{13}{c}{\it{\textbf{Proprietary}}} \\
\midrule
Gemini-3-Pro & 99.07 & 68.90 & 87.67 & 73.10 & 79.67 & 66.95 & 65.81 & 62.38 & 85.55 & 70.78 & 71.12 & 32.85 \\
Claude-Sonnet-4 & 93.52 & 61.55 & 76.18 & 70.60 & 71.90 & 60.85 & 58.95 & 60.25 & 85.70 & 65.27 & 65.99 & 26.44 \\
GPT-5.2 & 99.07 & 68.91 & 89.04 & 71.47 & 79.98 & 67.74 & 64.29 & 62.38 & 83.96 & 73.02 & 71.42 & 35.40 \\
Seed1.5-VL & 79.63 & 58.59 & 78.27 & 65.15 & 72.35 & 55.35 & 61.29 & 60.47 & 78.06 & 65.58 & 64.19 & 19.53 \\
Seed1.6-VL & 79.63 & 56.37 & 78.75 & 67.18 & 72.62 & 56.73 & 63.93 & 58.69 & 71.67 & 66.22 & 63.12 & 25.51 \\
\midrule
\multicolumn{13}{c}{\it{\textbf{Open-Source LMMs (non-thinking)}}} \\
\midrule
LLaVA-OV-Qwen2-7B-SI & 0.00 & - & - & - & - & - & - & - & - & - & - & 0.00 \\
LLaVA-OV-Qwen2-7B-OV & 0.00 & - & - & - & - & - & - & - & - & - & - & 0.00 \\
DeepSeek-VL-7B & 4.63 & 28.00 & 56.00 & 32.00 & 42.00 & 19.00 & 0.00 & 21.00 & 26.00 & 30.37 & 25.20 & 5.60 \\
kimi-VL-A3B & 64.81 & 45.64 & 68.50 & 48.00 & 55.00 & 40.86 & 32.50 & 47.00 & 37.93 & 53.62 & 45.36 & 16.41 \\
Qwen2-VL-7B & 31.48 & 45.74 & 71.03 & 39.41 & 51.62 & 35.88 & 26.03 & 37.50 & 31.32 & 51.02 & 41.03 & 8.65 \\
Qwen2-VL-72B & 64.81 & 52.14 & 75.29 & 57.21 & 67.00 & 48.57 & 48.79 & 53.07 & 61.14 & 59.66 & 56.41 & 15.81 \\
InternVL-2.5-8B & 37.04 & 38.97 & 66.28 & 42.82 & 41.92 & 36.28 & 26.92 & 43.33 & 35.64 & 51.58 & 38.81 & 12.85 \\
InternVL-2.5-38B & 77.78 & 51.49 & 75.65 & 59.23 & 65.65 & 47.98 & 45.48 & 51.07 & 61.55 & 59.86 & 55.51 & 22.88 \\
InternVL-3-8B & 12.96 & 40.00 & 62.86 & 41.43 & 45.00 & 37.86 & 9.64 & 35.71 & 33.93 & 50.73 & 35.61 & 3.14 \\
InternVL-3-38B & 36.11 & 59.10 & 79.74 & 61.28 & 66.28 & 52.69 & 54.62 & 55.00 & 61.79 & 60.17 & 60.42 & 18.56 \\
InternVL-3.5-38B & 24.07 & 55.58 & 77.31 & 54.23 & 68.27 & 50.00 & 65.00 & 52.69 & 63.27 & 61.15 & 60.69 & 21.85 \\
GLM-4V-9B & 41.67 & 15.68 & 37.95 & 11.02 & 21.48 & 11.02 & 1.02 & 7.16 & 9.32 & 31.64 & 12.84 & 1.00 \\
GLM-4.6v-Flash & 55.56 & 60.08 & 77.67 & 62.50 & 69.92 & 50.25 & 51.92 & 53.00 & 65.00 & 65.64 & 60.23 & 22.57 \\
Intern-VL-3.5-8B & 49.07 & 53.30 & 73.02 & 50.85 & 58.58 & 45.85 & 30.09 & 51.42 & 41.98 & 53.18 & 48.85 & 10.87 \\
MiMo-VL-7B-RL & 42.59 & 61.16 & 80.70 & 61.40 & 71.16 & 56.40 & 49.65 & 55.47 & 70.35 & 64.39 & 57.68 & 21.44 \\
MiMo-VL-7B-SFT & 32.41 & 57.00 & 69.86 & 57.57 & 66.43 & 46.86 & 40.00 & 49.86 & 62.86 & 64.68 & 54.74 & 22.83 \\
Qwen2.5-VL-7B & 44.44 & 39.90 & 71.88 & 44.79 & 53.23 & 38.44 & 27.29 & 40.83 & 34.27 & 50.30 & 41.78 & 7.81 \\
Qwen2.5-VL-72B & 47.22 & 53.82 & 79.41 & 61.08 & 70.39 & 52.45 & 58.24 & 55.69 & 70.59 & 63.95 & 61.37 & 24.48 \\
Molmo-7B-D & 1.85 & 30.00 & 80.00 & 22.50 & 20.00 & 22.50 & 10.00 & 15.00 & 0.00 & 45.58 & 24.00 & 0.00 \\
Qwen3-VL-30B-A3B & 74.07 & 55.77 & 72.88 & 63.65 & 72.95 & 52.82 & 58.40 & 60.00 & 67.12 & 63.02 & 60.23 & 26.34 \\
Qwen3-Dense-32B & 80.56 & 63.88 & 82.00 & 66.82 & 72.00 & 58.76 & 61.80 & 63.65 & 73.47 & 66.64 & 65.27 & 25.84 \\
\midrule
\multicolumn{13}{c}{\it{\textbf{Open-Source LMMs (thinking)}}} \\
\midrule
MiMo-VL-7B-RL & 37.96 & 60.73 & 78.17 & 62.80 & 73.90 & 52.93 & 61.95 & 55.98 & 74.63 & 68.05 & 64.38 & 29.82 \\
MiMo-VL-7B-SFT & 37.96 & 51.38 & 74.62 & 60.00 & 67.62 & 50.00 & 52.88 & 53.00 & 60.62 & 64.04 & 56.04 & 23.29 \\
Qwen3-VL-30B-A3B & 80.56 & 59.24 & 80.35 & 68.20 & 69.83 & 55.70 & 55.81 & 55.06 & 76.69 & 66.57 & 62.86 & 23.56 \\
\bottomrule
\end{tabular}}
\end{table}

\begin{table}[p]
\small
\vspace{-1.2cm}
\centering
\caption{Details Evaluation results on level-2 result(Base-Score Details)}
\label{tab:level2_details_base}
\resizebox{\linewidth}{!}{
\renewcommand\arraystretch{1.1}
\begin{tabular}{l|r|cccccccccc|c}
\toprule
\multirow{1}{*}{\textbf{Model}} & \multirow{1}{*}{\textbf{\shortstack{Exec.\\Rate}}} & \multicolumn{10}{c}{\textbf{Code-Level}} & \multicolumn{1}{c}{\textbf{Chart-Level}}  \\

 &  & Color & Grid & Layout & Legend & Visual & Data & Text & Type & Base & LLM & LMM \\
\midrule
\multicolumn{13}{c}{\it{\textbf{Proprietary}}} \\
\midrule
Gemini-3-Pro & 97.23 & 52.32 & 75.31 & 86.45 & 63.33 & 81.58 & 62.75 & 77.16 & 93.86 & 70.78 & 72.21 & 33.41 \\
Claude-Sonnet-4 & 90.20 & 47.17 & 65.29 & 55.32 & 56.51 & 81.50 & 54.88 & 80.52 & 93.29 & 63.65 & 66.45 & 25.40 \\
GPT-5.2 & 96.04 & 58.44 & 80.83 & 61.51 & 58.16 & 84.65 & 64.66 & 83.77 & 94.52 & 70.93 & 75.66 & 33.03 \\
Seed1.5-VL & 60.20 & 44.39 & 69.37 & 52.06 & 52.80 & 79.85 & 52.02 & 77.21 & 92.66 & 61.67 & 65.45 & 18.30 \\
Seed1.6-VL & 70.00 & 43.61 & 69.37 & 52.25 & 51.32 & 79.70 & 53.11 & 79.46 & 92.32 & 61.77 & 65.40 & 17.43 \\
\midrule
\multicolumn{13}{c}{\it{\textbf{Open-Source LMMs (non-thinking)}}} \\
\midrule
LLaVA-OV-Qwen2-7B-SI & 2.57 & 15.35 & 59.38 & 53.85 & 30.79 & 56.34 & 22.59 & 52.70 & 55.33 & 38.43 & 37.40 & 6.40 \\
LLaVA-OV-Qwen2-7B-OV & 1.68 & 15.37 & 63.48 & 38.24 & 64.56 & 50.15 & 18.13 & 48.81 & 58.43 & 39.07 & 35.44 & 7.35 \\
DeepSeek-VL-7B & 30.46 & 15.32 & 42.58 & 34.44 & 26.24 & 52.95 & 22.36 & 51.56 & 68.43 & 35.16 & 26.87 & 4.20 \\
kimi-VL-A3B & 50.69 & 26.22 & 64.50 & 40.68 & 39.93 & 66.08 & 34.05 & 64.97 & 83.03 & 47.96 & 44.43 & 9.01 \\
Qwen2-VL-7B & 24.55 & 18.72 & 57.60 & 41.87 & 32.73 & 58.65 & 25.41 & 55.07 & 76.38 & 41.06 & 34.82 & 5.27 \\
Qwen2-VL-72B & 57.23 & 27.82 & 64.07 & 42.86 & 40.59 & 69.53 & 36.83 & 64.35 & 84.76 & 49.55 & 39.88 & 10.02 \\
InternVL-2.5-8B & 24.06 & 23.86 & 62.25 & 41.21 & 36.70 & 65.03 & 32.72 & 62.46 & 81.14 & 46.20 & 41.14 & 7.47 \\
InternVL-2.5-38B & 30.30 & 30.40 & 68.25 & 50.34 & 48.97 & 72.92 & 37.38 & 70.78 & 87.96 & 53.48 & 52.90 & 10.67 \\
InternVL-3-8B & 4.55 & 23.22 & 62.68 & 57.03 & 31.34 & 66.79 & 29.34 & 58.63 & 84.49 & 46.61 & 38.02 & 5.48 \\
InternVL-3-38B & 69.80 & 35.87 & 68.36 & 48.66 & 49.30 & 75.86 & 45.33 & 72.28 & 90.53 & 56.74 & 56.98 & 13.32 \\
GLM-4V-9B & 10.50 & 19.71 & 61.79 & 48.11 & 34.54 & 59.98 & 26.53 & 57.80 & 65.80 & 42.05 & 34.78 & 3.87 \\
GLM-4.6v-flash & 18.22 & 48.06 & 69.23 & 50.13 & 51.20 & 81.56 & 54.78 & 77.90 & 91.86 & 62.76 & 63.25 & 17.91 \\
Intern-VL-3.5-8B & 37.52 & 29.55 & 68.21 & 43.87 & 41.56 & 69.85 & 39.58 & 67.41 & 83.82 & 51.30 & 49.52 & 9.71 \\
InternVL-3.5-38B & 4.26 & 44.11 & 92.54 & 67.22 & 60.24 & 72.35 & 40.38 & 76.53 & 88.56 & 62.64 & 57.48 & 8.28 \\
MiMo-VL-7B-RL & 17.23 & 42.72 & 70.76 & 54.30 & 46.81 & 74.50 & 49.23 & 74.57 & 89.75 & 59.42 & 56.95 & 14.54 \\
MiMo-VL-7B-SFT & 30.40 & 37.36 & 67.37 & 48.40 & 47.24 & 75.98 & 46.53 & 70.80 & 88.04 & 56.56 & 53.97 & 15.23 \\
Qwen2.5-VL-7B & 31.98 & 26.87 & 67.22 & 45.41 & 38.75 & 65.34 & 37.21 & 63.68 & 83.65 & 49.22 & 44.35 & 9.44 \\
Qwen2.5-VL-72B & 74.55 & 41.32 & 67.44 & 52.16 & 50.04 & 77.74 & 48.65 & 74.72 & 91.12 & 59.31 & 49.02 & 16.69 \\
Molmo-7B-D & 0.59 & 16.53 & 44.44 & 41.67 & 50.00 & 49.70 & 29.14 & 41.57 & 54.44 & 37.32 & 29.92 & 15.50 \\
Qwen3-VL-30B-A3B & 38.91 & 40.07 & 70.02 & 51.69 & 50.06 & 76.81 & 48.01 & 75.52 & 92.39 & 59.26 & 45.77 & 17.46 \\
\midrule
\multicolumn{13}{c}{\it{\textbf{Open-Source LMMs (thinking)}}} \\
\midrule
MiMo-VL-7B-RL & 26.14 & 42.47 & 69.05 & 48.47 & 45.96 & 80.76 & 50.41 & 74.80 & 90.56 & 59.53 & 61.04 & 17.81 \\
MiMo-VL-7B-SFT & 30.50 & 40.56 & 68.22 & 48.10 & 46.65 & 78.42 & 48.68 & 73.48 & 89.86 & 58.32 & 59.33 & 16.38 \\
Qwen3-VL-30B-A3B & 13.76 & 44.00 & 65.02 & 48.92 & 49.19 & 78.19 & 50.18 & 76.18 & 89.60 & 59.55 & 63.04 & 14.37 \\
Qwen3-32B & 63.76 & 42.90 & 71.98 & 50.76 & 52.24 & 77.96 & 52.35 & 77.43 & 92.09 & 61.30 & 64.15 & 18.27 \\
\bottomrule
\end{tabular}}
\end{table}

\begin{table}[p]
\vspace{0.2cm}
\small
\centering
\caption{Details Evaluation results on level-2 result(LLM-Scores Details)}
\label{tab:level2_details_llm}
\resizebox{\linewidth}{!}{
\renewcommand\arraystretch{1.1}
\begin{tabular}{l|r|cccccccccc|c}
\toprule
\multirow{1}{*}{\textbf{Model}} & \multirow{1}{*}{\textbf{\shortstack{Exec.\\Rate}}} & \multicolumn{10}{c}{\textbf{Code-Level}} & \multicolumn{1}{c}{\textbf{Chart-Level}}  \\

 &  & Color & Grid & Layout & Legend & Visual & Data & Text & Type & Base & LLM & LMM \\
\midrule
\multicolumn{13}{c}{\it{\textbf{Proprietary}}} \\
\midrule
Gemini-3-Pro & 97.23 & 85.21 & 89.26 & 86.91 & 78.32 & 69.72 & 48.24 & 73.06 & 87.00 & 70.78 & 72.21 & 33.41 \\
Claude-Sonnet-4 &  90.20  & 79.15 & 81.38 & 81.27 & 75.23 & 61.89 & 31.96 & 68.13 & 75.81 & 63.65 & 66.45 & 25.40 \\
GPT-5.2 &  96.04  & 86.34 & 90.05 & 87.25 & 80.06 & 71.65 & 52.00 & 74.57 & 89.89 & 70.93 & 75.66 & 33.03 \\
Seed1.5-VL &  60.20  & 78.37 & 80.30 & 79.64 & 72.17 & 60.46 & 31.42 & 67.87 & 75.55 & 61.67 & 65.45 & 18.30 \\
Seed1.6-VL &  70.00  & 78.48 & 80.57 & 80.88 & 71.56 & 61.88 & 30.12 & 66.99 & 74.90 & 61.77 & 65.40 & 17.43 \\
\midrule
\multicolumn{13}{c}{\it{\textbf{Open-Source LMMs (non-thinking)}}} \\
\midrule
LLaVA-OV-Qwen2-7B-SI &  2.57  & 47.88 & 66.73 & 46.15 & 44.04 & 33.85 & 11.92 & 40.58 & 23.08 & 38.43 & 37.40 & 6.40 \\
LLaVA-OV-Qwen2-7B-OV &  1.68  & 42.94 & 60.59 & 38.82 & 68.82 & 25.88 & 10.00 & 33.24 & 21.18 & 39.07 & 35.44 & 7.35 \\
DeepSeek-VL-7B &  30.46  & 36.13 & 52.15 & 34.04 & 36.15 & 23.12 & 2.83 & 33.42 & 11.85 & 35.16 & 26.87 & 4.20 \\
kimi-VL-A3B &  50.69  & 56.89 & 70.88 & 57.94 & 52.95 & 39.54 & 13.04 & 52.48 & 32.36 & 47.96 & 44.43 & 9.01 \\
Qwen2-VL-7B &  24.55  & 46.13 & 64.17 & 44.53 & 45.61 & 29.86 & 6.30 & 42.09 & 18.50 & 41.06 & 34.82 & 5.27 \\
Qwen2-VL-72B &  57.23  & 64.54 & 71.67 & 68.41 & 56.92 & 44.60 & 15.20 & 55.77 & 43.17 & 49.55 & 39.88 & 10.02 \\
InternVL-2.5-8B &  24.06  & 55.87 & 68.90 & 53.80 & 49.46 & 35.35 & 8.97 & 50.23 & 25.70 & 46.20 & 41.14 & 7.47 \\
InternVL-2.5-38B &  30.30  & 65.96 & 75.74 & 70.42 & 65.08 & 48.55 & 13.48 & 59.56 & 50.74 & 53.48 & 52.90 & 10.67 \\
InternVL-3-8B &  4.55  & 47.98 & 65.32 & 44.68 & 51.06 & 30.74 & 6.60 & 41.70 & 37.55 & 46.61 & 38.02 & 5.48 \\
InternVL-3-38B &  69.80  & 71.81 & 76.30 & 74.05 & 66.20 & 51.89 & 19.55 & 61.08 & 57.55 & 56.74 & 56.98 & 13.32 \\
InternVL-3.5-38B & 4.26 & 68.84 & 82.79 & 71.40 & 73.95 & 51.51 & 15.35 & 68.37 & 58.37 & 62.64 & 57.48 & 8.28 \\
GLM-4V-9B &  10.50  & 49.20 & 64.43 & 42.08 & 40.80 & 29.76 & 6.46 & 41.37 & 18.07 & 42.05 & 34.78 & 3.87 \\
GLM-4.6v-flash & 18.22 & 76.55 & 81.66 & 76.66 & 71.58 & 57.31 & 27.72 & 66.74 & 70.03 & 62.76 & 63.25 & 17.91 \\
Intern-VL-3.5-8B &  37.52  & 63.75 & 74.44 & 62.32 & 60.65 & 43.82 & 16.29 & 56.16 & 40.38 & 51.30 & 49.52 & 9.71 \\
MiMo-VL-7B-RL &  17.23  & 70.29 & 79.54 & 69.17 & 60.46 & 52.39 & 26.84 & 61.67 & 52.07 & 59.42 & 56.95 & 14.54 \\
MiMo-VL-7B-SFT &  30.40  & 69.11 & 74.66 & 66.56 & 61.03 & 49.00 & 22.66 & 58.90 & 49.59 & 56.56 & 53.97 & 15.23 \\
Qwen2.5-VL-7B &  31.98  & 55.09 & 70.80 & 57.89 & 54.77 & 39.49 & 12.23 & 50.76 & 35.12 & 49.22 & 44.35 & 9.44 \\
Qwen2.5-VL-72B &  74.55  & 75.34 & 76.72 & 79.06 & 68.19 & 56.79 & 26.03 & 65.49 & 66.16 & 59.31 & 49.02 & 16.69 \\
Molmo-7B-D &  0.59  & 38.33 & 55.00 & 40.00 & 60.00 & 25.00 & 0.83 & 30.83 & 10.00 & 37.32 & 29.92 & 15.50 \\
Qwen3-VL-30B-A3B &  38.91  & 74.04 & 80.26 & 78.01 & 66.96 & 56.33 & 25.10 & 64.51 & 59.31 & 59.26 & 45.77 & 17.46 \\
Qwen3-VL-32B & 63.76 & 75.96 & 82.61 & 78.22 & 70.49 & 58.98 & 30.53 & 67.80 & 70.47 & 61.30 & 64.15 & 18.27 \\
\midrule
\multicolumn{13}{c}{\it{\textbf{Open-Source LMMs (thinking)}}} \\
\midrule
MiMo-VL-7B-RL &  26.14  & 76.98 & 78.08 & 75.50 & 66.06 & 57.53 & 25.57 & 63.68 & 66.81 & 59.53 & 61.04 & 17.81 \\
MiMo-VL-7B-SFT &  30.50  & 75.96 & 78.18 & 74.14 & 65.65 & 54.04 & 23.84 & 61.16 & 62.49 & 58.32 & 59.33 & 16.38 \\
Qwen3-VL-30B-A3B &  13.76  & 74.93 & 76.80 & 78.17 & 72.12 & 57.68 & 26.55 & 66.14 & 76.51 & 59.55 & 63.04 & 14.37 \\
\bottomrule
\end{tabular}}
\end{table}

\begin{table}[p]
\small
\centering
\caption{Details Evaluation results on level-3 details result(Base-Score Details)}
\label{tab:level3_base}
\resizebox{\linewidth}{!}{
\renewcommand\arraystretch{1.1}
\begin{tabular}{l|r|cccccccccc|c}
\toprule
\multirow{1}{*}{\textbf{Model}} & \multirow{1}{*}{\textbf{\shortstack{Exec.\\Rate}}} & \multicolumn{10}{c}{\textbf{Code-Level}} & \multicolumn{1}{c}{\textbf{Chart-Level}}  \\

 &  & Color & Grid & Layout & Legend & Visual & Data & Text & Type & Base & LLM & LMM \\
\midrule
\multicolumn{12}{c}{\it{\textbf{Proprietary}}} \\
\midrule
Gemini-3-Pro & 30.03 & 56.29 & 83.71 & 92.55 & 68.39 & 84.00 & 62.26 & 80.45 & 96.60 & 74.28 & 77.90 & 35.97 \\
Claude-Sonnet-4 & 46.65 & 35.43 & 68.97 & 84.08 & 61.77 & 74.44 & 46.73 & 67.77 & 89.95 & 61.13 & 62.04 & 16.60 \\
GPT-5.2 & 20.13 & 34.46 & 82.14 & 80.87 & 52.64 & 78.34 & 43.56 & 76.09 & 93.81 & 61.99 & 68.40 & 16.29 \\
Seed1.5-VL & 10.86 & 44.55 & 76.89 & 93.94 & 69.30 & 73.52 & 51.75 & 73.57 & 95.96 & 67.58 & 75.59 & 22.85 \\
Seed1.6-VL & 26.52 & 44.26 & 82.28 & 95.48 & 74.73 & 75.42 & 52.62 & 77.82 & 86.47 & 68.60 & 76.35 & 25.77 \\
\bottomrule
\end{tabular}}
\end{table}

\begin{table}[p]
\vspace{0.2cm}
\small
\centering
\caption{Details Evaluation results on level-3 detail result(LLM-Score Details)}
\label{tab:level3_llm}
\resizebox{\linewidth}{!}{
\renewcommand\arraystretch{1.1}
\begin{tabular}{l|r|cccccccccc|c}
\toprule
\multirow{1}{*}{\textbf{Model}} & \multirow{1}{*}{\textbf{\shortstack{Exec.\\Rate}}} & \multicolumn{10}{c}{\textbf{Code-Level}} & \multicolumn{1}{c}{\textbf{Chart-Level}}  \\

 &  & Color & Grid & Layout & Legend & Visual & Data & Text & Type & Base & LLM & LMM \\
\midrule
\multicolumn{12}{c}{\it{\textbf{Proprietary}}} \\
\midrule
Gemini-3-Pro & 30.03 & 82.91 & 91.70 & 83.33 & 86.65 & 78.69 & 54.68 & 74.70 & 88.72 & 74.28 & 77.90 & 35.97 \\
Claude-Sonnet-4 & 46.65 & 65.68 & 78.97 & 71.95 & 78.97 & 62.71 & 32.16 & 57.78 & 74.32 & 61.13 & 62.04 & 16.60 \\
GPT-5.2 & 20.13 & 70.00 & 84.60 & 78.10 & 81.35 & 68.10 & 42.51 & 69.10 & 77.78 & 61.99 & 68.40 & 16.29 \\
Seed1.5-VL & 10.86 & 77.65 & 87.06 & 81.74 & 90.88 & 72.94 & 57.06 & 69.79 & 84.12 & 67.58 & 75.59 & 22.85 \\
Seed1.6-VL & 26.52 & 78.77 & 88.07 & 86.45 & 88.55 & 75.43 & 52.77 & 77.12 & 84.82 & 68.60 & 76.35 & 25.77 \\

\bottomrule
\end{tabular}}
\end{table}

\begin{figure}[htbp]
    \centering
    \includegraphics[width=.95\linewidth]{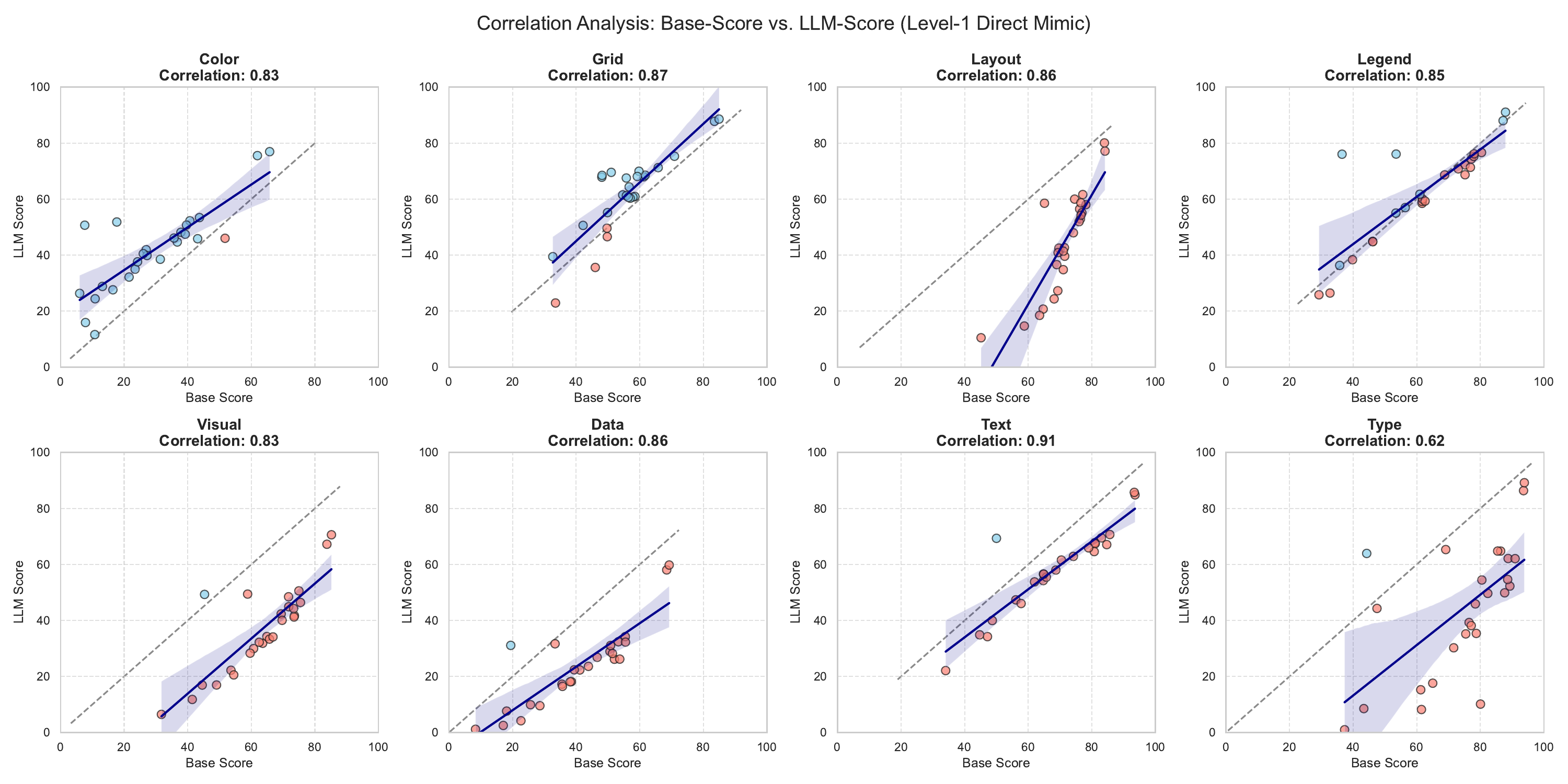}
    \caption{Correlation Analysis on Level-1 DR: Base-Score vs. LLM-Score (Code-Level Metrics).}
    \label{fig:DR_Base-Score vs. LLM-Score}
\end{figure}

\begin{figure}[htbp]
    \centering
    \includegraphics[width=.95\linewidth]{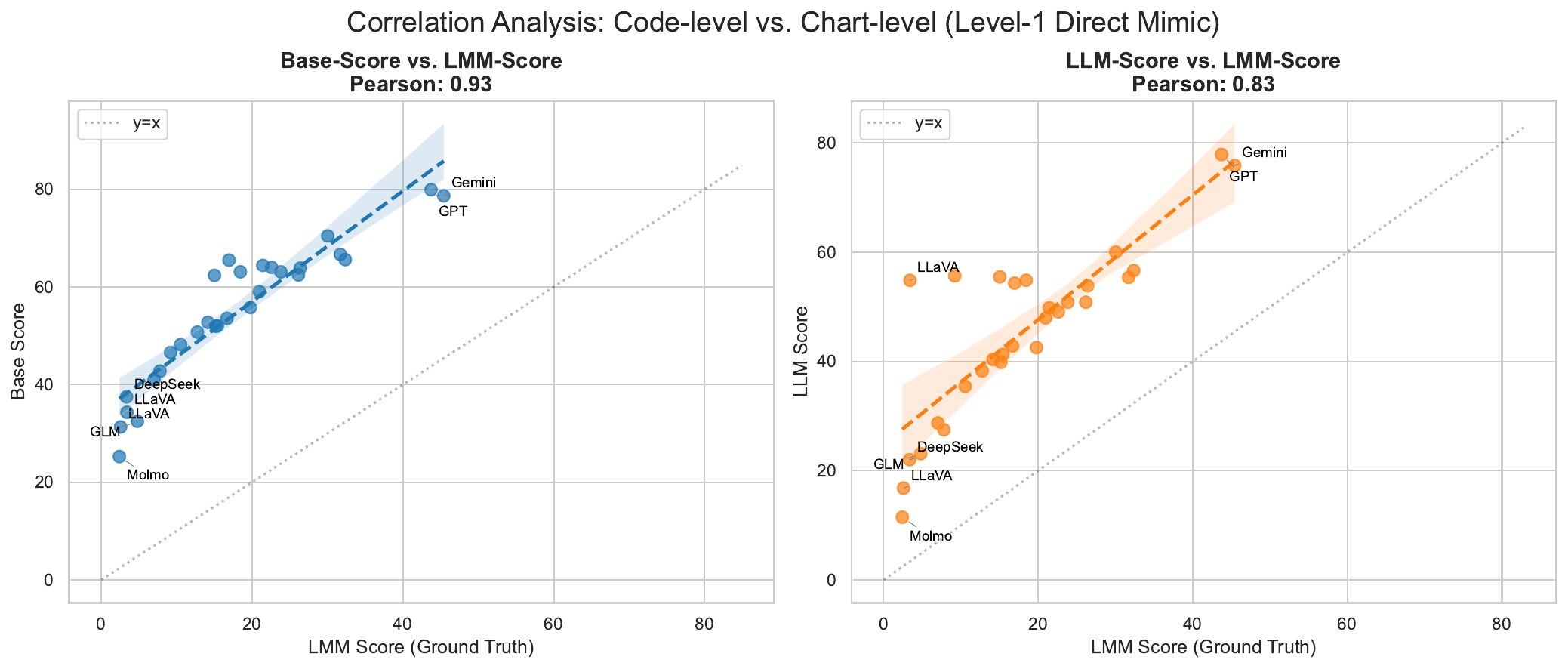}
    \caption{Correlation between Code-level and Chart-level on Level-1 DR}
    \label{fig:DR_code-level vs. chart-level}
\end{figure}

\paragraph{Discrepancy Between base-Score and LLM-Score.}
\label{Base vs LLM}
Since the Chart-Level LMM-Score serves as the closest  proxy to human evaluation (reflecting the actual rendered reality), we re-examine the results in Tab.~\ref{tab:level1_customize} and Tab.~\ref{tab:level1_customize_LLM}. 
While both the rule-based Base-Score and the semantic-aware LLM-Score at the Code-Level demonstrate a degree of correlation with visual similarity—effectively capturing the model's \textit{generative intent}—neither metric successfully approximates the lower, more realistic values of the LMM-Score. 
For instance, Gemini-3-pro achieves impressive Code-Level scores (Base: 94.44\% Type; LLM: 73.75\% Type), yet its Chart-Level LMM score remains significantly lower at 40.72\%. 
This consistent disparity highlights a critical \textbf{semantic modality gap} between the textual code domain and the visual chart domain. 
Code-Level metrics primarily evaluate the syntactic correctness and logical structure of the generation script (i.e., the instructions), but they fail to account for execution-time artifacts, such as occlusion, poor scaling, or color clashing, which only manifest during the rendering process. ~Fig. \ref{fig:DR_Base-Score vs. LLM-Score}  illustrates the similarity between Base-Score and LLM-Score across different evaluation dimensions. Both metrics adopt the same weighting scheme, where the data-parameter and color dimensions are assigned higher weights of 0.2, while all other dimensions are weighted at 0.1.
Consequently, high Code-Level scores imply that models are proficient at writing valid instructions ("code intent"), but the substantial drop in Chart-Level scores reveals that accurate syntax does not guarantee high-fidelity visual alignment ("visual reality").  However, as shown in Fig. \ref{fig:DR_code-level vs. chart-level}, the code-level metrics can still, to some extent, reflect the overall trend of visual fidelity.


\begin{figure}[htbp]
    \centering
    \includegraphics[width=.95\linewidth]{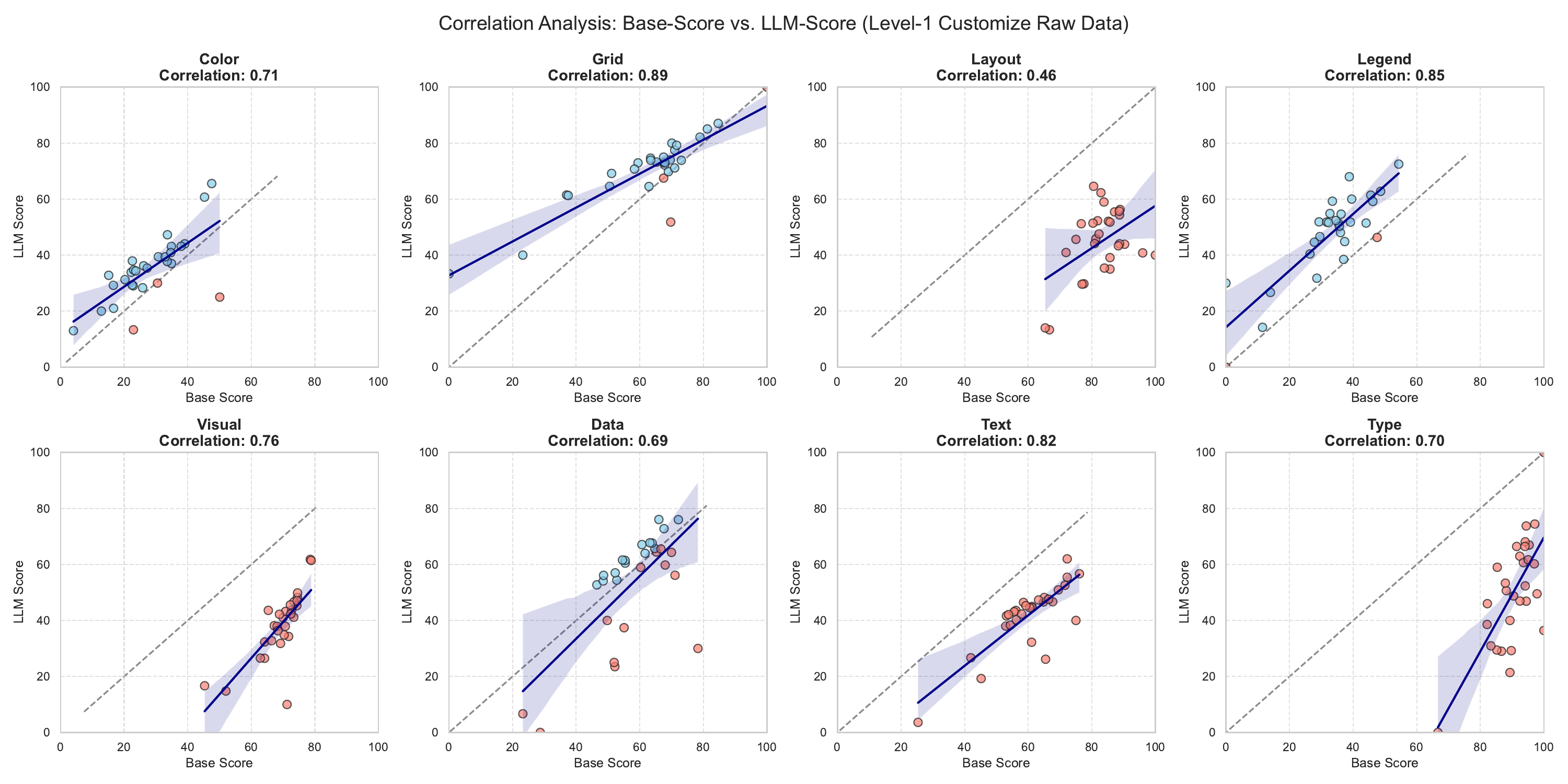}
    \caption{Correlation Analysis on Level-1 CRD: Base-Score vs. LLM-Score (Code-Level Metrics).}
    \label{fig:CRD_Base-Score vs. LLM-Score}
\end{figure}
\begin{figure}[htbp]
    \centering
    \includegraphics[width=.95\linewidth]{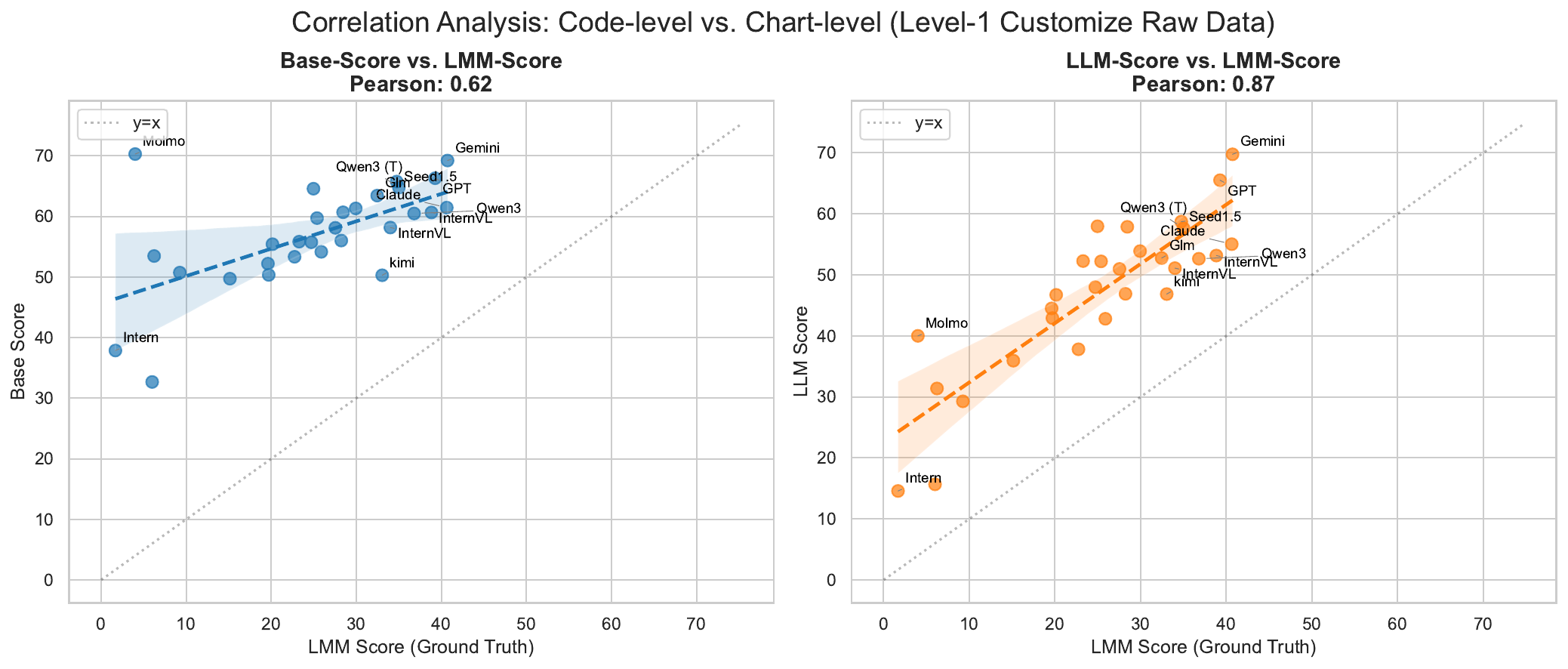}
    \caption{Correlation between Code-level and Chart-level on Level-1 CRD}
    \label{fig:CRD_code-level vs. chart-level}
\end{figure}

\begin{figure}[htbp]
    \centering
    \includegraphics[width=.95\linewidth]{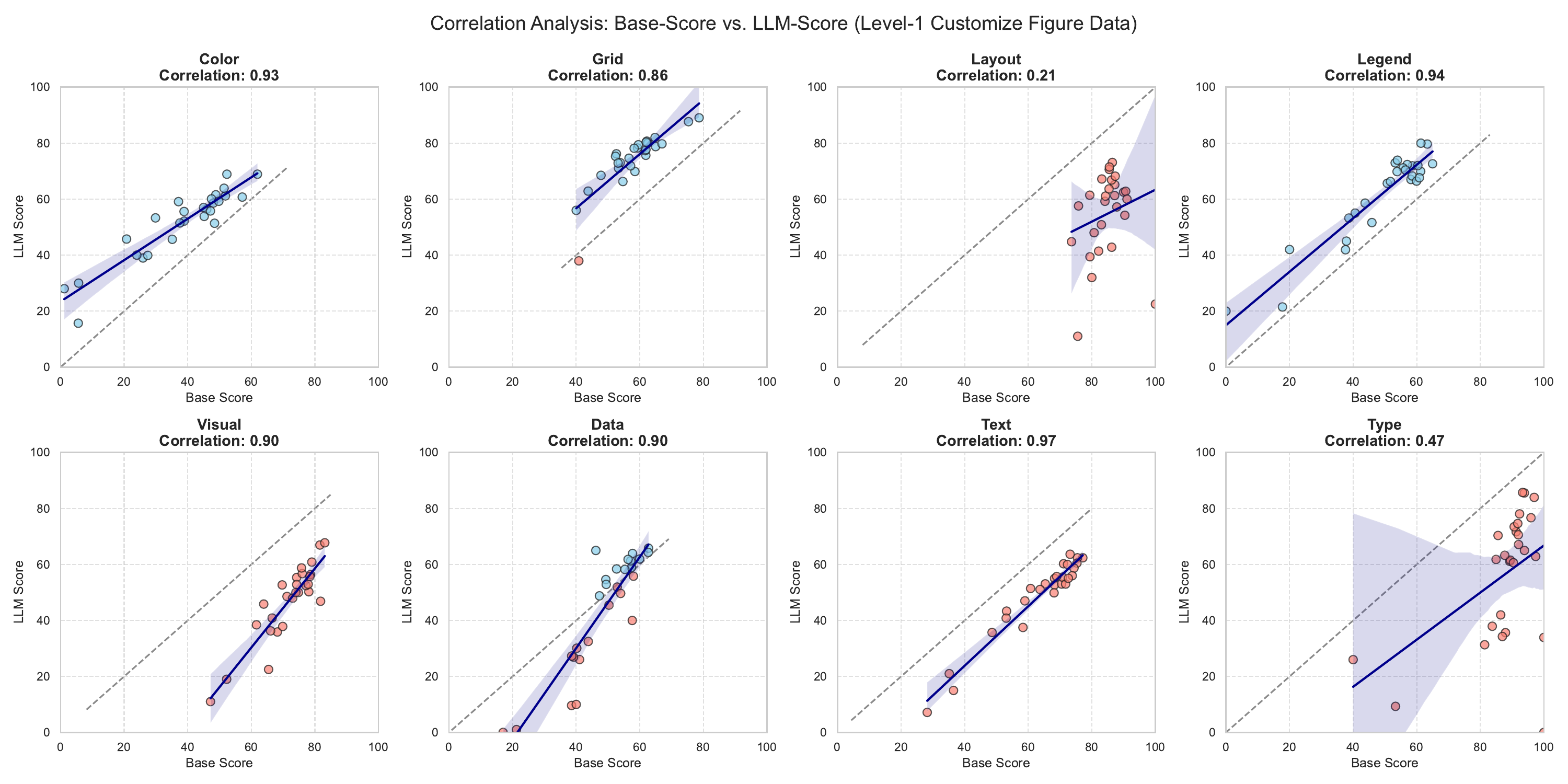}
    \caption{Correlation Analysis on Level-1 CFD: Base-Score vs. LLM-Score (Code-Level Metrics).}
    \label{fig:CFD_Base-Score vs. LLM-Score}
\end{figure}
\begin{figure}[htbp]
    \centering
    \includegraphics[width=.95\linewidth]{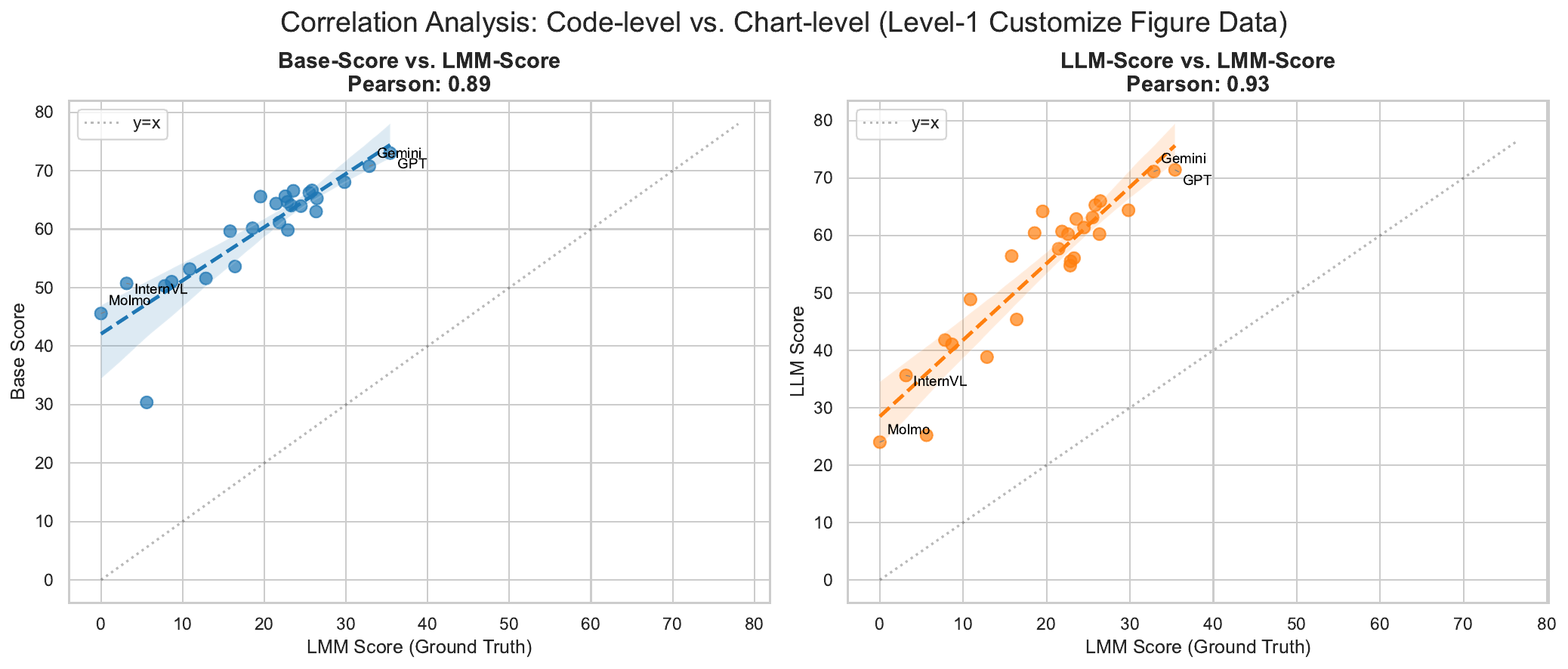}
    \caption{Correlation between Code-level and Chart-level on Level-1 CFD}
    \label{fig:CFD_code-level vs. chart-level}
\end{figure}




\paragraph{Discrepancy Between LLM-Score and LMM-Score.}
\label{LLM vs LMM}
Fig.\ref{fig:llmlmmscoreheatmap} illustrates model performance across ten representative task cases, evaluated by both LLM-score for code quality (left) and LMM-score for rendered chart fidelity (right). 
A clear discrepancy emerges: proprietary models such as GPT-5.2, Gemini-3-Pro, and Claude-Sonnet-4 achieve consistently high LLM-scores across most tasks (often $\geq$0.2), indicating strong code-level compliance. 
However, their corresponding LMM-scores are much lower (typically $\leq$0.35), suggests that code-level evaluation by LLMs lacks precision and tends to yield inflated scores, whereas chart-level assessments align more closely with human visual perception. This contrast highlights that although current models achieve high scores at the code level, they face fundamental limitations in attaining pixel-level chart fidelity, particularly in diverse and challenging task scenarios.
Open-source models, in contrast, underperform on both metrics, with particularly low LMM-scores across all tasks. 
This reveals, on one hand, that relying solely on code-level evaluation leads to a severe text-visual informational discrepancy that fails to accurately reflect actual performance; on the other hand, it underscores a significant performance gap between open-source and proprietary models in actual chart visual fidelity.

\subsection{Correlation Between Code Generation Quality and Visual Fidelity}
To validate the effectiveness of our evaluation framework, we analyzed the correlation between Code-Level Metrics (assessed via both Base-Score and LLM-Score) and the final Visual Fidelity of the reproduced charts, using the LMM-Score as the ground truth benchmark.

\textbf{Positive Correlation Between Code Quality and Visual fidelity.} A cross-comparison of the Base-Score (Tab.~\ref{tab:level2_details_base}) and LLM-Score (Tab.~\ref{tab:level2_details_llm}) against the LMM-Score reveals a strong positive correlation, confirming that code-level metrics serve as reliable proxies for visual fidelity. 

Models that excelled in generating structurally and semantically correct code consistently achieved higher visual fidelity scores. For instance, proprietary models such as Gemini-3-pro and GPT-5.2 dominated the code-level evaluations, achieving high scores in critical dimensions like Grid and Text across both scoring methods (e.g., Gemini-3-pro: Base-Score Text 93.59, LLM-Score Text 84.86). This code proficiency directly translated into the highest LMM-Scores in the benchmark (45.42 and 43.73, respectively). 

Conversely, models with weaker code generation capabilities, such as the LLaVA series, exhibited low scores across all code dimensions and correspondingly failed to produce visually coherent charts (LMM-Scores < 3.0). This trend substantiates the hypothesis that high-quality code generation is a prerequisite for, and a strong predictor of, high visual fidelity.

\textbf{Sensitivity of LLM-Score in Reflecting Visual Nuances.} While both code-level metrics correlate with visual outcomes, the LLM-Score demonstrates higher sensitivity to visual defects than the rule-based Base-Score, offering a more accurate reflection of the "visual logic" embedded in the code.

Correction of Score Inflation: The Base-Score often awards high marks based on rigid keyword matching, potentially masking visual alignment issues. For example, Claude-Sonnet-4 achieved a high Base-Score of 85.66 in the Text dimension. However, its LLM-Score for Text dropped to 70.69, aligning more closely with its moderate LMM-Score of 32.36. This suggests that the LLM-based evaluator effectively penalizes code that is syntactically correct but visually suboptimal (e.g., resulting in overlapping text or misaligned labels), thereby reducing the "false positive" rate of rule-based evaluation.

Dimensional Impact: The drop in LLM-Scores for dimensions like Layout and Grid in mid-tier models (e.g., Seed1.5-VL dropping from a Base Layout of 76.16 to an LLM Layout of 56.46) mirrors the steep decline seen in their LMM-Scores. This indicates that the LLM-Score captures the structural integrity of the chart more effectively, serving as a tighter upper bound for potential visual performance.

\textbf{The "Thinking" Mechanism and Structural Fidelity.} The data from "Thinking" models further reinforces the link between code logic and visual fidelity. The introduction of reasoning capabilities (e.g., in MiMo-VL-7B-RL) resulted in significant improvements in structural code dimensions such as Grid (Base-Score improving from 61.70 in the non-thinking variant to 74.48). This enhancement in code structure directly contributed to a notable increase in the LMM-Score (rising from 21.43 to 30.04). This finding suggests that improving a model's ability to reason about code logic—specifically in defining coordinate systems and layouts—is a critical pathway to bridging the gap between code generation and pixel-perfect visual reproduction.

Conclusion In summary, our experiments demonstrate that code-level metrics are not merely syntactic checks but act as significant indicators of visual fidelity. While the code-visual gap remains, particularly for open-source models, the alignment between robust code scores (especially LLM-Scores) and high LMM-Scores confirms that optimizing for code-level accuracy is essential for achieving high-fidelity chart reproduction.

\subsection{Evaluation Case}
We present a set of representative case studies generated by Claude Sonnet 4 on the DR task to provide a fine-grained analysis of our multi-dimensional evaluation framework, as shown in Fig. 11–Fig. 16 and Tab. 17–Tab. 19. These cases are carefully selected to illustrate how different evaluation paradigms assess the same chart instance from complementary perspectives.

Specifically, our framework consists of three distinct evaluation protocols: (1) a rule-based method (\texttt{Base}) that quantifies structural and parameter-level consistency directly from code, (2) an LLM-based evaluator (\texttt{LLM}) that performs semantic-level assessment over the generated code, and (3) an LMM-based evaluator (\texttt{LMM}) that measures visual similarity directly at the rendered chart level. For each case, we report not only the final scores from these three evaluators but also detailed sub-dimension scores and corresponding explanations, providing a transparent view of how discrepancies arise.

As illustrated in the example in Fig. 11–12 and Tab. 17, although \texttt{Base} and \texttt{LLM} exhibit relatively consistent scoring trends—indicating strong agreement at the code and structural levels—they both show noticeable divergence from \texttt{LMM}, which evaluates pixel-level visual fidelity. This discrepancy highlights that code-level correctness, even when semantically validated by LLMs, does not necessarily translate into high visual similarity.

Across multiple cases, we consistently observe that the three evaluation methods exhibit a certain degree of correlation, particularly in capturing coarse-grained model ranking. However, the differences in their scoring behaviors reveal complementary strengths: rule-based and LLM-based methods are more sensitive to structural and logical correctness, while LMM-based evaluation better captures perceptual and rendering-level discrepancies such as layout shifts, axis misalignment, and visual styling differences.

Overall, these case studies demonstrate that our multi-dimensional evaluation framework provides a more comprehensive and interpretable assessment by jointly considering code-level correctness and chart-level visual fidelity. This further suggests that relying solely on rule-based or code-based evaluation may lead to incomplete conclusions, whereas incorporating visual-level assessment is crucial for faithfully measuring real-world chart generation quality.

\begin{evaluationcase}
    \centering
     \appendixplatetitle{Evaluation Cases}
    \vspace{1.5mm}
    \begin{minipage}[t]{0.47\linewidth}
        \centering
        \includegraphics[width=\linewidth,height=0.30\textheight,keepaspectratio]{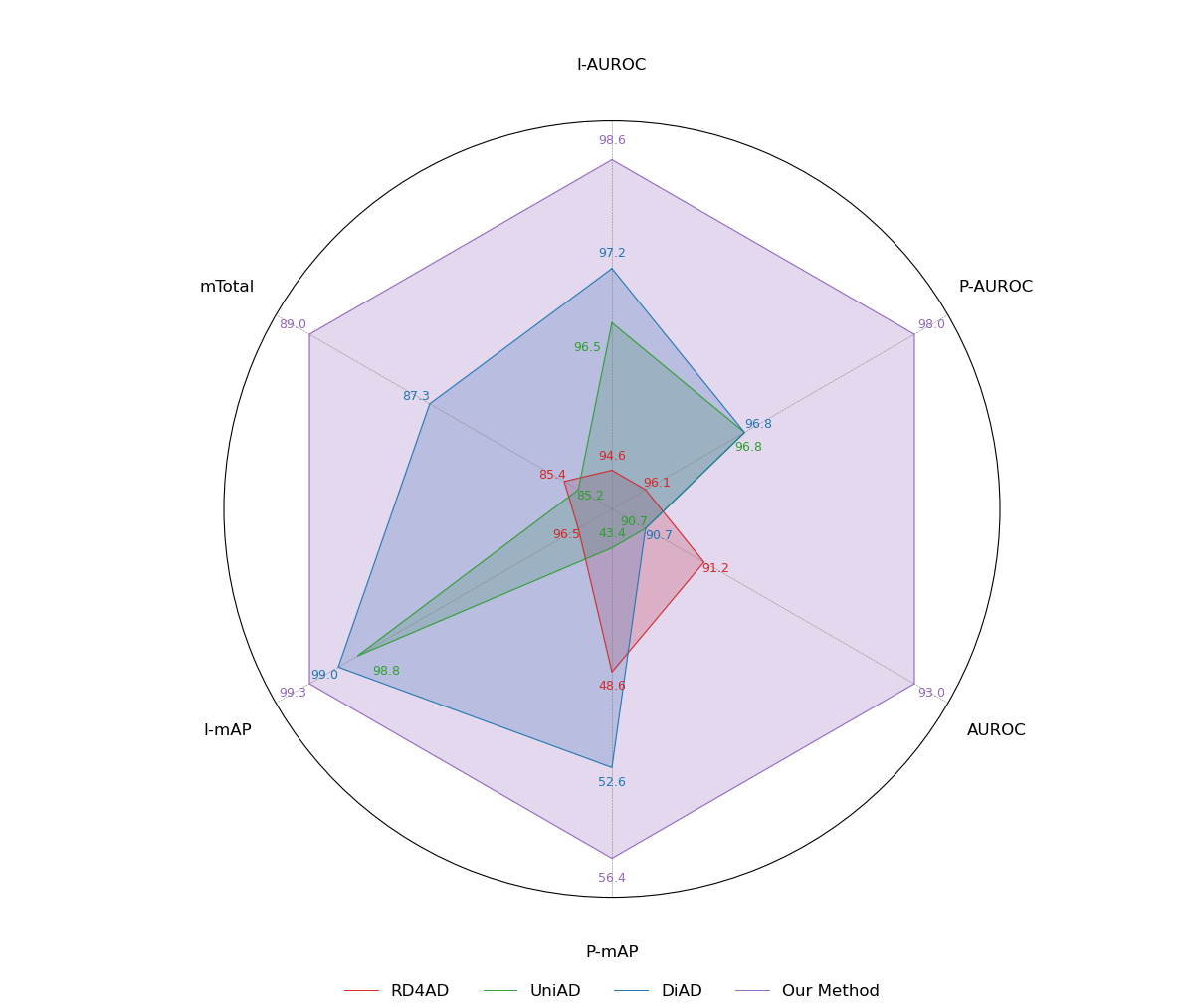}
        \caption{\textbf{GT\_Figure (Left).}} 
        \label{fig:GT_radar36}
    \end{minipage}
    \hfill 
    \begin{minipage}[t]{0.47\linewidth}
        \centering
        \includegraphics[width=\linewidth,height=0.30\textheight,keepaspectratio]{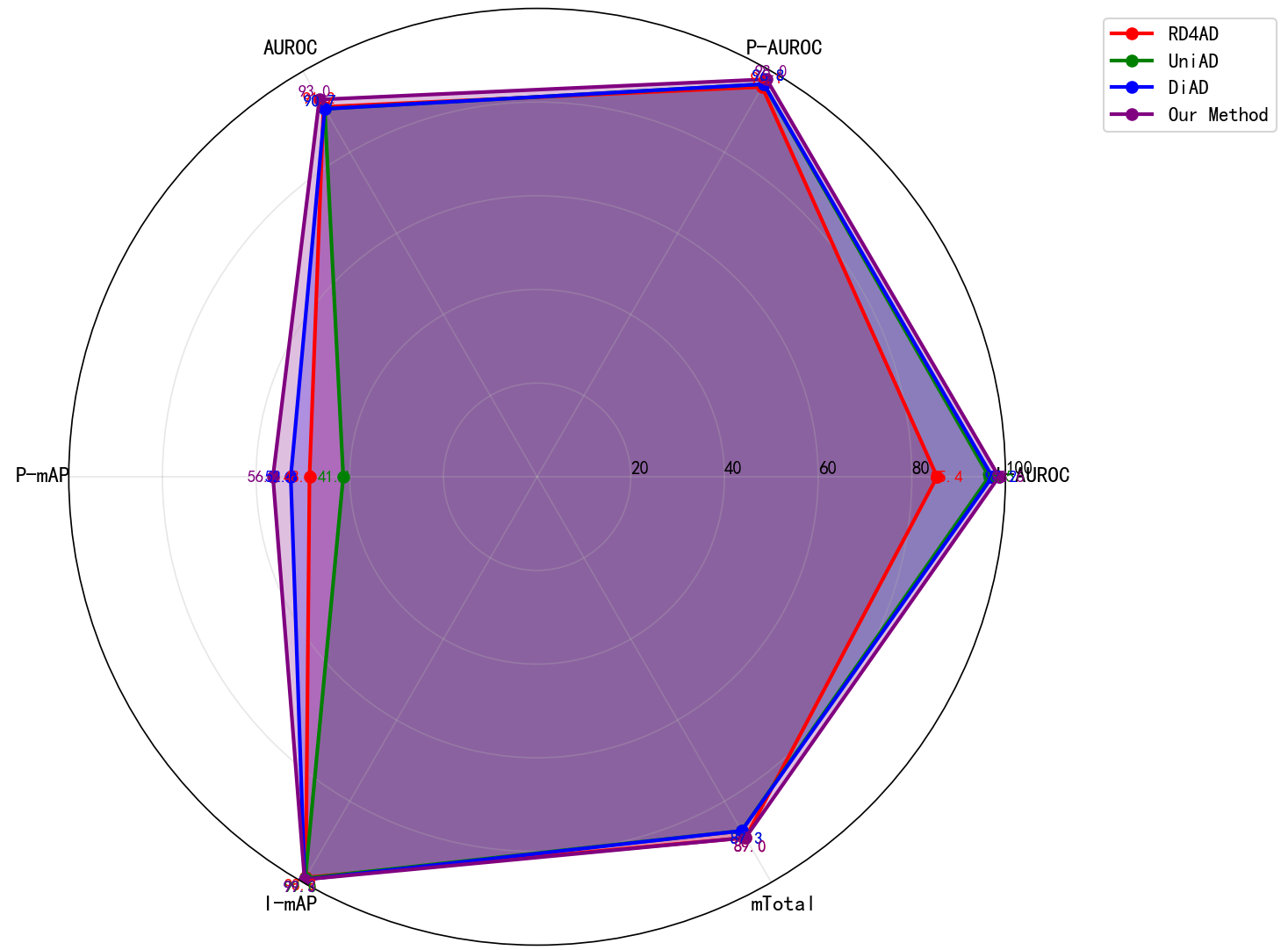} 
        \caption{\textbf{Generation\_Figure (Right).}} 
        \label{fig:Gen_radar36}
    \end{minipage}

    \vspace{6mm} 

    \captionsetup{type=table}
    \caption{\textbf{Detailed multi-dimensional evaluation statistics.} By pairing related metrics, we observe comprehensive performance across API, Base, and standard Evaluation reports.} 
    \label{tab:datastattable_radar36}
    \centering
    \setlength{\tabcolsep}{2.8pt}\renewcommand{\arraystretch}{1.00}
    
    {\small
    \begin{tabularx}{0.98\linewidth}{>{\bfseries}p{2.15cm} >{\bfseries}p{2.15cm} X}
        \toprule
        Metric 1 (Score) & Metric 2 (Score) & Combined Reason / Key Insight \\
        \midrule
        
        \multicolumn{3}{l}{\cellcolor{gray!15}\textit{\textbf{LLM Report (Overall Score: 38.50)}}} \\
        \midrule
        Chart Type (0.0) & Data Sim (0.0) & GT is normalized radial radar (radius [0.1,0.9]); GEN plots raw absolute values (0-100). Fundamentally different data mapping. \\
        Visual Params (30.0) & Color (80.0) & Similar basic colors used, but GEN uses thicker lines, different markers ('o'), and varying radius scaling. \\
        Layout (50.0) & Grid (40.0) & Figure size, theta offset, and limits differ. GT uses subtle dashed grey grid without y-ticks; GEN shows visible default radial ticks. \\
        Legend (60.0) & Text Content (45.0)& Legend locations differ (bottom vs. upper-right). Numeric annotation formats, offsets, and tick labels differ significantly. \\
        
        \midrule
        \multicolumn{3}{l}{\cellcolor{gray!15}\textit{\textbf{Base Report (Overall Weighted Score: 45.84)}}} \\
        \midrule
        Type F1 (1.00) & Layout F1 (1.00) & Perfect match in basic layout container and chart type detection. \\
        Legend F1 (0.78) & Text F1 (0.78) & High similarity in legend content and overall text OCR results. \\
        Visual Param (0.74) & Data Param (0.14) & Visual parameters closely match, but substantial errors exist in data extraction and mapping. \\
        Color F1 (0.00) & Grid F1 (0.00) & Complete mismatch in color precision and grid configuration. \\

        \midrule
        \multicolumn{3}{l}{\cellcolor{gray!15}\textit{\textbf{LMM Report (Final Similarity Score: 18.0)}}} \\
        \midrule
        Chart Type (100) & Color Style (80) & Both use radar charts. Colors are semantically consistent (red/green/blue/purple) with only minor hue/opacity differences. \\
        Data (25) & Layout (30) & \textbf{Major Errors:} Axis alignment/mapping differs drastically; numeric labels do not match GT. Legend moved, title changed, and axis orientation rotated. \\
        \midrule
        
        \multicolumn{3}{p{0.94\linewidth}}{\small \textbf{Summary:} Wharts share the same radar chart type and overall high/low ordering by series, but there are multiple clear visual mismatches: major data-axis alignment/annotation differences and substantial layout/annotation shifts (legend, ticks, orientation), plus minor color/style and numeric-label placement differences.} \\
        \bottomrule
        \label{tab:radar36_results}
    \end{tabularx}
    }
\end{evaluationcase}

\begin{evaluationcase}
    \centering
    \appendixplatetitle{Evaluation Cases}
    \vspace{1.5mm}
    
    \begin{minipage}[t]{0.47\linewidth}
        \centering
        \includegraphics[width=\linewidth,height=0.30\textheight,keepaspectratio]{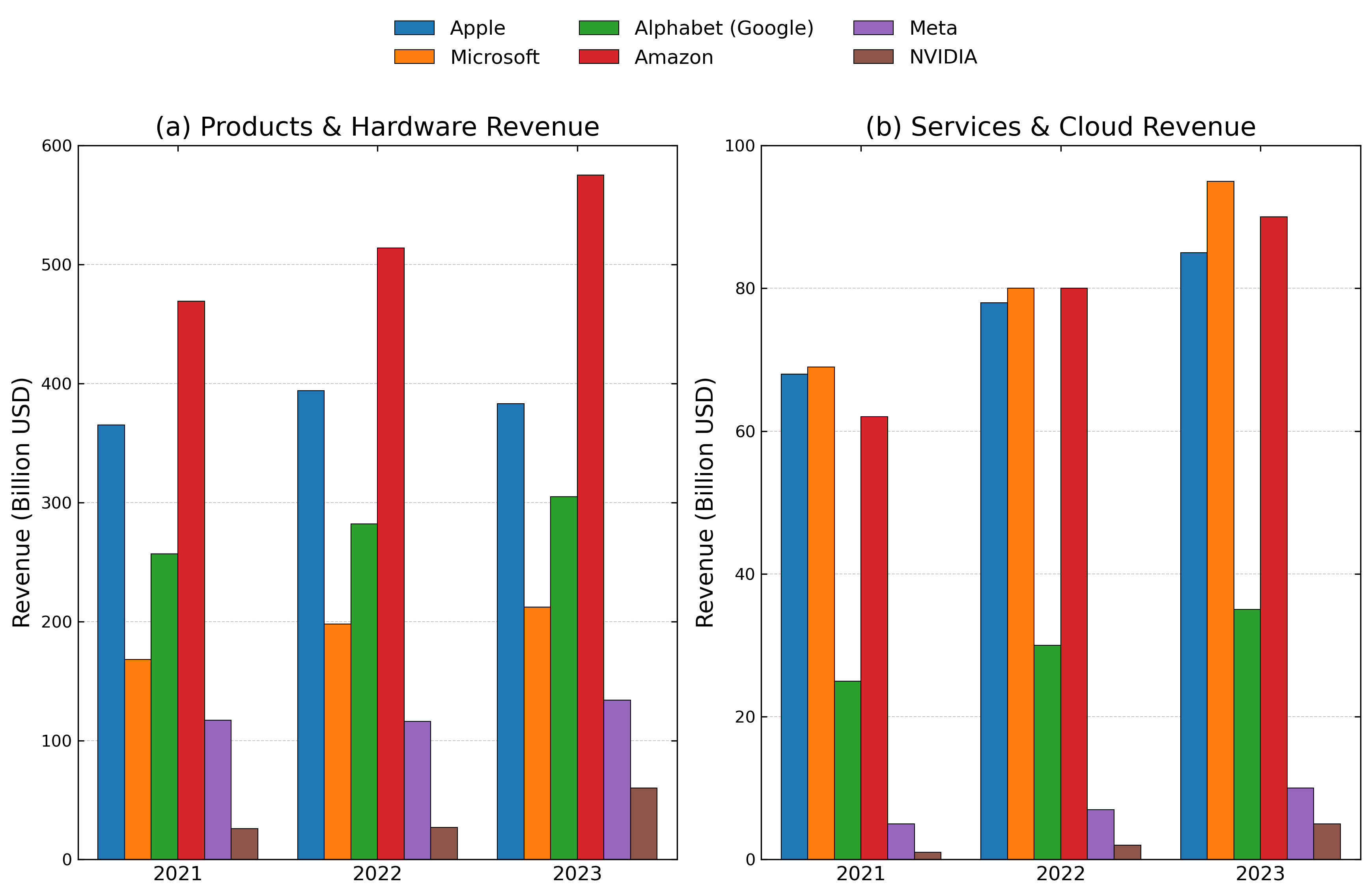}
        \caption{\textbf{GT\_Figure (Left).}} 
        \label{fig:GT_bar62}
    \end{minipage}
    \hfill 
    \begin{minipage}[t]{0.47\linewidth}
        \centering
        \includegraphics[width=\linewidth,height=0.30\textheight,keepaspectratio]{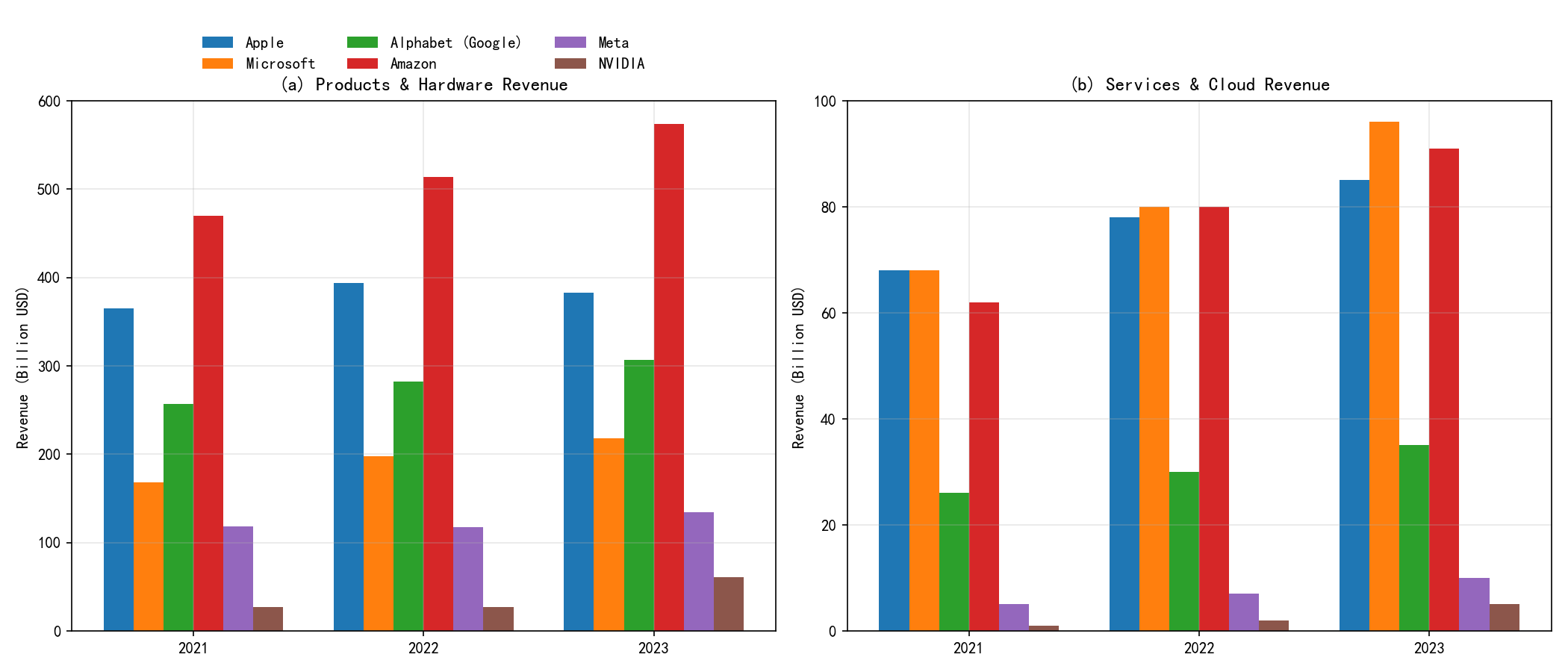} 
        \caption{\textbf{Generation\_Figure (Right).}} 
        \label{fig:Gen_bar62}
    \end{minipage}

    \vspace{6mm} 

    \captionsetup{type=table}
    \caption{\textbf{Detailed multi-dimensional evaluation statistics.} By pairing related metrics, we observe comprehensive performance across API, Base, and standard Evaluation reports.} 
    \centering
    \setlength{\tabcolsep}{2.8pt}\renewcommand{\arraystretch}{1.00}
    
    {\small
    \begin{tabularx}{0.98\linewidth}{>{\bfseries}p{2.15cm} >{\bfseries}p{2.15cm} X}
        \toprule
        Metric 1 (Score) & Metric 2 (Score) & Combined Reason / Key Insight \\
        \midrule
        
        \multicolumn{3}{l}{\cellcolor{gray!15}\textit{\textbf{LLM Report (Overall Score: 85.00)}}} \\
        \midrule
        Chart Type (100.0) & Data Sim (65.0) & Both render side-by-side grouped vertical bar charts. Most values match, but several numeric differences exist (e.g., Microsoft 212 vs 218), producing perceptible height differences. \\
        Visual Params (75.0) & Color (100.0) & Same color arrays used. Exact bar and figure sizes differ (13x8 vs 14x6), which slightly changes spacing and relative bar thickness. \\
        Layout (80.0) & Grid (85.0) & Both have 1x2 subplots with horizontal y-axis grids. GT's grid is dashed and more prominent; placement methods for legends cause slight margin differences. \\
        Legend (85.0) & Text Content (95.0)& Legends are centered above plots with matching entries. Titles and labels match exactly; GT explicitly sets larger font sizes. \\
        
        \midrule
        \multicolumn{3}{l}{\cellcolor{gray!15}\textit{\textbf{Base Report (Overall Weighted Score: 77.67)}}} \\
        \midrule
        Type F1 (1.00) & Layout F1 (1.00) & Perfect structural match in basic layout container and chart type detection. \\
        Data Param (0.99) & Color F1 (0.96) & Exceptional performance in extracting data values and precise color matching. \\
        Text F1 (0.88) & Visual Param (0.85) & Strong alignment in text OCR results and visual parameter rendering. \\
        Grid F1 (0.12) & Legend F1 (0.00) & Model struggles significantly with extracting legend configurations and grid properties. \\

        \midrule
        \multicolumn{3}{l}{\cellcolor{gray!15}\textit{\textbf{LMM Report (Final Similarity Score: 85.0)}}} \\
        \midrule
        Chart Type (100) & Color Style (80) & Both are grouped vertical bar charts. Color hues are consistent; minor style differences exist in stroke thickness and bar edges. \\
        Data (95) & Layout (80) & Heights, ordering, and trends match with negligible pixel differences. Minor shifts in font sizes, legend marker spacing, and subplot margins. \\
        \midrule
        
        \multicolumn{3}{p{0.94\linewidth}}{\small \textbf{Summary:} Charts are visually very similar with matching chart types, data trends, axes, and layout; only minor differences in styling (font weight/size, slight legend marker spacing) and extremely small numeric/text rendering variations are present.} \\
        \bottomrule
        \label{tab:bar62_results}
    \end{tabularx}
    }
\end{evaluationcase}

is a direct indicator of multimodal interaction [16, 54, 83],
it can reveal whether a text token successfully grounds in
a relevant visual region [82, 93]. Therefore, by visualizing
cross-attention maps for object-related words of Qwen2.5-
VL [2] in Fig. 2, we aim to uncover alignment patterns and
weaknesses not apparent from standard accuracy metrics.
Interestingly, our cross-attention analysis reveals a systematic discrepancy: Attribute words [27, 36, 56] produce
sharper and more localized activations in the visual modality, while object nouns [52] result in diffuse and scattered
attention patterns. We attribute this discrepancy to biases
in semantic reference. In large-scale multimodal corpora,
attribute words, such as colors or textures, correspond to
specific and spatially localized visual patterns, while object
nouns occur in diverse contexts, diluting their spatial association. Moreover, attribute words naturally map to low-level
visual features, while object nouns rely on high-level semantic representations often varied across instances, which
leads to poor alignment without explicit supervision.
This finding suggests that improving fine-grained object understanding requires explicitly strengthening crossmodal correspondence for object nouns, and thus inspires
us to pursue direct supervision between object words and
their associated visual regions. To provide such supervision signals, an enriched video understanding dataset that
pairs object-level visual annotations with natural language
prompts containing clear references is required. Thanks to
earlier visual-prompt-based approaches [50, 89], collecting
training data with mask annotations is not difficult. We start
from VideoRefer [89], a video fine-grained object understanding dataset providing frame-level object masks aligned
with textual prompts via placeholder tokens used for visual
prompting. While these masks are valuable, the associated
text does not contain clear natural language references to the
objects. We therefore design a GPT-4o-powered [49] data
refinement pipeline and construct the NL-Refer dataset.
Specifically, for each placeholder, we automatically replace
it with a concise natural language description of the specific
instance, informed by the context.
Based on the NL-Refer dataset, we propose SWIM
(See What I Mean), a simple yet effective training strategy
that explicitly aligns vision and language representations
to strengthen fine-grained object understanding in MLLMs.
Specifically, during supervised fine-tuning, SWIM extracts
cross-attention maps for object nouns from multiple intermediate layers and aligns them with ground-truth object
masks, enforcing spatial consistency between textual identity and visual grounding. By providing explicit alignment
signal throughout training, SWIM guides the model to preserve and utilize fine-grained object-level information, enabling more precise visual localization from purely textual
prompts at inference. Extensive experiments across finegrained object understanding benchmarks demonstrate that
SWIM enhances text–visual alignment and outperforms
visual-prompt-dependent approaches.
We summarize our contributions as follows.
• We point out design limitations in existing fine-grained
object understanding models and the insufficient vision–language alignment in general MLLMs. Based on
the observed systematic discrepancy, we introduce NLRefer dataset, in which each object is referred with explicit natural language expressions.
• We propose a novel training strategy, SWIM (See What I
Mean), which explicitly enforces alignment between visual content and object nouns during training, resulting in
a model that requires no visual prompt inputs at inference.
• Through experiments on fine-grained object understanding benchmarks, SWIM demonstrates consistent improvements over visual-prompt-based approaches. Quantitative and qualitative analyses of text–visual alignment
further corroborate our claim.\begin{evaluationcase}
    \centering
     \appendixplatetitle{Evaluation Cases}
    \vspace{1.5mm}
    \begin{minipage}[t]{0.47\linewidth}
        \centering
        \includegraphics[width=\linewidth,height=0.30\textheight,keepaspectratio]{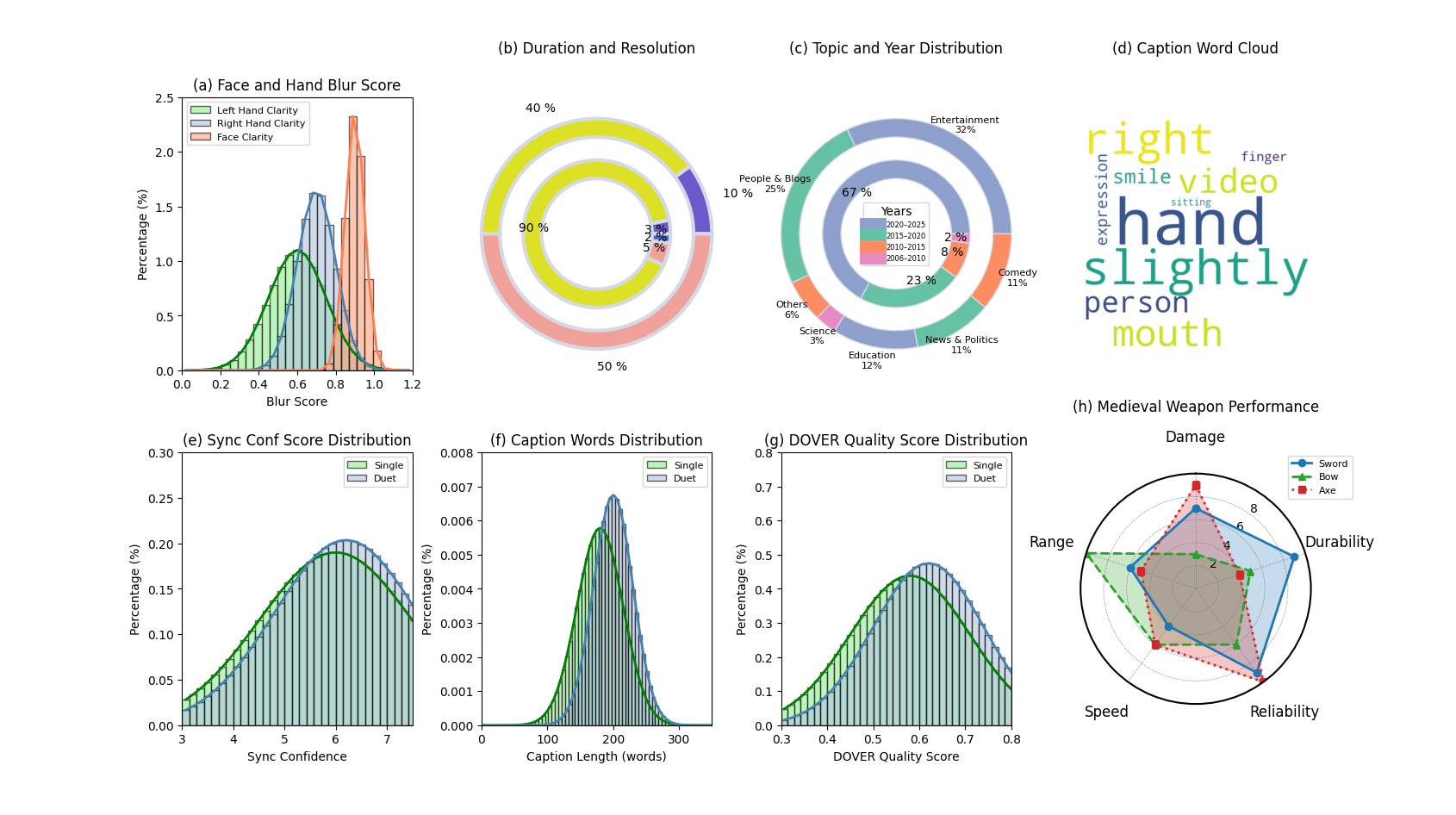}
        \caption{\textbf{GT\_Figure (Left).}} 
        \label{fig:GT_comb45}
    \end{minipage}
    \hfill 
    \begin{minipage}[t]{0.47\linewidth}
        \centering
        \includegraphics[width=\linewidth,height=0.30\textheight,keepaspectratio]{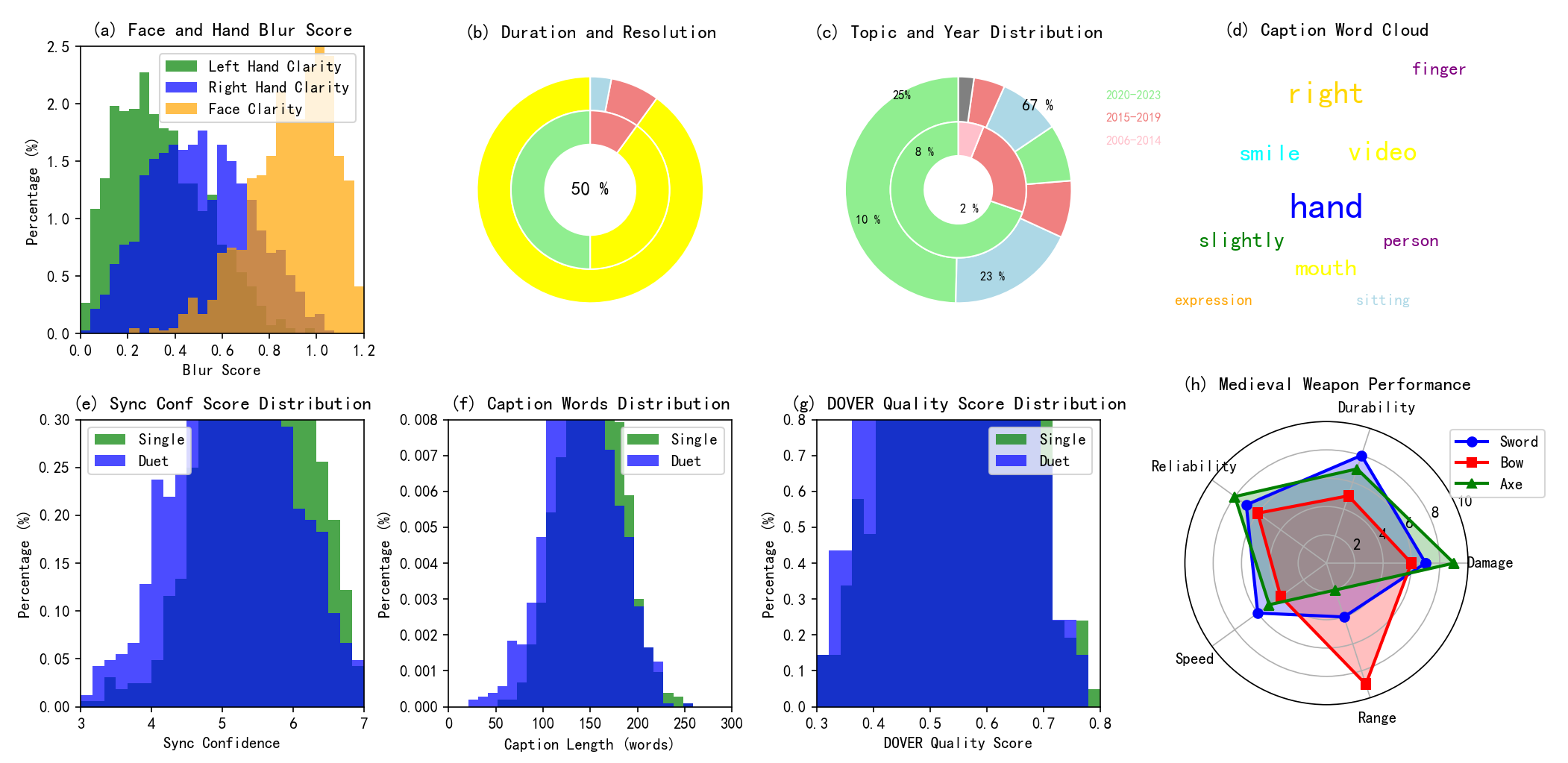} 
        \caption{\textbf{Generation\_Figure (Right).}} 
        \label{fig:Gen_comb45}
    \end{minipage}

    \vspace{6mm} 

    \captionsetup{type=table}
    \caption{\textbf{Detailed multi-dimensional evaluation statistics (Combination).} By pairing related metrics, we observe comprehensive performance across API, Base, and standard Evaluation reports.} 
    \label{tab:datastattable_comb45}
    \centering
    \setlength{\tabcolsep}{2.8pt}\renewcommand{\arraystretch}{1.00}
    
    {\small
    \begin{tabularx}{0.98\linewidth}{>{\bfseries}p{2.15cm} >{\bfseries}p{2.15cm} X}
        \toprule
        Metric 1 (Score) & Metric 2 (Score) & Combined Reason / Key Insight \\
        \midrule
        
        \multicolumn{3}{l}{\cellcolor{gray!15}\textit{\textbf{LLM Report (Overall Score: 39.50)}}} \\
        \midrule
        Chart Type (20.0) & Data Sim (25.0) & GT uses deterministic analytic plots and specific categorical values; GEN samples random data (np.random) and uses histogram-based density plots, completely altering the geometry and underlying distributions. \\
        Visual Params (40.0) & Color (30.0) & GT applies specific bar widths, lw=2 overlays, and a light color palette (e.g., lightgreen); GEN uses default histogram styles and a mismatched green/blue/orange palette. \\
        Layout (50.0) & Grid (70.0) & GT arranges a 2x4 grid (17x10) with specific polar dashed grids; GEN uses a 3x4 gridspec (16x12) with default Cartesian histogram grids, causing noticeable aspect differences. \\
        Legend (60.0) & Text Content (45.0) & GT features precise legend locations and annotation texts (exact percentages); GEN uses default placements and simulates a word cloud with different text. \\
        
        \midrule
        \multicolumn{3}{l}{\cellcolor{gray!15}\textit{\textbf{Base Report (Overall Weighted Score: 46.02)}}} \\
        \midrule
        Text F1 (83.92) & Type F1 (75.00) & Strong performance in text OCR and reasonable detection of the multiple basic chart types present. \\
        Legend F1 (71.50) & Data Param (62.10) & Moderate alignment in legend extraction and data parameter identification, though overall recall remains insufficient. \\
        Visual Param (54.71) & Color F1 (25.41) & Model struggles significantly with visual rendering parameters and shows severe color mismatches. \\
        Layout F1 (0.00) & Grid F1 (0.00) & Complete failure in capturing the complex multi-subplot arrangement and grid configurations. \\

        \midrule
        \multicolumn{3}{l}{\cellcolor{gray!15}\textit{\textbf{LMM Report (Final Similarity Score: 0.0)}}} \\
        \midrule
        Chart Type (100) & Color Style (0) & Successfully identifies the complex chart mix. However, \textbf{Major Errors} exist: missing KDE lines, hatching, bar outlines, and severe donut/wordcloud color mismatches. \\
        Data (0) & Layout (0) & \textbf{Major Errors:} Fundamental data deviations (donut rings collapsed from 90\%/40\% to a single 50\% ring). Subplot composition, aspect ratios, and KDE curve presence are entirely altered. \\
        \midrule
        
        \multicolumn{3}{p{0.94\linewidth}}{\small \textbf{Summary:} Automatic failure (0 score) triggered by multiple major errors. The model fails fundamentally in data presentation (e.g., random data sampling instead of deterministic plotting, collapsed donut rings) and layout composition (missing KDE curves, altered grid arrangement).} \\
        \bottomrule
        \label{tab:comb45_results}
    \end{tabularx}
    }
\end{evaluationcase}

\subsection{Error Case Study:}
\label{error_case_study}
To improve interpretability, we conduct a fine-grained analysis of model failures across the three hierarchical levels of our benchmark. Building upon the preliminary observations in Section ~\ref{sec:Analysis} and Appendix ~\ref{sec:appendix_case_study}, we further perform a granular metadata-based study on failure cases to refine our taxonomy and better characterize the dominant error sources at each level. Selected qualitative examples are provided in following pages. Fig.~\ref{fig:color errors 1}--Fig.~\ref{fig:visual_style errors 2} show the common error types in three tasks. These include color consistency errors, component position errors, data extraction and encoding errors, and axis scale errors, among others. These error types are shared across the three tasks and constitute the main errors related to the model’s visual perception and chart code reconstruction abilities.

\textbf{Level 1: Perception and Data Extraction Failures.} At Level 1, failures are primarily attributable to insufficient visual perception and weak data extraction capabilities. Models frequently overlook fine-grained visual attributes—such as grid configurations, marker styles, and line patterns—resulting in degraded visual fidelity. In addition, inaccurate or incomplete data extraction, particularly in chart-from-data (CFD) tasks, leads to erroneous or missing plotted values. Although the generated code is often syntactically correct and executable, the rendered charts fail to faithfully reproduce the target visualization.

\textbf{Level 2: Reasoning and Instruction Alignment Failures.} While Level 2 inherits the perception and extraction limitations observed in Level 1, the dominant bottleneck shifts to reasoning capacity. The editing tasks at this level involve substantially more complex and compositional instructions. Limited reasoning ability prevents models from executing valid edits that simultaneously respect the original visual context and satisfy the specified transformation requirements. This explains the consistent performance gap between Level 2 and Level 1 across evaluated models.

\textbf{Level 3: Long-Context and Instruction Comprehension Failures.} The severe performance degradation observed at Level 3 is mainly driven by long-context processing limitations. Handling extensive raw tabular data consumes a substantial portion of the model’s available context window, leading to the well-documented “lost-in-the-middle” phenomenon. Consequently, models exhibit incomplete data extraction and weakened understanding of plotting instructions. This results in a pronounced decline in both execution success rate and visual fidelity.

\begin{compacterrorpage}
    \centering
    \appendixplatetitle{Common Error Category: Color Errors}
    \vspace{1.0mm}
    \begin{minipage}{\linewidth}
        \centering
        \textbf{Case 1: Box Chart (level3)}\\[0.08cm]
        
        \begin{minipage}[t]{0.48\linewidth}
            \centering
            \includegraphics[width=\linewidth,height=0.160\textheight,keepaspectratio]{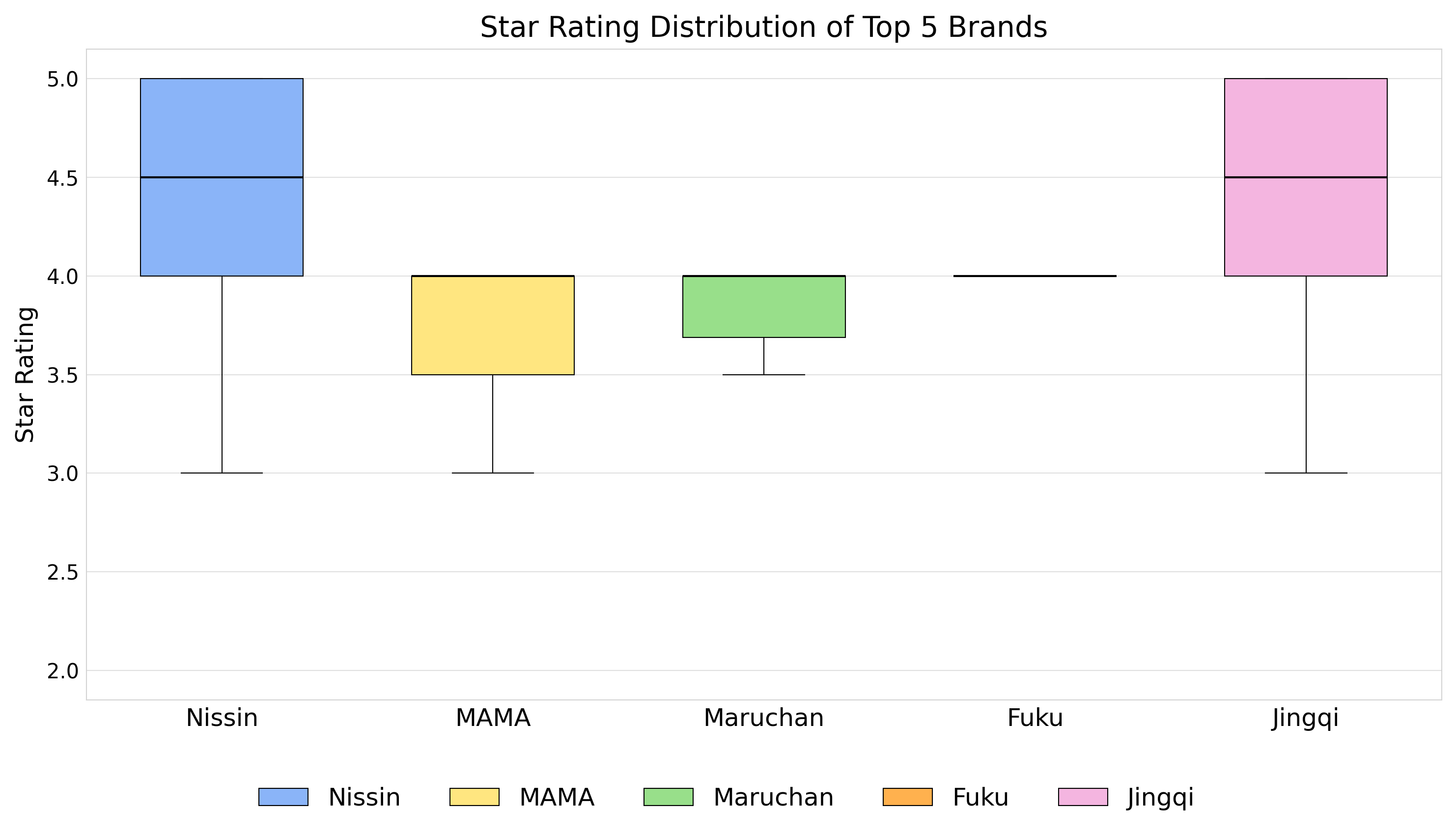}
            \\[0.02cm] \scriptsize (a) Ground Truth
        \end{minipage}
        \hfill
        \begin{minipage}[t]{0.48\linewidth}
            \centering
            \includegraphics[width=\linewidth,height=0.160\textheight,keepaspectratio]{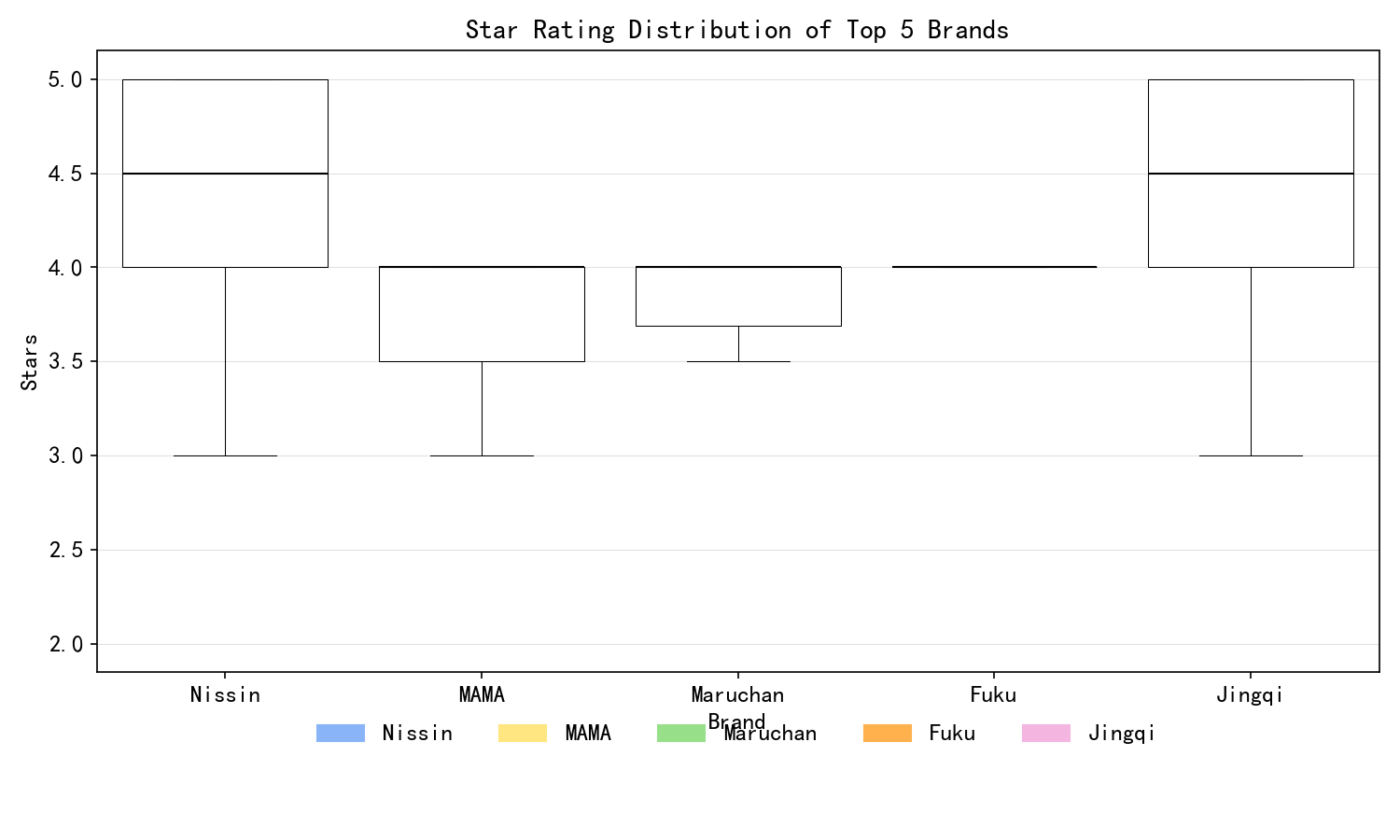}
            \\[0.02cm] \scriptsize (b) Model Output
        \end{minipage}
        
        \vspace{0.08cm}
        \begin{minipage}{0.96\linewidth}
            \footnotesize \textbf{Analysis:} Although the model identified the color sequence in the legend area, this perception failed to transfer to the plot objects within the coordinate system, reflecting a logical inconsistency between local feature recognition and global mapping.
        \end{minipage}
    \end{minipage}

    \errorcaseseparator

    \begin{minipage}{\linewidth}
        \centering
        \textbf{Case 2: Heatmap Chart (level1)}\\[0.08cm]
        
        \begin{minipage}[t]{0.48\linewidth}
            \centering
            \includegraphics[width=\linewidth,height=0.160\textheight,keepaspectratio]{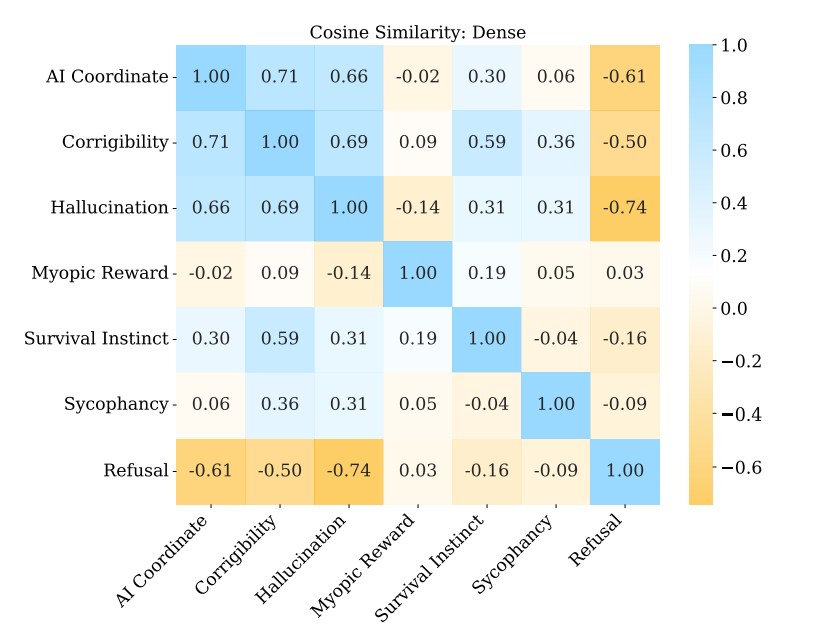}
            \\[0.02cm] \scriptsize (c) Ground Truth
        \end{minipage}
        \hfill
        \begin{minipage}[t]{0.48\linewidth}
            \centering
            \includegraphics[width=\linewidth,height=0.160\textheight,keepaspectratio]{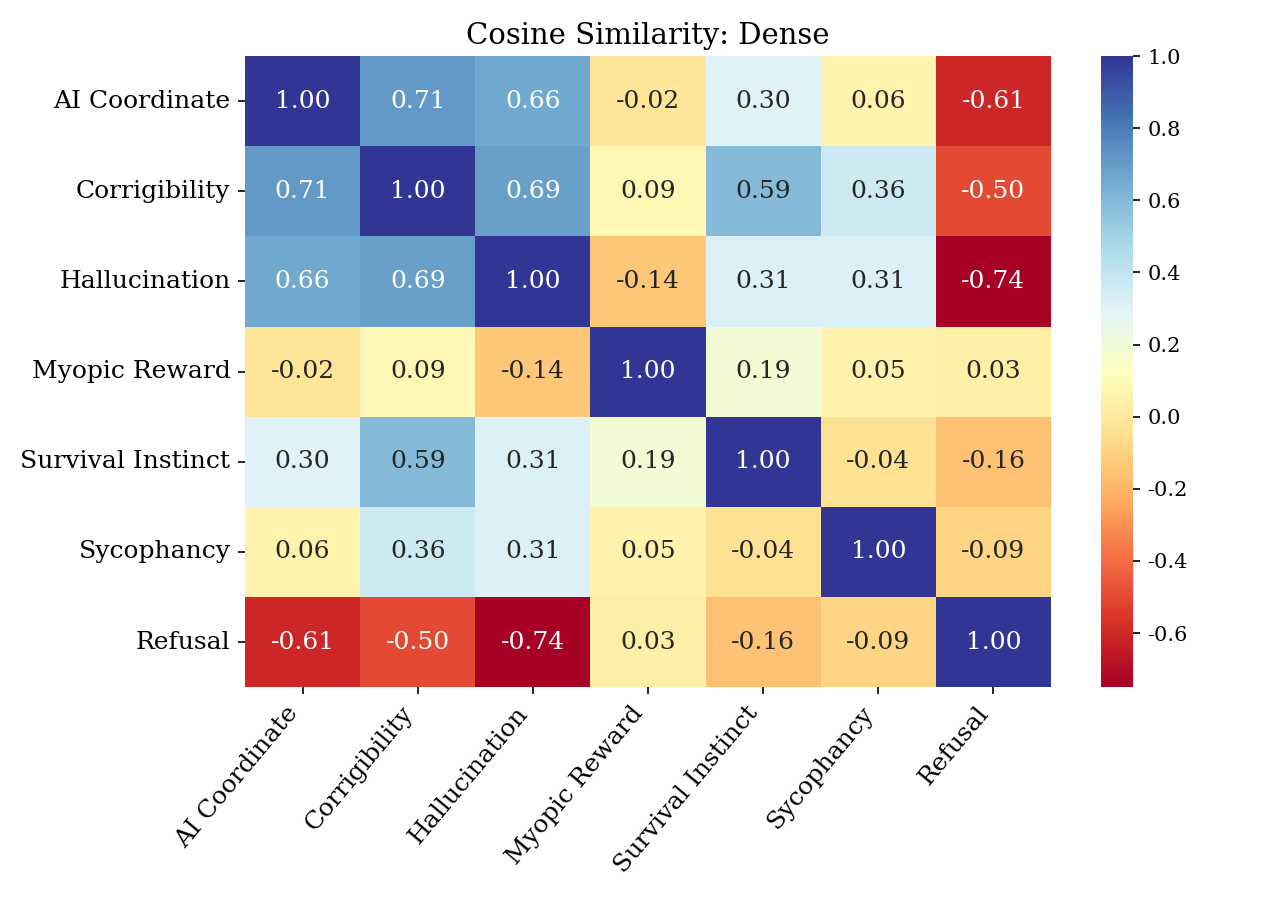}
            \\[0.02cm] \scriptsize (d) Model Output
        \end{minipage}
        
        \vspace{0.08cm}
        \begin{minipage}{0.96\linewidth}
            \footnotesize \textbf{Analysis:} Significant discrepancies in color transitions are observed between the model-generated output (right) and the ground-truth (GT) figure, indicating limitations in the model’s ability to accurately perceive and reproduce subtle variations in color.
        \end{minipage}
    \end{minipage}

    \errorcaseseparator
    
    \begin{minipage}{\linewidth}
        \centering
        \textbf{Case 3: Combination Chart (level2)}\\[0.08cm]
        
        \begin{minipage}[t]{0.48\linewidth}
            \centering
            \includegraphics[width=\linewidth,height=0.160\textheight,keepaspectratio]{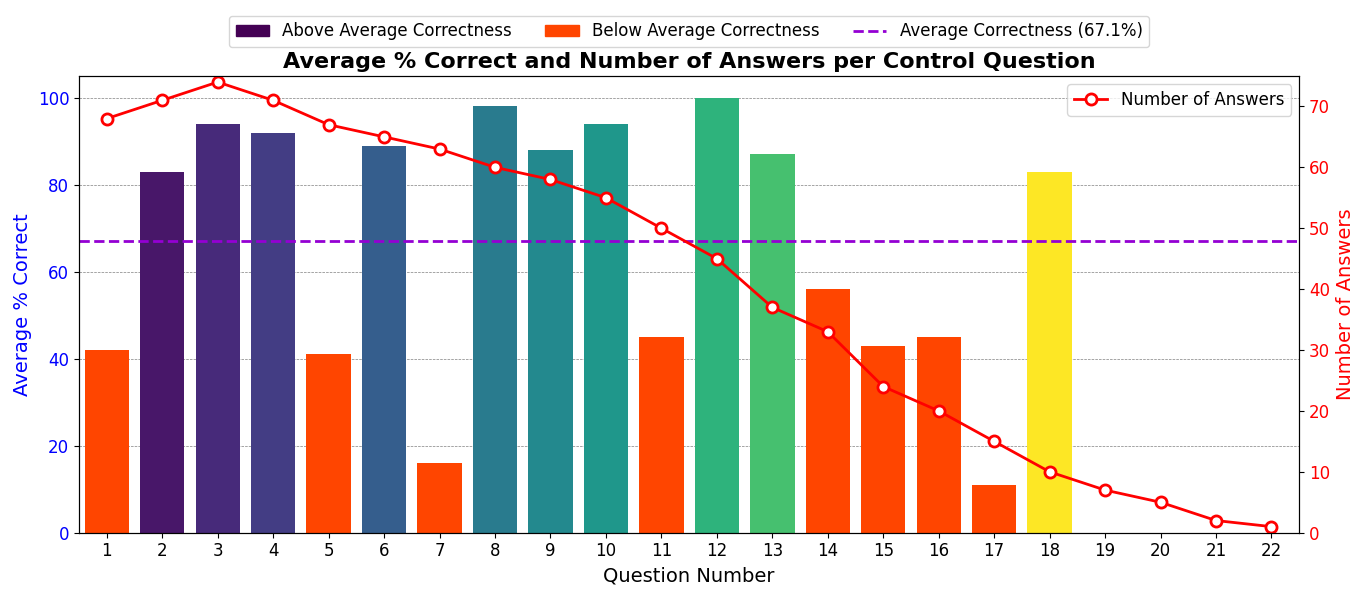}
            \\[0.02cm] \scriptsize (e) Ground Truth
        \end{minipage}
        \hfill
        \begin{minipage}[t]{0.48\linewidth}
            \centering
            \includegraphics[width=\linewidth,height=0.160\textheight,keepaspectratio]{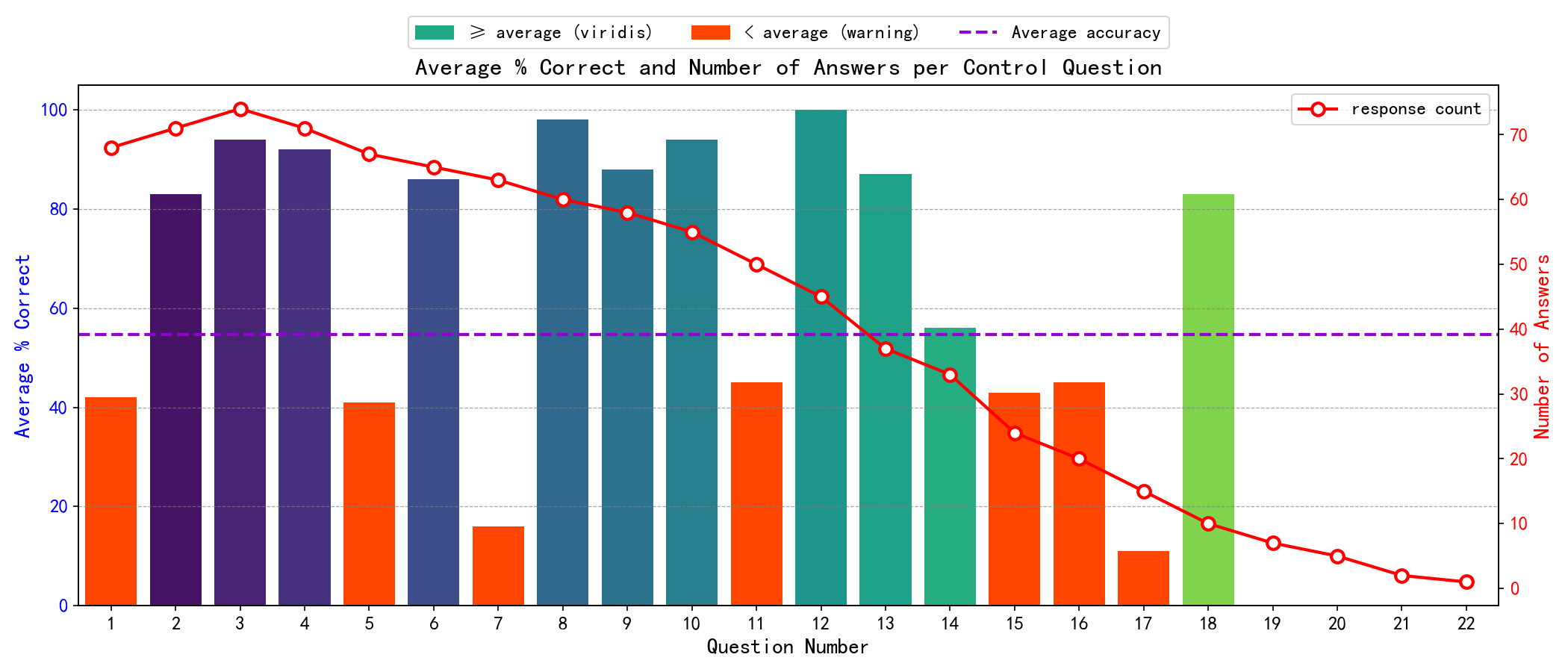}
            \\[0.02cm] \scriptsize (f) Model Output
        \end{minipage}
        
        \vspace{0.08cm}
        \begin{minipage}{0.96\linewidth}
            \footnotesize \textbf{Analysis:} The legend colors in the model output on the right do not match those in the GT Figure, and there is a noticeable color discrepancy at X=18. This indicates that the model lacks fine-grained image perception capabilities in complex charts.
        \end{minipage}
    \end{minipage}
    
    \vfill
    \caption{\textbf{Qualitative Error Analysis.} Color inconsistency issues across different chart types, including mismatches among axis colors, legend entries, and the main plot elements.
}
    \label{fig:color errors 1}
\end{compacterrorpage}

\begin{compacterrorpage}
    \centering
    \appendixplatetitle{Common Error Category: Color Errors}
    \vspace{1.0mm}
    \begin{minipage}{\linewidth}
        \centering
        \textbf{Case 1: Box Chart(level1)}\\[0.08cm]
        
        \begin{minipage}[t]{0.48\linewidth}
            \centering
            \includegraphics[width=\linewidth,height=0.160\textheight,keepaspectratio]{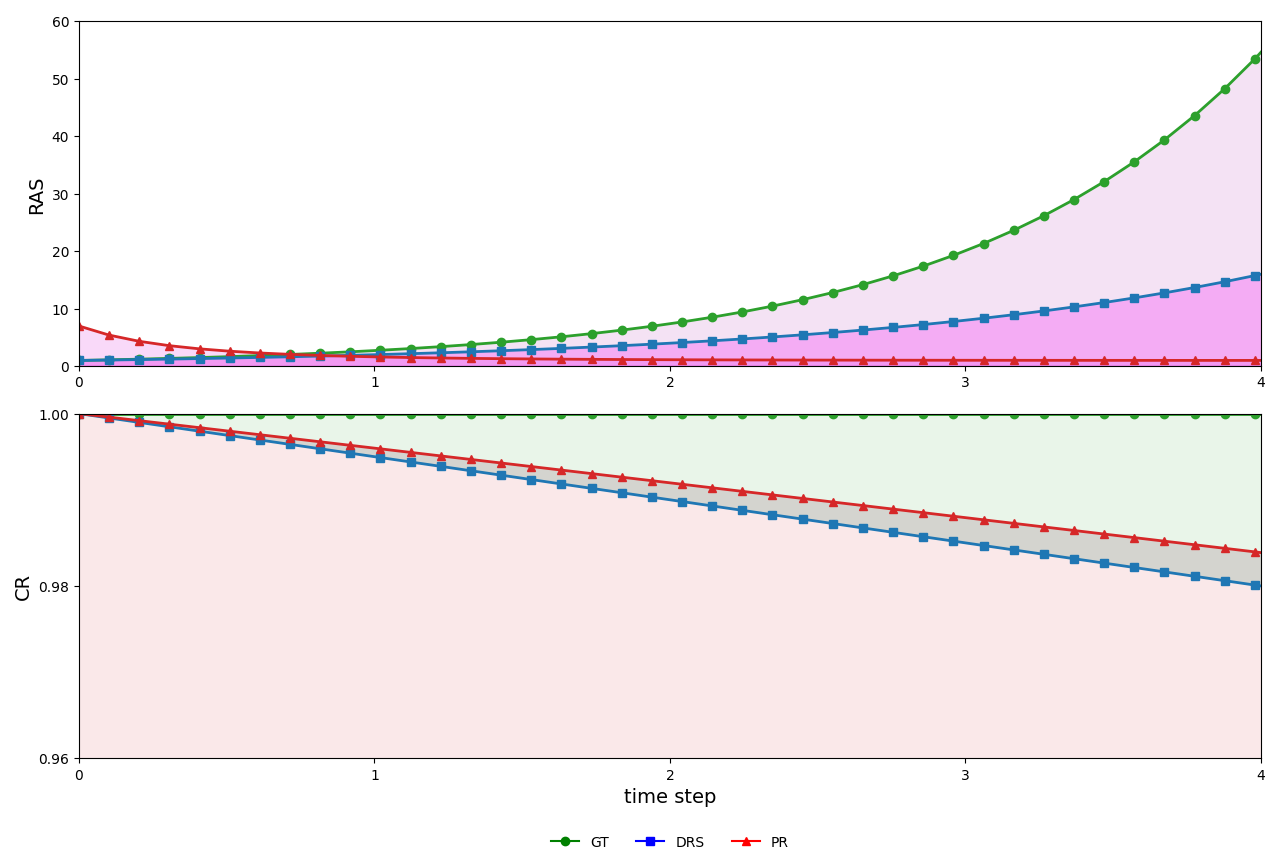}
            \\[0.02cm] \scriptsize (a) Ground Truth
        \end{minipage}
        \hfill
        \begin{minipage}[t]{0.48\linewidth}
            \centering
            \includegraphics[width=\linewidth,height=0.160\textheight,keepaspectratio]{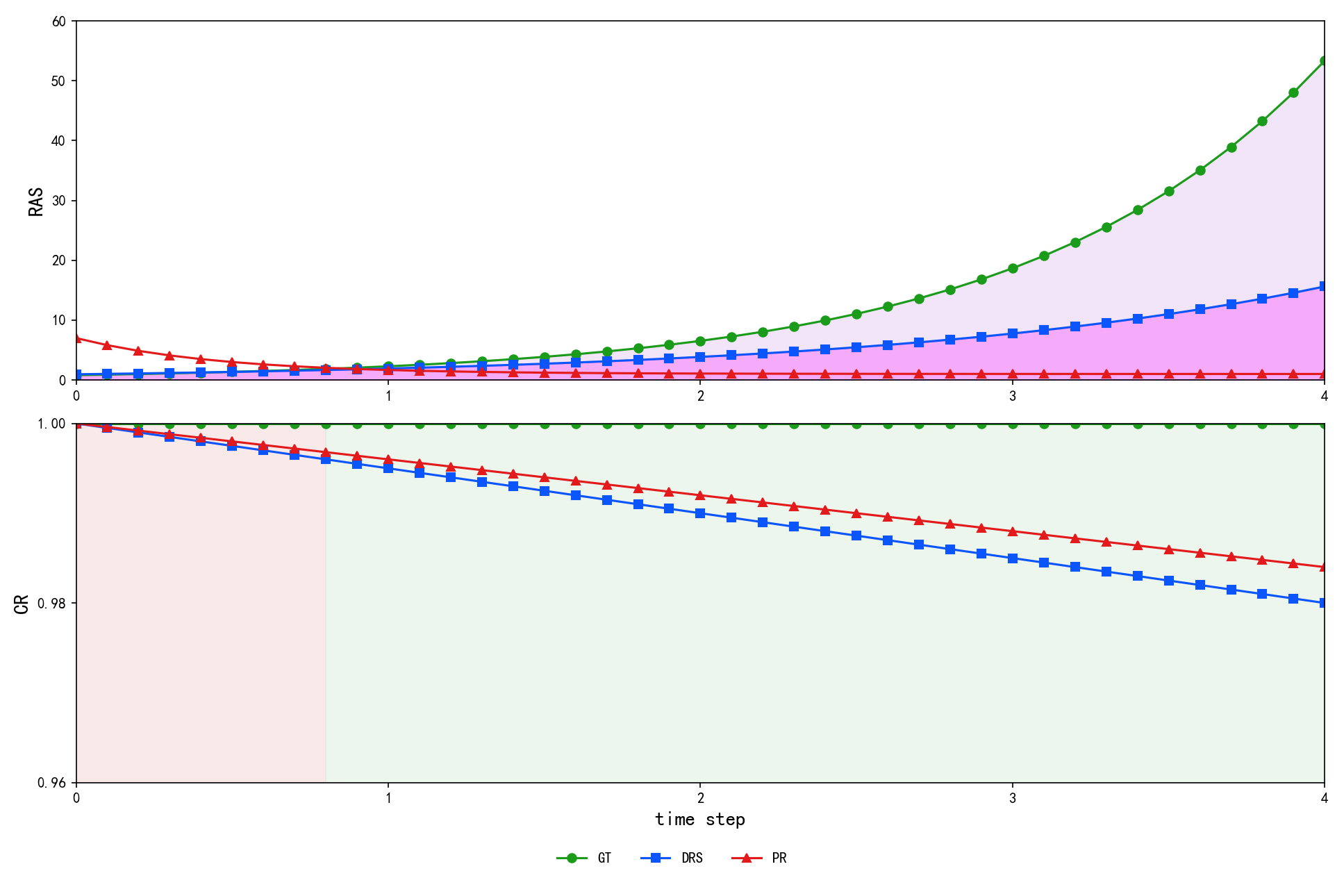}
            \\[0.02cm] \scriptsize (b) Model Output
        \end{minipage}
        
        \vspace{0.08cm}
        \begin{minipage}{0.96\linewidth}
            \footnotesize \textbf{Analysis:} The color of the bottom subplot in the model output on the right does not match the GT Figure; the model failed to understand the color-coded regions in the area chart.
        \end{minipage}
    \end{minipage}

    \errorcaseseparator

    \begin{minipage}{\linewidth}
        \centering
        \textbf{Case 2: Heatmap Chart(level2)}\\[0.08cm]
        
        \begin{minipage}[t]{0.48\linewidth}
            \centering
            \includegraphics[width=\linewidth,height=0.160\textheight,keepaspectratio]{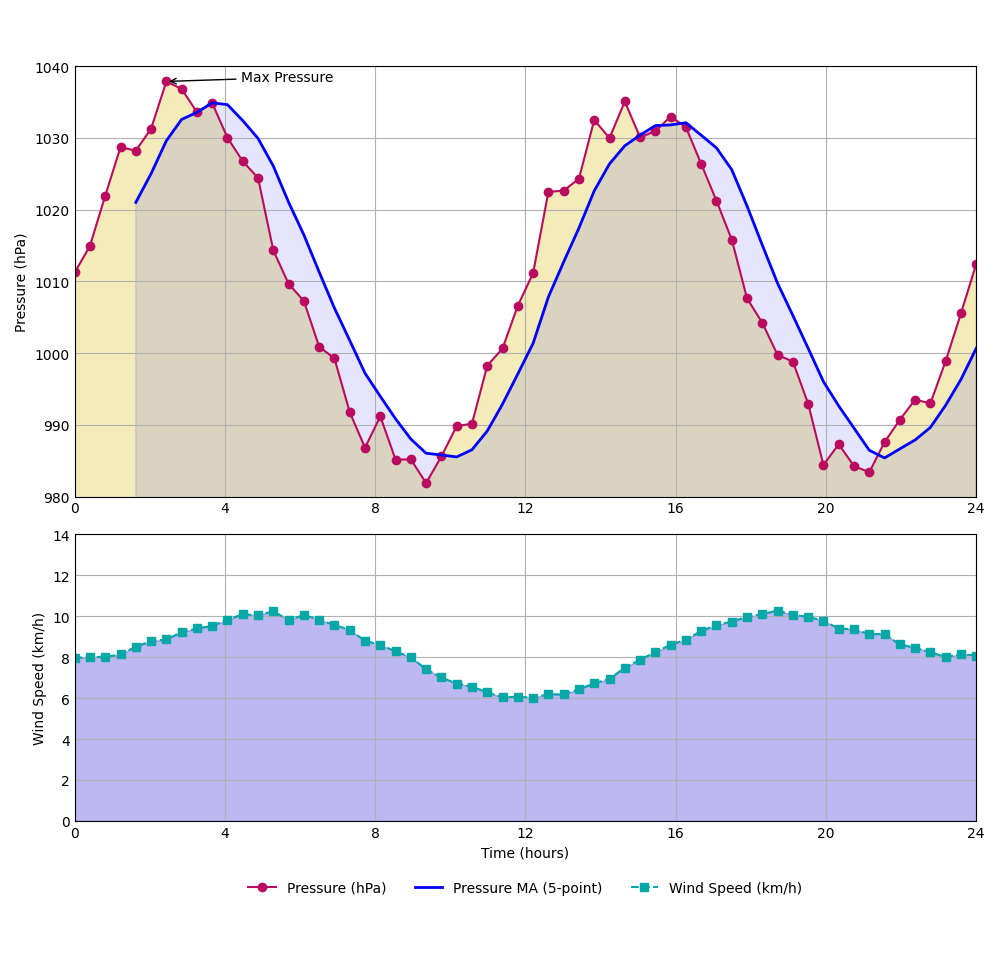}
            \\[0.02cm] \scriptsize (c) Ground Truth
        \end{minipage}
        \hfill
        \begin{minipage}[t]{0.48\linewidth}
            \centering
            \includegraphics[width=\linewidth,height=0.160\textheight,keepaspectratio]{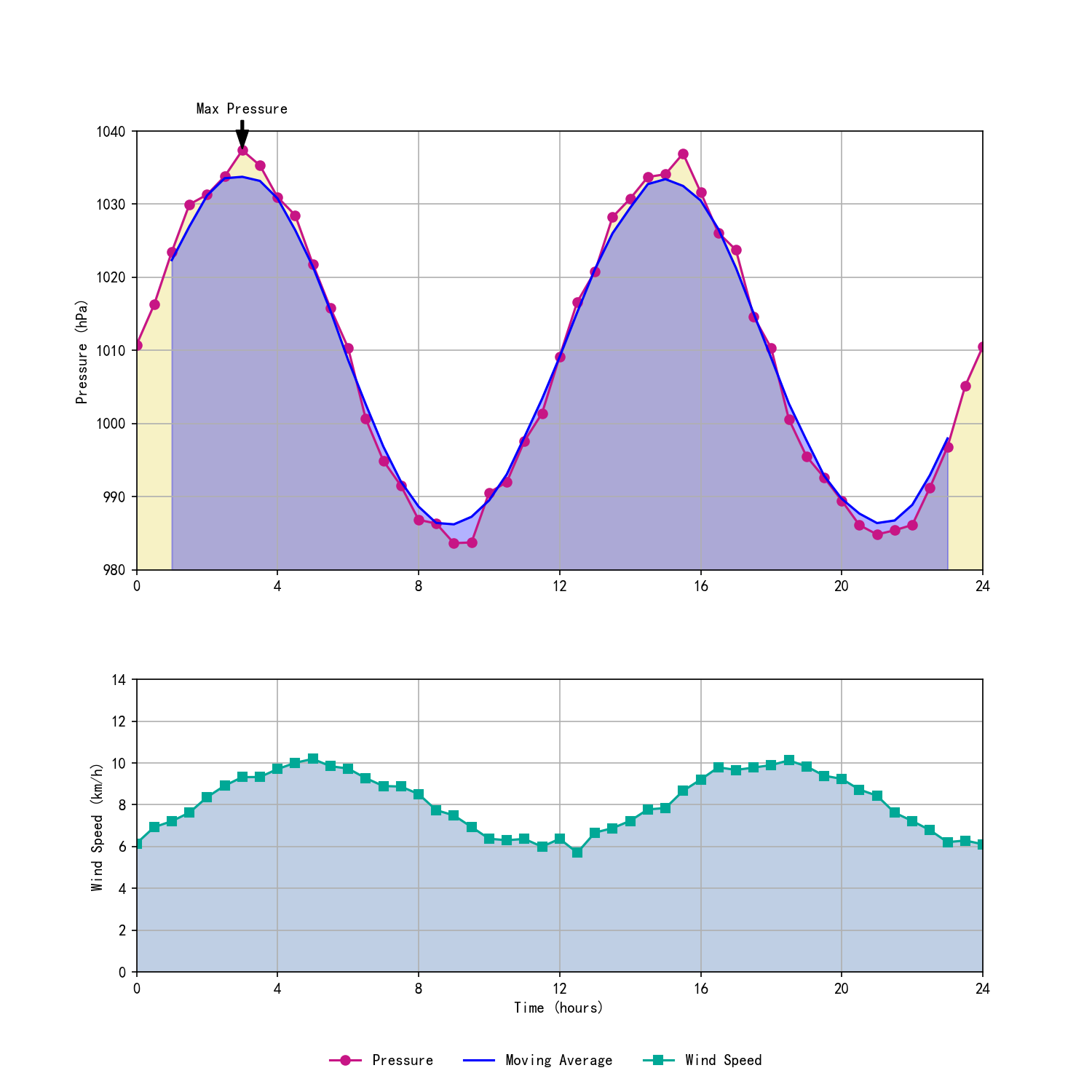}
            \\[0.02cm] \scriptsize (d) Model Output
        \end{minipage}
        
        \vspace{0.08cm}
        \begin{minipage}{0.96\linewidth}
            \footnotesize \textbf{Analysis:} The color filling in the model output on the right is inconsistent with the GT Figure and lacks visual appeal; the transparency of the filled colors in the area regions has not been appropriately adjusted.
        \end{minipage}
    \end{minipage}

    \errorcaseseparator
    
    \begin{minipage}{\linewidth}
        \centering
        \textbf{Case 3: Combination Chart(level1)}\\[0.08cm]
        
        \begin{minipage}[t]{0.48\linewidth}
            \centering
            \includegraphics[width=\linewidth,height=0.160\textheight,keepaspectratio]{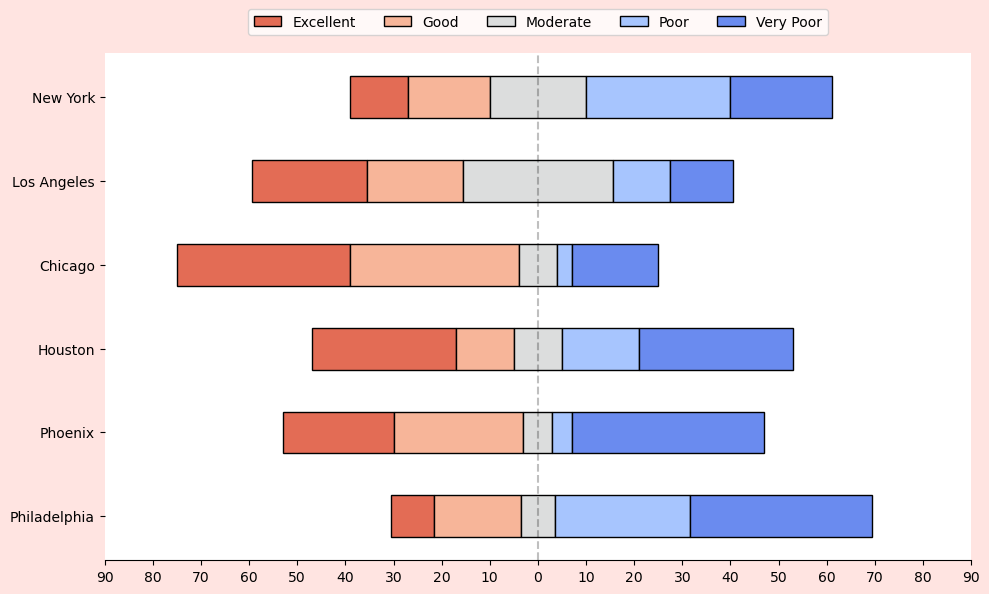}
            \\[0.02cm] \scriptsize (e) Ground Truth
        \end{minipage}
        \hfill
        \begin{minipage}[t]{0.48\linewidth}
            \centering
            \includegraphics[width=\linewidth,height=0.160\textheight,keepaspectratio]{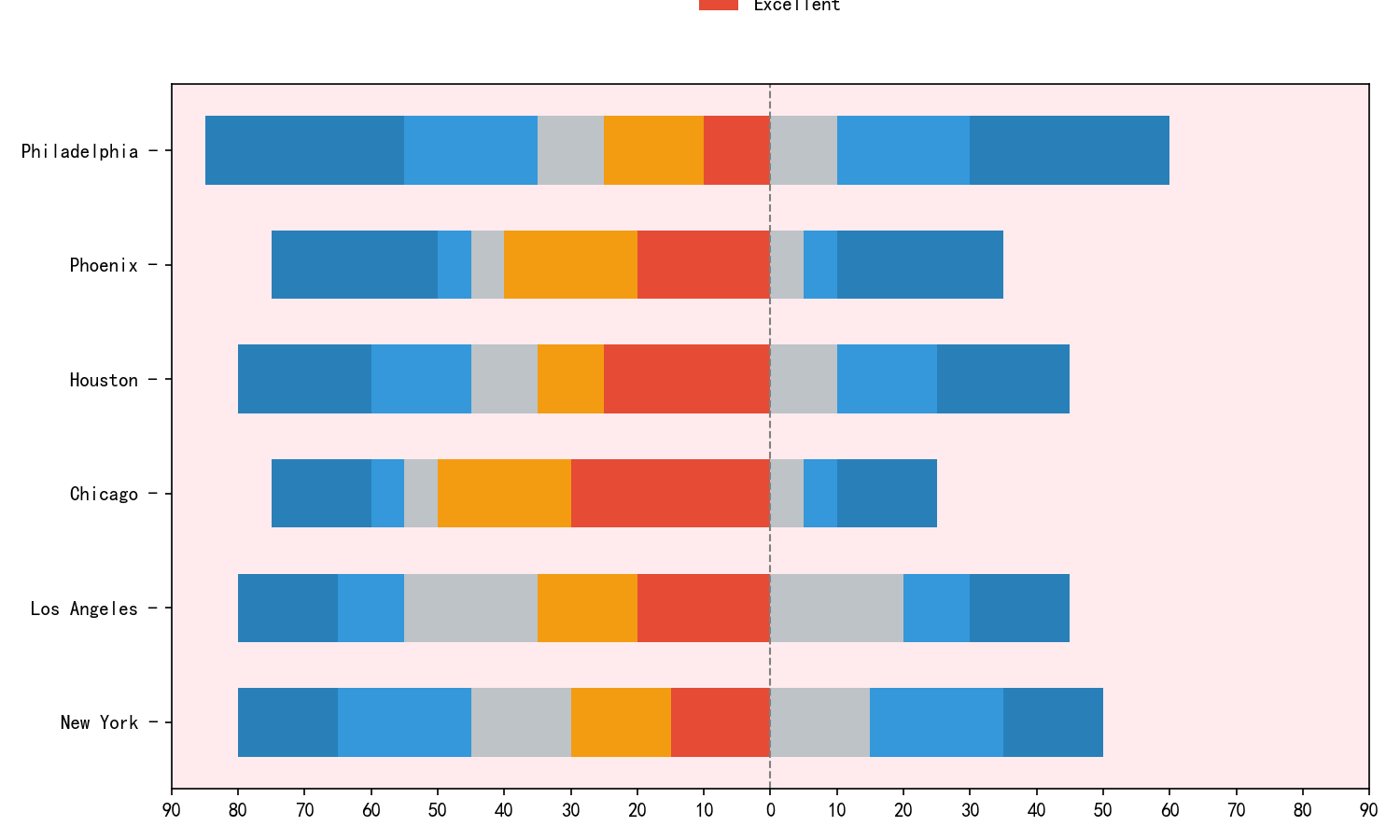}
            \\[0.02cm] \scriptsize (f) Model Output
        \end{minipage}
        
        \vspace{0.08cm}
        \begin{minipage}{0.96\linewidth}
            \footnotesize \textbf{Analysis:} The color filling in the model output on the right is completely opposite to that of the GT Figure, indicating the model's inability to accurately recognize colors. Additionally, there are issues with the recognition of the legend.
        \end{minipage}
    \end{minipage}
    
    \vfill
    \caption{\textbf{Qualitative Error Analysis.} Illustration of color consistency errors across various chart types, highlighting mismatches between axes, legend labels, and primary visual elements.}
    \label{fig:color errors 2}
\end{compacterrorpage}                  

\begin{compacterrorpage}
    \centering
    \appendixplatetitle{Common Error Category: Component Position Errors}
    \vspace{1.0mm}
    \begin{minipage}{\linewidth}
        \centering
        \textbf{Case 1: Bar Chart (level1)}\\[0.08cm]
        
        \begin{minipage}[t]{0.48\linewidth}
            \centering
            \includegraphics[width=\linewidth,height=0.160\textheight,keepaspectratio]{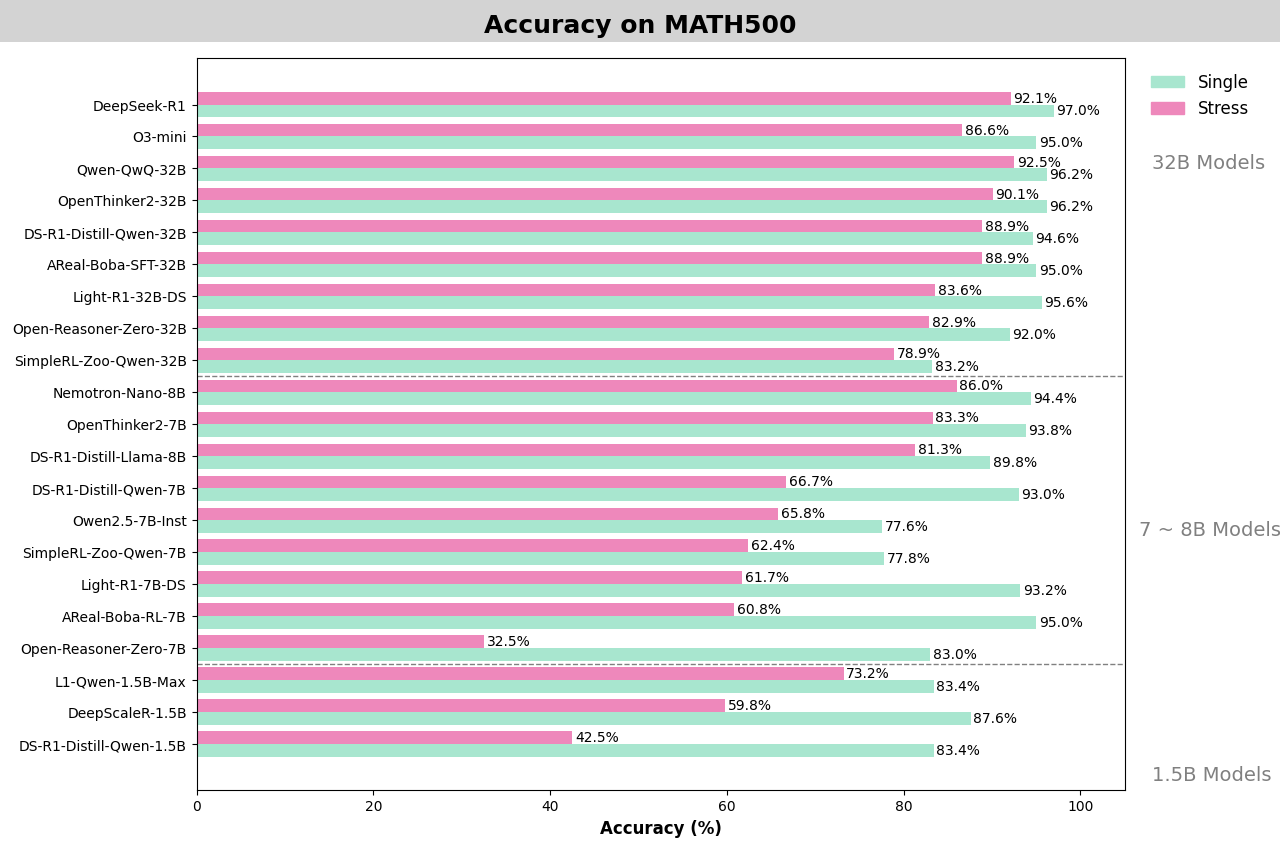}
            \\[0.02cm] \scriptsize (a) Ground Truth
        \end{minipage}
        \hfill
        \begin{minipage}[t]{0.48\linewidth}
            \centering
            \includegraphics[width=\linewidth,height=0.160\textheight,keepaspectratio]{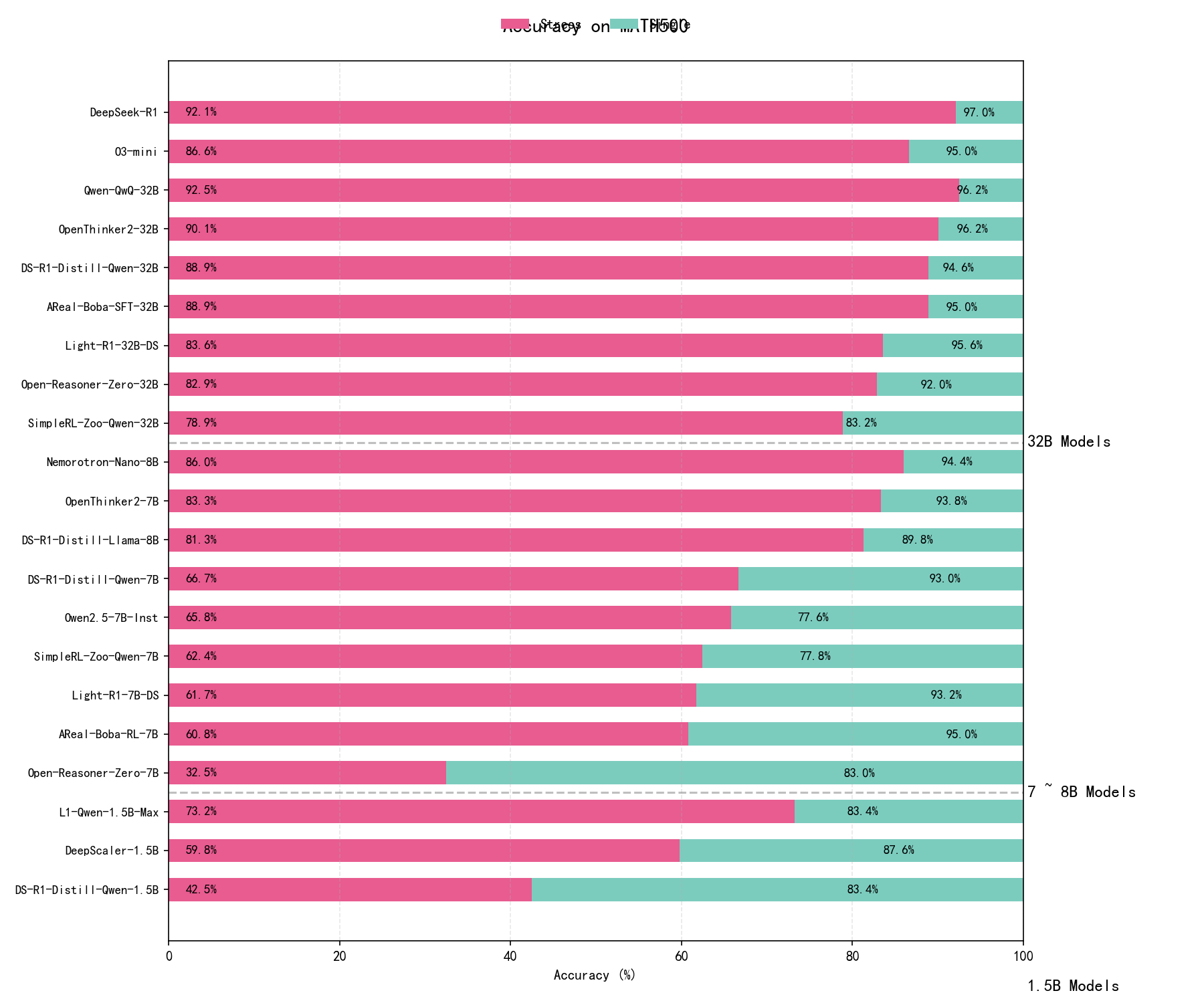}
            \\[0.02cm] \scriptsize (b) Model Output
        \end{minipage}
        
        \vspace{0.08cm}
        \begin{minipage}{0.96\linewidth}
            \footnotesize \textbf{Analysis:} The legend position in the model output on the right does not match that of the GT Figure, and the placement of the numerical values is also completely different.
        \end{minipage}
    \end{minipage}

    \errorcaseseparator

    \begin{minipage}{\linewidth}
        \centering
        \textbf{Case 2: Bar Chart (level1)}\\[0.08cm]
        
        \begin{minipage}[t]{0.48\linewidth}
            \centering
            \includegraphics[width=\linewidth,height=0.160\textheight,keepaspectratio]{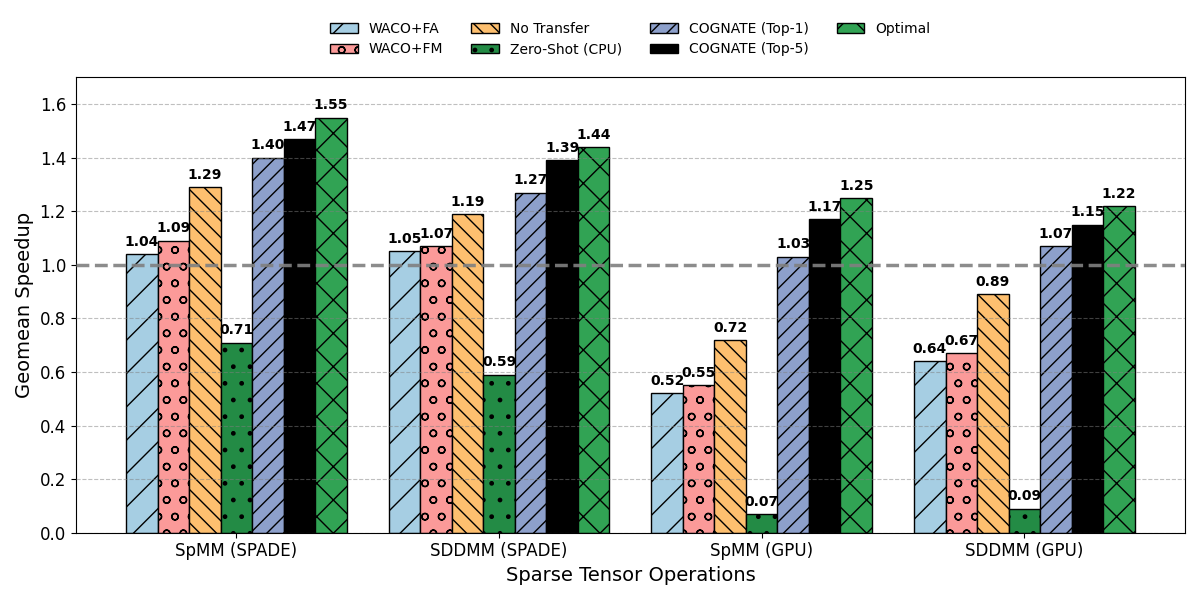}
            \\[0.02cm] \scriptsize (c) Ground Truth
        \end{minipage}
        \hfill
        \begin{minipage}[t]{0.48\linewidth}
            \centering
            \includegraphics[width=\linewidth,height=0.160\textheight,keepaspectratio]{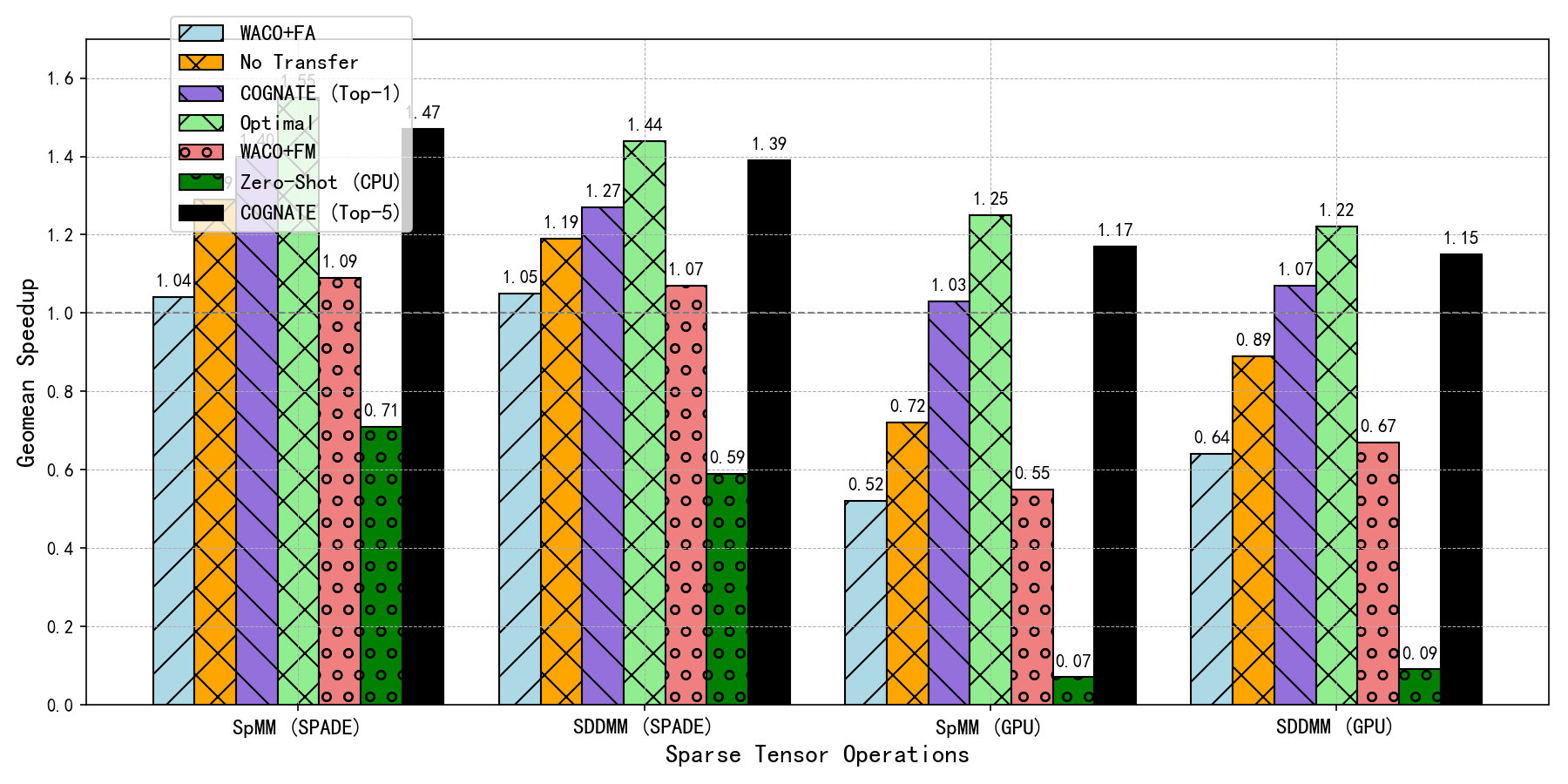}
            \\[0.02cm] \scriptsize (d) Model Output
        \end{minipage}
        
        \vspace{0.08cm}
        \begin{minipage}{0.96\linewidth}
            \footnotesize \textbf{Analysis:} The legend position in the model output on the right does not match that of the GT Figure, severely obscuring the data portion of the bar chart and causing significant visibility issues.
        \end{minipage}
    \end{minipage}

    \errorcaseseparator
    
    \begin{minipage}{\linewidth}
        \centering
        \textbf{Case 3: Combination Chart (level2)}\\[0.08cm]
        
        \begin{minipage}[t]{0.48\linewidth}
            \centering
            \includegraphics[width=\linewidth,height=0.160\textheight,keepaspectratio]{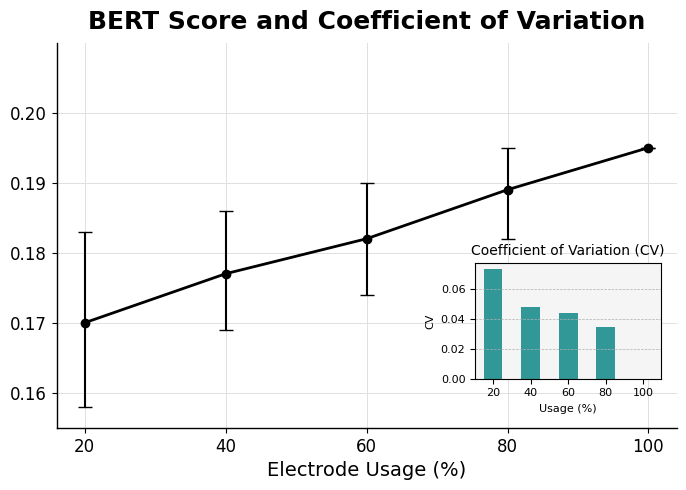}
            \\[0.02cm] \scriptsize (e) Ground Truth
        \end{minipage}
        \hfill
        \begin{minipage}[t]{0.48\linewidth}
            \centering
            \includegraphics[width=\linewidth,height=0.160\textheight,keepaspectratio]{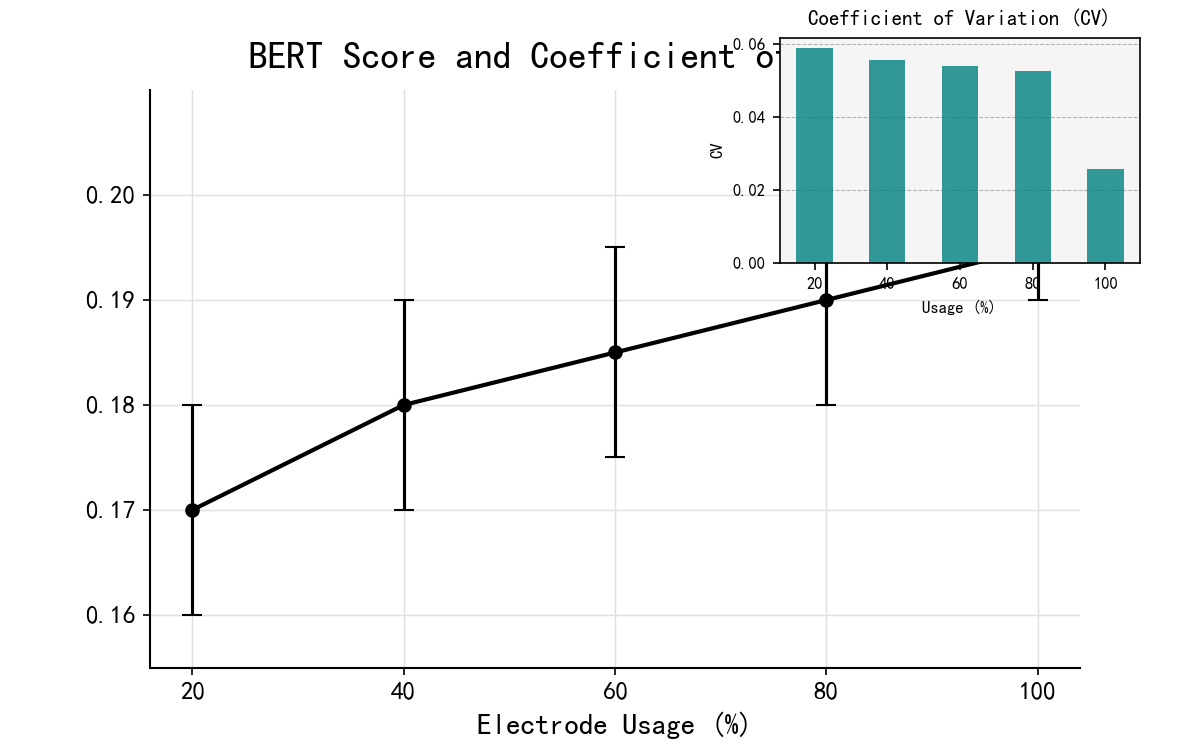}
            \\[0.02cm] \scriptsize (f) Model Output
        \end{minipage}
        
        \vspace{0.08cm}
        \begin{minipage}{0.96\linewidth}
            \footnotesize \textbf{Analysis:} The subplot position in the model output on the right does not match that of the GT Figure, severely obscuring the data portion of the error points and causing significant visibility issues.
        \end{minipage}
    \end{minipage}
    
    \vfill
    \caption{\textbf{Qualitative Error Analysis.} Common positional mistakes in model outputs, including overlap between the legend and chart elements, overlapping visual elements within the chart, and text overlapping with graphical components.
}
    \label{fig:component position errors 1}
\end{compacterrorpage}

\begin{compacterrorpage}
    \centering
    \appendixplatetitle{Common Error Category: Component Position Errors}
    \vspace{1.0mm}
    \begin{minipage}{\linewidth}
        \centering
        \textbf{Case 1: Errorpoint Chart (level2)}\\[0.08cm]
        
        \begin{minipage}[t]{0.48\linewidth}
            \centering
            \includegraphics[width=\linewidth,height=0.160\textheight,keepaspectratio]{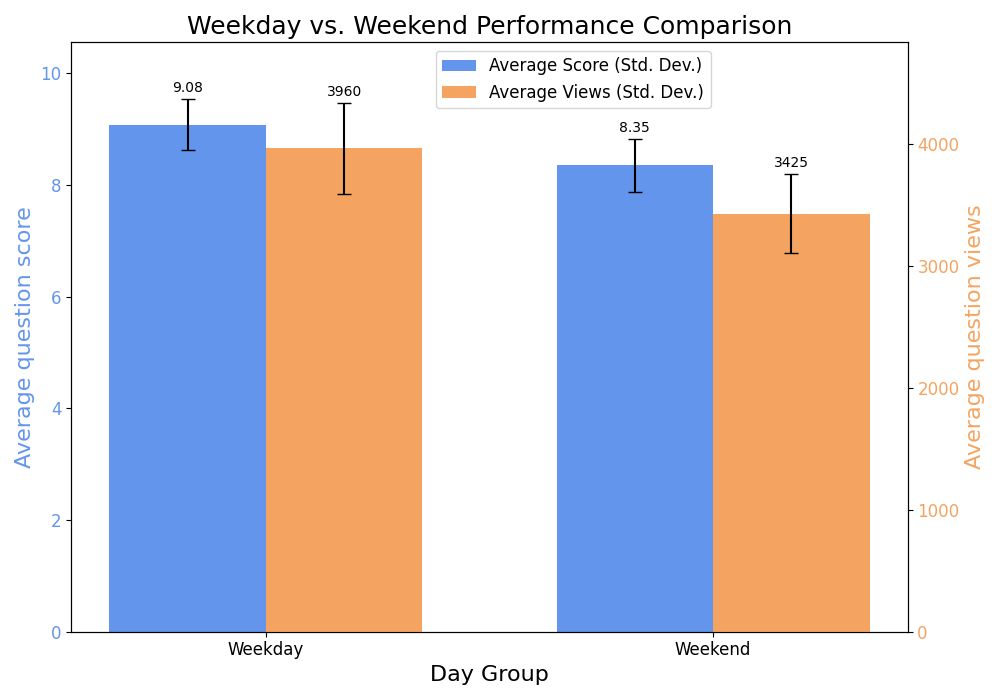}
            \\[0.02cm] \scriptsize (a) Ground Truth
        \end{minipage}
        \hfill
        \begin{minipage}[t]{0.48\linewidth}
            \centering
            \includegraphics[width=\linewidth,height=0.160\textheight,keepaspectratio]{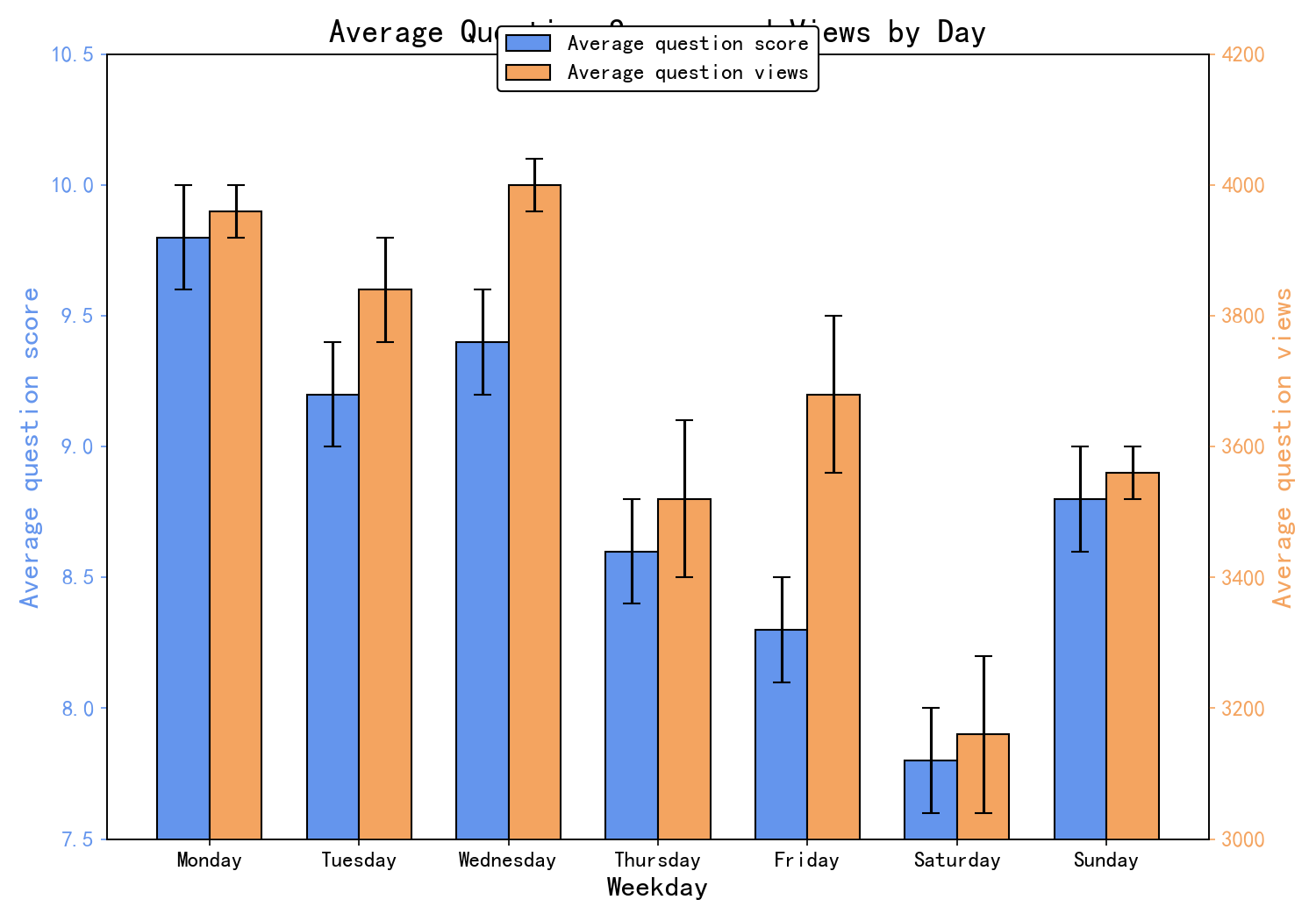}
            \\[0.02cm] \scriptsize (b) Model Output
        \end{minipage}
        
        \vspace{0.08cm}
        \begin{minipage}{0.96\linewidth}
            \footnotesize \textbf{Analysis:} The legend position in the model output on the right does not match that of the GT Figure, severely obscuring the chart title and making the letters in the title illegible.
        \end{minipage}
    \end{minipage}

    \errorcaseseparator

    \begin{minipage}{\linewidth}
        \centering
        \textbf{Case 2: Pie Chart (level3)}\\[0.08cm]
        
        \begin{minipage}[t]{0.48\linewidth}
            \centering
            \includegraphics[width=\linewidth,height=0.160\textheight,keepaspectratio]{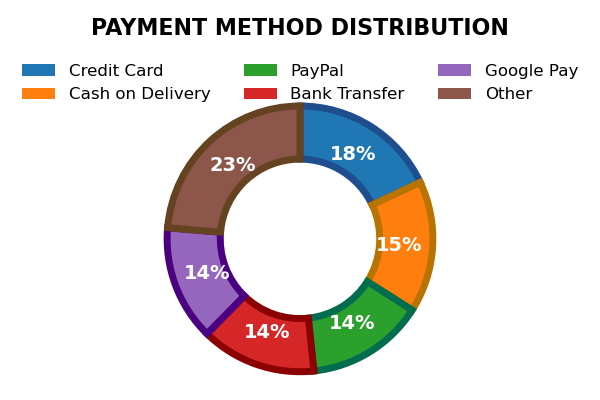}
            \\[0.02cm] \scriptsize (c) Ground Truth
        \end{minipage}
        \hfill
        \begin{minipage}[t]{0.48\linewidth}
            \centering
            \includegraphics[width=\linewidth,height=0.160\textheight,keepaspectratio]{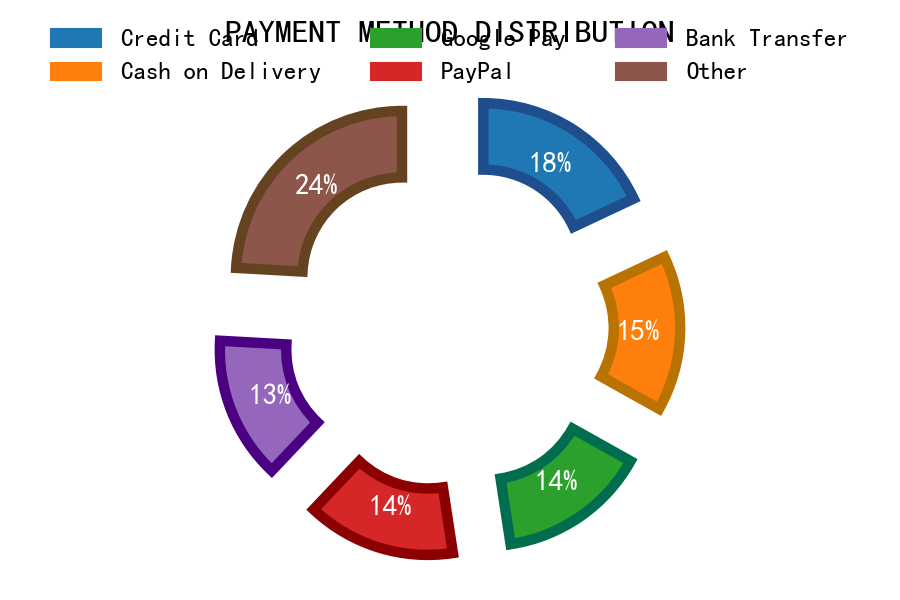}
            \\[0.02cm] \scriptsize (d) Model Output
        \end{minipage}
        
        \vspace{0.08cm}
        \begin{minipage}{0.96\linewidth}
            \footnotesize \textbf{Analysis:} The model failed to accurately extract the absolute spatial coordinates of the legend, leading to positional displacement and visual overlapping with the title, reflecting a failure in spatial perception for complex layouts.
        \end{minipage}
    \end{minipage}

    \errorcaseseparator
    
    \begin{minipage}{\linewidth}
        \centering
        \textbf{Case 3: Combination Chart(level3)}\\[0.08cm]
        
        \begin{minipage}[t]{0.48\linewidth}
            \centering
            \includegraphics[width=\linewidth,height=0.160\textheight,keepaspectratio]{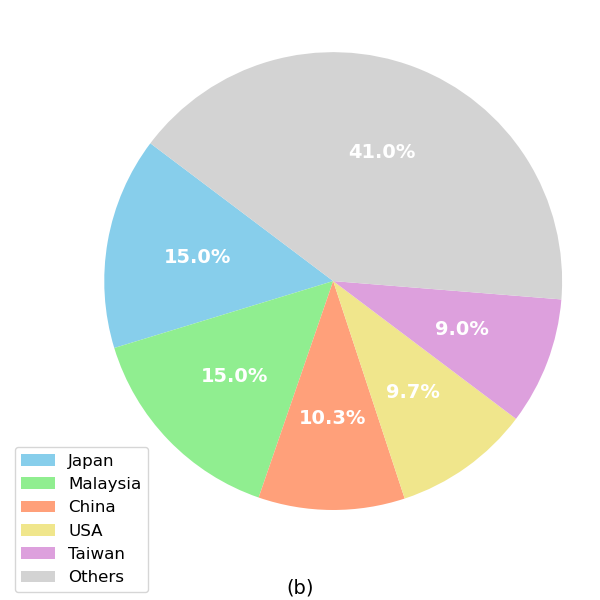}
            \\[0.02cm] \scriptsize (e) Ground Truth
        \end{minipage}
        \hfill
        \begin{minipage}[t]{0.48\linewidth}
            \centering
            \includegraphics[width=\linewidth,height=0.160\textheight,keepaspectratio]{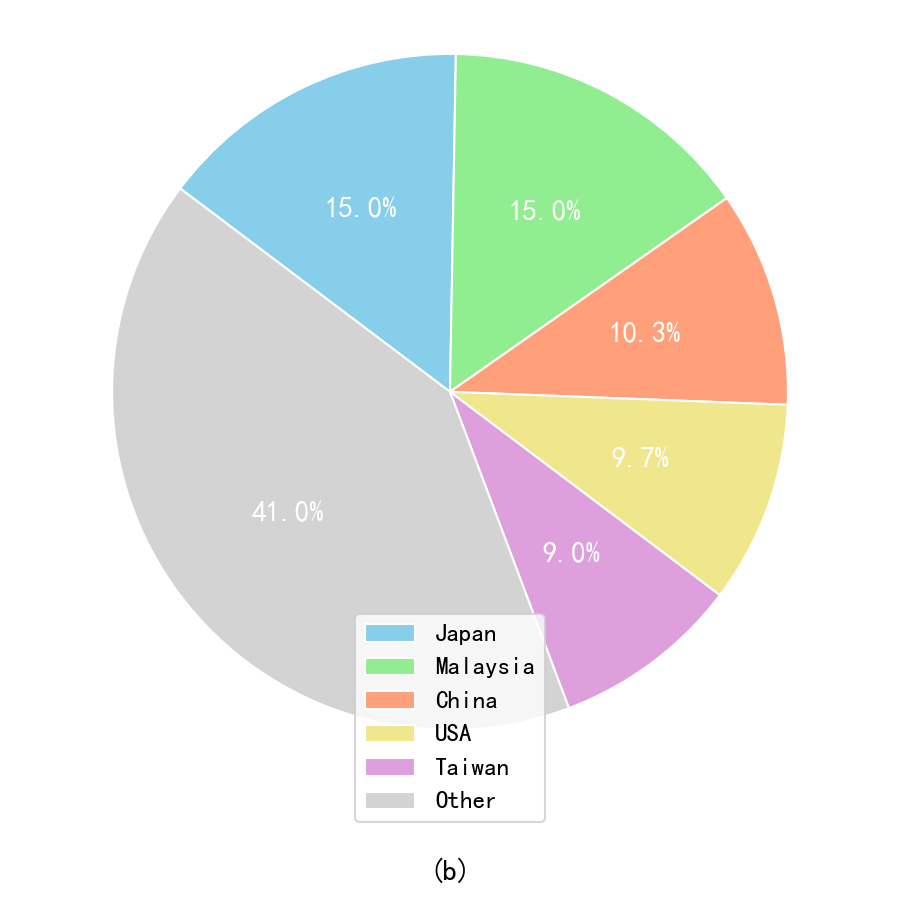}
            \\[0.02cm] \scriptsize (f) Model Output
        \end{minipage}
        
        \vspace{0.08cm}
        \begin{minipage}{0.96\linewidth}
            \footnotesize \textbf{Analysis:} The model exhibits a clear visual perception bottleneck in spatial reasoning, failing to recognize the physical boundaries between the legend, title, and chart body. This leads to severe visual overlapping and positional misalignment, as the model overlooks the layout constraints required to prevent component occlusion.
        \end{minipage}
    \end{minipage}
    
    \vfill
    \caption{\textbf{Qualitative Error Analysis.} Examples of frequent positional inconsistencies in model-generated charts, such as legend–plot overlap, internal element collisions, and text–graphic occlusion.}
    \label{fig:component position errors 2}
\end{compacterrorpage}

\begin{compacterrorpage}
    \centering
    \appendixplatetitle{Common Error Category: Data Encoding Errors}
    \vspace{1.0mm}
    \begin{minipage}{\linewidth}
        \centering
        \textbf{Case 1: Combination Chart (level2)}\\[0.08cm]
        
        \begin{minipage}[t]{0.48\linewidth}
            \centering
            \includegraphics[width=\linewidth,height=0.160\textheight,keepaspectratio]{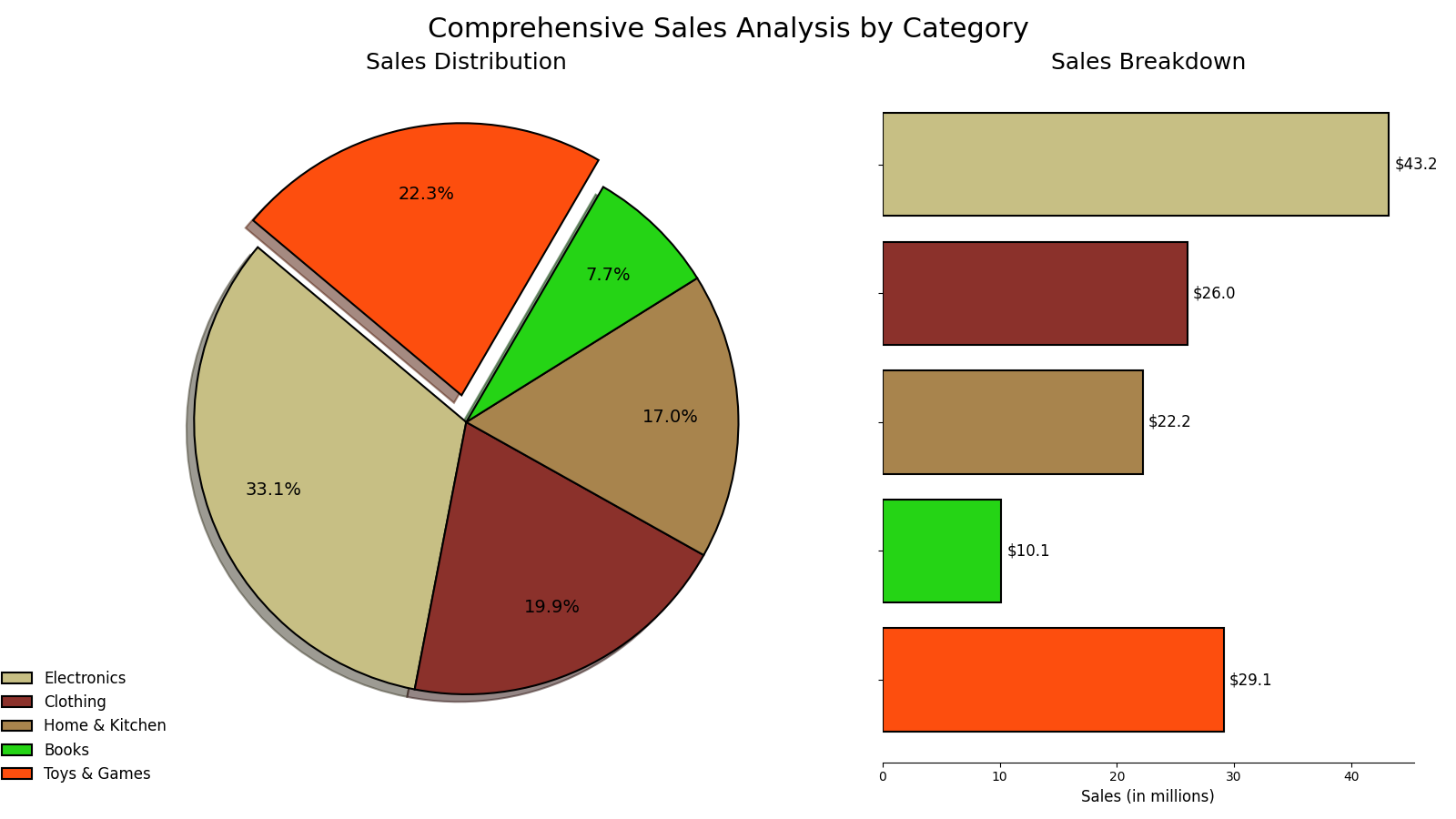}
            \\[0.02cm] \scriptsize (a) Ground Truth
        \end{minipage}
        \hfill
        \begin{minipage}[t]{0.48\linewidth}
            \centering
            \includegraphics[width=\linewidth,height=0.160\textheight,keepaspectratio]{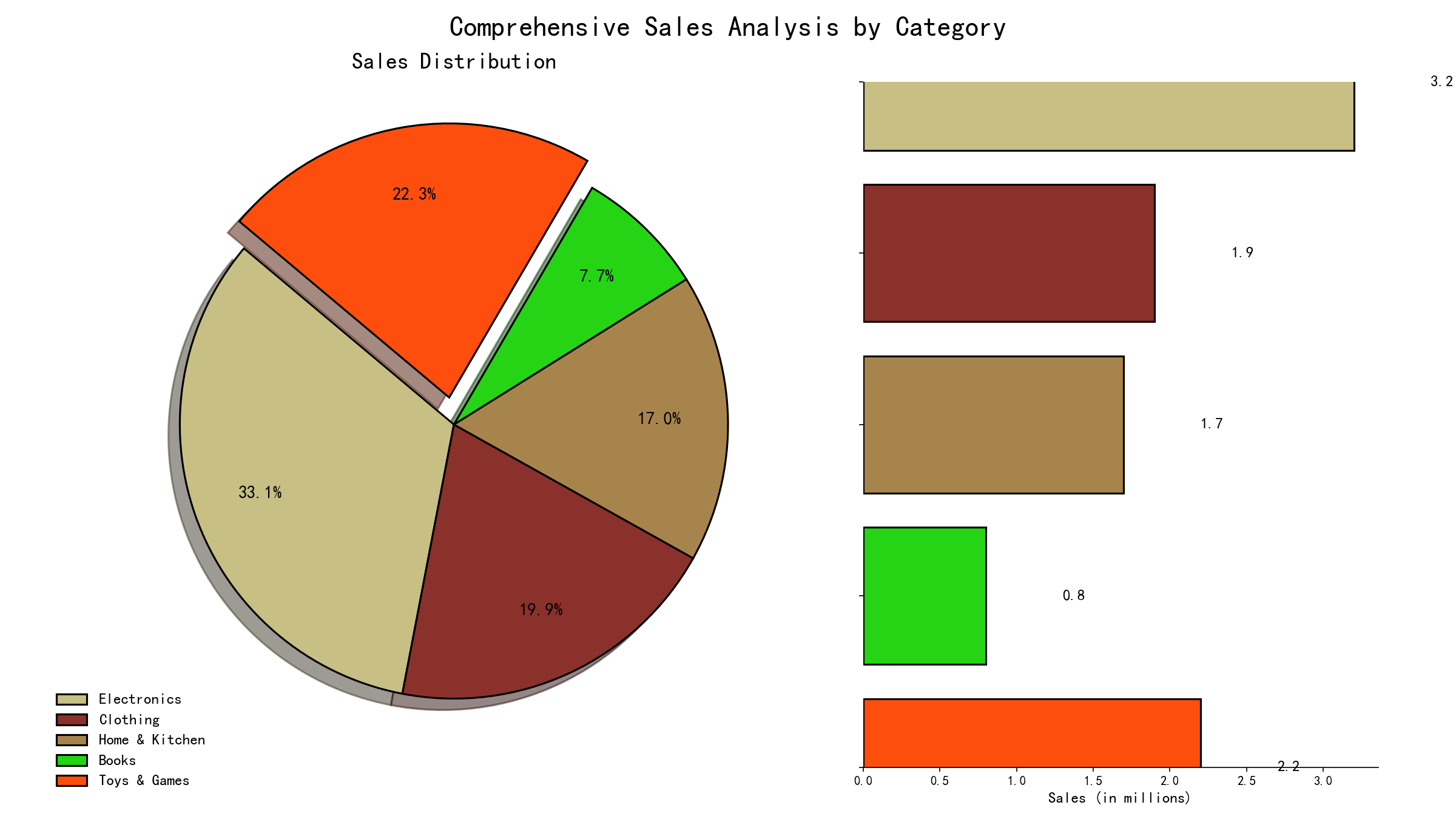}
            \\[0.02cm] \scriptsize (b) Model Output
        \end{minipage}
        
        \vspace{0.08cm}
        \begin{minipage}{0.96\linewidth}
            \footnotesize \textbf{Analysis:} The data distribution in the model output on the right is inconsistent with that of the GT Figure, indicating insufficient data accuracy and a deficiency in the model's ability to recognize and analyze the original chart's data.
        \end{minipage}
    \end{minipage}

    \errorcaseseparator

    \begin{minipage}{\linewidth}
        \centering
        \textbf{Case 2: Heatmap Chart (level3)}\\[0.08cm]
        
        \begin{minipage}[t]{0.48\linewidth}
            \centering
            \includegraphics[width=\linewidth,height=0.160\textheight,keepaspectratio]{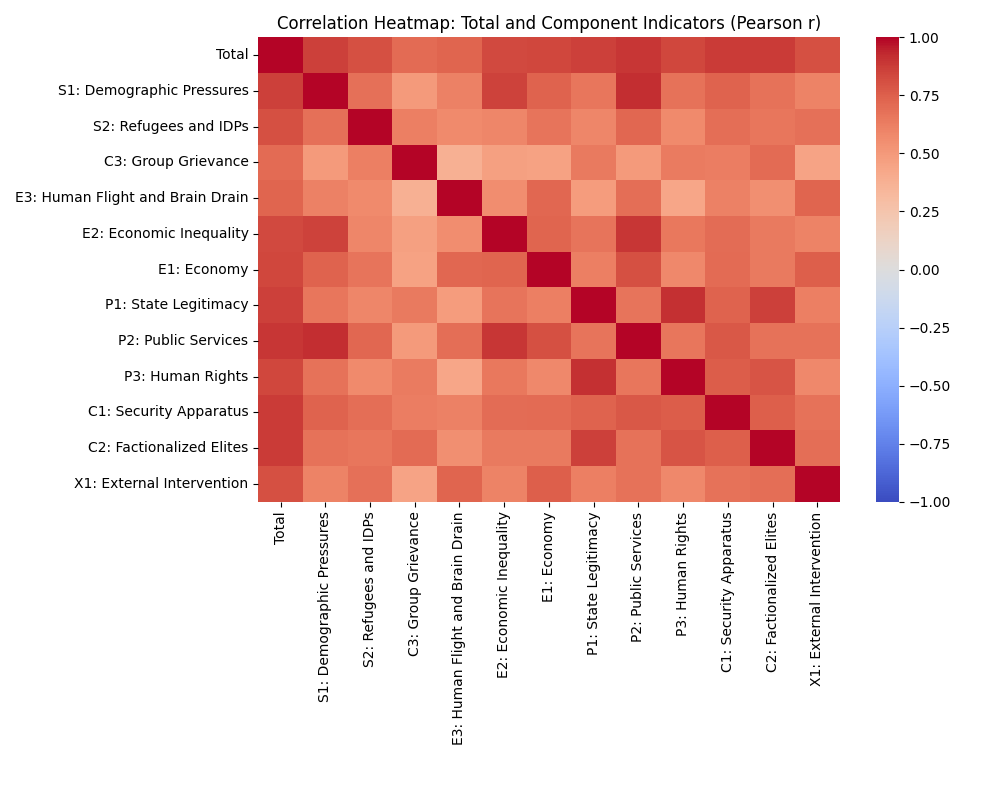}
            \\[0.02cm] \scriptsize (c) Ground Truth
        \end{minipage}
        \hfill
        \begin{minipage}[t]{0.48\linewidth}
            \centering
            \includegraphics[width=\linewidth,height=0.160\textheight,keepaspectratio]{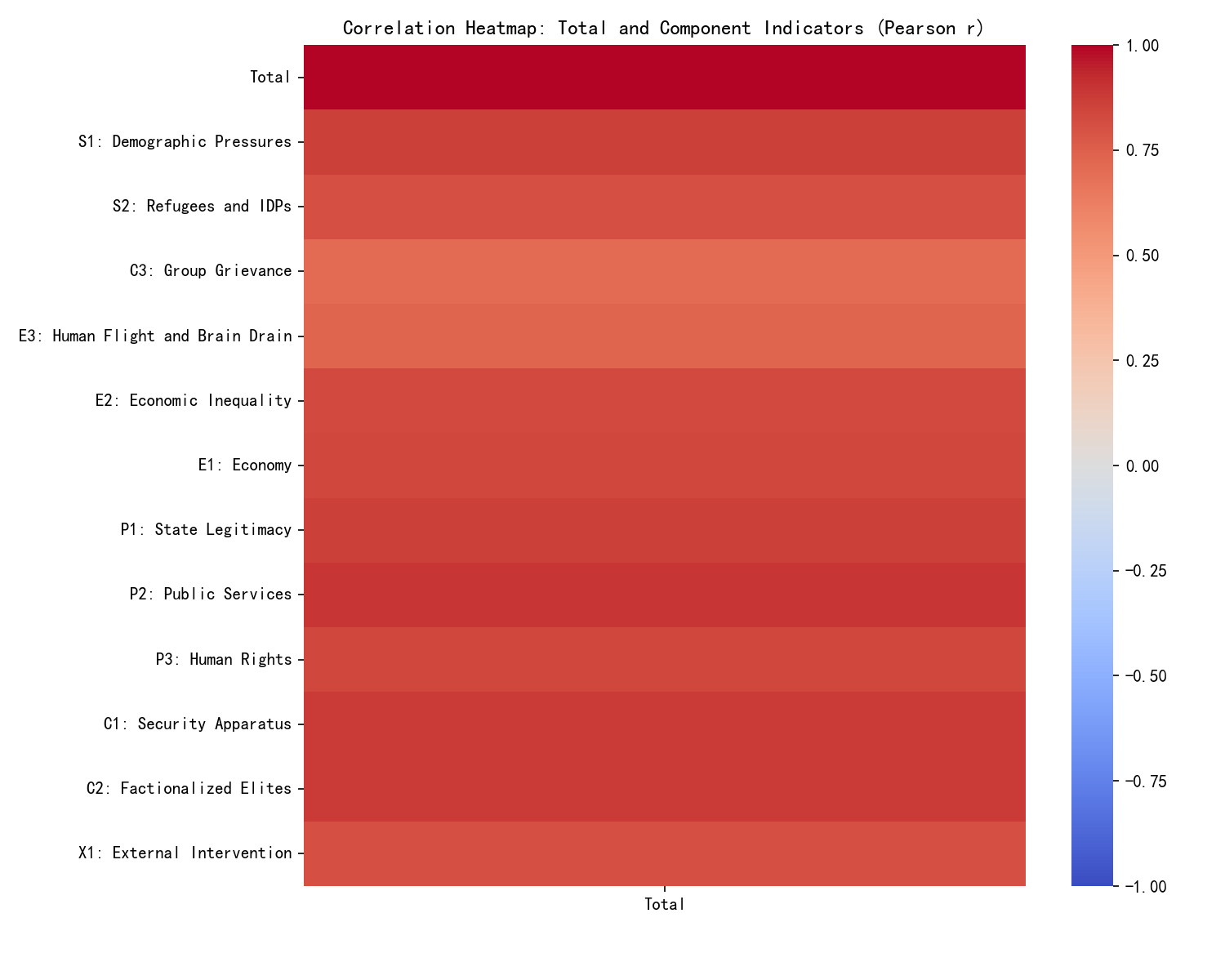}
            \\[0.02cm] \scriptsize (d) Model Output
        \end{minipage}
        
        \vspace{0.08cm}
        \begin{minipage}{0.96\linewidth}
            \footnotesize \textbf{Analysis:} In the generated heatmap, the model only reproduced the data for the "Total" indicator, completely omitting all other component indicators (e.g., S1, S2, C3, E3, etc.), which reflects a critical failure in data extraction and comprehensive information reproduction capabilities.
        \end{minipage}
    \end{minipage}

    \errorcaseseparator
    
    \begin{minipage}{\linewidth}
        \centering
        \textbf{Case 3: Combination Chart (level3)}\\[0.08cm]
        
        \begin{minipage}[t]{0.48\linewidth}
            \centering
            \includegraphics[width=\linewidth,height=0.160\textheight,keepaspectratio]{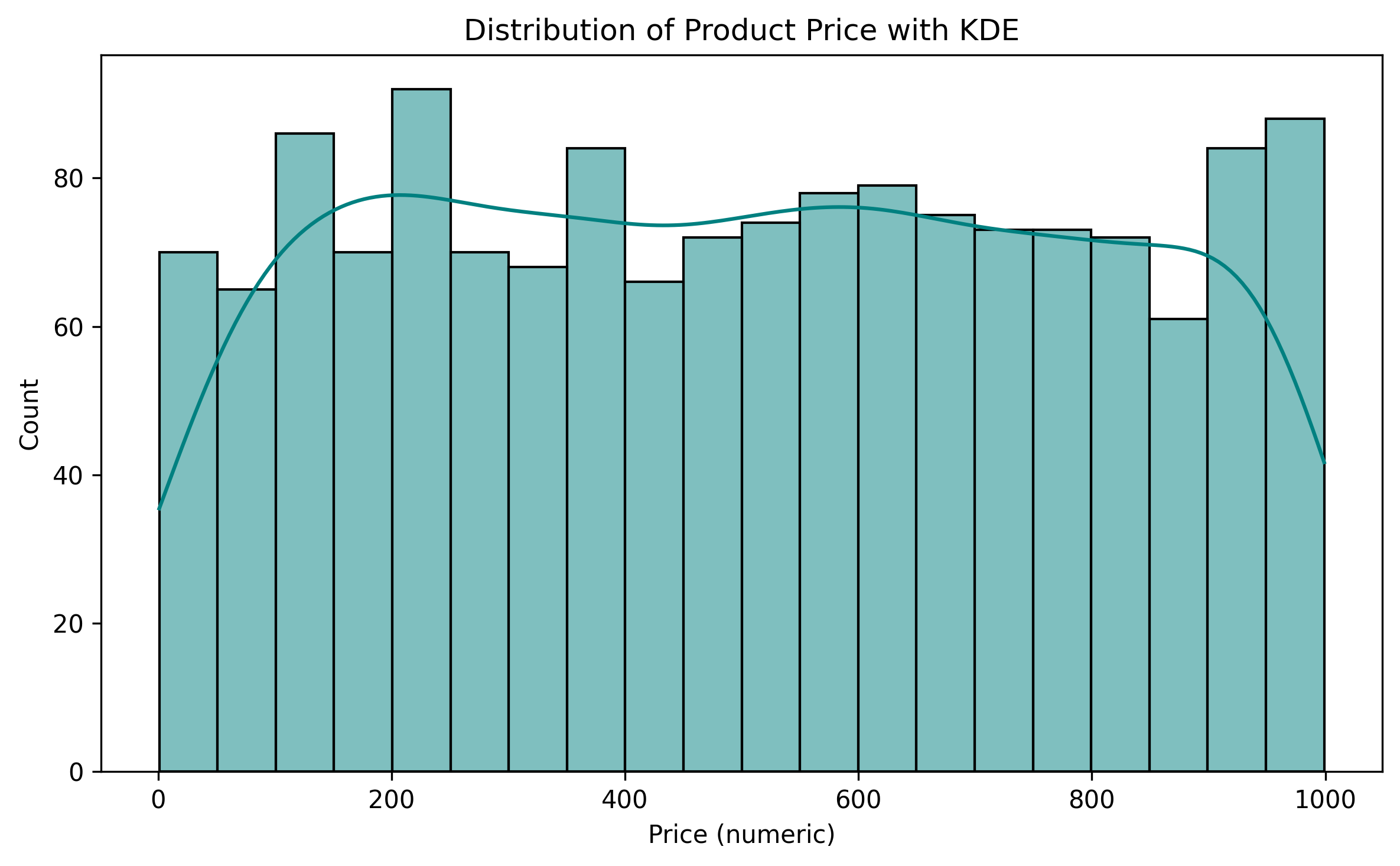}
            \\[0.02cm] \scriptsize (e) Ground Truth
        \end{minipage}
        \hfill
        \begin{minipage}[t]{0.48\linewidth}
            \centering
            \includegraphics[width=\linewidth,height=0.160\textheight,keepaspectratio]{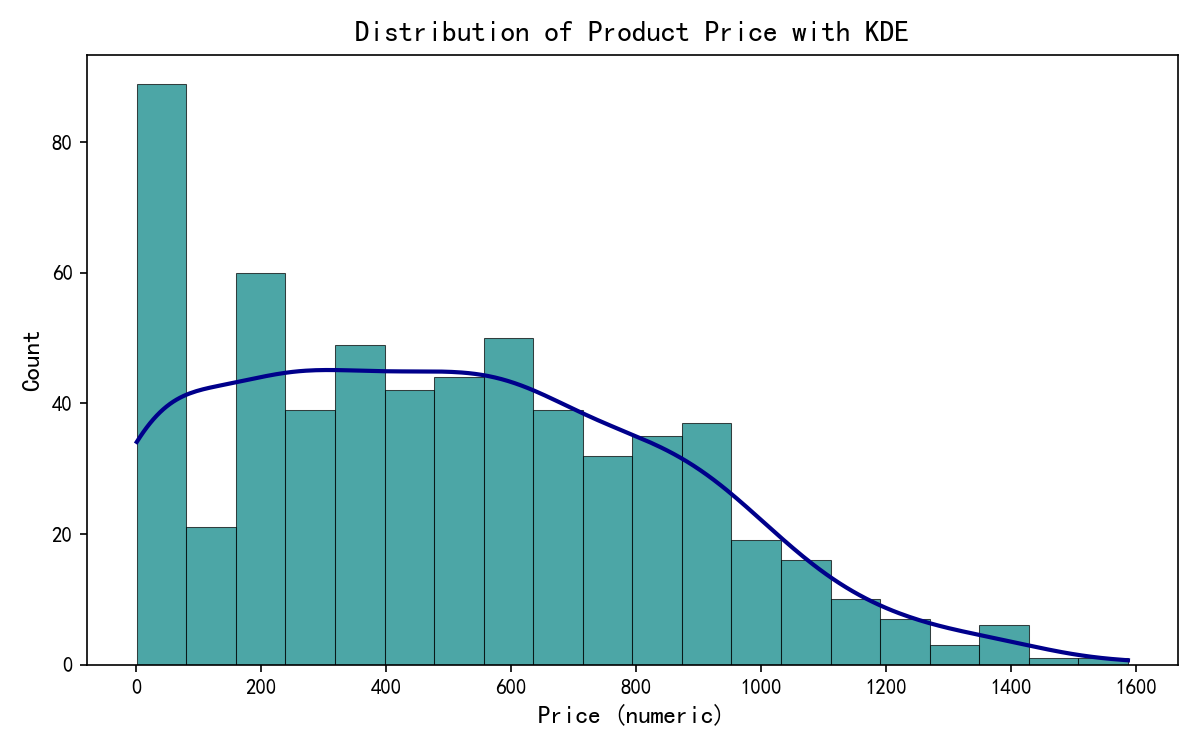}
            \\[0.02cm] \scriptsize (f) Model Output
        \end{minipage}
        
        \vspace{0.08cm}
        \begin{minipage}{0.96\linewidth}
            \footnotesize \textbf{Analysis:} The model made critical errors in data extraction, failing to accurately reproduce the histogram bar heights, KDE curve shape, and X-axis range from the ground truth, resulting in a fundamentally incorrect data distribution.
        \end{minipage}
    \end{minipage}
    
    \vfill
    \caption{\textbf{Qualitative Error Analysis.} Model outputs exhibiting data-related inaccuracies, such as incorrect data extraction and erroneous data distributions, leading to poor visual fidelity.
}
    \label{fig:data encoding errors 1}
\end{compacterrorpage}

\begin{compacterrorpage}
    \centering
    \appendixplatetitle{Error Category: Data Encoding Errors}
    \vspace{1.0mm}
    \begin{minipage}{\linewidth}
        \centering
        \textbf{Case 1: Radar Chart (level1)}\\[0.08cm]
        
        \begin{minipage}[t]{0.48\linewidth}
            \centering
            \includegraphics[width=\linewidth,height=0.160\textheight,keepaspectratio]{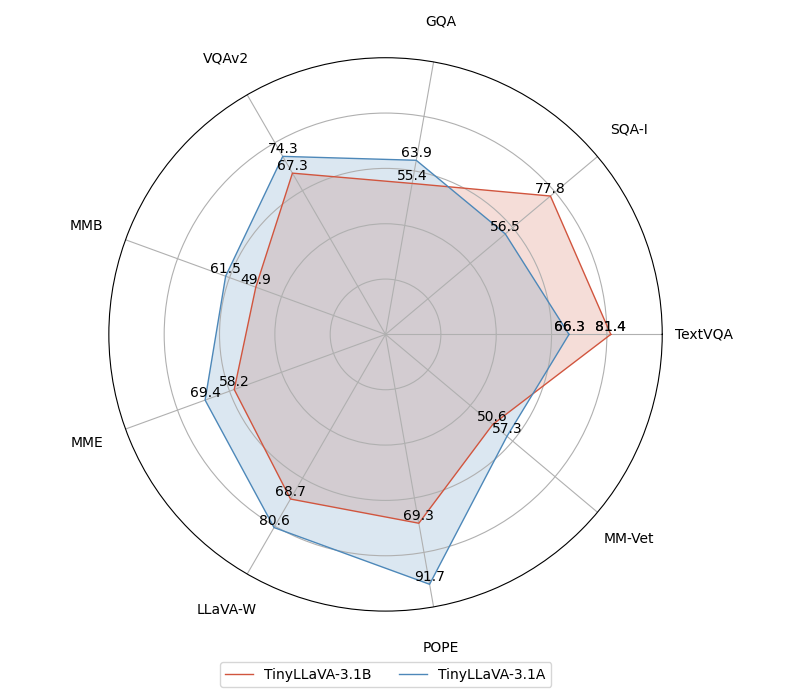}
            \\[0.02cm] \scriptsize (a) Ground Truth
        \end{minipage}
        \hfill
        \begin{minipage}[t]{0.48\linewidth}
            \centering
            \includegraphics[width=\linewidth,height=0.160\textheight,keepaspectratio]{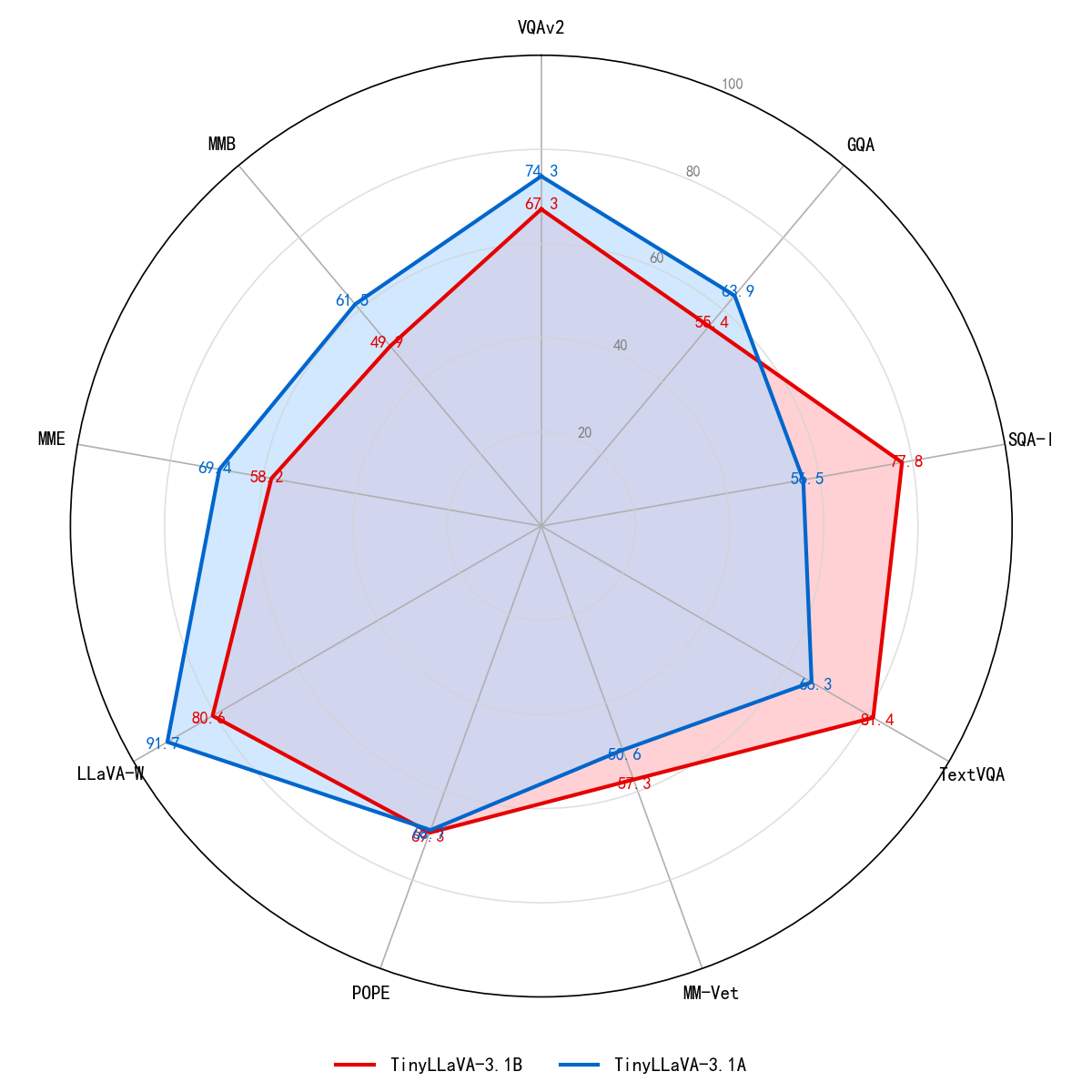}
            \\[0.02cm] \scriptsize (b) Model Output
        \end{minipage}
        
        \vspace{0.08cm}
        \begin{minipage}{0.96\linewidth}
            \footnotesize \textbf{Analysis:} The data distribution in the model output on the right is inconsistent with that of the GT Figure, indicating insufficient data accuracy and a deficiency in the model's ability to recognize and analyze the original chart's data.
        \end{minipage}
    \end{minipage}

    \errorcaseseparator

    \begin{minipage}{\linewidth}
        \centering
        \textbf{Case 2: Combination Chart (level1)}\\[0.08cm]
        
        \begin{minipage}[t]{0.48\linewidth}
            \centering
            \includegraphics[width=\linewidth,height=0.160\textheight,keepaspectratio]{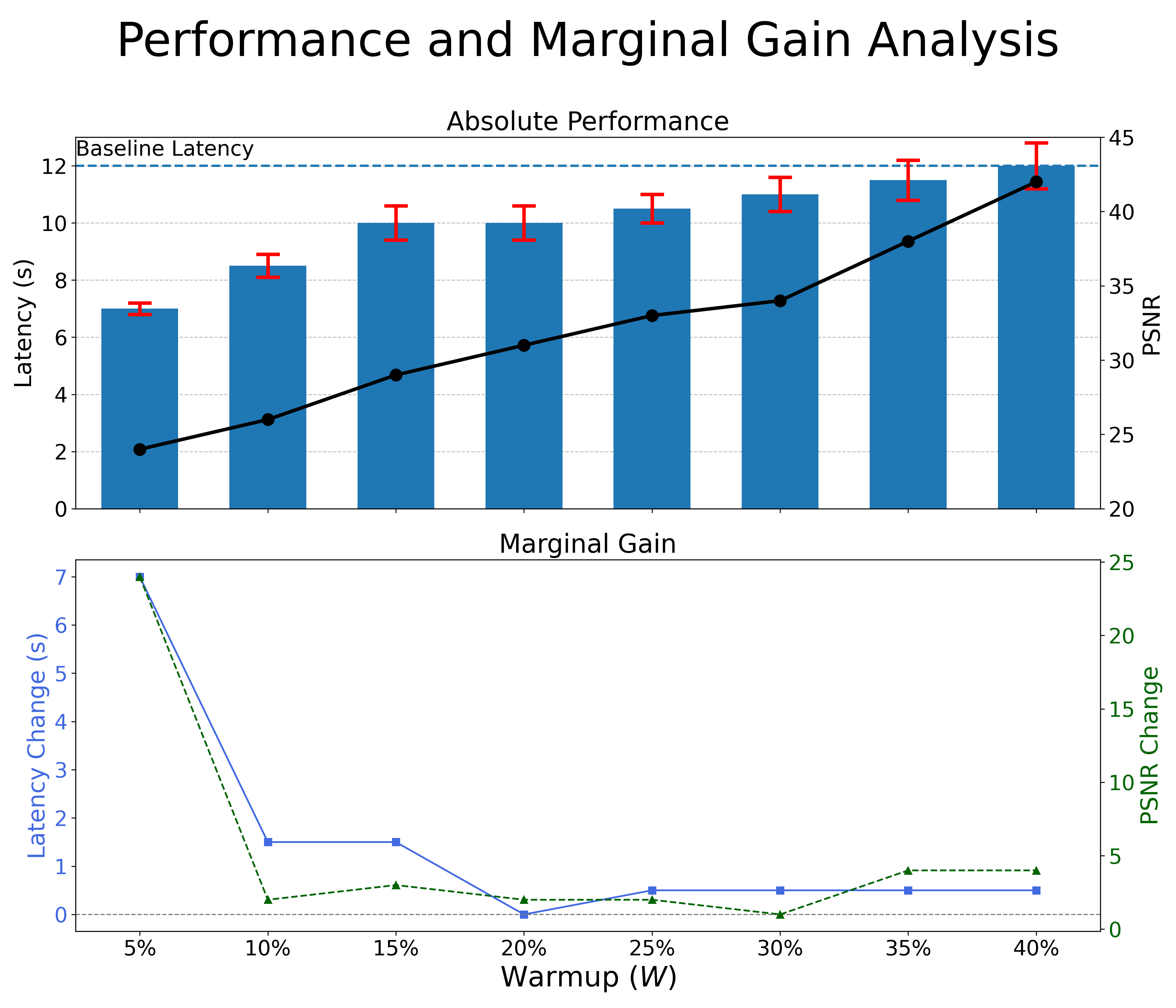}
            \\[0.02cm] \scriptsize (c) Ground Truth
        \end{minipage}
        \hfill
        \begin{minipage}[t]{0.48\linewidth}
            \centering
            \includegraphics[width=\linewidth,height=0.160\textheight,keepaspectratio]{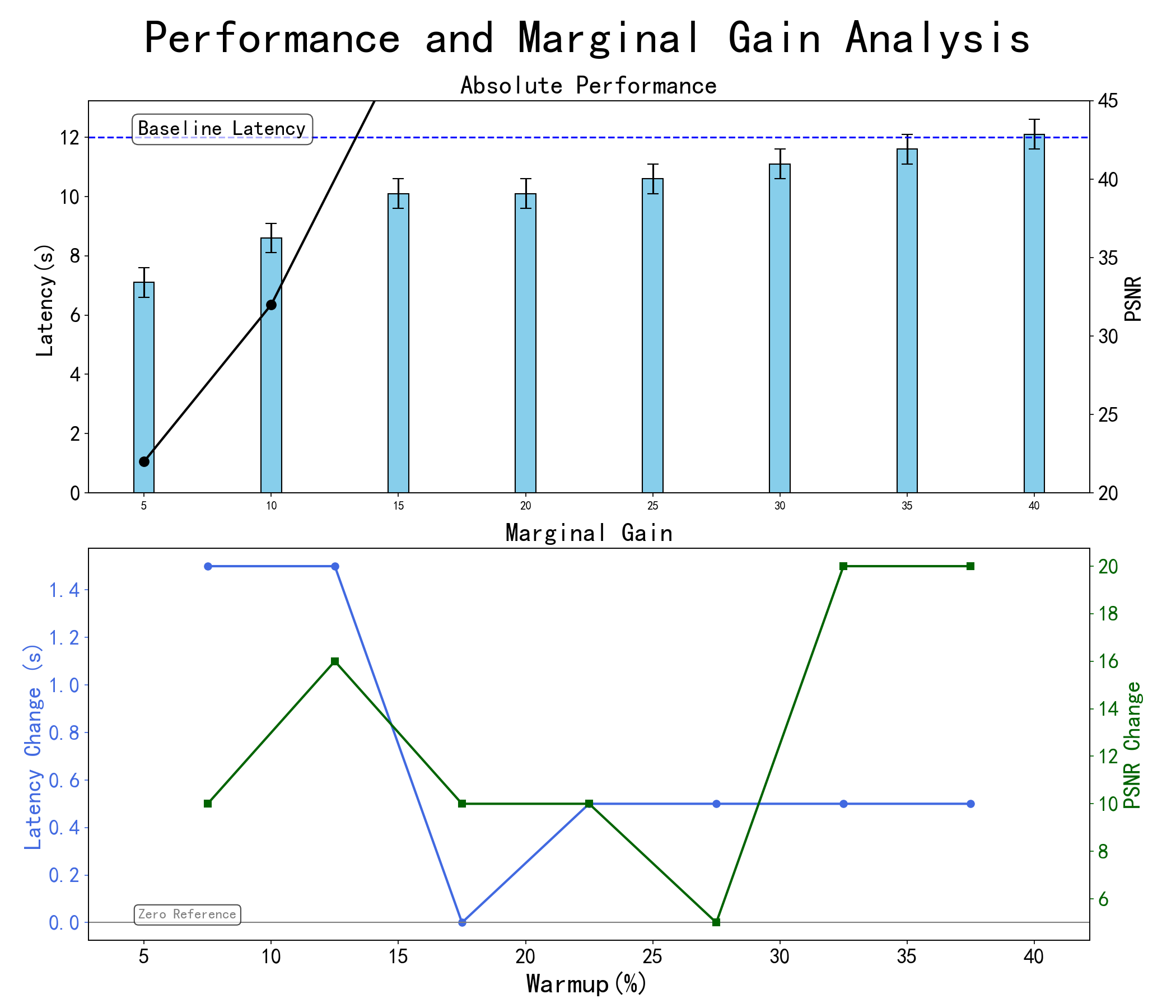}
            \\[0.02cm] \scriptsize (d) Model Output
        \end{minipage}
        
        \vspace{0.08cm}
        \begin{minipage}{0.96\linewidth}
            \footnotesize \textbf{Analysis:} The data distribution in the model output on the right is inconsistent with that of the GT Figure, indicating insufficient data accuracy and a deficiency in the model's ability to recognize and analyze the original chart's data.
        \end{minipage}
    \end{minipage}

    \errorcaseseparator
    
    \begin{minipage}{\linewidth}
        \centering
        \textbf{Case 3: Combination Chart (level1)}\\[0.08cm]
        
        \begin{minipage}[t]{0.48\linewidth}
            \centering
            \includegraphics[width=\linewidth,height=0.160\textheight,keepaspectratio]{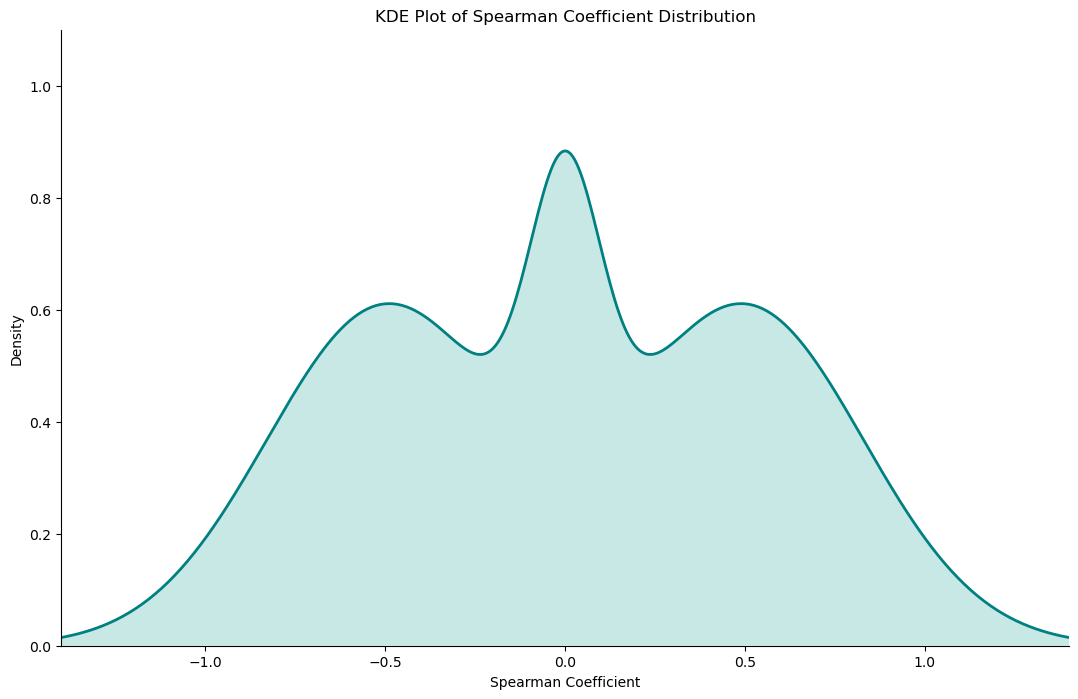}
            \\[0.02cm] \scriptsize (e) Ground Truth
        \end{minipage}
        \hfill
        \begin{minipage}[t]{0.48\linewidth}
            \centering
            \includegraphics[width=\linewidth,height=0.160\textheight,keepaspectratio]{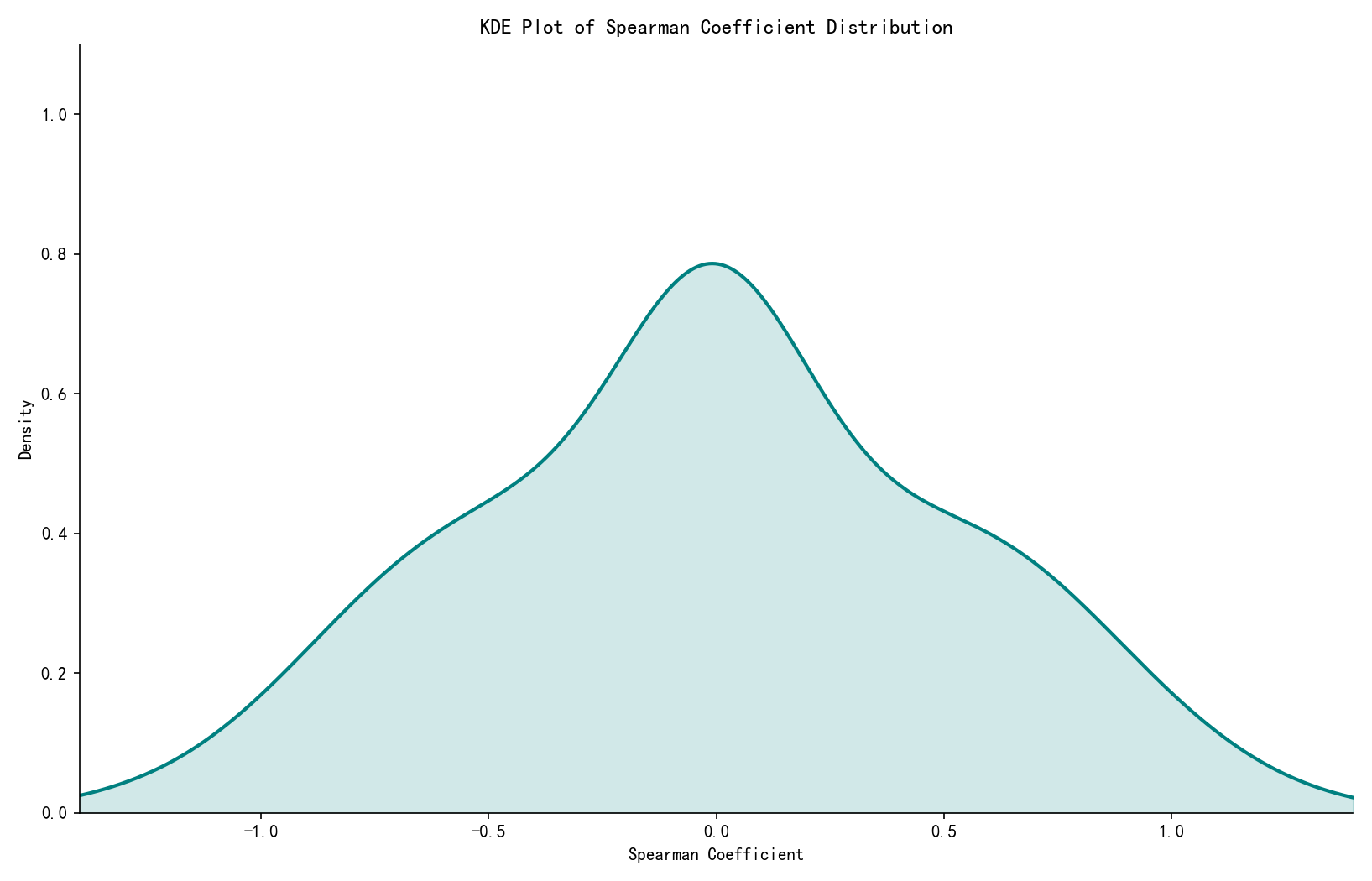}
            \\[0.02cm] \scriptsize (f) Model Output
        \end{minipage}
        
        \vspace{0.08cm}
        \begin{minipage}{0.96\linewidth}
            \footnotesize \textbf{Analysis:} The data distribution in the model output on the right is inconsistent with that of the GT Figure, indicating insufficient data accuracy and a deficiency in the model's ability to recognize and analyze the original chart's data.
        \end{minipage}
    \end{minipage}
    
    \vfill
    \caption{\textbf{Qualitative Error Analysis.} Examples of data misrepresentation in model-generated charts, including extraction mistakes and incorrect value distributions, resulting in degraded visual fidelity.}
    \label{fig:data encoding errors 2}
\end{compacterrorpage}

\begin{compacterrorpage}
    \centering
    \appendixplatetitle{Common Error Category: Axis Scale Errors}
    \vspace{1.0mm}
    \begin{minipage}{\linewidth}
        \centering
        \textbf{Case 1: Bar Chart (level1)}\\[0.08cm]
        
        \begin{minipage}[t]{0.48\linewidth}
            \centering
            \includegraphics[width=\linewidth,height=0.160\textheight,keepaspectratio]{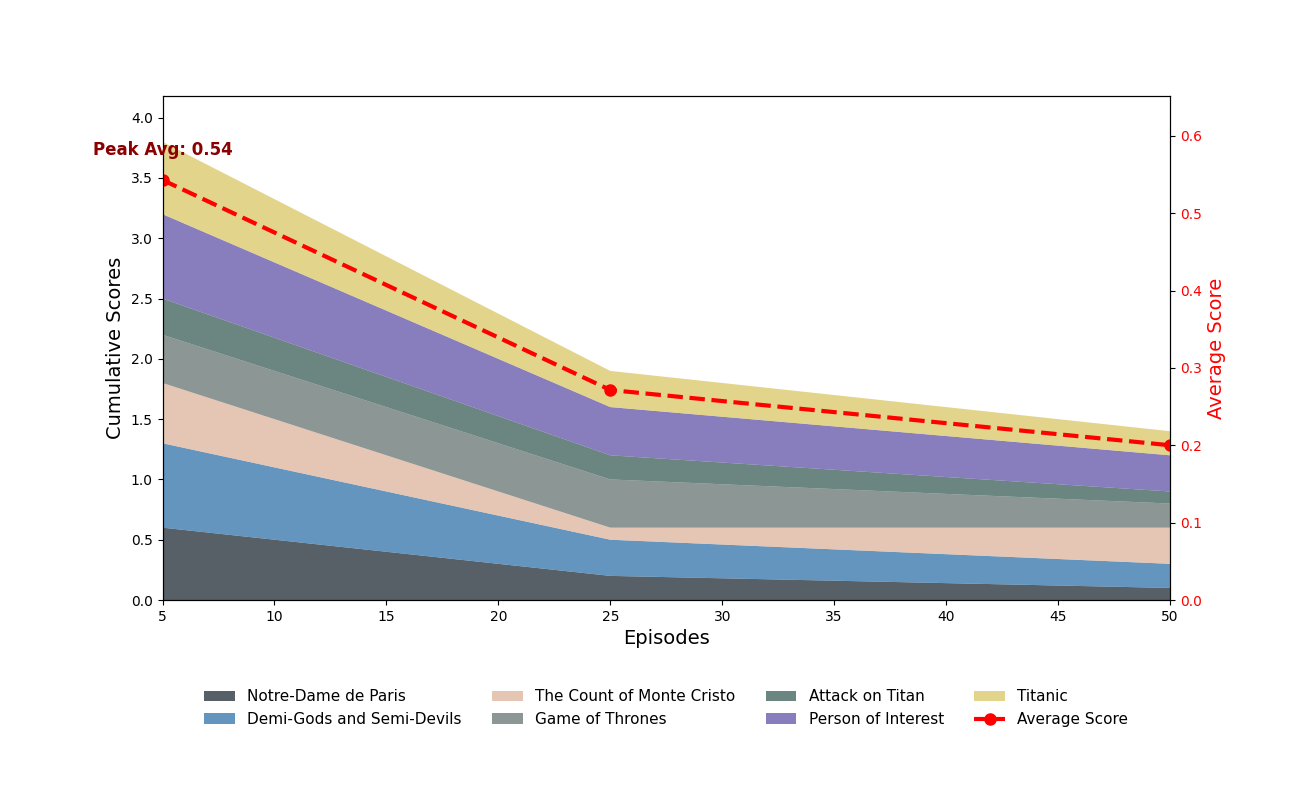}
            \\[0.02cm] \scriptsize (a) Ground Truth
        \end{minipage}
        \hfill
        \begin{minipage}[t]{0.48\linewidth}
            \centering
            \includegraphics[width=\linewidth,height=0.160\textheight,keepaspectratio]{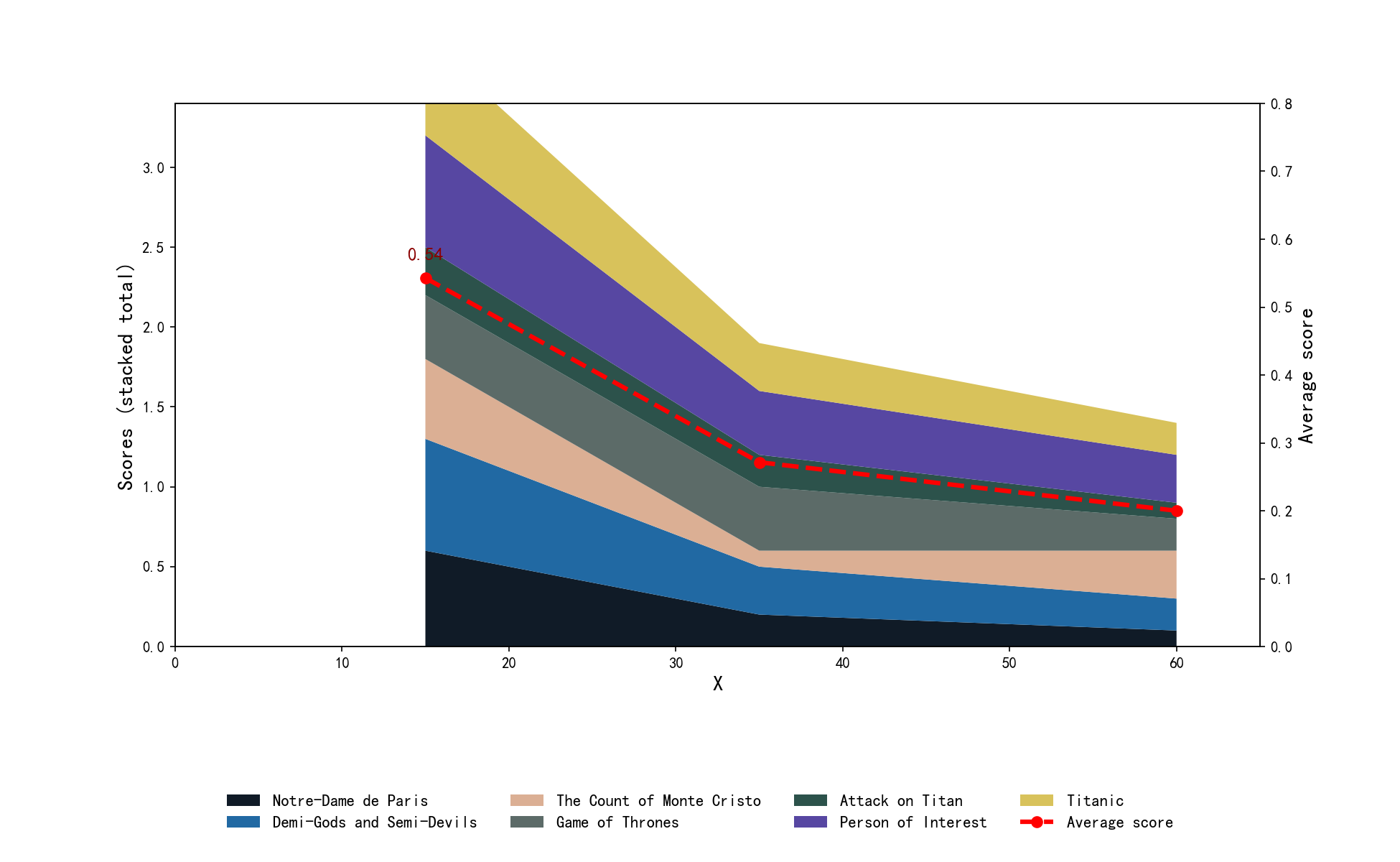}
            \\[0.02cm] \scriptsize (b) Model Output
        \end{minipage}
        
        \vspace{0.08cm}
        \begin{minipage}{0.96\linewidth}
            \footnotesize \textbf{Analysis:} In the model-generated chart, the y-axis range is incorrectly set to 0–3.5, which is narrower than the 0–4 range in the ground truth, resulting in the upper portion of the stacked area chart being truncated. Additionally, the leftmost segment of the chart is missing. These issues indicate deficiencies in the model’s data extraction capability and its ability to accurately infer the underlying numerical scale.
        \end{minipage}
    \end{minipage}

    \errorcaseseparator

    \begin{minipage}{\linewidth}
        \centering
        \textbf{Case 2: Heatmap Chart (level2)}\\[0.08cm]
        
        \begin{minipage}[t]{0.48\linewidth}
            \centering
            \includegraphics[width=\linewidth,height=0.160\textheight,keepaspectratio]{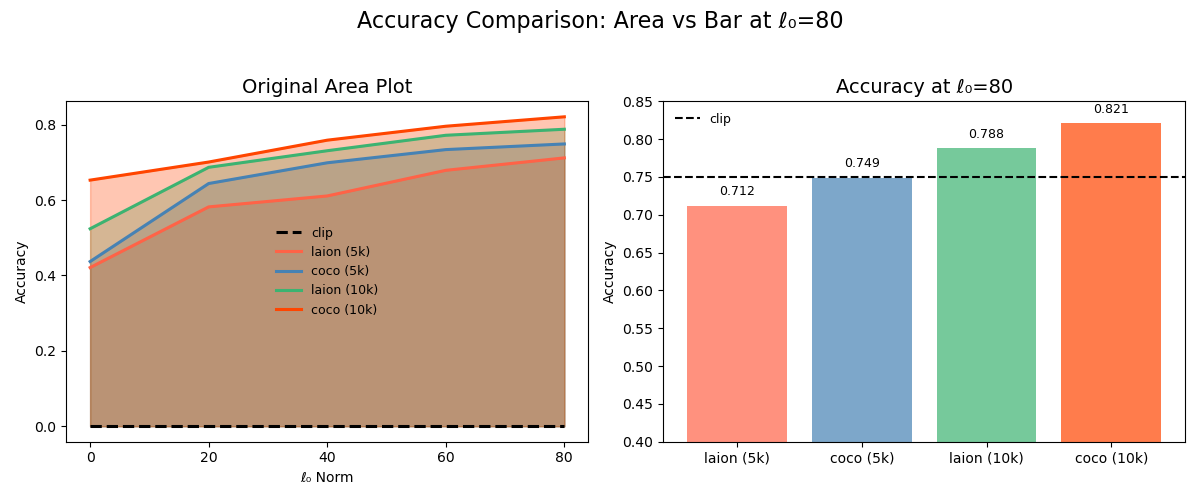}
            \\[0.02cm] \scriptsize (c) Ground Truth
        \end{minipage}
        \hfill
        \begin{minipage}[t]{0.48\linewidth}
            \centering
            \includegraphics[width=\linewidth,height=0.160\textheight,keepaspectratio]{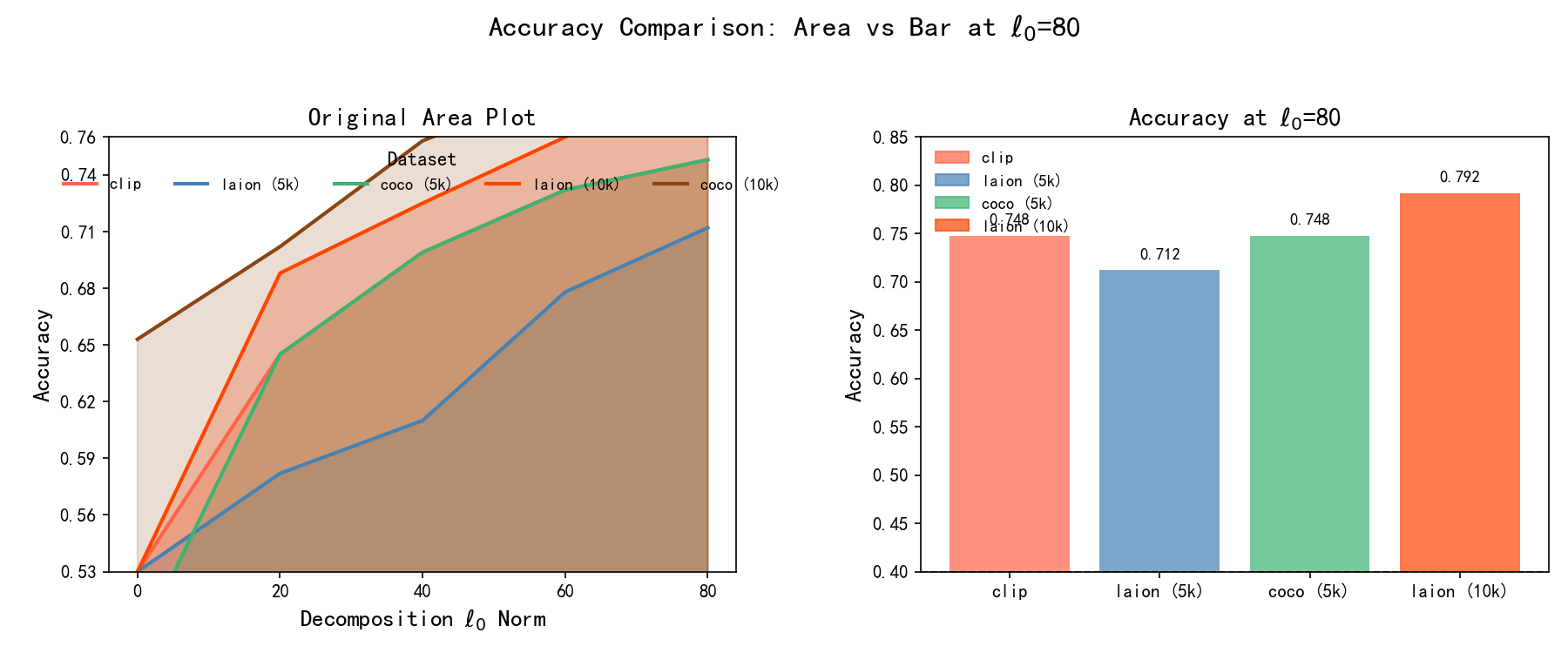}
            \\[0.02cm] \scriptsize (d) Model Output
        \end{minipage}
        
        \vspace{0.08cm}
        \begin{minipage}{0.96\linewidth}
            \footnotesize \textbf{Analysis:} In the generated chart, the Y-axis range of the left subplot is incorrectly compressed to 0.53–0.76, and the legend is misplaced over the data area, both of which severely distort the data representation and visual clarity compared to the ground truth.
        \end{minipage}
    \end{minipage}

    \errorcaseseparator
    
    \begin{minipage}{\linewidth}
        \centering
        \textbf{Case 3: Combination Chart (level2)}\\[0.08cm]
        
        \begin{minipage}[t]{0.48\linewidth}
            \centering
            \includegraphics[width=\linewidth,height=0.160\textheight,keepaspectratio]{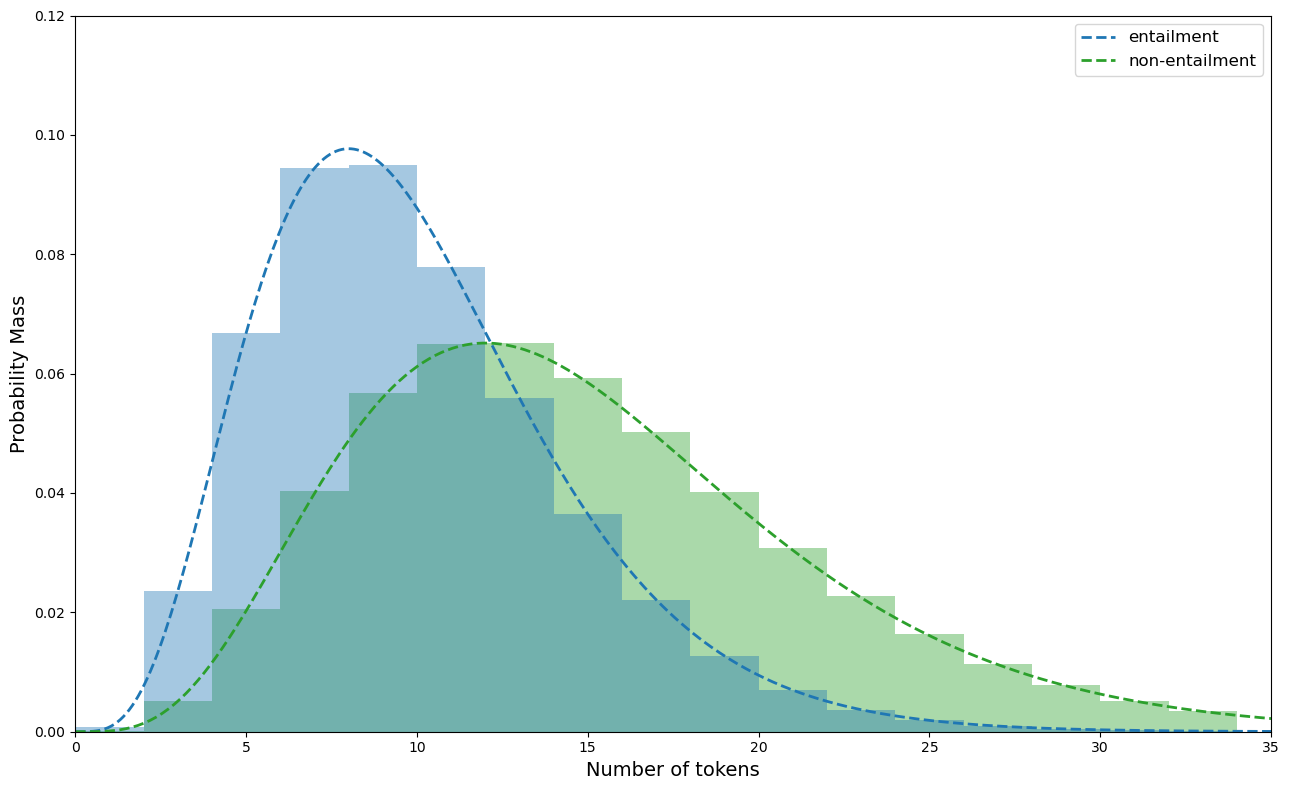}
            \\[0.02cm] \scriptsize (e) Ground Truth
        \end{minipage}
        \hfill
        \begin{minipage}[t]{0.48\linewidth}
            \centering
            \includegraphics[width=\linewidth,height=0.160\textheight,keepaspectratio]{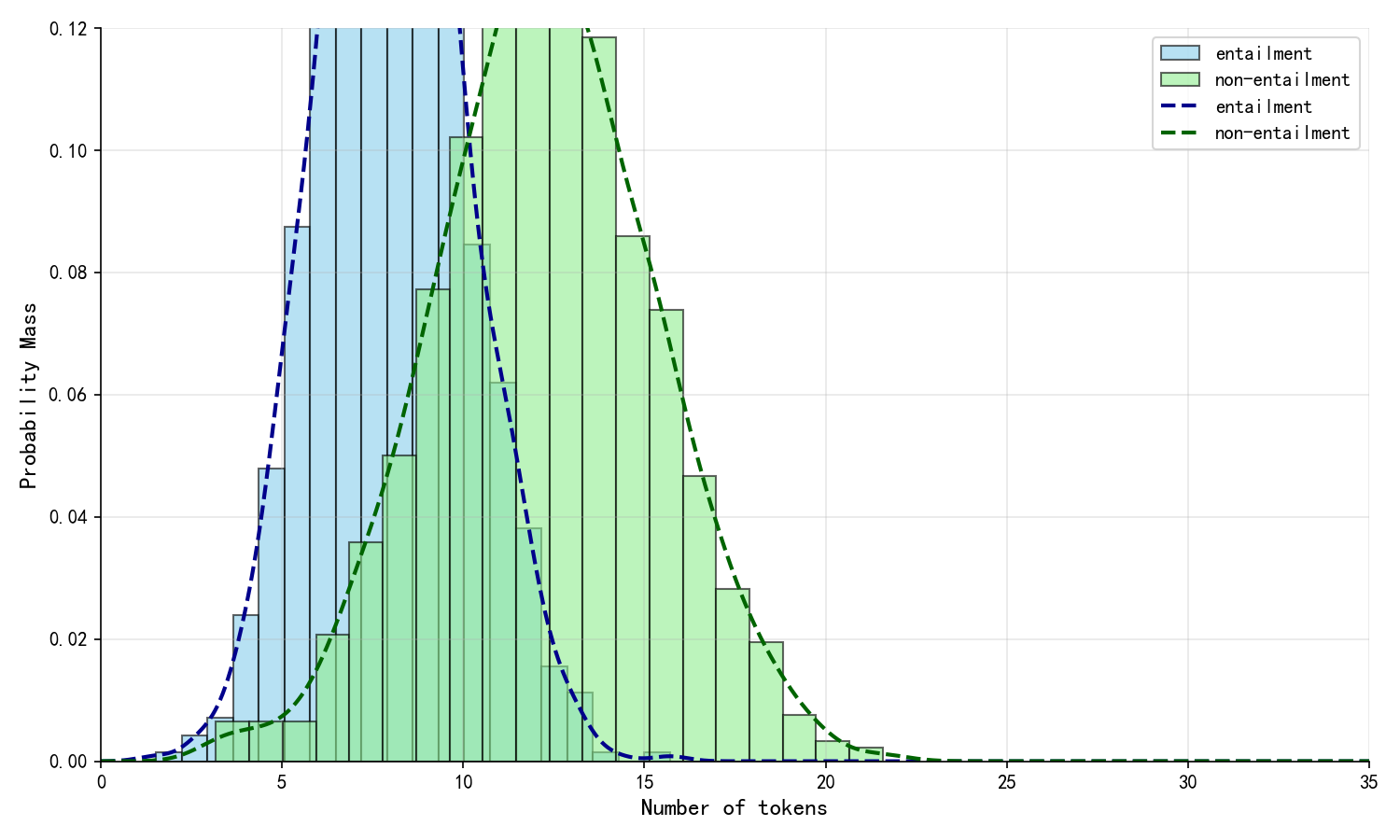}
            \\[0.02cm] \scriptsize (f) Model Output
        \end{minipage}
        
        \vspace{0.08cm}
        \begin{minipage}{0.96\linewidth}
            \footnotesize \textbf{Analysis:} The y‑axis range in the model output on the right is problematic, causing part of the data to be invisible. This indicates an insufficient match between the axis range and the data. Additionally, errors occurred in data recognition and analysis.
        \end{minipage}
    \end{minipage}
    
    \vfill
    \caption{\textbf{Qualitative Error Analysis.} Scaling inconsistencies in the model outputs, leading to issues such as the disappearance of primary chart elements or disproportionate visual layouts, thereby reducing overall visual fidelity.
}
    \label{fig:axis scale errors}
\end{compacterrorpage}

\begin{compacterrorpage}
    \centering
    \appendixplatetitle{Common Error Category: Type Errors}
    \vspace{1.0mm}
    \begin{minipage}{\linewidth}
        \centering
        \textbf{Case 1: Combination Chart (level1)}\\[0.08cm]
        
        \begin{minipage}[t]{0.48\linewidth}
            \centering
            \includegraphics[width=\linewidth,height=0.160\textheight,keepaspectratio]{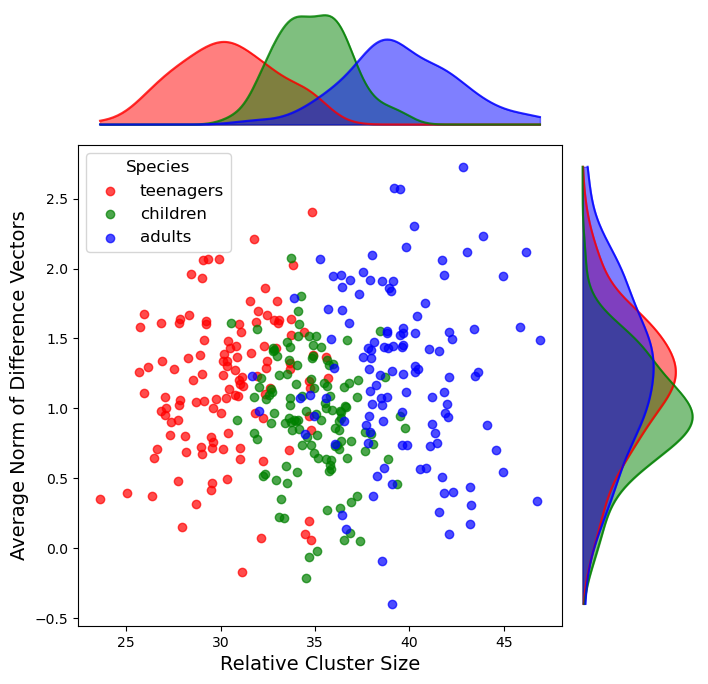}
            \\[0.02cm] \scriptsize (a) Ground Truth
        \end{minipage}
        \hfill
        \begin{minipage}[t]{0.48\linewidth}
            \centering
            \includegraphics[width=\linewidth,height=0.160\textheight,keepaspectratio]{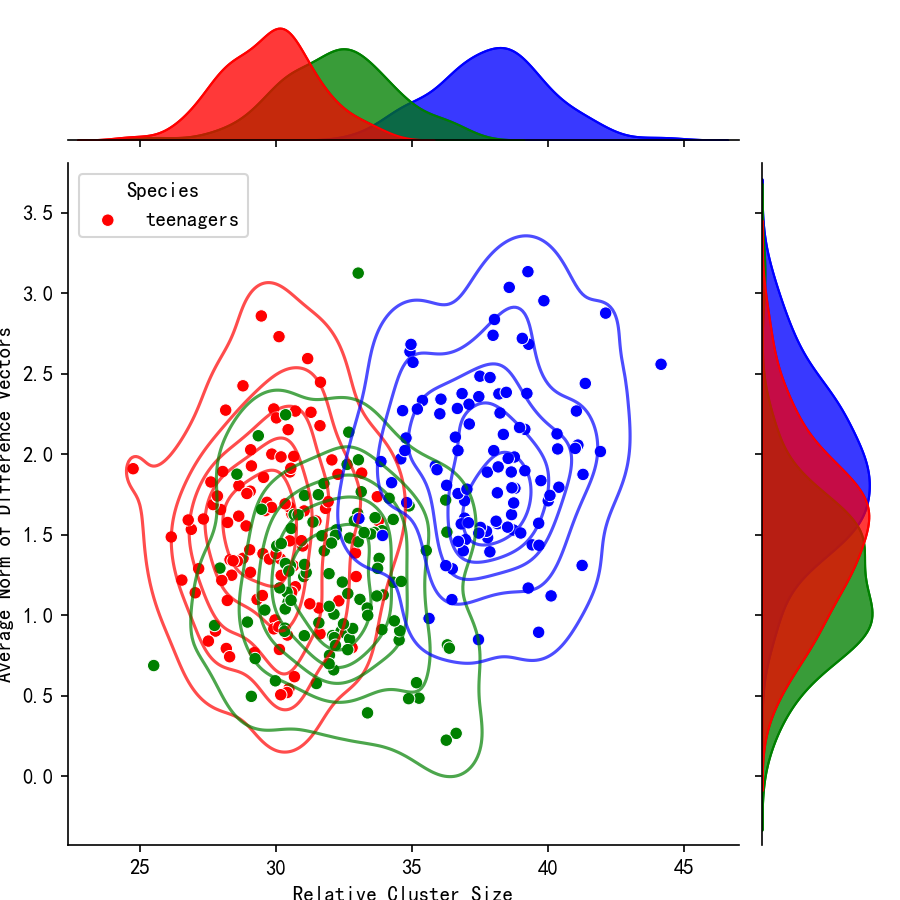}
            \\[0.02cm] \scriptsize (b) Model Output
        \end{minipage}
        
        \vspace{0.08cm}
        \begin{minipage}{0.96\linewidth}
            \footnotesize \textbf{Analysis:} The chart type in the model output on the right is inconsistent with that of the GT Figure; the scatter plot has been misidentified as a contour plot.
        \end{minipage}
    \end{minipage}

    \errorcaseseparator

    \begin{minipage}{\linewidth}
        \centering
        \textbf{Case 2: Bar Chart (level1)}\\[0.08cm]
        
        \begin{minipage}[t]{0.48\linewidth}
            \centering
            \includegraphics[width=\linewidth,height=0.160\textheight,keepaspectratio]{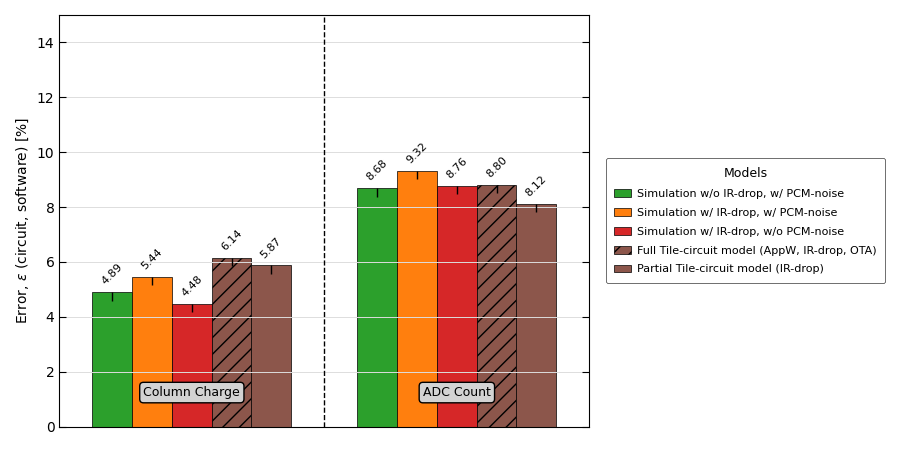}
            \\[0.02cm] \scriptsize (c) Ground Truth
        \end{minipage}
        \hfill
        \begin{minipage}[t]{0.48\linewidth}
            \centering
            \includegraphics[width=\linewidth,height=0.160\textheight,keepaspectratio]{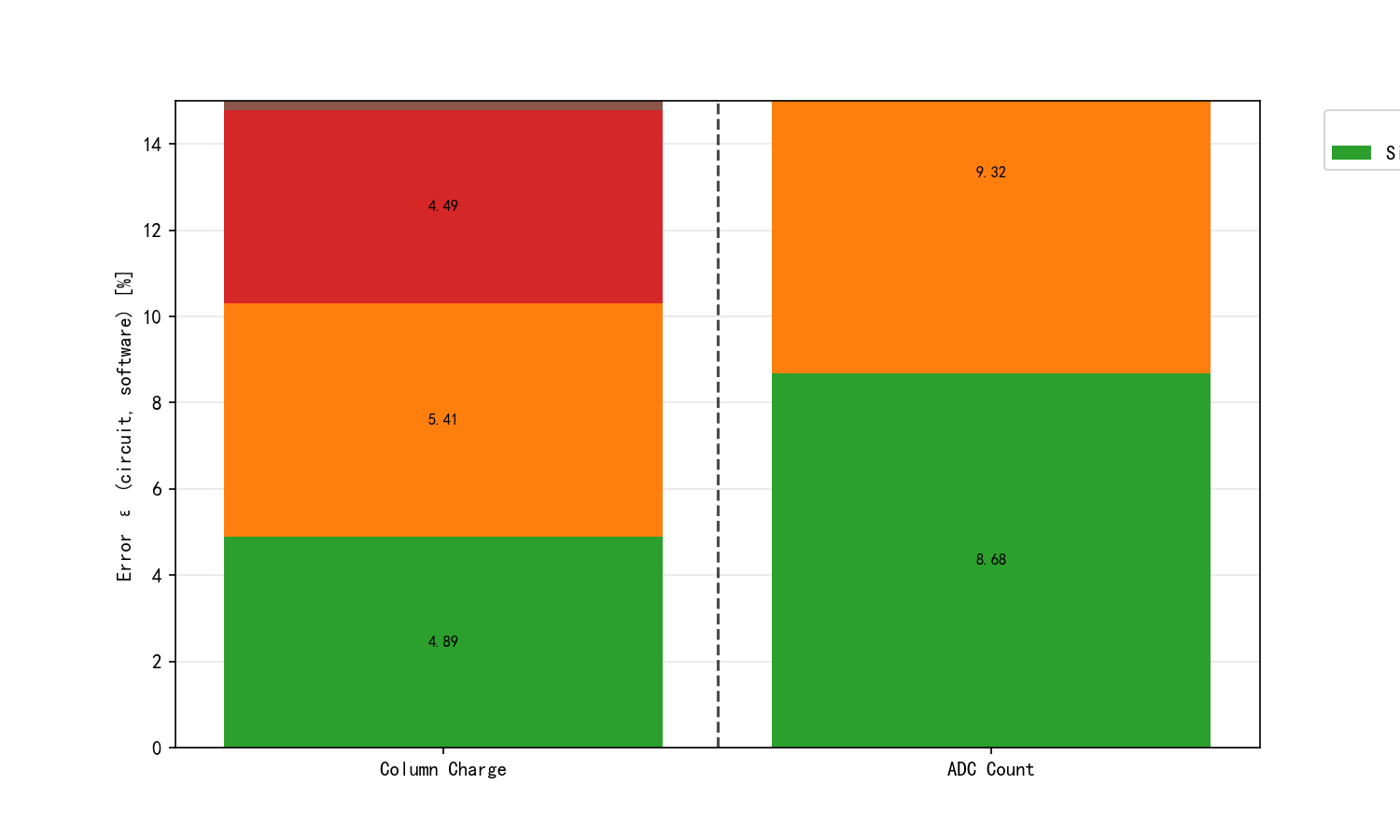}
            \\[0.02cm] \scriptsize (d) Model Output
        \end{minipage}
        
        \vspace{0.08cm}
        \begin{minipage}{0.96\linewidth}
            \footnotesize \textbf{Analysis:} The chart type in the model output on the right is inconsistent with the GT Figure; the regular bar chart has been misidentified as a stacked bar chart, which deviates significantly from the original.
        \end{minipage}
    \end{minipage}

    \errorcaseseparator
    
    \begin{minipage}{\linewidth}
        \centering
        \textbf{Case 3: Combination Chart (level2)}\\[0.08cm]
        
        \begin{minipage}[t]{0.48\linewidth}
            \centering
            \includegraphics[width=\linewidth,height=0.160\textheight,keepaspectratio]{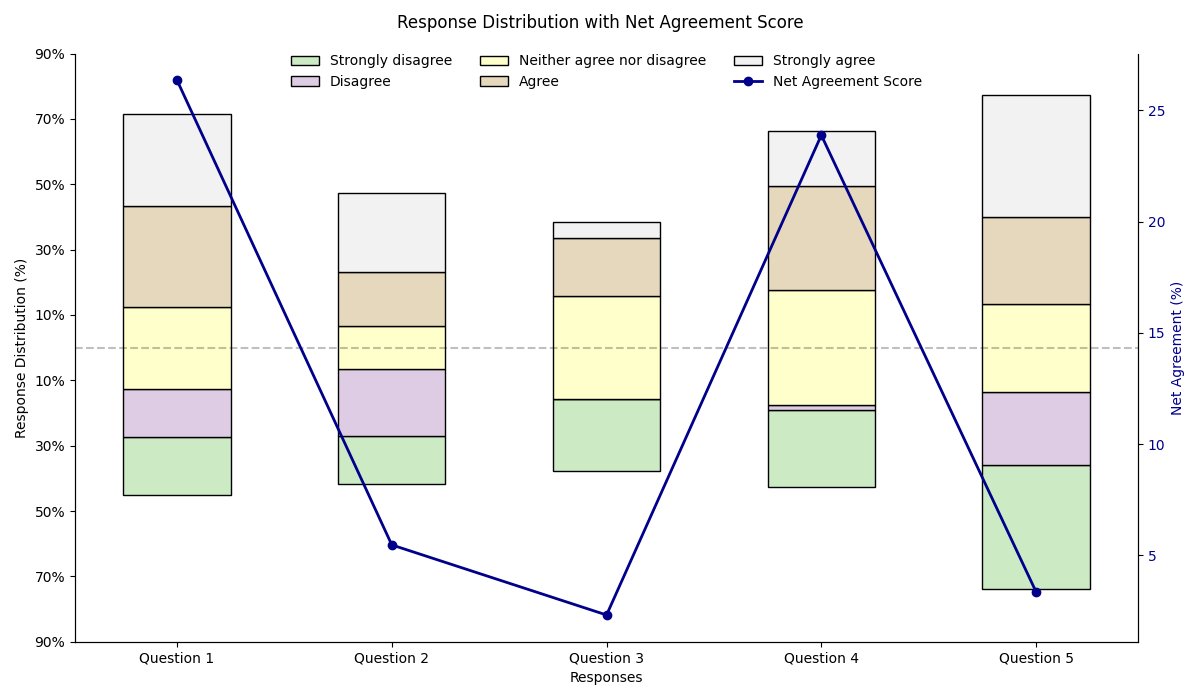}
            \\[0.02cm] \scriptsize (e) Ground Truth
        \end{minipage}
        \hfill
        \begin{minipage}[t]{0.48\linewidth}
            \centering
            \includegraphics[width=\linewidth,height=0.160\textheight,keepaspectratio]{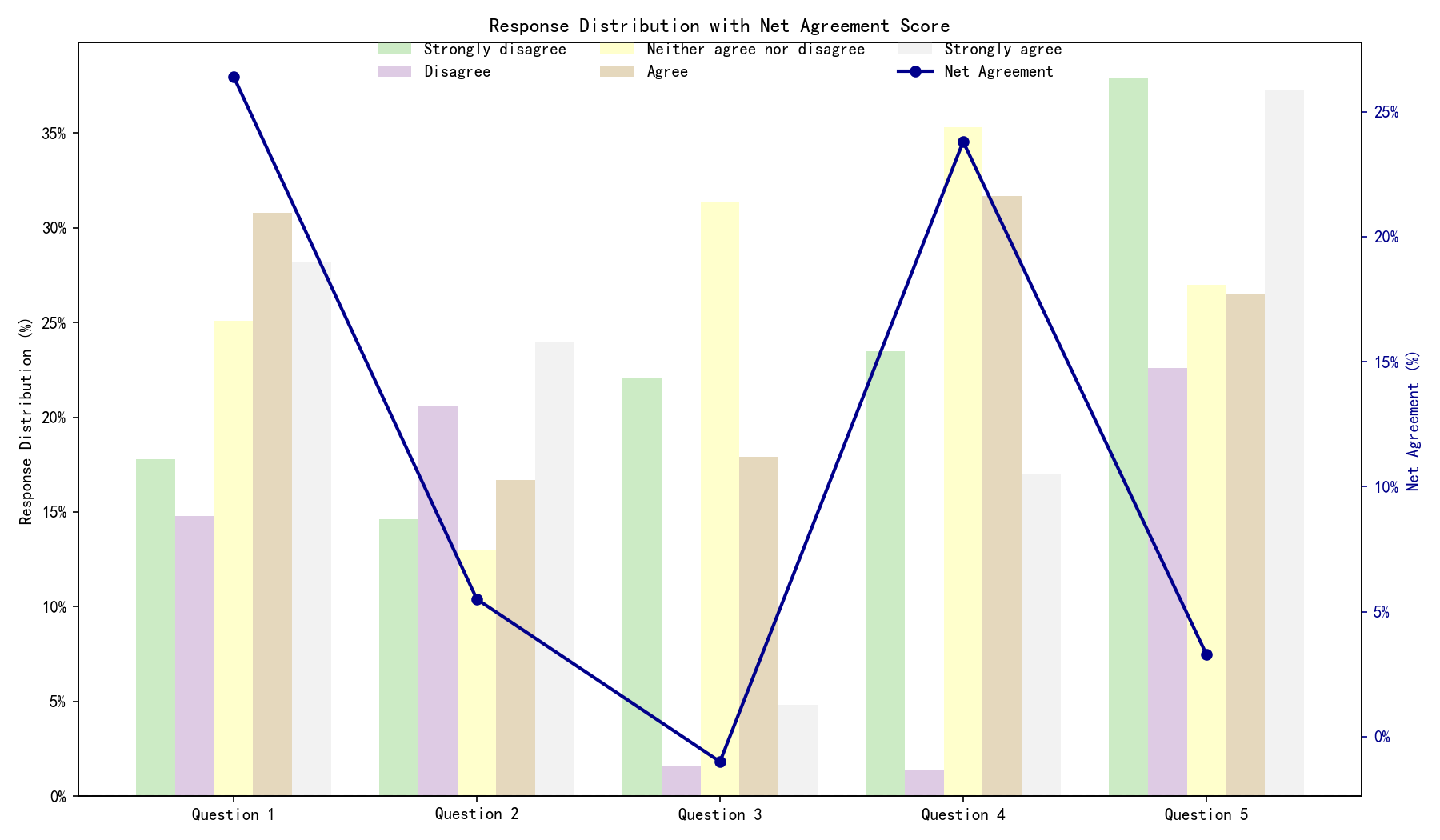}
            \\[0.02cm] \scriptsize (f) Model Output
        \end{minipage}
        
        \vspace{0.08cm}
        \begin{minipage}{0.96\linewidth}
            \footnotesize \textbf{Analysis:} The chart type in the model output on the right is inconsistent with the GT Figure; the bar chart on the right differs significantly from the stacked bar chart in the GT Figure.
        \end{minipage}
    \end{minipage}
    
    \vfill
    \caption{\textbf{Qualitative Error Analysis.} Examples of chart-type misclassification in generated results, including confusion between stacked and standard bar charts, horizontal and vertical orientations, and other structural type mismatches.
}
    \label{fig:type errors 1}
\end{compacterrorpage}

\begin{compacterrorpage}
    \centering
    \appendixplatetitle{Common Error Category: Type Errors}
    \vspace{1.0mm}
    \begin{minipage}{\linewidth}
        \centering
        \textbf{Case 1: Combination Chart (level2)}\\[0.08cm]
        
        \begin{minipage}[t]{0.48\linewidth}
            \centering
            \includegraphics[width=\linewidth,height=0.160\textheight,keepaspectratio]{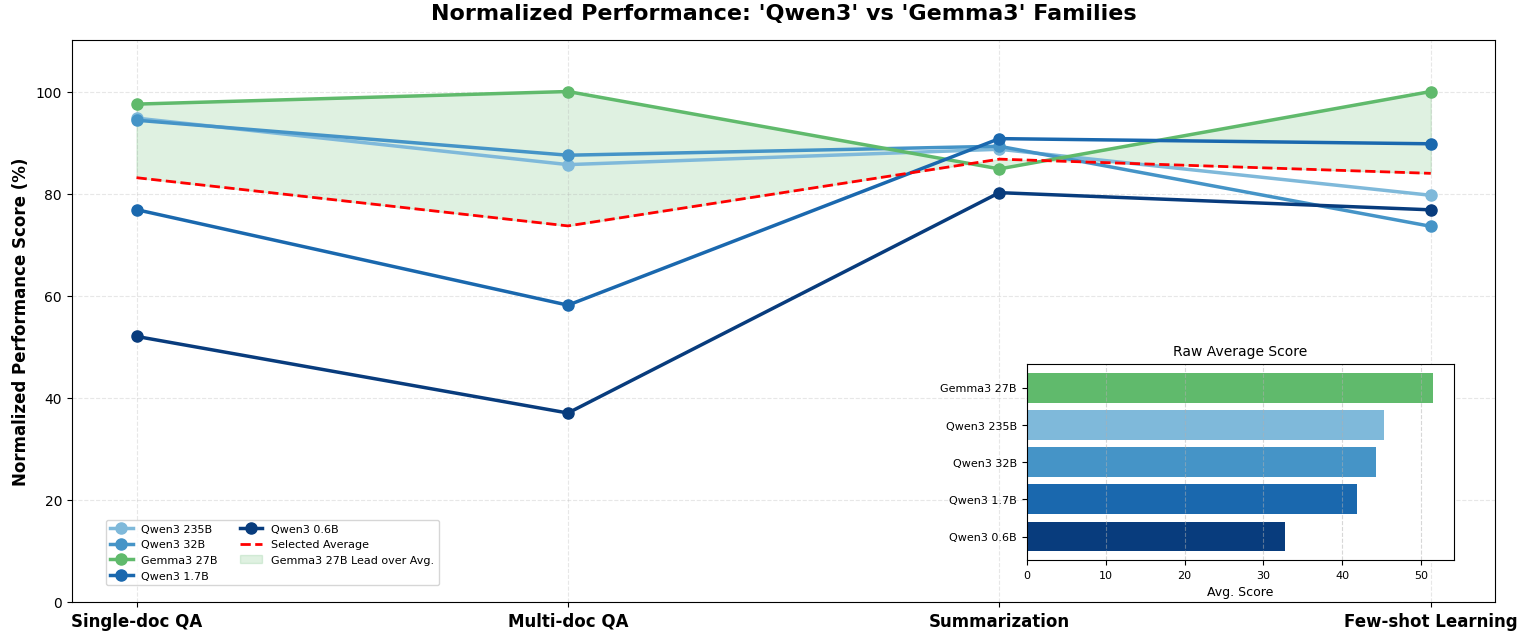}
            \\[0.02cm] \scriptsize (a) Ground Truth
        \end{minipage}
        \hfill
        \begin{minipage}[t]{0.48\linewidth}
            \centering
            \includegraphics[width=\linewidth,height=0.160\textheight,keepaspectratio]{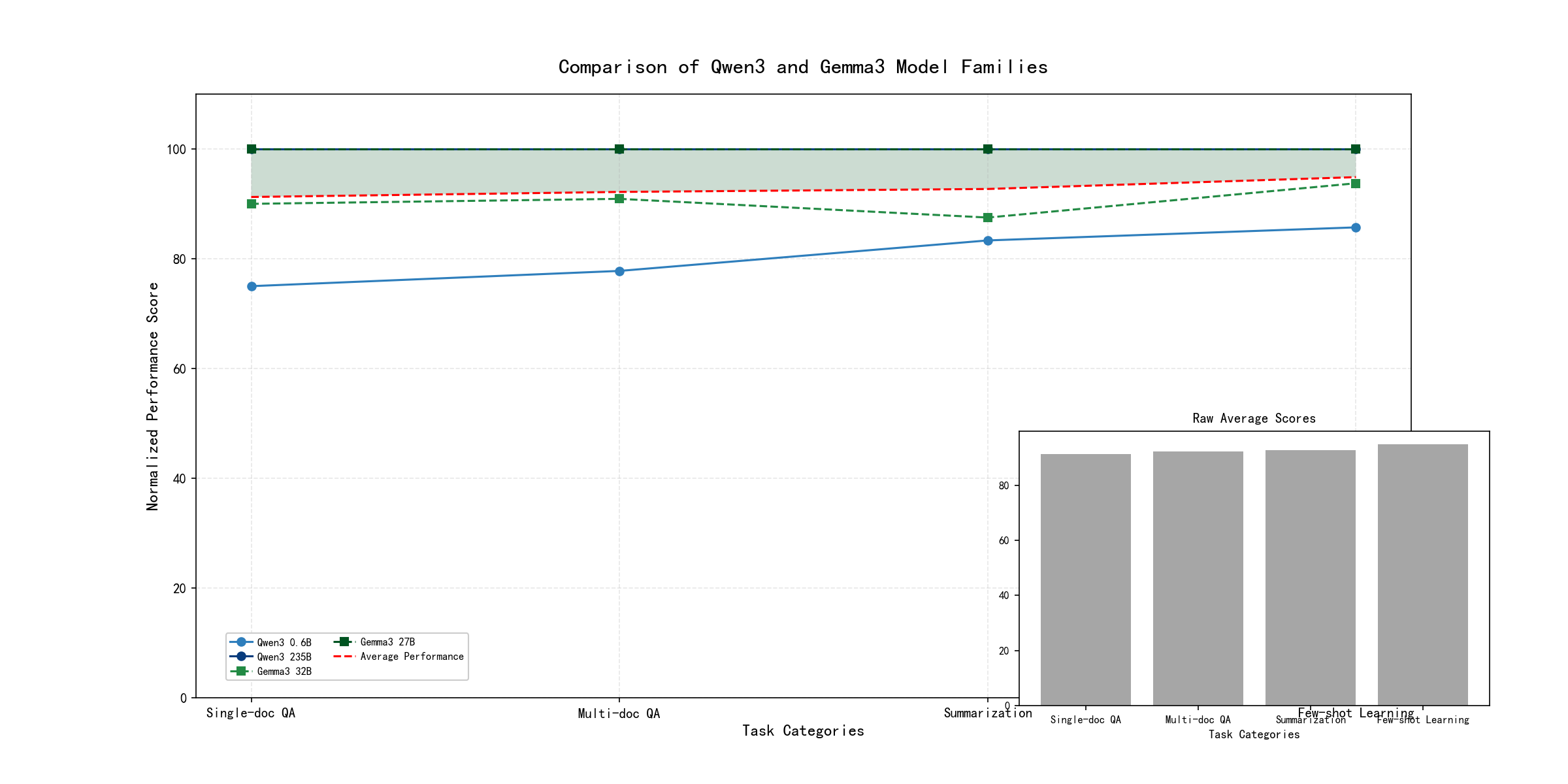}
            \\[0.02cm] \scriptsize (b) Model Output
        \end{minipage}
        
        \vspace{0.08cm}
        \begin{minipage}{0.96\linewidth}
            \footnotesize \textbf{Analysis:} The chart type in the model output on the right is inconsistent with that of the GT Figure; the right subplot differs significantly from the corresponding subplot in the GT Figure. The orientation of the bars in the bar chart is inconsistent, and there are notable differences in the remaining parts as well.
        \end{minipage}
    \end{minipage}

    \errorcaseseparator

    \begin{minipage}{\linewidth}
        \centering
        \textbf{Case 2: Violin Chart (level3)}\\[0.08cm]
        
        \begin{minipage}[t]{0.48\linewidth}
            \centering
            \includegraphics[width=\linewidth,height=0.160\textheight,keepaspectratio]{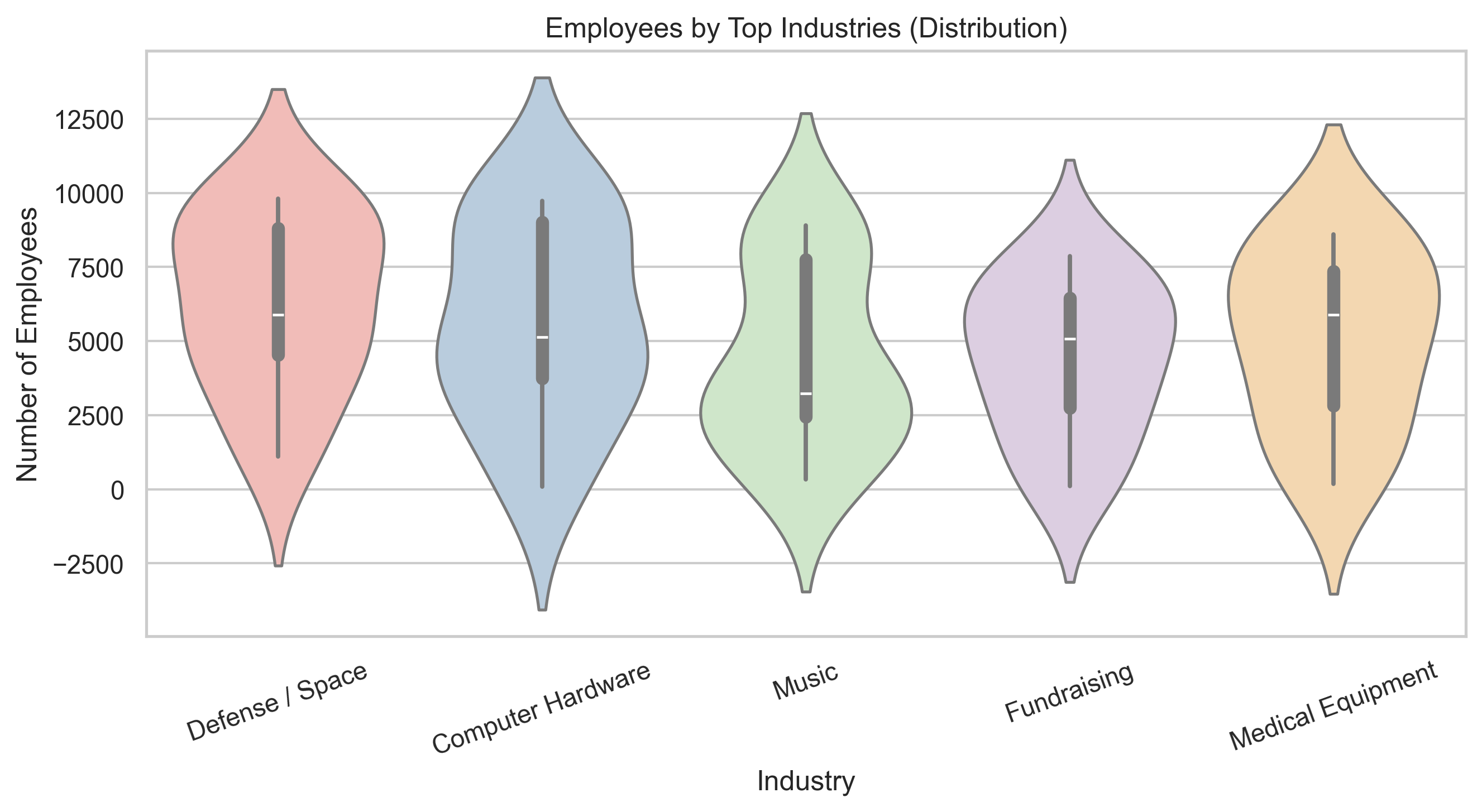}
            \\[0.02cm] \scriptsize (c) Ground Truth
        \end{minipage}
        \hfill
        \begin{minipage}[t]{0.48\linewidth}
            \centering
            \includegraphics[width=\linewidth,height=0.160\textheight,keepaspectratio]{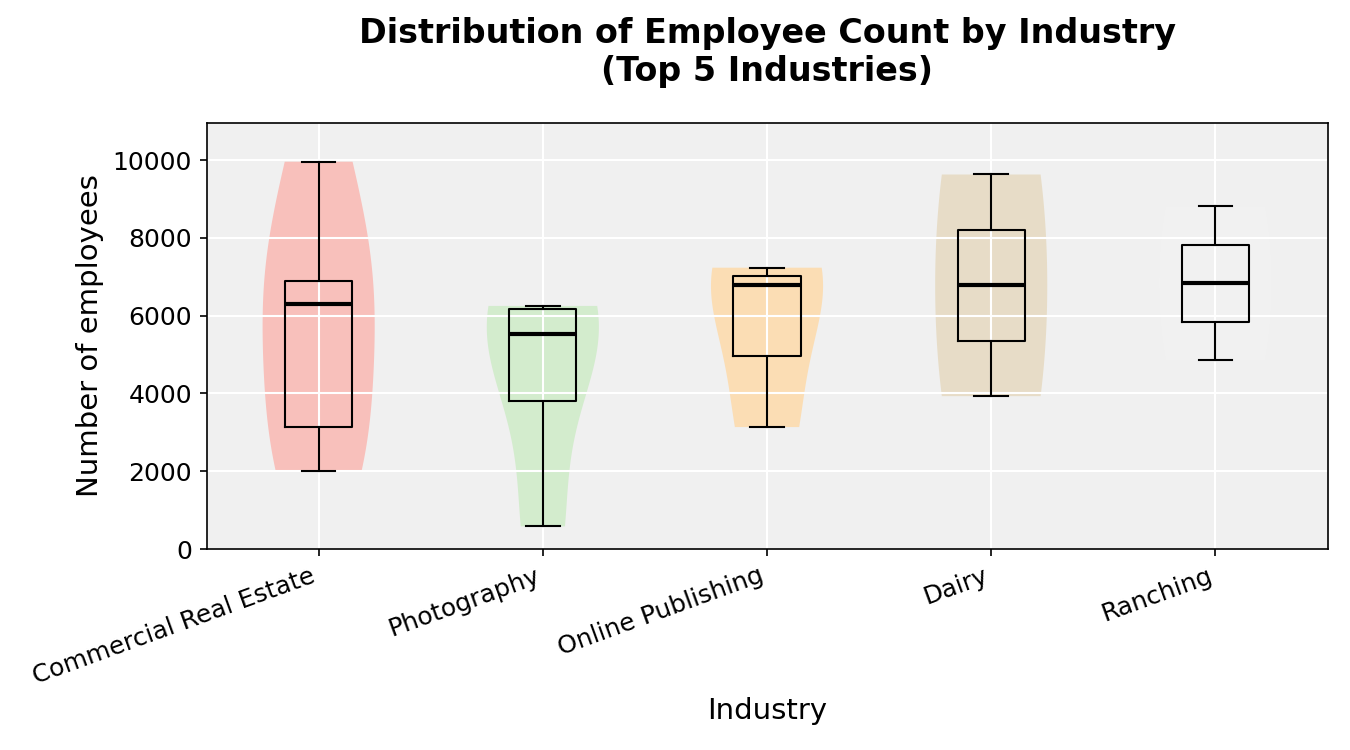}
            \\[0.02cm] \scriptsize (d) Model Output
        \end{minipage}
        
        \vspace{0.08cm}
        \begin{minipage}{0.96\linewidth}
            \footnotesize \textbf{Analysis:} The chart type in the model output on the right differs from that of the GT Figure. The GT Figure is a violin chart, but the model generated a box chart.
        \end{minipage}
    \end{minipage}

    \errorcaseseparator
    
    \begin{minipage}{\linewidth}
        \centering
        \textbf{Case 3: Polar Chart (level3)}\\[0.08cm]
        
        \begin{minipage}[t]{0.48\linewidth}
            \centering
            \includegraphics[width=\linewidth,height=0.160\textheight,keepaspectratio]{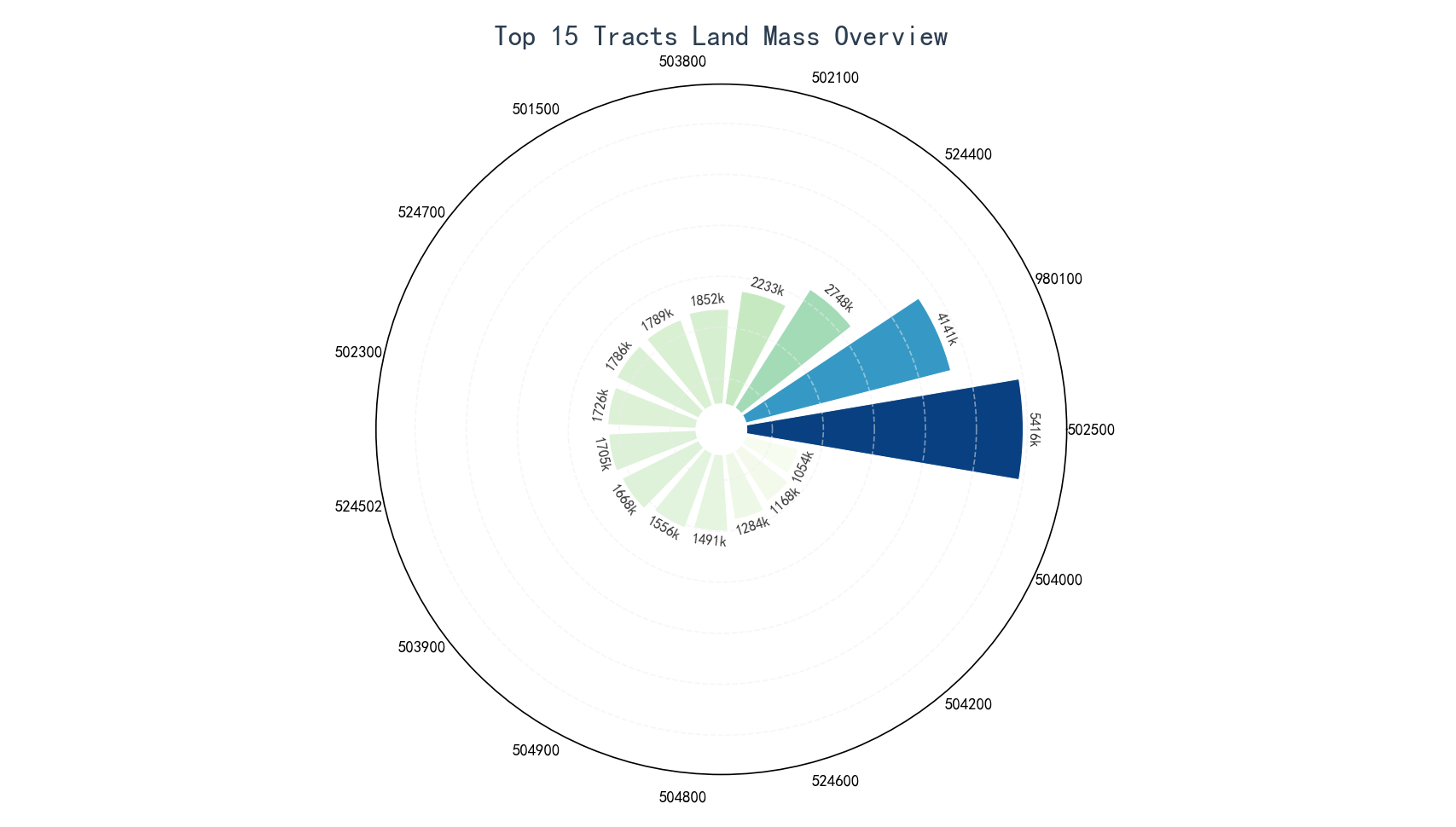}
            \\[0.02cm] \scriptsize (e) Ground Truth
        \end{minipage}
        \hfill
        \begin{minipage}[t]{0.48\linewidth}
            \centering
            \includegraphics[width=\linewidth,height=0.160\textheight,keepaspectratio]{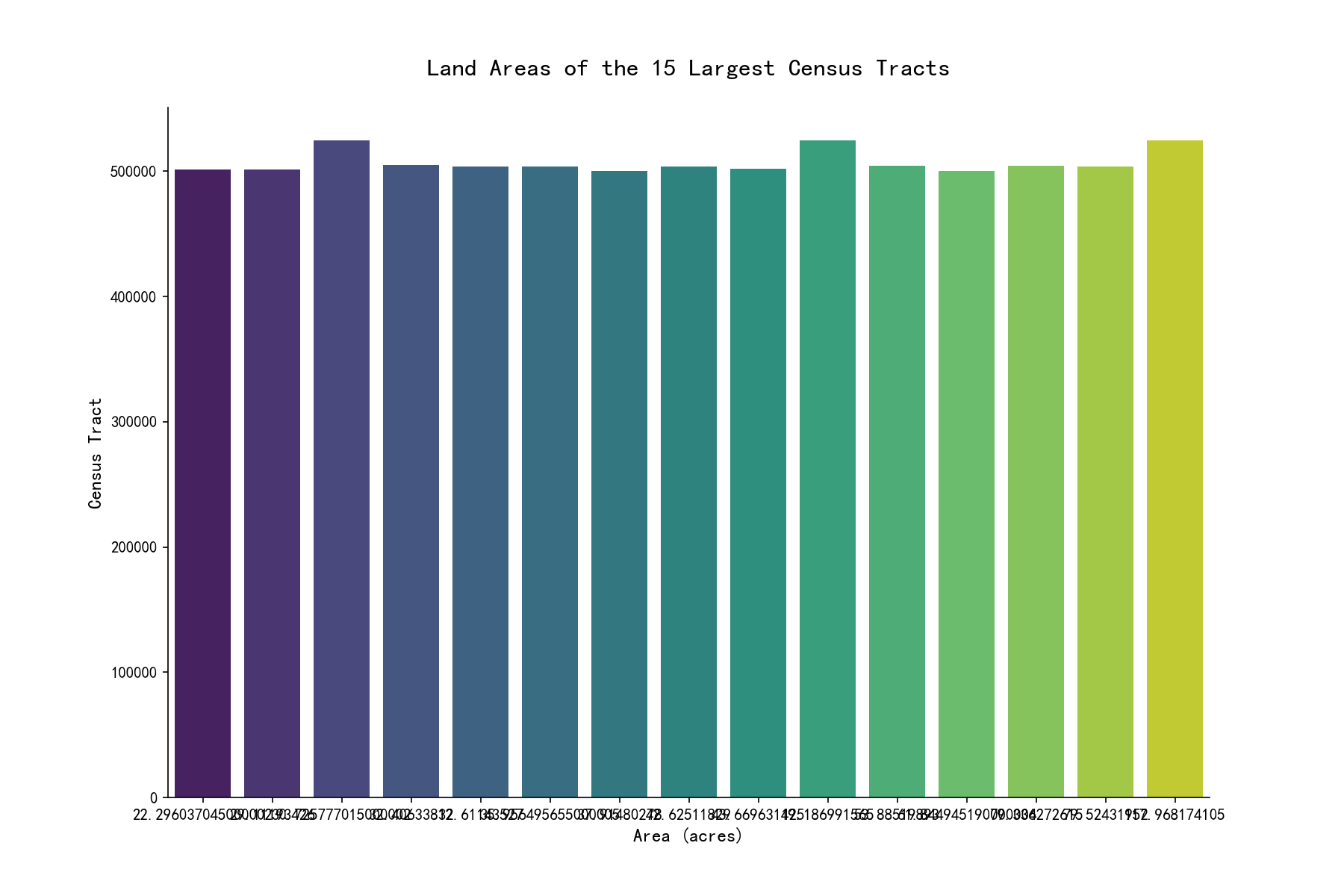}
            \\[0.02cm] \scriptsize (f) Model Output
        \end{minipage}
        
        \vspace{0.08cm}
        \begin{minipage}{0.96\linewidth}
            \footnotesize \textbf{Analysis:} The chart type in the model output on the right differs from that of the GT Figure. The GT Figure is a circular bar chart, but the model generated a regular bar chart.
        \end{minipage}
    \end{minipage}
    
    \vfill
    \caption{\textbf{Qualitative Error Analysis.} Incorrect chart-type predictions in model outputs, such as confusing stacked bar charts with standard bar charts, or misclassifying horizontal and vertical bar charts, among other type-related mistakes.}
    \label{fig:type errors 2}
\end{compacterrorpage}

\begin{compacterrorpage}
    \centering
    \appendixplatetitle{Common Error Category: Grid Errors}
    \vspace{1.0mm}
    \begin{minipage}{\linewidth}
        \centering
        \textbf{Case 1: Line Chart (level1)}\\[0.08cm]
        
        \begin{minipage}[t]{0.48\linewidth}
            \centering
            \includegraphics[width=\linewidth,height=0.160\textheight,keepaspectratio]{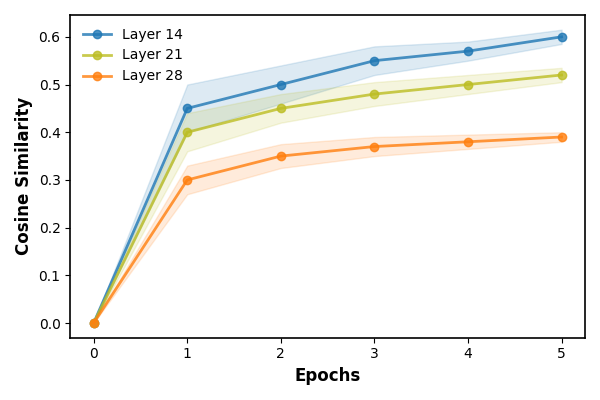}
            \\[0.02cm] \scriptsize (a) Ground Truth
        \end{minipage}
        \hfill
        \begin{minipage}[t]{0.48\linewidth}
            \centering
            \includegraphics[width=\linewidth,height=0.160\textheight,keepaspectratio]{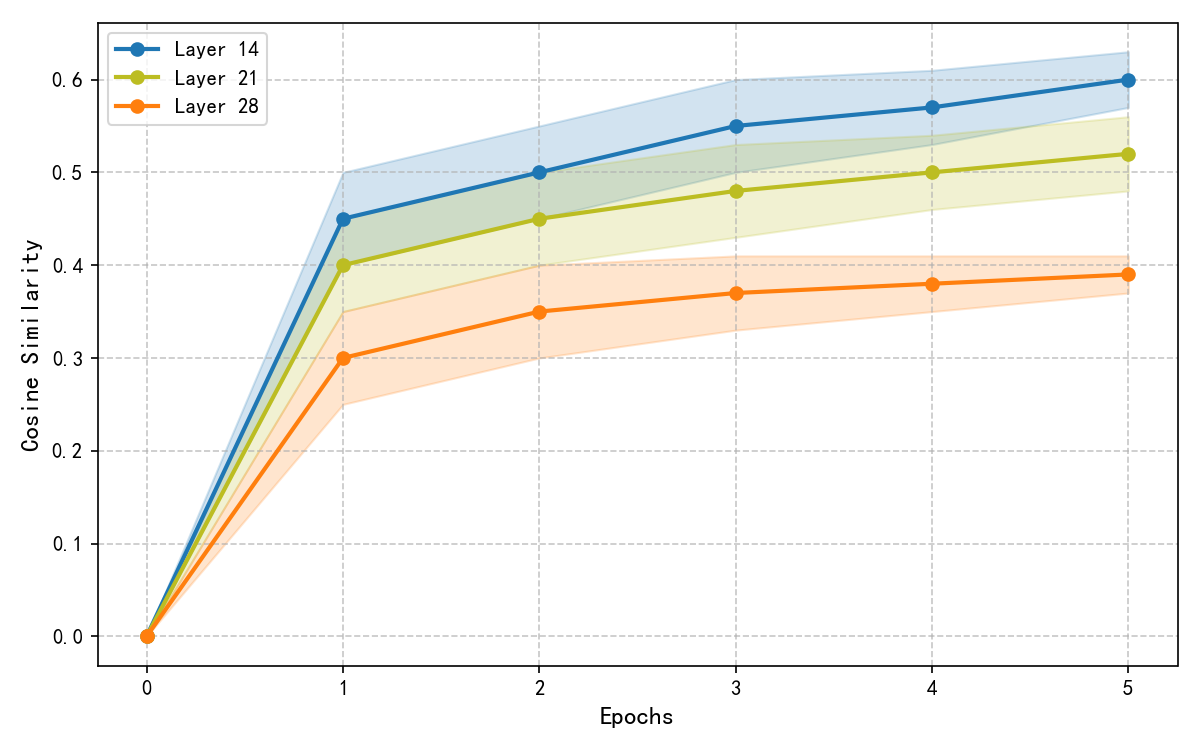}
            \\[0.02cm] \scriptsize (b) Model Output
        \end{minipage}
        
        \vspace{0.08cm}
        \begin{minipage}{0.96\linewidth}
            \footnotesize \textbf{Analysis:} The grid line style in the model-generated output (right) is inconsistent with that of the ground-truth (GT) figure, with dashed lines appearing on the right side where they are not present in the reference. This discrepancy indicates a deficiency in the model’s ability to perceive and faithfully reproduce fine-grained visual details.
        \end{minipage}
    \end{minipage}

    \errorcaseseparator

    \begin{minipage}{\linewidth}
        \centering
        \textbf{Case 2: Line Chart (level1)}\\[0.08cm]
        
        \begin{minipage}[t]{0.48\linewidth}
            \centering
            \includegraphics[width=\linewidth,height=0.160\textheight,keepaspectratio]{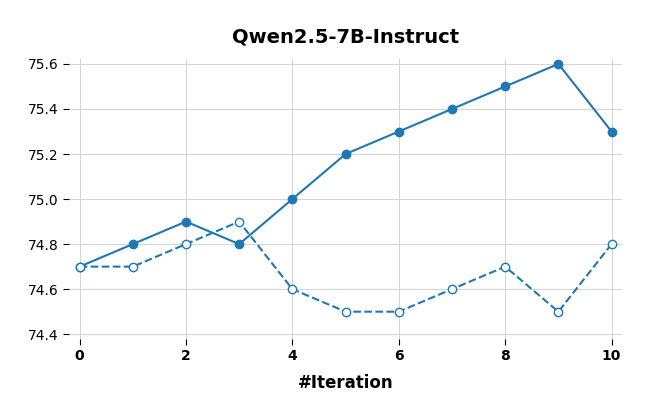}
            \\[0.02cm] \scriptsize (c) Ground Truth
        \end{minipage}
        \hfill
        \begin{minipage}[t]{0.48\linewidth}
            \centering
            \includegraphics[width=\linewidth,height=0.160\textheight,keepaspectratio]{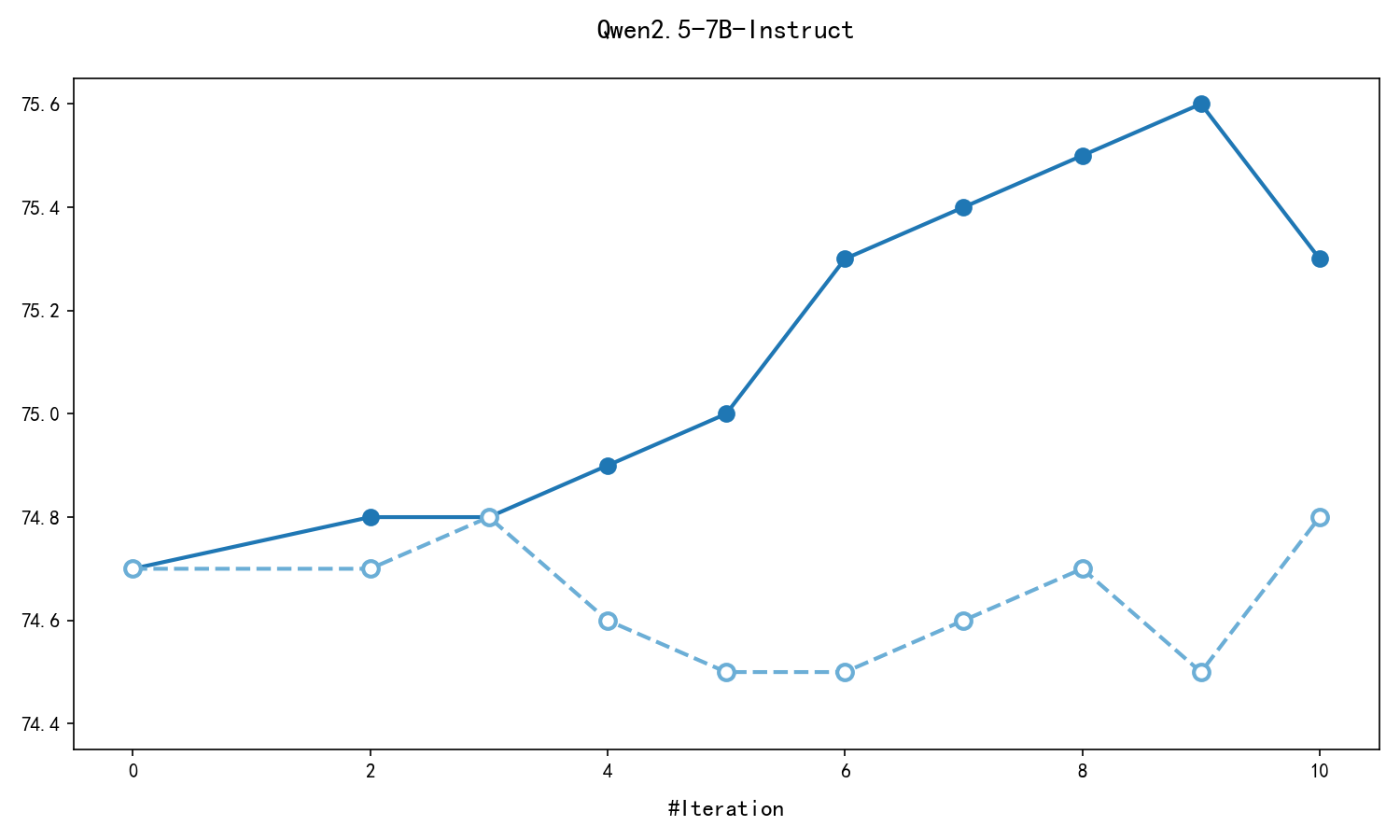}
            \\[0.02cm] \scriptsize (d) Model Output
        \end{minipage}
        
        \vspace{0.08cm}
        \begin{minipage}{0.96\linewidth}
            \footnotesize \textbf{Analysis:} The grid line style in the model output on the right is inconsistent with that of the GT Figure, indicating a deficiency in the model's ability to perceive fine-grained visual details.
        \end{minipage}
    \end{minipage}

    \errorcaseseparator
    
    \begin{minipage}{\linewidth}
        \centering
        \textbf{Case 3: Combination Chart (level2)}\\[0.08cm]
        
        \begin{minipage}[t]{0.48\linewidth}
            \centering
            \includegraphics[width=\linewidth,height=0.160\textheight,keepaspectratio]{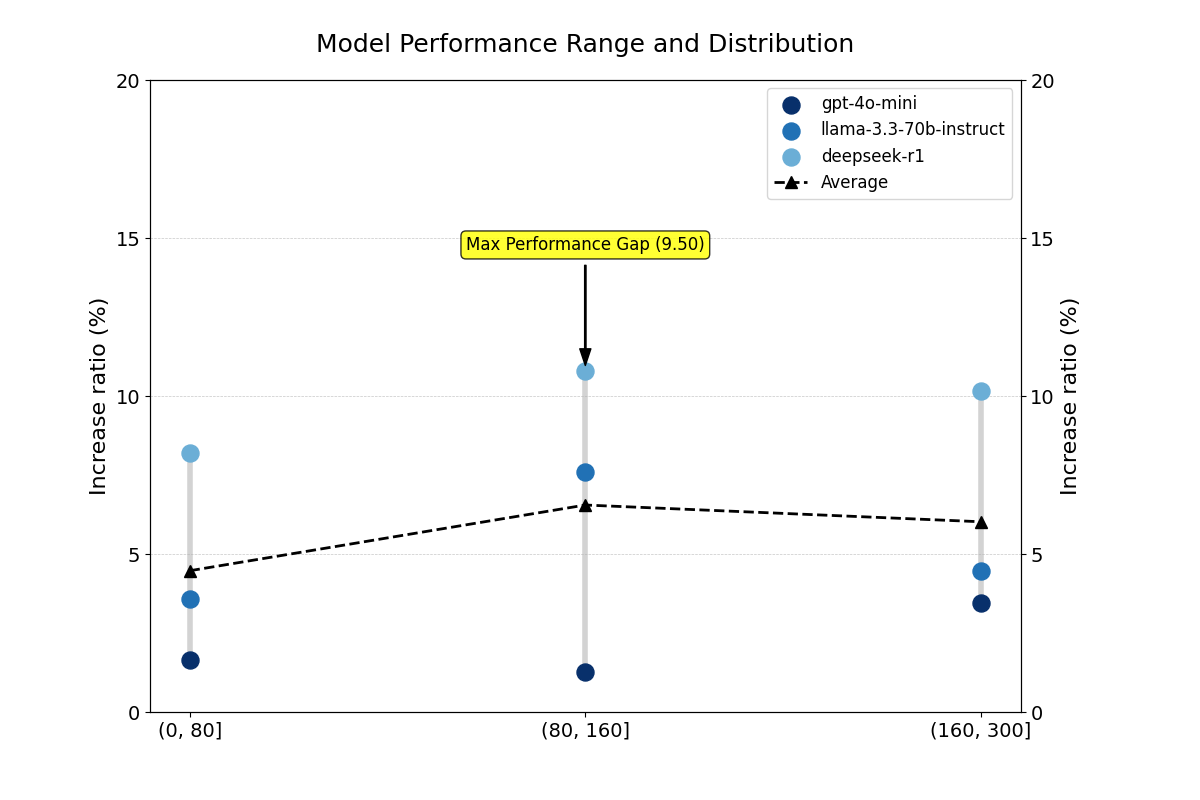}
            \\[0.02cm] \scriptsize (e) Ground Truth
        \end{minipage}
        \hfill
        \begin{minipage}[t]{0.48\linewidth}
            \centering
            \includegraphics[width=\linewidth,height=0.160\textheight,keepaspectratio]{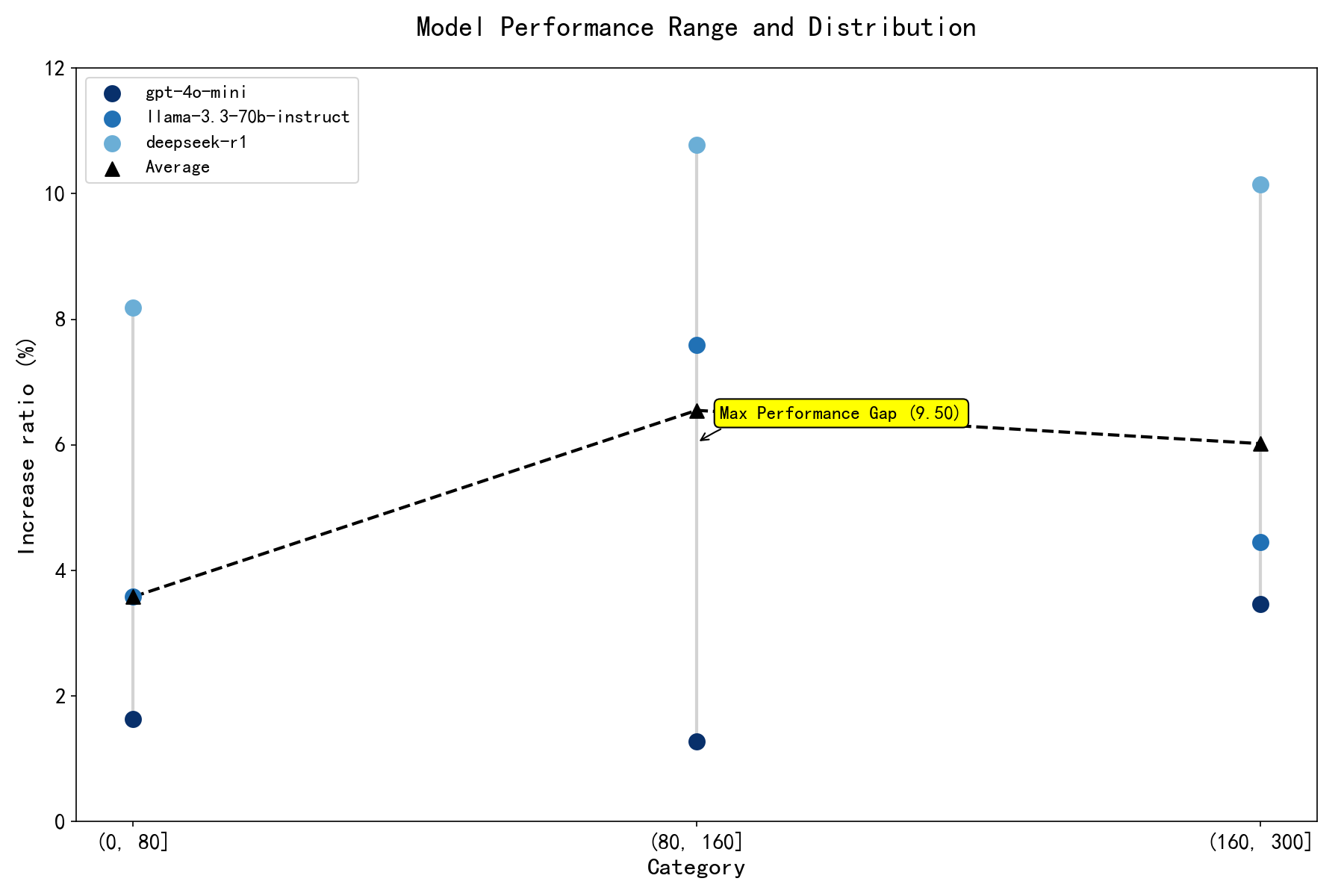}
            \\[0.02cm] \scriptsize (f) Model Output
        \end{minipage}
        
        \vspace{0.08cm}
        \begin{minipage}{0.96\linewidth}
            \footnotesize \textbf{Analysis:} The grid line style in the model output on the right is inconsistent with that of the GT Figure, indicating a deficiency in the model's ability to perceive fine-grained visual details.
        \end{minipage}
    \end{minipage}
    
    \vfill
    \caption{\textbf{Qualitative Error Analysis.} Discrepancies in grid configuration within model-generated charts, including mismatches in horizontal and vertical gridlines, missing or extraneous grids, and incorrect gridline styles or thickness.
}
    \label{fig:grid errors 1}
\end{compacterrorpage}

\begin{compacterrorpage}
    \centering
    \appendixplatetitle{Common Error Category: Grid Errors}
    \vspace{1.0mm}
    \begin{minipage}{\linewidth}
        \centering
        \textbf{Case 1: Combination Chart (level2)}\\[0.08cm]
        
        \begin{minipage}[t]{0.48\linewidth}
            \centering
            \includegraphics[width=\linewidth,height=0.160\textheight,keepaspectratio]{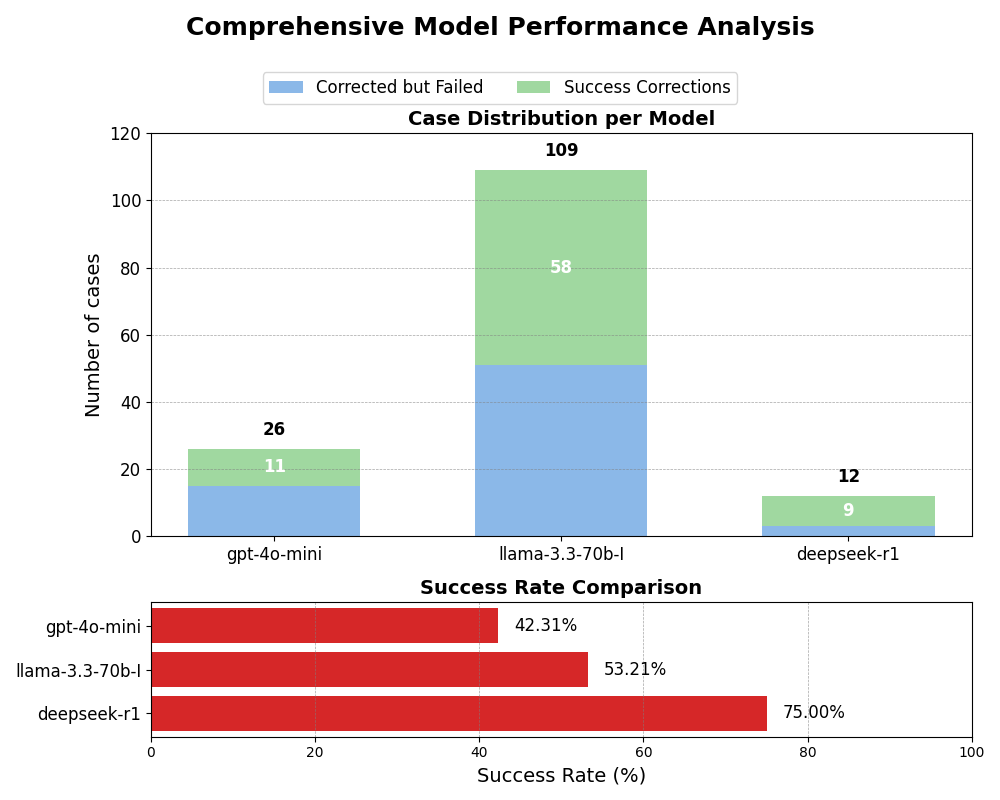}
            \\[0.02cm] \scriptsize (a) Ground Truth
        \end{minipage}
        \hfill
        \begin{minipage}[t]{0.48\linewidth}
            \centering
            \includegraphics[width=\linewidth,height=0.160\textheight,keepaspectratio]{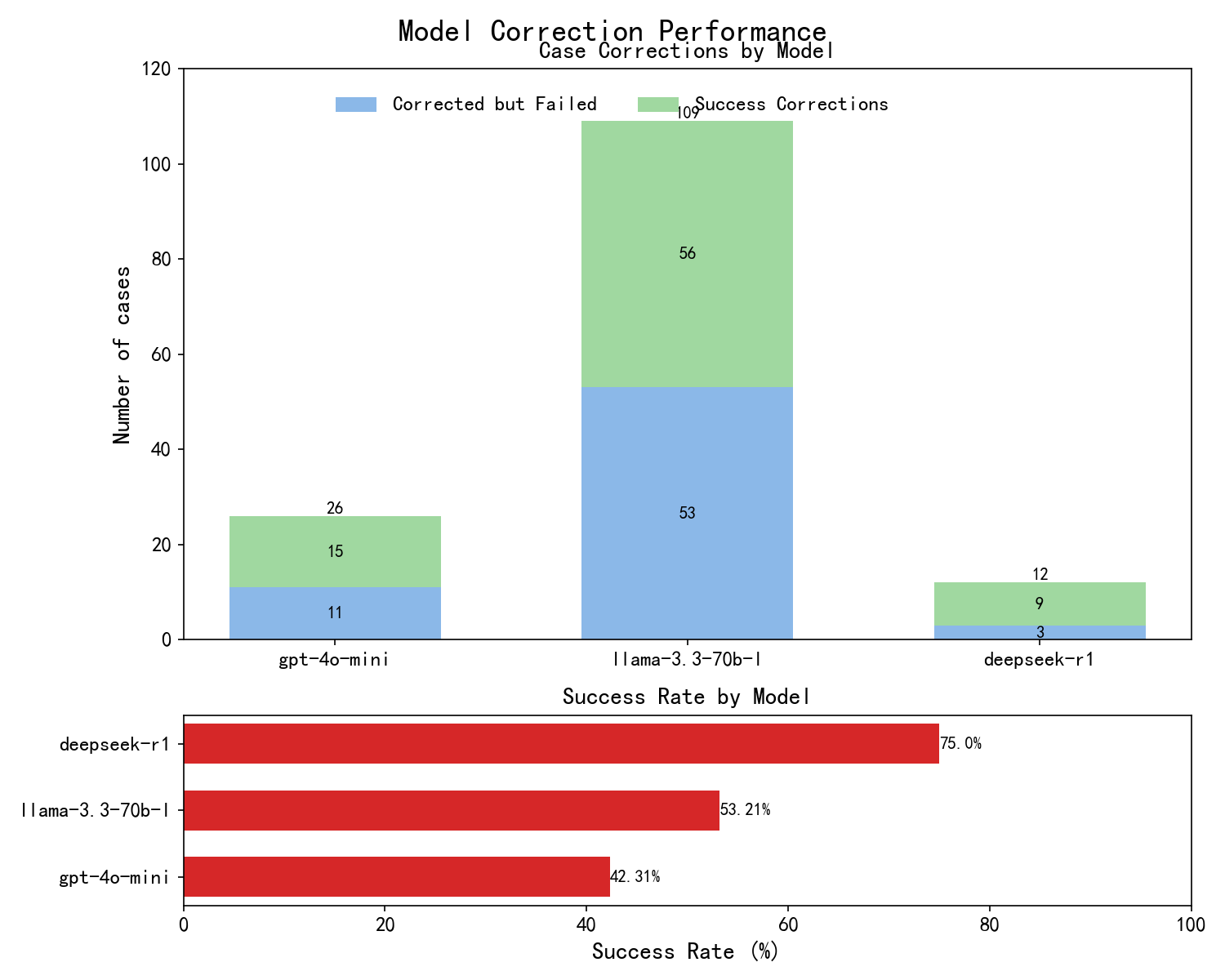}
            \\[0.02cm] \scriptsize (b) Model Output
        \end{minipage}
        
        \vspace{0.08cm}
        \begin{minipage}{0.96\linewidth}
            \footnotesize \textbf{Analysis:} The grid line style in the model output on the right is inconsistent with that of the GT Figure, indicating a deficiency in the model's ability to perceive fine-grained visual details.
        \end{minipage}
    \end{minipage}

    \errorcaseseparator

    \begin{minipage}{\linewidth}
        \centering
        \textbf{Case 2: Line Chart (level2)}\\[0.08cm]
        
        \begin{minipage}[t]{0.48\linewidth}
            \centering
            \includegraphics[width=\linewidth,height=0.160\textheight,keepaspectratio]{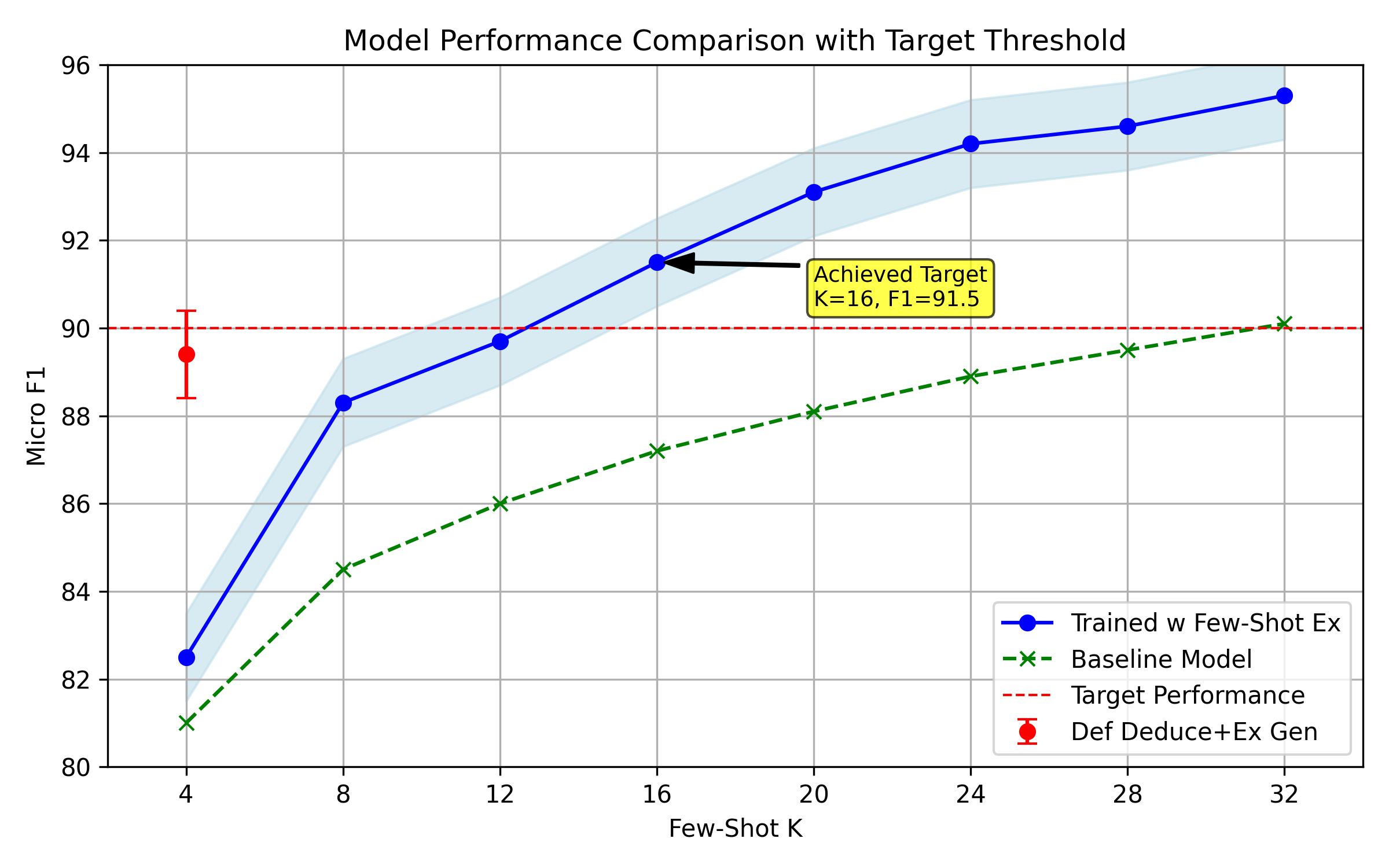}
            \\[0.02cm] \scriptsize (c) Ground Truth
        \end{minipage}
        \hfill
        \begin{minipage}[t]{0.48\linewidth}
            \centering
            \includegraphics[width=\linewidth,height=0.160\textheight,keepaspectratio]{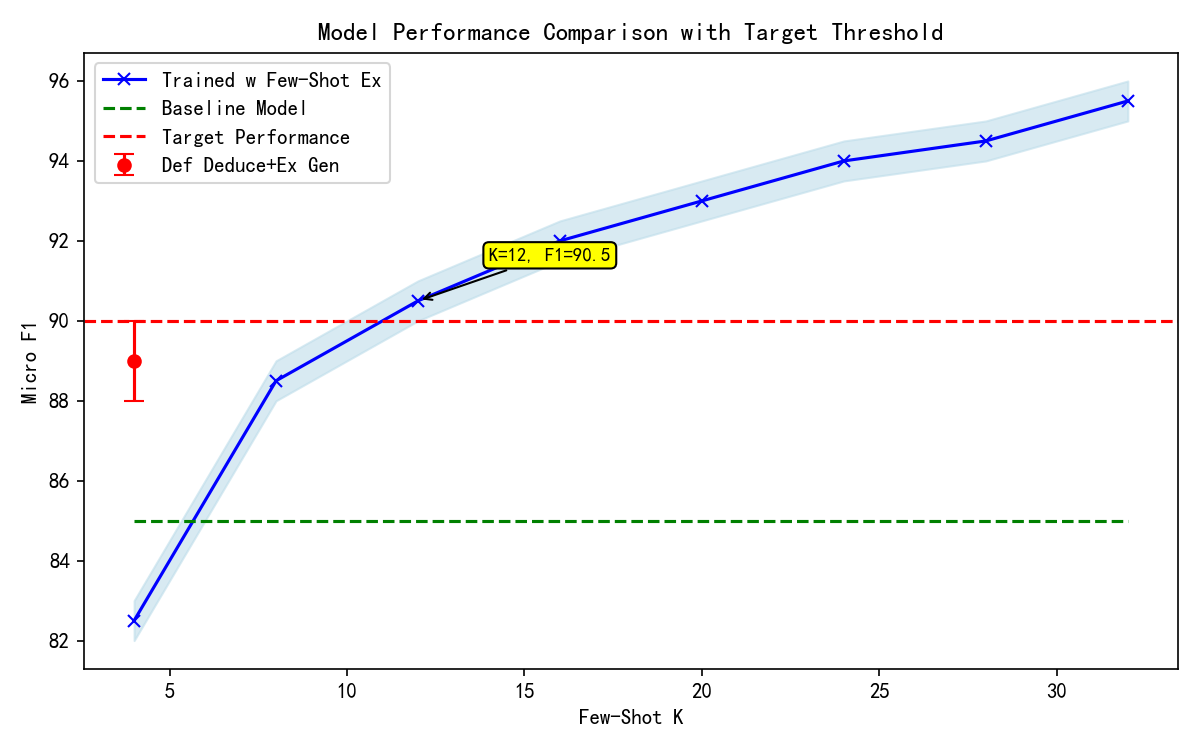}
            \\[0.02cm] \scriptsize (d) Model Output
        \end{minipage}
        
        \vspace{0.08cm}
        \begin{minipage}{0.96\linewidth}
            \footnotesize \textbf{Analysis:} The grid line style in the model output on the right is inconsistent with that of the GT Figure, indicating a deficiency in the model's ability to perceive fine-grained visual details.
        \end{minipage}
    \end{minipage}

    \errorcaseseparator
    
    \begin{minipage}{\linewidth}
        \centering
        \textbf{Case 3: Polar Chart (level3)}\\[0.08cm]
        
        \begin{minipage}[t]{0.48\linewidth}
            \centering
            \includegraphics[width=\linewidth,height=0.160\textheight,keepaspectratio]{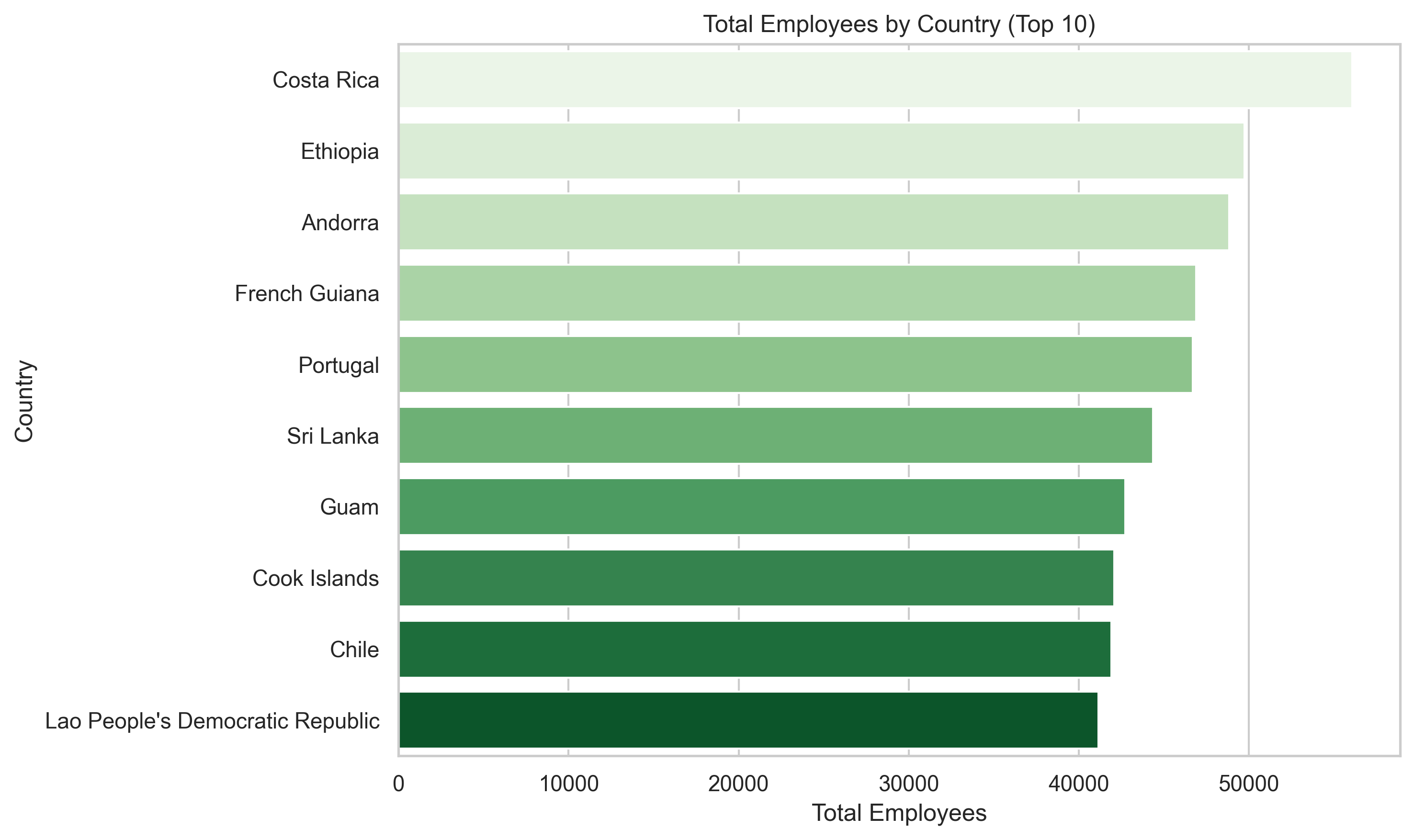}
            \\[0.02cm] \scriptsize (e) Ground Truth
        \end{minipage}
        \hfill
        \begin{minipage}[t]{0.48\linewidth}
            \centering
            \includegraphics[width=\linewidth,height=0.160\textheight,keepaspectratio]{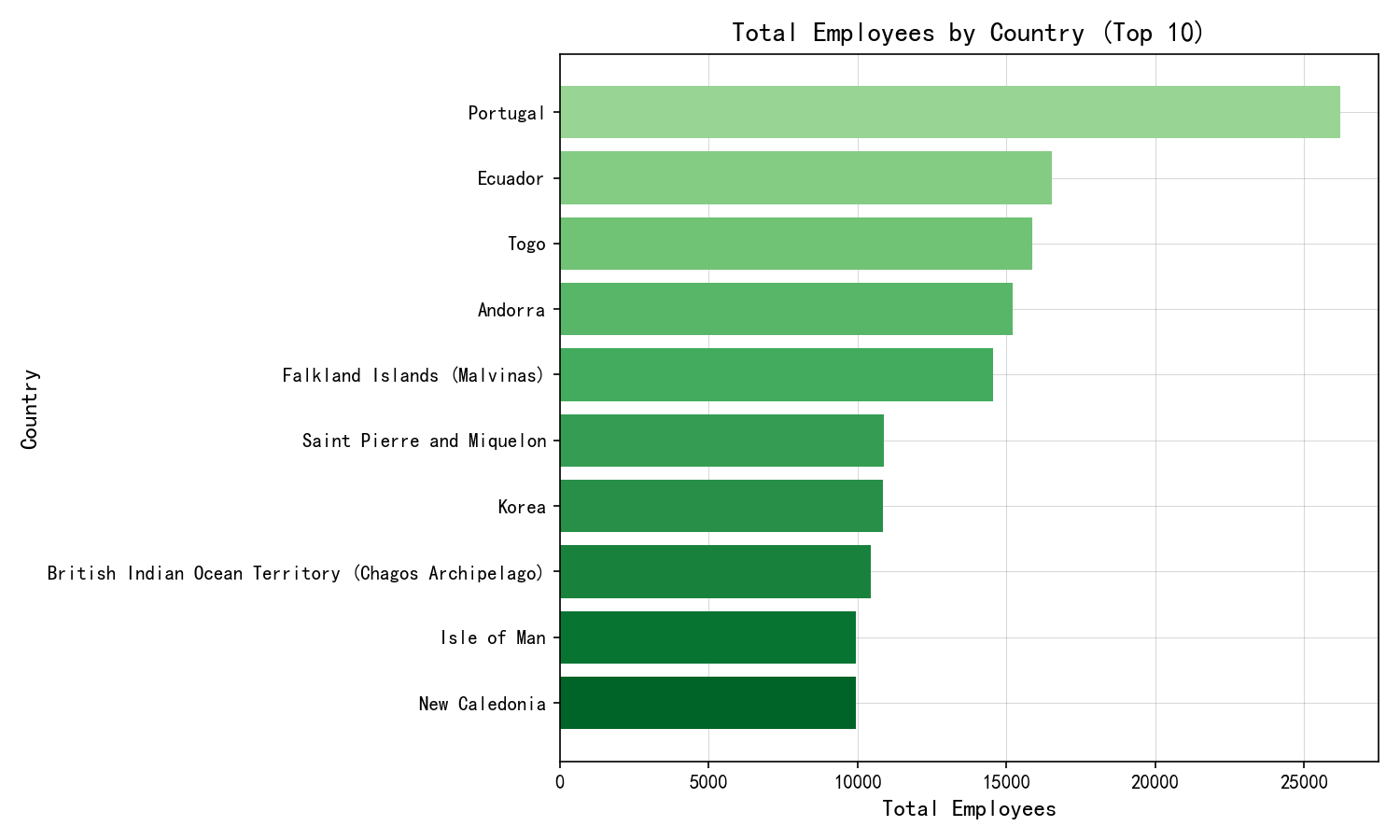}
            \\[0.02cm] \scriptsize (f) Model Output
        \end{minipage}
        
        \vspace{0.08cm}
        \begin{minipage}{0.96\linewidth}
            \footnotesize \textbf{Analysis:} The grid line style in the model output on the right differs from that of the GT Figure. In the original figure, there are no horizontal grid lines. Additionally, there are issues with the data in the chart.
        \end{minipage}
    \end{minipage}
    
    \vfill
    \caption{\textbf{Qualitative Error Analysis.} Examples of incorrect grid settings in generated charts, such as improper horizontal/vertical gridlines, grid presence inconsistencies, and errors in line style or thickness.}
    \label{fig:grid errors 2}
\end{compacterrorpage}

\begin{compacterrorpage}
    \centering
    \appendixplatetitle{Common Error Category: Text Occlusion Errors}
    \vspace{1.0mm}
    \begin{minipage}{\linewidth}
        \centering
        \textbf{Case 1: Bar Chart (level2)}\\[0.08cm]
        
        \begin{minipage}[t]{0.48\linewidth}
            \centering
            \includegraphics[width=\linewidth,height=0.160\textheight,keepaspectratio]{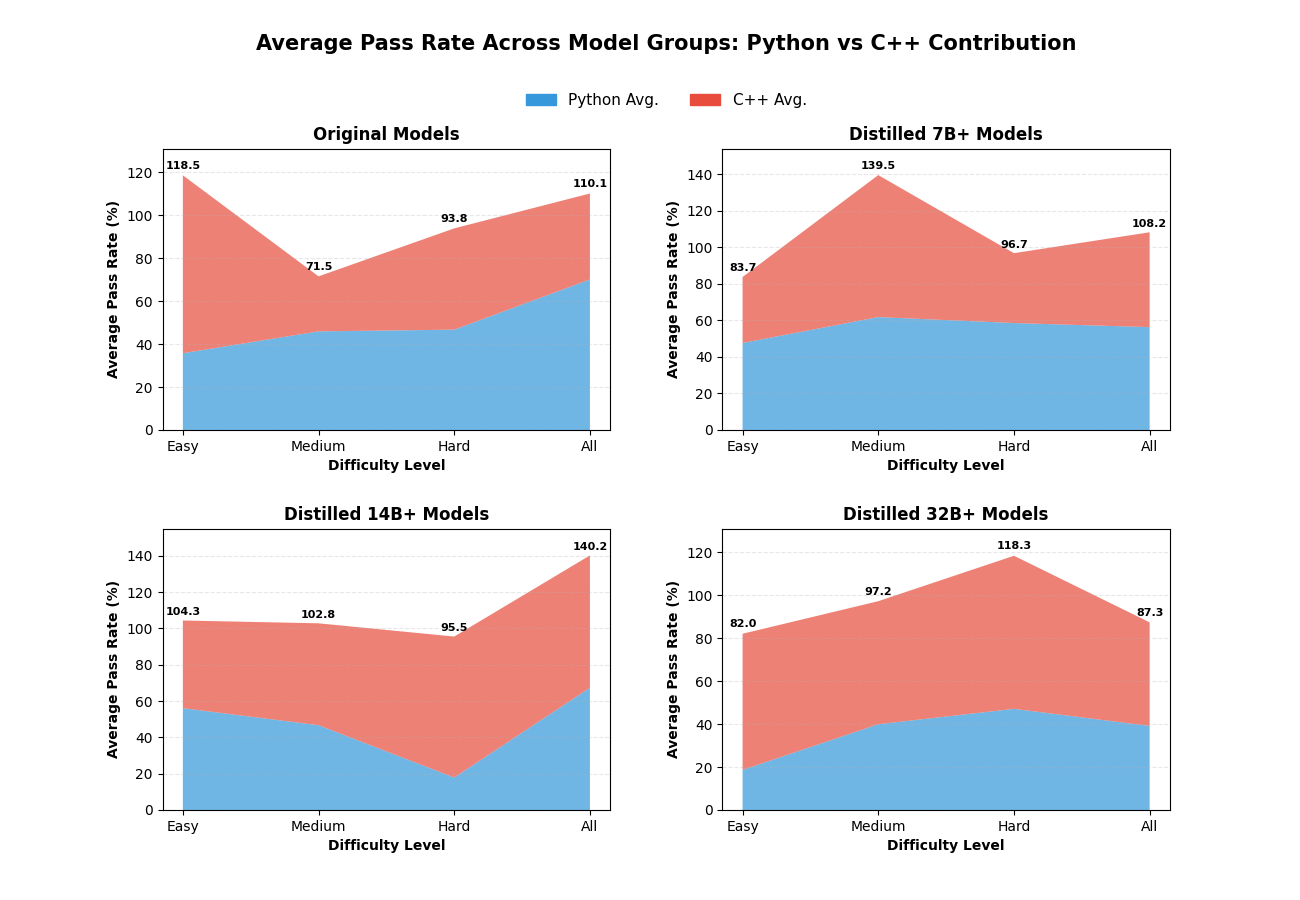}
            \\[0.02cm] \scriptsize (a) Ground Truth
        \end{minipage}
        \hfill
        \begin{minipage}[t]{0.48\linewidth}
            \centering
            \includegraphics[width=\linewidth,height=0.160\textheight,keepaspectratio]{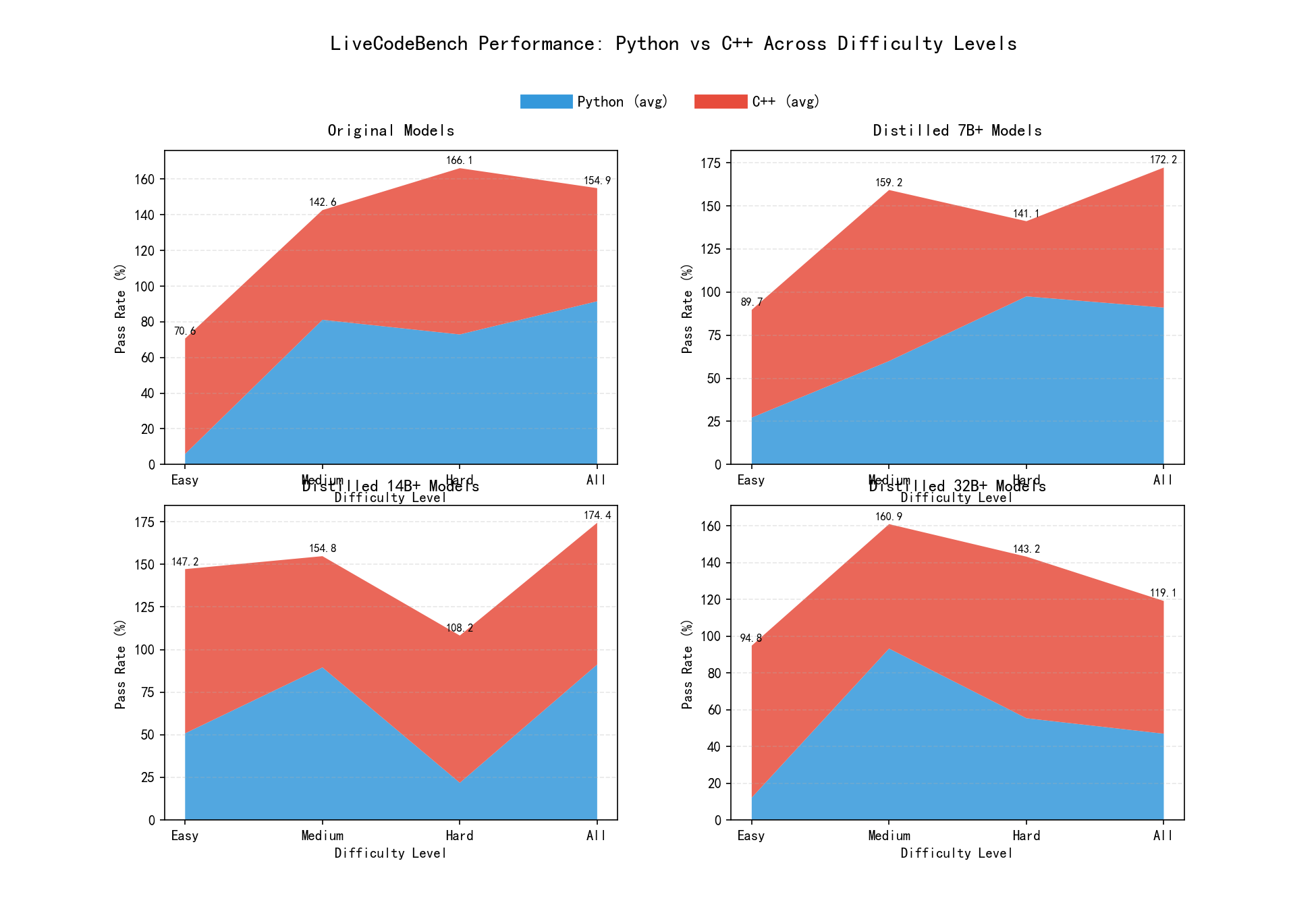}
            \\[0.02cm] \scriptsize (b) Model Output
        \end{minipage}
        
        \vspace{0.08cm}
        \begin{minipage}{0.96\linewidth}
            \footnotesize \textbf{Analysis:} There is a serious text overlap issue in the model output on the right, with overlapping text between the upper and lower subplots, which severely affects the visual aesthetics.
        \end{minipage}
    \end{minipage}

    \errorcaseseparator

    \begin{minipage}{\linewidth}
        \centering
        \textbf{Case 2: Bar Chart (level2)}\\[0.08cm]
        
        \begin{minipage}[t]{0.48\linewidth}
            \centering
            \includegraphics[width=\linewidth,height=0.160\textheight,keepaspectratio]{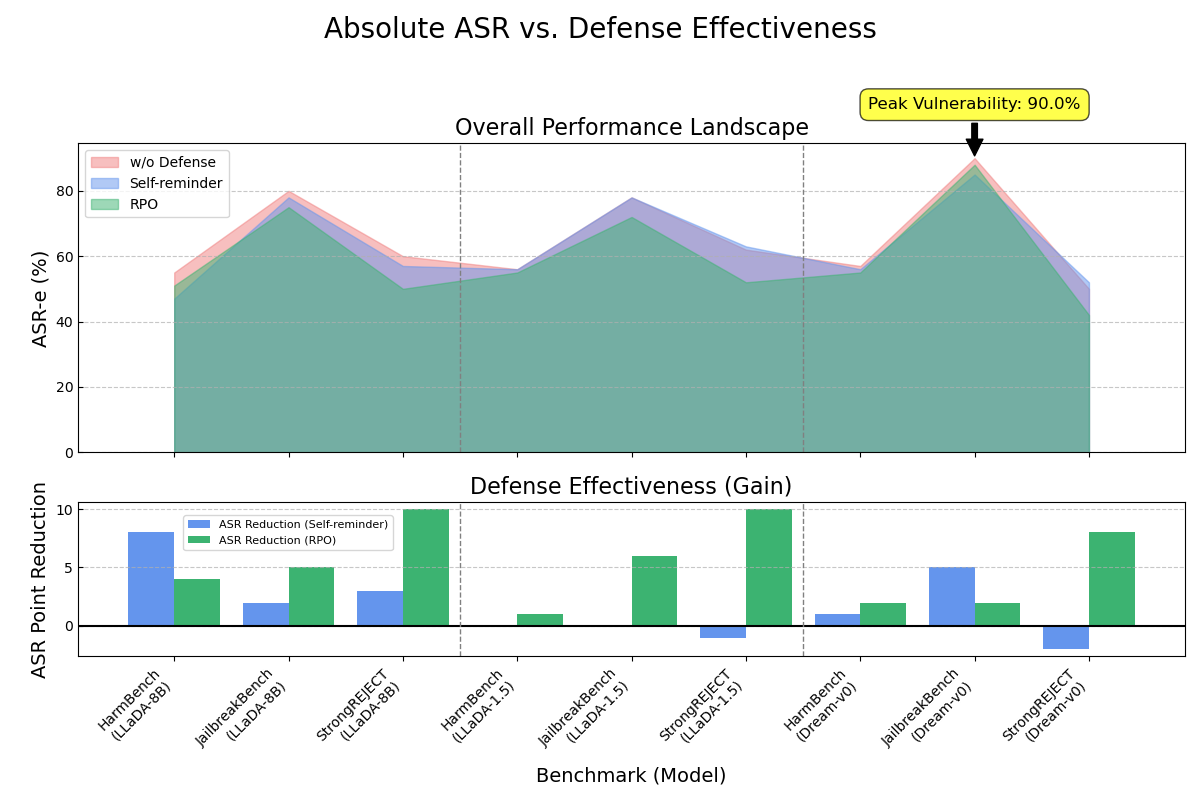}
            \\[0.02cm] \scriptsize (c) Ground Truth
        \end{minipage}
        \hfill
        \begin{minipage}[t]{0.48\linewidth}
            \centering
            \includegraphics[width=\linewidth,height=0.160\textheight,keepaspectratio]{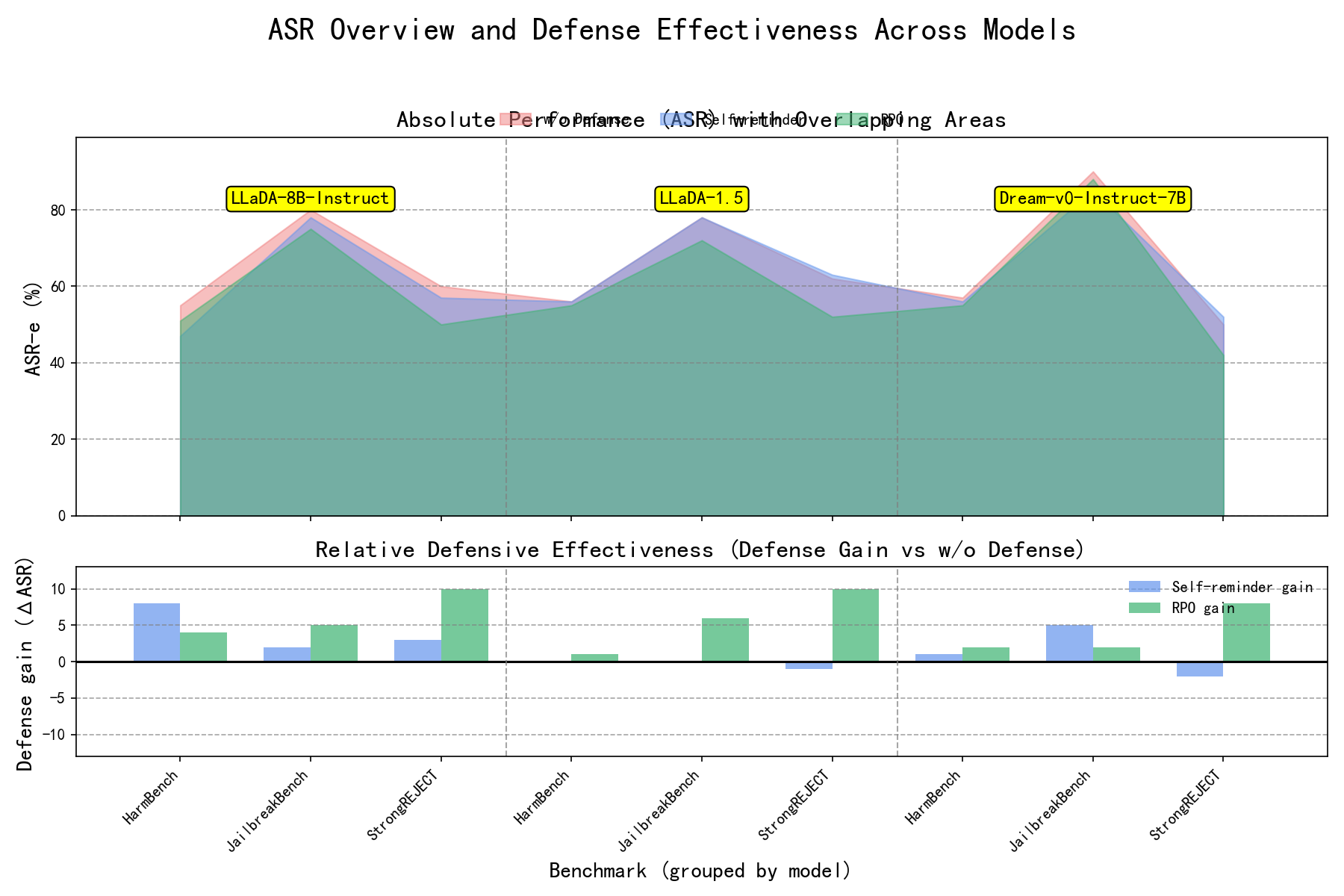}
            \\[0.02cm] \scriptsize (d) Model Output
        \end{minipage}
        
        \vspace{0.08cm}
        \begin{minipage}{0.96\linewidth}
            \footnotesize \textbf{Analysis:} The legend and title in the model output on the right severely overlap, significantly impacting the visual aesthetics.
        \end{minipage}
    \end{minipage}

    \errorcaseseparator
    
    \begin{minipage}{\linewidth}
        \centering
        \textbf{Case 3: Combination Chart (level2)}\\[0.08cm]
        
        \begin{minipage}[t]{0.48\linewidth}
            \centering
            \includegraphics[width=\linewidth,height=0.160\textheight,keepaspectratio]{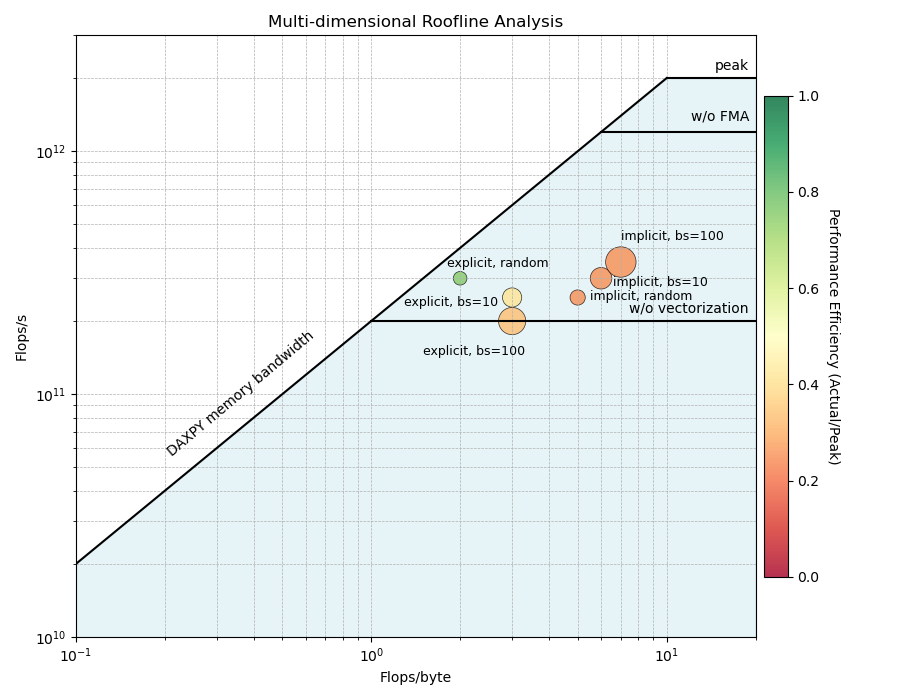}
            \\[0.02cm] \scriptsize (e) Ground Truth
        \end{minipage}
        \hfill
        \begin{minipage}[t]{0.48\linewidth}
            \centering
            \includegraphics[width=\linewidth,height=0.160\textheight,keepaspectratio]{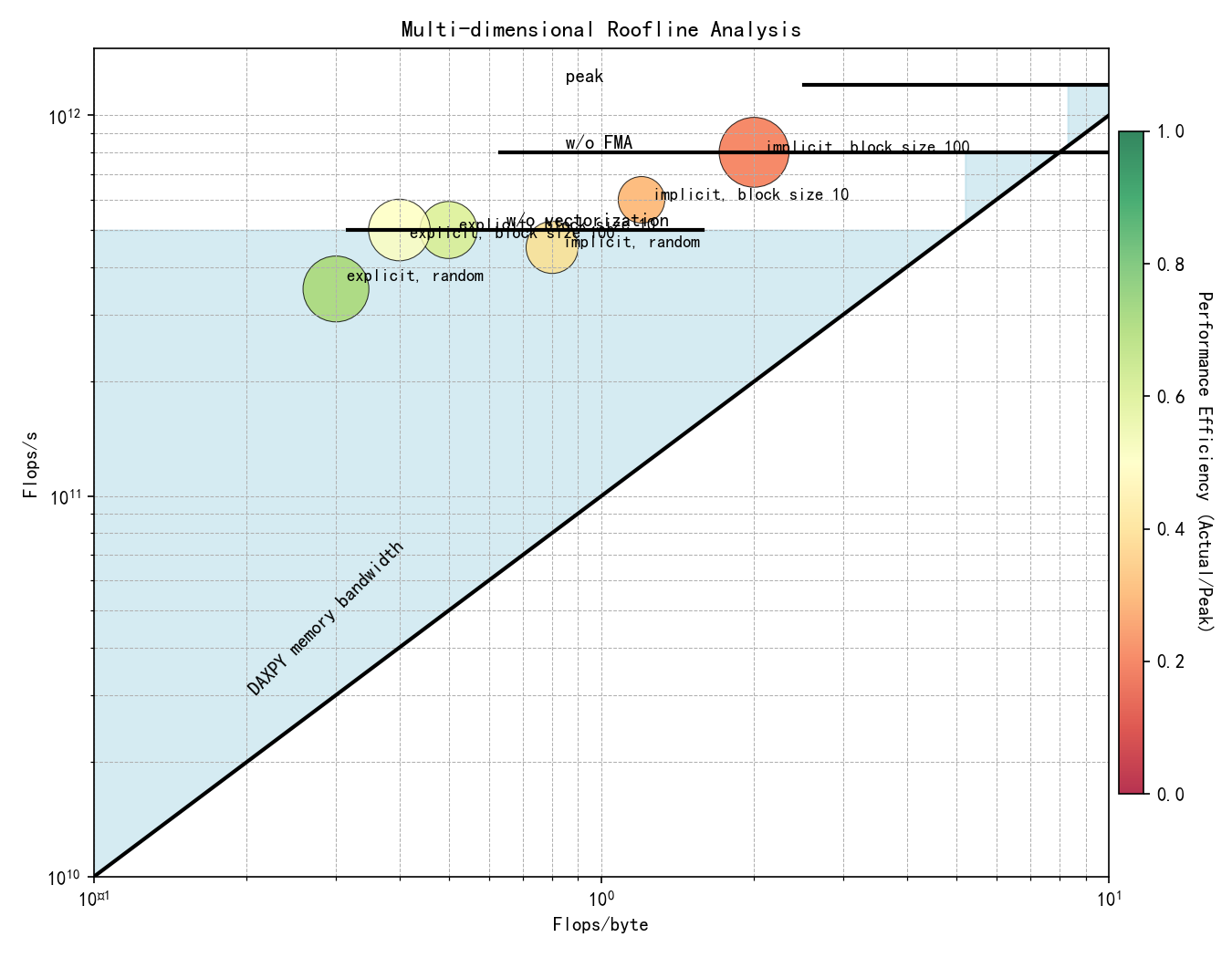}
            \\[0.02cm] \scriptsize (f) Model Output
        \end{minipage}
        
        \vspace{0.08cm}
        \begin{minipage}{0.96\linewidth}
            \footnotesize \textbf{Analysis:} There is severe overlap among the labels in the model output on the right, which greatly impacts the visual aesthetics. Additionally, there is a noticeable deficiency in the recognition of numerical values.
        \end{minipage}
    \end{minipage}
    
    \vfill
    \caption{\textbf{Qualitative Error Analysis.} Frequent text overlap issues in model-generated charts, occurring across legends, titles, x-axis labels, y-axis labels, and other textual elements.
}
    \label{fig:text_occlusion errors}
\end{compacterrorpage}

\begin{compacterrorpage}
    \centering
    \appendixplatetitle{Common Error Category: Visual Style Errors}
    \vspace{1.0mm}
    \begin{minipage}{\linewidth}
        \centering
        \textbf{Case 1: Combination Chart (level2)}\\[0.08cm]
        
        \begin{minipage}[t]{0.48\linewidth}
            \centering
            \includegraphics[width=\linewidth,height=0.160\textheight,keepaspectratio]{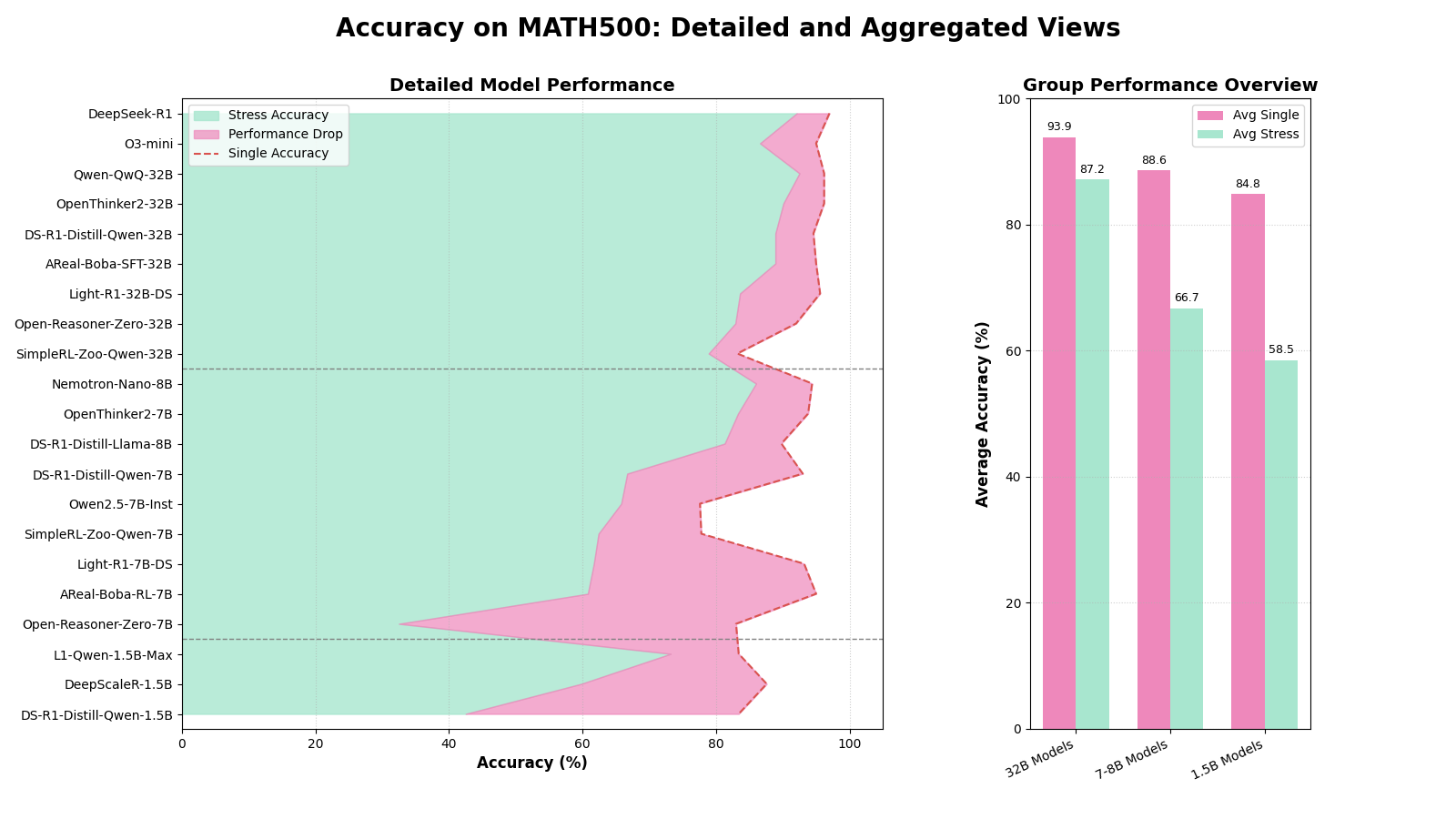}
            \\[0.02cm] \scriptsize (a) Ground Truth
        \end{minipage}
        \hfill
        \begin{minipage}[t]{0.48\linewidth}
            \centering
            \includegraphics[width=\linewidth,height=0.160\textheight,keepaspectratio]{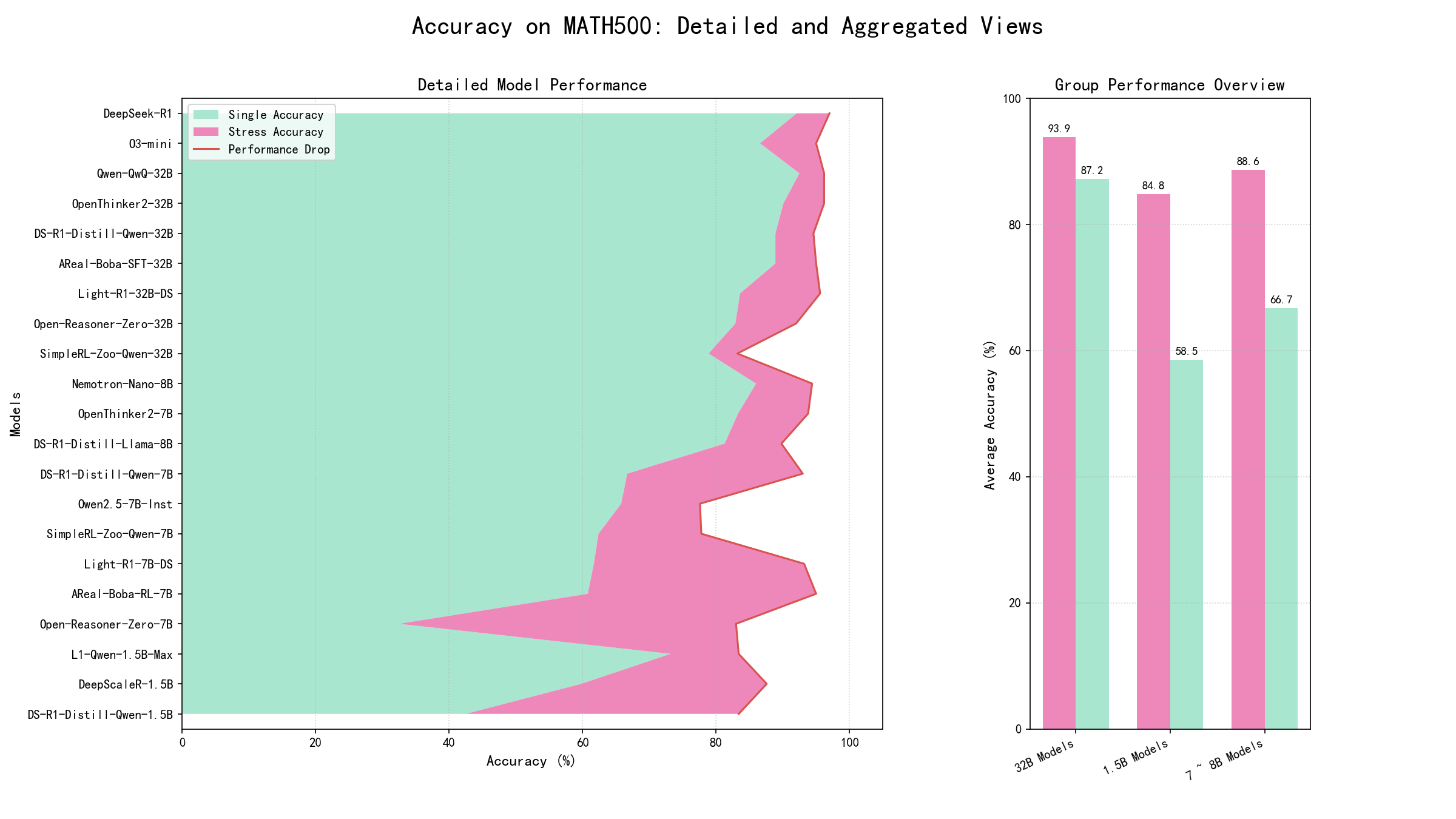}
            \\[0.02cm] \scriptsize (b) Model Output
        \end{minipage}
        
        \vspace{0.08cm}
        \begin{minipage}{0.96\linewidth}
            \footnotesize \textbf{Analysis:} The transparency in the left plot of the model output on the right is inconsistent with that of the GT Figure. Additionally, the legend style differs—the GT Figure uses a dashed line, while the model output uses a solid line. Furthermore, the text bolding also deviates from the GT Figure.
        \end{minipage}
    \end{minipage}

    \errorcaseseparator

    \begin{minipage}{\linewidth}
        \centering
        \textbf{Case 2: Combination Chart (level2)}\\[0.08cm]
        
        \begin{minipage}[t]{0.48\linewidth}
            \centering
            \includegraphics[width=\linewidth,height=0.160\textheight,keepaspectratio]{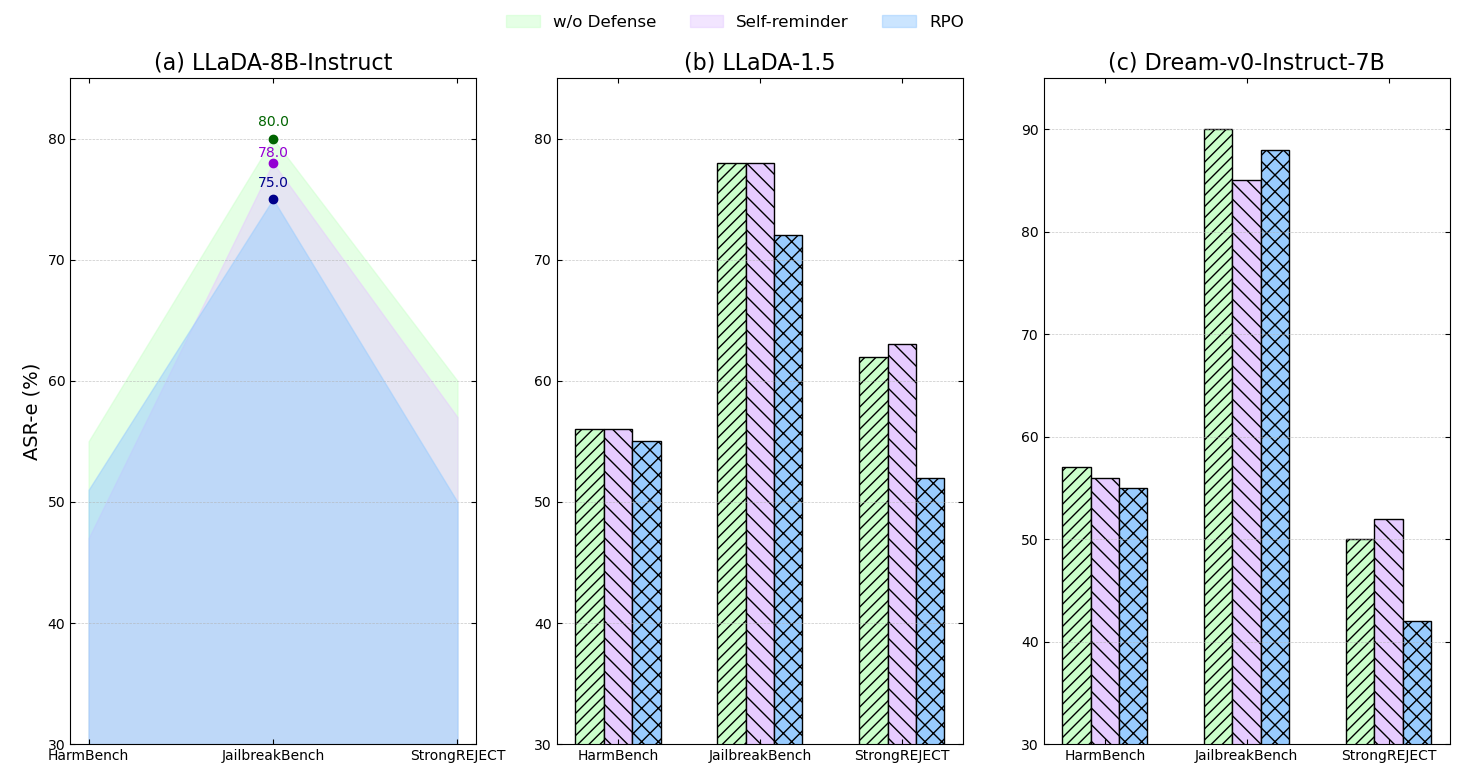}
            \\[0.02cm] \scriptsize (c) Ground Truth
        \end{minipage}
        \hfill
        \begin{minipage}[t]{0.48\linewidth}
            \centering
            \includegraphics[width=\linewidth,height=0.160\textheight,keepaspectratio]{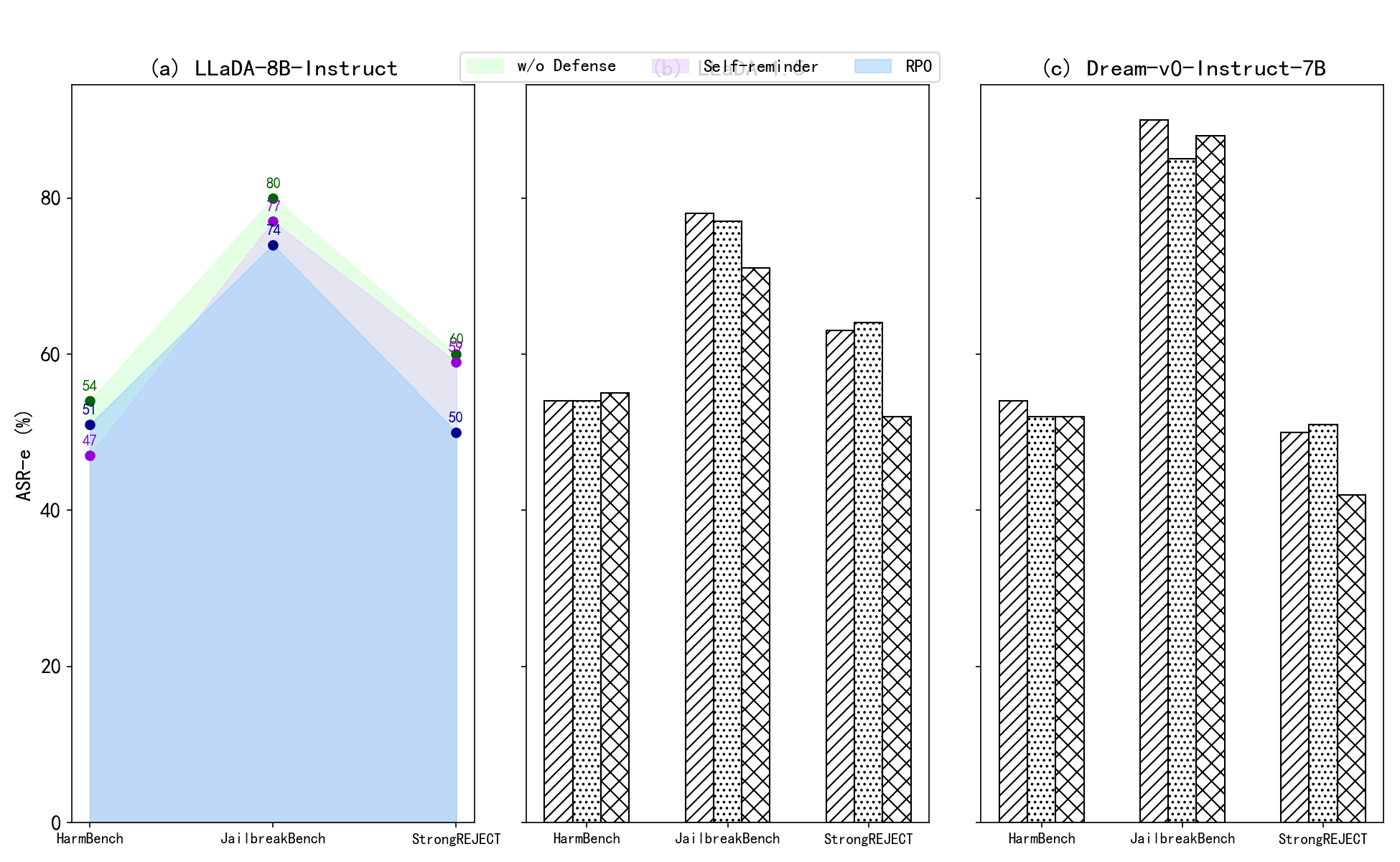}
            \\[0.02cm] \scriptsize (d) Model Output
        \end{minipage}
        
        \vspace{0.08cm}
        \begin{minipage}{0.96\linewidth}
            \footnotesize \textbf{Analysis:} The colors in the left plot of the model output on the right are inconsistent with those in the GT Figure, and the legend positions also differ. In the GT Figure, the bar chart on the right uses different colors, whereas in the model output, it is white.
        \end{minipage}
    \end{minipage}

    \errorcaseseparator
    
    \begin{minipage}{\linewidth}
        \centering
        \textbf{Case 3: Bar Chart (level1)}\\[0.08cm]
        
        \begin{minipage}[t]{0.48\linewidth}
            \centering
            \includegraphics[width=\linewidth,height=0.160\textheight,keepaspectratio]{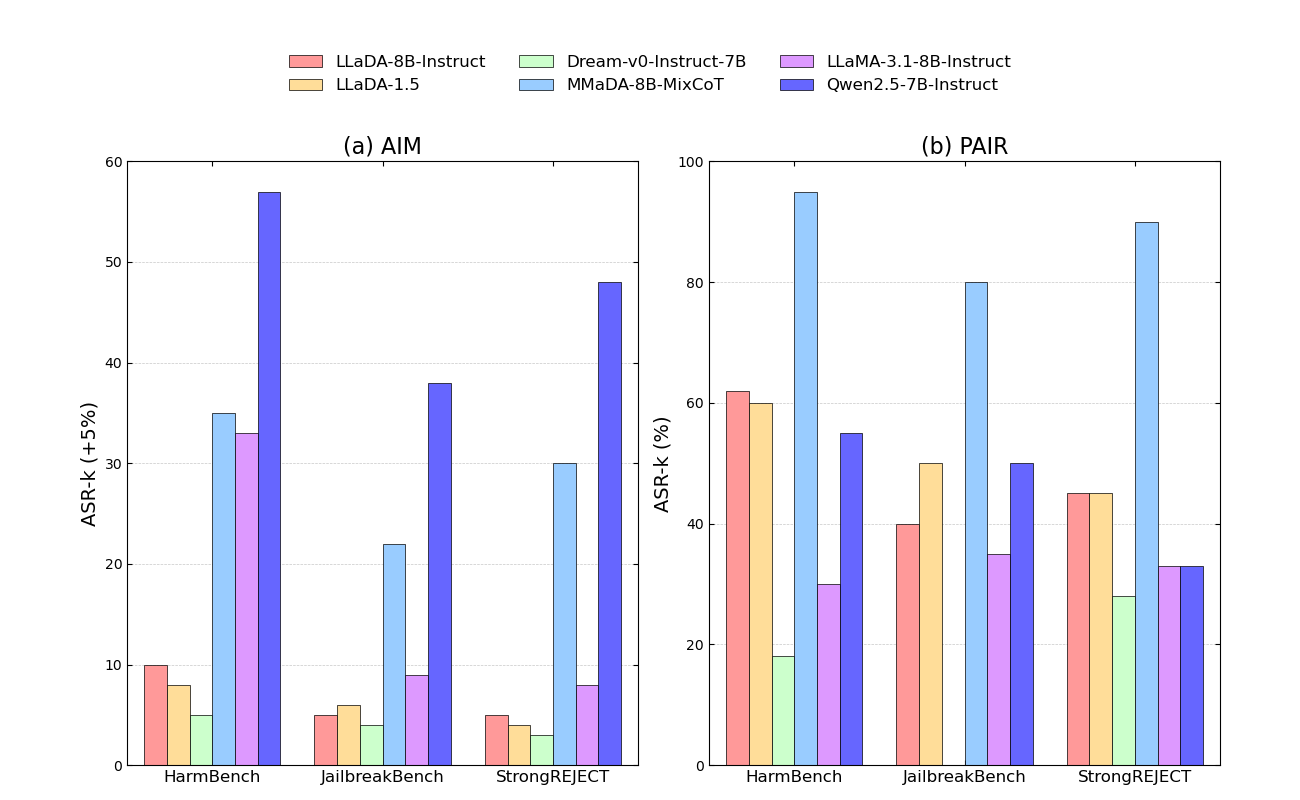}
            \\[0.02cm] \scriptsize (e) Ground Truth
        \end{minipage}
        \hfill
        \begin{minipage}[t]{0.48\linewidth}
            \centering
            \includegraphics[width=\linewidth,height=0.160\textheight,keepaspectratio]{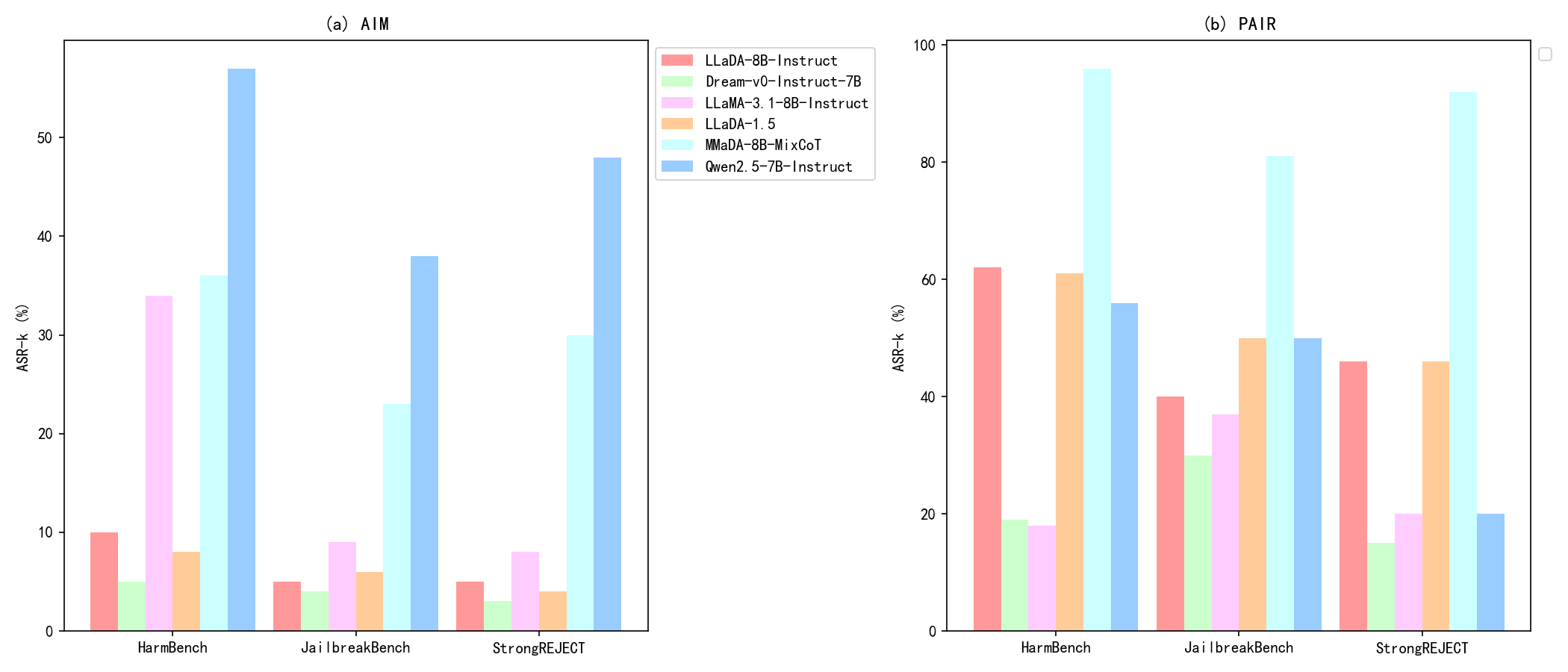}
            \\[0.02cm] \scriptsize (f) Model Output
        \end{minipage}
        
        \vspace{0.08cm}
        \begin{minipage}{0.96\linewidth}
            \footnotesize \textbf{Analysis:} In the generated chart, the dashed grid lines that were present in the ground truth are completely missing, which directly reflects the model's failure to capture and reproduce fine-grained visual attributes.
        \end{minipage}
    \end{minipage}
    
    \vfill
    \caption{\textbf{Qualitative Error Analysis.}Deficiencies in visual styling of model-generated charts, including improper control of line width, transparency, color gradients, edge outlines, and other visual parameters that affect overall appearance.
}
    \label{fig:visual_style errors 1}
\end{compacterrorpage}

\begin{compacterrorpage}
    \centering
    \appendixplatetitle{Common Error Category: Visual Style Errors}
    \vspace{1.0mm}
    \begin{minipage}{\linewidth}
        \centering
        \textbf{Case 1: Bar Chart (level2)}\\[0.08cm]
        
        \begin{minipage}[t]{0.48\linewidth}
            \centering
            \includegraphics[width=\linewidth,height=0.160\textheight,keepaspectratio]{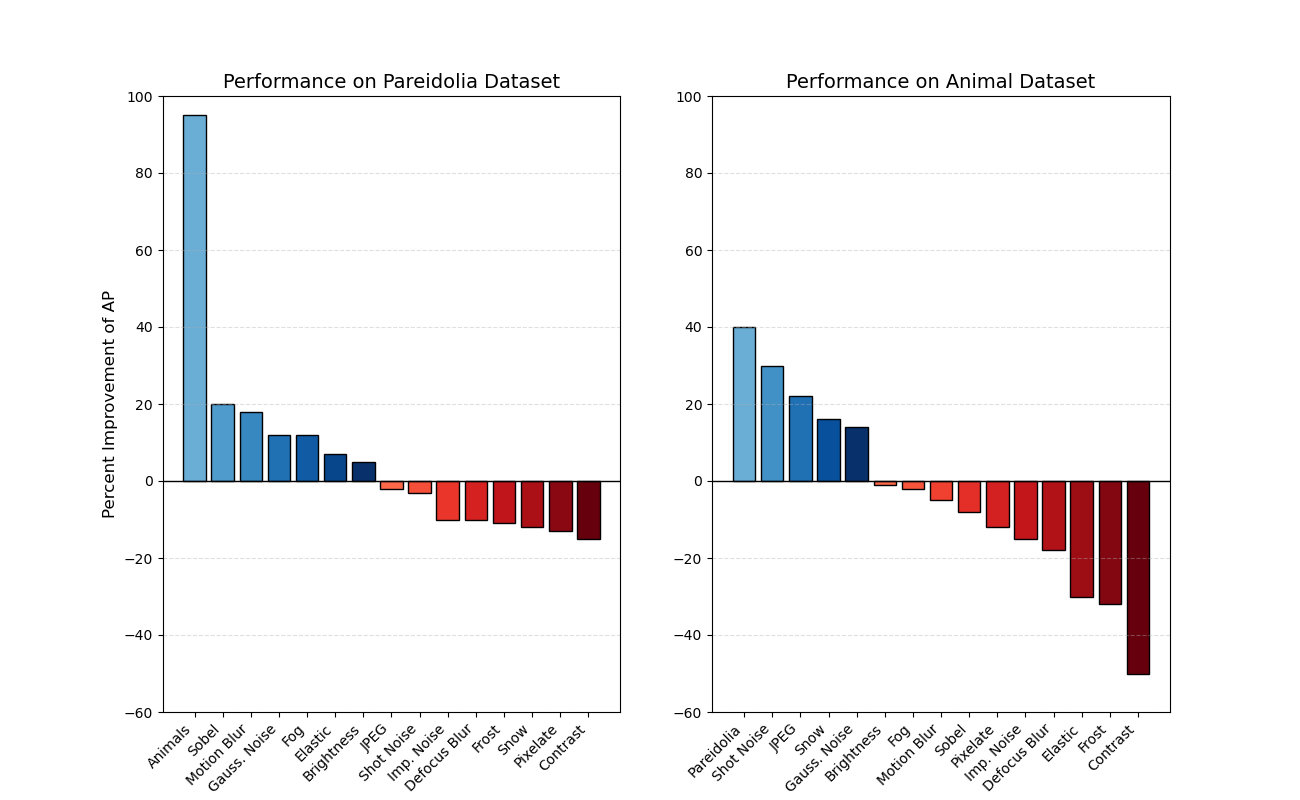}
            \\[0.02cm] \scriptsize (a) Ground Truth
        \end{minipage}
        \hfill
        \begin{minipage}[t]{0.48\linewidth}
            \centering
            \includegraphics[width=\linewidth,height=0.160\textheight,keepaspectratio]{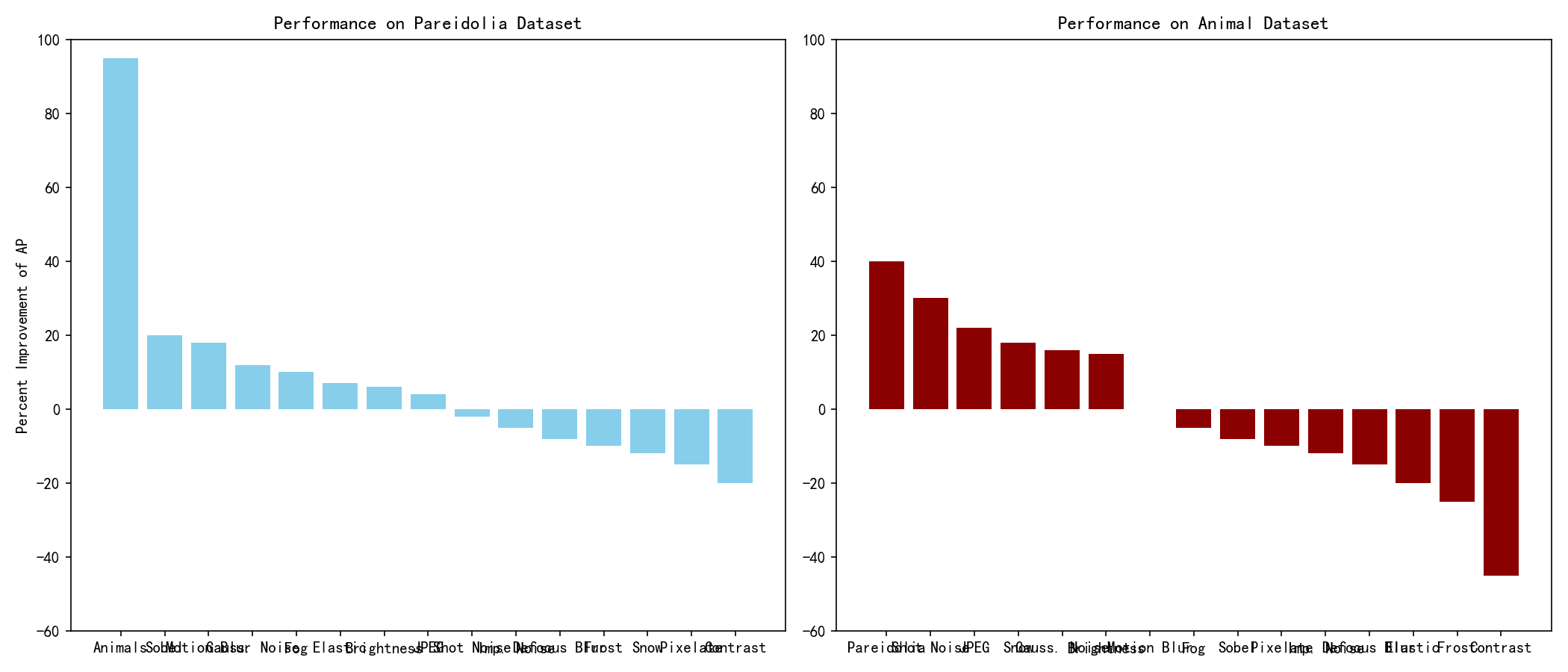}
            \\[0.02cm] \scriptsize (b) Model Output
        \end{minipage}
        
        \vspace{0.08cm}
        \begin{minipage}{0.96\linewidth}
            \footnotesize \textbf{Analysis:} In the generated chart, the dashed grid lines present in the ground truth are completely missing, and the color gradients are not accurately replicated. This directly reflects the model’s failure to capture and reproduce fine-grained visual attributes.
        \end{minipage}
    \end{minipage}

    \errorcaseseparator

    \begin{minipage}{\linewidth}
        \centering
        \textbf{Case 2: Line Chart (level3)}\\[0.08cm]
        
        \begin{minipage}[t]{0.48\linewidth}
            \centering
            \includegraphics[width=\linewidth,height=0.160\textheight,keepaspectratio]{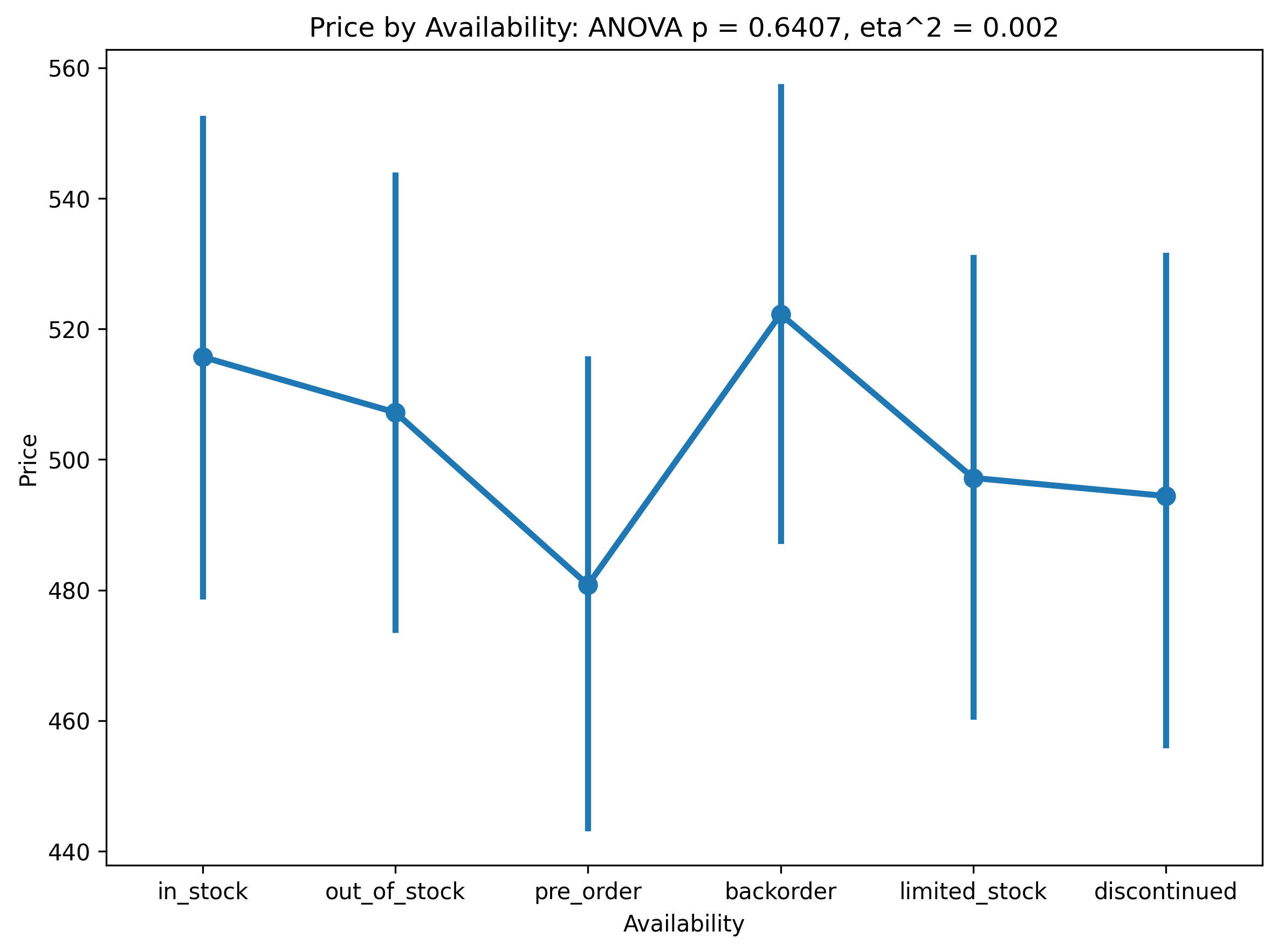}
            \\[0.02cm] \scriptsize (c) Ground Truth
        \end{minipage}
        \hfill
        \begin{minipage}[t]{0.48\linewidth}
            \centering
            \includegraphics[width=\linewidth,height=0.160\textheight,keepaspectratio]{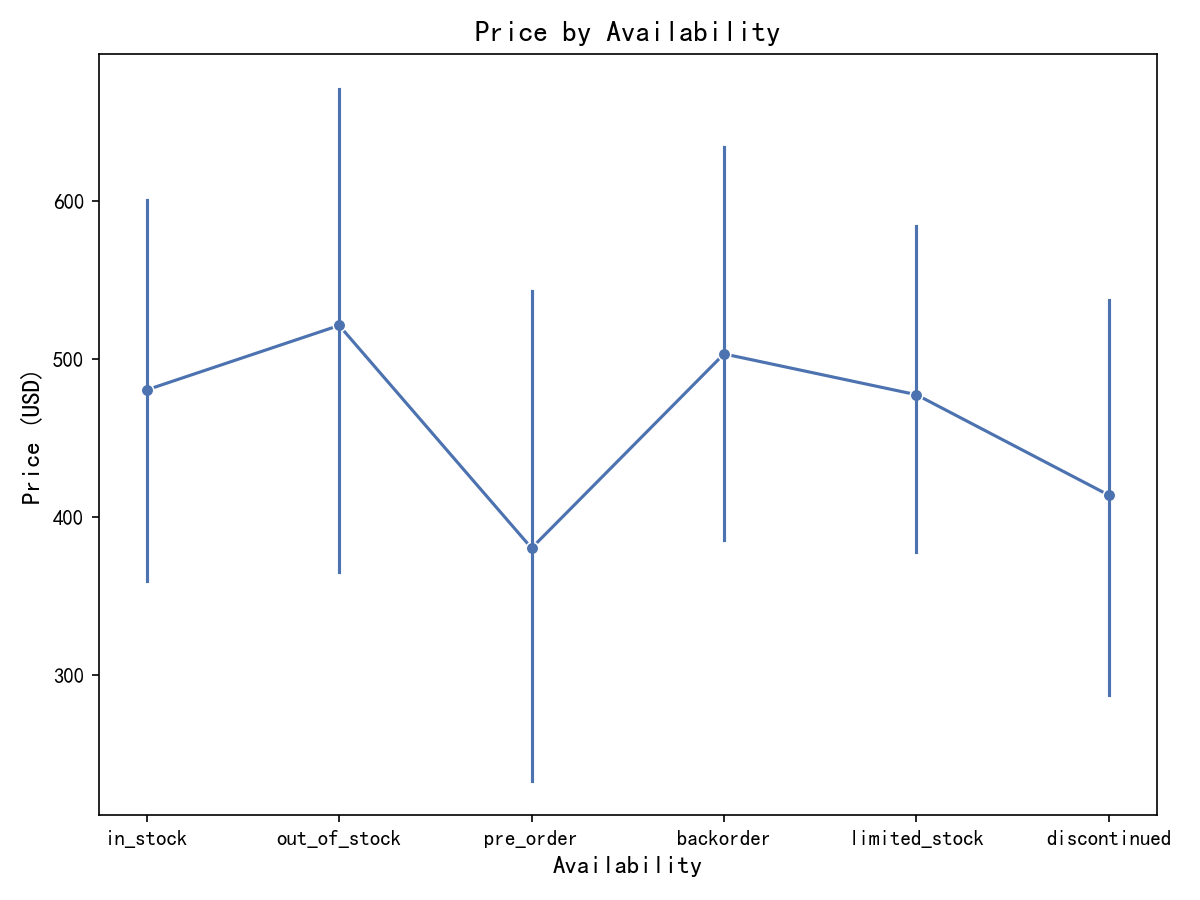}
            \\[0.02cm] \scriptsize (d) Model Output
        \end{minipage}
        
        \vspace{0.08cm}
        \begin{minipage}{0.96\linewidth}
            \footnotesize \textbf{Analysis:} During the visual perception stage, the model failed to capture key stylistic attributes, overlooking the bolded line width and prominent marker sizes present in the original chart
        \end{minipage}
    \end{minipage}

    \errorcaseseparator
    
    \begin{minipage}{\linewidth}
        \centering
        \textbf{Case 3: Combination Chart (level3)}\\[0.08cm]
        
        \begin{minipage}[t]{0.48\linewidth}
            \centering
            \includegraphics[width=\linewidth,height=0.160\textheight,keepaspectratio]{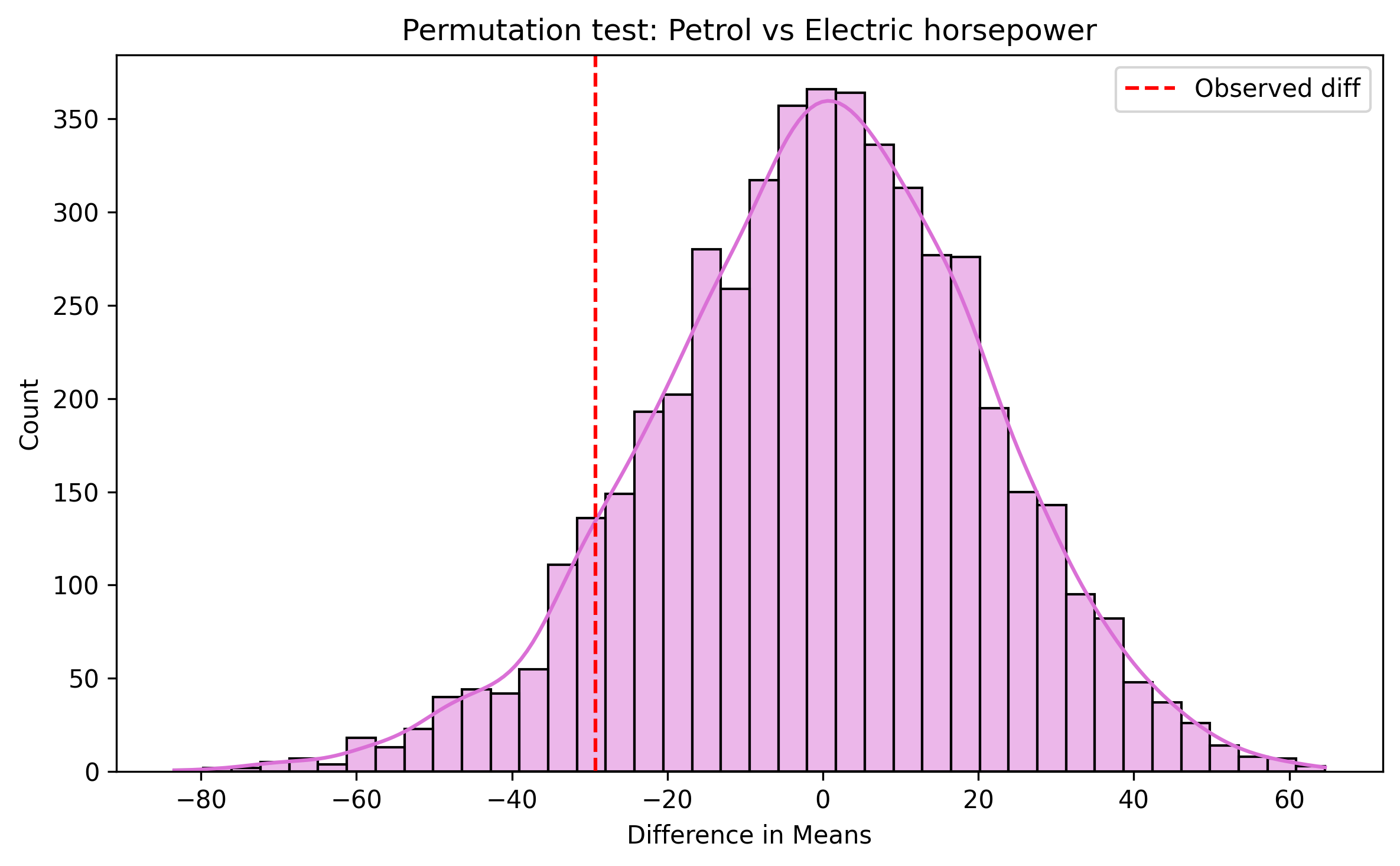}
            \\[0.02cm] \scriptsize (e) Ground Truth
        \end{minipage}
        \hfill
        \begin{minipage}[t]{0.48\linewidth}
            \centering
            \includegraphics[width=\linewidth,height=0.160\textheight,keepaspectratio]{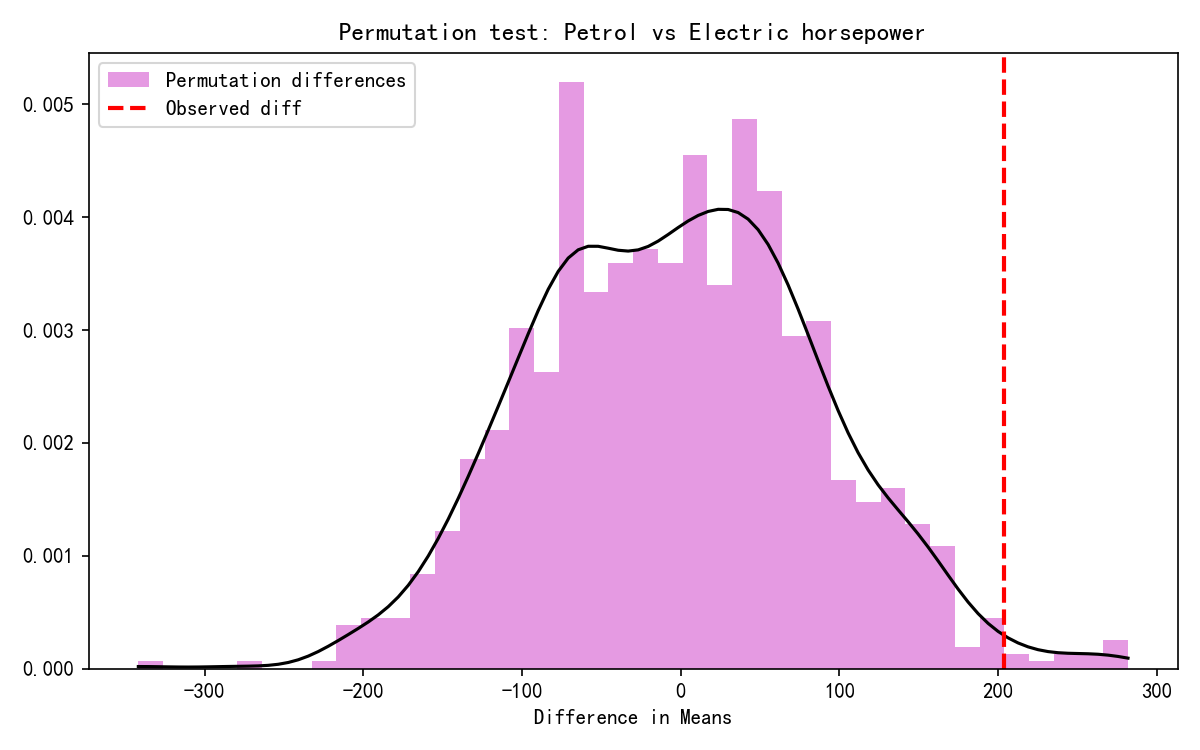}
            \\[0.02cm] \scriptsize (f) Model Output
        \end{minipage}
        
        \vspace{0.08cm}
        \begin{minipage}{0.96\linewidth}
            \footnotesize \textbf{Analysis:} In the generated chart, the histogram bars lack the distinct black borders that were present in the ground truth, which reduces the visual clarity and contrast of the bar distribution, reflecting the model's failure to reproduce fine-grained visual details.
        \end{minipage}
    \end{minipage}
    
    \vfill
    \caption{\textbf{Qualitative Error Analysis.} Errors in visual parameterization, such as incorrect line thickness, opacity, gradient schemes, and edge styling, leading to suboptimal visual presentation.}
    \label{fig:visual_style errors 2}
\end{compacterrorpage}

\begin{compacterrorpage}
    \centering
    \appendixplatetitle{Special Error Category for level2}
    \vspace{1.0mm}
    \begin{minipage}{\linewidth}
        \centering
        \textbf{Case 1: Bar Chart (level2)}\\[0.08cm]
        
        \begin{minipage}[t]{0.48\linewidth}
            \centering
            \includegraphics[width=\linewidth,height=0.160\textheight,keepaspectratio]{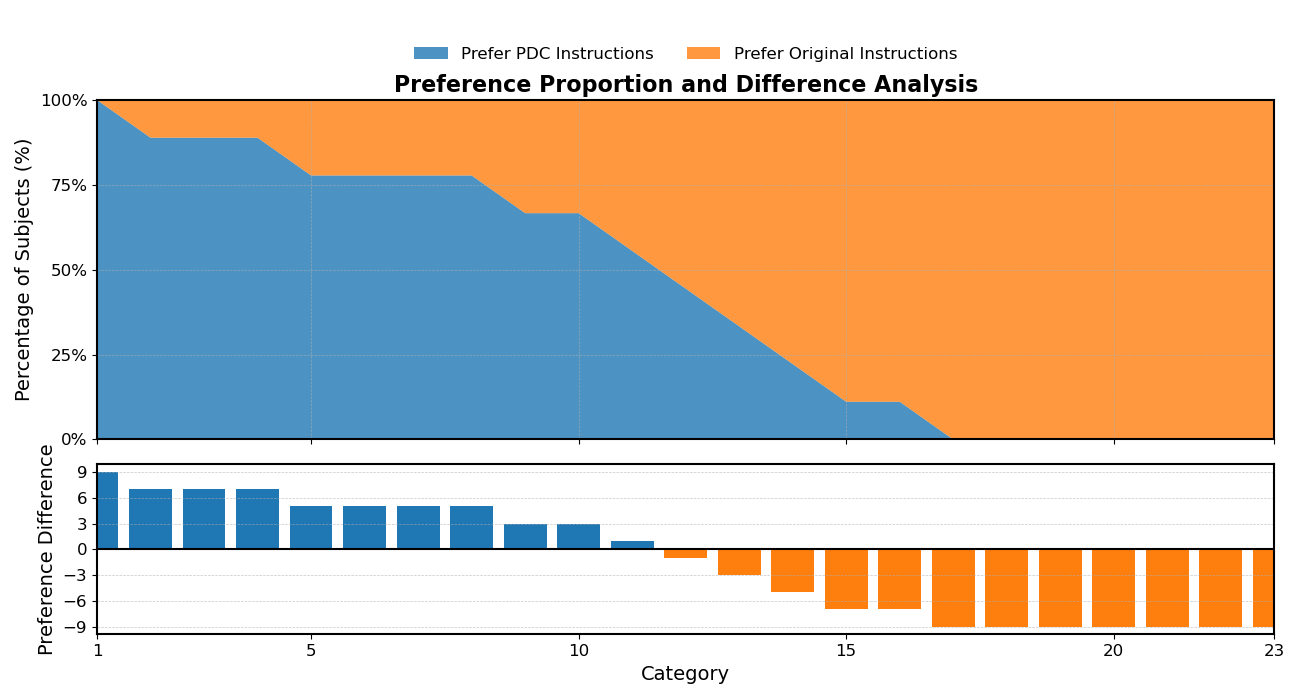}
            \\[0.02cm] \scriptsize (a) Ground Truth
        \end{minipage}
        \hfill
        \begin{minipage}[t]{0.48\linewidth}
            \centering
            \includegraphics[width=\linewidth,height=0.160\textheight,keepaspectratio]{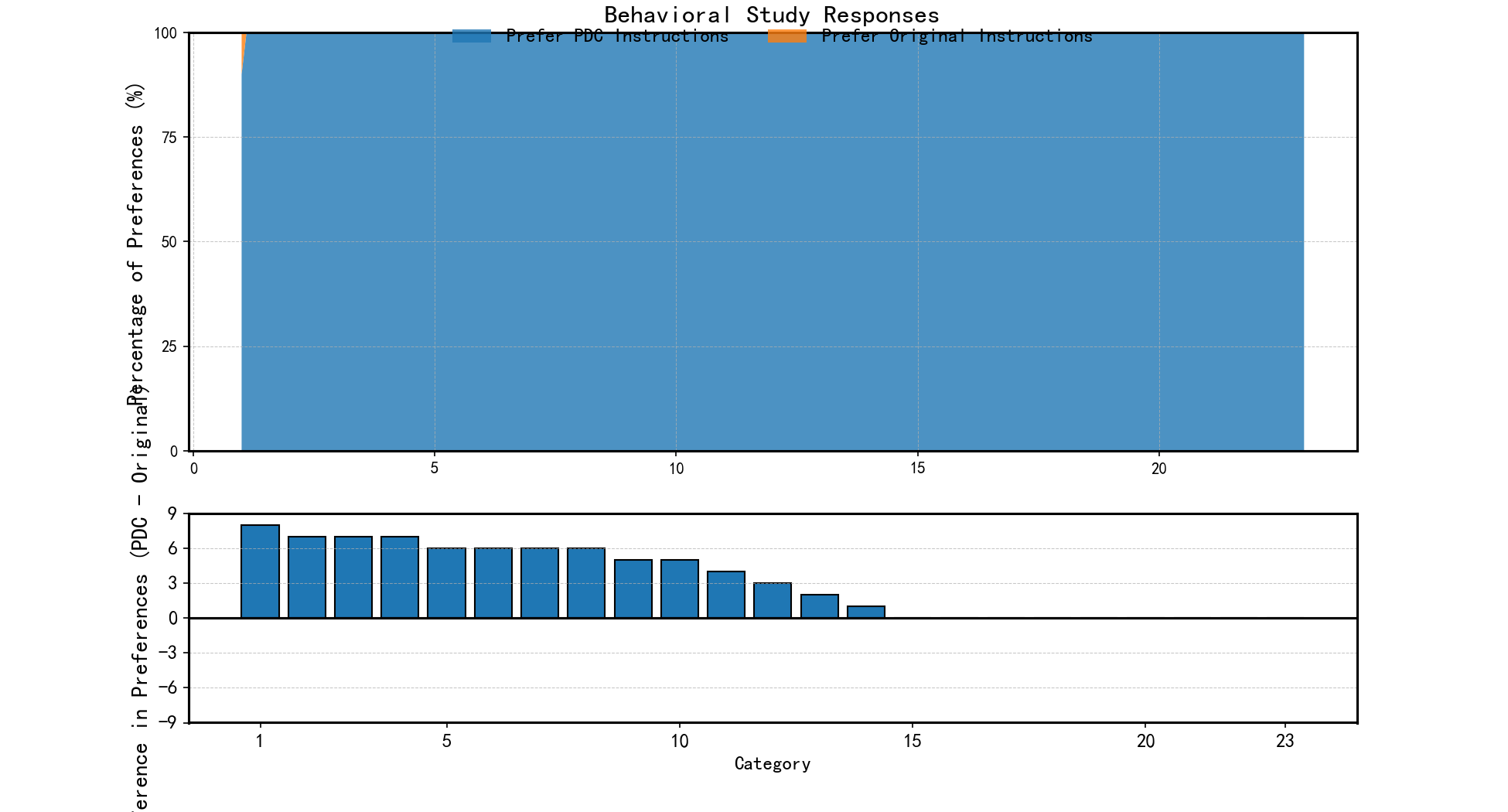}
            \\[0.02cm] \scriptsize (b) Model Output
        \end{minipage}
        
        \vspace{0.08cm}
        \begin{minipage}{0.96\linewidth}
            \footnotesize \textbf{Analysis:} The generated output exhibits a severe logical discrepancy compared to the ground truth. It misrepresents the 100\% stacked proportions in the top subplot and completely omits the corresponding negative difference bars in the bottom subplot, failing to reflect the required compositional mathematical constraints.
        \end{minipage}
    \end{minipage}

    \errorcaseseparator

    \begin{minipage}{\linewidth}
        \centering
        \textbf{Case 2: Line Chart (level2)}\\[0.08cm]
        
        \begin{minipage}[t]{0.48\linewidth}
            \centering
            \includegraphics[width=\linewidth,height=0.160\textheight,keepaspectratio]{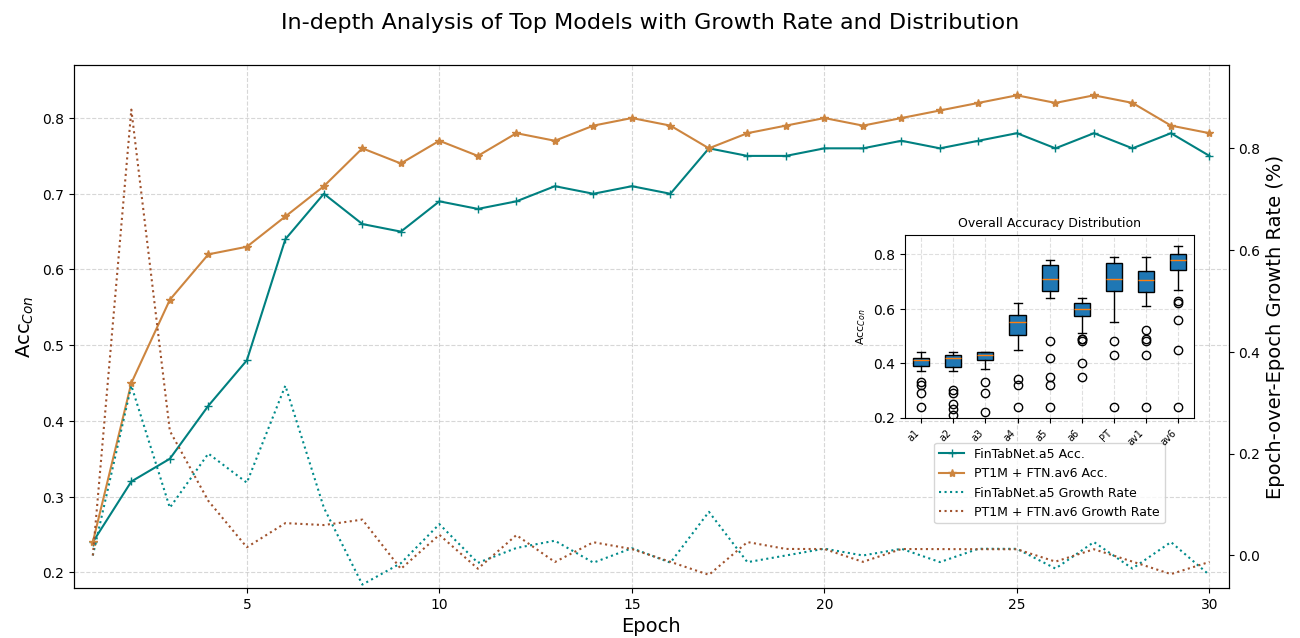}
            \\[0.02cm] \scriptsize (c) Ground Truth
        \end{minipage}
        \hfill
        \begin{minipage}[t]{0.48\linewidth}
            \centering
            \includegraphics[width=\linewidth,height=0.160\textheight,keepaspectratio]{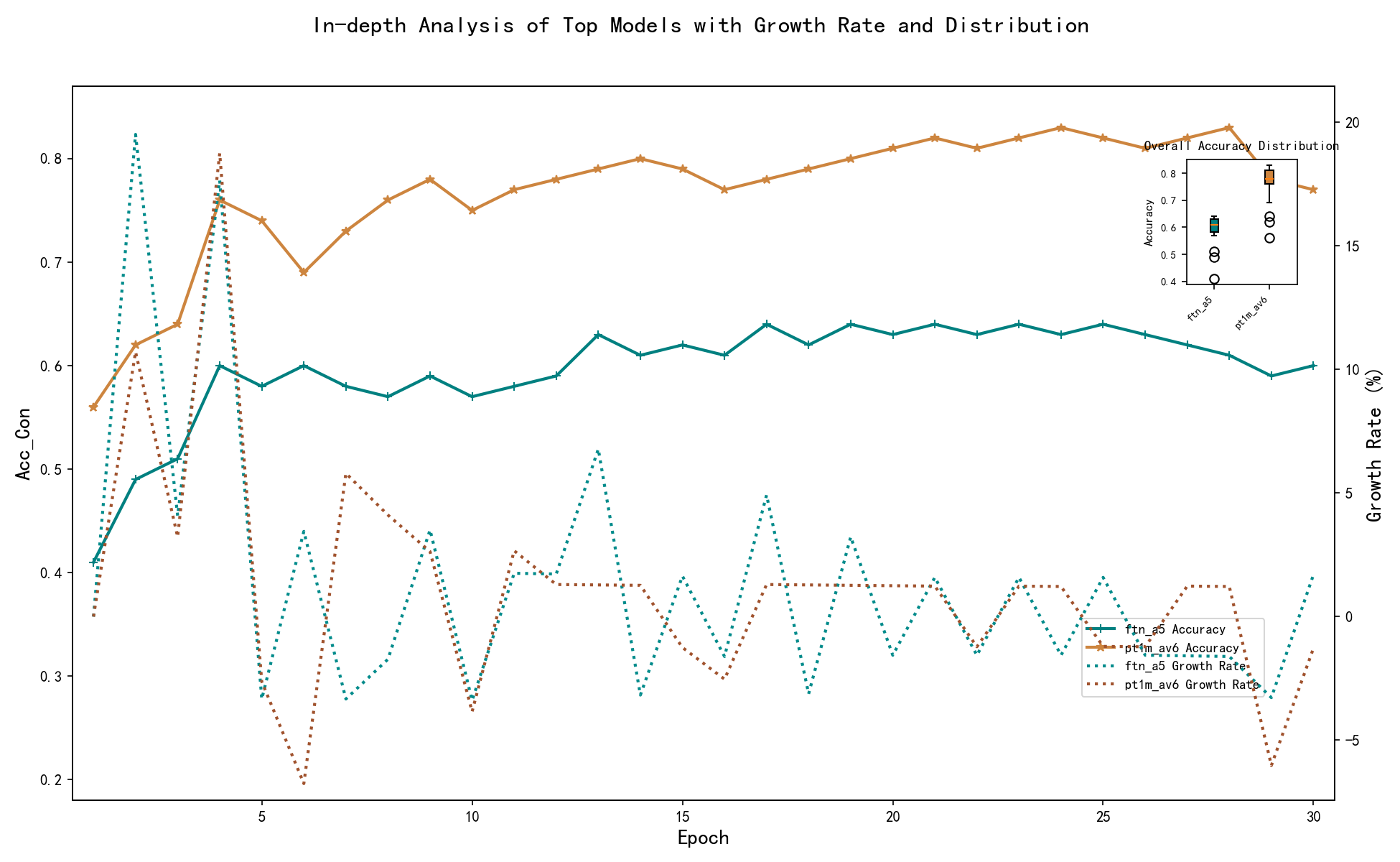}
            \\[0.02cm] \scriptsize (d) Model Output
        \end{minipage}
        
        \vspace{0.08cm}
        \begin{minipage}{0.96\linewidth}
            \footnotesize \textbf{Analysis:} As illustrated, the significantly complex instruction of embedding a box plot inset within a dual-axis line chart exceeds the model's reasoning capacity; this limitation prevents it from performing valid edits that satisfy the visual context, ultimately resulting in a structurally collapsed inset and severely distorted secondary axis data mapping.
        \end{minipage}
    \end{minipage}

    \errorcaseseparator
    
    \begin{minipage}{\linewidth}
        \centering
        \textbf{Case 3: Combination Chart (level2)}\\[0.08cm]
        
        \begin{minipage}[t]{0.48\linewidth}
            \centering
            \includegraphics[width=\linewidth,height=0.160\textheight,keepaspectratio]{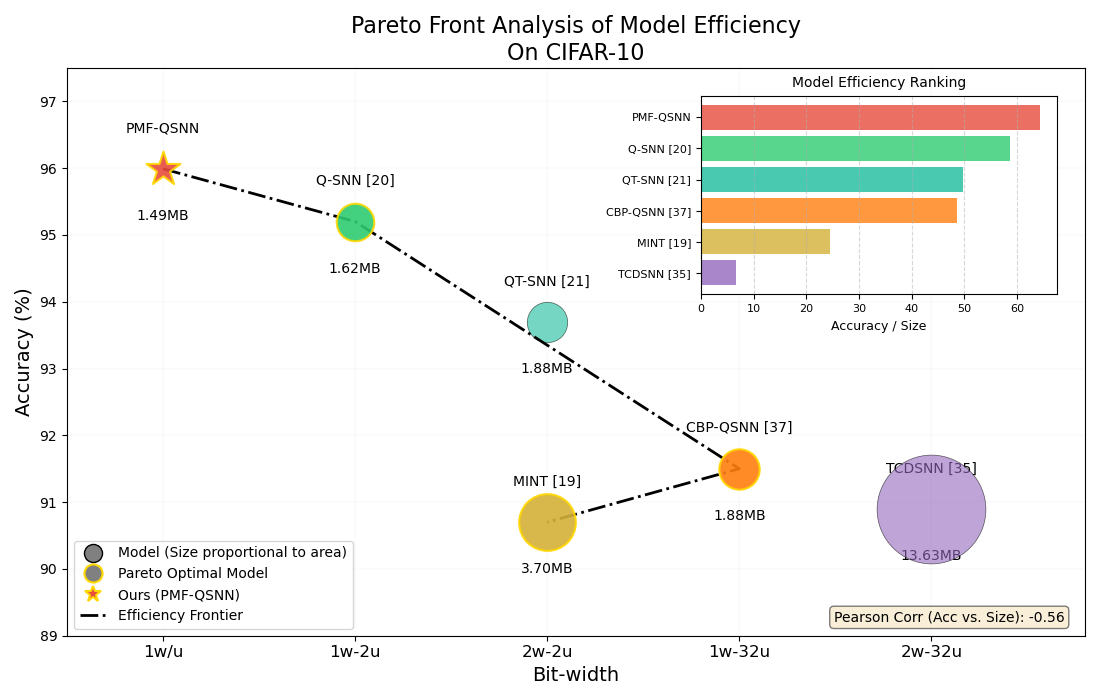}
            \\[0.02cm] \scriptsize (e) Ground Truth
        \end{minipage}
        \hfill
        \begin{minipage}[t]{0.48\linewidth}
            \centering
            \includegraphics[width=\linewidth,height=0.160\textheight,keepaspectratio]{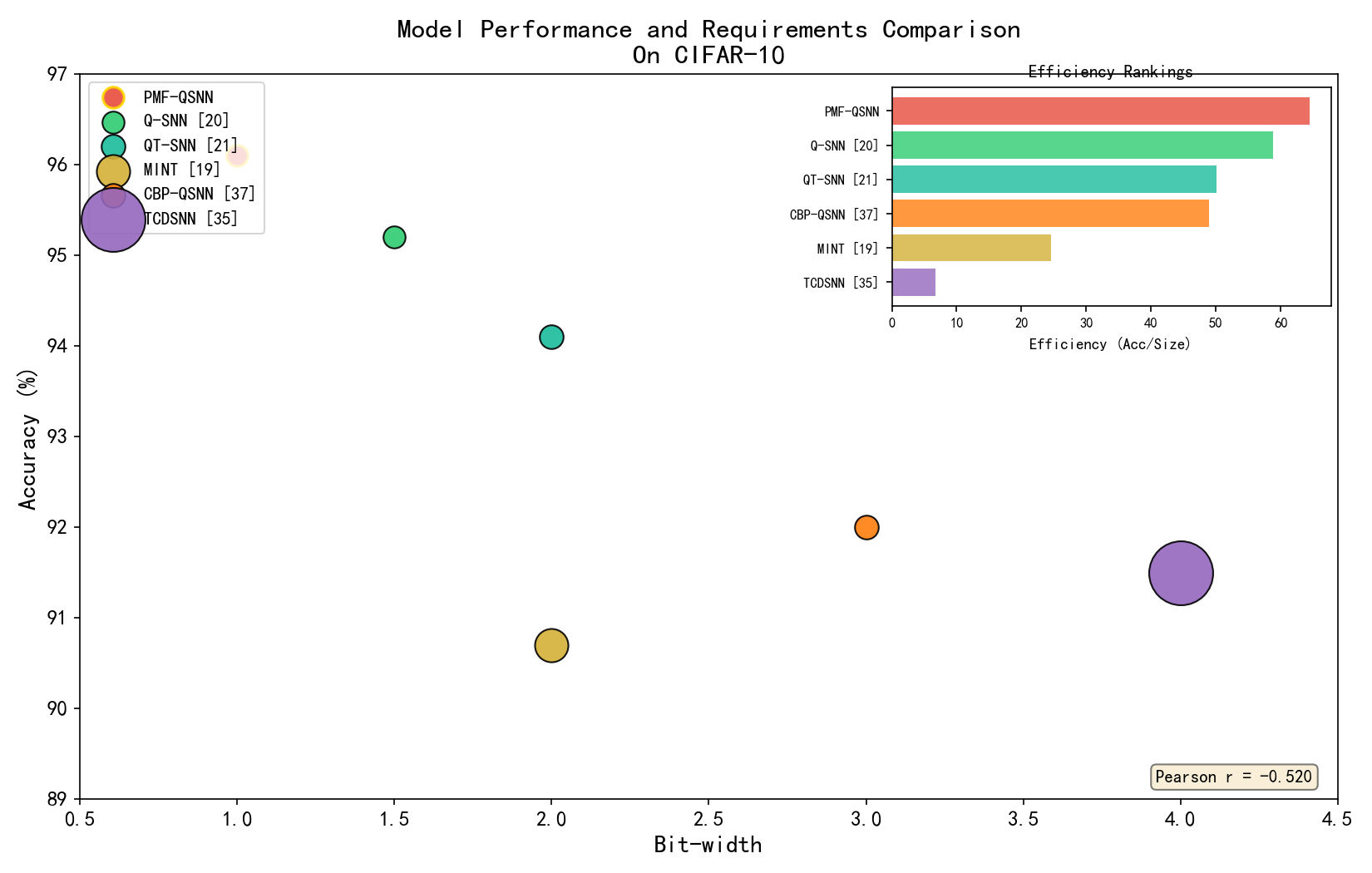}
            \\[0.02cm] \scriptsize (f) Model Output
        \end{minipage}
        
        \vspace{0.08cm}
        \begin{minipage}{0.96\linewidth}
            \footnotesize \textbf{Analysis:} The highly complex compound plotting instructions—requiring an inset chart, a Pareto front line, custom markers, and text annotations—overwhelmed the model's execution capacity, resulting in unimplemented key features (e.g., missing lines and text labels) and partial data loss (e.g., the categorical X-axis being erroneously replaced by arbitrary numerical values).
        \end{minipage}
    \end{minipage}
    
    \vfill
    \caption{\textbf{Level 2–Specific Errors:}  Due to the substantially more complex editing instructions, reasoning becomes the primary bottleneck. Limited reasoning capacity prevents the model from producing edits that simultaneously satisfy the visual context and the given instructions, leading to consistently lower performance than Level 1.
}
    \label{fig:special_level2_1}
\end{compacterrorpage}

\begin{compacterrorpage}
    \centering
    \appendixplatetitle{Special Error Category for level2}
    \vspace{1.0mm}
    \begin{minipage}{\linewidth}
        \centering
        \textbf{Case 1: Combination Chart (level2)}\\[0.08cm]
        
        \begin{minipage}[t]{0.48\linewidth}
            \centering
            \includegraphics[width=\linewidth,height=0.160\textheight,keepaspectratio]{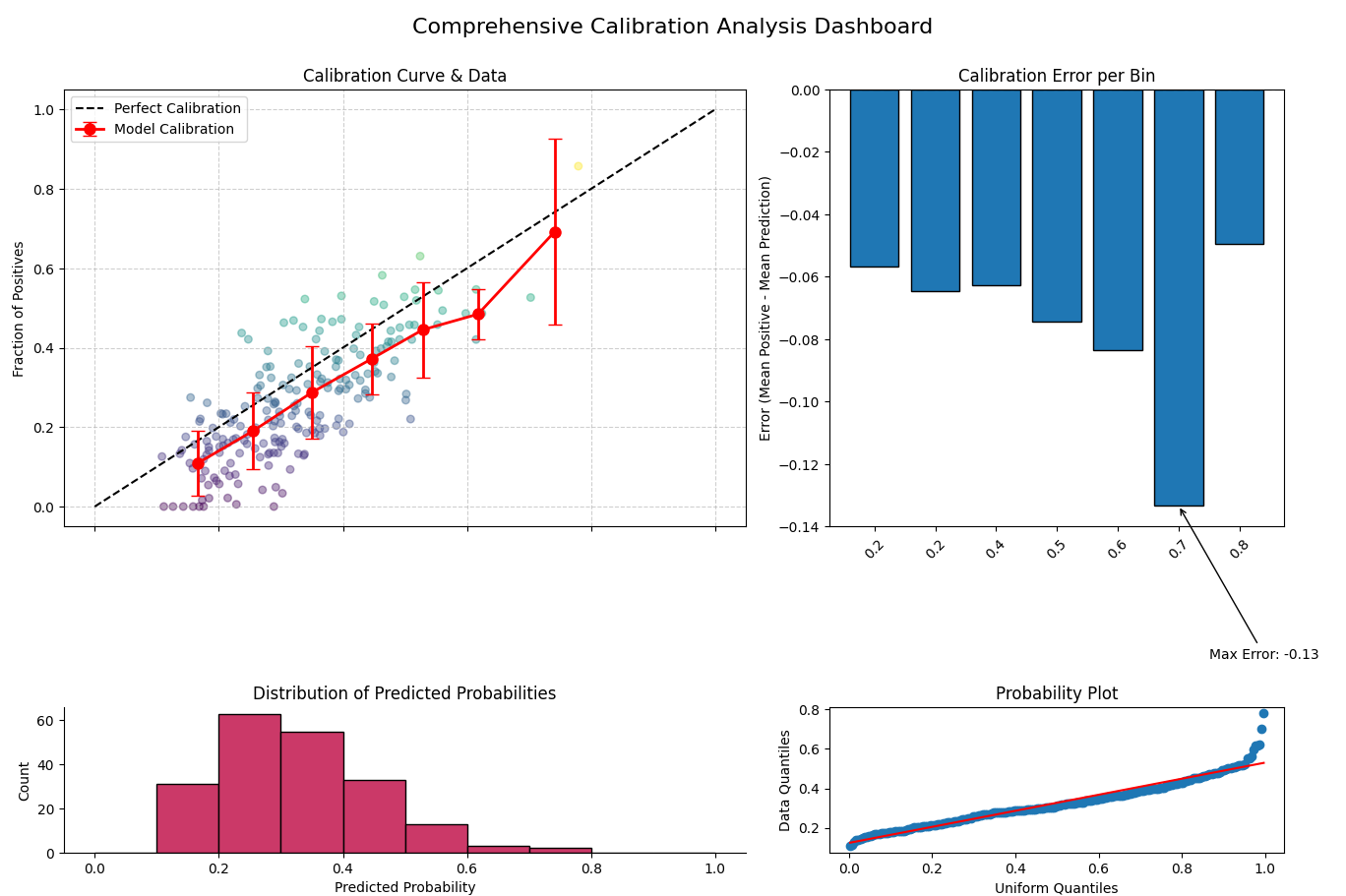}
            \\[0.02cm] \scriptsize (a) Ground Truth
        \end{minipage}
        \hfill
        \begin{minipage}[t]{0.48\linewidth}
            \centering
            \includegraphics[width=\linewidth,height=0.160\textheight,keepaspectratio]{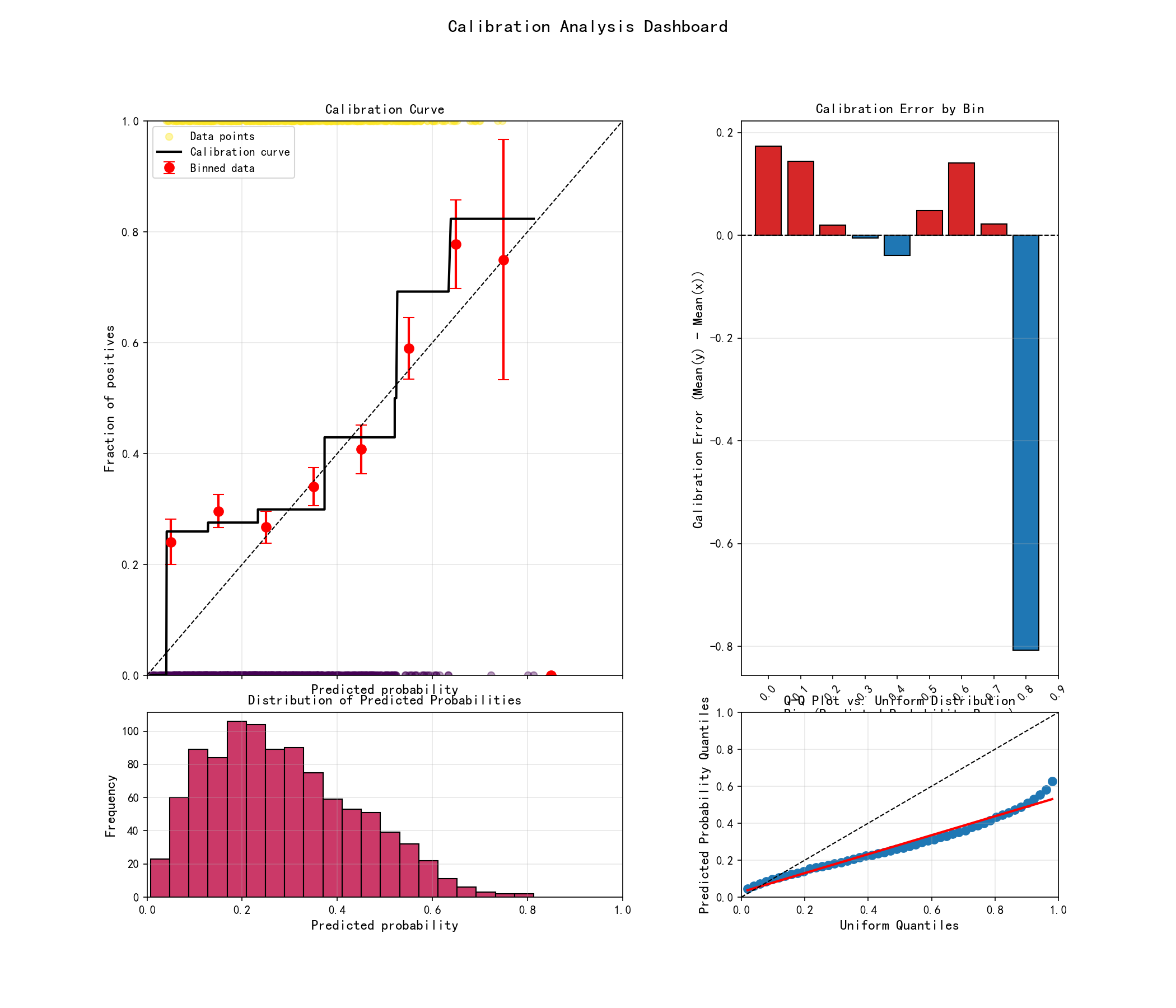}
            \\[0.02cm] \scriptsize (b) Model Output
        \end{minipage}
        
        \vspace{0.08cm}
        \begin{minipage}{0.96\linewidth}
            \footnotesize \textbf{Analysis:} As illustrated by the calibration dashboard, the highly complex composite instructions—which demand cross-plot axis sharing, and data-conditional annotations—overwhelm the model's logical reasoning and multi-task scheduling capacities; this not only causes a complete collapse of structural layout constraints but also leads to the total omission of crucial analytical visual elements, such as the maximum error highlight.
        \end{minipage}
    \end{minipage}

    \errorcaseseparator

    \begin{minipage}{\linewidth}
        \centering
        \textbf{Case 2: Combination Chart (level2)}\\[0.08cm]
        
        \begin{minipage}[t]{0.48\linewidth}
            \centering
            \includegraphics[width=\linewidth,height=0.160\textheight,keepaspectratio]{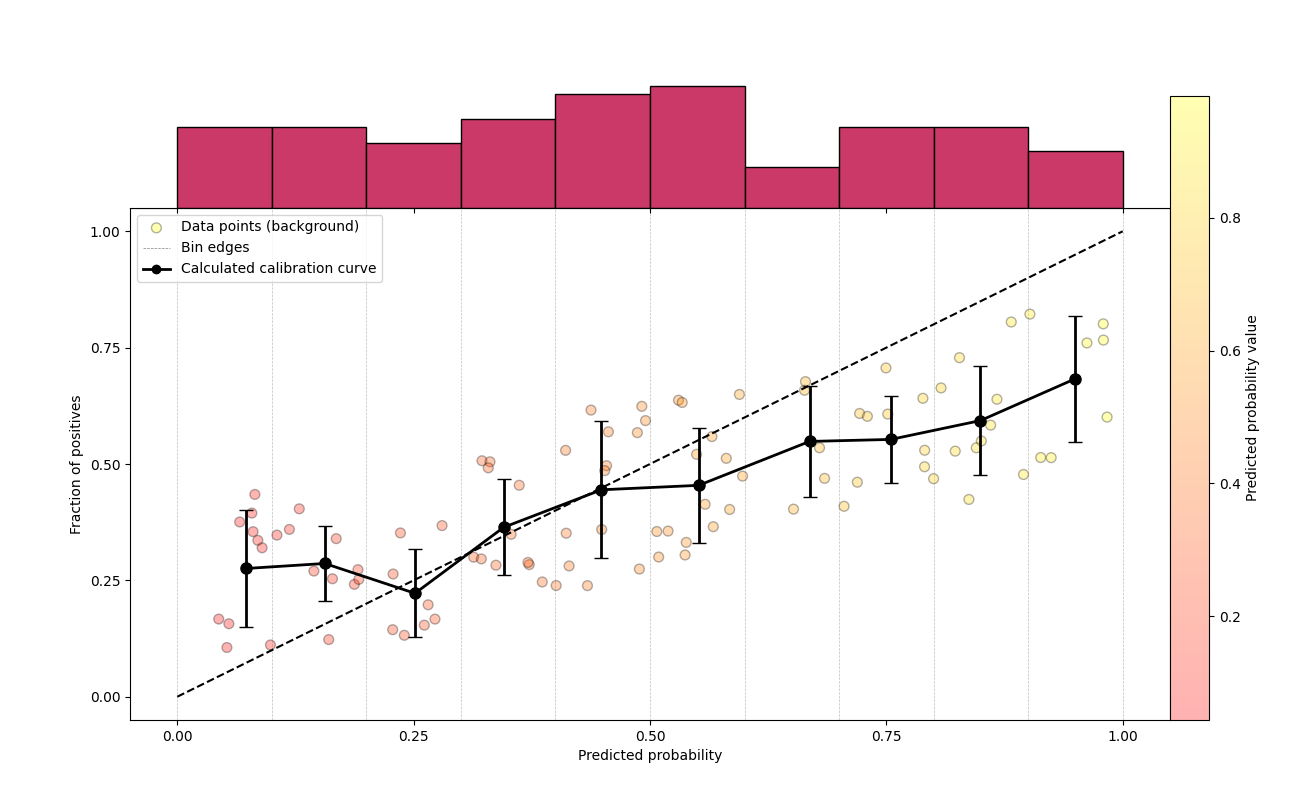}
            \\[0.02cm] \scriptsize (c) Ground Truth
        \end{minipage}
        \hfill
        \begin{minipage}[t]{0.48\linewidth}
            \centering
            \includegraphics[width=\linewidth,height=0.160\textheight,keepaspectratio]{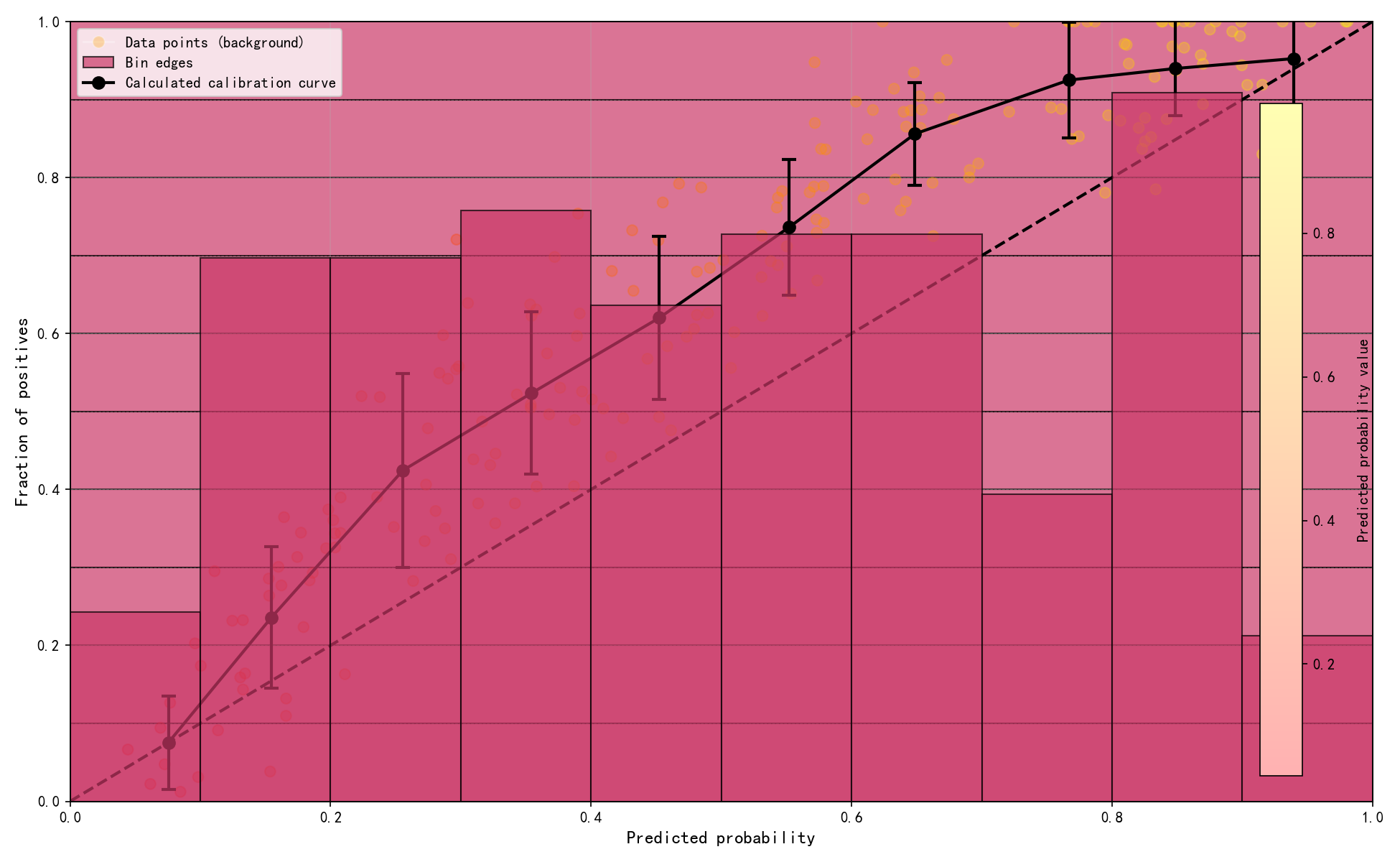}
            \\[0.02cm] \scriptsize (d) Model Output
        \end{minipage}
        
        \vspace{0.08cm}
        \begin{minipage}{0.96\linewidth}
            \footnotesize \textbf{Analysis:} When instructions simultaneously demand dynamic statistical calculations (e.g., binning for means and standard deviations) and rigorous spatial/layering constraints (e.g., absolute coordinate positioning and multiple Z-order configurations), the model suffers a complete layout collapse due to insufficient spatial reasoning. This is evidenced by the wildly disproportionate histogram that obfuscates the core plotting area, highlighting the model's limitations in complex, multi-task visual rendering.
        \end{minipage}
    \end{minipage}

    \errorcaseseparator
    
    \begin{minipage}{\linewidth}
        \centering
        \textbf{Case 3: Combination Chart (level2)}\\[0.08cm]
        
        \begin{minipage}[t]{0.48\linewidth}
            \centering
            \includegraphics[width=\linewidth,height=0.160\textheight,keepaspectratio]{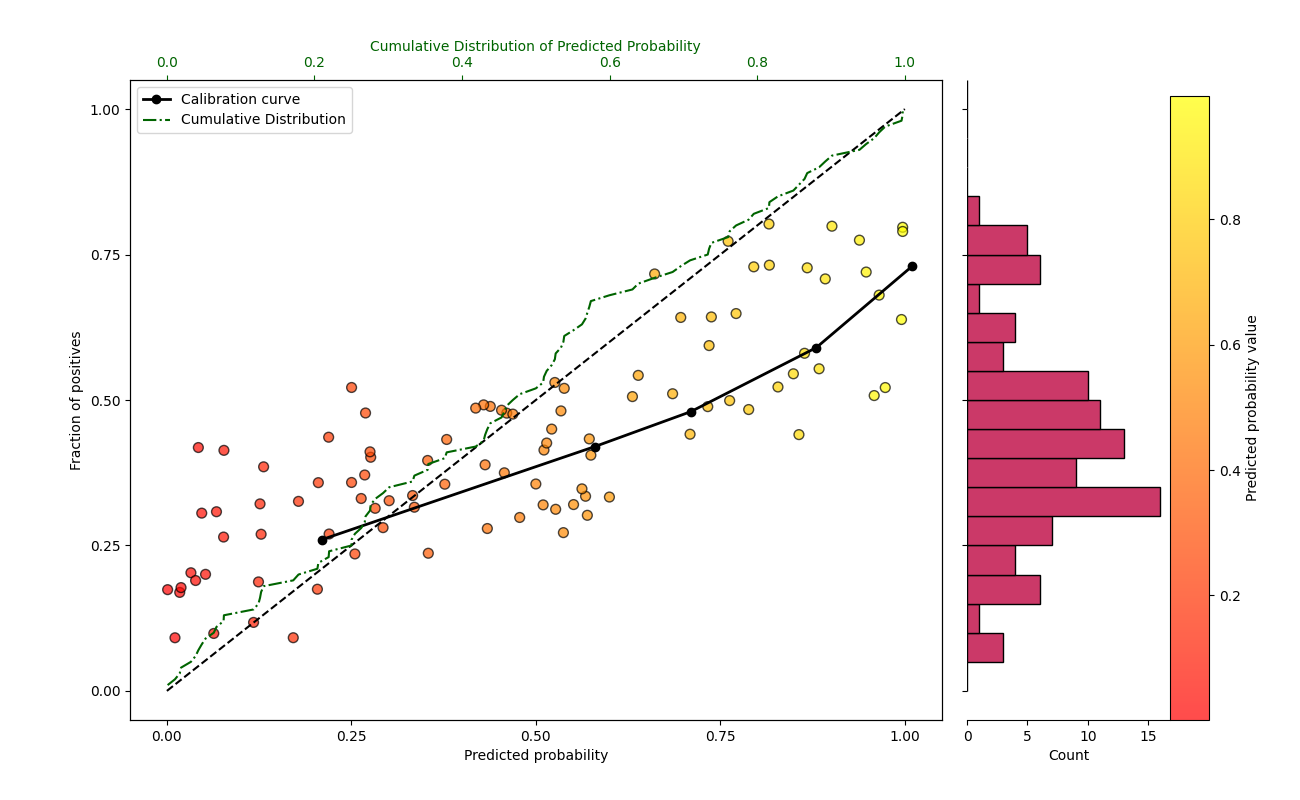}
            \\[0.02cm] \scriptsize (e) Ground Truth
        \end{minipage}
        \hfill
        \begin{minipage}[t]{0.48\linewidth}
            \centering
            \includegraphics[width=\linewidth,height=0.160\textheight,keepaspectratio]{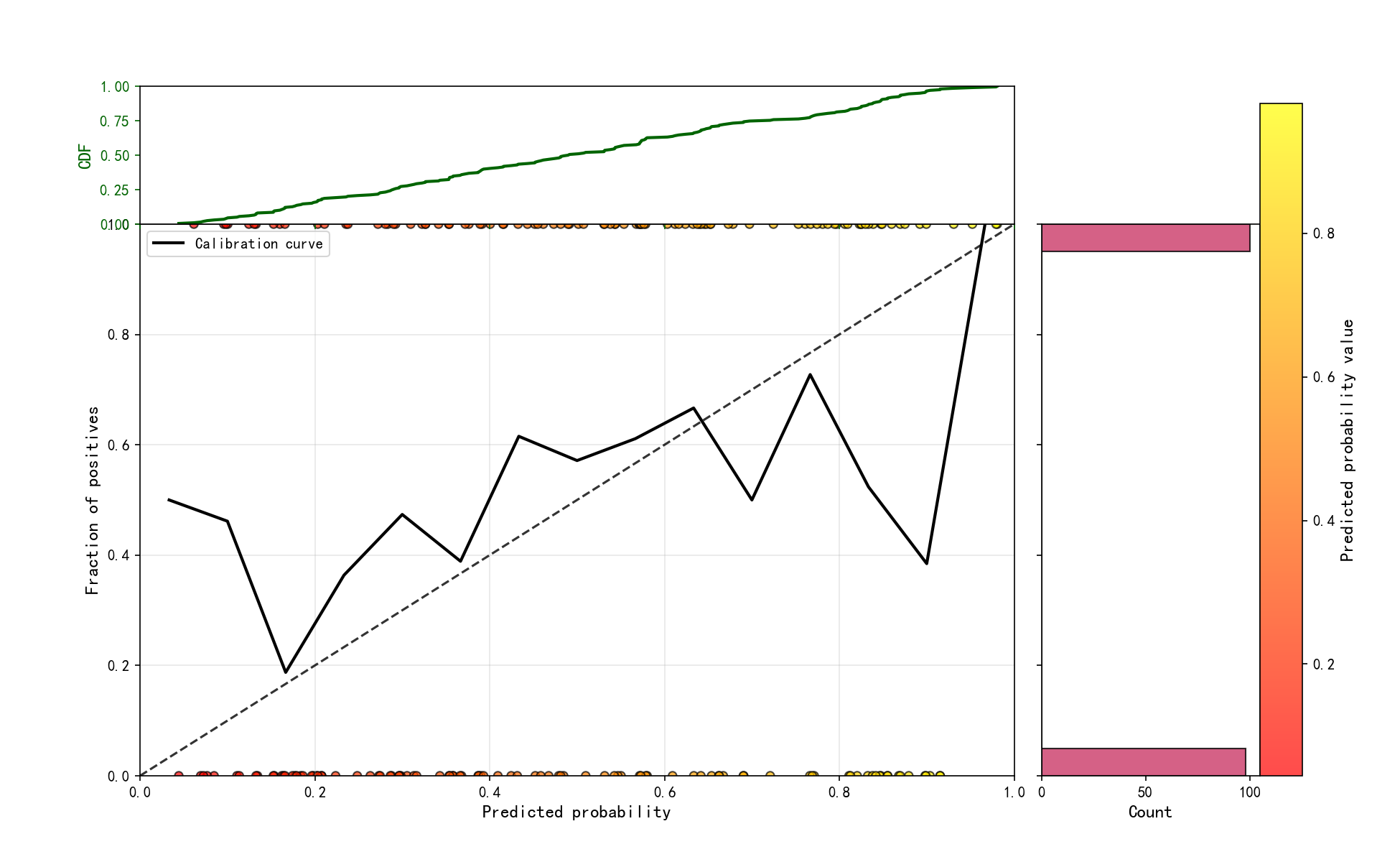}
            \\[0.02cm] \scriptsize (f) Model Output
        \end{minipage}
        
        \vspace{0.08cm}
        \begin{minipage}{0.96\linewidth}
            \footnotesize \textbf{Analysis:} The catastrophic layout failures shown in the figure, including the disconnected CDF axis and missing histogram data, highlight a critical reasoning bottleneck. When confronted with highly reasoning-dependent tasks like reformatting the chart into a complex left-right layout with shared axes, the model fails completely, resulting in an unusable output.
        \end{minipage}
    \end{minipage}
    
    \vfill
    \caption{\textbf{Level 2–Specific Errors:}  Due to the substantially more complex editing instructions, reasoning becomes the primary bottleneck. Limited reasoning capacity prevents the model from producing edits that simultaneously satisfy the visual context and the given instructions, leading to consistently lower performance than Level 1.
}
    \label{fig:special_level2_2}
\end{compacterrorpage}

\begin{compacterrorpage}
    \centering
    \appendixplatetitle{Special Error Category for level3}
    \vspace{1.0mm}
    \begin{minipage}{\linewidth}
        \centering
        \textbf{Case 1: Heatmap Chart (level3)}\\[0.08cm]
        
        \begin{minipage}[t]{0.48\linewidth}
            \centering
            \includegraphics[width=\linewidth,height=0.160\textheight,keepaspectratio]{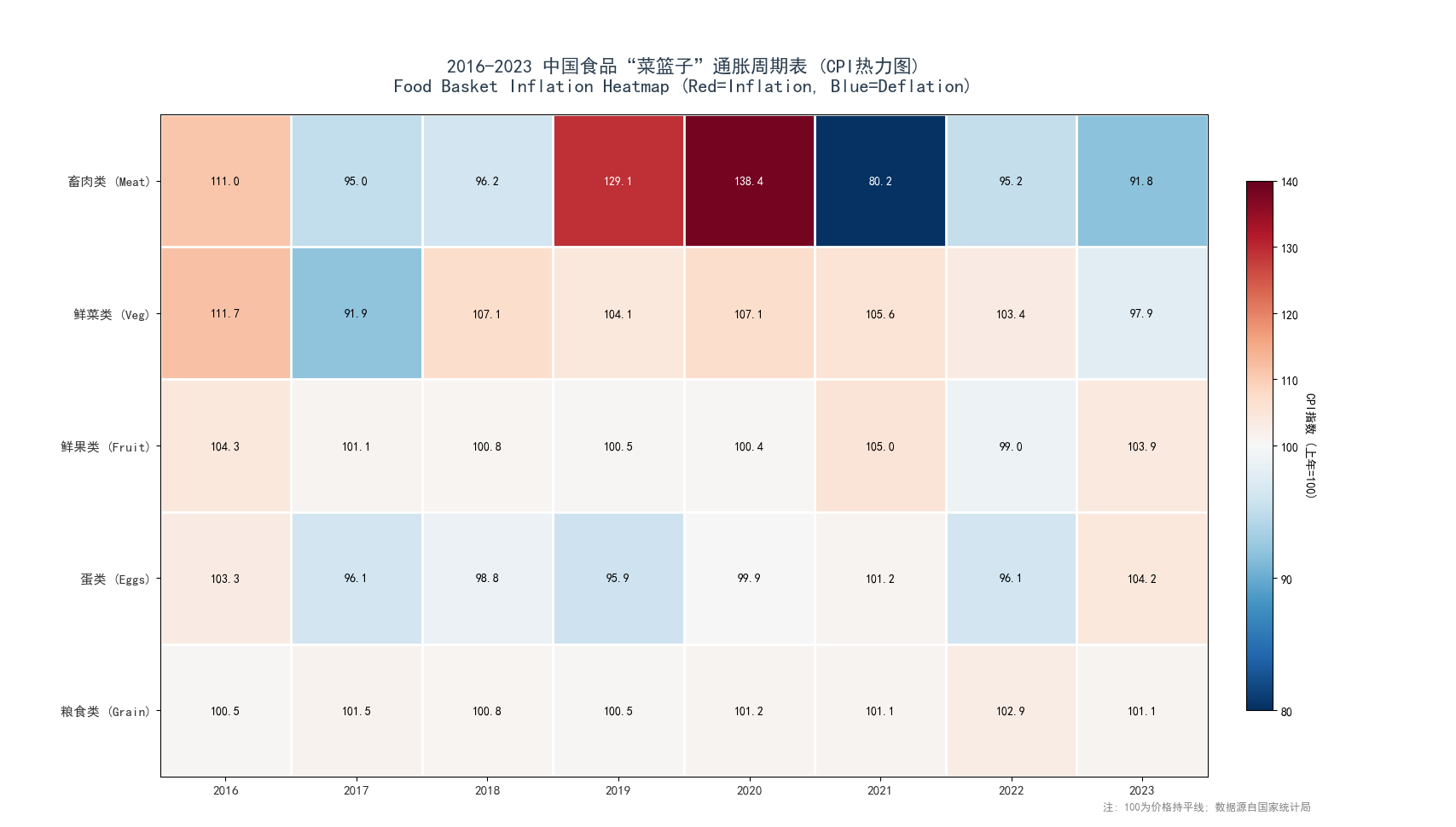}
            \\[0.02cm] \scriptsize (a) Ground Truth
        \end{minipage}
        \hfill
        \begin{minipage}[t]{0.48\linewidth}
            \centering
            \includegraphics[width=\linewidth,height=0.160\textheight,keepaspectratio]{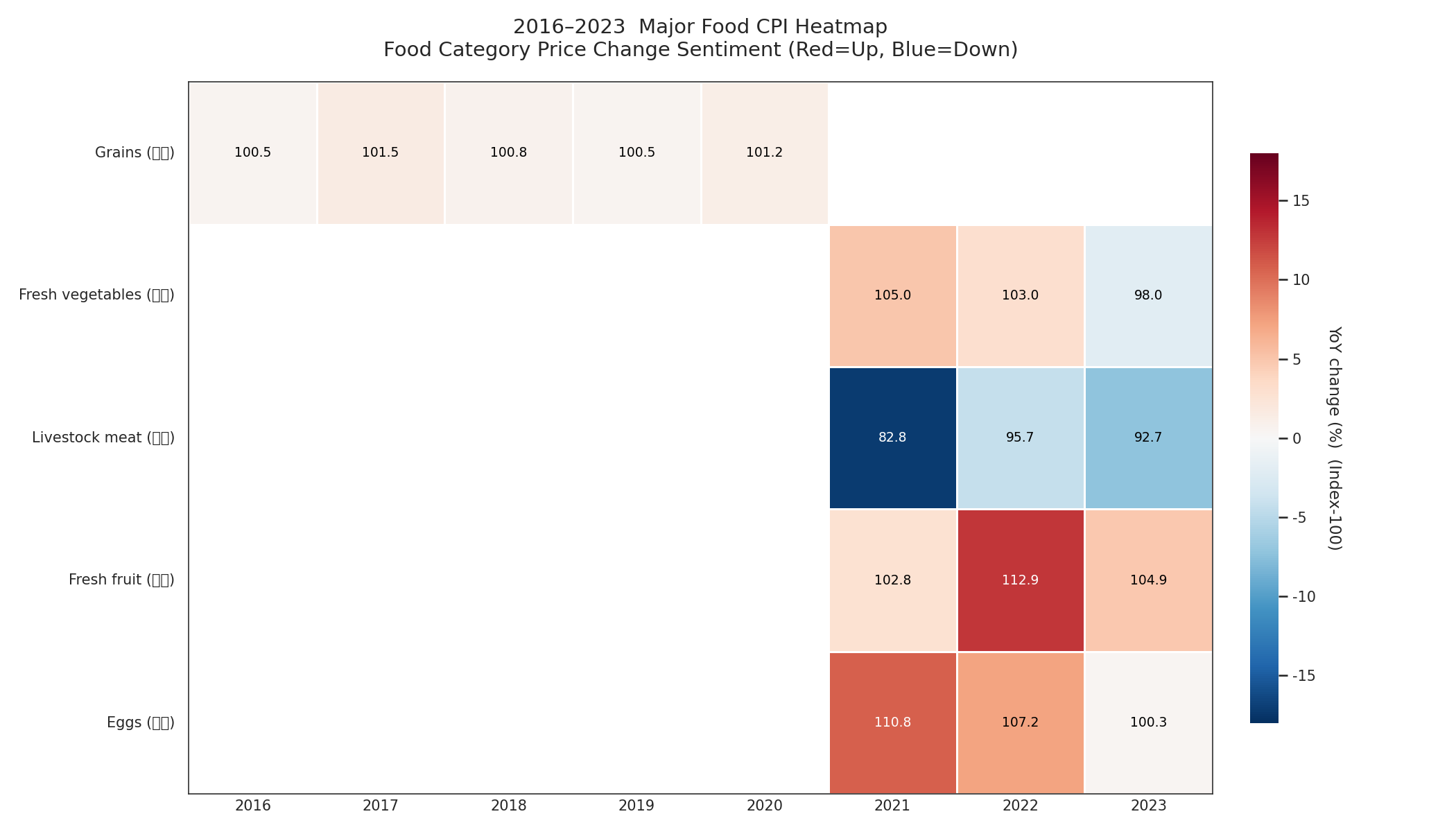}
            \\[0.02cm] \scriptsize (b) Model Output
        \end{minipage}
        
        \vspace{0.08cm}
        \begin{minipage}{0.96\linewidth}
            \footnotesize \textbf{Analysis:} As evidenced by the extensive data blanks in the generated heatmap, processing lengthy raw tabular data heavily occupies the context window capacity, inducing the "lost-in-the-middle" phenomenon and ultimately resulting in highly incomplete data extraction.
        \end{minipage}
    \end{minipage}

    \errorcaseseparator

    \begin{minipage}{\linewidth}
        \centering
        \textbf{Case 2: Combination Chart (level3)}\\[0.08cm]
        
        \begin{minipage}[t]{0.48\linewidth}
            \centering
            \includegraphics[width=\linewidth,height=0.160\textheight,keepaspectratio]{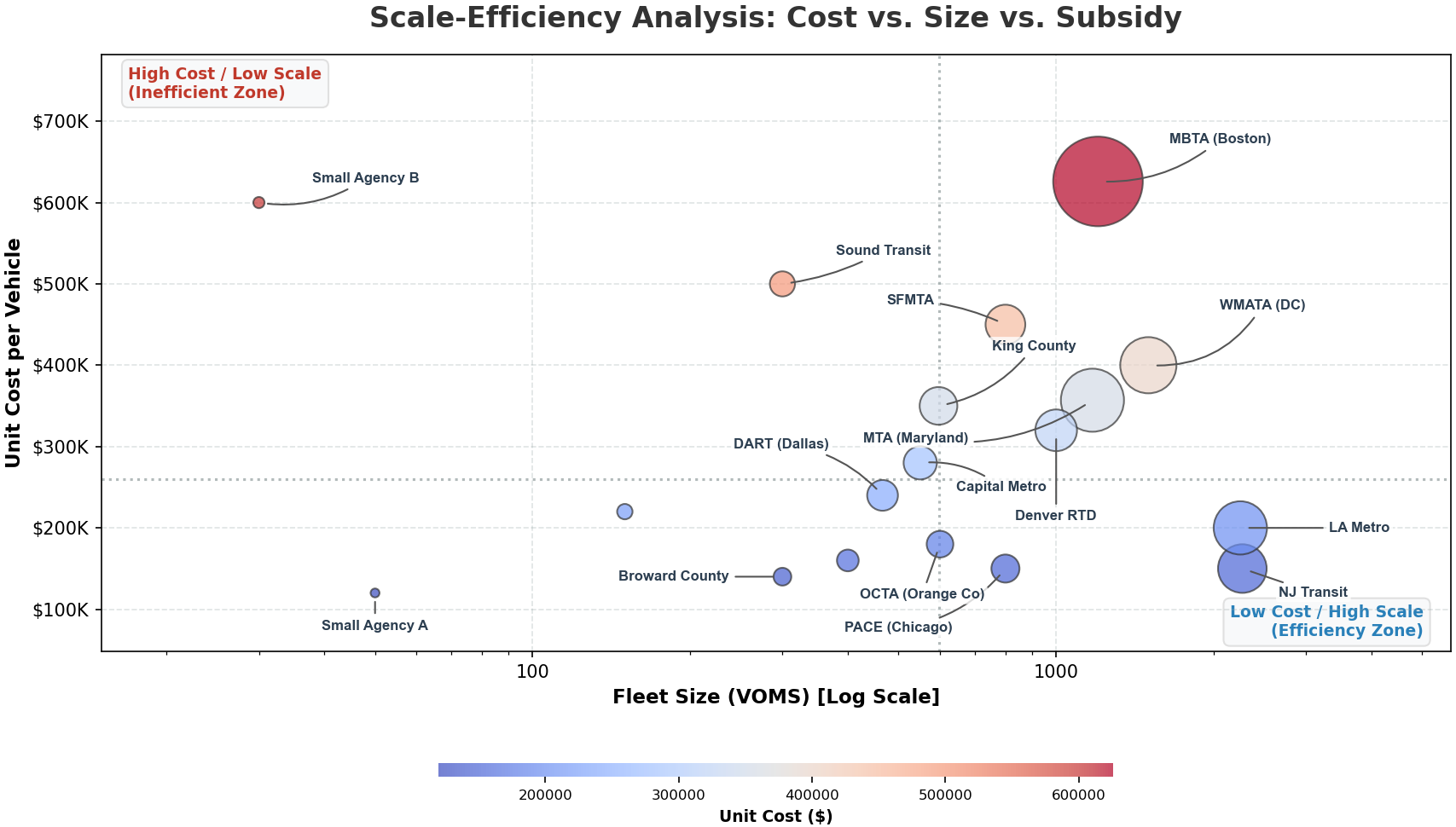}
            \\[0.02cm] \scriptsize (c) Ground Truth
        \end{minipage}
        \hfill
        \begin{minipage}[t]{0.48\linewidth}
            \centering
            \includegraphics[width=\linewidth,height=0.160\textheight,keepaspectratio]{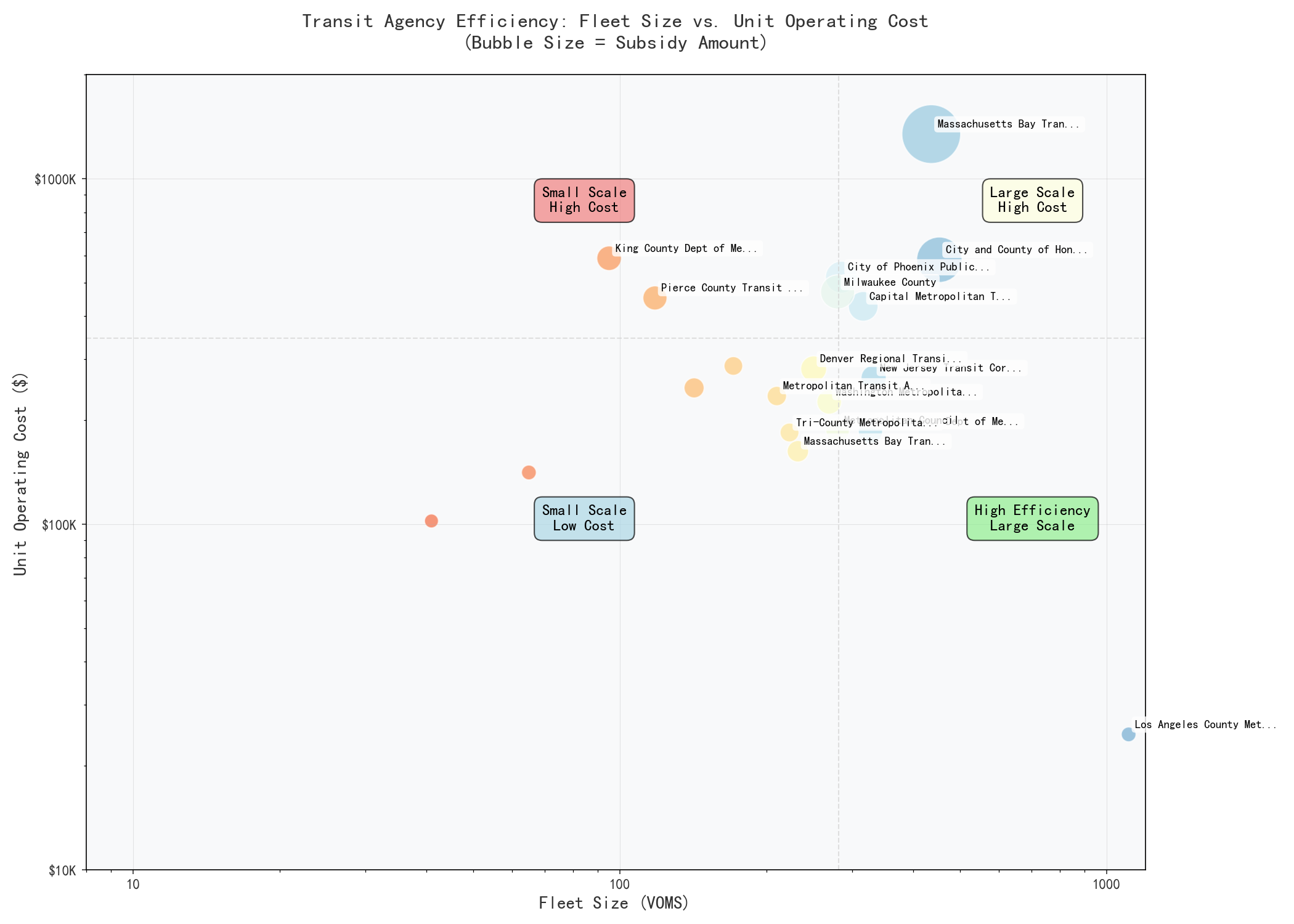}
            \\[0.02cm] \scriptsize (d) Model Output
        \end{minipage}
        
        \vspace{0.08cm}
        \begin{minipage}{0.96\linewidth}
            \footnotesize \textbf{Analysis:} In the generated bubble chart, the model suffered from severe multi-dimensional mapping failure when processing complex data (X, Y, bubble size, and color), not only erroneously altering the linear axis to a logarithmic scale—causing extreme distortion in data distribution—but also completely losing the continuous color mapping.
        \end{minipage}
    \end{minipage}

    \errorcaseseparator
    
    \begin{minipage}{\linewidth}
        \centering
        \textbf{Case 3: Combination Chart (level3)}\\[0.08cm]
        
        \begin{minipage}[t]{0.48\linewidth}
            \centering
            \includegraphics[width=\linewidth,height=0.160\textheight,keepaspectratio]{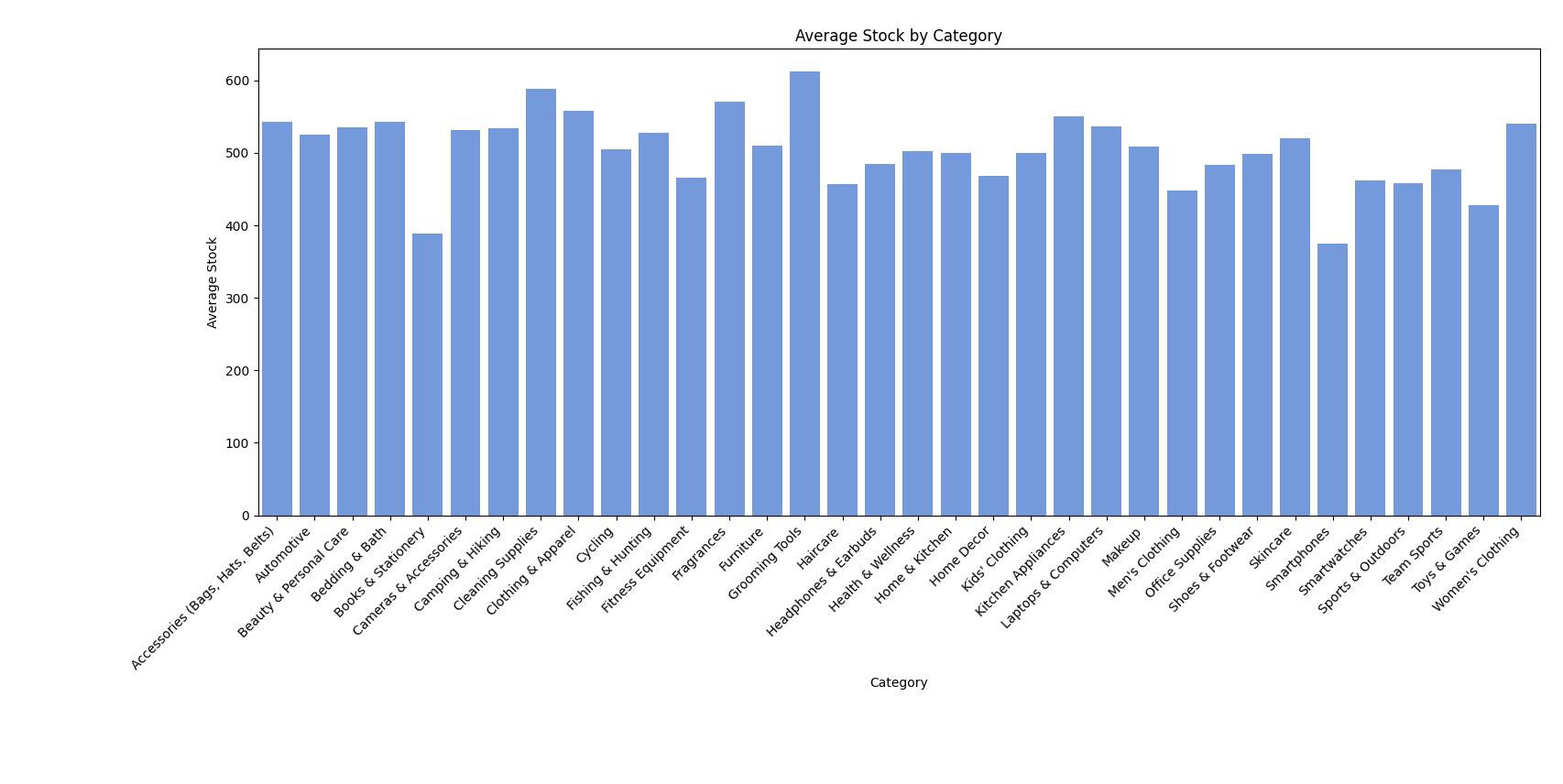}
            \\[0.02cm] \scriptsize (e) Ground Truth
        \end{minipage}
        \hfill
        \begin{minipage}[t]{0.48\linewidth}
            \centering
            \includegraphics[width=\linewidth,height=0.160\textheight,keepaspectratio]{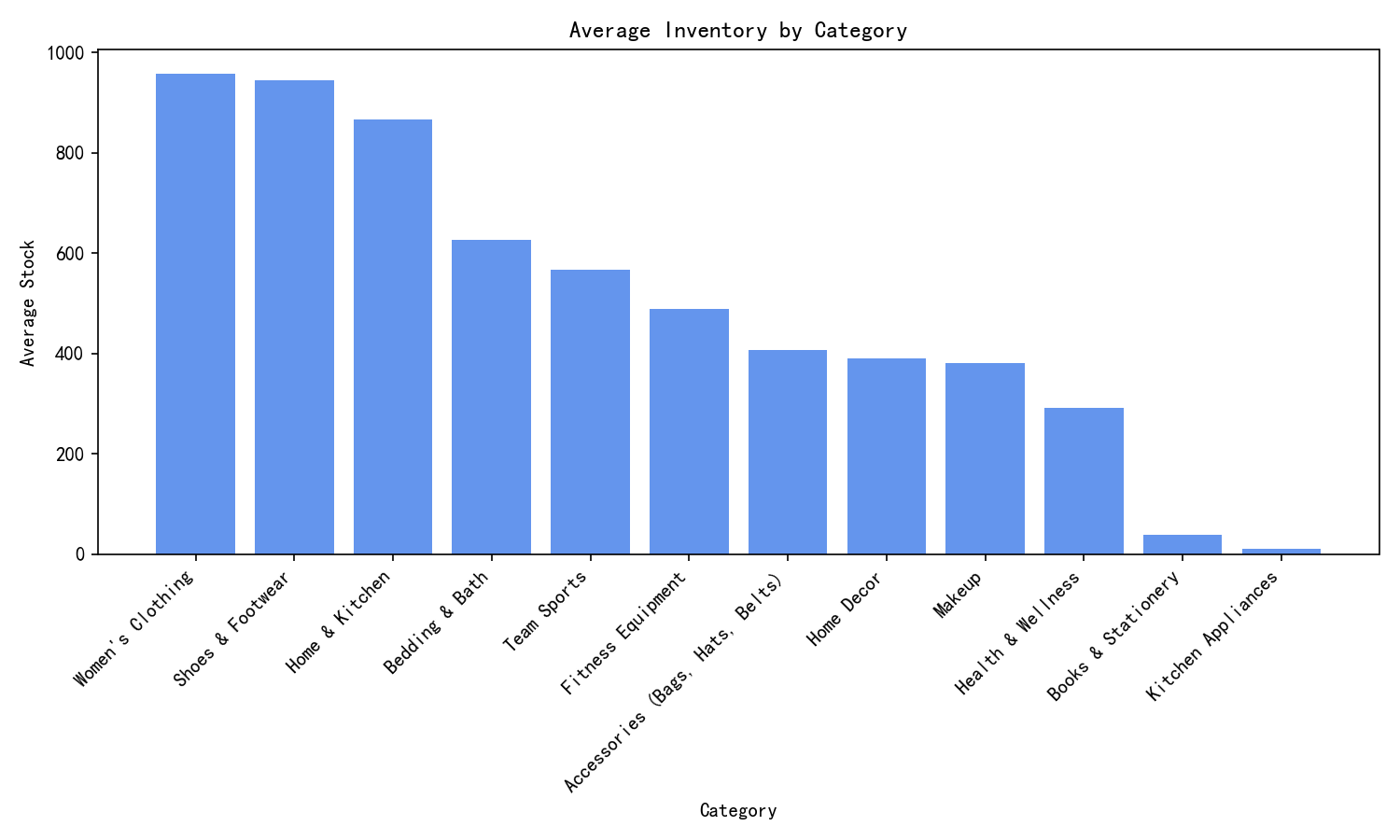}
            \\[0.02cm] \scriptsize (f) Model Output
        \end{minipage}
        
        \vspace{0.08cm}
        \begin{minipage}{0.96\linewidth}
            \footnotesize \textbf{Analysis:} The sheer volume of the raw tabular data causes severe information attenuation within the context window, resulting in significant categorical omissions and distorted numerical values in the generated chart.
        \end{minipage}
    \end{minipage}
    
    \vfill
    \caption{\textbf{Level 3–Specific Errors: }  Performance degradation primarily stems from long-context limitations. Processing extensive tabular data exhausts the context window, leading to incomplete data extraction and weakened instruction comprehension, which significantly reduces execution success rate and visual fidelity.
}
    \label{fig:special_level3_1}
\end{compacterrorpage}

\begin{compacterrorpage}
    \centering
    \appendixplatetitle{Special Error Category for level3}
    \vspace{1.0mm}
    \begin{minipage}{\linewidth}
        \centering
        \textbf{Case 1: Combination Chart (level3)}\\[0.08cm]
        
        \begin{minipage}[t]{0.48\linewidth}
            \centering
            \includegraphics[width=\linewidth,height=0.160\textheight,keepaspectratio]{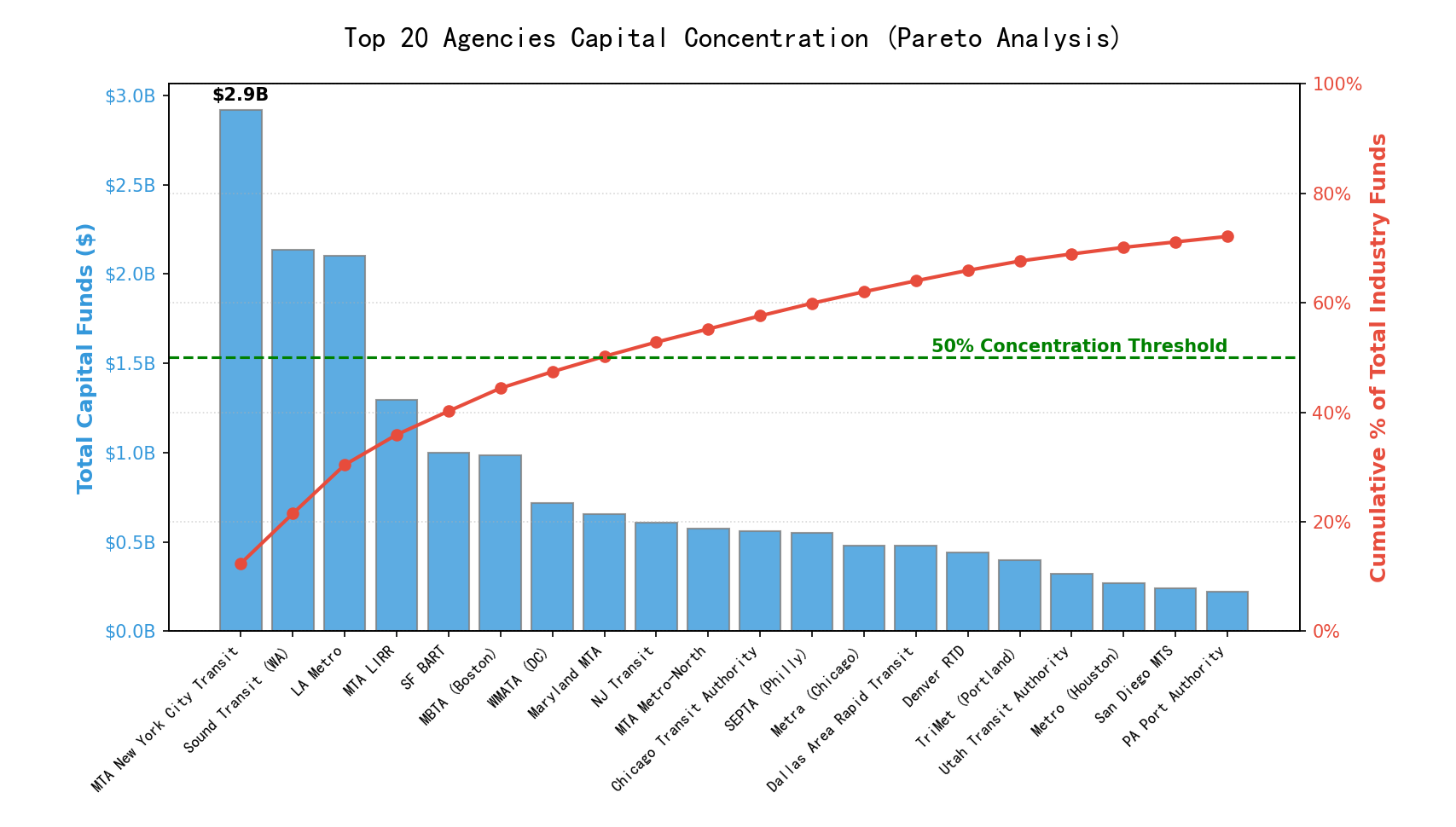}
            \\[0.02cm] \scriptsize (a) Ground Truth
        \end{minipage}
        \hfill
        \begin{minipage}[t]{0.48\linewidth}
            \centering
            \includegraphics[width=\linewidth,height=0.160\textheight,keepaspectratio]{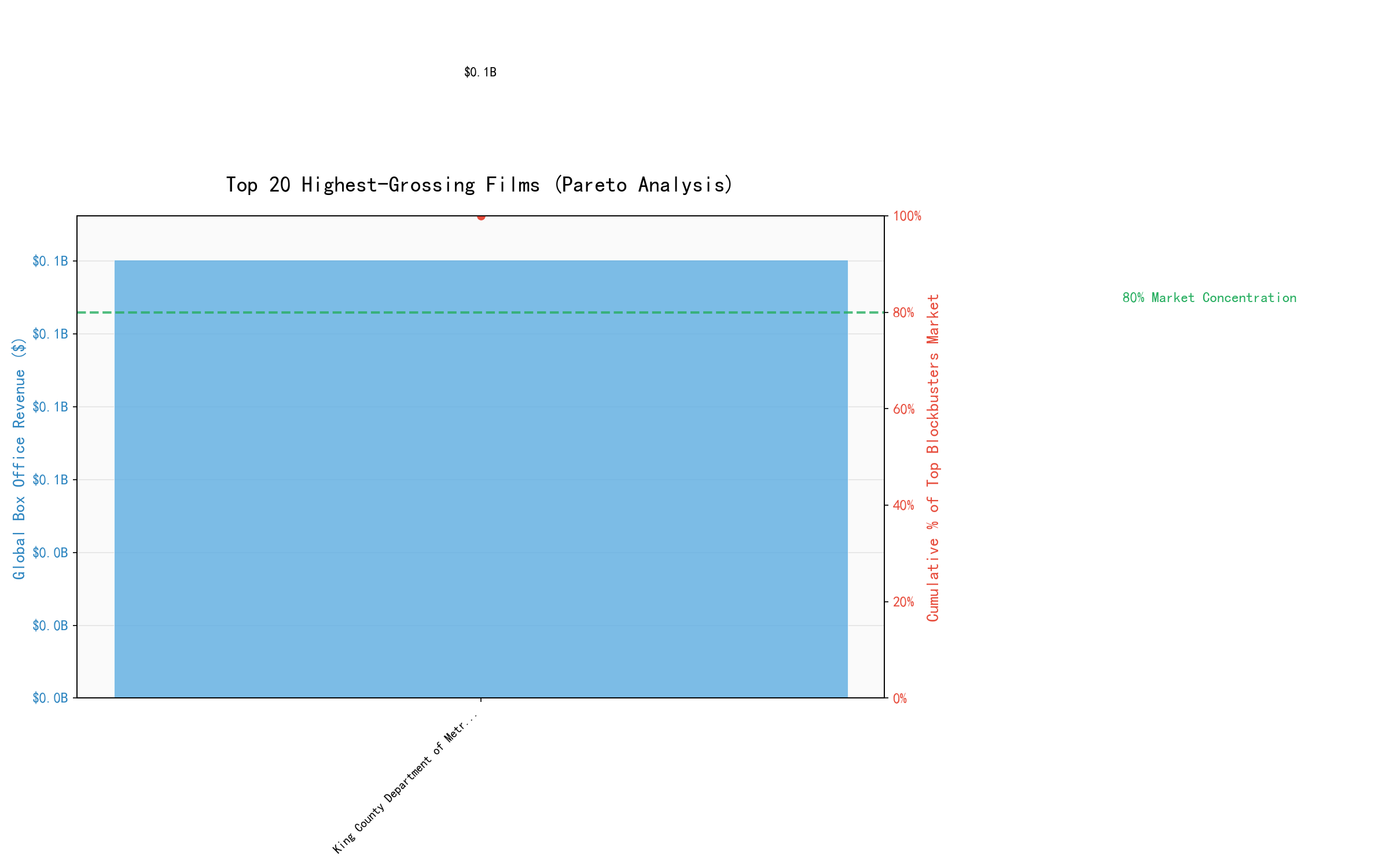}
            \\[0.02cm] \scriptsize (b) Model Output
        \end{minipage}
        
        \vspace{0.08cm}
        \begin{minipage}{0.96\linewidth}
            \footnotesize \textbf{Analysis:} The extraordinarily large tabular input coupled with complex plotting instructions caused severe context overload in the model, leading not only to catastrophic data truncation and massive omissions but also to instruction forgetting and complete semantic hallucinations.
        \end{minipage}
    \end{minipage}

    \errorcaseseparator

    \begin{minipage}{\linewidth}
        \centering
        \textbf{Case 2: Combination Chart (level3)}\\[0.08cm]
        
        \begin{minipage}[t]{0.48\linewidth}
            \centering
            \includegraphics[width=\linewidth,height=0.160\textheight,keepaspectratio]{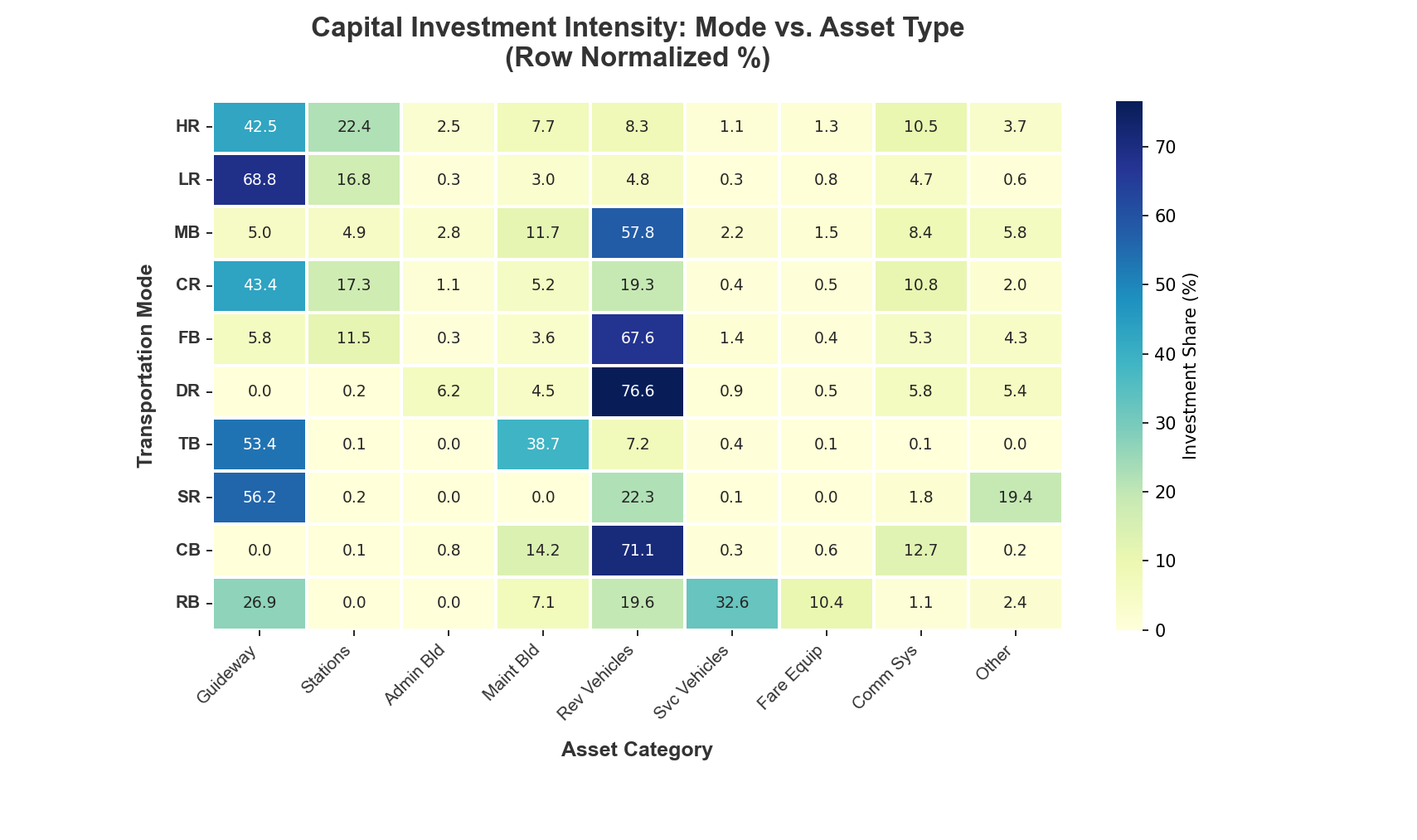}
            \\[0.02cm] \scriptsize (c) Ground Truth
        \end{minipage}
        \hfill
        \begin{minipage}[t]{0.48\linewidth}
            \centering
            \includegraphics[width=\linewidth,height=0.160\textheight,keepaspectratio]{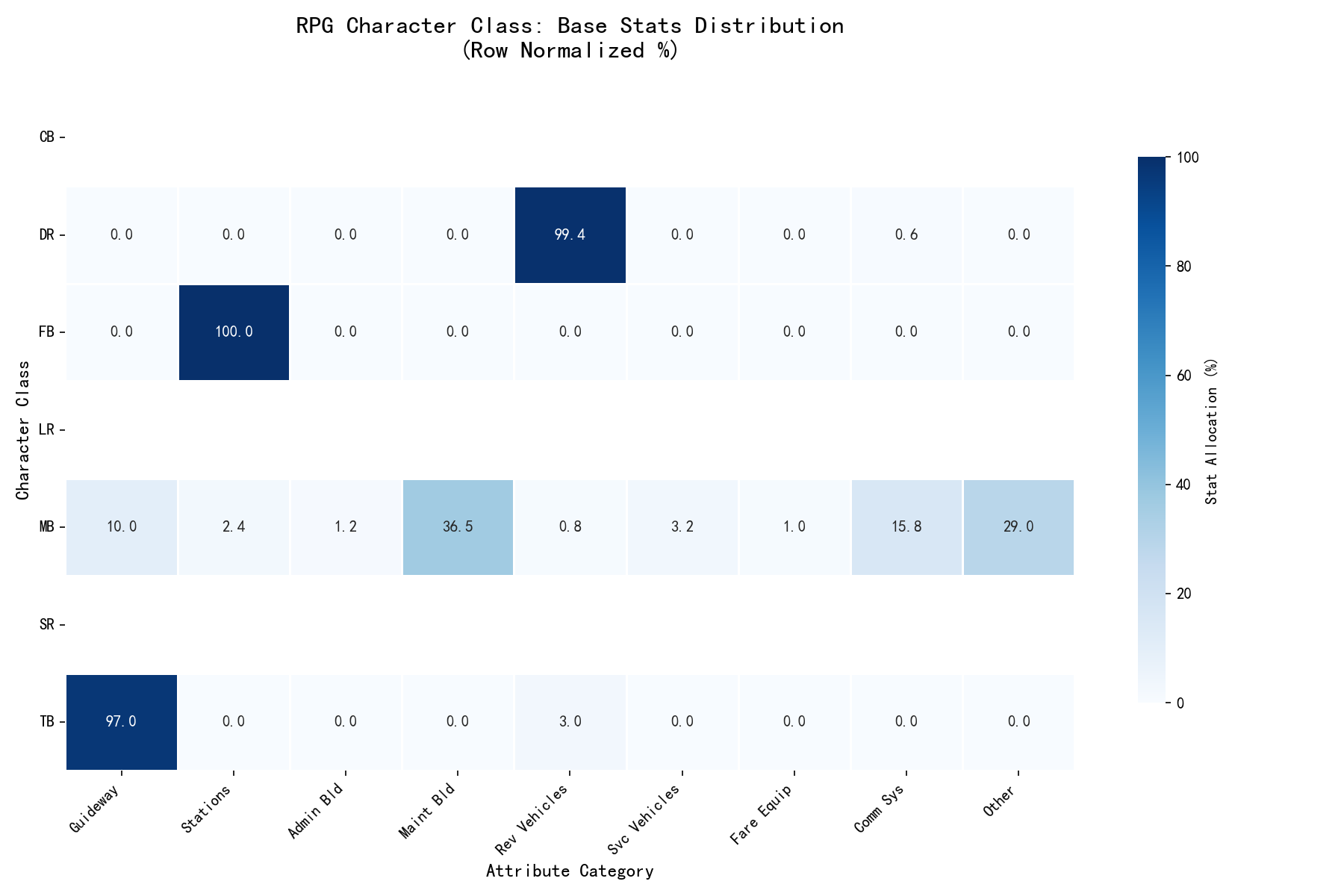}
            \\[0.02cm] \scriptsize (d) Model Output
        \end{minipage}
        
        \vspace{0.08cm}
        \begin{minipage}{0.96\linewidth}
            \footnotesize \textbf{Analysis:} As illustrated in the generated heatmap, the complex two-dimensional tabular data exceeds the model's effective context capacity, triggering a severe 'lost-in-the-middle' phenomenon that ultimately results in massive row-level data extraction failures and extensive blank areas.
        \end{minipage}
    \end{minipage}

    \errorcaseseparator
    
    \begin{minipage}{\linewidth}
        \centering
        \textbf{Case 3: Combination Chart (level2)}\\[0.08cm]
        
        \begin{minipage}[t]{0.48\linewidth}
            \centering
            \includegraphics[width=\linewidth,height=0.160\textheight,keepaspectratio]{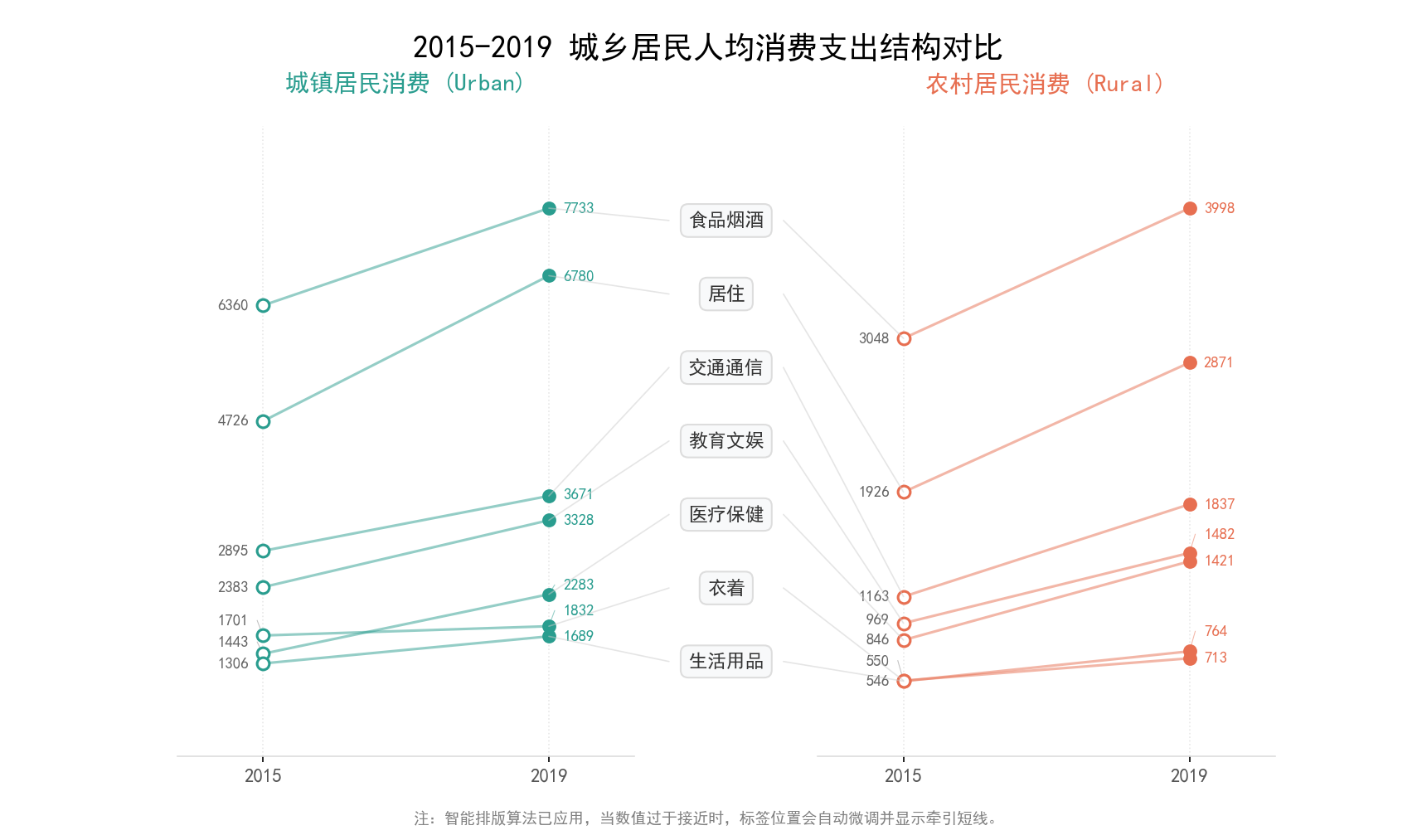}
            \\[0.02cm] \scriptsize (e) Ground Truth
        \end{minipage}
        \hfill
        \begin{minipage}[t]{0.48\linewidth}
            \centering
            \includegraphics[width=\linewidth,height=0.160\textheight,keepaspectratio]{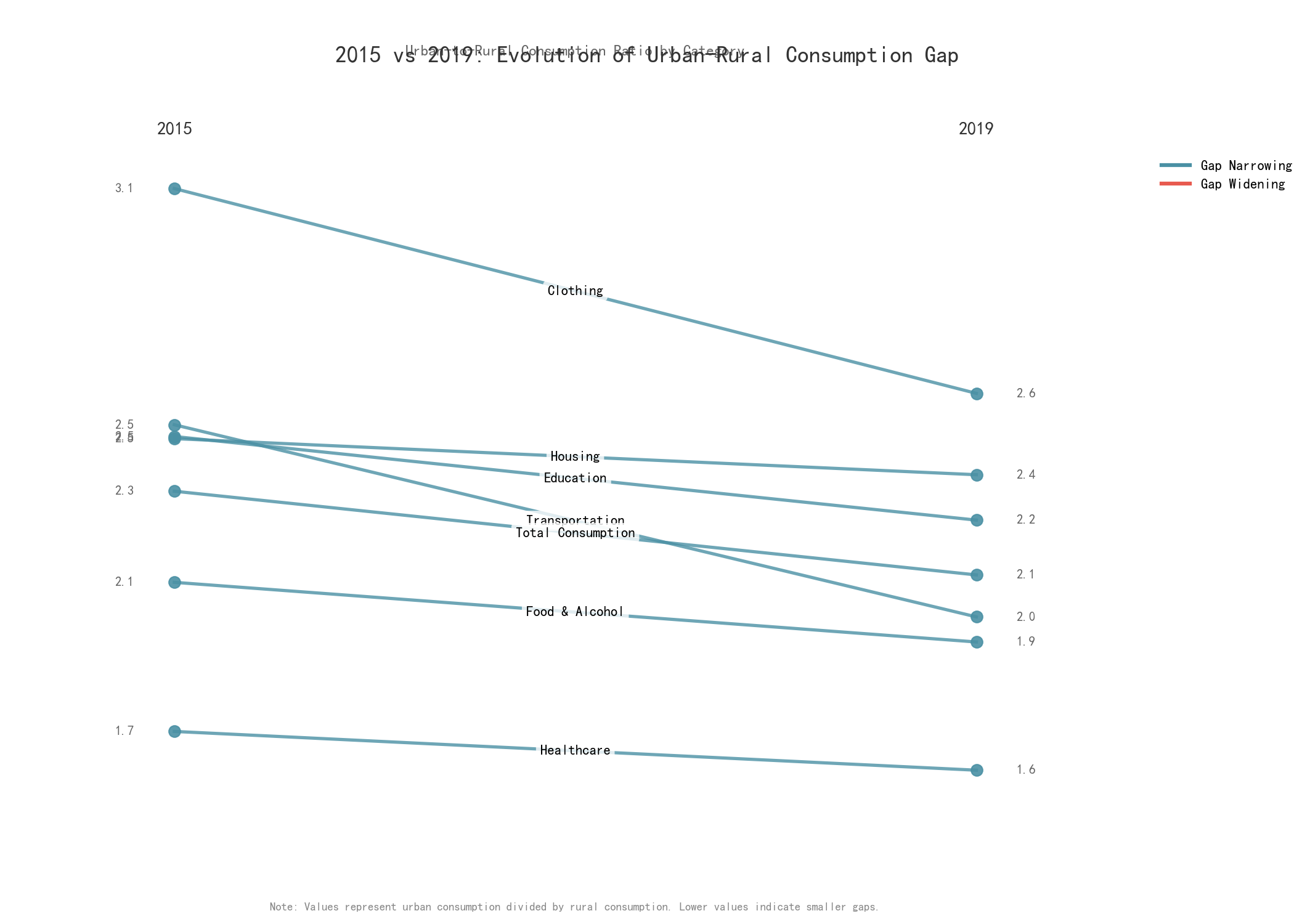}
            \\[0.02cm] \scriptsize (f) Model Output
        \end{minipage}
        
        \vspace{0.08cm}
        \begin{minipage}{0.96\linewidth}
            \footnotesize \textbf{Analysis:} The generated output exhibits a severe structural degradation compared to the ground truth. While the GT presents a complex dual-panel layout simultaneously comparing two distinct demographics (Urban and Rural), the generated chart collapses this into a single, simplified slope graph plotting a derived ratio. This stark contrast highlights a critical failure in processing extensive data context and comprehending complex layout instructions, resulting in incomplete visual extraction.
        \end{minipage}
    \end{minipage}
    
    \vfill
    \caption{\textbf{Level 3–Specific Errors: }  Performance degradation primarily stems from long-context limitations. Processing extensive tabular data exhausts the context window, leading to incomplete data extraction and weakened instruction comprehension, which significantly reduces execution success rate and visual fidelity.
}
    \label{fig:special_level3_2}
\end{compacterrorpage}

\section{Limitations}

While Chart2Code serves as a robust benchmark, we identify two limitations to be addressed in future work:

\begin{itemize}
    \item \textbf{Linguistic Diversity and Generalization}: The dataset annotations and task prompts are exclusively in English. This design choice may overestimate the capabilities of current models when applied to multilingual contexts. Real-world data visualization often involves diverse languages and cultural formatting standards. Consequently, the current benchmark does not fully evaluate how well LMMs can generalize to non-English queries or preserve text rendering accuracy in multi-byte character sets (e.g., CJK languages).
    
    \item \textbf{Evaluation Bias in LMM-based Judges}: Our evaluation protocol relies on strong LMMs to assess code correctness and visual fidelity. While this ensures scalability, it poses a risk of \textit{evaluation bias}. LMM-based judges might exhibit leniency towards generated code patterns similar to their own training data, or struggle to accurately penalize subtle visual artifacts (such as layout misalignment) that do not trigger code execution errors. Developing more objective, reference-free visual metrics remains an open challenge.
\end{itemize}

\begin{figure}[htbp]
    \centering
    \includegraphics[width=.95\linewidth]{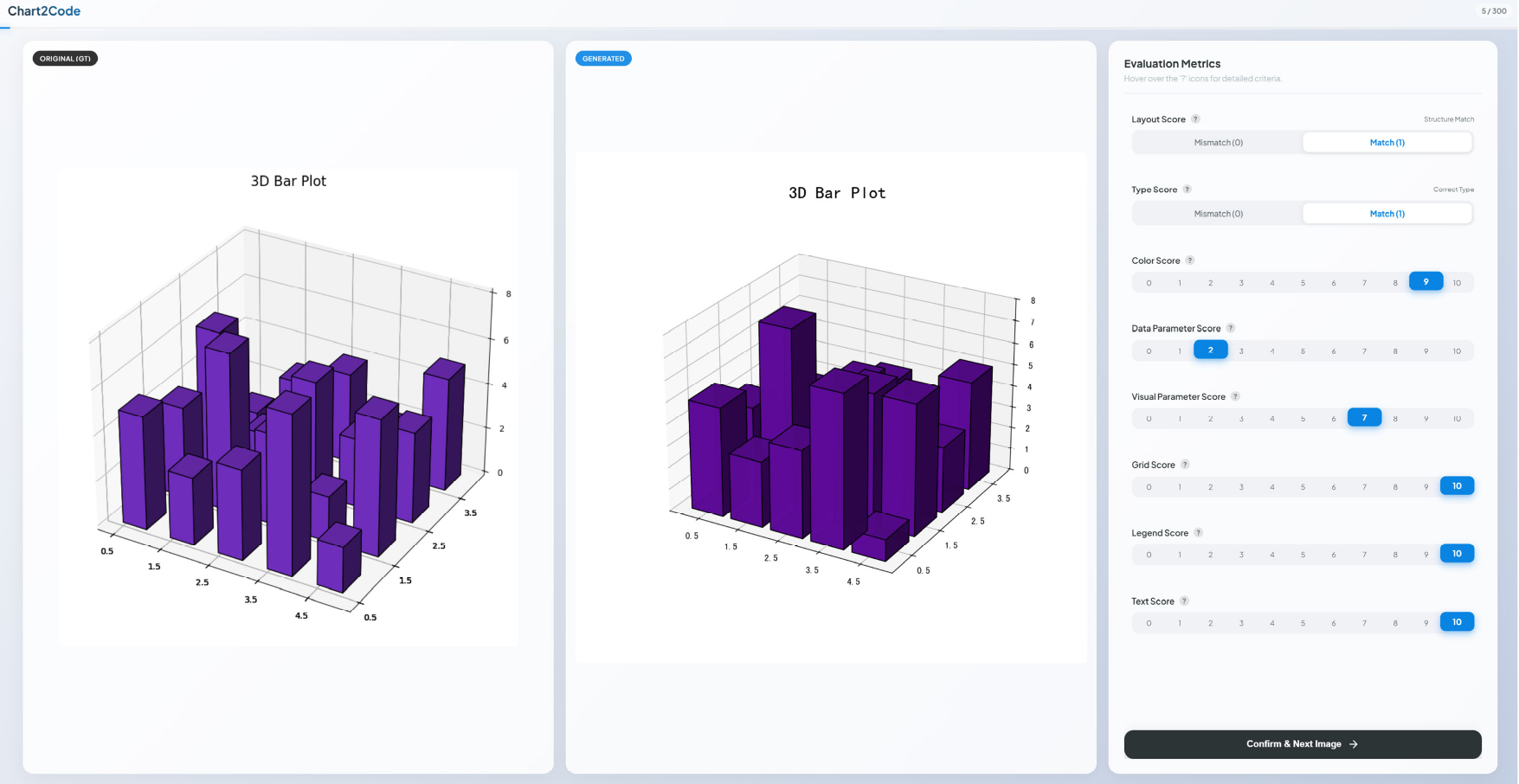}
    \caption{Human vs model evaluate system.}
    \label{fig:human_eval}
\end{figure}

\section{Human Evaluation}
\label{human_evaluation}
\subsection{Correlation between Automated Metrics and Human Evaluation}
\begin{figure}[htbp]
    \centering
    \begin{subfigure}[t]{0.48\linewidth}
        \centering
        \includegraphics[width=\linewidth]{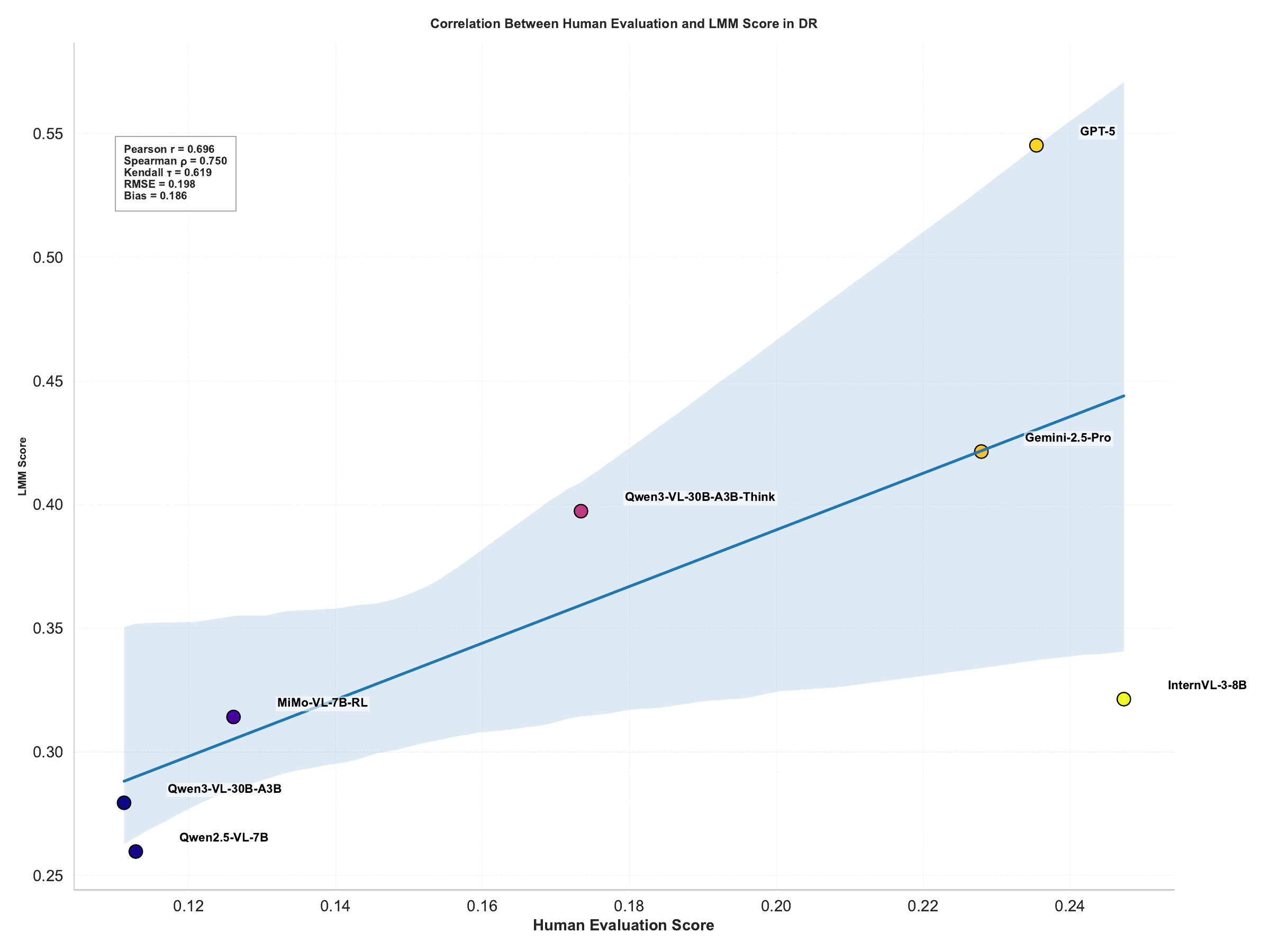}
        \caption{Human vs. model evaluation: correlation between human evaluation and LMM Score in DR.}
        \label{fig:human_evaluate_level1_direct}
    \end{subfigure}
    \hfill
    \begin{subfigure}[t]{0.48\linewidth}
        \centering
        \includegraphics[width=\linewidth]{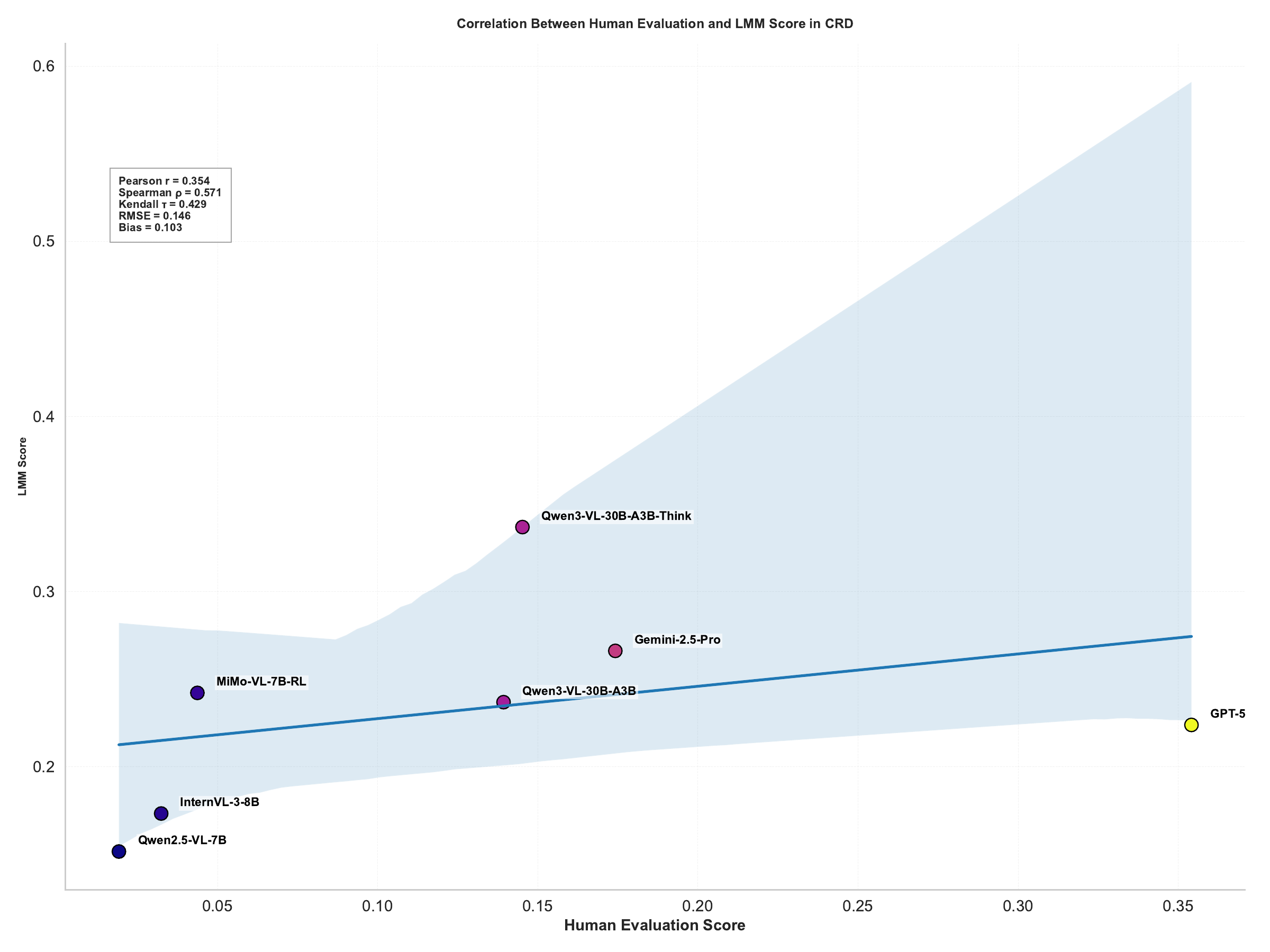}
        \caption{Human vs. model evaluation: correlation between human evaluation and LMM Score in CRD.}
        \label{fig:human_evaluate_level1_customize}
    \end{subfigure}

    \medskip

    \begin{subfigure}[t]{0.48\linewidth}
        \centering
        \includegraphics[width=\linewidth]{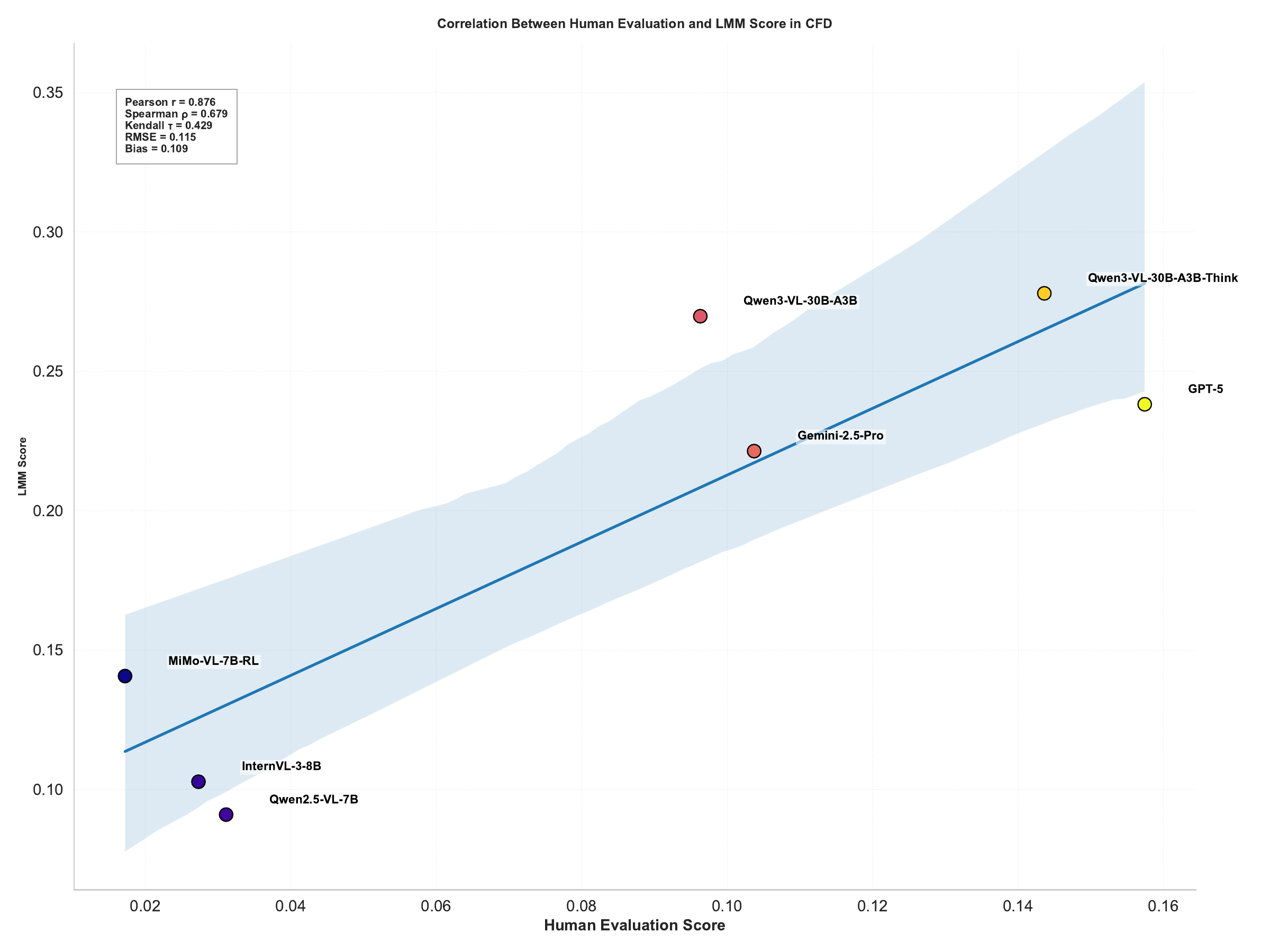}
        \caption{Human vs. model evaluation: correlation between human evaluation and LMM Score in CFD.}
        \label{fig:human_evaluate_level1_figure}
    \end{subfigure}
    \hfill
    \begin{subfigure}[t]{0.48\linewidth}
        \centering
        \includegraphics[width=\linewidth]{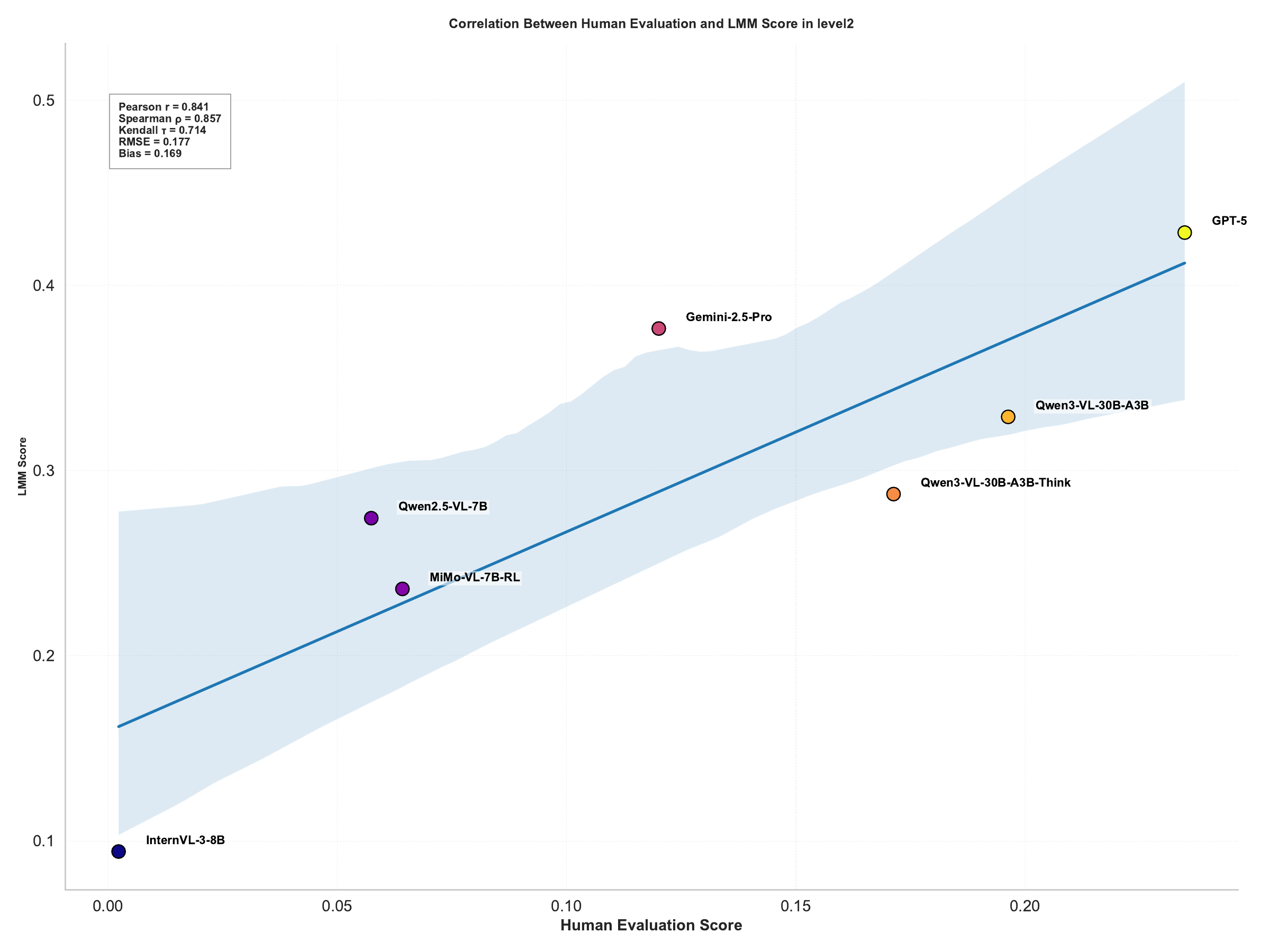}
        \caption{Human vs. model evaluation: correlation between human evaluation and LMM Score in Level 2.}
        \label{fig:human_evaluate_level2}
    \end{subfigure}

    \caption{Human vs. model evaluation across all tasks: correlation between human evaluation and LMM Score.}
    \label{fig:human_evaluate_all}
\end{figure}

\begin{figure}[htbp]
    \centering
    \includegraphics[width=.9\linewidth]{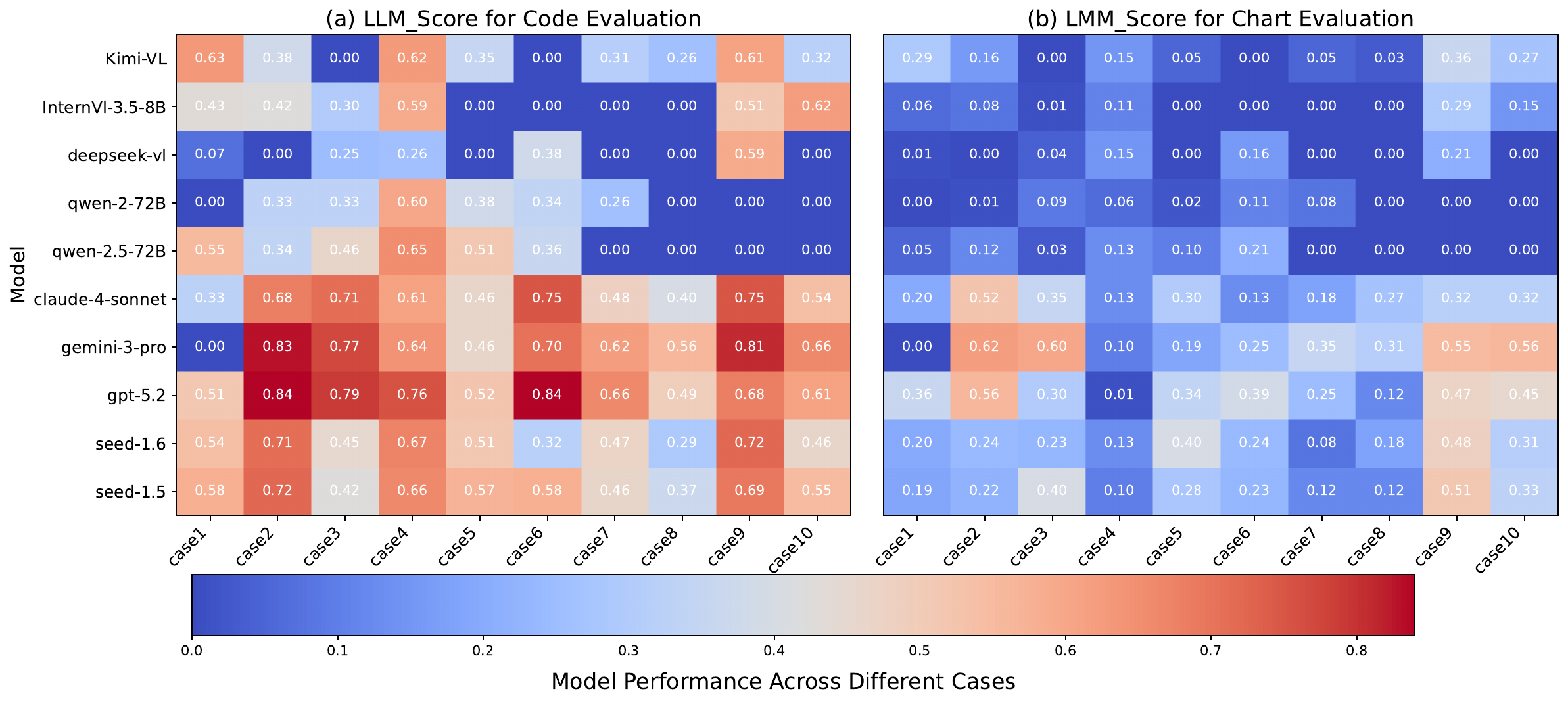}
    \caption{Analysis of model performance on different task cases with LLM-score and LMM-score.}
    \label{fig:llmlmmscoreheatmap}
\end{figure}

\begin{figure}[t]
  \centering
    \begin{minipage}[t]{0.31\linewidth}
    \centering
    \vspace{0pt}
    \includegraphics[width=\linewidth]{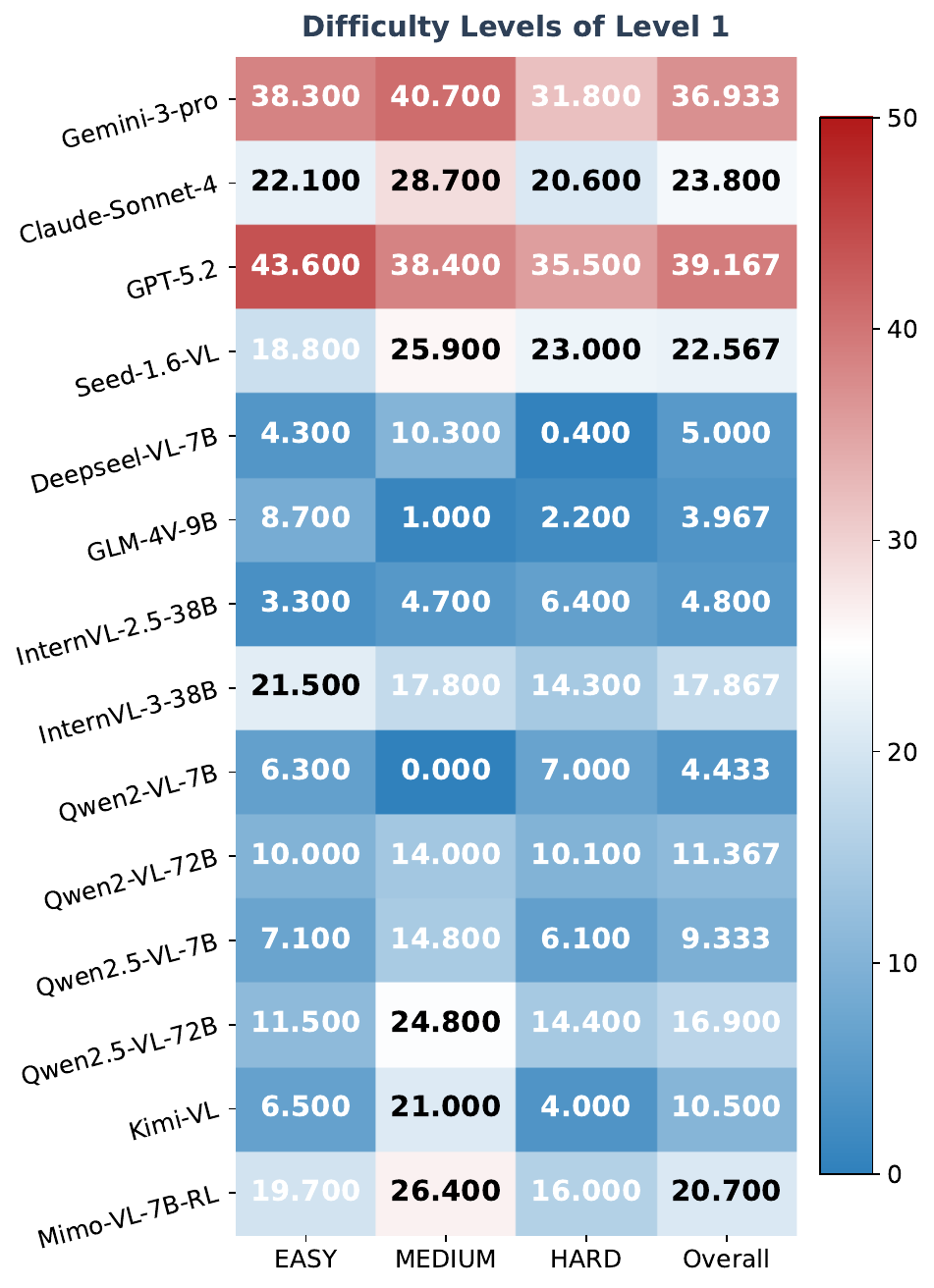}
    \end{minipage}%
  \begin{minipage}[t]{0.31\linewidth}
    \centering
    \vspace{0pt}
    \includegraphics[width=\linewidth]{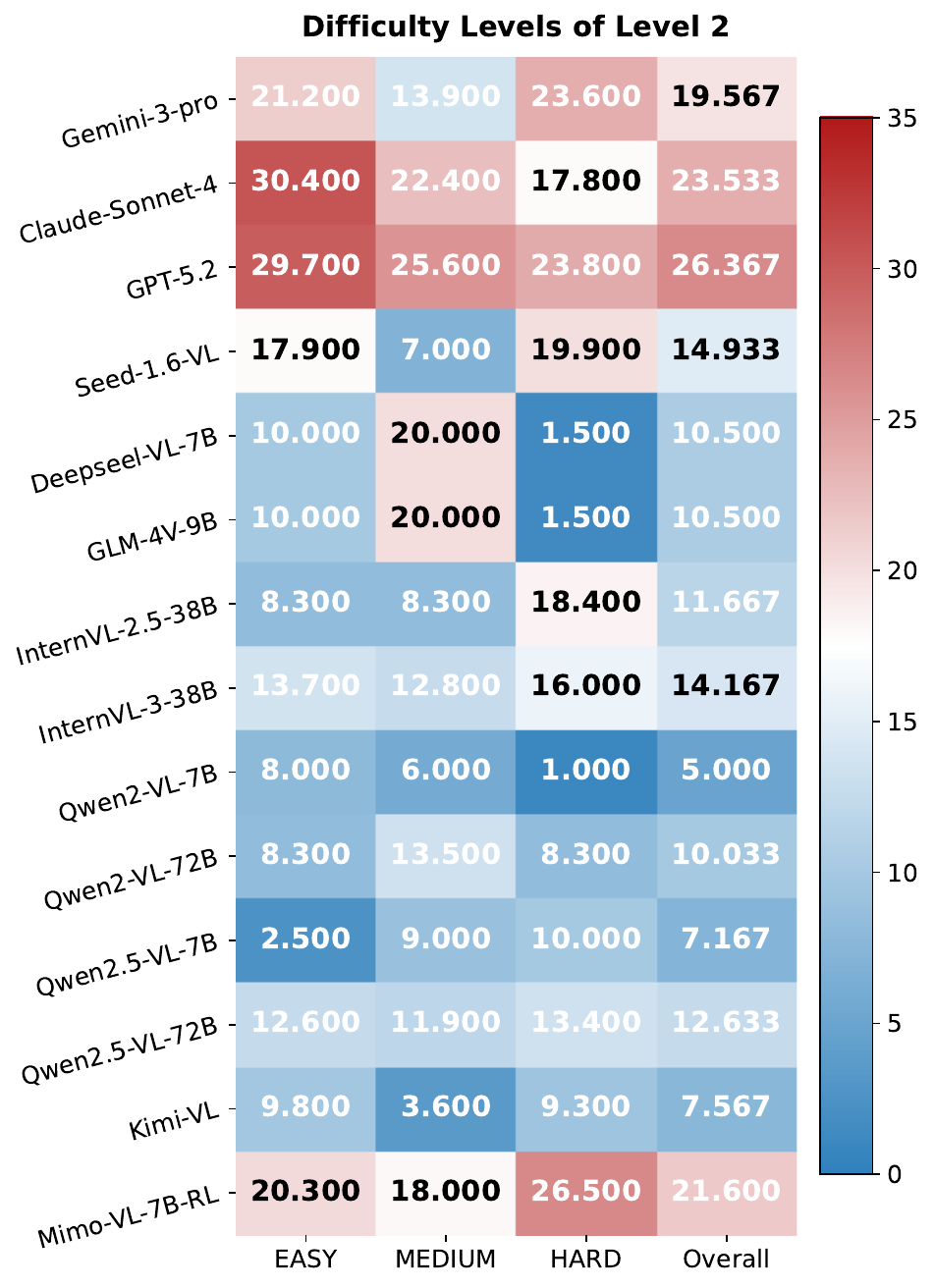}
    \end{minipage}%
    \begin{minipage}[t]{0.31\linewidth}
    \centering
    \vspace{0pt}
    \includegraphics[width=\linewidth]{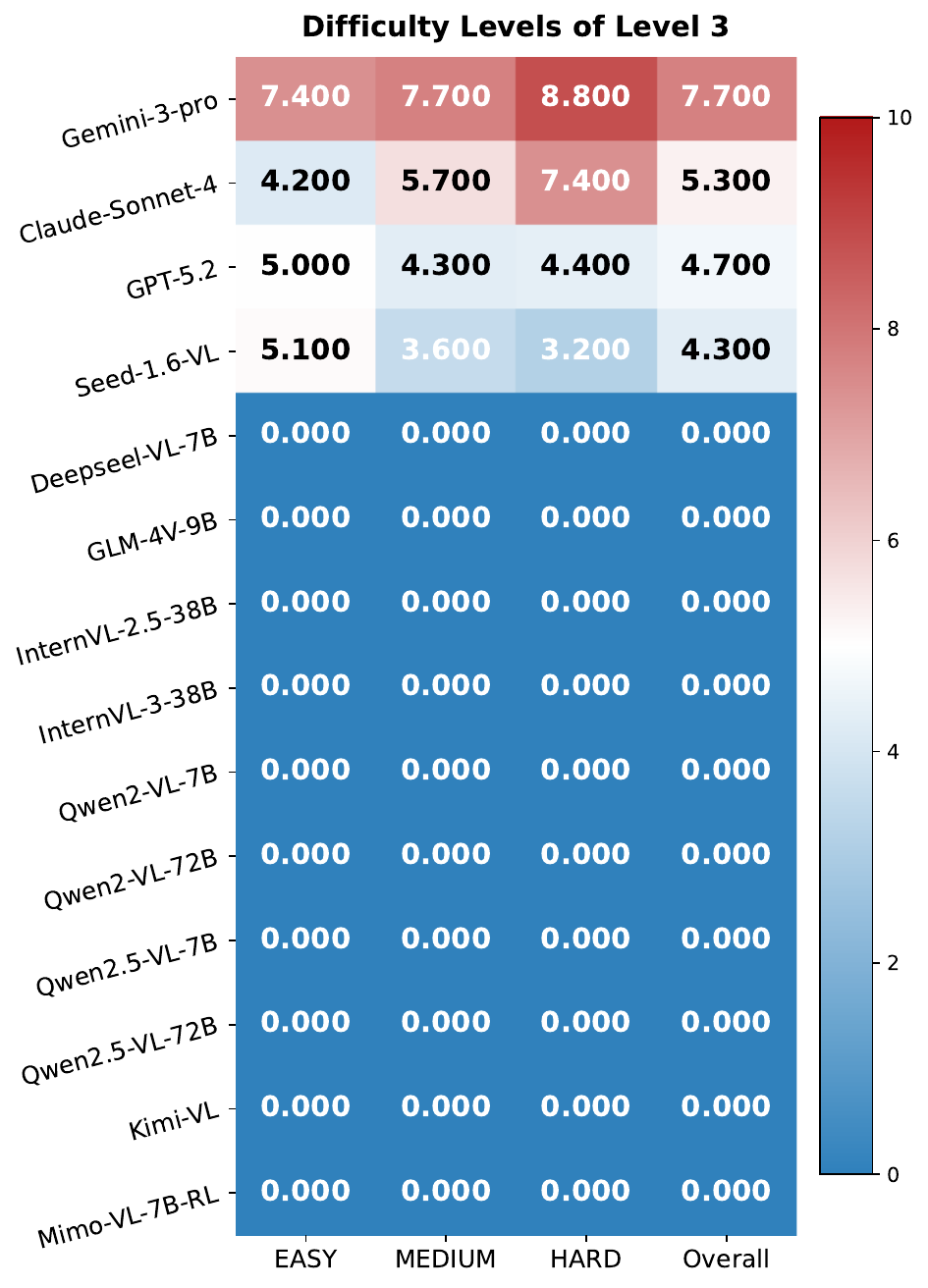}
    \end{minipage}%
     \caption{Correlation of the model performance (i.e, LMM-score) on different manually annotated difficulty levels (i.e., Easy, Medium, Hard) on Level 1, 2, 3, respectively.}
     \label{fig:difficultscoreheatmap}
\end{figure}

To validate the reliability of our proposed automated metrics and their alignment with human perception, we conducted a human evaluation study involving 20 undergraduate students from the School of Computer Science. Participants assessed both the ground truth (GT) charts and the generated charts across multiple dimensions, aiming for a comprehensive evaluation of the reproduction quality, the evaluation system is illustrated in Fig.~\ref{fig:human_eval}. Regarding the sampling strategy, we randomly selected 300 result images for the DR task and the Level 2 task, respectively. For the CRD and CFD tasks, we evaluated the complete set of output results. We assessed the outputs generated by seven models: \texttt{Qwen3-VL-30b\_A3B-Instruct}, \texttt{Qwen3-VL-30b\_A3B-Think}, \texttt{MiMo-Vl-7B-RL}, \texttt{Gemini-3-pro}, \texttt{InternVL-3-8B}, \texttt{Qwen2.5-VL-7B}, and \texttt{GPT-5.2}. The detailed results for each task, along with the calculated correlation coefficients, are presented in Fig.~\ref{fig:human_evaluate_level1_direct}, Fig.~\ref{fig:human_evaluate_level1_customize}, Fig.~\ref{fig:human_evaluate_level1_figure}, Fig.~\ref{fig:human_evaluate_level2} .

From the results shown in the Fig.~\ref{fig:human_evaluate_level1_direct}, Fig.~\ref{fig:human_evaluate_level1_customize}, Fig.~\ref{fig:human_evaluate_level1_figure}, Fig.~\ref{fig:human_evaluate_level2}, we observe that the correlation between LMM scores and human ratings is generally strong across most tasks, with the exception of the CRD setting, where the linear correlation is comparatively weaker. This suggests that LMMs can effectively evaluate the quality differences between two charts, and that their evaluation criteria broadly align with those used by human annotators—leading to similar scoring trends. The higher Spearman and Kendall correlation coefficients further indicate that, rather than achieving precise numerical agreement, the models are more reliable at capturing the relative ranking of chart quality. In other words, although there are notable discrepancies between the absolute LMMs scores and human ratings, LMMs remain capable of distinguishing higher-quality charts from lower-quality ones.

Regarding error metrics, both the RMSE and Bias values reveal a consistent upward deviation of LMM predictions from human evaluations, indicating a tendency for LMMs to overestimate chart quality. This systematic bias highlights an important limitation: while LMM scores are correlated with human assessments, they are not yet calibrated to the same level of strictness as human judgment. Among different models, proprietary models such as GPT-5.2 and Gemini-3-Pro tend to align more closely with the regression trend line—demonstrating smaller deviations from human evaluations—whereas many open-source models exhibit higher variance.

In summary, these findings demonstrate that LMM-based automatic scoring provides a useful, though not perfect, signal for chart-quality evaluation. Despite the remaining challenges in fine-grained assessment for complex chart-understanding tasks, LMM scores still offer sufficiently reliable trends to serve as effective auxiliary indicators in large-scale benchmarking.





\subsection{Performance Comparison}
\begin{figure}[htbp]
    \centering
    \includegraphics[width=.9\linewidth]{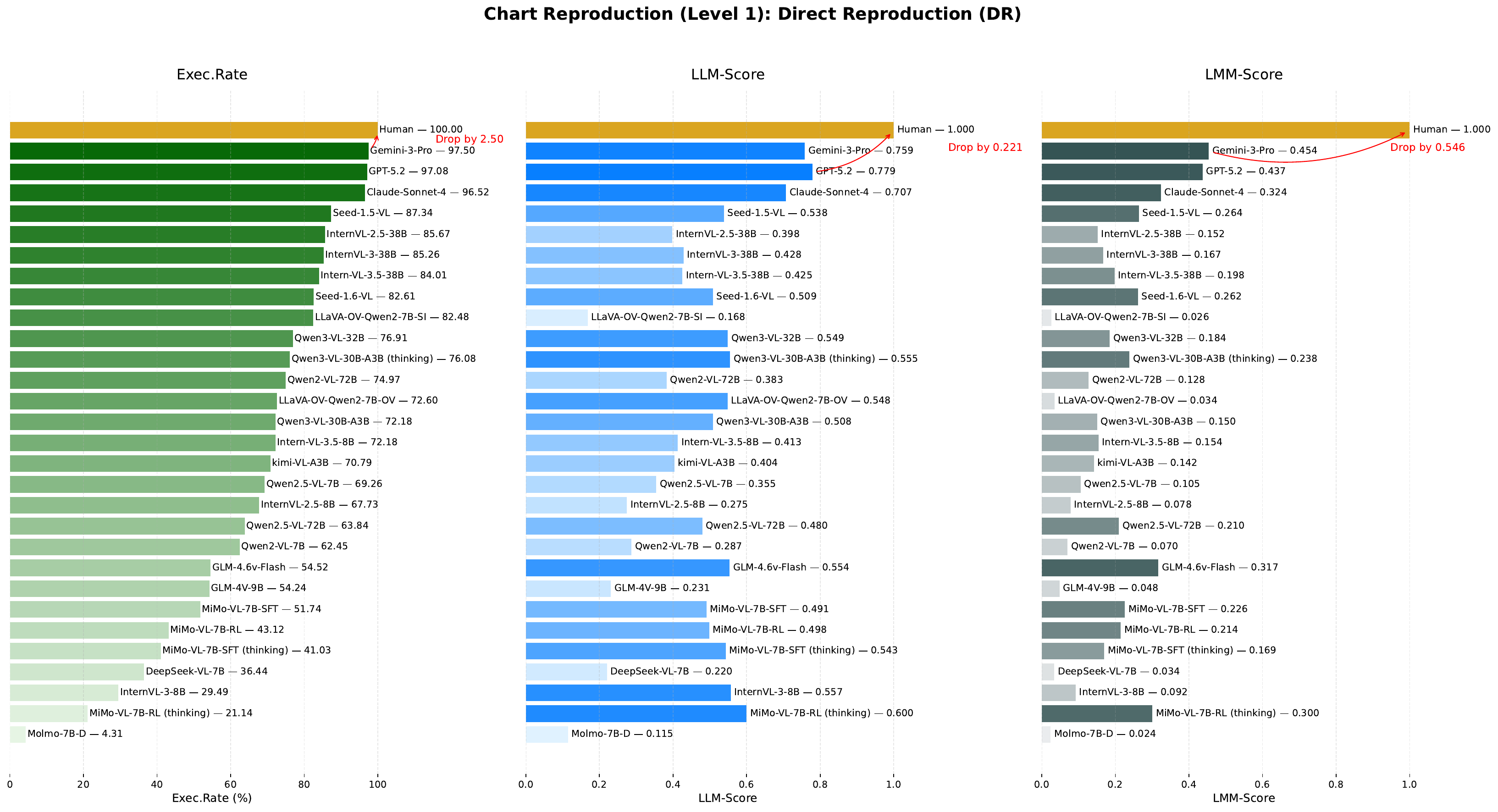}
    \caption{Human vs model performance: LLM-score and LMM-score across level 1 direct tasks.}
    \label{fig:level 1 direct}
\end{figure}

\begin{figure}[htbp]
    \centering 
    \includegraphics[width=.9\linewidth]{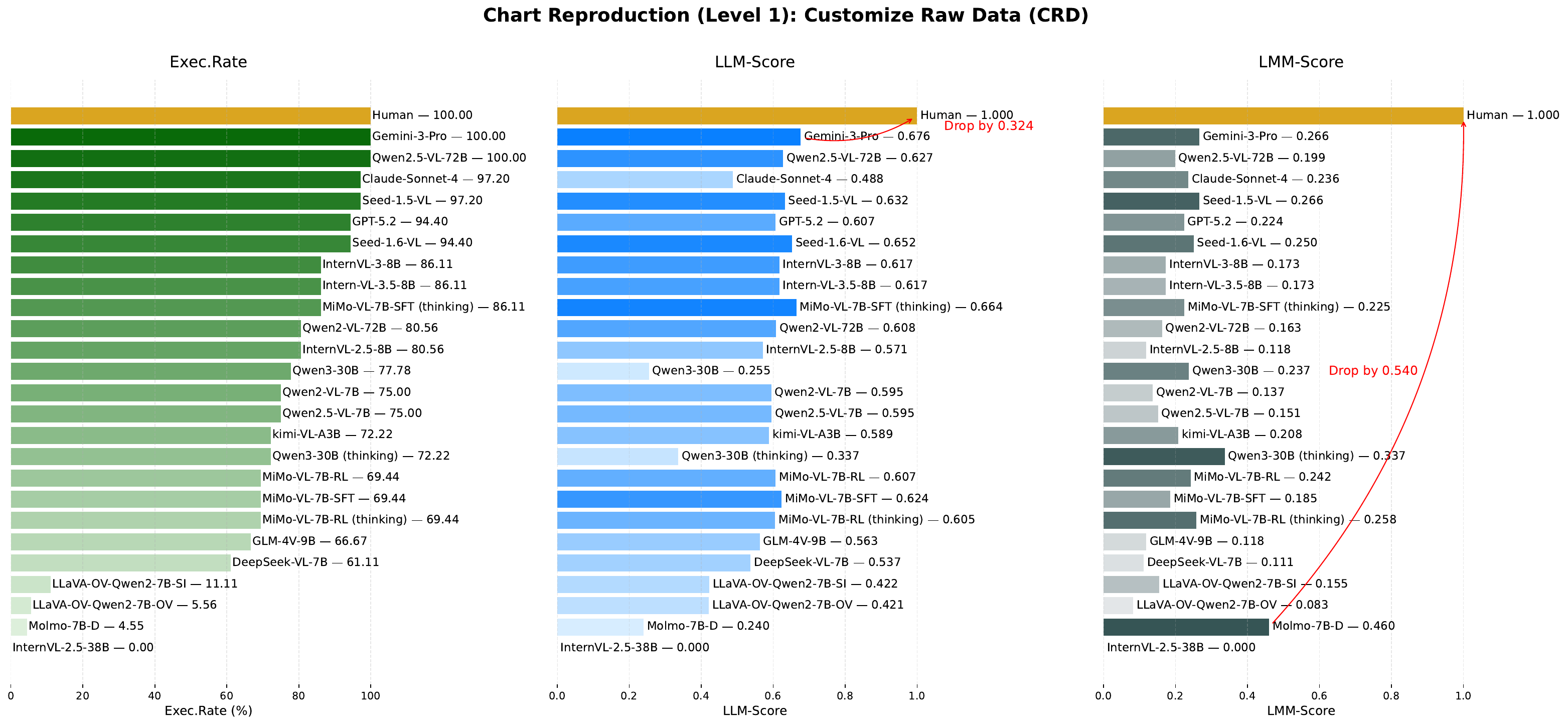}
    \caption{Human vs model performance: LLM-score and LMM-score across level 1 customize tasks.}
    \label{fig:level 1 customize}
\end{figure}

\begin{figure}[htbp]
    \centering
    \includegraphics[width=.9\linewidth]{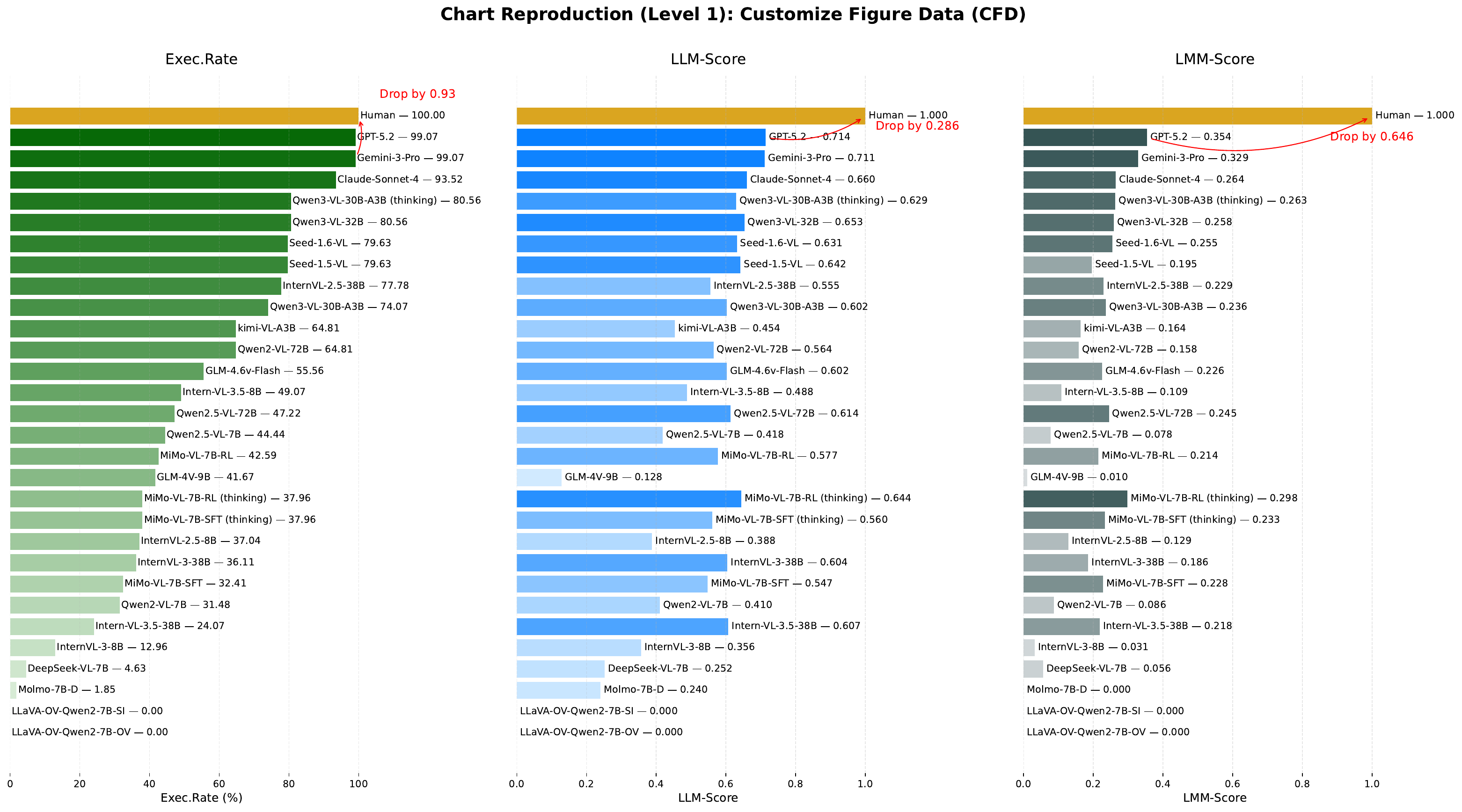}
    \caption{Human vs model performance: LLM-score and LMM-score across level 1 figure tasks.}
    \label{fig:level 1 figure}
\end{figure}

\begin{figure}[htbp]
    \centering
    \includegraphics[width=.9\linewidth]{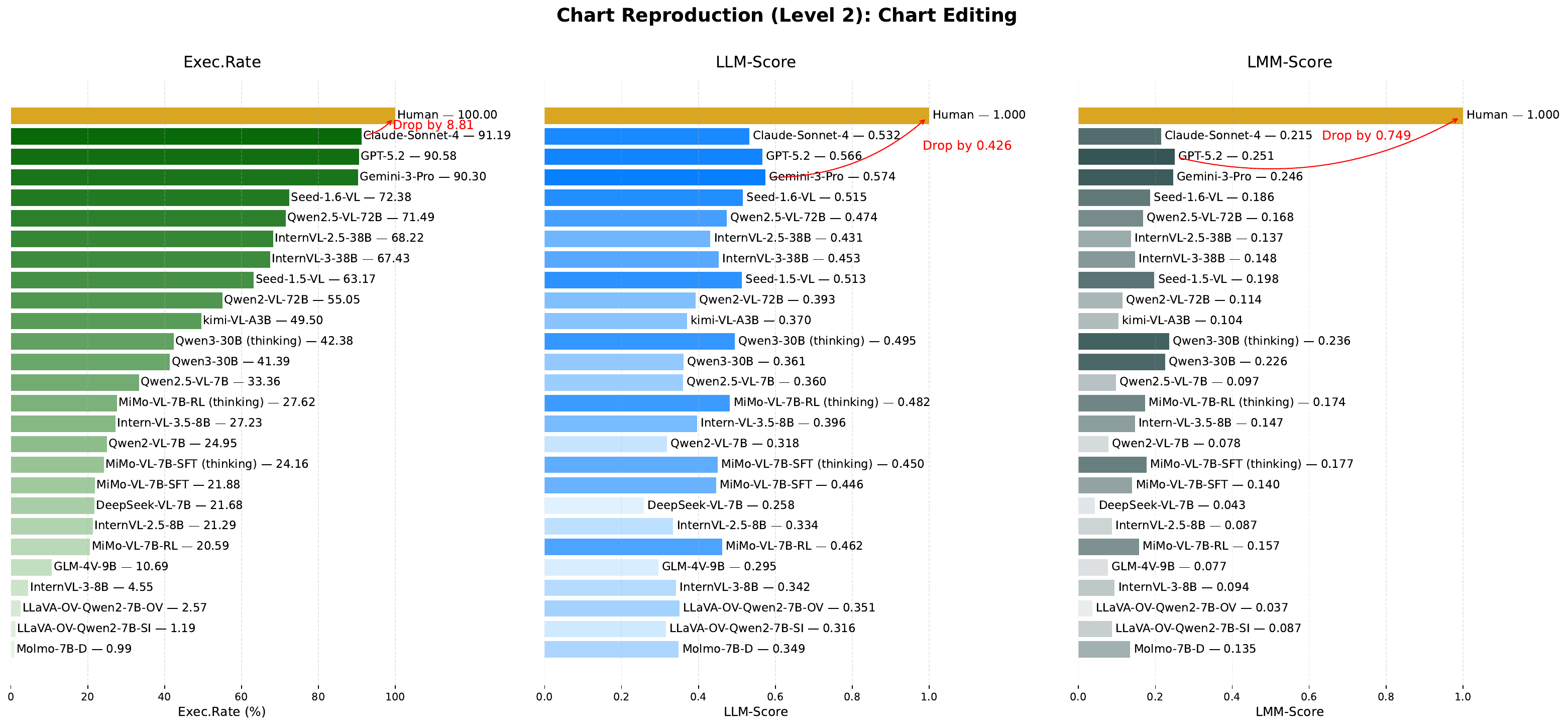}
    \caption{Human vs model performance: LLM-score and LMM-score across level 2 tasks.}
    \label{fig:level 2}
\end{figure}

\begin{figure}[htbp]
    \centering
    \includegraphics[width=.9\linewidth]{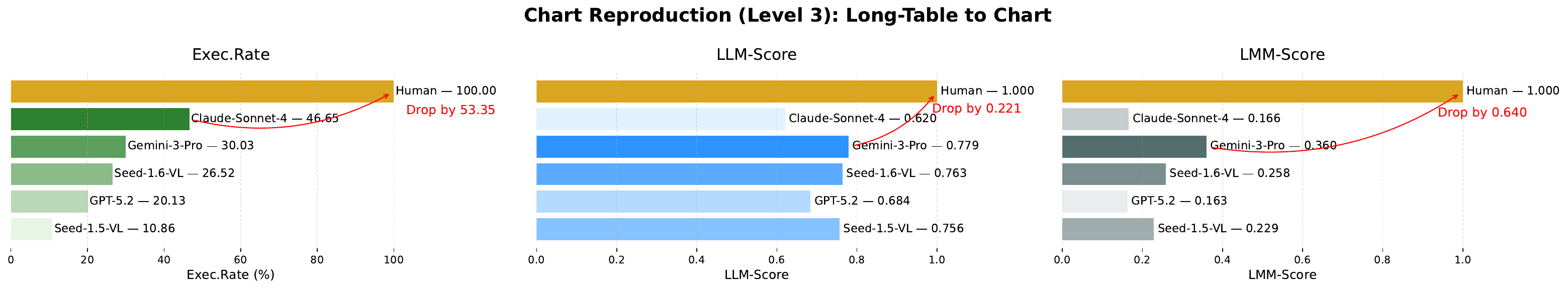}
    \caption{Human vs model performance: LLM-score and LMM-score across level 3 tasks.}
    \label{fig:level 3}
\end{figure}

To provide a more intuitive comparison of LMM performance across our hierarchical tasks, Fig.~\ref{fig:level 1 direct}, Fig.~\ref{fig:level 1 customize}, Fig.~\ref{fig:level 1 figure}, Fig.~\ref{fig:level 2}, Fig.~\ref{fig:level 3} presents the Executability Rate, LLM Score, and LMM Score, benchmarked against human performance.

\textbf{High Executability in Proprietary Models}, Approaching Human Levels In terms of Execution Rate (Exec.Rate), proprietary models demonstrate exceptional stability. This indicates that current top-tier LMMs have largely solved the challenge of generating syntactically correct and runnable code for simple chart reproduction.

\textbf{Visual Fidelity Remains the Major Bottleneck The LMM-Score}, which measures the visual similarity between the generated chart and the ground truth, reveals the biggest challenge.
Most models show a sharp decline in this metric compared to their Code-Level scores.

\textbf{The Gap Between Code Generation and Visual Perception.} There is a noticeable discrepancy between the blue bars (LLM-Score) and grey bars (LMM-Score).
This indicates that while models understand how to write code, they often fail to perceive or reproduce the fine-grained visual nuances (such as exact color hex codes, marker styles, or specific layout constraints) required for pixel-perfect reproduction.

While top-tier models have mastered the syntax required to generate charts (High Exec.Rate), the semantic understanding of the data (LLM-Score) is being revolutionized by "thinking" models. However, achieving high visual fidelity (LMM-Score) remains a difficult frontier for most models.

\FloatBarrier
\section{Evaluation Metric Details}

\subsection{Overall}
\label{evaluation overall}

To better evaluate the performance of different models, we conduct comparative assessments from two levels: the \textbf{code-level} and the \textbf{chart-level}.  
Throughout the evaluation process, we first examine the executability of the generated code. The execution rate is defined as the ratio between the number of executable code snippets that successfully generate images ($s$) and the total number of tasks ($t$). Formally, the execution rate is expressed as:  

\begin{equation}\
\label{exce_rate}
    \text{exec\_rate} = \frac{s}{t}.
\end{equation}

The execution rate is reported as a percentage.  

At the \textbf{code-level}, we first extract plotting elements from the \texttt{matplotlib.Figure} object and propose eight evaluation dimensions as the \textbf{base evaluation}. The detailed specifications are given in \ref{base evaluation}. Subsequently, we leverage \texttt{gpt-5-mini} to perform a holistic similarity assessment of the code's visualization results, thereby providing a more reliable confidence score at the code level. We refer this as \textbf{LLM-Score}.

At the \textbf{chart-level}, we input the executable code into \texttt{gpt-5-mini} for image-based evaluation. By designing specific prompts, the large multimodal model (LMM) assesses multiple dimensions and produces an aggregated score. This chart-level evaluation offers an intuitive similarity measure of the visual outputs, thereby serving as a direct indicator of model performance.  We refer this as \textbf{LMM-Score}.
 The implementation details of these two evaluation mechanisms are described as follows.
\subsection{Base Evaluation}
\label{base evaluation}
To evaluate visualization effects from the code perspective\citep{chartmimic} we investigated commonly used Python plotting libraries and found that Seaborn, Matplotlib, NetworkX, and WordCloud all rely on Matplotlib’s underlying plotting functions. When using these libraries for plotting, a \texttt{Figure} object is generated in memory, which contains all the elements of the plot. This implies that we can extract all visualization-related elements from the \texttt{Figure} object and compare the GT\_code with the generated\_code to evaluate their visualization effects.

\subsubsection{Comparison Between Chartmimic}
\textbf{More Efficient.} Unlike ChartMimic \citep{chartmimic}, which depends on code tracers and code injection, our evaluation method is substantially more efficient. In practice, ChartMimic must execute both the GT\_code and generated\_code for each evaluation dimension, resulting in up to twelve executions for a single generated\_code. This process incurs significant computational overhead in both time and memory. By contrast, our method executes the GT\_code and generated\_code only once, caches their corresponding Figure objects, and then evaluates multiple dimensions directly on these objects, thereby greatly reducing execution cost.

\textbf{More General.} In comparison to ChartMimic’s \citep{chartmimic} hard-coded rules, which exhibit limited adaptability and strong dependence on specific Matplotlib versions, our evaluation method is inherently more general. ChartMimic enforces rule-based matching of plotting elements, which not only imposes strict version constraints but also leaves many elements unsupported. Our approach instead parses the \texttt{Figure} object directly, which comprehensively encapsulates all elements in memory, ensuring greater robustness and version independence.

\textbf{More Versatile.} Whereas ChartMimic \citep{chartmimic} is restricted to a narrow set of functions from specific libraries, our method offers broad applicability. By operating directly on core Matplotlib objects, our approach seamlessly extends to all visualization libraries that build upon Matplotlib’s primitives, thereby achieving substantially stronger cross-library generalization.

\textbf{More Precise.} Unlike ChartMimic \citep{chartmimic}, which evaluates function call patterns rather than visual outputs, our method emphasizes the visualization results themselves. ChartMimic leaves a gap between code execution and rendered charts, while our approach directly inspects visual objects such as \texttt{Line} and \texttt{Patch}. This enables a more faithful and precise evaluation of visualization quality at the code-to-visualization level.

\textbf{More Dimensions.} 
Our evaluation framework introduces a richer set of dimensions compared to ChartMimic. Specifically, we incorporate grid alignment, legend consistency, visual parameters, and data parameters, enabling a more fine-grained assessment of code-level similarity. By aggregating multiple complementary dimensions, our approach provides a more comprehensive evaluation of similarity from the code perspective.
Through analysis of LLM-Score and Base-Score, we observe a high degree of consistency between the two metrics, indicating that the proposed hard-rule matching strategy aligns well with LLM-based assessments to a certain extent. Moreover, Base-Score also reflects visual similarity at the trend level, as illustrated in Fig.~\ref{fig:DR_code-level vs. chart-level}, Fig.~\ref{fig:CRD_code-level vs. chart-level} where we analyze the relationship between code-level and chart-level similarity.

\subsubsection{Color Score}

Traditional approaches typically treat all colors in a chart as an unordered set, neglecting the binding relationship between colors and specific data items\citep{chartmimic}. To address this issue, we propose an efficient and more professional method for color extraction strategy designed to parse colors and their corresponding semantic information from Matplotlib's graphical objects \texttt{Figure}. This strategy decomposes the chart into different types of visual elements and organizes the extracted color information into a structured mapping, which can be expressed as:
\begin{equation}
    \{ \text{ElementType} \rightarrow \{ \text{DataKey} \rightarrow \text{HexColor} \} \}
\end{equation}

where: 
\begin{itemize}
    \item  ElementType: Refers to the object to which the color is applied, such as the fill color of a bar chart (\texttt{patch\_face}), the line color of a line chart (\texttt{line\_color}), the color of a scatter plot (\texttt{scatter\_color}), or the background of the axes (\texttt{axes\_bg}).
    \item   DataKey: Refers to the specific data entity bound to the color. This is typically the label in the legend, the tick label on the axis, or the content of a text element.
    \item  HexColor: The standardized hexadecimal color code.
\end{itemize}

After obtaining the structured color data, we design a set of weighted evaluation metrics to quantify the color fidelity between generated\_code and GT\_code. The core principle of this evaluation is that not all colors are equally important. For example, errors in the colors of data series are more severe than errors in the colors of axis grid lines.

To this end, we introduce element-type weights ($w_t$), assigning a predefined weight to each \texttt{ElementType} $t$. Core data elements (e.g., \texttt{patch\_face}, \texttt{line\_color}) are assigned high weights (e.g., 1.0), whereas auxiliary or decorative elements (e.g., \texttt{figure\_bg}, \texttt{spine}) are assigned lower weights (e.g., 0.01).

The evaluation is performed only on the element types and data keys shared by both generated\_code(gen) and GT\_code(gt). This ensures a valid comparison, avoiding mismatches such as comparing a line color in generated with a bar color in gt\_code.

The total weighted similarity $S_{\text{total}}$ serves as the core of our model, and is computed as:
\begin{equation}
    S_{\text{total}} = \sum_{t \in T_{\cap}} \sum_{k \in K_{\cap, t}} w_t \cdot \sigma \big( C_{\text{gen},t,k}, \, C_{\text{gt},t,k} \big),
\end{equation}

where:
\begin{itemize}
    \item $T_gen$ and $T_{gt}$ denote the sets of all element types present in the generated chart and the ground-truth chart, respectively.
    \item $K_{gen,t}$ and $K_{gt,t}$ denote the sets of all data keys under element type $t$ in the generated and ground-truth charts, respectively.
    \item $w_t$ is the predefined weight for element type $t$.
    \item $C_{gen,t,k}$ and $C_{gt,t,k}$ are the colors corresponding to element type $t$ and key $k$ in the generated and ground-truth charts, respectively.
    \item $\sigma(C_1, C_2)$ is a function measuring the similarity between two hexadecimal colors.
\end{itemize}

The color similarity function $\sigma(C_1, C_2)$ is used to quantify the visual closeness between two colors. In our implementation, we adopt a normalized reversed Euclidean distance in the RgenB color space to compute similarity.

First, the hexadecimal color $C$ is converted into its RGB representation $(R, G, B)$. The Euclidean distance between two colors $C_1$ and $C_2$ is defined as:
\begin{equation}
    d(C_1, C_2) = (R_1 - R_2)^2 + (G_1 - G_2)^2 + (B_1 - B_2)^2.
\end{equation}

The maximum possible distance in the RGB space corresponds to the distance between $(0,0,0)$ and $(255,255,255)$, i.e.,
\begin{equation}
    d_{\max} = 3 \cdot 255^2.
\end{equation}

We then normalize the distance $d$ and transform it into a similarity score $\sigma$ within the range $[0,1]$:
\begin{equation}
    \sigma(C_1, C_2) = 1 - \frac{d(C_1, C_2)}{d_{\max}}.
\end{equation}

When two colors are identical, $\sigma = 1.0$; when they differ maximally, $\sigma = 0.0$.

To provide comprehensive and interpretable evaluation results, we map the computed total weighted similarity ($S_{\text{total}}$) to three standard metrics widely used in the information retrieval domain: Precision, Recall, and F1-Score.

\textbf{Total Weight}: We first compute the total weights of the generated chart and the ground-truth chart, representing the maximum theoretically achievable similarity score.

\begin{gather}
    W_{\text{gen}} = \sum_{t \in T_{\text{gen}}} \sum_{k \in K_{\text{gen},t}} w_t \\
    W_{\text{gt}} = \sum_{t \in T_{\text{gt}}} \sum_{k \in K_{\text{gt},t}} w_t
\end{gather}

\textbf{Precision}: Measures the accuracy of all color elements in the generated chart. It answers the question: “Among all generated colors, what proportion is correct?”
\begin{equation}
    \text{Precision} = \frac{S_{\text{total}}}{W_gen}.
\end{equation}

\textbf{Recall}: Measures the extent to which all color elements in the ground-truth chart are correctly reproduced in the generated chart. It answers the question: “Among all required colors, what proportion has been correctly generated?”\
\begin{equation}
    \text{Recall} = \frac{S_{\text{total}}}{W_{gt}}.
\end{equation}

\textbf{F1-Score}: The harmonic mean of Precision and Recall, providing a single comprehensive evaluation score.
\begin{equation}
    \text{F1-Score} = \frac{2 \cdot \text{Precision} \cdot \text{Recall}}{\text{Precision} + \text{Recall}}.
\end{equation}

\subsubsection{grid Score}

We define a structured \textbf{Grid State Descriptor}. For each subplot $ax$ in a chart, we extract the visibility of its X-axis and Y-axis grid lines, and encode them as a Boolean dictionary:
\begin{equation}
    \mathcal{S}_{\text{grid}} = 
    \begin{cases}
        \texttt{'x\_grid\_visible'} : \text{bool} \\
        \texttt{'y\_grid\_visible'} : \text{bool}
    \end{cases}
\end{equation}

We traverse all \texttt{Axes} objects within a \texttt{Figure}, and for each subplot where at least one grid line (X-axis or Y-axis) is visible, we generate a grid state descriptor. Ultimately, the grid configuration of an entire chart is abstracted as a list of such descriptors, which can be mathematically regarded as a multiset.

For example, in a \texttt{Figure} with two subplots, where the first subplot has only Y-axis grid lines and the second subplot has both X-axis and Y-axis grid lines, the grid configuration is represented as:
\begin{equation}
\begin{aligned}
    1.~ & \{ \texttt{'x\_grid'}: \texttt{False}, \texttt{'y\_grid'}: \texttt{True} \} \\
    2.~ & \{ \texttt{'x\_grid'}: \texttt{True}, \texttt{'y\_grid'}: \texttt{True} \}
\end{aligned}
\end{equation}

This structured representation is not only precise but also completely ignores the specific styles of grid lines (e.g., color, linewidth). Instead, it focuses solely on their presence, which captures the core semantics and makes the evaluation more robust. 

After extracting the multisets of grid state descriptors from the generated figure ($G_{\text{gen}}$) and the ground-truth figure ($G_{\text{gt}}$), we further use the F1 metric to measure the accuracy of this parameter.

We define the following notations:

\begin{itemize}
    \item $G_{\text{gen}}$: the multiset of grid state descriptors extracted from the generated figure.
    \item $G_{\text{gt}}$: the multiset of grid state descriptors extracted from the ground-truth figure.
\end{itemize}

The number of true positives (TP) is defined as the cardinality of the intersection between the two multisets:
\begin{equation}
    TP = | G_{\text{gen}} \cap G_{\text{gt}} |.
\end{equation}

\textbf{ True Positives (TP)}  
A true positive is defined as a grid state descriptor that appears in $G_{gen}$ and exactly matches one in $G_{gt}$. The total number of true positives is given by the size of the intersection of these two multisets:
\begin{equation}
    TP = |G_{gen} \cap G_{gt}|.
\end{equation}

\textbf{ Precision}  
Precision measures the proportion of correctly activated grid configurations among all grid configurations in the generated figure (i.e., those that also exist in the ground-truth figure):
\begin{equation}
    \text{Precision} = \frac{TP}{|G_{gen}|} = \frac{|G_{gen} \cap G_{gt}|}{|G_{gen}|}.
\end{equation}

If $|G_{gen}| = 0$, we define $\text{Precision} = 1.0$.  

\textbf{ Recall}  
Recall measures the proportion of required grid configurations in the ground-truth figure that are successfully reproduced in the generated figure:
\begin{equation}
    \text{Recall} = \frac{TP}{|G_{gt}|} = \frac{|G_{gen} \cap G_{gt}|}{|G_{gt}|}.
\end{equation}

If $|G_{gt}| = 0$, we define $\text{Recall} = 1.0$.  

\textbf{F1-Score}  
The F1-score, as the harmonic mean of precision and recall, provides a single comprehensive metric:
\begin{equation}
    \text{F1-Score} = 2 \cdot \frac{\text{Precision} \cdot \text{Recall}}{\text{Precision} + \text{Recall}}.
\end{equation}

\subsubsection{Layout score}

For each individual subplot (i.e., an \texttt{Axes} object) in a chart, we create a unique and quantitative \textbf{Layout Descriptor}. This descriptor fully defines the size and position of the subplot within a virtual grid (\texttt{GridSpec}). Instead of relying on pixel coordinates, we extract the underlying structural information from Matplotlib's \texttt{SubplotSpec} object.

For each subplot $ax$ in a \texttt{Figure}, we extract the following six key parameters to construct its layout descriptor $D$:

\begin{itemize}
    \item $nrows$ ($R$): the total number of rows in the corresponding \texttt{GridSpec}.
    \item $ncols$ ($C$): the total number of columns in the corresponding \texttt{GridSpec}.
    \item $row\_start$ ($r_s$): the starting row index of the grid cells occupied by the subplot.
    \item $row\_end$ ($r_e$): the ending row index of the grid cells occupied by the subplot.
    \item $col\_start$ ($c_s$): the starting column index of the grid cells occupied by the subplot.
    \item $col\_end$ ($c_e$): the ending column index of the grid cells occupied by the subplot.
\end{itemize}

Thus, the layout of each subplot can be precisely represented as a 6-tuple:
\begin{equation}
    D = (R, C, r_s, r_e, c_s, c_e).
\end{equation}

By traversing all \texttt{Axes} objects in a \texttt{Figure}, the overall layout can be abstracted as a multiset of these layout descriptors $D$, denoted as $L$.

We define the following notation:
\begin{itemize}
    \item $L_{gen}$: the multiset of layout descriptors extracted from the generated figure.
    \item $L_{GT}$: the multiset of layout descriptors extracted from the ground-truth figure.
\end{itemize}

\textbf{ True Positives (TP)}  
A true positive represents a layout descriptor that exists in $L_{gen}$ and exactly matches one in $L_{gt}$. The total number of true positives is defined as the size of the intersection of these two multisets:
\begin{equation}
    TP = |L_{gen} \cap L_{gt}|
\end{equation}

This indicates the number of subplots that are correctly generated and placed in the correct positions.

\textbf{Precision}  
Precision measures the proportion of correctly generated subplots among all generated subplots:
\begin{equation}
    \text{Precision} = \frac{TP}{|L_{gen}|} = \frac{|L_{gen} \cap L_{gt}|}{|L_{gen}|}
\end{equation}

Here, $|L_{gen}|$ denotes the total number of subplots in the generated figure. A low precision indicates that the model produced redundant or incorrectly placed subplots.

\textbf{Recall}  
Recall measures the proportion of required subplots in the ground-truth figure that were successfully generated:
\begin{equation}
    \text{Recall} = \frac{TP}{|L_{gt}|} = \frac{|L_{gen} \cap L_{gt}|}{|L_{gt}|}
\end{equation}

Here, $|L_{gt}|$ denotes the total number of subplots in the ground-truth figure. A low recall suggests that the model failed to generate all required subplots.

\textbf{F1-Score}  
The F1-score, as the harmonic mean of precision and recall, provides a single balanced metric for evaluating the overall quality of the layout:
\begin{equation}
    \text{F1-Score} = 2 \cdot \frac{\text{Precision} \cdot \text{Recall}}{\text{Precision} + \text{Recall}}
\end{equation}

\subsubsection{Legend score}
We propose a Dual-Constraint Matching Framework for Legend Evaluation. This framework decomposes legend evaluation into independent assessments of the semantic and spatial properties of each individual legend entry, and quantifies the consistency between the generated and ground-truth figures through a flexible matching algorithm. Consequently, it provides a more comprehensive and robust evaluation scheme.

Our method does not treat the legend as a single entity but decomposes it into a collection of independent legend entries. For each visible legend object in the chart, we traverse all its text labels and create an atomic, structured \textbf{Legend Descriptor} for each label.

The descriptor $D$ is defined as a 2-tuple that captures both semantic and spatial information:
\begin{equation}
    D = (t, B)
\end{equation}

where:

\begin{itemize}
    \item $t$ is a string representing the textual content of the legend entry. This element captures the semantic correctness of the legend.
    \item $B$ is a 4-tuple $(x_0, y_0, x_1, y_1)$ representing the bounding box of the entire legend object containing the text entry, expressed in the screen rendering coordinate system. This element captures the spatial correctness of the legend.
\end{itemize}

By traversing all legends from both the \texttt{Axes} objects and the \texttt{Figure} object itself, we can extract all visible legend entries of a chart and represent them collectively as a multiset of descriptors $D$, denoted as $L$.

After extracting the multisets of legend descriptors $L_{gen}$ and $L_{gt}$ from the generated and ground-truth figures, respectively, we design a dual-constraint matching algorithm to compute their similarity. The algorithm can flexibly operate in two modes: semantic-only matching or combined semantic and spatial matching.

A descriptor $D_{gen} = (t_{gen}, B_{gen})$ from $L_{gen}$ matches a descriptor $D_{gt} = (t_{gt}, B_{gt})$ from $L_{gt}$ if and only if one or both of the following constraints are satisfied:

\noindent\textbf{Semantic Constraint:} The text content of the two descriptors must be identical:
\begin{equation}
    t_{gen} = t_{gt}.
\end{equation}

\textbf{Positional Constraint:} The bounding boxes of the legend objects containing the descriptors must have a positive intersection area:
\begin{equation}
    \text{Area}_{intersection}(B_{gen}, B_{gt}) > 0.
\end{equation}

For two bounding boxes $B_1 = (x_{1,0}, y_{1,0}, x_{1,1}, y_{1,1})$ and $B_2 = (x_{2,0}, y_{2,0}, x_{2,1}, y_{2,1})$, the intersection area is computed as:
\begin{equation}
    \begin{aligned}
        x_A &= \max(x_{1,0}, x_{2,0}), \quad y_A = \max(y_{1,0}, y_{2,0}) \\
        x_B &= \min(x_{1,1}, x_{2,1}), \quad y_B = \min(y_{1,1}, y_{2,1}) \\
        &\text{Area}_{\text{int}} = \max(0, x_B - x_A) \\
        &\quad \cdot \max(0, y_B - y_A)
    \end{aligned}
\end{equation}

The algorithm finds unique matching pairs that satisfy the above constraints (removing matched descriptors from the pool) and computes the total number of true positives (TP). Based on TP, we perform the final quantitative evaluation using standard precision, recall, and F1-score metrics:

\begin{equation}
\begin{split} 
    \text{Precision} = \frac{TP}{|L_{gen}|}, \quad
    \text{Recall} = \frac{TP}{|L_{gt}|}, \\ 
    \text{F1-Score} = 2 \cdot \frac{\text{Precision} \cdot \text{Recall}}{\text{Precision} + \text{Recall}}.
\end{split}
\end{equation}
\subsubsection{data parameter score}
\label{data_parameter}
The primary goal of data visualization is to faithfully and accurately convey the underlying data. We introduce an evaluation framework designed to quantify the fidelity of a chart's \textit{data parameters}. This framework inspects the chart at a deep level, directly verifying the correctness of its underlying data. 

The first step of the framework is to identify and extract the \textit{data parameters} that directly define the data representation of the chart. Through introspection of Matplotlib plotting elements, we categorize these parameters into distinct types. The set of data parameters, denoted as $K_{data}$, is explicitly defined as:


\begin{equation}
\begin{split}
    K_{\text{data}} = \{ & \texttt{'xdata'}, \texttt{'ydata'}, \texttt{'offsets'}, \texttt{'xy'}, \\
                         & \texttt{'verts'}, \texttt{'width'}, \texttt{'height'}, \texttt{'sizes'} \}.
\end{split}
\end{equation}
These parameters directly correspond to the geometric and positional properties of chart elements:

\begin{itemize}
    \item For line plots (\texttt{Line2D}), we extract \texttt{xdata} and \texttt{ydata}.
    \item For bar charts (\texttt{Rectangle}), we extract the lower-left corner coordinates \texttt{xy}, as well as \texttt{width} and \texttt{height}.
    \item For filled plots (\texttt{Polygon}), we extract all vertex coordinates \texttt{verts}.
    \item For scatter plots (\texttt{Collection}), we extract the center coordinates \texttt{offsets} and the point sizes \texttt{sizes}.
\end{itemize}

Through this process, each chart is decomposed into a multiset $E$ of element-parameter dictionaries. 

Data parameters, especially those represented as arrays, cannot be compared using simple equality operators. To robustly handle variations in data point ordering or floating-point precision, we define a dedicated similarity function $S(v_1, v_2)$. The core logic for data parameters is as follows:

\textbf{Numeric Type:} For scalar values, we use \texttt{numpy.isclose} to determine whether two floating-point numbers are approximately equal within a tolerance $\epsilon$:
\begin{equation}
    \label{eq_s1}
    S(v_1, v_2) =
    \begin{cases}
    1 & \text{if } |v_1 - v_2| \leq \epsilon \\
    0 & \text{otherwise}
    \end{cases} 
\end{equation}

\textbf{Array-like Type:} For array data, which is crucial for evaluating data parameters, we adopt the Jaccard similarity coefficient to measure the overlap between the contents of two arrays. Let $V_1$ and $V_2$ denote the sets of elements in $v_1$ and $v_2$, respectively:
\begin{equation}
    \label{eq_s2}
    S(v_1, v_2) = \frac{|V_1 \cap V_2|}{|V_1 \cup V_2|}
\end{equation}

This method is insensitive to the order of data points and accurately reflects the true content overlap between two datasets.

After quantifying the similarity between parameters, we employ a two-stage algorithm to compute the final evaluation metrics.

\textbf{Element Matching:} To address differences in element order and quantity across charts, we use a greedy optimal matching algorithm. For each element $e_{gt}$ in the ground-truth chart, the algorithm searches among elements of the same type in the generated chart to find the best match $e^*_{gen}$ that maximizes the total similarity across all parameters. This matching is performed globally, considering all parameter types. The result is a set of successful matches:
\begin{equation}
    M = \{ (e_{gen}, e_{gt}) \}.
\end{equation}

\textbf{Data Metric Computation:} Once the matching set $M$ is obtained, we focus exclusively on data parameters to aggregate the scores. The total true positive score for the data dimension, $TP_{data}$, is computed as the sum of similarities across all matched pairs. We iterate over the union of keys to ensure penalties for missing or extra parameters:

\begin{equation}
    \text{TP} = \sum_{(e_{\text{gen}}, e_{\text{gt}}) \in M} \sum_{k \in \mathcal{K}_{e}} S(e_{\text{gen}}[k], e_{\text{gt}}[k])
\end{equation}

$\mathcal{K}_{e} = (\text{keys}(e_{\text{gen}}) \cup \text{keys}(e_{\text{gt}})) \cap K_{\text{data}}$, Next, we count the total number of data parameters in the generated chart and the ground-truth chart, denoted as $N_{data,gen}$ and $N_{data,gt}$, respectively. Finally, we compute the precision, recall, and F1-score for the data dimension:

\begin{equation} 
    \begin{aligned} 
    \text{Precision} &= \frac{TP}{N_{data,gen}}, \\ \text{Recall} &= \frac{TP}{N_{data,gt}}, \\ \text{F1-Score} &= 2 \cdot \frac{\text{Precision} \cdot \text{Recall}} {\text{Precision}+ \text{Recall}}. 
    \end{aligned} 
\end{equation}

\subsubsection{visual parameter score}
The visual style of a chart is also an important component of chart reproduction quality. Visual style is governed by a set of \textit{visual parameters}, such as line styles, marker shapes, element transparency, and so on. Correct usage of these parameters not only affects the aesthetic quality and professionalism of the chart, but also directly determines whether it adheres to specific design guidelines or user instructions. We propose a framework, running in parallel with the data parameter evaluation, specifically designed to quantify the consistency of a chart with respect to its \textit{visual parameters}.

This framework builds upon the parameterized representation established in \ref{data_parameter}. After extracting all parameters of an element, we identify the set of \textit{visual parameters} ($K_{visual}$) by exclusion. A parameter key $k$ is classified as a visual parameter if it satisfies:

\begin{equation}
    k \notin K_{data} \quad \text{and} \quad k \notin K_{ignore}
\end{equation}

where $K_{data}$ is the predefined set of data parameters, and $K_{ignore}$ is the set of parameters handled by other evaluators (e.g., color). Typical visual parameters include: \texttt{'linestyle'}, \texttt{'linewidth'}, \texttt{'marker'}, \texttt{'markersize'}, \texttt{'alpha'}, and so on. The extraction process is performed in parallel with that of the data parameters, but subsequent evaluation computations focus exclusively on this subset of parameters.

We employ the same general similarity function $S(v_1, v_2)$ introduced in the \eqref{eq_s1} and \eqref{eq_s2} to compare the values of visual parameters. Its robustness is equally applicable to various data types of visual parameters:

\begin{itemize}
    \item \textbf{String type:} For parameters such as \texttt{linestyle} (e.g., '-' vs '--') or \texttt{marker} (e.g., 'o' vs 'x'), the function performs a direct string equality comparison.
    \item \textbf{Numeric type:} For parameters such as \texttt{linewidth} (e.g., 1.5 vs 2.0) or \texttt{alpha} (e.g., 0.8 vs 1.0), the function uses \texttt{numpy.isclose} to perform a tolerance-based comparison.
\end{itemize}

This consistent definition of similarity ensures intrinsic coherence across different evaluation dimensions.

\textbf{Element Matching:} We reuse the set of matched element pairs $M = \{(e_{gen}, e_{gt})\}$ obtained through the greedy optimal matching algorithm. This implies that the matching of elements is determined based on their overall similarity (data + visual), consistent with human perception — we always perceive an element as a whole. Establishing a match indicates that both the data and visual aspects will be evaluated for that pair.

\textbf{Visual Metric Computation:} Given the set of matched pairs $M$, we focus exclusively on the visual parameters to aggregate the scores. We compute the total true positive score for the visual dimension ($TP_{visual}$), defined as the sum of visual parameter similarities across all matched pairs:

\begin{equation}
\begin{split}
    \text{TP}_{\text{visual}} = & \sum_{(e_{\text{gen}}, e_{\text{gt}}) \in M} \\
    & \sum_{k \in \mathcal{K}_{\text{v}}} S(e_{\text{gen}}[k], e_{\text{gt}}[k])
\end{split}
\end{equation}

Similarly, we count the total number of visual parameters in the generated and ground-truth charts, denoted as $N_{visual,gen}$ and $N_{visual,gt}$, respectively. Finally, the precision, recall, and F1-score for the visual dimension are computed as:

\begin{equation}
\begin{aligned}
    \text{Precision}_{visual} &= \frac{TP_{visual}}{N_{visual,gen}}, \\
    \text{Recall}_{visual}    &= \frac{TP_{visual}}{N_{visual,gt}}, \\
    \text{F1-Score}_{visual} &= 2 \cdot \frac{\text{Precision}_{visual} \cdot \text{Recall}_{visual}}
                                        {\text{Precision}_{visual} + \text{Recall}_{visual}}.
\end{aligned}
\end{equation}

\subsubsection{type score}
We propose an evaluation framework based on \textit{Artist Class Introspection}. Unlike methods that rely on the visual rendering of charts, this framework directly inspects the object model constructed in memory by the plotting library (Matplotlib). By examining the core drawing \textit{artists} (i.e., primitive graphical objects) and their associated classes, the framework deterministically and robustly infers the composition of a chart. The key idea is that Matplotlib employs different classes of artist objects for different types of plots. For example, a line plot is rendered using \texttt{Line2D} objects, whereas a bar chart is rendered using \texttt{Rectangle} objects. Leveraging this intrinsic correspondence, we can infer the chart types present in a figure by identifying which classes of artist objects it contains.

Our algorithm operates by traversing all subplots (Axes) within a \texttt{matplotlib.Figure} object and inspecting the list of artists contained in each subplot (e.g., \texttt{ax.lines}, \texttt{ax.patches}, \texttt{ax.collections}, etc.). 

The algorithm aggregates all detected chart types within a figure into a \textit{set}. This set-based representation has a significant advantage: it naturally supports the identification and evaluation of \textit{composite charts}. For example, a chart that overlays a line plot on top of a bar chart will be recognized as containing both \texttt{bar\_or\_hist} and \texttt{line}.
  
The number of true positives is defined as the size of the intersection between the two sets, that is, the number of chart types present in both the generated chart and the reference chart:
\begin{equation}
    TP = \lvert T_{\text{gen}} \cap T_{\text{gt}} \rvert
\end{equation}

Precision measures the proportion of correct chart types among all generated chart types:
\begin{equation}
    \text{Precision} = \frac{TP}{\lvert T_{\text{gen}} \rvert} 
= \frac{\lvert T_{\text{gen}} \cap T_{\text{gt}} \rvert}{\lvert T_{\text{gen}} \rvert}
\end{equation}

where \(\lvert T_{\text{gen}} \rvert\) denotes the total number of distinct chart types detected in the generated chart.

Recall measures the proportion of reference chart types that are successfully generated:
\begin{equation}
    \text{Recall} = \frac{TP}{\lvert T_{\text{gt}} \rvert} 
    = \frac{\lvert T_{\text{gen}} \cap T_{\text{gt}} \rvert}{\lvert T_{\text{gt}} \rvert}
\end{equation}

where \(\lvert T_{\text{gt}} \rvert\) denotes the total number of distinct chart types in the reference chart.

The F1-Score is the harmonic mean of precision and recall, providing a comprehensive evaluation metric:
\begin{equation}
    \text{F1-Score} = \frac{2 \cdot \text{Precision} \cdot \text{Recall}}{\text{Precision} + \text{Recall}}
\end{equation}

\subsubsection{text score}

We propose a text evaluation framework based on \emph{semantic categorization} and \emph{fuzzy matching}. 
In this framework, all textual elements in a chart are categorized according to their functional roles, 
and a fuzzy matching algorithm based on edit distance is applied among texts within the same category. 
This enables a quantitative evaluation of chart text that is both strict and robust.

To achieve precise evaluation of textual roles, we first design an extractor 
(\_extract\_texts\_from\_figure) that introspects the \texttt{matplotlib} 
\texttt{Figure} object to identify and classify all visible textual elements. 
Instead of treating all texts as an undifferentiated set, we categorize them 
into predefined semantic classes.

Through this process, the entire textual content of a chart is transformed into 
a structured \textit{Text Map}, denoted as $T$. Its form is a dictionary that 
maps each category name to the list of text strings belonging to that category:  
$T = \{c \rightarrow [t_{1}, t_{2}, \ldots]\}$. For example, $T_{\text{title}}$ 
represents the list of all subplot titles in the figure. This categorization 
mechanism ensures context-aware evaluation and prevents, for instance, an axis 
label from being incorrectly compared with a title.

After obtaining the text maps of the generated chart and the reference chart, 
$T_{\text{gen}}$ and $T_{\text{gt}}$, we designed an evaluation algorithm to 
quantify their consistency. To tolerate minor textual differences, we adopt the 
Levenshtein Ratio as the similarity function between two strings $s_{1}$ and 
$s_{2}$, denoted as $S_{L}(s_{1}, s_{2})$. This function is based on computing 
the minimum number of single-character edits (insertions, deletions, or 
substitutions) required to transform one string into the other (i.e., the 
Levenshtein Distance), and normalizes the value to the interval $[0,1]$:

\begin{equation}
    S_{L}(s_{1}, s_{2}) = 1 - 
\frac{\text{LevenshteinDistance}(s_{1}, s_{2})}
     {\max(|s_{1}|, |s_{2}|)}
\end{equation}

A higher value of $S_{L}$ indicates greater similarity between the two strings. 
Identical strings achieve a similarity of 1.

Our evaluation algorithm operates independently within each semantic category. 
For each category $c$, the algorithm searches for the best match $t_{gt}^{*}$ 
for every generated text $t_{\text{gen}} \in T_{\text{gen},c}$ from the available 
reference texts $T_{\text{gt},c}$, such that $S_{L}(t_{\text{gen}}, t_{gt})$ is 
maximized. To prevent one-to-many matches, once a reference text is matched, 
it is removed from the candidate pool.

We then accumulate the similarity scores of all best matches across all 
categories to obtain a total similarity score ($TP_{\text{score}}$), which can 
be regarded as a weighted sum of ``true positives'':

\begin{equation}
    TP_{\text{score}} = 
    \sum_{c \in C} \;
    \sum_{t_{\text{gen}} \in T_{\text{gen},c}} \;
    \max_{t_{gt} \in T_{\text{gt},c}^{\prime}}
    S_{L}(t_{\text{gen}}, t_{gt})
\end{equation}

where $C$ denotes the union of all text categories present in both charts, 
and $T_{\text{gt},c}^{\prime}$ is the set of unmatched reference texts in 
category $c$.

Finally, we compute the total number of generated and reference texts 
($N_{\text{gen}}$ and $N_{\text{gt}}$), and derive the Precision, Recall, 
and F1-Score as follows:

\begin{equation}
   \text{Precision} = \frac{TP_{\text{score}}}{N_{\text{gen}}}, 
    \quad 
    N_{\text{gen}} = \sum_{c} |T_{\text{gen},c}|
\end{equation}

\begin{equation}
    \text{Recall} = \frac{TP_{\text{score}}}{N_{\text{gt}}}, 
    \quad 
    N_{\text{gt}} = \sum_{c} |T_{\text{gt},c}|
\end{equation}

\begin{equation}
    \text{F1-Score} = 
 \frac{2 \cdot \text{Precision} \cdot \text{Recall}}
     {\text{Precision} + \text{Recall}}
\end{equation}

\subsection{LLM-Evaluation}
\label{LLM_Evaluation}

This study designs and implements a multi-dimensional visualization code evaluation framework based on Large Language Models (LLMs). The framework does not execute code or render images; instead, it leverages the powerful code understanding and reasoning capabilities of LLMs to perform static analysis directly on the source code of both the generated and reference scripts. By decomposing the complex problem of ``visual similarity'' into a series of well-defined and mutually orthogonal evaluation dimensions, and by designing strict scoring instructions for each, our framework provides a comprehensive, in-depth, and interpretable quantitative assessment of chart code quality.

We deconstruct the ambiguous task of ``code quality'' assessment into eight specific and independent evaluation dimensions, denoted as $D_i$. This approach makes the LLM's evaluation task more focused and renders the final results more diagnostic and interpretable. The eight dimensions are defined as follows:

\begin{itemize}
    \item \textbf{Data Logic ($D_{data}$):} Evaluates the integrity of data presentation, including raw values, sorting, filtering, aggregation, and mathematical transformations, ensuring the quantitative information conveyed is identical to the ground truth.
    \item \textbf{Chart Type and Geometry ($D_{type}$):} Evaluates the correctness of the core plotting primitive (e.g., scatter, bar, contour) and the dimensionality of the visualization.
    \item \textbf{Color and Color Mapping ($D_{color}$):} Evaluates the final rendered colors, colormap families, direction, and normalization ranges, distinguishing between data-driven mappings and fixed style choices.
    \item \textbf{Visual Parameters ($D_{vis}$):} Evaluates non-color stylistic attributes such as line width, marker styles, transparency (alpha), and line styles.
    \item \textbf{Layout and Structure ($D_{layout}$):} Evaluates the overall figure configuration, including canvas size, aspect ratio, subplot grid structures, and spacing adjustments (e.g., tight layout).
    \item \textbf{Legend Configuration ($D_{legend}$):} Evaluates the presence, content, ordering, and positioning of the legend, ensuring it accurately reflects the data series.
    \item \textbf{Grid and Axes ($D_{grid}$):} Evaluates the configuration of coordinate systems, including grid line visibility/style, axis limits, spines, and tick specifications.
    \item \textbf{Text Content ($D_{text}$):} Evaluates the accuracy of all textual annotations, including titles, axis labels, and specific data point annotations.
\end{itemize}

\subsection{LMM-Evaluation}
\label{LMM-Evaluation}
The ultimate criterion for evaluating automatically generated charts should be human visual perception. Although programmatic evaluation and source code analysis can technically ensure the correctness of chart components and parameters, they may not fully capture all visual details, artifacts, or the overall aesthetic coherence in the final rendered image. To establish an evaluation system that more closely approximates a "gold standard," we argue for the necessity of directly assessing the final visual output---the chart image itself.

To this end, this study designs and implements a holistic chart image evaluation framework based on Vision-Language Models (VLMs). This framework utilizes advanced multimodal large models by simultaneously providing them with both the reference and the generated images, supplemented by a set of rigorous evaluation instructions, to directly quantify the visual similarity between the two. This end-to-end visual evaluation method can capture a wide range of discrepancies, from macroscopic layout to microscopic pixel-level differences, thereby providing a comprehensive and holistic quality score. Here, we adopt a holistic evaluation approach, assessing all visual aspects in a single call. To ensure rigor, we extend and reinforce the philosophy of a \textbf{deduction-based scoring system}. The instructions require the model to assume a perfect score of 100, and then to deduct points for every visual discrepancy it finds between the two images.

The evaluation prompt is in \ref{LMM_evaluation}


\begin{table}[p]
\centering
\caption{Run configurations for all models. Unset values indicate that their default values are being used. For Proprietary models, we are unable to use a Top-P of exactly 1 due to their API settings, and we end up using a value of $0.99999$. Temp. denotes temperature. We use model pages' code to set up the run configurations whenever possible.}
\resizebox{\linewidth}{!}{
\renewcommand{\arraystretch}{1.35} 
\begin{tabular}{ll|ccccc@{}}
\hline
\toprule
\small \textbf{Model} & \small \textbf{Version/HF Checkpoint} & \small \textbf{Do Sample} & \small \textbf{level 1 2 Max} & \textbf{level 3 Max} & \small \textbf{Temp.} & \small \textbf{Top-P}   \\
\midrule
\multicolumn{7}{c}{\textbf{Proprietary Multimodal Large Language Models}} \\
\midrule
GPT-5.2 \cite{gpt5.2} & \texttt{gpt-5.2-2025-12-11} &  & default & 55000& 0.1 & 1   \\
Claude 4 Sonnet \cite{Claude4} & \texttt{claude-4-sonnet-20250523} &  & default & 55000 & 0.1 & 1    \\
Gemini-3-Pro \cite{gemini2.5} & \texttt{gemini-3-pro-20251118} &  & default &55000 & 0.1 & 1   \\
doubao-seed-1-5 \cite{seed1.5} & \texttt{seed1.5-VL-20250513} &  & default & 16000&0.1 & 1  \\
doubao-seed-1-6 \cite{seed1.6} & \texttt{seed1.6-VL-20250625} &  & default & 32768&0.1 & 1     \\
\midrule
\multicolumn{7}{c}{\textbf{Open-Source Multimodal Large Language Models}} \\
\midrule
Qwen2-VL-7B \cite{qwen2vl} & \texttt{Qwen/Qwen2-VL-7B-Instruct} & \texttt{True} & 8192 & 32768 &0.1 & 0.95    \\
Qwen2-VL-72B \cite{qwen2vl} & \texttt{Qwen/Qwen2-VL-72B-Instruct} & \texttt{True} & 8192& 32768 &0.1 & 0.95    \\
Qwen2.5-VL-7B \cite{qwen2.5vl} & \texttt{Qwen/Qwen2.5-VL-7B-Instruct} & \texttt{True} & 8192 & 32768 & 0.1 & 0.95   \\
qwen2.5-VL-72B \cite{qwen2.5vl} & \texttt{Qwen/Qwen2.5-VL-72B-Instruct} & \texttt{True} & 8192 & 32768& 0.1 & 0.95   \\
deepseek-VL-7B \cite{deepseekvl} & \texttt{deepseek-ai/deepseek-vl-7b-base} & \texttt{True} &  8192& 32768 & 0.1 &0.95     \\
kimi-VL-A3B \cite{kimiteam2025kimivltechnicalreport} & \texttt{moonshotai/Kimi-VL-A3B-Thinking} & \texttt{True} & 8192 & 32768& 0.1 & 0.95    \\
MiMo-VL-7B-RL \cite{mimo} & \texttt{XiaomiMiMo/MiMo-VL-7B-RL-2508} & \texttt{True} &  8192 &32768 &0.1 &0.95     \\
MiMo-VL-7B-SFT \cite{mimo} & \texttt{XiaomiMiMo/MiMo-VL-7B-SFT-2508} & \texttt{True} & 8192 &32768 &  0.1&0.95    \\
GLM-4-9b \cite{GLM-4V} & \texttt{zai-org/glm-4-9b} & \texttt{True} & 8192  & 32768 &  0.1&0.95   \\
GLM-4.6V-Flash \cite{glm4.6} & \texttt{zai-org/glm-4.6v-Flash} & \texttt{True} & 8192  & 32768 &  0.1&0.95   \\
Intern-VL 2.5 8B \cite{internvl2.5} & \texttt{OpenGVLab/InternVL2\_5-8B} & \texttt{True} & 8192 &32768 &0.1 &0.95     \\
Intern-VL 2.5 38B \cite{internvl2.5} & \texttt{OpenGVLab/InternVL2\_5-38B} & \texttt{True} &  8192 &32768& 0.1 & 0.95    \\
Intern-VL 3 8B \cite{internvl3} & \texttt{OpenGVLab/InternVL3-8B} & \texttt{True} &  8192 & 32768&0.1 & 0.95    \\
Intern-VL 3 38B \cite{internvl3} & \texttt{OpenGVLab/InternVL3-38B} & \texttt{True} & 8192 & 32768& 0.1 &  0.95    \\

Intern-VL 3.5 8B \cite{internvl3_5} & \texttt{OpenGVLab/InternVL3\_5-8B} & \texttt{True} &  8192 & 32768& 0.1 & 0.95   \\
Intern-VL 3.5 38B \cite{internvl3_5} & \texttt{OpenGVLab/InternVL3\_5-38B} & \texttt{True} & 8192  &32768& 0.1 &  0.95 \\

llava-onevision-qwen2-7b-si \cite{llavaonevision} & \texttt{lmms-lab/llava-onevision-qwen2-7b-si} & \texttt{True} & 8192 & 32768&0.1 & 0.95   \\
llava-onevision-qwen2-7b-ov \cite{llavaonevision} & \texttt{lmms-lab/llava-onevision-qwen2-7b-ov} & \texttt{True}  & 8192 & 32768& 0.1 & 0.95    \\

Qwen3-VL-30B-A3B \cite{qwen3} & \texttt{Qwen/Qwen3-VL-30B-A3B-Instruct} & \texttt{True} &  8192 & 32768& 0.1 & 0.95   \\
Qwen3-VL-30B-A3B 3.5 38B \cite{qwen3} & \texttt{Qwen/Qwen3-VL-30B-A3B-Think} & \texttt{True} & 8192  &32768& 0.1 &  0.95 \\
Qwen3-VL-32B \cite{qwen3} & \texttt{Qwen/Qwen3-VL-32B-Instruct} & \texttt{True} & 8192  &32768& 0.1 &  0.95 \\
\bottomrule
\hline
\end{tabular}
}
\label{tab:run_configurations}
\end{table}
\FloatBarrier
\section{Run configurations}

We summarize the run configurations for all evaluated models in Table~\ref{tab:run_configurations}. To ensure a fair and consistent comparison across different systems, we adopt a unified configuration strategy whenever possible.

For proprietary models, we primarily rely on their official APIs and follow the default settings provided by the platforms, with minimal modifications. In particular, due to API constraints, we set the Top-P value to $0.99999$ instead of exactly $1$. The temperature is fixed at $0.1$ to reduce randomness and improve output stability.

For open-source models, we enable stochastic decoding (\texttt{do\_sample=True}) and use a consistent configuration with a context length of 8192 tokens for Level 1 and Level 2 tasks, and up to 32768 tokens for Level 3 tasks. These settings are chosen to balance generation quality and computational efficiency.

Whenever possible, we follow the official implementation or model documentation to configure inference parameters. This ensures that each model is evaluated under conditions that are both reproducible and representative of its intended usage.

\FloatBarrier
\section{Open-Source Model Components}

We summarize the key components of all evaluated open-source multimodal models in Table~\ref{tab:os_model_components}, including their vision encoders, language backbones, and input resolutions.

The evaluated models exhibit diverse architectural designs. Most models adopt variants of Vision Transformers (ViT) or CLIP-based encoders (e.g., Qwen-VL, InternVL, and GLM-4V), while others incorporate hybrid designs such as SigLIP combined with SAM for enhanced visual representation (e.g., DeepSeek-VL). On the language side, models are typically built upon large language model backbones such as Qwen, InternLM, or custom proprietary architectures.

Regarding input resolution, we follow the default settings of each model whenever possible, denoted as \textit{original} in the table. This is particularly important for models that support dynamic resolution inputs. For models with stricter input requirements (e.g., DeepSeek-VL-7B and GLM-4-9B), we apply a maximum resolution constraint to ensure compatibility with their architectures.

This diversity in model components reflects the current design space of multimodal systems and provides a comprehensive basis for evaluating their performance under realistic settings.
\begin{table}[p]
\centering
\caption{We summarize the visual and language components of the open-source models evaluated in our benchmark, along with the input resolutions used in our evaluation. Here, \textit{original} denotes that we use the default image size, as the corresponding models support dynamic resolution inputs. Note that for DeepSeekVL-7B and GLM-4-9B , we apply a maximum input size constraint to accommodate their requirements.}
\scalebox{0.8}{
\begin{tabular}{llll@{}}
\hline
\toprule
\small \textbf{Model} & \small \textbf{Vision} & \small \textbf{Language} & \small \textbf{Resolu-} \\
\textbf{} & \small \textbf{Encoder} & \small \textbf{Model} & \small \textbf{tion}  \\
\midrule
Qwen2-VL-7B & Qwen2-VL ViT-14-224  &  Qwen2-VL-LLM-7B & \textit{origianl} \\
\midrule
Qwen2-VL-72B & Qwen2-VL ViT-14-224 &  Qwen2-VL-LLM-72B & \textit{origianl} \\
\midrule
Qwen2.5-VL-7B & Qwen2.5-VL ViT-14-224 & Qwen2.5-VL-LLM-7B & \textit{origianl}\\
 \midrule
Qwen2.5-VL-72B & Qwen2.5-VL ViT-14-224 &  Qwen2.5-VL-LLM-72B & \textit{origianl} \\
\midrule
Deepseek-VL-7B & SigLIP-384-SO400M \&  & DeepSeek-LLM-7B & $1152\times1152$* \\
&SAM-ViT-Base&& \\
\midrule
Kimi-VL-A3B & MoonViT  &  Moonlight Model & \textit{origianl}\\
\midrule
MiMo-VL & Qwen2.5-ViT & MiMo-7B & \textit{origianl} \\
\midrule
GLM-4-9B  & CLIP ViT-L-14-336 & InternLM-7B & $1120\times1120$* \\
\midrule
InternVL-2.5-8B  & InternViT-6B-448px-V2\_5 & internlm2\_5-7b-chat & \textit{origianl} \\
\midrule
InternVL-2.5-38B & InternViT-6B-448px-V2\_5 & Qwen2.5-32B-Instruct & \textit{origianl} \\
\midrule
InternVL-3-8B & InternViT-300M-448px-V2\_5 & Qwen2.5-7B & \textit{origianl} \\
\midrule
InternVL-3-38B & InternViT-6B-448px-V2\_5 & Qwen2.5-32B & \textit{origianl} \\
\midrule
InternVL-3.5-8B & InternViT-300M \&  & Qwen3-8B & \textit{origianl} \\
& InternViT-6B & & \\
\midrule
InternVL-3.5-38B & InternViT-300M \&  & Qwen3-38B & \textit{origianl} \\
& InternViT-6B & & \\
\midrule
llava-onevision-qwen2-7b-si & SigLIP-384-SO400M  & Qwen2-7B & \textit{origianl} \\
\midrule
llava-onevision-qwen2-7b-ov & SigLIP-384-SO400M & Qwen2-7B & \textit{origianl} \\

\bottomrule
\hline
\end{tabular}
}
\label{tab:os_model_components}
\end{table}

\begin{table}[p]
\caption{The release time and model source of LMMs used in our benchmark.}
\label{tab:modelsource}
\centering
\vspace{1mm}
\scalebox{0.68}{
\begin{tabular}{l|l|l}
\toprule
Model & Release Time & \multicolumn{1}{c}{Source} \\
\midrule
\multicolumn{3}{c}{\textit{Closed-source Models}} \\
\cmidrule{1-3}
GPT-5.2  & 2025-12-11 & \url{https://openai.com/zh-Hans-CN/index/introducing-gpt-5-2/}\\
Claude 4 Sonnet & 2025-05-23 & \url{https://www.anthropic.com/news/claude-4} \\
Gemini-3-Pro  & 2025-11-18 & \url{https://deepmind.google/models/gemini/pro/} \\
doubao-seed-1.5  & 2025-05-11 & \url{https://www.volcengine.com/product/doubao}\\
doubao-seed-1.6 & 2025-06-11 & \url{https://www.volcengine.com/product/doubao} \\
\midrule
\multicolumn{3}{c}{\textit{Open-source Models}} \\
\cmidrule{1-3}
Qwen2-VL-7B  & 2024-09-18 & \url{https://huggingface.co/Qwen/Qwen2-VL-7B-Instruct} \\
Qwen2-VL-72B  & 2024-09-18 & \url{https://huggingface.co/Qwen/Qwen2-VL-72B-Instruct}\\
qwen2.5-VL-7B & 2025-01-26 & \url{https://huggingface.co/Qwen/Qwen2.5-VL-7B-Instruct} \\
qwen2.5-VL-72B  & 2025-01-26 & \url{https://huggingface.co/Qwen/Qwen2.5-VL-72B-Instruct} \\
deepseek-VL-7B & 2024-03-09 & \url{https://huggingface.co/deepseek-ai/deepseek-vl-7b-base}\\
kimi-VL-A3B & 2024-08-20 & \url{https://huggingface.co/moonshotai/Kimi-VL-A3B-Thinking}\\
MiMo-VL-7B-RL & 2025-08-10 & \url{https://huggingface.co/XiaomiMiMo/MiMo-VL-7B-RL-2508}\\
MiMo-VL-7B-SFT & 2025-08-10 & \url{https://huggingface.co/XiaomiMiMo/MiMo-VL-7B-SFT-2508} \\
GLM-4-9B & 2024-06-19 & \url{https://huggingface.co/zai-org/glm-4-9b}\\
Intern-VL 2.5 8B  & 2024-11-21 & \url{https://huggingface.co/OpenGVLab/InternVL2_5-8B}\\
Intern-VL 2.5 38B   & 2024-11-21 & \url{https://huggingface.co/OpenGVLab/InternVL2_5-38B}\\
Intern-VL 3 8B   & 2025-04-10 & \url{https://huggingface.co/OpenGVLab/InternVL3-8B}\\
Intern-VL 3 38B   & 2025-04-10 & \url{https://huggingface.co/OpenGVLab/InternVL3-38B}\\
Intern-VL 3.5 8B  & 2025-08-25 & \url{https://huggingface.co/OpenGVLab/InternVL3_5-8B}\\
Intern-VL 3.5 38B  & 2024-08-25 & \url{https://huggingface.co/OpenGVLab/InternVL3_5-38B}\\
llava-onevision-qwen2-7b-si  & 2024-07-29 & \url{https://huggingface.co/lmms-lab/llava-onevision-qwen2-7b-si}\\
llava-onevision-qwen2-7b-ov   & 2024-07-25 & \url{https://huggingface.co/lmms-lab/llava-onevision-qwen2-7b-ov}\\
Qwen3-VL-30B-A3B-Instruct   & 2025-10-03 & \url{https://huggingface.co/Qwen/Qwen3-VL-30B-A3B-Instruct}\\
Qwen3-VL-30B-A3B-Thinking   & 2025-10-03 & \url{https://huggingface.co/Qwen/Qwen3-VL-30B-A3B-Thinking}\\
Qwen3-VL-32B-Instruction &  2025-10-23 & \url{https://huggingface.co/Qwen/Qwen3-VL-32B-Instruct}\\
GLM-4.6v-Flash  & 2025-12-08 & \url{https://huggingface.co/zai-org/GLM-4.6V-Flash}\\
\bottomrule
\end{tabular}
}
\end{table}

\FloatBarrier
\section{Model License Information}

To ensure transparency and clarify the usage conditions of all models involved in our experiments, we provide a summary of their licensing information in Table~\ref{tab:model_license}. This table reports both the \textit{model license} and the \textit{code license} for each evaluated model.

The \textbf{model license} refers to the terms governing the distribution and usage of the pretrained model weights or API services, while the \textbf{code license} corresponds to the licensing of the associated implementation, including training or inference code when available. For proprietary models, both licenses are typically governed by platform-specific terms of service rather than open-source licenses.

We include a diverse set of models, ranging from proprietary multimodal large language models (e.g., GPT-5.2, Claude 4 Sonnet, Gemini-3-Pro) to open-source models released under permissive licenses such as Apache 2.0 and MIT. This diversity reflects the practical landscape of current multimodal systems and enables a comprehensive evaluation across different licensing regimes.

Entries marked with ``Not Applicable'' indicate that no explicit license information is publicly available in the corresponding repository or model documentation. In such cases, users should refer to the official platform or provider for the most up-to-date usage terms.
\begin{table}[p]
\centering

\caption{Summary of licenses in models that are evaluated in \bench{}. Entries marked with ``Not Applicable'' indicate that authors do not have an explicit code license displayed within the codebase or model checkpoint page.}
\begin{tabular}{lll}
\hline
\toprule
\textbf{Name}        & \textbf{Model License} & \textbf{Code License} \\
\midrule
GPT-5.2               & Proprietary            & Proprietary           \\
Claude 4 Sonnet      & Proprietary            & Proprietary           \\
Gemini-3-Pro       & Proprietary            & Proprietary           \\
doubao-seed-1.6            & Proprietary            & Proprietary           \\
doubao-seed-1.5            & Proprietary            & Proprietary           \\
Qwen2-VL-7B          & qwen            & Apache 2.0           \\
Qwen2-VL-72B           & qwen            & Apache 2.0           \\
qwen2.5-VL-7B      & qwen            & Apache 2.0           \\
qwen2.5-VL-72B      & qwen            & Apache 2.0           \\
qwen3-VL-30B-A3B      & qwen            & Apache 2.0           \\
qwen3-VL-30B-Instruct      & qwen            & Apache 2.0           \\
deepseek-VL-7B         & deepseek            & MIT           \\
kimi-VL-A3B   & MIT                    & MIT                   \\
MiMo-VL-7B-RL          & MIT            & Apache 2.0             \\
MiMo-VL-7B-SFT        & MIT            & Apache 2.0            \\
GLM-4-9B     & glm-4             & Apache 2.0            \\
GLM-4.6v-Flash     & glm-4            & Apache 2.0            \\
Intern-VL 2.5 8B                  & Apache-2.0         & MIT   \\
Intern-VL 2.5 38B                     & Apache-2.0   & MIT     \\
Intern-VL 3 8B                    & Apache-2.0        & MIT           \\
Intern-VL 3 38B                  & Apache-2.0      & MIT  \\
Intern-VL 3.5 8B                                 & Apache-2.0      & MIT      \\
Intern-VL 3.5 38B                           & Apache 2.0       & MIT     \\
llava-onevision-qwen2-7b-si & Apache 2.0             & Apache 2.0            \\
llava-onevision-qwen2-7b-ov                  & Apache 2.0             & Apache 2.0            \\
\bottomrule
\hline
\end{tabular}
\label{tab:model_license}
\end{table}

\FloatBarrier
\section{Model Source}
We summarize the release dates and official sources of all evaluated models in Table~\ref{tab:modelsource}. This information provides transparency regarding the provenance of each model and ensures the reproducibility of our benchmark.

Our evaluation includes both proprietary and open-source models. For proprietary systems, we reference their official product pages or announcements (e.g., OpenAI, Anthropic, and Google DeepMind). For open-source models, we provide links to their corresponding repositories or checkpoints, primarily hosted on platforms such as Hugging Face.

The release timeline spans from early 2024 to late 2025, covering multiple generations of multimodal models. This temporal diversity allows us to evaluate both earlier and more recent architectures, providing a more comprehensive view of the current landscape.

By explicitly documenting model sources and release dates, we aim to ensure that all results in our benchmark are traceable and verifiable.

\FloatBarrier
\section{Sample of Data}

To provide a clearer understanding of the structure and characteristics of our benchmark, we present representative examples from our dataset in this appendix.

\begin{wideexample}{Level 1 Direct sample 1}\tiny

\textbf{Instruction}: You are a Python developer proficient in data visualization, with expertise in using libraries such as Matplotlib, NetworkX, Seaborn, and others.I have a plot generated by Python code, but I don't have the corresponding code that generated this plot. Your task is to generate the Python code that can perfectly reproduce the picture based on the image I provide.

Here are the requirements for the task:
1. Data Extraction: Extract the actual data from the provided image. Based on the visual features of the plot, you must infer the data and recreate the plot.
2. Recreate the Image: Generate the Matplotlib code that reproduces the image exactly as it appears, including all elements such as:
   - Plot type (scatter, line, bar, etc.)
   - Axis labels and titles
   - Colors, markers, line styles, and other visual styles
   - Any legends, annotations, or gridlines present in the image
3. Self-contained Code: The Python code should be complete, executable, and self-contained. It should not require any external data files or variables not already present in the code.
Your objective is to extract the any necessary details from the image and generate a Python script that accurately reproduces the plot.

Now, please generate the Python code to reproduce the picture below.

\textbf{Reference figure}:

\begin{center}
    \includegraphics[width=0.8\linewidth]{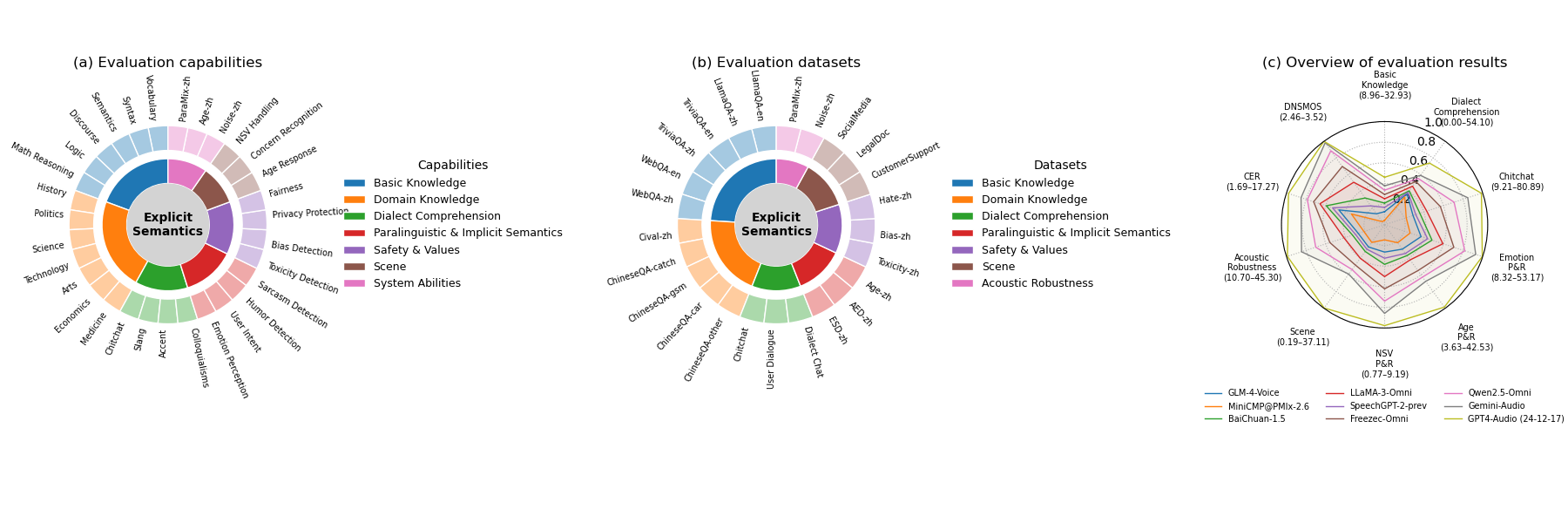}
\end{center}

\textbf{GT Code}: 
\begin{lstlisting}[
  language=Python,
  firstline=1,      % 想分段时随时改行号
  basicstyle=\ttfamily\tiny
]
# == CB_38 figure code ==
import matplotlib.pyplot as plt
import numpy as np
import matplotlib.colors as mcolors

# == CB_38 figure data ==
capabilities = {
    'Basic Knowledge': [
        'Vocabulary', 'Syntax', 'Semantics', 'Discourse', 'Logic', 'Math Reasoning'
    ],...}
datasets = ...
def lighten_color(color, amount=0.5):
    rgb = mcolors.to_rgb(color)
    return tuple(rgb[i] + (1.0 - rgb[i]) * amount for i in range(3))
...

inner_a, size_a, col_a, outer_a, osize_a, ocol_a = prepare_sunburst(capabilities)
inner_b, size_b, col_b, outer_b, osize_b, ocol_b = prepare_sunburst(datasets)
# == figure plot ==
fig = plt.figure(figsize=(18.0, 6.0))
plt.subplots_adjust(left=0.05, right=0.85, wspace=0.7)

# -- (a) Evaluation capabilities sunburst --
ax1 = fig.add_subplot(1, 3, 1)
...
wedges1, _ = ax1.pie(size_a, radius=0.8, labels=None, startangle=90, colors=col_a, wedgeprops=dict(width=0.3, edgecolor='white'))
centre = plt.Circle((0, 0), 0.5, color='lightgray', linewidth=0)
ax1.add_artist(centre)
ax1.text(0, 0, 'Explicit\nSemantics',
         ha='center', va='center', fontsize=10, weight='bold')
ax1.set(aspect='equal')
ax1.set_title('(a) Evaluation capabilities', fontsize=12, pad=45)
ax1.legend(wedges1, inner_a, title='Capabilities', loc='center left', bbox_to_anchor=(1.3, 0.5), fontsize=9, frameon=False)
# -- (b) Evaluation datasets sunburst --
...
centre2 = plt.Circle((0, 0), 0.5, color='lightgray', linewidth=0)
ax2.add_artist(centre2)
# -- (c) Overview of evaluation results (radar) --
ax3 = fig.add_subplot(1, 3, 3, projection='polar')
N = len(categories)
angles = np.linspace(0, 2*np.pi, N, endpoint=False).tolist()
angles += angles[:1]
...
ax3.xaxis.set_ticks(angles[:-1])
ax3.set_xticklabels([])     
ax3.grid(True, linestyle=':')

ax3.set_yticks([0.2, 0.4, 0.6, 0.8, 1.0])
ax3.set_ylim(0, 1)
...
ax3.set_title('(c) Overview of evaluation results', fontsize=12, pad=45)
ax3.legend(loc='lower center', bbox_to_anchor=(0.5, -0.5), ncol=3, fontsize=7, frameon=False)
\end{lstlisting}
\end{wideexample}
\begin{wideexample}{Level 1 Direct sample 2}\tiny

\textbf{Instruction}: You are a Python developer proficient in data visualization, with expertise in using libraries such as Matplotlib, NetworkX, Seaborn, and others.I have a plot generated by Python code, but I don't have the corresponding code that generated this plot. Your task is to generate the Python code that can perfectly reproduce the picture based on the image I provide.

Here are the requirements for the task:
1. Data Extraction: Extract the actual data from the provided image. Based on the visual features of the plot, you must infer the data and recreate the plot.
2. Recreate the Image: Generate the Matplotlib code that reproduces the image exactly as it appears, including all elements such as:
   - Plot type (scatter, line, bar, etc.)
   - Axis labels and titles
   - Colors, markers, line styles, and other visual styles
   - Any legends, annotations, or gridlines present in the image
3. Self-contained Code: The Python code should be complete, executable, and self-contained. It should not require any external data files or variables not already present in the code.
Your objective is to extract the any necessary details from the image and generate a Python script that accurately reproduces the plot.

Now, please generate the Python code to reproduce the picture below.

\textbf{Reference figure}:
\begin{center}
    \includegraphics[width=0.8\linewidth]{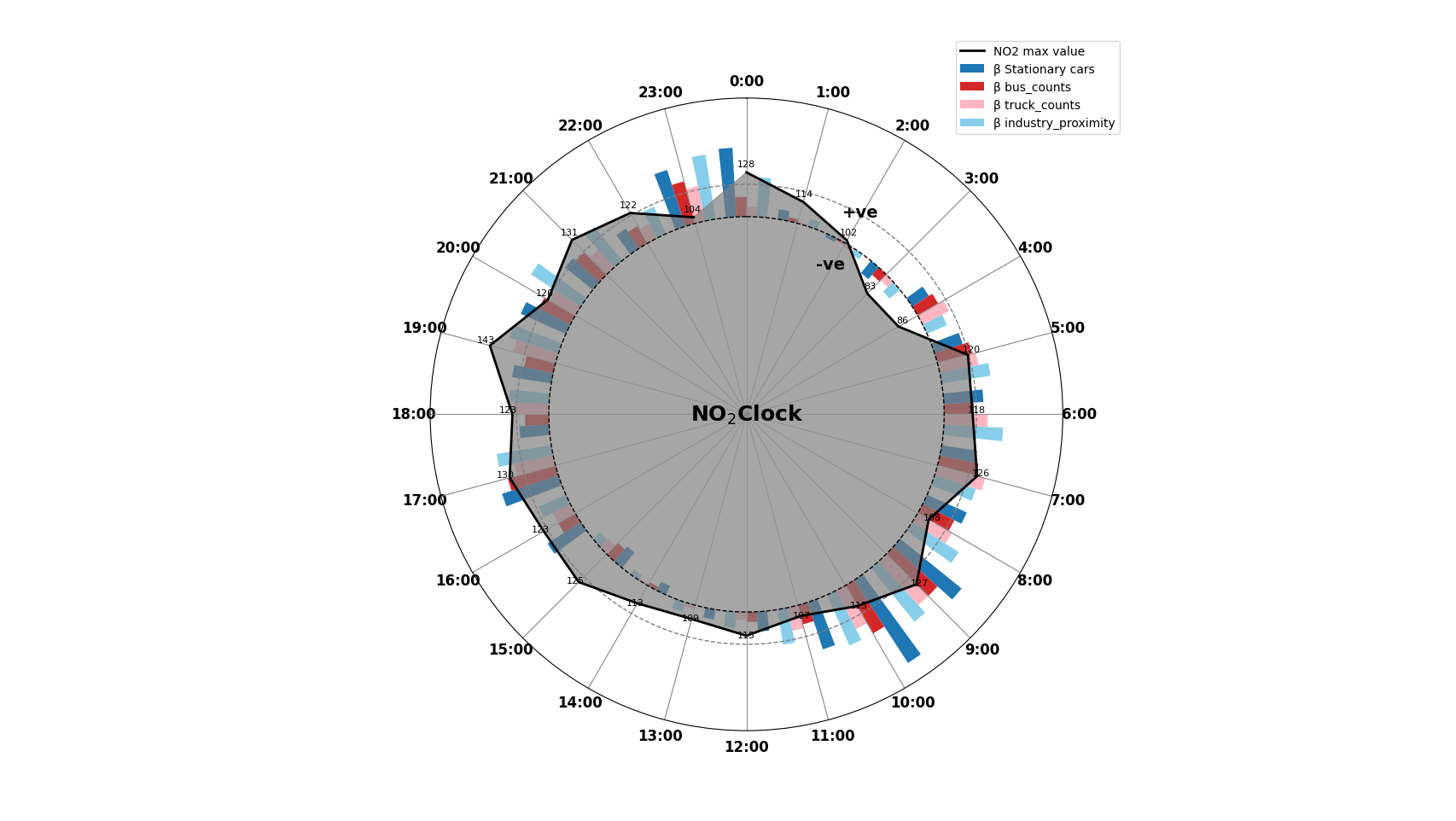}
\end{center}
\textbf{GT Code}: 
\begin{lstlisting}[
  language=Python,
  firstline=1,      % 想分段时随时改行号
  % lastline=100,   % 例如先显示前 100 行
]
import numpy as np
import matplotlib.pyplot as plt

hours = np.arange(24)
angles = 2 * np.pi * hours / 24

stationary = ...
bus_counts = np.array(...

fig = plt.figure(figsize=(10,10))
...
for th in angles:
    ax.plot([th, th], [0, 160], color='grey', linewidth=0.5)

baseline = 100
theta = np.linspace(0, 2*np.pi, 360)
ax.plot(theta, np.full_like(theta, baseline), linestyle='--', color='black', linewidth=1)
inner_circle = np.mean(no2)
ax.plot(theta, np.full_like(theta, inner_circle), linestyle='--', color='grey', linewidth=1)

ax.text(0, 0, r'NO$_2$Clock', fontsize=18, fontweight='bold', ha='center', va='center')

bar_width = 2*np.pi/24 * 0.2
offsets = np.array([-1.5, -0.5, 0.5, 1.5]) * bar_width
for vals, off, color, label in zip(
        [stationary, bus_counts, truck_counts, industry_proximity],
        offsets,
        ['tab:blue','tab:red','lightpink','skyblue'],..
    ax.bar(angles + off, vals * 100, bottom=baseline, width=bar_width, color=color, label=label)

scale = 0.8
no2_scaled = baseline + (no2 - baseline) * scale
ln, = ax.plot(angles, no2_scaled, color='black', linewidth=2, label='NO2 max value')
ax.fill(angles, no2_scaled, color='grey', alpha=0.7)

for ang, orig_val, r in zip(angles, no2, no2_scaled):
    ax.text(ang, r + 2, f'{orig_val}', ha='center', va='bottom', fontsize=8, color='black')

ax.text(np.deg2rad(30), baseline+15, '+ve', fontsize=14, fontweight='bold', ha='center')
ax.text(np.deg2rad(30), baseline-15, '-ve', fontsize=14, fontweight='bold', ha='center')

ax.legend(loc='upper right', bbox_to_anchor=(1.1,1.1), fontsize=10)

plt.show()
\end{lstlisting}
\end{wideexample}
\begin{wideexample}{Level 1 Direct sample 3}\tiny

\textbf{Instruction}: You are a Python developer proficient in data visualization, with expertise in using libraries such as Matplotlib, NetworkX, Seaborn, and others.I have a plot generated by Python code, but I don't have the corresponding code that generated this plot. Your task is to generate the Python code that can perfectly reproduce the picture based on the image I provide.

Here are the requirements for the task:
1. Data Extraction: Extract the actual data from the provided image. Based on the visual features of the plot, you must infer the data and recreate the plot.
2. Recreate the Image: Generate the Matplotlib code that reproduces the image exactly as it appears, including all elements such as:
   - Plot type (scatter, line, bar, etc.)
   - Axis labels and titles
   - Colors, markers, line styles, and other visual styles
   - Any legends, annotations, or gridlines present in the image
3. Self-contained Code: The Python code should be complete, executable, and self-contained. It should not require any external data files or variables not already present in the code.
Your objective is to extract the any necessary details from the image and generate a Python script that accurately reproduces the plot.

Now, please generate the Python code to reproduce the picture below.

\textbf{Reference figure}:
\begin{center}
    \includegraphics[width=0.8\linewidth]{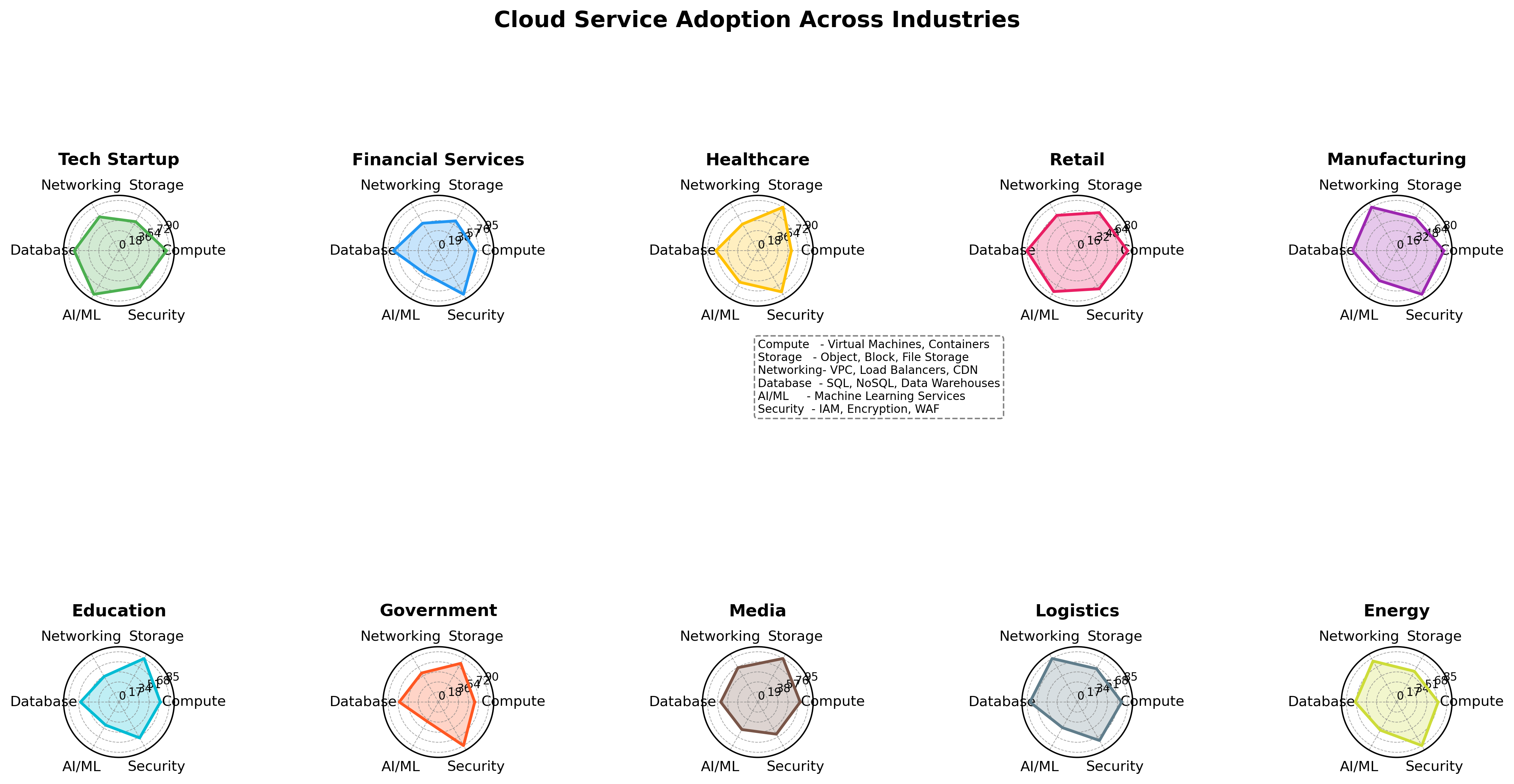}
\end{center}

\textbf{GT Code}: 
\begin{lstlisting}[
  language=Python,
  firstline=1,      % 想分段时随时改行号
  % lastline=100,   % 例如先显示前 100 行
]
import matplotlib.pyplot as plt
import numpy as np

# == New radar figure data ==

labels = ['Compute', 'Storage', 'Networking', 'Database', 'AI/ML', 'Security']
num_metrics = len(labels)
# angle of each axis in the plot (in radians)
angles = np.linspace(0, 2 * np.pi, num_metrics, endpoint=False).tolist()
# complete the loop
angles += angles[:1]

# Values for each industry's cloud service adoption (0-100 scale)
data = ...

industries = list(data.keys())

# New modern color scheme
colors = ...
# == figure plot ==

fig, axes = plt.subplots(2, 5,
                         figsize=(15.0, 9.0), # Slightly larger for readability
                         subplot_kw=dict(polar=True))
axes = axes.ravel()

for ax, name in zip(axes, industries):
    vals = data[name]
    # close the loop
    vals_loop = vals + vals[:1]
    i = industries.index(name)
    ....    
    ax.set_yticks(rticks)
    ax.set_yticklabels([f"{int(x)}" for x in rticks], fontsize=8)
    ax.set_ylim(0, max_val * 1.1) # Add a small buffer to max_val
    
    # title
    ax.set_title(name, fontsize=12, fontweight='bold', pad=10)
    
    # light grid
    ax.grid(color='gray', linestyle='--', linewidth=0.5, alpha=0.7)
    ax.spines['polar'].set_linewidth(1.0)
    ...
...
plt.tight_layout(rect=[0, 0, 1, 0.96]) # Adjust layout to make space for a potential suptitle
plt.suptitle('Cloud Service Adoption Across Industries', fontsize=16, fontweight='bold', y=0.99)
plt.savefig("./datasets_level2/radar_15.png", bbox_inches="tight", dpi=300)  # Save the figure
plt.show()
\end{lstlisting}
\end{wideexample}
\begin{wideexample}{Level 1 Customized (raw data) sample 1}\tiny

\textbf{Instruction}: I want to use a heatmap to show the variation range of each category for each month, with the horizontal axis representing time and the vertical axis representing the three categories: Energy, Metals, and Food. The color intensity represents the magnitude of the variation. Please refer to the uploaded image style to generate runnable Python code.  

\textbf{Reference figure}:
\begin{center}
    \includegraphics[width=0.5\linewidth]{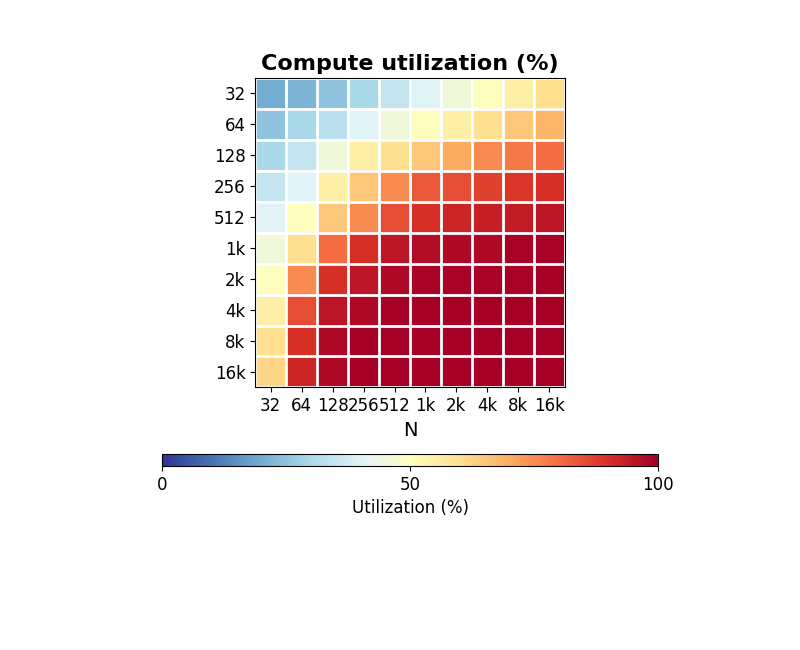}
\end{center}

\textbf{Raw data}: {
  "dates": [
    "2020-01-01",
    "2020-02-01", ...
    "2024-08-01",
    "2024-09-01"
  ],
  "commodities": [
    "Energy",
    "Metals",
    "Food"
  ],
  "values": [
    [
      4.7, ...
      -8.1
    ],
    [
      1.6, ...
      -4.7
    ],
    [
      8.8, ...
      -0.3
    ]
  ]
}

\textbf{GT Code}: 
\begin{lstlisting}[
  language=Python,
  firstline=1,      % 想分段时随时改行号
  % lastline=100,   % 例如先显示前 100 行
]
import numpy as np
import matplotlib.pyplot as plt

# Data
dates = ..
commodities = ["Energy", "Metals", "Food"]
values = ..

data = np.array(values)

# Plot
fig, ax = plt.subplots(figsize=(14, 6))
fig.subplots_adjust(bottom=0.25)

# Determine symmetric range around zero
max_abs = np.max(np.abs(data))
im = ax.imshow(data, cmap='RdYlBu_r', aspect='auto', vmin=-max_abs, vmax=max_abs)

...

# Labels and title
ax.set_xlabel('Month', fontsize=14)
ax.set_title('Monthly Commodity Price Change (%)', fontsize=16, fontweight='bold')

# Gridlines
ax.set_xticks(np.arange(data.shape[1] + 1) - 0.5, minor=True)
ax.set_yticks(np.arange(data.shape[0] + 1) - 0.5, minor=True)
ax.grid(which='minor', color='white', linestyle='-', linewidth=2)
ax.tick_params(which='minor', bottom=False, left=False)

# Colorbar
cbar = fig.colorbar(im, ax=ax, orientation='horizontal', pad=0.3, aspect=40, shrink=0.8)
cbar.set_label('Change (%)', fontsize=12)
cbar.ax.tick_params(labelsize=12)

plt.show()
\end{lstlisting}
\textbf{GT Figure}: 
\begin{center}
   \includegraphics[width=0.8\linewidth]{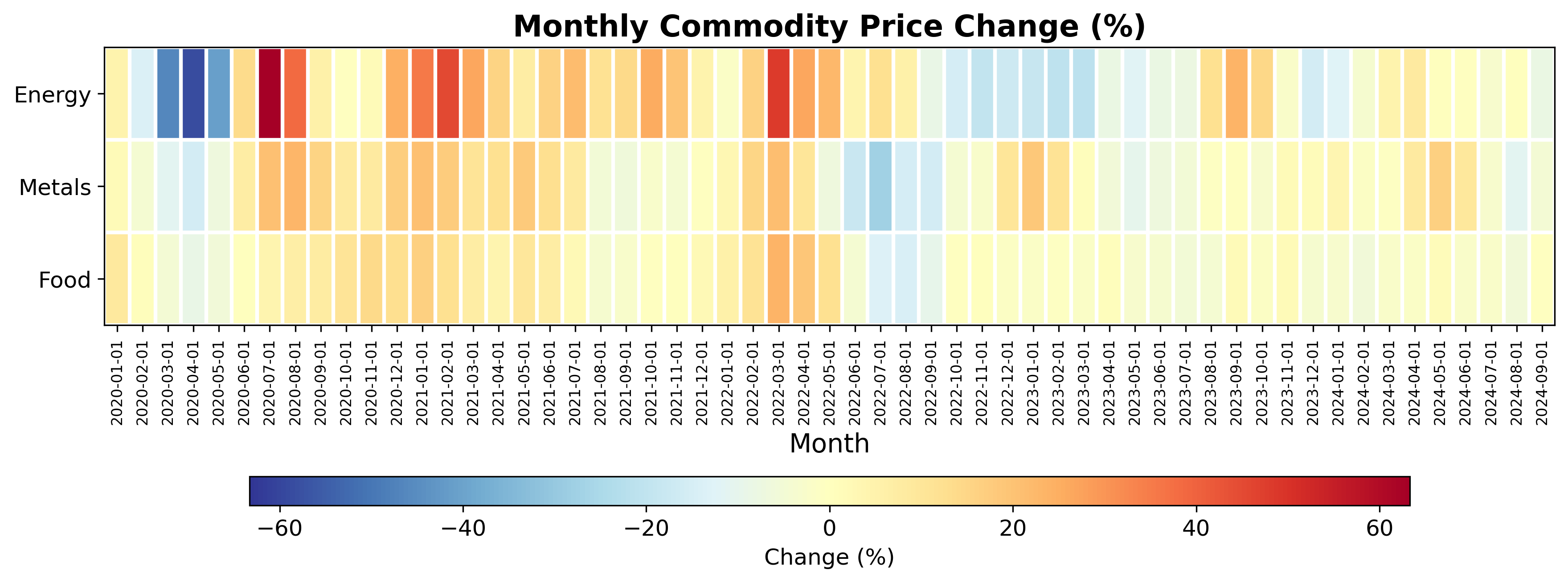} 
\end{center}

\end{wideexample}

\begin{wideexample}{Level 1 Customized (table figure) sample 1}\tiny

\textbf{Instruction}: I want to use the data from the uploaded director compensation table (PNG) and create a combination chart based on the style of the reference combination chart: the horizontal axis represents the names of the directors, the bar chart displays cash compensation, stock awards, and total compensation respectively, and a dashed line chart highlights the trends of these three items. Thank you!  
Adjust the image size to match the aspect ratio of the reference image; use the dark blue, cyan, and light gray tones from the reference image; for the x-axis labels, tilt them 45 degrees and align them to the right, mimicking the text style of the reference image; add a title centered at the top, with font effects similar to the reference image; set the y-axis scale range and intervals according to the reference image; keep the legend position consistent with the reference image, arranged horizontally at the top; apply dashed line styles as in the reference image, and mimic the marker shapes from the reference image.

\textbf{Reference figure}:

\begin{center}
\includegraphics[width=0.5\linewidth]{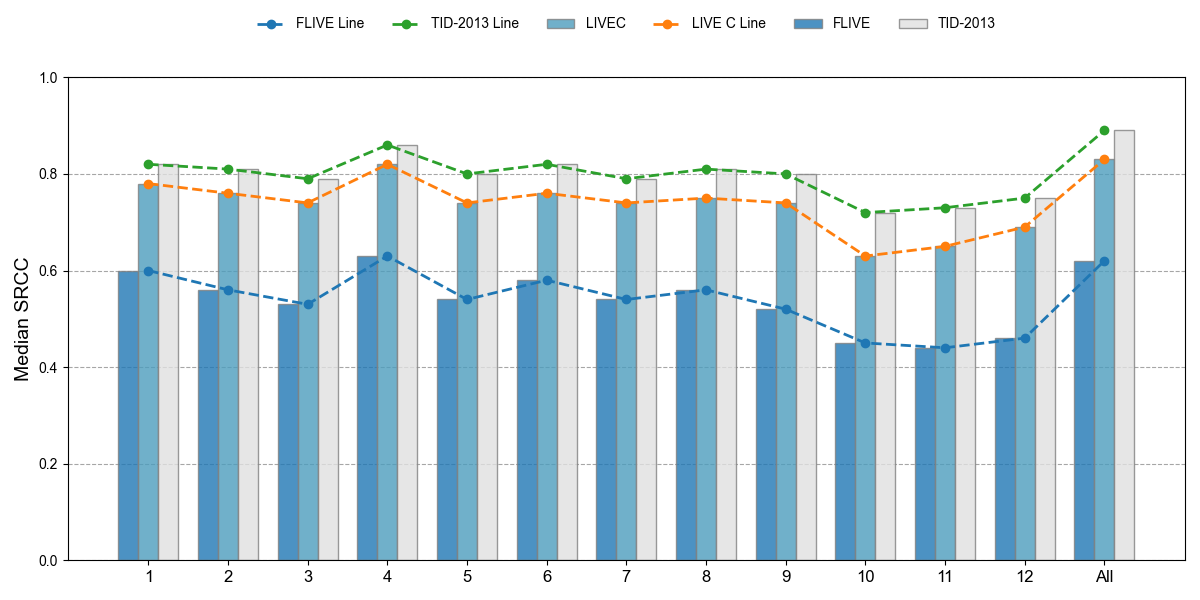}
\end{center}

\textbf{Data figure}:

\begin{center}
\includegraphics[width=0.5\linewidth]{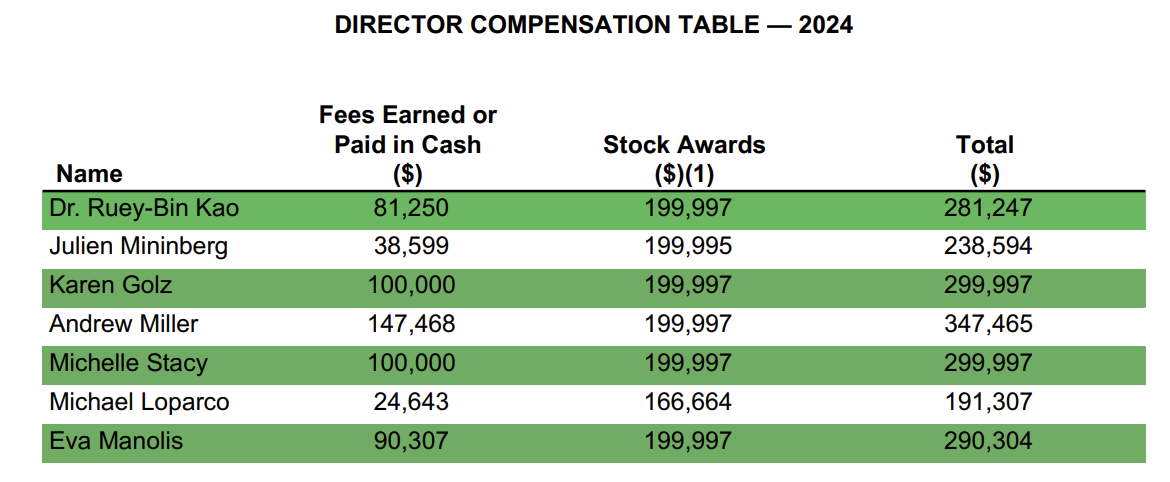}
\end{center}

\textbf{GT Code}: 
\begin{lstlisting}[
  language=Python,
  firstline=1,      % 想分段时随时改行号
  % lastline=100,   % 例如先显示前 100 行
]
import numpy as np
import matplotlib.pyplot as plt

plt.rcParams.update({
    'font.family': 'sans-serif',
    'font.sans-serif': ['Arial']
})

names = ['Dr. Ruey-Bin Kao', 'Julien Mininberg', ..., 'Eva Manolis']
fees = [81250, 38599, ..., 90307]
stock_awards = [199997, 199995, ..., 199997]
total = [281247, 238594, ..., 290304]

x = np.arange(len(names))
fig, ax = plt.subplots(figsize=(12, 6))

ax.grid(axis='y', linestyle='--', alpha=0.7)
ax.bar(x - 0.25, fees, 0.25, label='Fees Earned', color='#1f77b4', alpha=0.8)
ax.bar(x, stock_awards, 0.25, label='Stock Awards', color='#4c9dbd', alpha=0.8)
ax.bar(x + 0.25, total, 0.25, label='Total Compensation', color='#e0e0e0', alpha=0.8)

ax.plot(x, fees, '--o', color='#1f77b4', label='Fees Line')
ax.plot(x, stock_awards, '--o', color='#ff7f0e', label='Stock Awards Line')
ax.plot(x, total, '--o', color='#2ca02c', label='Total Line')
ax.set_xticks(x)
ax.set_xticklabels(names, rotation=45, ha='right')
ax.set_ylabel('Compensation ($)')

handles, labels = ax.get_legend_handles_labels()
ax.legend(handles, labels, loc='upper center', bbox_to_anchor=(0.5, 1.15), ncol=3)

plt.tight_layout()
plt.show()
\end{lstlisting}
\textbf{GT Figure}: 
\begin{center}
   \includegraphics[width=0.5\linewidth]{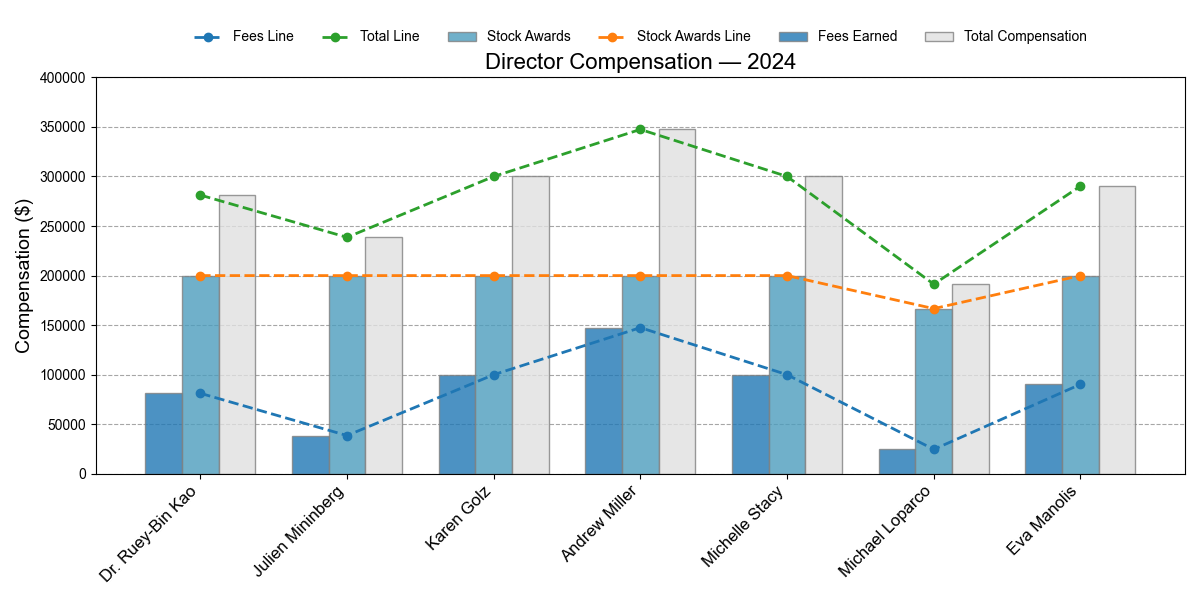} 
\end{center}
\end{wideexample}
\begin{wideexample}{Level 2 sample 1}\tiny

\textbf{Instruction}: Use GridSpec to create a complex 1+2 layout. The top section will feature a large subplot (spanning the entire width) to display "raincloud plots" (half-violin plots + box plots + scatter plots) for all four categories... enabling an in-depth comparison of these two distinctly different distributions.

On this basis:
- Set the overall canvas size to 14 inches wide × 10 inches high.
- Continue using four fixed colors: light orange `\#FFC0A0`, light green `\#B0E0B0`, light purple `\#B9A0E0`, and beige `\#FFE4C4`. Use a red line to mark the mean value in the histograms.
- Use a GridSpec layout with two rows and two columns. The first row spans both columns for the top plot, while the second row places the two histograms side by side, one in each column. The row height ratio should be explicitly set to 2:1. ....
- Rotate the X-axis tick labels of the top subplot counterclockwise by 20 degrees.
- Maintain a white background and gray grid lines (`\#D3D3D3`).

\textbf{Reference figure}:

\begin{center}
\includegraphics[width=0.5\linewidth]{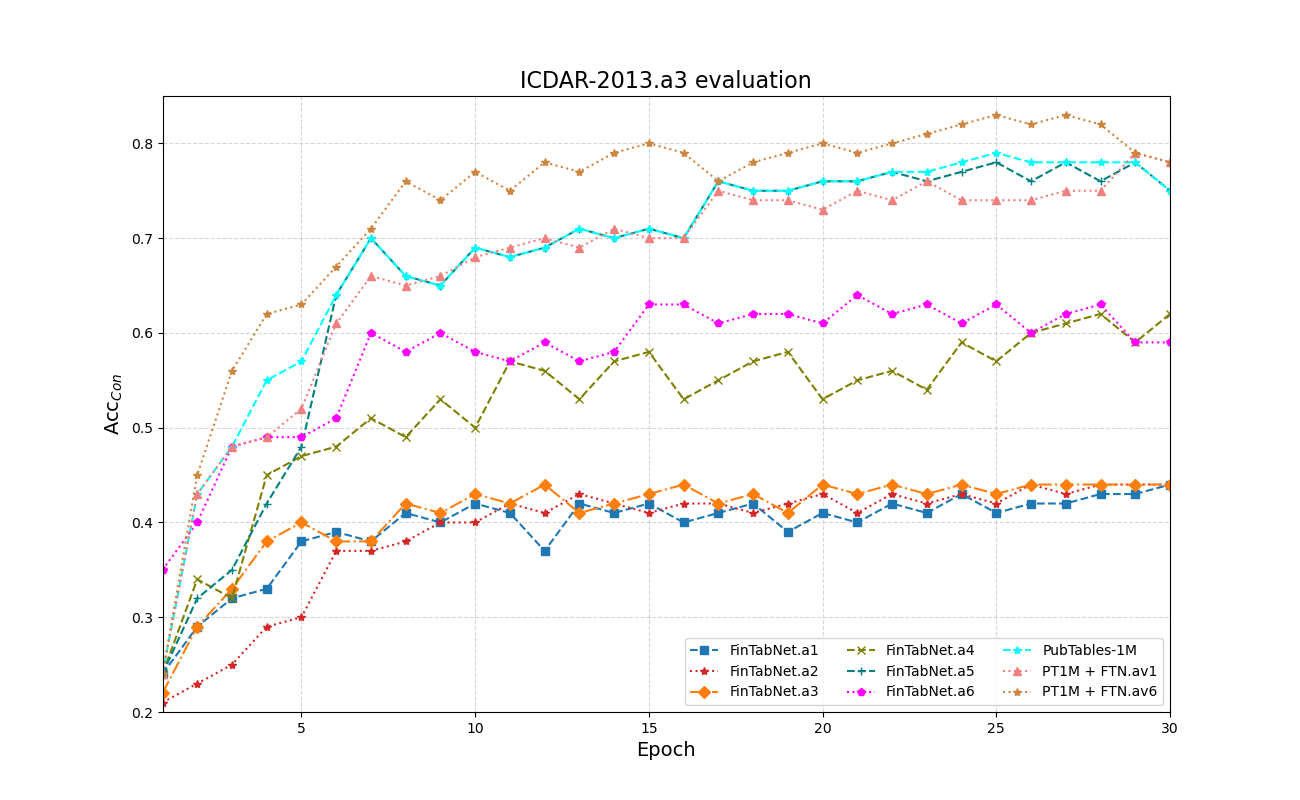}

\end{center}

\textbf{GT Code}: 
\begin{lstlisting}[
  language=Python,
  firstline=1,      % 想分段时随时改行号
  % lastline=100,   % 例如先显示前 100 行
]
# == line_19 figure code ==
import matplotlib.pyplot as plt
import numpy as np
import matplotlib.gridspec as gridspec

# == line_19 figure data ==
epochs = np.arange(1, 31)

# FinTabNet variants
ftn_a1 = np.array([...
...

# == Data Processing for Dashboard ==
# 1. Group data
fintabnet_group_data = np.array([ftn_a1, ftn_a2, ftn_a3, ftn_a4, ftn_a5, ftn_a6])
pt1m_based_group_data = np.array([pubtables, pt1m_av1, pt1m_av6])
all_models_data = np.vstack([fintabnet_group_data, pt1m_based_group_data])
all_models_labels = ['FTN.a1', 'FTN.a2', 'FTN.a3', 'FTN.a4', 'FTN.a5', 'FTN.a6', 'PubTables', 'PT1M+FTN.av1', 'PT1M+FTN.av6']
...
# 3. Final performance data
...
# 4. Significant surpass point
diff = pt1m_av6 - pubtables
surpass_margin = 0.05
surpass_epoch_idx = np.where(diff > surpass_margin)[0]
first_surpass_epoch = epochs[surpass_epoch_idx[0]] if len(surpass_epoch_idx) > 0 else None

...
# Plot 3: Key Model Showdown

...
plt.tight_layout(rect=[0, 0.03, 1, 0.95])
# plt.savefig("./datasets/line_19.png")
plt.show()
\end{lstlisting}

\textbf{GT figure}:

\begin{center}
\includegraphics[width=0.8\linewidth]{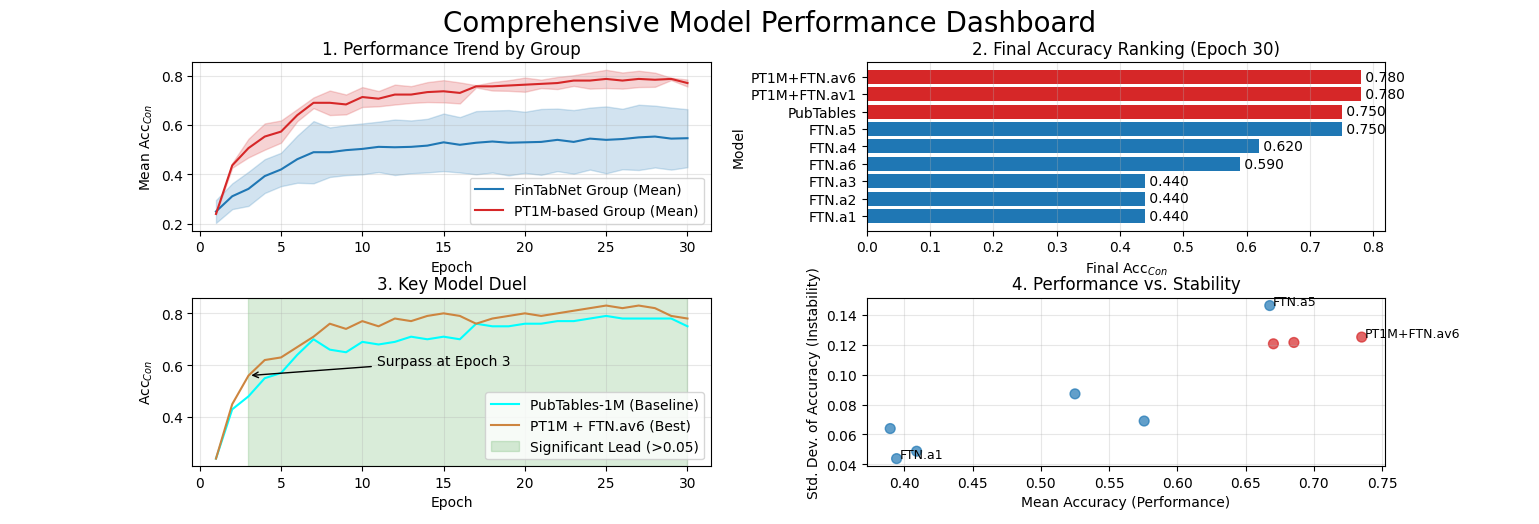}
\end{center}

\end{wideexample}

\begin{wideexample}{Level 2 sample 2}\tiny

\textbf{Instruction}: Create a comprehensive, dashboard-style analytical view that juxtaposes raw data trends, statistical distributions, and localized details.

1. Layout Modifications: Use `GridSpec` to create a complex 2x2 grid layout.  
     The top-left main plot (spanning the 1st row and 1st column) is a composite chart (three CCA lines + CKA bar chart).  
     The top-right subplot (spanning the 1st row and 2nd column) is a box plot, used to display the overall data distribution of four data series (\texttt{cca\_top1}, \texttt{cca\_top3}, \texttt{cca\_top10}, \texttt{cka}).
     The large bottom plot (spanning the 2nd row and all columns) is a "zoomed-in" view of the main plot, specifically focusing on the "Center Layer" in the range of 10 to 20 for the CCA line chart details.  

2. Chart Type Conversion and Combination:
     In the top-right subplot, create a box plot for each of the four datasets and set appropriate labels.  
     In the bottom zoomed-in plot, only draw the three CCA line charts and omit the CKA bar chart to emphasize the localized CCA dynamics. ...

Additional Requirements:  
– Set the canvas size to 15×10 inches.  
– Use a 2×2 `GridSpec` layout with width ratios `[2,1]` and height ratios `[1,1]`. The top-left main plot occupies the 1st row and 1st column, the top-right box plot occupies the 1st row and 2nd column, and the bottom zoomed-in plot spans the 2nd row across all columns...
– For the box plots, use a fill color of `\#d3d3d3`, black borders, and red median lines.  
– For the zoomed-in region rectangle, use a gray fill with transparency 0.2, a red dashed border, and red dashed connecting lines.

\textbf{Reference figure}:

\begin{center}
\includegraphics[width=0.5\linewidth]{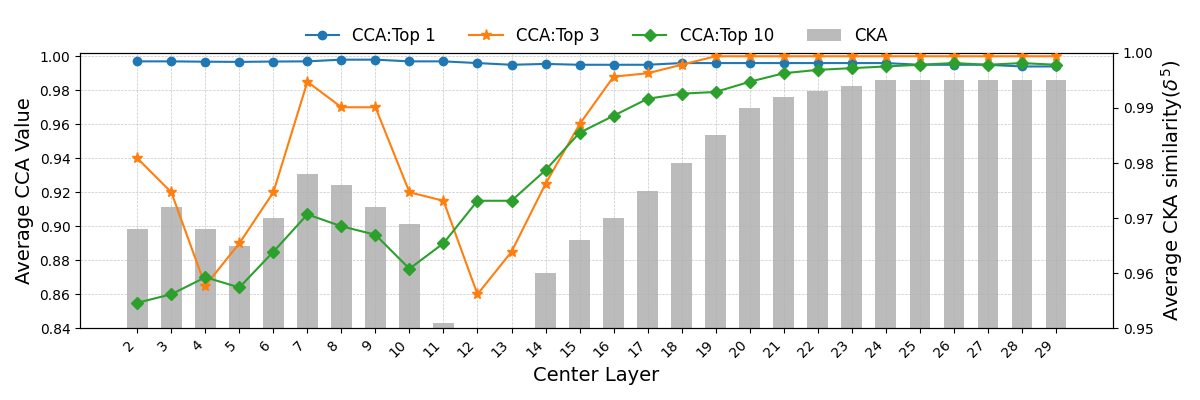}
\end{center}

\textbf{GT Code}: 
\begin{lstlisting}[
  language=Python,
  firstline=1,      % 想分段时随时改行号
  % lastline=100,   % 例如先显示前 100 行
]
import matplotlib.pyplot as plt
import matplotlib.gridspec as gridspec
from matplotlib.patches import Rectangle, ConnectionPatch
import numpy as np

layers = list(range(2, 30))
cca_top1 = [0.997, 0.997, ...
# Create figure with constrained layout
fig = plt.figure(figsize=(15, 10), constrained_layout=True)
gs = gridspec.GridSpec(2, 2, width_ratios=[2, 1], height_ratios=[1, 1], figure=fig)
...
ax_main_twin = ax_main.twinx()

# --- Main Plot (Top-Left) ---
bar_width = 0.6
...
labels = [h.get_label() for h in handles]
ax_main.legend(handles, labels, loc='lower center', ncol=4, fontsize=10, bbox_to_anchor=(0.5, -0.25))

# --- Box Plot (Top-Right) ---
data_for_boxplot = [cca_top1, cca_top3, cca_top10, cka]
box_labels = ['CCA:Top 1', 'CCA:Top 3', 'CCA:Top 10', 'CKA']
...
ax_box.grid(True, axis='y', linestyle='--', linewidth=0.5, alpha=0.7)

# --- Zoomed Plot (Bottom) ---
zoom_range = (10, 20)
ax_zoom.plot(layers, cca_top1, color='#1f77b4', marker='o', markersize=6, lw=1.5)
...
ax_zoom.grid(True, linestyle='--', linewidth=0.5, alpha=0.7)

# --- Visual Connection ---
rect = Rectangle((zoom_range[0], 0.84), zoom_range[1] - zoom_range[0], 1.002 - 0.84,
                 facecolor='grey', alpha=0.2, edgecolor='red', linestyle='--')
...
fig.add_artist(con2)

plt.show()
\end{lstlisting}

\textbf{GT figure}:

\begin{center}
\includegraphics[width=0.5\linewidth]{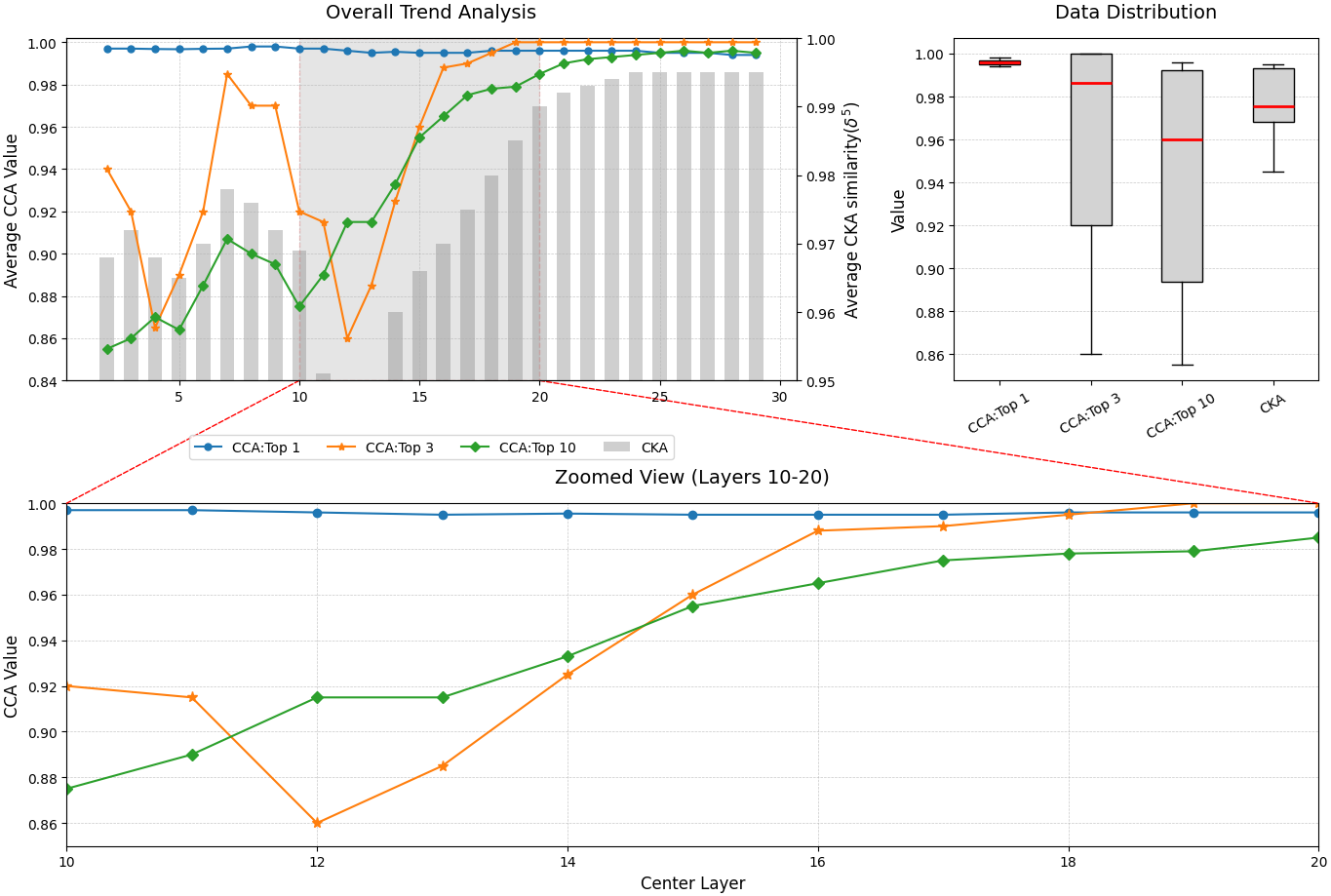}
\end{center}

\end{wideexample}
\begin{wideexample}{Level 2 sample 3}\tiny
\textbf{Instruction}: Create a comprehensive, dashboard-style multi-panel analysis plot to deeply explore the relationships between model performance, tool wear growth, and model comparisons. The specific requirements are as follows:

\textbf{Reference figure}:
\begin{center}
    \includegraphics[width=0.4\linewidth]{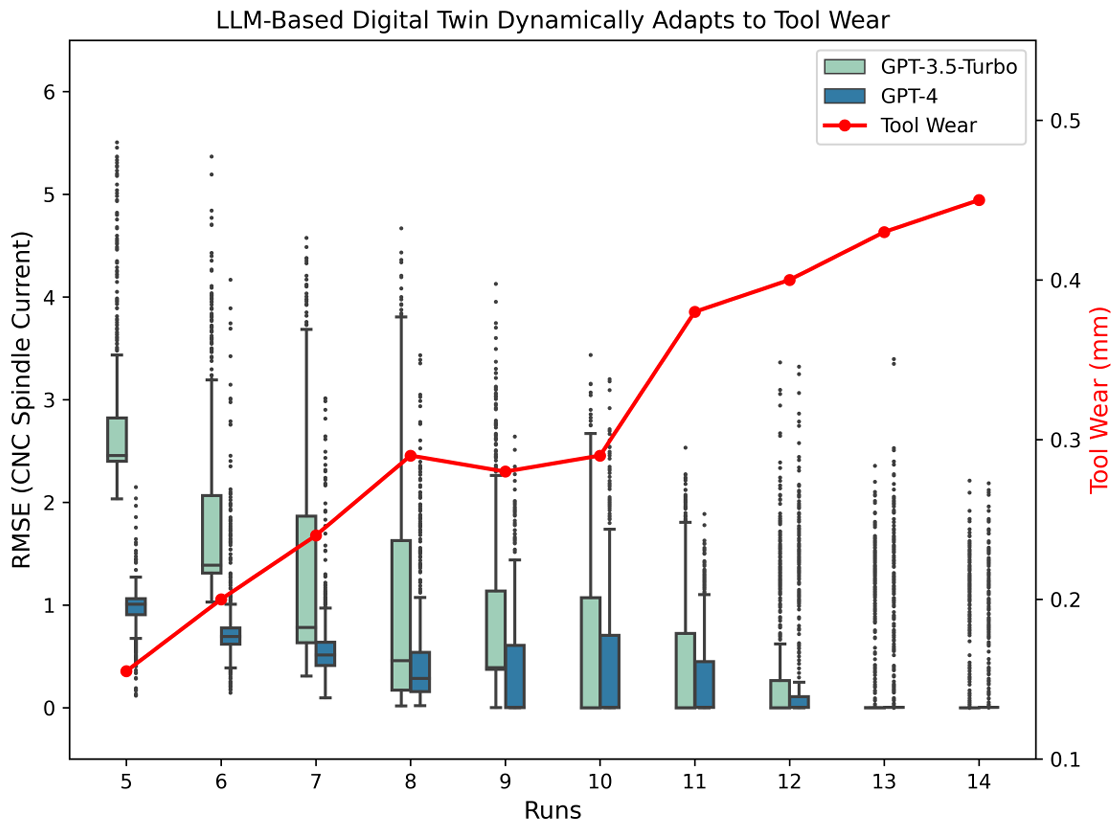}
\end{center}

\textbf{GT Code}: 
\begin{lstlisting}[
  language=Python,
  firstline=1,      % 想分段时随时改行号
  % lastline=100,   % 例如先显示前 100 行
]
import numpy as np
import matplotlib.pyplot as plt
import matplotlib.patches as mpatches
import matplotlib.gridspec as gridspec

np.random.seed(0)
runs = np.arange(5, 15)
mean35 = [2.5, ..
std35  = [0.5,...

fig = plt.figure(figsize=(18, 10))
gs = gridspec.GridSpec(2, 2, width_ratios=[3, 2], height_ratios=[1, 1])


pos1 = runs - 0.2
pos2 = runs + 0.2
..
vp1 = ax_main.violinplot(data35, positions=pos1, widths=0.4, showmedians=True)
ax_main.grid(True, linestyle="--", alpha=0.6)
ax_main.set_title("A) Model Performance Distribution vs. Tool Wear", fontsize=16, loc='left')

ax_wear = ax_main.twinx()
ax_wear.plot(runs, tool_wear, color="red", marker="o", markersize=6, linewidth=2)
ax_wear.set_ylabel("Tool Wear (mm)", color="red", fontsize=14)
...
ax_growth.grid(axis='y', linestyle='--', alpha=0.6)

median35 = [np.median(d) for d in data35]
median4 = [np.median(d) for d in data4]

highlight_run_idx = max_growth_idx + 1 
ax_compare.scatter(median35, median4, c=runs, cmap='viridis', s=60, alpha=0.8)
...
ax_compare.grid(True, linestyle='--', alpha=0.6)
ax_compare.text(0.95, 0.05, 'GPT-4 Better', transform=ax_compare.transAxes,
                ha='right', va='bottom', fontsize=12, color='green', style='italic')
...
ax_main.annotate('Max Wear Growth', xy=(max_growth_run, 4.0), xytext=(max_growth_run, 5.0),
                 arrowprops=dict(facecolor='#e31a1c', shrink=0.05, width=1.5, headwidth=8),
                 fontsize=12, color='#e31a1c', ha='center', bbox=dict(boxstyle="round,pad=0.3", fc="white", ec="#e31a1c", lw=1))

fig.suptitle("Comprehensive Analysis of LLM-based Digital Twin Performance", fontsize=20, y=0.98)
plt.tight_layout(rect=[0, 0, 1, 0.95])
plt.show()
\end{lstlisting}

\textbf{GT figure}:
\begin{center}
    \includegraphics[width=0.5\linewidth]{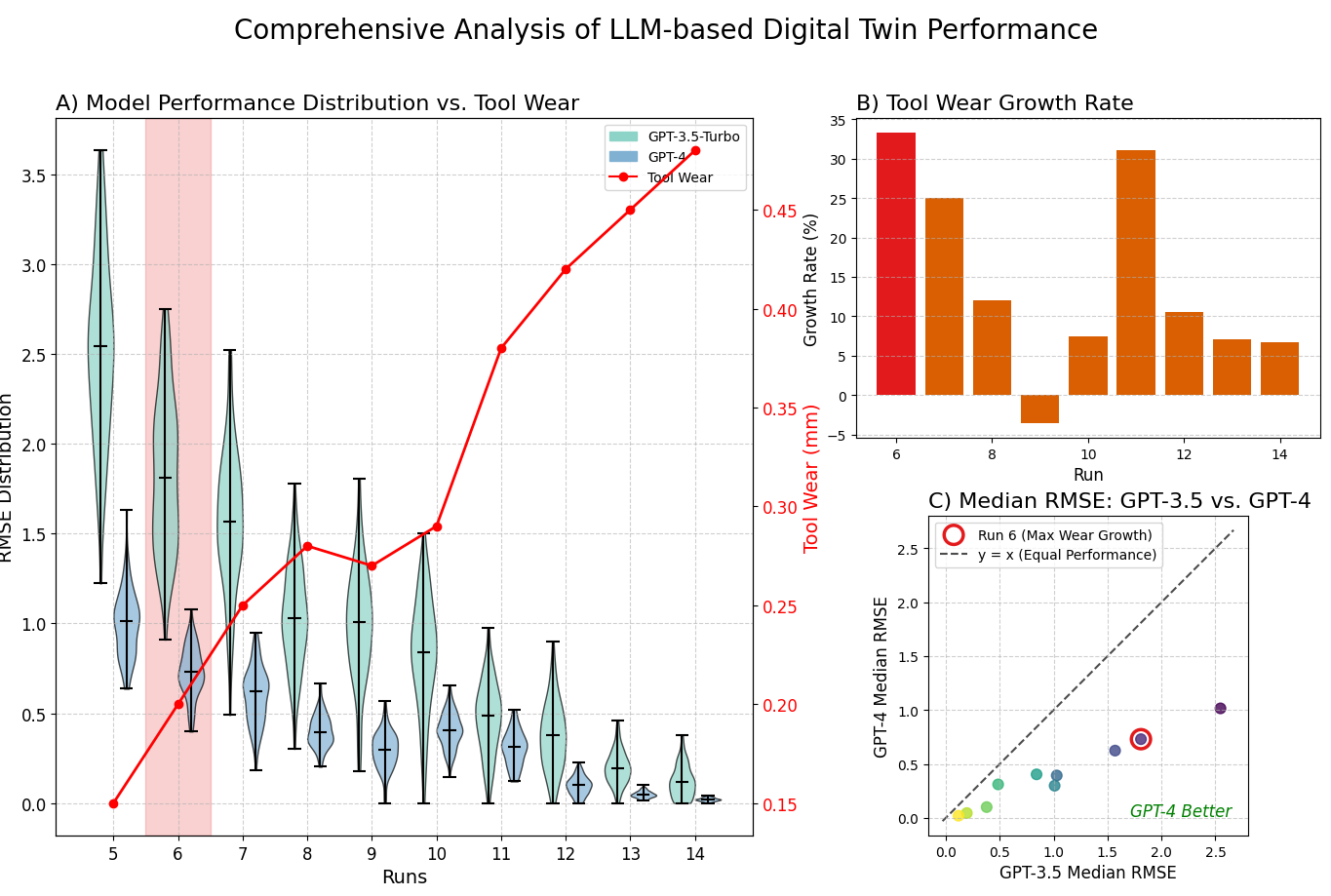}
\end{center}

\end{wideexample}
\begin{wideexample}{Level 2 sample 4}\tiny

\textbf{Instruction}:

1. Use `GridSpec` to create a complex dashboard-style layout:
   - The left side contains a main plot occupying a 2x2 space.
   - The right side contains two subplots, each occupying a 1x1 space.

2. **Main Plot (Left Side)**:
   - Retain the original bar chart and exponential trend line.
   - Display the absolute values and trends of the annual research count.

3. **Top-Right Subplot**:
   - Convert the original data into an area chart.
   - Show the cumulative total of research counts to analyze the expansion of overall scale.

4. **Bottom-Right Subplot**:
   - Use a donut chart to display the proportion of research counts from the last three years (2022–2024) relative to their total.
   - Highlight the distribution of recent contributions.

5. Add titles to all subplots and ensure a unified visual style for clear communication and coordinated layout.

**Additional Modifications**:
- Adjust the overall canvas size to 16 inches × 9 inches.
- Configure the layout as `GridSpec(2,3)`:
  - The main plot occupies the first and second columns of all rows.
  - The top-right subplot is placed in the first row, third column.
  - The bottom-right subplot is placed in the second row, third column.
- **Styling**:
  - Main plot bar color: `'\#1a5276'`.
  - Main plot trend line color: `red`.
  - Area chart fill color: `'\#5dade2'`, line color: `'\#1a5276'`.
  - Donut chart colors: `['\#1abc9c', '\#f1c40f', '\#e74c3c']`.
  - Donut chart percentage text: white and bold.
  - Overall title font: size 22, bold.
  - Subplot titles font: size 16.
  - Axis titles font: size 14.
  - Tick labels font: size 12.
  - Top-right chart annotations font: size 12, bold.
  - Donut chart center text font: size 14, bold.
  - Pie chart percentage text font: size 8, bold.

\textbf{Reference figure}:
\begin{center}
    \includegraphics[width=0.5\linewidth]{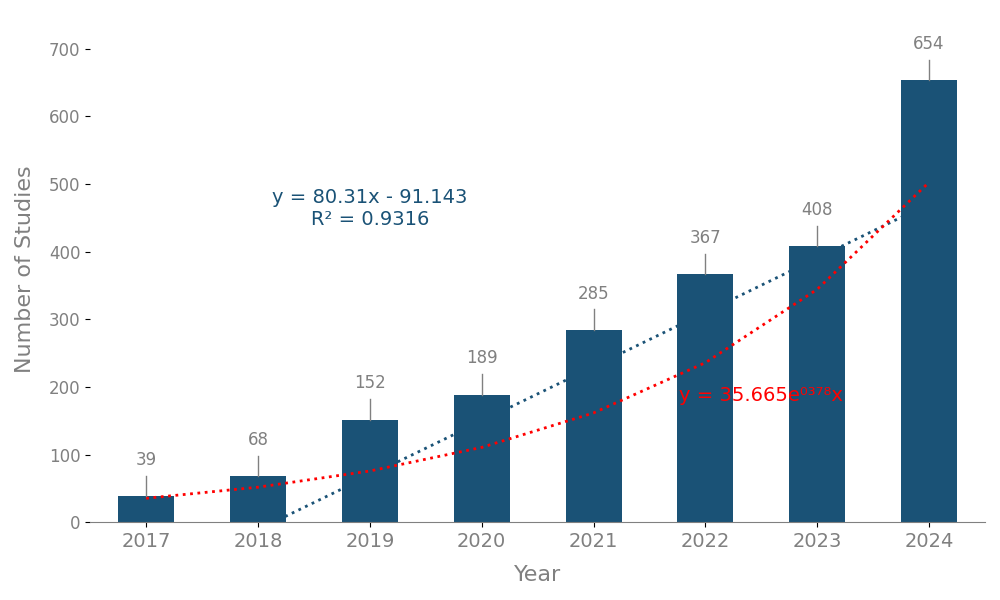}

\end{center}

\textbf{GT Code}: 
\begin{lstlisting}[
  language=Python,
  firstline=1,      % 想分段时随时改行号
  % lastline=100,   % 例如先显示前 100 行
]
import numpy as np
import matplotlib.pyplot as plt
import matplotlib.gridspec as gridspec

years = np.array([2017, 2018, 2019, 2020, 2021, 2022, 2023, 2024])
x = np.arange(len(years))
...
gs = gridspec.GridSpec(2, 3, figure=fig)
ax1 = fig.add_subplot(gs[:, 0:2])
...
ax1.set_xlabel('Year', fontsize=14, color='grey')
ax1.set_ylabel('Number of Studies', fontsize=14, color='grey')
...
for spine in ['top', 'right']:
    ax2.spines[spine].set_visible(False)
ax2.grid(axis='y', linestyle='--', alpha=0.7)
ax2.text(years[-1], cumulative_y[-1], f' Total:\n {cumulative_y[-1]}', ha='right', va='top', fontsize=12, fontweight='bold')


colors = ['#1abc9c', '#f1c40f', '#e74c3c']
wedges, texts, autotexts = ax3.pie(last_3_years_data,...
ax3.add_artist(centre_circle)
ax3.set_title('Contribution in Last 3 Years', fontsize=16, pad=10)
ax3.text(0, 0, f'Total:\n{sum(last_3_years_data)}', ha='center', va='center', fontsize=14, fontweight='bold')
plt.setp(autotexts, size=8, weight="bold", color="white")

plt.tight_layout(rect=[0, 0, 1, 0.95])
plt.show()
\end{lstlisting}

\textbf{GT figure}:
\begin{center}
    \includegraphics[width=0.5\linewidth]{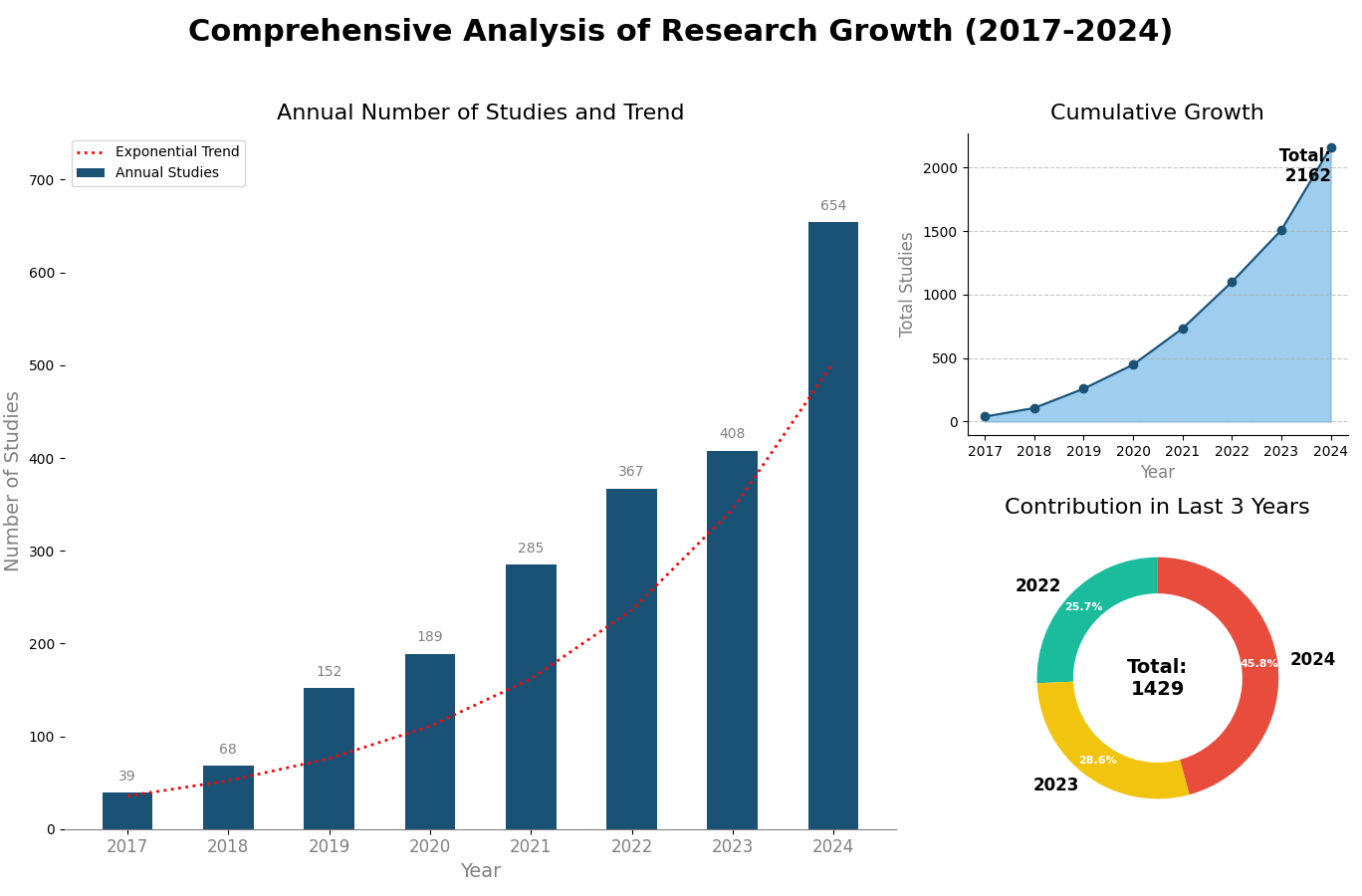}

\end{center}

\end{wideexample}

\begin{wideexample}{Level 2 sample 5}\tiny

\textbf{Instruction}: 

Create a 2x2 dashboard to comprehensively compare model performance.  

1. **Top-left plot (Performance Trend Comparison):** Divide the models into two groups: 'FinTabNet' and 'PT1M-based'. ...  

2. **Top-right plot (Final Performance Ranking):** Use a horizontal bar chart to show the final accuracy of all 9 models at the last epoch..y.

3. **Bottom-left plot (Key Model Showdown):** Plot the performance curves of the best model `pt1m\_av6` and the baseline model `pubtables` separately. Identify the epoch where `pt1m\_av6` first surpasses `pubtables` by more than 0.05 in accuracy, and use `axvspan` to highlight the region from ...

4. **Bottom-right plot (Performance vs. Stability):** Create a scatter plot where the X-axis represents the average accuracy of each model (mean over 30 epochs), and the Y-axis represents the standard deviation of accuracy. This plot evaluates whether high performance is accompanied by high instability. Add text labels to the best-performing, most stable, and most unstable models on the plot.  

— Additional Modifications:  
- Set the overall canvas size to 16×12 inches.  
- Use a 2-row, 2-column `GridSpec` layout with row spacing of 0.4 and column spacing of 0.3.  
- Use a bold font size of 20 for the main title, regular font size of 12 for subplot titles, axis labels, and tick marks, and font size of 10 for legends... and semi-transparency. Use font size 9 for labels and adjust them horizontally by 0.002.  
- Use dashed grid lines with approximately 30


\textbf{Reference figure}:
\begin{center}
    \includegraphics[width=0.5\linewidth]{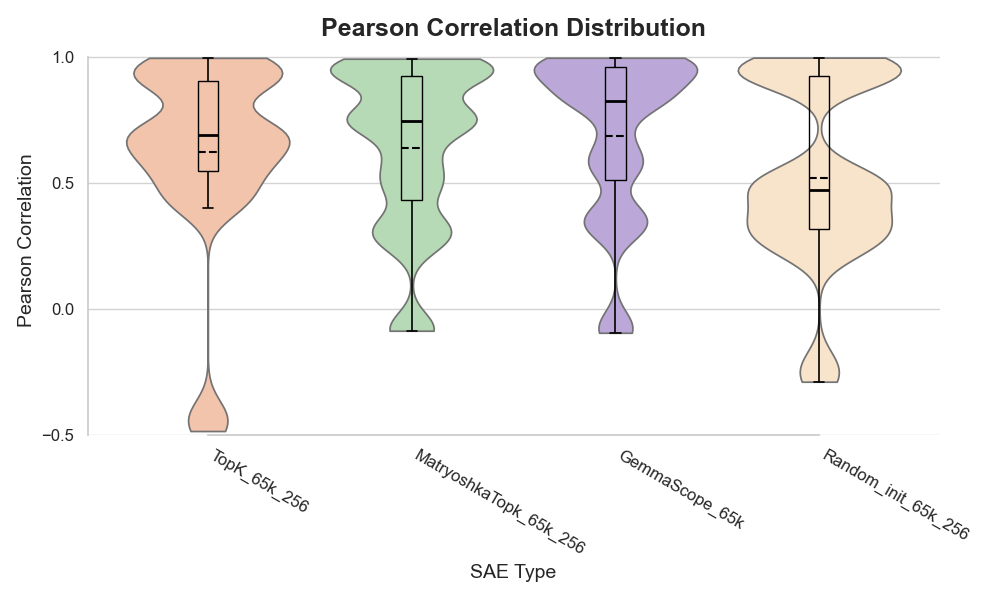}

\end{center}

\textbf{GT Code}: 
\begin{lstlisting}[
  language=Python,
  firstline=1,      % 想分段时随时改行号
  % lastline=100,   % 例如先显示前 100 行
]
import pandas as pd
import matplotlib.pyplot as plt
import seaborn as sns
import matplotlib.gridspec as gridspec

data = {
    "TopK_65k_256": [-0.4625, -0.4049, ...
clean_data = {k: [x for x in v if x is not None] for k, v in data.items()}
...
fig.suptitle("Comprehensive Analysis of Pearson Correlation", fontsize=20, fontweight='bold')

# --- Top Plot: Raincloud Plot ---
order = ["TopK_65k_256", "MatryoshkaTopk_65k_256", "GemmaScope_65k", "Random_init_65k_256"]
colors = ["#FFC0A0", "#B0E0B0", "#B9A0E0", "#FFE4C4"]
...
# Jittered points
sns.stripplot(x="SAE Type", y="Pearson Correlation", data=df, order=order, ax=ax_main,..

# Boxplot
sns.boxplot(x="SAE Type", y="Pearson Correlation", data=df, order=order, ax=ax_main,
...
ax_main.tick_params(axis='x', labelsize=12, labelrotation=-20)
ax_main.tick_params(axis='y', labelsize=12)

# --- Bottom-Right Plot: Histogram for Random_init ---
random_data = df[df["SAE Type"] == "Random_init_65k_256"]["Pearson Correlation"]
...
sns.despine(fig=fig)
plt.tight_layout(rect=[0, 0, 1, 0.96])
plt.show()
\end{lstlisting}

\textbf{GT figure}:
\begin{center}
    \includegraphics[width=0.5\linewidth]{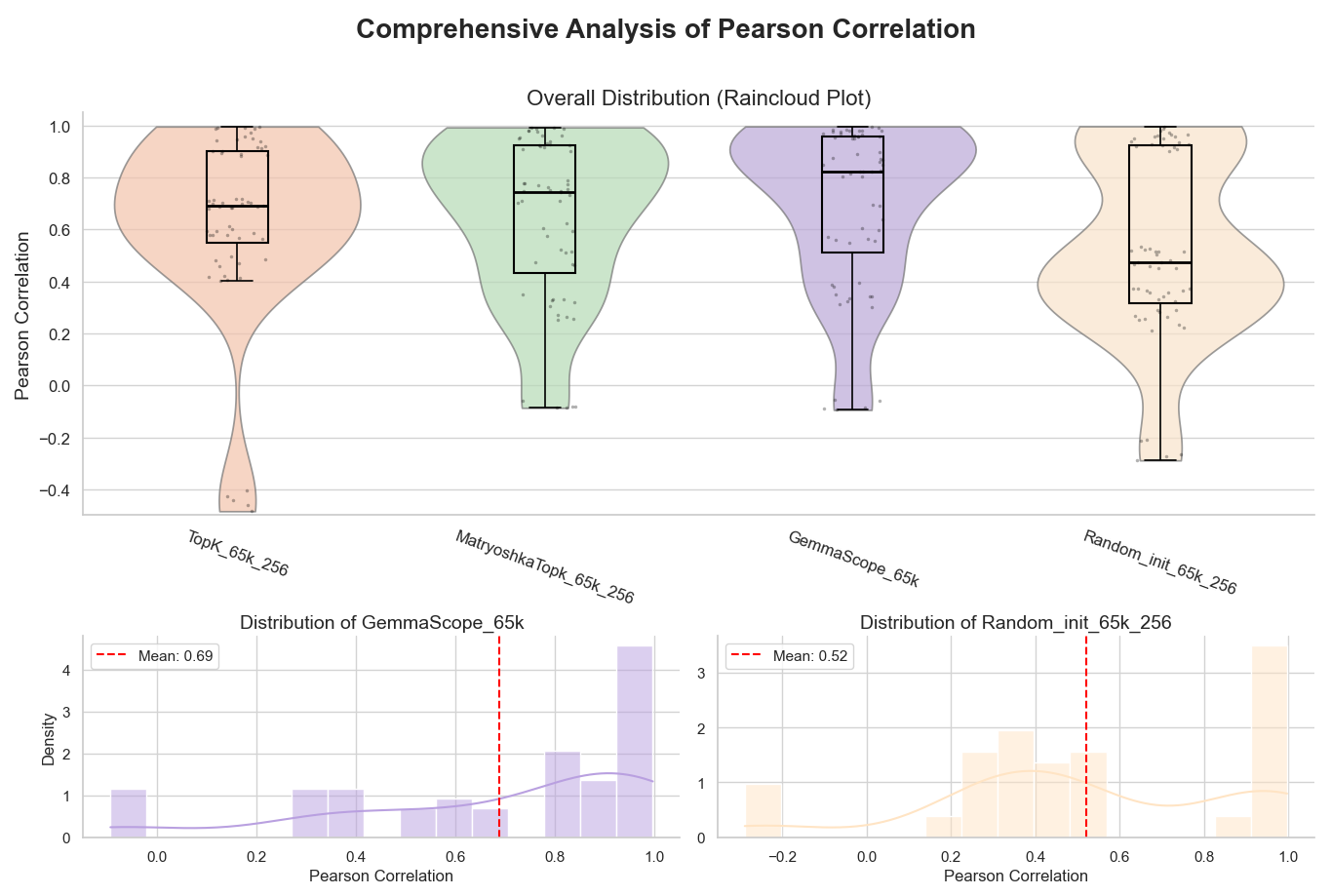}
\end{center}

\end{wideexample}

\begin{wideexample}{Level 3 sample 1}\tiny

\textbf{Instruction}: I have an Excel spreadsheet to analyze, which contains fuel types and corresponding horsepower values. Please generate a plotting code based on the style of the grouped box plot I uploaded to display the horsepower distribution for different fuel types. Use a canvas size precisely 13 inches wide and 8 inches high, with the color scheme set to Set3. The entire chart should contain only one subplot, without complex layouts like GridSpec. The title should be "Horsepower by Fuel Type," the X-axis label should be "Fuel Type," and the Y-axis label should be "Horsepower (hp)." Keep all text at Matplotlib's default font size and style; rotate the X-axis tick labels 45 degrees; finally, apply a tight layout to ensure there is no excess whitespace between elements.

\textbf{Reference Figure}: 

\begin{center}
   \includegraphics[width=0.6\linewidth]{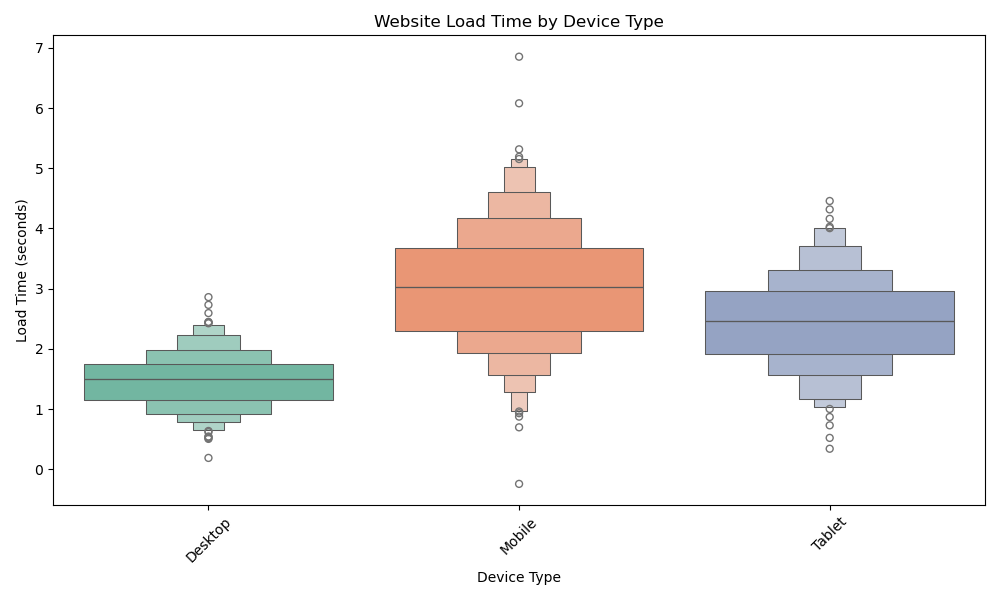} 
\end{center}

\textbf{GT Code}: 
\begin{lstlisting}[
  language=Python,
  firstline=1,      % 想分段时随时改行号
  % lastline=100,   % 例如先显示前 100 行
]
import matplotlib.pyplot as plt
import seaborn as sns
import pandas as pd

data_x_groups = ['plug in hyrbrid', 'Petrol', 'Diesel', 'Hybrid', ...]
data_y_values = [963.0, 563.0, 381.0, 1160.0, ...]

df = pd.DataFrame({
    'fuel_types': data_x_groups,
    'horsepower_num': data_y_values
})

plt.figure(figsize=(13, 8))

sns.boxenplot(data=df, x='fuel_types', y='horsepower_num', palette='Set3')

plt.title('Horsepower by Fuel Type')
plt.xlabel('Fuel Type')
plt.ylabel('Horsepower (hp)')
plt.xticks(rotation=45)
plt.tight_layout()
plt.show()
\end{lstlisting}

\textbf{GT Figure}: 
\begin{center}
   \includegraphics[width=0.6\linewidth]{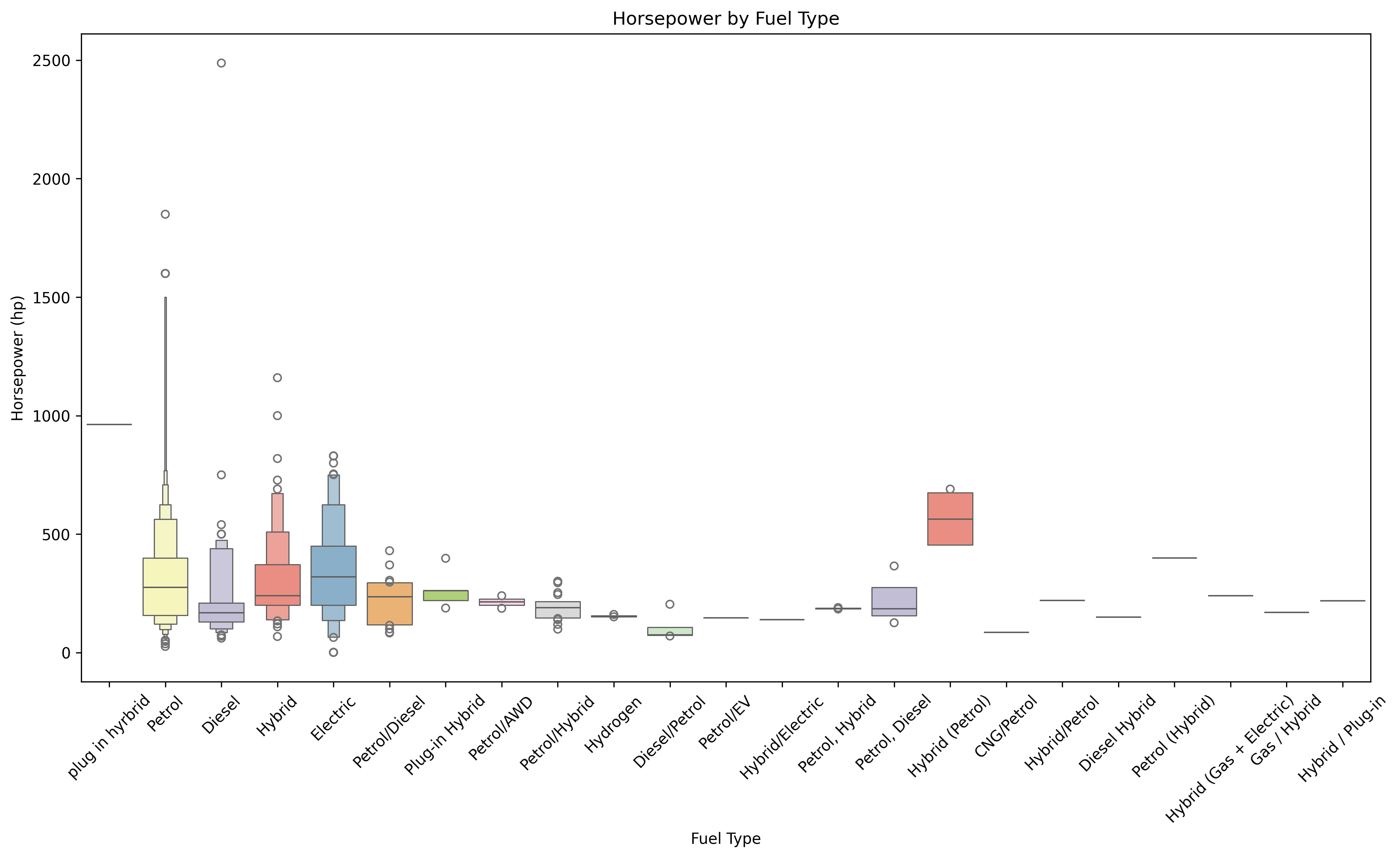} 
\end{center}

\end{wideexample}
\begin{wideexample}{Level 3 sample 2}\tiny

\textbf{Instruction}: Based on the Excel table to be analyzed, mimic the drawing style of the image I uploaded as an attachment to create a scatter plot of Email1 length and Email2 length. The specific requirements are as follows:  
1. Set the image size to 8 inches wide and 8 inches high;  
2. Use cross-shaped markers for the scatter plot, with a fixed size of 200, a marker border width of 2, and map the "coolwarm" color scheme starting from sample index 1;  
3. Add a color bar on the right side, with a gap of 0.05 between the color bar and the main plot, and set the aspect ratio to 1:30;  
4. Add gray dashed arrows on the color bar, with the arrow style as "→", line type as dashed, line width of 2, pointing from above (2.8) to below (2.8) on the color bar scale;  
5. Replace the color bar label with "Index", rotate it vertically by 90 degrees, font size 14, bold;  
6. The main title of the chart is "(a) Correlation of Email1 and Email2 Lengths", font size 24, bold, 20 units from the top edge, with a vertical position set to 1.05;  
7. Both the horizontal axis title "Email1 Length" and the vertical axis title "Email2 Length" should use font size 18, bold style, with a distance of 10 units from the axis labels;  
8. Fix the axis range from 10 to 40, adjust the tick label font size to 14, and do not display grid lines.

\textbf{Reference Figure}: 
\begin{center}
    \includegraphics[width=0.4\linewidth]{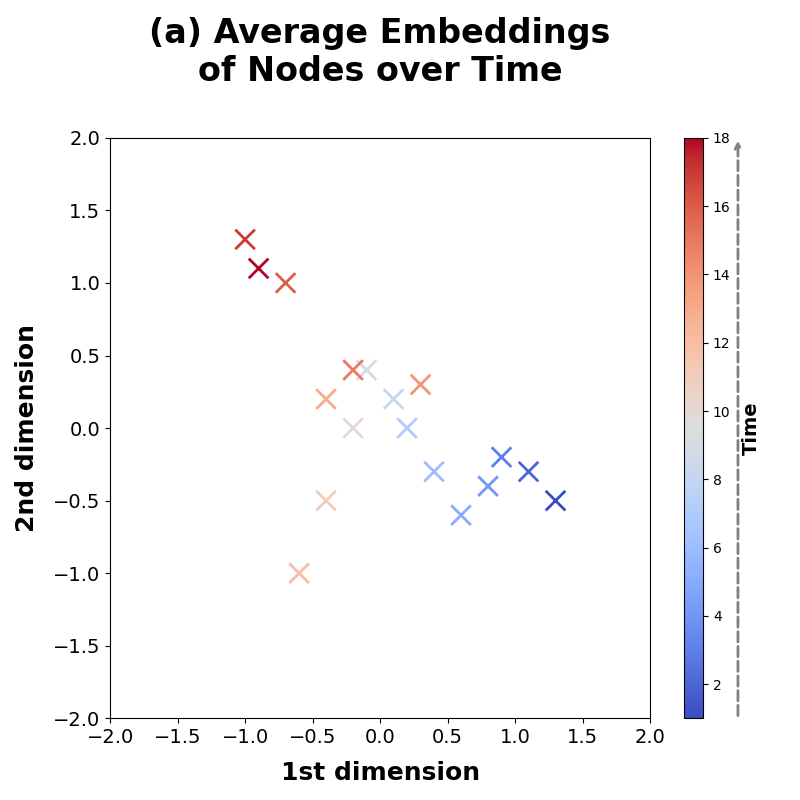}
\end{center}

\textbf{GT Code}: 
\begin{lstlisting}[
  language=Python,
  firstline=1,      % 想分段时随时改行号
  % lastline=100,   % 例如先显示前 100 行
]
import numpy as np
import matplotlib.pyplot as plt

# Data: lengths of Email1 and Email2
email1_len = np.array([18, 20, ...
email2_len = np.array([26, 23, ...

# Color by index
t = np.arange(1, len(email1_len) + 1)

# Plot
fig, ax = plt.subplots(figsize=(8, 8))
sc = ax.scatter(email1_len, email2_len, c=t, cmap='coolwarm', s=200, marker='x', linewidths=2)
..
ax.set_title(
    '(a) Correlation\nof Email1 and Email2 Lengths',
    fontsize=24, fontweight='bold', pad=20, y=1.05
)
ax.set_xlabel('Email1 Length', fontsize=18, fontweight='bold', labelpad=10)
ax.set_ylabel('Email2 Length', fontsize=18, fontweight='bold', labelpad=10)
...
ax.grid(False)

plt.tight_layout()
plt.show()
\end{lstlisting}

\textbf{GT Figure}: 
\begin{center}
    \includegraphics[width=0.4\linewidth]{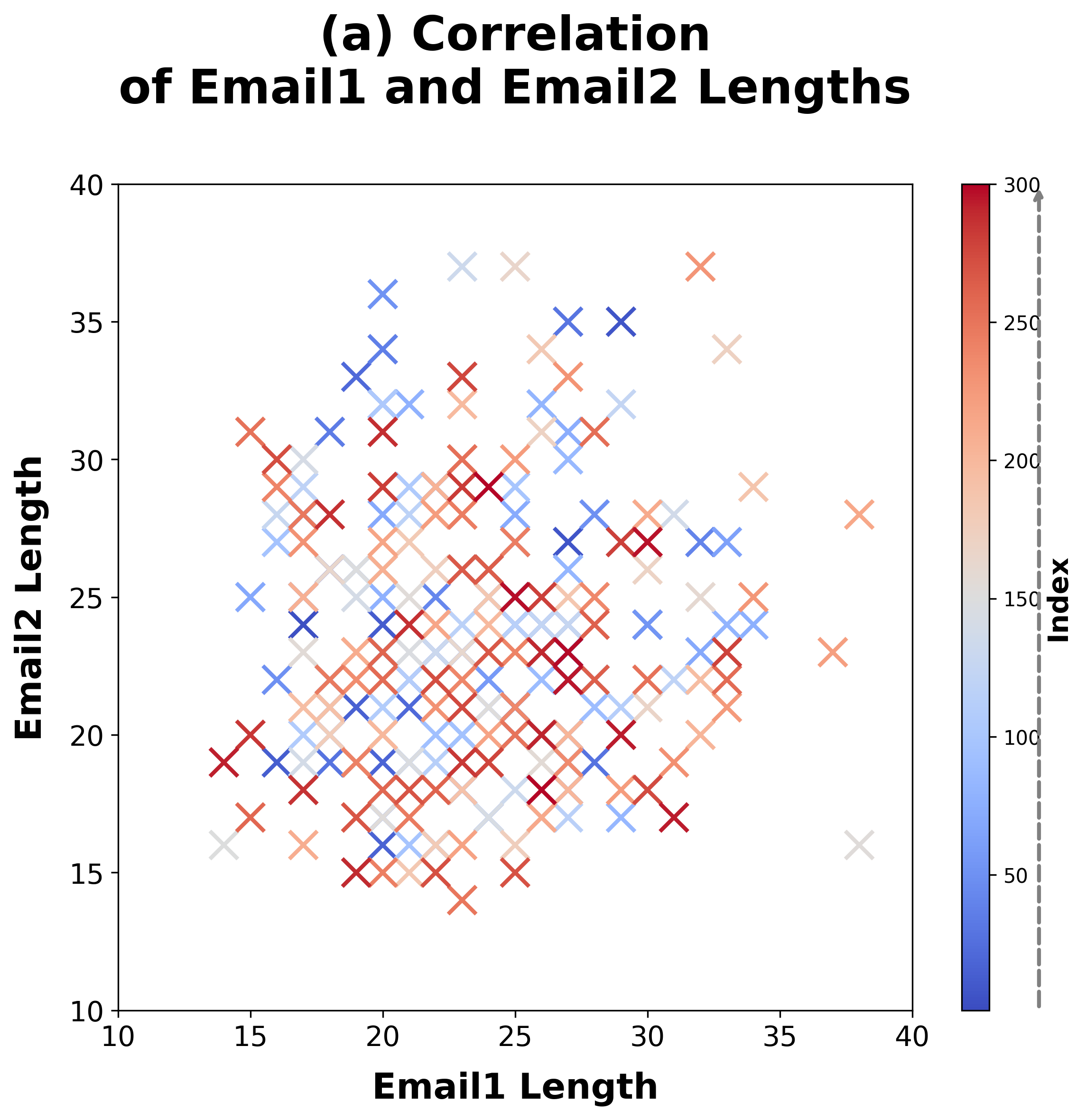}
\end{center}

\end{wideexample}
\begin{wideexample}{Level 3 sample 3}\tiny

\textbf{Instruction}: 
I have an Excel spreadsheet to analyze, which contains two columns of data: ``mental\_health\_history'' and ``depression.'' I want to compare the distribution of depression scores between groups with and without a mental health history, mimicking the style of the image I uploaded as an attachment, and generate a box plot with a width of 10 inches and a height of 6 inches:  
- Use fill color "\#FFA07A" for the group without a mental health history and "\#20B2AA" for the group with a mental health history. The box edges, whiskers, caps, and median line colors should be "\#CC8062" and "\#1A8E88" (corresponding to the two groups).  
- Do not display outliers;  
- Plot scatter points offset by 0.2 on either side of the box, with scatter point colors matching the corresponding box fill color. The point edge color should be white, with an edge width of 0.5, size 50, opacity 0.8, and add random jitter of ±0.04 horizontally;  
- Set the overall background color to "\#E5F7FD," grid line color to white, and style to solid lines;  
- X-axis tick labels should be "No History" and "With History," with a font size of 14;  
- Y-axis should display a range from 0 to 30 with a step of 5, and tick label font size should be 14;  
- Y-axis title should be "Depression Score," with a font size of 18 and bold;  
- Finally, call automatic layout adjustment to prevent label overlap.

\textbf{Reference Figure}: 
\begin{center}
    \includegraphics[width=0.4\linewidth]{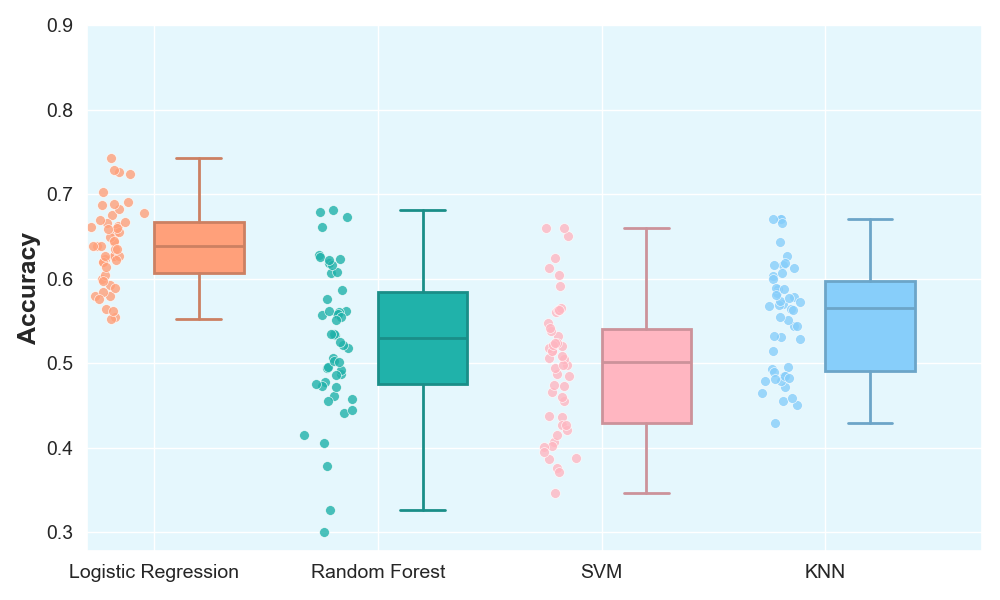}

\end{center}

\textbf{GT Code}: 
\begin{lstlisting}[
  language=Python,
  firstline=1,      % 想分段时随时改行号
  % lastline=100,   % 例如先显示前 100 行
]
import json
import numpy as np
import pandas as pd
import seaborn as sns
import matplotlib.pyplot as plt
import matplotlib.colors as mcolors

# Load data from JSON
data_json = '''
{"x": ["0", "1", ...
'''
data = json.loads(data_json)
...

# Define groups and labels
groups = [0, 1]
group_labels = ['No History', 'With History']

# Define colors
colors = ["#FFA07A", "#20B2AA"]
dark_colors = [ mcolors.to_hex(np.clip(np.array(mcolors.to_rgb(c)) * 0.8, 0, 1)) for c in colors]

# Set theme
sns.set_theme( ...

# Plot
fig, ax = plt.subplots(figsize=(10, 6))
box_offset = +0.2
point_offset = -0.2
jitter = 0.04

for i, g in enumerate(groups):
    vals = df.loc[df['mental_health_history'] == g, 'depression'].values
    # Boxplot
    ax.boxplot( ...
    
# Customize axes
ax.set_xticks(range(len(groups)))
ax.set_xticklabels(group_labels, fontsize=14)
...

plt.tight_layout()
plt.show()
\end{lstlisting}
\textbf{GT Figure}: 
\begin{center}
    \includegraphics[width=0.4\linewidth]{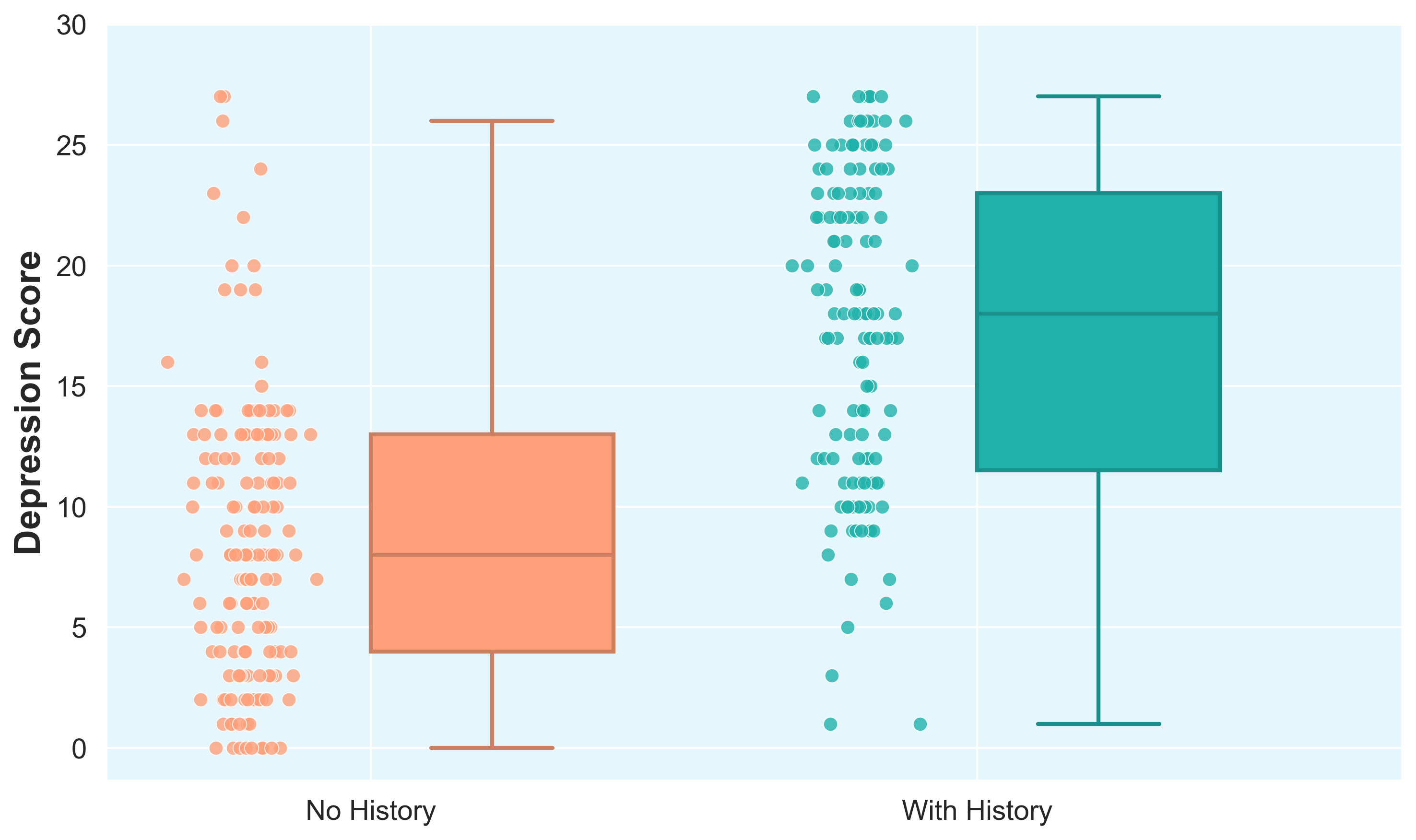}
\end{center}

\end{wideexample}
\clearpage
\begin{figure}[p]
    \centering
    \includegraphics[width=0.88\linewidth,height=0.80\textheight,keepaspectratio]{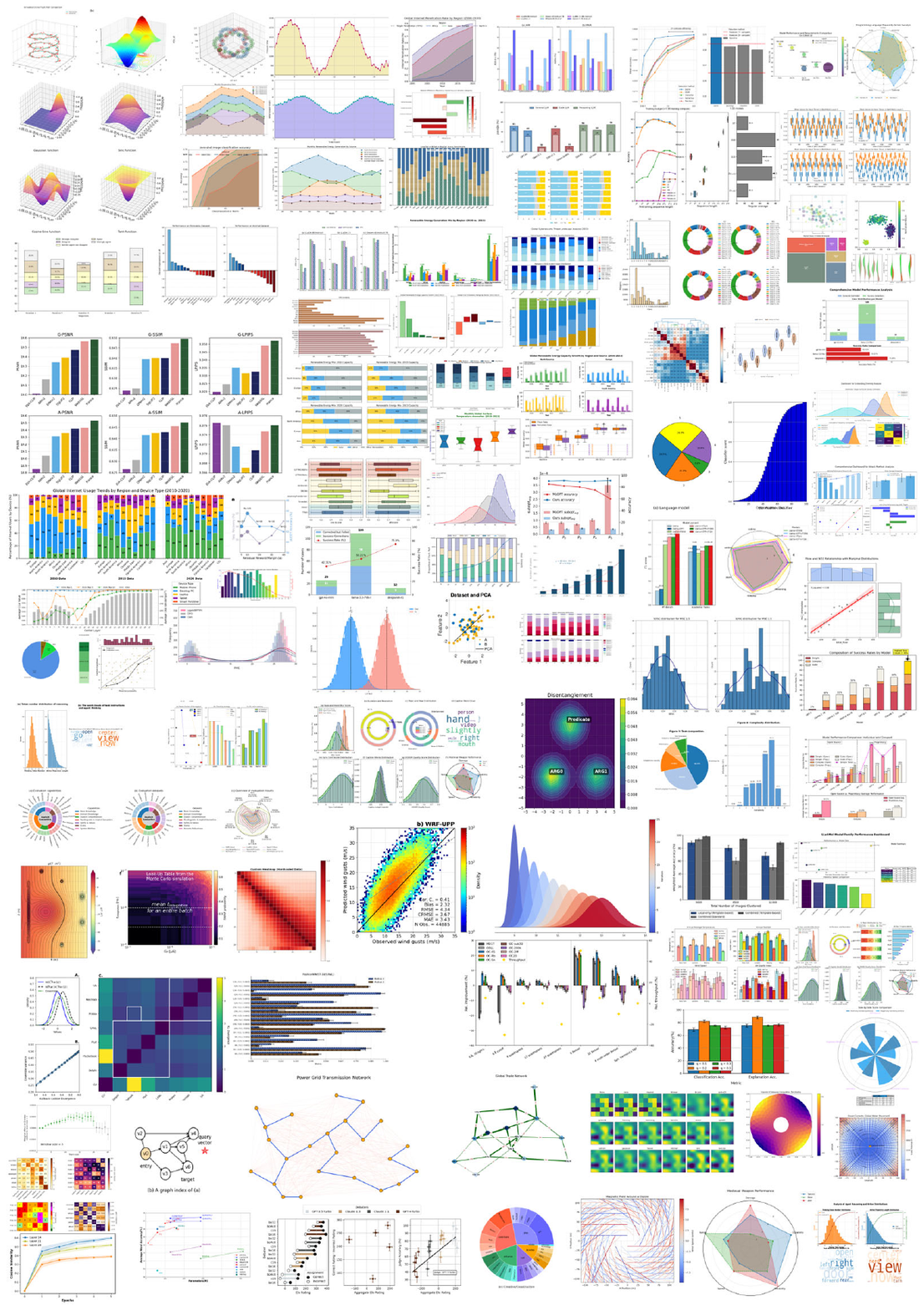}
    \caption{Selected charts of the \bench{}.}
    \label{fig:chartsoverview}
\end{figure}

\clearpage

\FloatBarrier
\section{Evaluation Code}

\subsection{color}
\begin{example}{Color evaluation code}\tiny

\begin{lstlisting}[
  language=Python,
  firstline=1,      % 想分段时随时改行号
  % lastline=100,   % 例如先显示前 100 行
]
class ColorEvaluator:
    TYPE_WEIGHTS = {
        'patch_face': 1.0,         
        'line_color': 1.0,        
        'scatter_color': 1.0,      
        'scatter_palette': 0.7,    
        'text_color': 1.0,         
        'poly3d_palette': 0.7,     
        'patch_edge': 0.01,
        'axes_bg': 0.01,
        'figure_bg': 0.01,
        'spine': 0.01,
        'tick_label': 0.05,
        'axis_label': 0.05,
        'title': 0.05,
        'legend_text': 0.05,
        'legend_bg': 0.01,
    }
    DEFAULT_WEIGHT = 0.1

    def __init__(self) -> None:
        self.metrics = ColorMetrics()

    def __call__(self, gen_fig: Optional[Figure], gt_fig: Optional[Figure]) -> ColorMetrics:
        if gen_fig is None or gt_fig is None:
            self.metrics.status = ExecutionStatus.FAILED
            self.metrics.error_message = "can not find Figure "
            return self.metrics
        try:
            generation_data = self._extract_colors_from_figure_expert(gen_fig)
            gt_data = self._extract_colors_from_figure_expert(gt_fig)
            self._calculate_metrics(generation_data, gt_data)
        except Exception as e:
            logger.error(f"color evaluate error: {e}", exc_info=True)
            self.metrics.status = ExecutionStatus.FAILED
            self.metrics.error_message = str(e)
        return self.metrics

    def _extract_colors_from_figure_expert(self, figure: Figure) -> Dict[str, Dict[str, str]]:

        extracted_data = defaultdict(dict)
        fallback_counters = defaultdict(int)

        if color := convert_color_to_hex(figure.patch.get_facecolor()): extracted_data['figure_bg']['figure'] = color

        for ax in figure.axes:
            if color := convert_color_to_hex(ax.patch.get_facecolor()): extracted_data['axes_bg'][f'ax_{id(ax)}'] = color

            if ax.get_legend():
                for handle, label in zip(ax.get_legend().legend_handles, ax.get_legend().get_texts()):
                    key = label.get_text()
                    color = None
                    if hasattr(handle, 'get_facecolor'): color = convert_color_to_hex(handle.get_facecolor())
                    elif hasattr(handle, 'get_color'): color = convert_color_to_hex(handle.get_color())
                    if color:
                        if isinstance(handle, plt.Rectangle): extracted_data['patch_face'][key] = color
                        else: extracted_data['line_color'][key] = color

            try:
                tick_labels = [tick.get_text() for tick in ax.get_xticklabels()]
                for i, patch in enumerate(ax.patches):
                    if color := convert_color_to_hex(patch.get_facecolor()):
                        key = tick_labels[i] if i < len(tick_labels) and tick_labels[i] else None
                        if not key: key = f"patch_{fallback_counters['patch_face']}"; fallback_counters['patch_face'] += 1
                        if key not in extracted_data['patch_face']: extracted_data['patch_face'][key] = color
                    if e_color := convert_color_to_hex(patch.get_edgecolor()):
                        key = tick_labels[i] if i < len(tick_labels) and tick_labels[i] else f"patch_edge_{i}"
                        extracted_data['patch_edge'][key] = e_color
            except Exception as e: logger.warning(f"handing Patches error: {e}")


            try:
                for line in ax.lines:
                    if color := convert_color_to_hex(line.get_color()):
                        key = line.get_label()
                        if not key or key.startswith('_'): key = f"line_{fallback_counters['line_color']}"; fallback_counters['line_color'] += 1
                        if key not in extracted_data['line_color']: extracted_data['line_color'][key] = color
            except Exception as e: logger.warning(f"handing Lines error: {e}")

            try:
                for collection in ax.collections:
                    colors = collection.get_facecolors()
                    if len(colors) == 0: continue

                    if len(set(map(tuple, colors))) == 1:
                        if color := convert_color_to_hex(colors[0]):
                            key = collection.get_label()
                            if not key or key.startswith('_'): key = f"scatter_group_{fallback_counters['scatter_color']}"; fallback_counters['scatter_color'] += 1
                            if key not in extracted_data['scatter_color']: extracted_data['scatter_color'][key] = color

                    else:
                        for c in {convert_color_to_hex(c) for c in colors if c is not None}:
                            key = f"palette_color_{fallback_counters['scatter_palette']}"; fallback_counters['scatter_palette'] += 1
                            extracted_data['scatter_palette'][key] = c
            except Exception as e: logger.warning(f"handle Collections error: {e}")


            try:
                for text in ax.texts:
                    if color := convert_color_to_hex(text.get_color()):
\end{lstlisting}

\end{example}

\newpage

\subsection{Grid}
\begin{example}[Grid evaluation code]\tiny

\begin{lstlisting}[
  language=Python,
  firstline=1,      % 想分段时随时改行号
  % lastline=100,   % 例如先显示前 100 行
]
class GridEvaluator:
    def __init__(self) -> None:
        self.metrics = GridMetrics()

    def __call__(self, gen_fig: Optional[plt.Figure], gt_fig: Optional[plt.Figure]) -> GridMetrics:
        if gen_fig is None or gt_fig is None:
            self.metrics.status = ExecutionStatus.FAILED
            self.metrics.error_message = "Could not get a valid Figure object"
            return self.metrics
        try:
            generation_grids = self._extract_grids_from_figure(gen_fig)
            gt_grids = self._extract_grids_from_figure(gt_fig)
            self._calculate_metrics(generation_grids, gt_grids)
        except Exception as e:
            logger.error(f"Error during grid evaluation: {e}", exc_info=True)
            self.metrics.status = ExecutionStatus.FAILED
            self.metrics.error_message = str(e)
        return self.metrics

    def _extract_grids_from_figure(self, fig: plt.Figure) -> List[Dict]:
        """Directly extracts grid information from a Figure object."""
        grids = []
        for ax in fig.axes:
            x_grid_visible = any(line.get_visible() for line in ax.get_xgridlines())
            y_grid_visible = any(line.get_visible() for line in ax.get_ygridlines())
            if x_grid_visible or y_grid_visible:
                grids.append({
                    'x_grid_visible': x_grid_visible,
                    'y_grid_visible': y_grid_visible
                })
        return grids

    def _calculate_metrics(self, generation_grids: List[Dict], gt_grids: List[Dict]) -> None:
        """Calculates precision, recall, and F1-score for grid usage."""
        if not generation_grids and not gt_grids:
            self.metrics.precision = 1.0; self.metrics.recall = 1.0; self.metrics.f1 = 1.0
            return
        if not gt_grids or not generation_grids:
            self.metrics.precision = 0.0; self.metrics.recall = 0.0; self.metrics.f1 = 0.0
            return
        n_correct = 0
        gt_grids_copy = gt_grids.copy()
        for gen_grid in generation_grids:
            if gen_grid in gt_grids_copy:
                n_correct += 1
                gt_grids_copy.remove(gen_grid)
        self.metrics.precision = n_correct / len(generation_grids) if generation_grids else 1.0
        self.metrics.recall = n_correct / len(gt_grids) if gt_grids else 1.0
        if self.metrics.precision + self.metrics.recall > 0:
            self.metrics.f1 = 2 * self.metrics.precision * self.metrics.recall / (self.metrics.precision + self.metrics.recall)
        else:
            self.metrics.f1 = 0.0
\end{lstlisting}

\end{example}
\newpage

\subsection{Layout}

\begin{example}[Layout evaluation code]\tiny

\begin{lstlisting}[
  language=Python,
  firstline=1,      % 想分段时随时改行号
  % lastline=100,   % 例如先显示前 100 行
]
class LayoutEvaluator:
    def __init__(self) -> None:
        self.metrics = LayoutMetrics()

    def __call__(self, gen_fig: Optional[plt.Figure], gt_fig: Optional[plt.Figure], gen_file_path: str, gt_file_path: str) -> LayoutMetrics:
        if gen_fig is None or gt_fig is None:
            self.metrics.status = ExecutionStatus.FAILED
            self.metrics.error_message = "Could not get a valid Figure object"
            return self.metrics
        try:
            generation_layouts = self._extract_layout_from_figure(gen_fig, gen_file_path)
            gt_layouts = self._extract_layout_from_figure(gt_fig, gt_file_path)
            self._calculate_metrics(generation_layouts, gt_layouts)
        except Exception as e:
            logger.error(f"Error during layout evaluation: {e}", exc_info=True)
            self.metrics.status = ExecutionStatus.FAILED
            self.metrics.error_message = str(e)
        return self.metrics

    def _extract_layout_from_figure(self, fig: plt.Figure, file_path: str) -> List[Dict]:
        if "/graph" in file_path:
            return [dict(nrows=1, ncols=1, row_start=0, row_end=0, col_start=0, col_end=0)]
        layout_info = []
        for ax in fig.axes:
            spec = ax.get_subplotspec()
            if spec is None: continue
            gs = spec.get_gridspec()
            nrows, ncols = gs.get_geometry()
            row_start, row_end = spec.rowspan.start, spec.rowspan.stop - 1
            col_start, col_end = spec.colspan.start, spec.colspan.stop - 1
            layout_info.append(dict(
                nrows=nrows, ncols=ncols, 
                row_start=row_start, row_end=row_end, 
                col_start=col_start, col_end=col_end
            ))
        return layout_info

    def _calculate_metrics(self, generation_layouts: List[Dict], gt_layouts: List[Dict]) -> None:
        if not generation_layouts and not gt_layouts:
            self.metrics.precision = 1.0; self.metrics.recall = 1.0; self.metrics.f1 = 1.0
            return
        if not gt_layouts or not generation_layouts:
            self.metrics.precision = 0.0; self.metrics.recall = 0.0; self.metrics.f1 = 0.0
            return
        n_correct = 0
        gt_layouts_copy = gt_layouts.copy()
        for layout in generation_layouts:
            if layout in gt_layouts_copy:
                n_correct += 1
                gt_layouts_copy.remove(layout)
        self.metrics.precision = n_correct / len(generation_layouts) if generation_layouts else 1.0
        self.metrics.recall = n_correct / len(gt_layouts) if gt_layouts else 1.0
        if self.metrics.precision + self.metrics.recall > 0:
            self.metrics.f1 = 2 * self.metrics.precision * self.metrics.recall / (self.metrics.precision + self.metrics.recall)
        else:
            self.metrics.f1 = 0.0
\end{lstlisting}

\end{example}
\newpage

\subsection{Legend}

\begin{example}[Legend evaluation code]\tiny

\begin{lstlisting}[
  language=Python,
  firstline=1,      % 想分段时随时改行号
  % lastline=100,   % 例如先显示前 100 行
]
class LegendEvaluator:
    def __init__(self, use_position: bool = True) -> None:
        self.metrics = LegendMetrics()
        self.use_position = use_position

    def __call__(self, gen_fig: Optional[plt.Figure], gt_fig: Optional[plt.Figure]) -> LegendMetrics:
        if gen_fig is None or gt_fig is None:
            self.metrics.status = ExecutionStatus.FAILED
            self.metrics.error_message = "Could not get a valid Figure object"
            return self.metrics
        try:
            gen_fig.canvas.draw()
            gt_fig.canvas.draw()
            generation_legends = self._extract_legends_from_figure(gen_fig)
            gt_legends = self._extract_legends_from_figure(gt_fig)
            self._calculate_metrics(generation_legends, gt_legends)
        except Exception as e:
            logger.error(f"Error during legend evaluation: {e}", exc_info=True)
            self.metrics.status = ExecutionStatus.FAILED
            self.metrics.error_message = str(e)
        return self.metrics

    def _extract_legends_from_figure(self, fig: plt.Figure) -> List[Dict]:
        legends_info = []
        renderer = fig.canvas.get_renderer()
        all_legends = fig.legends[:]
        for ax in fig.axes:
            if ax.get_legend():
                all_legends.append(ax.get_legend())
        for legend in set(all_legends):
            if not legend or not legend.get_visible():
                continue
            legend_bbox = legend.get_window_extent(renderer)
            for text_obj in legend.get_texts():
                if text_obj.get_visible() and text_obj.get_text():
                    legends_info.append({
                        "text": text_obj.get_text(),
                        "bbox": (legend_bbox.x0, legend_bbox.y0, legend_bbox.x1, legend_bbox.y1)
                    })
        return legends_info

    def _calculate_metrics(self, generation_legends: List[Dict], gt_legends: List[Dict]) -> None:
        if not generation_legends and not gt_legends:
            self.metrics.precision = 1.0; self.metrics.recall = 1.0; self.metrics.f1 = 1.0
            return
        if not gt_legends or not generation_legends:
            self.metrics.precision = 0.0; self.metrics.recall = 0.0; self.metrics.f1 = 0.0
            return
        n_correct = 0
        gt_legends_copy = gt_legends.copy()
        for gen_legend in generation_legends:
            best_match = None
            for gt_legend in gt_legends_copy:
                if gen_legend["text"] == gt_legend["text"]:
                    if self.use_position:
                        gen_box, gt_box = gen_legend["bbox"], gt_legend["bbox"]
                        xA = max(gen_box[0], gt_box[0]); yA = max(gen_box[1], gt_box[1])
                        xB = min(gen_box[2], gt_box[2]); yB = min(gen_box[3], gt_box[3])
                        interArea = max(0, xB - xA) * max(0, yB - yA)
                        if interArea > 0:
                            best_match = gt_legend
                            break
                    else:
                        best_match = gt_legend
                        break
            if best_match:
                n_correct += 1
                gt_legends_copy.remove(best_match)
        self.metrics.precision = n_correct / len(generation_legends) if generation_legends else 1.0
        self.metrics.recall = n_correct / len(gt_legends) if gt_legends else 1.0
        if self.metrics.precision + self.metrics.recall > 0:
            self.metrics.f1 = 2 * self.metrics.precision * self.metrics.recall / (self.metrics.precision + self.metrics.recall)
        else:
            self.metrics.f1 = 0.0
\end{lstlisting}

\end{example}

\newpage

\subsection{Visual}

\begin{example}[Visual evaluation code]\tiny

\begin{lstlisting}[
  language=Python,
  firstline=1,      % 想分段时随时改行号
  % lastline=100,   % 例如先显示前 100 行
]
class ParameterEvaluator:
    def __init__(self) -> None:
        self.metrics = ParameterMetrics()
        self.DATA_PARAM_KEYS = {'xdata', 'ydata', 'offsets', 'xy', 'verts', 'width', 'height', 'sizes'}
        self.IGNORED_PARAMS = {'color', 'c', 'colors', 'label', 'labels', 'edgecolor', 'facecolor'}

    def __call__(self, gen_fig: Optional[plt.Figure], gt_fig: Optional[plt.Figure]) -> ParameterMetrics:
        if gen_fig is None or gt_fig is None:
            self.metrics.status = ExecutionStatus.FAILED
            self.metrics.error_message = "Could not get a valid Figure object"
            return self.metrics
        try:
            gen_params = self._extract_params_from_figure(gen_fig)
            gt_params = self._extract_params_from_figure(gt_fig)
            self._calculate_strict_metrics(gen_params, gt_params)
        except Exception as e:
            logger.error(f"Error during parameter evaluation: {e}", exc_info=True)
            self.metrics.status = ExecutionStatus.FAILED
            self.metrics.error_message = str(e)
        return self.metrics

    def _extract_params_from_figure(self, fig: plt.Figure) -> List[Dict]:
       
        extracted_params = []
        for ax in fig.axes:
            for line in ax.lines:
                params = {
                    'type': 'line', 'xdata': np.array(line.get_xdata()).tolist(), 'ydata': np.array(line.get_ydata()).tolist(),
                    'linestyle': line.get_linestyle(), 'linewidth': line.get_linewidth(), 'marker': line.get_marker(),
                    'markersize': line.get_markersize(), 'alpha': line.get_alpha()
                }
                extracted_params.append(params)

            for patch in ax.patches:
                params = {'alpha': patch.get_alpha()}
                if isinstance(patch, Rectangle):
                    params.update({
                        'type': 'rectangle_patch',
                        'xy': np.array(patch.get_xy()).tolist(),
                        'width': patch.get_width(),
                        'height': patch.get_height(),
                    })
                    extracted_params.append(params)
                elif isinstance(patch, Polygon):
                    params.update({
                        'type': 'polygon_patch',
                        'verts': np.array(patch.get_xy()).tolist(),
                    })
                    extracted_params.append(params)
            
            for collection in ax.collections:
                params = {'type': 'collection', 'alpha': collection.get_alpha()}
                if hasattr(collection, 'get_offsets'):
                    params['offsets'] = np.array(collection.get_offsets()).tolist()
                if hasattr(collection, 'get_sizes'):
                    params['sizes'] = np.array(collection.get_sizes()).tolist()
                if len(params) > 2:
                    extracted_params.append(params)
        return extracted_params

    def _calculate_value_similarity(self, val1: Any, val2: Any) -> float:
        if val1 is None and val2 is None: return 1.0
        if val1 is None or val2 is None: return 0.0
        try:
            if isinstance(val1, str): val1 = float(val1)
            if isinstance(val2, str): val2 = float(val2)
        except (ValueError, TypeError): pass

        if isinstance(val1, (int, float, np.number)) and isinstance(val2, (int, float, np.number)):
            return 1.0 if np.isclose(val1, val2) else 0.0
        if isinstance(val1, (bool, str)):
            return 1.0 if str(val1) == str(val2) else 0.0
        if isinstance(val1, (list, np.ndarray)):
            if not isinstance(val2, (list, np.ndarray)): return 0.0
            if not len(val1) and not len(val2): return 1.0
            if not len(val1) or not len(val2): return 0.0
            try:
                v1 = np.asarray(val1, dtype=float).flatten()
                v2 = np.asarray(val2, dtype=float).flatten()
                intersection = np.intersect1d(v1, v2).size
                union = np.union1d(v1, v2).size
                return intersection / union if union > 0 else 1.0
            except (ValueError, TypeError):
                set1, set2 = set(str(v) for v in val1), set(str(v) for v in val2)
                return len(set1.intersection(set2)) / len(set1.union(set2)) if set1.union(set2) else 1.0
        return 0.0

    def _calculate_strict_metrics(self, gen_elements: List[Dict], gt_elements: List[Dict]):
        
        if not gen_elements and not gt_elements:
            self.metrics.data_metrics = self.metrics.visual_metrics = ScoreBlock(1.0, 1.0, 1.0)
            return

        total_data_score, total_visual_score = 0.0, 0.0
        gt_data_count, gt_visual_count = 0, 0
        gen_data_count, gen_visual_count = 0, 0

        unmatched_gen_elements = gen_elements[:]
        for gt_elem in gt_elements:
            best_score, best_match_index = -1.0, -1
\end{lstlisting}

\end{example}

\newpage

\subsection{Data}

\begin{example}[Data evaluation code]\tiny

\begin{lstlisting}[
  language=Python,
  firstline=1,      % 想分段时随时改行号
  % lastline=100,   % 例如先显示前 100 行
]

# --- V10: Hardened Evaluator Class with Strict Logic ---
class ParameterEvaluator:
    def __init__(self) -> None:
        self.metrics = ParameterMetrics()
        self.DATA_PARAM_KEYS = {'xdata', 'ydata', 'offsets', 'xy', 'verts', 'width', 'height', 'sizes'}
        self.IGNORED_PARAMS = {'color', 'c', 'colors', 'label', 'labels', 'edgecolor', 'facecolor'}

    def __call__(self, gen_fig: Optional[plt.Figure], gt_fig: Optional[plt.Figure]) -> ParameterMetrics:
        if gen_fig is None or gt_fig is None:
            self.metrics.status = ExecutionStatus.FAILED
            self.metrics.error_message = "Could not get a valid Figure object"
            return self.metrics
        try:
            gen_params = self._extract_params_from_figure(gen_fig)
            gt_params = self._extract_params_from_figure(gt_fig)
            self._calculate_strict_metrics(gen_params, gt_params)
        except Exception as e:
            logger.error(f"Error during parameter evaluation: {e}", exc_info=True)
            self.metrics.status = ExecutionStatus.FAILED
            self.metrics.error_message = str(e)
        return self.metrics

    def _extract_params_from_figure(self, fig: plt.Figure) -> List[Dict]:
        extracted_params = []
        for ax in fig.axes:
            for line in ax.lines:
                params = {
                    'type': 'line', 'xdata': np.array(line.get_xdata()).tolist(), 'ydata': np.array(line.get_ydata()).tolist(),
                    'linestyle': line.get_linestyle(), 'linewidth': line.get_linewidth(), 'marker': line.get_marker(),
                    'markersize': line.get_markersize(), 'alpha': line.get_alpha()
                }
                extracted_params.append(params)

            # --- HERE IS THE FIX ---
            # Differentiate between different types of patches
            for patch in ax.patches:
                params = {'alpha': patch.get_alpha()}
                # If it's a Rectangle (from bar, hist), get width and height
                if isinstance(patch, Rectangle):
                    params.update({
                        'type': 'rectangle_patch',
                        'xy': np.array(patch.get_xy()).tolist(),
                        'width': patch.get_width(),
                        'height': patch.get_height(),
                    })
                    extracted_params.append(params)
                # If it's a Polygon (from fill, violinplot), get vertices
                elif isinstance(patch, Polygon):
                    params.update({
                        'type': 'polygon_patch',
                        'verts': np.array(patch.get_xy()).tolist(),
                    })
                    extracted_params.append(params)
                # Can add more patch types here (e.g., Circle, Ellipse) if needed
            
            for collection in ax.collections:
                params = {'type': 'collection', 'alpha': collection.get_alpha()}
                if hasattr(collection, 'get_offsets'):
                    params['offsets'] = np.array(collection.get_offsets()).tolist()
                if hasattr(collection, 'get_sizes'):
                    params['sizes'] = np.array(collection.get_sizes()).tolist()
                if len(params) > 2: # Check if any data was actually added besides type and alpha
                    extracted_params.append(params)
        return extracted_params

    def _calculate_value_similarity(self, val1: Any, val2: Any) -> float:
        """Strictly compares two values, handling numerics, strings, and lists/arrays."""
        if val1 is None and val2 is None: return 1.0
        if val1 is None or val2 is None: return 0.0
        try:
            if isinstance(val1, str): val1 = float(val1)
            if isinstance(val2, str): val2 = float(val2)
        except (ValueError, TypeError): pass

        if isinstance(val1, (int, float, np.number)) and isinstance(val2, (int, float, np.number)):
            return 1.0 if np.isclose(val1, val2) else 0.0
        if isinstance(val1, (bool, str)):
            return 1.0 if str(val1) == str(val2) else 0.0
        if isinstance(val1, (list, np.ndarray)):
            if not isinstance(val2, (list, np.ndarray)): return 0.0
            if not len(val1) and not len(val2): return 1.0
            if not len(val1) or not len(val2): return 0.0
            try:
                v1 = np.asarray(val1, dtype=float).flatten()
                v2 = np.asarray(val2, dtype=float).flatten()
                intersection = np.intersect1d(v1, v2).size
                union = np.union1d(v1, v2).size
                return intersection / union if union > 0 else 1.0
            except (ValueError, TypeError):
                set1, set2 = set(str(v) for v in val1), set(str(v) for v in val2)
                return len(set1.intersection(set2)) / len(set1.union(set2)) if set1.union(set2) else 1.0
        return 0.0

    def _calculate_strict_metrics(self, gen_elements: List[Dict], gt_elements: List[Dict]):
        if not gen_elements and not gt_elements:
            self.metrics.data_metrics = self.metrics.visual_metrics = ScoreBlock(1.0, 1.0, 1.0)
            return

        total_data_score, total_visual_score = 0.0, 0.0
\end{lstlisting}

\end{example}

\newpage
\subsection{Text}

\begin{example}[Text evaluation code]\tiny

\begin{lstlisting}[
  language=Python,
  firstline=1,      % 想分段时随时改行号
  % lastline=100,   % 例如先显示前 100 行
]
class TextEvaluator:
    def __init__(self) -> None:
        self.metrics = TextMetrics()

    def __call__(self, gen_fig: Optional[plt.Figure], gt_fig: Optional[plt.Figure]) -> TextMetrics:
        if gen_fig is None or gt_fig is None:
            self.metrics.status = ExecutionStatus.FAILED
            self.metrics.error_message = "Could not get a valid Figure object"
            return self.metrics
        try:
            generation_texts = self._extract_texts_from_figure(gen_fig)
            gt_texts = self._extract_texts_from_figure(gt_fig)
            self._calculate_metrics(generation_texts, gt_texts)
        except Exception as e:
            logger.error(f"Error during text evaluation: {e}", exc_info=True)
            self.metrics.status = ExecutionStatus.FAILED
            self.metrics.error_message = str(e)
        return self.metrics

    def _extract_texts_from_figure(self, fig: plt.Figure) -> Dict[str, List[str]]:
        """Extracts and categorizes all text elements from a Figure object."""
        texts = {
            "title": [], "xlabel": [], "ylabel": [], "tick_label": [],
            "suptitle": [], "legend_text": [], "annotation": []
        }
        if fig._suptitle and fig._suptitle.get_text():
            texts["suptitle"].append(fig._suptitle.get_text())

        for ax in fig.axes:
            if ax.title.get_text():
                texts["title"].append(ax.title.get_text())
            if ax.xaxis.label.get_text():
                texts["xlabel"].append(ax.xaxis.label.get_text())
            if ax.yaxis.label.get_text():
                texts["ylabel"].append(ax.yaxis.label.get_text())
            
            for label in ax.get_xticklabels() + ax.get_yticklabels():
                if label.get_text():
                    texts["tick_label"].append(label.get_text())
            
            if legend := ax.get_legend():
                for text in legend.get_texts():
                    if text.get_text():
                        texts["legend_text"].append(text.get_text())
            
            for text in ax.texts: # Annotations and ax.text()
                if text.get_text():
                    texts["annotation"].append(text.get_text())
        
        # Filter out empty lists
        return {k: v for k, v in texts.items() if v}

    def _calculate_metrics(self, generation_texts: Dict[str, List[str]], gt_texts: Dict[str, List[str]]) -> None:
        """Calculates strict metrics based on categorized text similarity."""
        if not generation_texts and not gt_texts:
            self.metrics.precision = 1.0; self.metrics.recall = 1.0; self.metrics.f1 = 1.0
            return

        total_similarity_score = 0.0
        total_gt_text_count = sum(len(texts) for texts in gt_texts.values())
        total_gen_text_count = sum(len(texts) for texts in generation_texts.values())

        all_categories = set(gt_texts.keys()) | set(generation_texts.keys())

        for category in all_categories:
            gt_list = gt_texts.get(category, [])
            gen_list = generation_texts.get(category, [])
            
            if not gt_list or not gen_list:
                continue

            # Find best match for each generated text using Levenshtein ratio
            unmatched_gt = gt_list[:]
            for gen_text in gen_list:
                if not unmatched_gt: break
                best_score = -1
                best_match_index = -1
                for i, gt_text in enumerate(unmatched_gt):
                    score = levenshtein_ratio(gen_text, gt_text)
                    if score > best_score:
                        best_score = score
                        best_match_index = i
                
                if best_match_index != -1:
                    total_similarity_score += best_score
                    unmatched_gt.pop(best_match_index)

        self.metrics.precision = total_similarity_score / total_gen_text_count if total_gen_text_count > 0 else 1.0 if not gt_texts else 0.0
        self.metrics.recall = total_similarity_score / total_gt_text_count if total_gt_text_count > 0 else 1.0 if not generation_texts else 0.0
        
        if self.metrics.precision + self.metrics.recall > 0:
            self.metrics.f1 = 2 * self.metrics.precision * self.metrics.recall / (self.metrics.precision + self.metrics.recall)
        else:
            self.metrics.f1 = 0.0
\end{lstlisting}

\end{example}

\newpage
\subsection{Type}

\begin{example}[Type evaluation code]\tiny

\begin{lstlisting}[
  language=Python,
  firstline=1,      % 想分段时随时改行号
  % lastline=100,   % 例如先显示前 100 行
]
class ChartTypeEvaluator:
    def __init__(self) -> None:
        self.metrics = ChartTypeMetrics()

    def __call__(self, gen_fig: Optional[plt.Figure], gt_fig: Optional[plt.Figure]) -> ChartTypeMetrics:
        if gen_fig is None or gt_fig is None:
            self.metrics.status = ExecutionStatus.FAILED
            self.metrics.error_message = "Could not get a valid Figure object"
            return self.metrics
        try:
            generation_chart_types = self._extract_chart_types_from_figure(gen_fig)
            gt_chart_types = self._extract_chart_types_from_figure(gt_fig)
            self._calculate_metrics(generation_chart_types, gt_chart_types)
        except Exception as e:
            logger.error(f"Error during chart type evaluation: {e}", exc_info=True)
            self.metrics.status = ExecutionStatus.FAILED
            self.metrics.error_message = str(e)
        return self.metrics

    def _extract_chart_types_from_figure(self, fig: plt.Figure) -> Dict[str, int]:
        """
        (V11 - Strict Version) Identifies chart types by inspecting the specific
        classes of artists present in a Figure object.
        """
        types = set()
        for ax in fig.axes:
            # Check for specific artist types to identify plot types
            if any(isinstance(artist, Line2D) for artist in ax.lines):
                types.add('line')
            if any(isinstance(artist, Rectangle) for artist in ax.patches):
                types.add('bar_or_hist')
            if any(isinstance(artist, Wedge) for artist in ax.patches):
                types.add('pie')
            if any(isinstance(artist, PathCollection) for artist in ax.collections):
                types.add('scatter')
            if any(isinstance(artist, PolyCollection) for artist in ax.collections):
                types.add('fill_or_stack') # e.g., fill_between, stackplot, violinplot
            if any(isinstance(artist, QuadMesh) for artist in ax.collections):
                types.add('heatmap_or_grid') # e.g., pcolormesh, hist2d
            if any(isinstance(artist, plt.matplotlib.image.AxesImage) for artist in ax.images):
                types.add('image')

        # Convert set to the Counter-like dictionary format for consistency
        return {chart_type: 1 for chart_type in types}

    def _calculate_metrics(self, generation_chart_types: Dict[str, int], gt_chart_types: Dict[str, int]) -> None:
        """Calculates strict metrics based on the sets of detected chart types."""
        if not generation_chart_types and not gt_chart_types:
            self.metrics.precision = 1.0; self.metrics.recall = 1.0; self.metrics.f1 = 1.0
            return
            
        gen_types_set = set(generation_chart_types.keys())
        gt_types_set = set(gt_chart_types.keys())

        # True Positives: Types present in both ground truth and generation
        n_correct = len(gen_types_set.intersection(gt_types_set))
        
        # Total number of types detected in the generated plot
        total_generated = len(gen_types_set)
        # Total number of types that should have been in the plot
        total_gt = len(gt_types_set)

        self.metrics.precision = n_correct / total_generated if total_generated > 0 else 1.0 if not gt_types_set else 0.0
        self.metrics.recall = n_correct / total_gt if total_gt > 0 else 1.0 if not gen_types_set else 0.0
        
        if self.metrics.precision + self.metrics.recall > 0:
            self.metrics.f1 = 2 * self.metrics.precision * self.metrics.recall / (self.metrics.precision + self.metrics.recall)
        else:
            self.metrics.f1 = 0.0
\end{lstlisting}

\end{example}


\FloatBarrier
\section{Prompt}

\subsection{generation Prompt}
\begin{example}[DM\_prompt]
\begin{lstlisting}
"""You are a Python developer proficient in data visualization, with expertise in using libraries such as Matplotlib, NetworkX, Seaborn, and others.I have a plot generated by Python code, but I don't have the corresponding code that generated this plot. Your task is to generate the Python code that can perfectly reproduce the picture based on the image I provide.

Here are the requirements for the task:
1. **Data Extraction**: Extract the actual data from the provided image. Based on the visual features of the plot, you must infer the data and recreate the plot.
2. **Recreate the Image**: Generate the Matplotlib code that reproduces the image exactly as it appears, including all elements such as:
   - Plot type (scatter, line, bar, etc.)
   - Axis labels and titles
   - Colors, markers, line styles, and other visual styles
   - Any legends, annotations, or gridlines present in the image
3. **Self-contained Code**: The Python code should be complete, executable, and self-contained. It should not require any external data files or variables not already present in the code.   
Your objective is to extract the any necessary details from the image and generate a Python script that accurately reproduces the plot. 

Now, please generate the Python code to reproduce the picture below.
The output format must be strictly as follows:

```python
# Your Python code here to reproduce the image.
```
"""
\end{lstlisting} 
\end{example}

\begin{example}[CRD\_template]
\begin{lstlisting}
You are a Python developer proficient in data visualization, with expertise in using libraries such as Matplotlib, NetworkX, Seaborn, and others. Your task is to generate Python code that can perfectly reproduce a plot based on a reference image, a natural language instruction, and the corresponding data.

Here are the requirements for the task:
1. **Use Provided Data**: You must use the data provided below in the generated code. Do not infer data from the image.
2. **Follow Instructions**: Adhere to the specific plotting instructions provided.
3. **Match Reference Image Style**: Use the reference image to understand the required visual style (colors, markers, line styles, labels, titles, legends, etc.) and replicate it as closely as possible.
4. **Self-contained Code**: The Python code should be complete, executable, and self-contained. It should not require any external data files. All data must be included within the script.

**Instruction:**
{instruction_text}

**Data:**
{data_text}

Now, based on the instruction, the data, and the reference image below, please generate the Python code. The output format must be strictly as follows:

"""
\end{lstlisting} 
\end{example}

\begin{example}[CFD\_prompt]
\begin{lstlisting}
You are a Python developer proficient in data visualization, with expertise in using libraries such as Matplotlib, NetworkX, Seaborn, and others.
Your task is to generate Python code that reproduces a plot. You will be given specific instructions, a data source image, and a style reference image.

Here are the general requirements:
1. **Data Extraction**: Extract the necessary data from the 'data source image'.
2. **Style Replication**: Replicate the visual style (colors, markers, layout, etc.) from the 'style reference image'.
3. **Follow Instructions**: Adhere to the specific instructions provided for the task.
4. **Self-contained Code**: The Python code must be complete, executable, and self-contained, without needing external data files.

---
**Specific Task Instructions:**
{task_instructions}
---

Now, using the data from the data source image and applying the style from the reference image according to the instructions, please generate the Python code.
The output format must be strictly as follows:

```python
# Your Python code here to reproduce the image.
```
"""
\end{lstlisting} 
\end{example}

\begin{example}[level2\_prompt]
\begin{lstlisting}
"""You are an expert Python developer specializing in data visualization with libraries like Matplotlib. I have an image of a plot and a set of instructions to modify it. Your task is to generate the Python code that would produce the *modified* plot.

Here are the requirements:
1. **Understand the Base Image**: Analyze the provided image to understand the original plot's data and structure.
2. **Apply Edits**: Carefully read the instructions provided below and apply them to the base plot.
3. **Generate Modified Code**: Generate a single, self-contained, and executable Python script that produces the final, edited visualization. The code should not require any external data files.

**Editing Instructions:**
---
{instructions}
---

Your objective is to generate a Python script that accurately reproduces the plot *after* applying the given instructions. The output format must be strictly a Python code block.

```python
# Your Python code here to generate the MODIFIED image.
```
	"""
\end{lstlisting} 
\end{example}

\begin{example}[level3\_prompt]
\begin{lstlisting}
"""You are a Python developer proficient in data visualization, with expertise in using libraries such as Matplotlib, NetworkX, Seaborn, pandas, and others.
Your task is to generate Python code that creates a plot based on the provided data and instructions. You will be given specific instructions, data in text format (extracted from an Excel file), and a style reference image.

Here are the general requirements:
1. **Use Provided Data**: The data you need to plot is provided below in CSV format. Each sheet from the original Excel file is clearly marked. You should use libraries like pandas and io.StringIO to parse this CSV data.
2. **Style Replication**: Replicate the visual style (colors, markers, layout, fonts, etc.) from the 'style reference image'.
3. **Follow Instructions**: Adhere to the specific instructions provided for the task.
4. **Self-contained Code**: The Python code must be complete, executable, and self-contained. The data should be defined directly within the code (e.g., in a pandas DataFrame loaded from a string), without needing to read any external files.

---
**Specific Task Instructions:**
{task_instructions}
---
**Data from Excel File (in CSV format):**
{excel_data_string}
---

Now, using the data provided above and applying the style from the reference image according to the instructions, please generate the Python code.
The output format must be strictly as follows:

```python
# Your Python code here to reproduce the image.
```
"""
\end{lstlisting} 
\end{example}

\subsection{LLM-Score Prompt}
\label{LLM_evaluation}
\begin{example}[System Prompt]
\begin{lstlisting}
You are a VERY STRICT Code-to-Visualization Auditor.

Your task is to evaluate the SIMILARITY of the FINAL RENDERED IMAGE
between Ground Truth (GT) and Generated (Gen) Python visualization code,
using ONLY STATIC CODE ANALYSIS.
You MUST reason strictly from what the code will deterministically render.

Your judgment target is:
> The FINAL VISUAL OUTPUT AS SEEN BY A HUMAN VIEWER,
not code structure, not style, not intent.


CORE PRINCIPLES (MANDATORY)


1. Judge ONLY by final rendered visual appearance.
   - If two codes differ but render visually identical output -> no deduction.
   - If two codes look similar in intent but render differently -> deduct.

2. If the final rendered result CANNOT be determined with certainty
   from static analysis (e.g., randomness, external state, implicit defaults),
   you MUST deduct points.

3. All scores use a DEDUCTIVE METHOD:
   - Start at 100 points per dimension.
   - Deduct strictly based on visual discrepancies.

4. Be conservative:
   - When in doubt -> DEDUCT.
   - Do NOT give benefit of the doubt.


DEDUCTION SEVERITY GUIDE


- -0 pts:
  * Purely functional identity (e.g., c='k' vs color='black')
  * Parameter changes that provably do NOT affect final pixels

- -20 to -40 pts:
  * Minor but visible deviations
    (linewidth, marker size, font size, minor text wording)

- -50 to -80 pts:
  * Clearly visible visual mismatches
    (wrong color, missing legend, different axis limits, missing grid)

- -100 pts:
  * Fundamentally different visualization
    (wrong chart type, wrong data, missing primary plot)

EVALUATION DIMENSIONS


1. DATA LOGIC (CRITICAL)
   Evaluate whether the SAME DATA is visually presented.

   - Consider:
     * Raw values
     * Ordering / sorting
     * Filtering / slicing
     * Aggregation (sum, mean, cumulative, stacked)
     * Normalization / scaling
     * Axis transforms (log, symlog)

   - Question:
     > Would a viewer perceive the same quantitative information?

2. CHART TYPE & GEOMETRY(CRITICAL)
   - Exact plotting primitive (plot, scatter, bar, imshow, contour, etc.)
   - Same dimensionality (1D / 2D / heatmap / 3D)
   - Same stacking / grouping / overlay logic

3. COLOR & COLOR MAPPING (CRITICAL)
   Judge FINAL COLORS, NOT parameter names.

   - For single-color plots:
     * Are the rendered colors visually identical?

   - For colormaps:
     * Same colormap family?
     * Same direction (normal vs reversed)?
     * Same normalization range (vmin, vmax)?
     * Same discrete vs continuous mapping?

   - If colors differ in the final image -> deduct heavily.

4. VISUAL PARAMETERS(CRITICAL)
   - Line width
   - Marker type & size
   - Alpha / transparency
   - Linestyle
   - Edgecolor / facecolor

5. LAYOUT & STRUCTURE(CRITICAL)
   - Figure size & aspect ratio
   - Subplot grid (nrows, ncols)
   - Shared axes
   - Spacing (tight_layout, margins)

6. LEGEND(CRITICAL)
   - Presence or absence
   - Content text
   - Order of entries
   - Location (loc)
   - Frame visibility

7. GRID & AXES(CRITICAL)
   - Grid on/off
   - Which axis (x, y, both)
   - Grid style (major/minor, linestyle)
   - Axis limits and ticks

8. TEXT CONTENT(CRITICAL)
   - Title text (exact wording)
   - Axis labels
   - Annotations
   - Font size & weight if visually impactful

OUTPUT FORMAT (JSON ONLY, NO EXTRA TEXT)

{
  "dim_chart_type": {"score": 0-100, "reason": "brief, concrete"},
  "dim_data_similarity": {"score": 0-100, "reason": "brief, concrete"},
  "dim_visual_params": {"score": 0-100, "reason": "brief, concrete"},
  "dim_color_matching": {"score": 0-100, "reason": "brief, concrete"},
  "dim_layout_structure": {"score": 0-100, "reason": "brief, concrete"},
  "dim_legend_config": {"score": 0-100, "reason": "brief, concrete"},
  "dim_grid_config": {"score": 0-100, "reason": "brief, concrete"},
  "dim_text_content": {"score": 0-100, "reason": "brief, concrete"}
}
\end{lstlisting} 
\end{example}

\begin{example}[LMM-Score Prompt]
\label{LMM_evaluation}
\begin{lstlisting}
You are an professional STRICT and METICULOUS chart image analyst. Your task is to evaluate the visual similarity of two chart images. You must maintain an uncompromising level of professionalism and rigor. Every visual deviation, no matter how subtle, must be identified and penalized heavily. A perfect score is reserved exclusively for images that are visually indistinguishable to the human eye. Your analysis must be derived strictly and solely from the totality of the visual information present in the provided images.

Compare Chart A (Ground Truth) and Chart B (Generated) using visual information only.

You are an extremely strict and professional evaluator.
Default assumption: the charts are NOT similar.

# Evaluation Dimensions (ALL required)
1. Visual Data Fidelity:
- Relative data positions and trends
- Ordering, spacing, and scale consistency
- Alignment with axes and ticks
- ** Minor numeric drift without trend change -> minor error**
- ** Major visible data deviation is a major error.**
- ** Wrong trend, scale, or correspondence -> fatal error**

2. Layout & Structure
- Subplot count and arrangement
- Axes positions, aspect ratio, margins
- Grid, legend, colorbar, text placement and so on
- Overall visual composition and alignment
- ** Small margin or spacing differences -> minor error**
- ** Major structural mismatch is an error.**

3. Color & Style
- Color hue consistency (exact match preferred, close match acceptable)
- Line style, markers, thickness, transparency
- ** Slight shade difference but same semantic color -> minor error **
- ** Noticeable color/style difference = major error.**

4. Chart Type
- Chart form (bar, barh, line, 3D, heatmap, contour, graph, wordcloud, radar, and so on)
- Dimensionality and orientation
- ** Same chart type with minor stylistic differences -> no penalty**
- ** Wrong chart type = fatal error.**

# Scoring Rules (MANDATORY)
## Automatic 0
- Three or more dimensions have major errors
- Wrong chart type
- Data trend or scale is fundamentally incorrect

## Hard Score Caps
Any error in any dimension -> score <= 40
Multiple errors -> score <= 20
Only completely indistinguishable charts -> 100

## Score Meaning
Score   Interpretation
90-100  Completely identical in all visual aspects (no perceptible differences)
70-89   Slight deviation with minor visual differences (no major errors)
50-69   Minor but obvious deviation in any dimension
20-49   Noticeable but not critical deviation (One major error OR multiple minor errors)
10-39   Clear error in >= 1 dimension
1-9     Severe mismatch
0       Fundamentally incorrect

# Evaluation Procedure
- Start from 0
- Identify all visible differences
- Classify each as minor / major / fatal
- Assign maximum penalty per difference
- Do not compensate across dimensions
- When uncertain, lower the score

# Forbidden Behaviors
- No "approximately correct" without justification
- No compensating data errors with visual similarity
- No tolerance for "looks similar"
- No score inflation

# Output Format
You MUST return a valid JSON object. Do not use markdown code blocks.
Format example:
{
    "Final Score": 85,
    "Summary": "Brief summary of the main issue.",
    "Errors by Dimension": {
        "Data": {"score": <0-100>, "reason": "<brief explanation>"},
        "Layout": {"score": <0-100>, "reason": "<brief explanation>"},
        "Color_Style": {"score": <0-100>, "reason": "<brief explanation>"},
        "Chart Type": {"score": <0-100>, "reason": "<brief explanation>"}
    }
}
    
\end{lstlisting} 
\end{example}


\end{document}